\documentclass[a4paper,12pt]{book}

\usepackage[english]{babel}
\usepackage[sectionbib]{natbib} 
\usepackage{url}
\usepackage[table]{xcolor}
\usepackage{graphicx,epic}
\usepackage{tikz} 
\usepackage{float}
\usepackage{amsmath,amssymb}
\usepackage{enumerate}
\usepackage{multirow}
\usepackage{indentfirst} 
\usepackage{ifthen}
\usepackage{calc}
\usepackage[T1]{fontenc} 
\usepackage[ansinew]{inputenc}
\usepackage[font=small,format=plain,labelfont=bf]{caption}
\usepackage{soul}

\usepackage{ifpdf} 
\ifpdf
   \usepackage{graphicx}
   \usepackage{epstopdf}
\else
   \usepackage{graphicx}
\fi

\newcounter{hours}
\newcounter{minutes}
\newcommand{\now}
	{
	\setcounter{hours}{\time/60}
	\setcounter{minutes}{\time-\value{hours}*60}
	\textcolor{red}{version: \today~(\thehours:\ifthenelse{\value{minutes}<10}{0}{}\theminutes)}
	}

\def\printerversion{0}

\definecolor{mycolordeco}{rgb}{0.801176,0.801176,0.801176}
\definecolor{mycolordeco2}{rgb}{0.92,0.92,0.92}
\definecolor{mycolorbox}{rgb}{0.93,0.93,0.93}
\definecolor{mycoloracrbox}{rgb}{0.9,0.9,0.9}
\definecolor{mycoloracrline}{rgb}{0.85,0.85,0.85}

\ifthenelse{\equal{\printerversion}{1}}{
	\definecolor{mycoloracrlink}{rgb}{0.0,0.0,0.0}
	\definecolor{mycolorlink}{rgb}{0.0,0.0,0.0}
	\definecolor{mycolorcite}{rgb}{0.0,0.0,0.0}
	\definecolor{mycolorbiblio}{rgb}{0.0,0.0,0.0}
	\definecolor{mycoloracr}{rgb}{0.0,0.0,0.0}
	}{
	\definecolor{mycoloracrlink}{rgb}{0.0,0.0,0.0}
	\definecolor{mycolorlink}{rgb}{0.0,0.0,0.6}
	\definecolor{mycolorcite}{rgb}{0.6,0.0,0.0}
	\definecolor{mycolorbiblio}{rgb}{0.0,0.0,0.6}
	\definecolor{mycoloracr}{rgb}{0.0,0.0,0.6}
	}

\newcommand{\cdp}{\clearpage{\pagestyle{plain}\cleardoublepage}}  

\newcommand{\linkup}[1]{%
	\protect\begin{picture}(0,0)%
	\protect\put(0,12){\protect\hypertarget{#1}{}}%
	\protect\end{picture}%
	}

\newcommand{\myparskip}{10pt}
\newcommand{\myparskiptoc}{2pt}

\newcommand{\localseclabel}{sec:}
\newcommand{\localsubseclabel}{sec:}
\newlength{\templength}
\newcounter{autotoc}
\newcounter{autolof}
\newcounter{autolot}

\def\thechapter{\Roman{chapter}}
\def\thesection{\arabic{section}}

\makeatletter

\def\@chapter[#1]#2#3{%
		\ifnum \c@secnumdepth >\m@ne
                       \if@mainmatter
				\refstepcounter{chapter}%
				\typeout{\@chapapp\space\thechapter.}%
				\addcontentsline{toc}{chapter}{\protect\numberline{\thechapter}#1}%
			\else
				\addcontentsline{toc}{chapter}{#1}%
			\fi
		\else
			\addcontentsline{toc}{chapter}{#1}%
		\fi
		\chaptermark{#1}%
		\addtocontents{lof}{\protect\addvspace{10\p@}}%
		\addtocontents{lot}{\protect\addvspace{10\p@}}%
		\pagestyle{normal}
		\thispagestyle{plain}
		\@makechapterhead{#2}{#3}%
		\@afterheading}

\def\@makechapterhead#1#2{%
	\vspace*{0\p@}%
	{\parindent \z@ \raggedleft \normalfont
	\ifnum \c@secnumdepth >\m@ne
		\if@mainmatter
			\begin{picture}(0,0)\put(0,25){\hypertarget{cha:#2}{}\hypertarget{autotoc:\theautotoc}{}\addtocounter{autotoc}{1}}\end{picture}%
			\label{cha:#2}%
			\large\textsc{\@chapapp} %
			\begin{picture}(40,0)
				\begin{tikzpicture}[overlay, mycolordeco2]
				\end{tikzpicture}
			\put(0,0){\makebox[40pt][c]{\Huge\textbf{\thechapter}}}
			\end{picture}\\
			\vskip 7\p@
		\fi
	\fi
	\Huge \centering \bfseries #1\par\nobreak
	\vskip 45\p@
	}}

\def\@schapter#1{%
	\if@twocolumn
		\@topnewpage[\@makeschapterhead{#1}]%
	\else
		\@makeschapterhead{#1}%
		\@afterheading
	\fi}

\def\@makeschapterhead#1{%
	\vspace*{29\p@}%
	{\parindent \z@ \raggedleft
	\normalfont
	\Huge \bfseries\begin{picture}(0,0)\put(0,25){\hypertarget{autotoc:\theautotoc}{}\addtocounter{autotoc}{1}}\end{picture} #1\\
	\vskip 45\p@}}

\def\@startsection#1#2#3#4#5#6{
	\if@noskipsec \leavevmode \fi
	\par
	\@tempskipa #4\relax
	\@afterindenttrue
	\ifdim \@tempskipa <\z@
		\@tempskipa -\@tempskipa \@afterindentfalse
	\fi
	\if@nobreak
		\everypar{}%
	\else
		\addpenalty\@secpenalty\addvspace\@tempskipa
	\fi
	\@ifstar
		{\@ssect{#2}{#3}{#4}{#5}{#6}}%
		{\@dblarg{\@sect{#1}{#2}{#3}{#4}{#5}{#6}}}%
	} 

\def\@sect#1#2#3#4#5#6[#7]#8{
	\ifnum #2>\c@secnumdepth
		\let\@svsec\@empty
	\else
		\refstepcounter{#1}%
		\protected@edef\@svsec{\@seccntformat{#1}\relax}%
	\fi
	\@tempskipa #5\relax
	\ifdim \@tempskipa>\z@
		\begingroup
			#6%
			\ifnum #2<2
				\settowidth{\mysecwidth}{\@hangfrom{\hskip #3\relax\@svsec}\large\bfseries#8}%
				\begin{tikzpicture}[overlay, mycolordeco]%
					\fill (9pt,-4.5pt) -- ++(\mysecwidth,0pt) -- ++(0,20pt) -- ++(-\mysecwidth,0pt) -- cycle;
					\fill (\mysecwidth+9pt, 5.5pt) circle (10pt);
					\fill (9pt, 5.5pt) circle (10pt);
					\fill[white] (9pt, 5.5pt) circle (8pt);
				\end{tikzpicture}%
				\hspace{5pt}%
			\fi%
			{\@hangfrom{\hskip #3\relax\@svsec}\interlinepenalty \@M #8\@@par}%
		\endgroup
		\csname #1mark\endcsname{#7}%
		\addcontentsline{toc}{#1}{%
		\ifnum #2>\c@secnumdepth \else
		\protect\numberline{\csname the#1\endcsname}%
		\fi
		#7}%
	\else
		\def\@svsechd{%
		#6{\hskip #3\relax
		\@svsec #8}%
		\csname #1mark\endcsname{#7}%
		\addcontentsline{toc}{#1}{%
		\ifnum #2>\c@secnumdepth \else
			\protect\numberline{\csname the#1\endcsname}%
		\fi
		#7}}%
	\fi
	\@xsect{#5}}

\def\@ssect#1#2#3#4#5#6{
	\@tempskipa #4\relax
	\ifdim \@tempskipa>\z@
		\begingroup
			#5%
			\ifnum #1<2
				\settowidth{\mysecwidth}{\@hangfrom{\hskip #2}large\bfseries#6}%
				\begin{tikzpicture}[overlay, mycolordeco]%
					\fill (10pt,-4.5pt) -- ++(\mysecwidth,0pt) -- ++(0pt,20pt) -- ++(-\mysecwidth,0pt) -- cycle;
					\fill (\mysecwidth+10pt, 5.5pt) circle (10pt);
					\fill (10pt, 5.5pt) circle (10pt);
				\end{tikzpicture}%
			\fi%
			{\@hangfrom{\hskip #2}\interlinepenalty \@M \hspace{10pt}#6\@@par}%
		\endgroup
	\else
		\def\@svsechd{#5{\hskip #2\relax #6}}%
	\fi
	\@xsect{#4}}

\renewcommand\section{\@startsection {section}{1}{\z@}%
	{-3.5ex \@plus -1ex \@minus -.2ex}%
	{2.3ex \@plus.2ex}%
	{\noindent\begin{picture}(0,0)\put(0,25){\hypertarget{\localseclabel}{}\hypertarget{autotoc:\theautotoc}{}\addtocounter{autotoc}{1}}\end{picture}\normalfont\large\bfseries\twospace}}
	
\renewcommand\subsection{\@startsection{subsection}{2}{\z@}%
	{-3.25ex\@plus -1ex \@minus -.2ex}%
	{1.5ex \@plus .2ex}%
	{\noindent\begin{picture}(0,0)\put(0,20){\hypertarget{\localsubseclabel}{}\hypertarget{autotoc:\theautotoc}{}\addtocounter{autotoc}{1}}\end{picture}\normalfont\normalsize\bfseries}}

\makeatother

\newcommand{\chanonumber}[1]{
	\cdp
	\chapter*{#1}
	\refstepcounter{chapter}
	\setcounter{chapter}{0}
	\pagestyle{nochapnumber}
	\thispagestyle{plain}
	\markboth{#1}{}
	\addcontentsline{toc}{chapter}{#1}
	\addtocontents{lof}{\protect\addvspace{10pt}}
	\addtocontents{lot}{\protect\addvspace{10pt}}
	}

\newlength{\mysecwidth}
\newcommand{\cha}[2]{\cdp\chapter{#2}{#1}} 
\newcommand{\sect}[2][]{\def\localseclabel{sec:#1}\section{#2}\label{sec:#1}}
\newcommand{\subsect}[2][]{\def\localsubseclabel{sec:#1}\subsection{#2}\label{sec:#1}} 

\newcommand{\refcha}[1]{\hyperlink{cha:#1}{\textcolor{mycolorlink}{Chapter~\ref{cha:#1}}}}
\newcommand{\refapp}[1]{\hyperlink{cha:#1}{\textcolor{mycolorlink}{Appendix~\ref{cha:#1}}}}

\newcommand{\refsec}[2][]{\hyperlink{sec:#2}{\textcolor{mycolorlink}{Section~\ifthenelse{\equal{#1}{}}{}{\ref{cha:#1}.}\ref{sec:#2}}}}
\newcommand{\refsect}[2][]{\hyperlink{sec:#2}{\textcolor{mycolorlink}{Section~\ifthenelse{\equal{#1}{}}{}{\ref{cha:#1}.}\ref{sec:#2}}}, page~\pageref{sec:#2}}
\newcommand{\refsecp}[2][]{\hyperlink{sec:#2}{\textcolor{mycolorlink}{Section~\ifthenelse{\equal{#1}{}}{}{\ref{cha:#1}.}\ref{sec:#2}}} (page~\pageref{sec:#2})}

\newcommand{\refsub}[2][]{\hyperlink{sub:#2}{\textcolor{mycolorlink}{Section~\ifthenelse{\equal{#1}{}}{}{\ref{cha:#1}.}\ref{sub:#2}}}}

\newcommand{\tab}[6]{
	\settowidth{\templength}{#6}
	\begin{table}
	\hypertarget{tab:#2}{}\hypertarget{autolot:\theautolot}{}\addtocounter{autolot}{1}
	\begin{center}
	\begin{minipage}{#1}
	\caption{#3}
	\label{tab:#2}
	\vspace{0.2cm}
	\begin{tabular*}{\linewidth}{#4}
	\hline\hline
	#5
	\hline
	\ifthenelse{\lengthtest{\templength=0pt}}
		{\end{tabular*}}
		{
		\vspace{-0.3cm}
		\end{tabular*}
		{\footnotesize #6}
		}
	\end{minipage}
	\end{center}
	\end{table}}

\newcommand{\eqn}[2][]{
	\begin{picture}(0,0)\put(0,0){\hypertarget{eqn:#1}{}}\end{picture}
	\begin{align}
	#2
	\label{eqn:#1}
	\end{align}
	}

\newcounter{myboxctr}[chapter]
\makeatletter \renewcommand \themyboxctr {\ifnum \c@chapter>\z@ \thechapter.\fi \@arabic\c@myboxctr} \makeatother 

\newcommand{\mybox}[2][]{
	\ignorespaces
	\begin{center}
	\begin{minipage}{15cm}
	\ifthenelse{\equal{#1}{nocnt}}
		{}
		{
		\refstepcounter{myboxctr} 
		$\blacksquare$\,Box \themyboxctr\linkup{box:#1}\\
		\vspace{-0.8cm}
		}
	\begin{center}
	\setbox0=\hbox\bgroup
	\begin{minipage}{13.9cm}
	\setlength{\parindent}{15pt}
	\setlength\parskip{10pt}
	\ifthenelse{\equal{#1}{nocnt}}
		{\begin{flushleft}\vspace{-5pt}$\cdots$\vspace{-5pt}\end{flushleft}}
		{}
	#2
	\ifthenelse{\equal{#1}{nocnt}}{}{\begin{picture}(0,0)\put(0,0){\label{box:#1}}\end{picture}}
	\end{minipage}\egroup
	\fcolorbox{black}{mycolorbox}{\box0}\end{center}\end{minipage}
	\end{center}
	}

\makeatletter
\newcommand{\testacr}[3]{%
	\@ifundefined{s@#1}{\global\@namedef{s@#1}{1}#2\linkup{acr:#1}}{\hyperlink{acr:#1}{\textcolor{mycoloracrlink}{#1#3}}}%
}
\makeatother

\newcommand{\acr}[4][]{
	\ifthenelse{\equal{#2}{notarg}}
		{\hyperlink{acr:#3}{\textcolor{mycoloracrlink}{#3#1}}}%
		{%
		\testacr{#3}
			{%
			\label{tempacr:#3}%
			\ifthenelse{\isodd{\pageref{tempacr:#3}}}
				{\begin{picture}(0,0)\put(0,-4){\textcolor{mycoloracrbox}{\rule{\widthof{#4}}{13pt}}}\color{mycoloracrline}\put(0,-4){\line(1,0){1000}}\end{picture}\hbox{#4}}
				{\begin{picture}(0,0)\put(0,-4){\textcolor{mycoloracrbox}{\rule{\widthof{#4}}{13pt}}}\end{picture}\hbox{#4}\begin{picture}(0,0)\color{mycoloracrline}\put(0,-4){\line(-1,0){1000}}\end{picture}}
			\marginpar
				[\raggedleft\textcolor{mycoloracr}{\textsf{#3}}\space\begin{picture}(12,12)\put(-80,-6){\textcolor{white}{\rule{80pt}{3pt}}}\color{mycoloracr}\put(0,-4){\line(0,1){14}}\end{picture}]%
				{\begin{picture}(12,12)\color{mycoloracr}\put(12,-4){\line(0,1){14}}\put(12.2,-6){\textcolor{white}{\rule{80pt}{3pt}}}\end{picture}\raggedright\space\textcolor{mycoloracr}{\textsf{#3}}}%
			}{#1}%
		}%
	}

\newcommand{\reffig}[2][]{\hyperlink{fig:#2}{\textcolor{mycolorlink}{Fig.~\ref{fig:#2}\ifthenelse{\equal{#1}{}}{}{.#1}}}}
\newcommand{\reffigt}[2][]{\hyperlink{fig:#2}{\textcolor{mycolorlink}{Fig.~\ref{fig:#2}\ifthenelse{\equal{#1}{}}{}{.#1}}}, page~\pageref{fig:#2}}
\newcommand{\reffigp}[2][]{\hyperlink{fig:#2}{\textcolor{mycolorlink}{Fig.~\ref{fig:#2}\ifthenelse{\equal{#1}{}}{}{.#1}}} (page~\pageref{fig:#2})}

\newcommand{\reftab}[1]{\hyperlink{table:#1}{\textcolor{mycolorlink}{Table~\ref{table:#1}}}}

\newcommand{\refeqn}[1]{\hyperlink{eqn:#1}{\textcolor{mycolorlink}{Equation~\ref{eqn:#1}}}}

\newcommand{\myciteauthor}[1]{\citeauthor{#1}}
\newcommand{\myciteyear}[1]{\hyperlink{autobib:#1}{\textcolor{mycolorcite}{\citeyear{#1}}}}

\newcommand{\mycitealt}[1]{\myciteauthor{#1} \myciteyear{#1}}

\newcounter{autofootnote}
\makeatletter
\def\@makefnmark{\hbox{\@textsuperscript{\hyperlink{autofnt:\theautofootnote}{\normalfont\@thefnmark}}}}

\renewcommand\footnoterule{%
	\kern-3\p@
	\hrule\@width.4\columnwidth
	\kern2.6\p@}
\makeatother

\let\oldfootnote=\footnote
\renewcommand{\footnote}[1]{\stepcounter{autofootnote}\oldfootnote{\samepage$\ $\linkup{autofnt:\theautofootnote}#1}} 

\let\oldtoc=\tableofcontents
\renewcommand{\tableofcontents}{\setcounter{autotoc}{-1}\setlength\parskip{\myparskiptoc}\oldtoc\setlength\parskip{\myparskip}\setcounter{autotoc}{0}}
\let\oldlof=\listoffigures
\renewcommand{\listoffigures}{\setcounter{autolof}{0}\setlength\parskip{\myparskiptoc}\oldlof\setlength\parskip{\myparskip}\setcounter{autolof}{0}\setcounter{autotoc}{0}}
\let\oldlot=\listoftables
\renewcommand{\listoftables}{\setcounter{autolot}{0}\setlength\parskip{\myparskiptoc}\oldlot\setlength\parskip{\myparskip}\setcounter{autolot}{0}\setcounter{autotoc}{0}}

\makeatletter
\newcommand{\tocdotfill}[1]{\leaders\hbox{$\m@th\mkern #1 mu\hbox{.}\mkern #1 mu$}\hfill}

\renewcommand*{\l@chapter}[2]{%
	\ifnum \c@tocdepth >\m@ne
		\addpenalty{-\@highpenalty}%
		\vskip 1.0em\@plus\p@
		{\leftskip 0em\relax
		\rightskip \@tocrmarg
		\parfillskip -\rightskip
		\parindent 0em\relax\@afterindenttrue
		\interlinepenalty\@M
		\leavevmode
		\@tempdima 2.5em\relax
		\advance\leftskip \@tempdima \null\nobreak\hskip -\leftskip
		{\bfseries %
		\begin{tikzpicture}[overlay, mycolordeco2]%
			\fill (4pt,-4pt) -- ++(\linewidth-8,0pt) -- ++(0pt,16pt) -- ++(-\linewidth+8,0pt) -- cycle;
			\fill (\linewidth-4pt, 4pt) circle (8pt);
			\fill (4pt, 4pt) circle (8pt);
		\end{tikzpicture}%
		\hyperlink{autotoc:\theautotoc}{#1}\addtocounter{autotoc}{1}}\nobreak{\bfseries\tocdotfill{10000}}\hb@xt@\@pnumwidth{\hfil\bfseries \hyperlink{pag:#2}{#2}}\par}%
	\fi}

\renewcommand*{\l@section}[2]{%
	\ifnum \c@tocdepth >\z@
		\vskip \z@ \@plus.2\p@
		{\leftskip 1.0em\relax
		\rightskip \@tocrmarg
		\parfillskip -\rightskip
		\parindent 2.5em\relax\@afterindenttrue
		\interlinepenalty\@M
		\leavevmode
		\@tempdima 1.5em\relax
		\advance\leftskip \@tempdima \null\nobreak\hskip -\leftskip
		{\normalfont \hyperlink{autotoc:\theautotoc}{#1}\addtocounter{autotoc}{1}}\nobreak{\normalfont\tocdotfill{4.5}}\nobreak\hb@xt@\@pnumwidth{\hfil\normalfont \hyperlink{pag:#2}{#2}}\par}%
	\fi}

\renewcommand*{\l@subsection}[2]{%
	\ifnum \c@tocdepth >\@ne
		\vskip \z@ \@plus.2\p@
		{\leftskip 3.8em\relax
		\rightskip \@tocrmarg
		\parfillskip -\rightskip
		\parindent 4.0em\relax\@afterindenttrue
		\interlinepenalty\@M
		\leavevmode
		\@tempdima 2.0em\relax
		\advance\leftskip \@tempdima \null\nobreak\hskip -\leftskip
		{\normalfont \hyperlink{autotoc:\theautotoc}{#1}\addtocounter{autotoc}{1}}\nobreak{\normalfont\tocdotfill{4.5}}\nobreak\hb@xt@\@pnumwidth{\hfil\normalfont \hyperlink{pag:#2}{#2}}\par}%
	\fi}

\renewcommand*{\l@figure}[2]{%
	\vskip \z@ \@plus.2\p@
	{\leftskip 1.5em\relax
	\rightskip \@tocrmarg
	\parfillskip -\rightskip
	\parindent 1.5em\relax\@afterindenttrue
	\interlinepenalty\@M
	\leavevmode
	\@tempdima 3.3em\relax
	\advance\leftskip \@tempdima \null\nobreak\hskip -\leftskip
	{\normalfont \hyperlink{autolof:\theautolof}{#1}\addtocounter{autolof}{1}}\nobreak{\normalfont\tocdotfill{4.5}}\nobreak\hb@xt@\@pnumwidth{\hfil\normalfont \hyperlink{pag:#2}{#2}}\par}%
	}

\renewcommand*{\l@table}[2]{%
	\vskip \z@ \@plus.2\p@
	{\leftskip 1.5em\relax
	\rightskip \@tocrmarg
	\parfillskip -\rightskip
	\parindent 1.5em\relax\@afterindenttrue
	\interlinepenalty\@M
	\leavevmode
	\@tempdima 2.3em\relax
	\advance\leftskip \@tempdima \null\nobreak\hskip -\leftskip
	{\normalfont \hyperlink{autolot:\theautolot}{#1}\addtocounter{autolot}{1}}\nobreak{\normalfont\tocdotfill{4.5}}\nobreak\hb@xt@\@pnumwidth{\hfil\normalfont \hyperlink{pag:#2}{#2}}\par}%
	}

\makeatother

\usepackage{fancyhdr}
\newcommand{\pagenumber}{\thepage}

\fancypagestyle{normal}
	{
	\fancyhf{} 
	\fancyhead[LE]{\mychapname\thechapter}
	\fancyhead[CE]{~~~~~~~~~~~\nouppercase{{\sc\leftmark}}}
	\fancyhead[CO]{\nouppercase{\rightmark}}
	\fancyhead[R]{\linkup{pag:\thepage}} 
	\fancyfoot[LE,RO]{\pagenumber}

	}

\fancypagestyle{nochapnumber}
	{
	\fancyhf{}
	\fancyhead[CE]{\nouppercase{{\sc\leftmark}}}
	\fancyhead[CO]{\nouppercase{\rightmark}}
	\fancyhead[R]{\linkup{pag:\thepage}}
	\fancyfoot[LE,RO]{\pagenumber}

	}

\fancypagestyle{frontmatter}
	{
	\fancyhf{}
	\fancyhead[C]{\nouppercase{{\sc\leftmark}}}
	\fancyhead[R]{\linkup{pag:\thepage}}
	\fancyfoot[LE,RO]{\pagenumber}

	}
		
\fancypagestyle{plain} 
	{
	\fancyhf{}
	\fancyhead[R]{\linkup{pag:\thepage}}
	\fancyfoot[LE,RO]{\pagenumber}

	}

\renewcommand{\chaptermark}[1]{\markboth{{#1}}{}} 

\newcommand{\linkads}[2][]{\ifthenelse{\equal{\printerversion}{1}}{\hphantom{~{\footnotesize\href{#2}{\ifthenelse{\equal{#1}{}}{[ADS}{[#1}~\includegraphics[scale=0.6]{figs/deco_link.eps}~]}}}}{~{\footnotesize\href{#2}{\ifthenelse{\equal{#1}{}}{[ADS}{[#1}~\includegraphics[scale=0.6]{figs/deco_link.eps}~]}}}}

\renewenvironment{thebibliography}[1]{
	\renewcommand{\bibname}{Bibliography} 
	\markboth{\bibname}{} 
	\phantomsection  
	\addcontentsline{toc}{chapter}{\bibname}
	\cdp
	\pagestyle{frontmatter}
	\begin{oldthebibliography}{#1}
	\setlength{\columnsep}{40pt}
	\setlength{\parskip}{10pt}
	\setlength{\itemsep}{0pt}
	}{
	\end{oldthebibliography}}

\newcommand{\aj}{AJ}
\newcommand{\araa}{ARA\&A}
\newcommand{\apj}{ApJ}
\newcommand{\apjl}{ApJ}
\newcommand{\apjs}{ApJS}
\newcommand{\apss}{Ap\&SS}
\newcommand{\aap}{A\&A}
\newcommand{\aaps}{A\&AS}
\newcommand{\mnras}{MNRAS}
\newcommand{\pasp}{PASP}
\newcommand{\pasj}{PASJ}
\newcommand{\nat}{Nature}
\newcommand{\physrep}{Phys.~Rep.}

\newcommand{\U}[1]{\ensuremath{\mathrm{~#1}}} 

\newcommand{\ergs}{\U{erg~s}}

\newcommand{\msun}{$\U{M}_{\odot}$~}

\newcommand{\hi}{H{\sc i} }
\newcommand{\hii}{H{\sc ii} }

\newcommand{\chaphead}[1]
	{
	\begin{flushleft}
	\begin{picture}(0,0)
		\linethickness{1.5pt}
		\put(127,3){\line(-1,0){300}}
		\put(138,0){$\mathcal{O}$}
		\put(147,0){\emph{\textsf{verview}}}
		\put(320,3){\line(-1,0){125}}
		\qbezier(320,3)(340,3)(340,-12)
		\put(340,-12){\line(0,-1){12}}
	\end{picture}
	
	\begin{minipage}{400pt}
		\setlength{\arrayrulewidth}{1.5pt}
		\begin{tabular}{@{}p{320pt}@{}p{19.25pt}@{}|}
		#1 & \\
		\end{tabular}
	\end{minipage}

	\begin{picture}(0,0)
		\linethickness{1.5pt}
		\put(340,12){\line(0,1){12}}
		\qbezier(340,12)(340,-3)(320,-3)
		\put(320,-3){\line(-1,0){400}}
	\end{picture}
	\end{flushleft}
	\vspace{30pt}
	}

\newcommand{\chapheadesp}[1]
	{
	\begin{flushleft}
	\begin{picture}(0,0)
		\linethickness{1.5pt}
		\put(127,3){\line(-1,0){300}}
		\put(138,0){$\mathcal{R}$}
		\put(147,0){\emph{\textsf{esumen}}}
		\put(320,3){\line(-1,0){125}}
		\qbezier(320,3)(340,3)(340,-12)
		\put(340,-12){\line(0,-1){12}}
	\end{picture}
	
	\begin{minipage}{400pt}
		\setlength{\arrayrulewidth}{1.5pt}
		\begin{tabular}{@{}p{320pt}@{}p{19.25pt}@{}|}
		#1 & \\
		\end{tabular}
	\end{minipage}

	\begin{picture}(0,0)
		\linethickness{1.5pt}
		\put(340,12){\line(0,1){12}}
		\qbezier(340,12)(340,-3)(320,-3)
		\put(320,-3){\line(-1,0){400}}
	\end{picture}
	\end{flushleft}
	\vspace{30pt}
	}

\newcommand{\myrule}{\rule{\linewidth}{1mm}}
\usepackage{enumerate}  
\usepackage{amssymb}
\usepackage{lscape}
\usepackage{rotating}
\usepackage{textcomp}
\usepackage{longtable}

\newcommand{\filteri}{\textit{F814W}~}
\newcommand{\filterb}{\textit{F435W}~}
\newcommand{\filterh}{\textit{F160W}~}
\newcommand{\mi}{$M_{F814W}$~}
\newcommand{\mb}{$M_{F435W}$~}
\newcommand{\mbi}{$M_{F435W}$-$M_{F814W}$~}
\newcommand{\mini}{M$_{\rm{ini}}$~}

\newcommand{\reff}{$r_{\rm{eff}}$~}

\newcommand{\reffknot}{$r^{\rm{knot}}_{\rm{eff}}$~}
\newcommand{\reffpsf}{$r^{\rm{PSF}}_{\rm{eff}}$~}

\newcommand{\lir}{$L_{\rm{IR}}$~}
\newcommand{\lirnorm}{$L_{\rm{IR}}$~}
\newcommand{\lha}{$L_{H \alpha}$}
\renewcommand{\ergs}{erg s$^{-1}$~}
\newcommand{\lsun}{$L_{\odot}$~}
\newcommand{\ld}{$D_{\rm{L}}$~}
\newcommand{\av}{$A_V$~}
\newcommand{\twospace}{$\!\!$}
\newcommand{\onespace}{\hspace{-3pt}}
\newcommand{\arcsec}{\mbox{\ensuremath{^{\prime\prime}}}}

\newcommand{\ciii}{[C{\scriptsize{III}}]~}
\newcommand{\civ}{[C{\scriptsize{IV}}]~}
\newcommand{\ha}{H$\alpha$~}
\newcommand{\hb}{H$\beta$~}
\renewcommand{\hii}{H{\scriptsize{II}}~}
\newcommand{\hiiminus}{H{\tiny{II}}~}
\renewcommand{\hi}{H{\scriptsize{I}}~}
\newcommand{\oiii}{[O{\scriptsize{III}}]~}

\newcommand{\nii}{[N{\scriptsize{II}}]~}

\newcommand{\agn}{\acr{notarg}{AGN}{}~}
\newcommand{\acs}{\acr{notarg}{ACS}{}~}
\newcommand{\agns}{\acr[s]{notarg}{AGN}{}~}
\newcommand{\ew}{\acr{notarg}{EW}{}~}
\newcommand{\fov}{\acr{notarg}{FoV}{}~}
\newcommand{\fwhm}{\acr{notarg}{FWHM}{}~}

\newcommand{\gcs}{\acr[s]{notarg}{GC}{}~}
\newcommand{\gmcs}{\acr[s]{notarg}{GMC}{}~}
\newcommand{\gmc}{\acr{notarg}{GMC}{}~}
\newcommand{\hst}{\acr{notarg}{$HST$}{}~}
\newcommand{\imf}{\acr{notarg}{IMF}{}~}
\newcommand{\icmf}{\acr{notarg}{ICMF}{}~}
\newcommand{\icmfs}{\acr[s]{notarg}{ICMF}{}~}
\newcommand{\ifs}{\acr{notarg}{IFS}{}~}
\newcommand{\ifu}{\acr{notarg}{IFU}{}~}
\newcommand{\ifus}{\acr[s]{notarg}{IFU}{}~}
\newcommand{\ir}{\acr{notarg}{IR}{}~}
\newcommand{\iras}{\acr{notarg}{$IRAS$}{}~}
\newcommand{\ism}{\acr{notarg}{ISM}{}~}

\newcommand{\lirg}{\acr{notarg}{LIRG}{}~}
\newcommand{\lirgs}{\acr[s]{notarg}{LIRG}{}~}
\newcommand{\lf}{\acr{notarg}{LF}{}~}
\newcommand{\lfs}{\acr[s]{notarg}{LF}{}~}
\newcommand{\mdyn}{M$_{\rm{dyn}}$~}
\newcommand{\mf}{\acr{notarg}{MF}{}~}
\newcommand{\mfs}{\acr[s]{notarg}{MF}{}~}
\newcommand{\nicmos}{\acr{notarg}{NICMOS}{}~}
\newcommand{\psf}{\acr{notarg}{PSF}{}~}
\newcommand{\nir}{\acr{notarg}{NIR}{}~}
\newcommand{\smgs}{\acr[s]{notarg}{SMG}{}~}
\newcommand{\sbnn}{\acr{notarg}{SB99}{}~}

\newcommand{\ssc}{\acr{notarg}{SSC}{}~}
\newcommand{\sscs}{\acr[s]{notarg}{SSC}{}~}

\newcommand{\sfr}{\acr{notarg}{SFR}{}~}
\newcommand{\sfrs}{\acr[s]{notarg}{SFR}{}~}
\newcommand{\tdg}{\acr{notarg}{TDG}{}~}
\newcommand{\tdgs}{\acr[s]{notarg}{TDG}{}~}
\newcommand{\ulirg}{\acr{notarg}{ULIRG}{}~}
\newcommand{\ulirgs}{\acr[s]{notarg}{ULIRG}{}~}
\newcommand{\uv}{\acr{notarg}{UV}{}~}
\newcommand{\sfh}{\acr{notarg}{SFH}{}~}

\newcommand{\wfpc}{\acr{notarg}{WFPC2}{}~}
\newcommand{\wolfr}{\acr{notarg}{WR}{}~}
\newcommand{\wolfrs}{\acr[s]{notarg}{WR}{}~}

\newcommand{\ymcs}{\acr[s]{notarg}{YMC}{}~}

\usepackage[implicit=false, colorlinks=true, linkcolor=black, urlcolor=mycolorlink]{hyperref}

\newcommand{\eprint}[1]{\href{http://arxiv.org/abs/#1}{#1}}

\newcommand{\adsurl}[1]{\href{#1}{ADS}}
\providecommand{\url}[1]{\href{#1}{#1}}

\stepcounter{secnumdepth}

\addto\captionsenglish{}

\begin{document}
\sloppy

\pagestyle{frontmatter}\setcounter{page}{1}\pagenumbering{roman}

\cdp
\chapter*{}
\refstepcounter{chapter}
\phantomsection
\addcontentsline{toc}{section}{Front Page}
\pagestyle{empty}
\thispagestyle{empty}

\vspace{-5cm}

\begin{center}

\myrule \\
\vspace{-0.95cm}
\rule{\linewidth}{0.2mm}
\textbf{\Huge{ \textsc{Optically Selected Compact}} \\
\Huge{\textsc{Stellar Regions and}} \\
\Huge{\textsc{Tidal Dwarf Galaxies in \\
\onespace(Ultra)Luminous\hspace{-1pt} Infrared\hspace{-1pt} Galaxies\onespace \\}}}

\vspace{0.2cm}
\myrule
\vspace{-0.7cm}
\rule{\linewidth}{0.2mm}

\vspace{1cm}

{\Large Universidad Aut\'onoma de Madrid \\
 Facultad de Ciencias \\
 \vspace{0.2cm}
 Departamento de F\'isica Te\'orica} \\

\vspace{1.5cm}


\begin{figure*}[h]
\hspace{0.1\textwidth}
\includegraphics[width=0.2\textwidth]{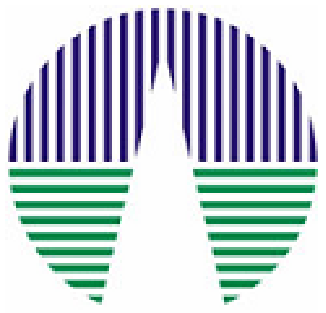}
\end{figure*}

\vspace{-3.2cm}
\begin{figure*}[h]
\hspace{0.53\textwidth}
\includegraphics[width=0.35\textwidth]{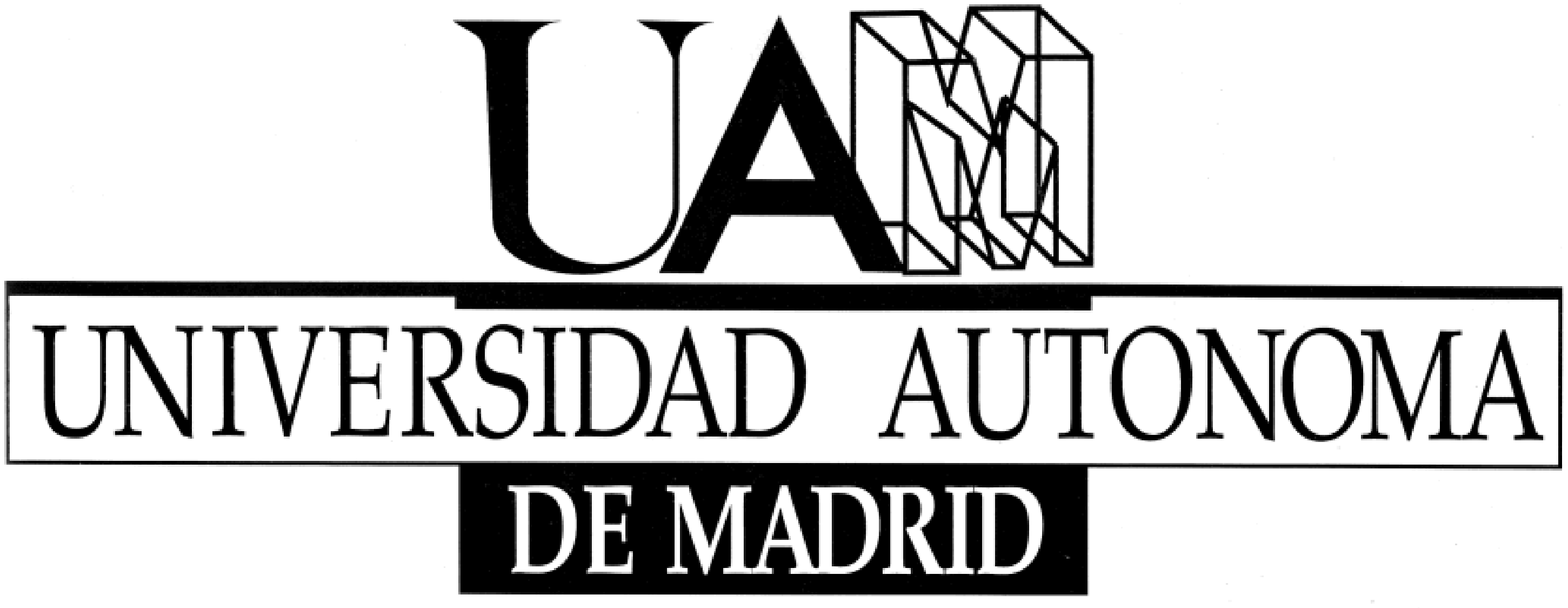}\\
\end{figure*}

\mbox{}

\vspace{-1cm}
\begin{figure*}[h!]
\hspace{0.3\textwidth}
\vspace{3cm}\includegraphics[width=0.39\textwidth]{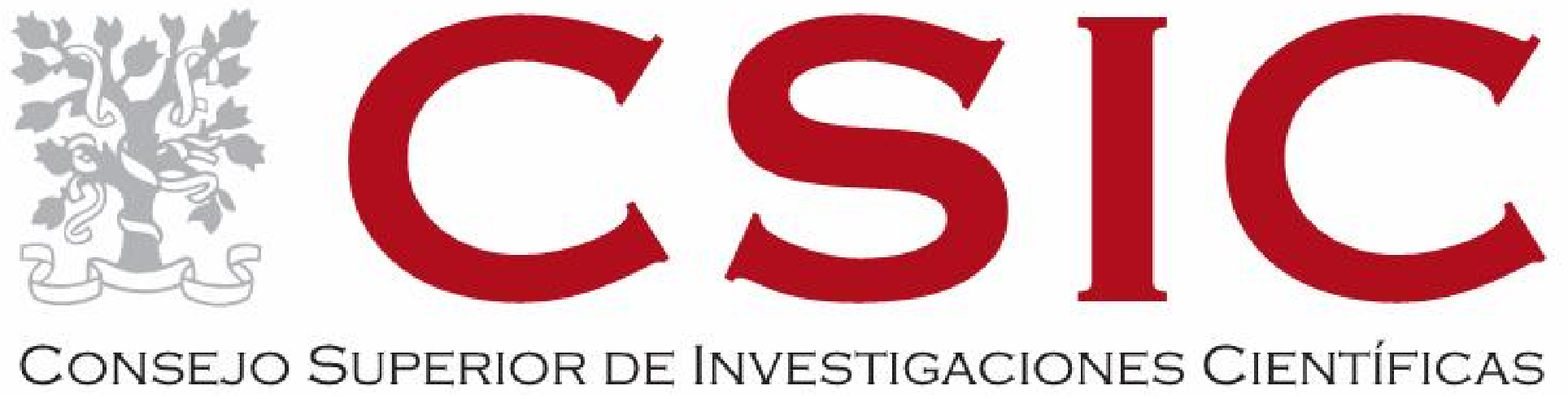}
\end{figure*}

\vspace{-2cm}
{\Large Consejo Superior de Investigaciones Cient\'ificas  \\
Centro de Astrobiolog\'ia \\
 \vspace{0.2cm}
Departamento de Astrof\'isica }  \\

\vspace{1.5cm}

{\LARGE \textbf{Daniel Miralles Caballero}}

\end{center}

\cleardoublepage


\mbox{}
\vspace{-1.2cm}

\begin{center}

\textbf{\Huge{ \textsc{Optically Selected Compact}} \\
\Huge{\textsc{Stellar Regions and}} \\
\Huge{\textsc{Tidal Dwarf Galaxies in \\
\onespace(Ultra)Luminous\hspace{-1pt} Infrared\hspace{-1pt} Galaxies\onespace \\}}}


\vspace{3cm} 

\Large PhD Thesis by \\
{\Large \textbf{Daniel \textsc{Miralles Caballero}}}

\vspace{1.5cm}

{\Large \textbf{Universidad Aut\'onoma de Madrid} \\
Facultad de Ciencias \\
Departamento de F\'isica Te\'orica} \\

\vspace{1.5cm}

{\Large \textbf{Consejo Superior de Investigaciones Cient\'ificas} } \\

{\Large Centro de Astrobiolog\'ia \\
Departamento de Astrof\'isica }  \\

\end{center}

\vspace{1.5cm}

\begin{tabular}{@{}l@{\ \ }lll}
\hline
   \noalign{\smallskip}
   \noalign{\smallskip}
Director:&	Dr. Luis &\textsc{Colina Robledo}&	CAB-CSIC, Madrid \\
Tutora:&	Dra. \'Angeles &\textsc{D\'iaz Beltr\'an }&	Universidad Aut\'onoma, Madrid \\
   \noalign{\smallskip}
\hline
\end{tabular}

\begin{flushright}
{\small Madrid, November 2011}
\end{flushright}


\cleardoublepage

\cdp
\begin{flushright}
\textit{A mi abuela}
\end{flushright}
\cleardoublepage

\chapter*{Acknowledgments}
\refstepcounter{chapter}
\markboth{{\sc Acknowledgments}}{}
\phantomsection
\addcontentsline{toc}{section}{Acknowledgments}
\pagestyle{frontmatter}

Como en un programa de televisi\'on, en donde la gente s\'olo conoce generalmente al presentador pero detr\'as hay una ingente cantidad de personas haciendo que el programa funcione, esta tesis no s\'olo es trabajo de uno. S\'i, presenta el trabajo de investigaci\'on que he realizado durante estos \'ultimos cuatro a\~nos. Sin embargo, no es \'unicamente fruto de mi completa dedicaci\'on, sino tambi\'en de la colaboraci\'on y apoyo de otras personas e instituciones. Sin ellos mi programa no habr\'ia funcionado. 

En primer lugar, quer\'ia agradecer el apoyo, dedicaci\'on y consejos que me ha propiciado mi director de tesis, Luis Colina. Gracias por el esp\'iritu cr\'itico que me ha transmitido, tan necesario en este campo. Agradezco tambi\'en el apoyo prestado por Almudena y Santiago. Asimismo, agradezco a mi tutora, Prof. \'Angeles D\'iaz, por su inter\'es y ayuda prestada en los momentos finales, y enlace con la UAM.

Quisiera expresar mi agradecimiento a todos los desarrolladores que, muchos de ellos sin \'animo de lucro, ponen a disposici\'on de la comunidad programas y herramientas de trabajo que simplifica muchos aspectos del trabajo. 

Todo empez\'o en una \textit{nave} ubicada en el CSIC de Serrano. Gracias a todas las personas que han pasado por ella, por las risas, las barritas, las sugerencias y muchas otras experiencias que hemos pasado juntos. En especial quer\'ia agradecer a Maca su ayuda prestada reci\'en llegado. Gracias tambi\'en a las otras personas del Departamento de Astrof\'isica Molecular Infrarroja del CSIC, trasaladadas ahora al Centro de Astrobiolog\'ia en el INTA. Lo cual me lleva tambi\'en a agradecer a los nuevos compa\~neros de la becar\'ia del mismo centro, donde he finalizado mi trabajo. Gracias a la gente que est\'a y a la gente que estuvo, por todas esas cosas que hacen que todo sea mucho m\'as llevadero y divertido. Gracias por los momentos que hemos pasado juntos (y espero pasar), sobre todo fuera de la becar\'ia. 

La asistencia a congresos internacionales y la realizaci\'on de estancias en centros de investigaci\'on extranjeros me ha ofrecido la posibilidad de conocer a muchas personas. A todas ellas les agradezco el tiempo compartido. En particular, me gustar\'ia agradecer la buena acogida, disposici\'on, interacci\'on y colaboraci\'on de Pierre-Alain en el IAP en Par\'is. Asimismo, a su estudiante Pierre-Emmanuel, a Fr\'ed\'eric y a la gente del telescopio espacial Herschel, que me han propiciado una estancia muy agradable. Y, c\'omo no, a mi casera Martine y a Gabriela. Gracias a todos vosotros la estancia en Par\'is ha sido inolvidable.

Gracias tambi\'en a mis amigos de Gand\'ia y Valencia, de por Espa\~na, de por el \mbox{mundo \ldots} Amigos de antes, de ahora y de siempre. Vosotros hab\'eis sido capaces de ``soportarme'' y me hab\'eis apoyado y animado principalmente en estos \'ultimos meses tan dif\'iciles e intensos en cuanto a trabajo que han culminado en el presente proyecto. Vuestra compa\~n\'ia y amistad han sido y ser\'an impagables.

La presente tesis no habr\'ia sido posible sin la financiaci\'on del Ministerio de Educaci\'on y Ciencia (hoy Ministerio de Ciencia e Innovaci\'on) a trav\'es de la beca BES-2007-16198 y la financiaci\'on del Consejo Superior de Investigaciones Cient\'ificas a trav\'es de los proyectos \mbox{ESP2005-01480}, \mbox{ESP2007-65475-C02-01} y \mbox{AYA2010-21161-C02-01}.

\textit{Last but not least}, agradezco profundamente a las personas que siempre han estado a mi lado, en los buenos y en los malos momentos. Gracias Javi, por aguantarme sobre todo desde que empec\'e a escribir la tesis. Gracias a toda mi familia, que en ning\'un momento ha dudado en apoyar y secundar todas las decisiones que he tomado. Gracias a mis padres, a mis hermanas, a mis prim@s, a mis t\'i@s  y, en especial, a mi abuela que, aunque no le gustara nada que me fuera tan lejos del pueblo, siempre ha estado pendiente de su nieto y le ha apoyado en todo.

\cdp
\chapter*{Abstract}
\refstepcounter{chapter}
\pagestyle{plain} 
\markboth{{\sc Abstract}}{}
\phantomsection
\addcontentsline{toc}{section}{Abstract}


Galaxy mergers can transform the type of the parent galaxy into another and, during their occurrence, drive phenomena that are extraordinary compared to the processes that take place in quiescent galaxies. Specifically, simulations and observations show that they trigger star formation events, and young objects at least as massive as globular clusters can be formed. Among the merger environments, luminous (\lirgs\twospace; L$_{\rm{bol}}$ $\sim$ \lir = L$_{[8-1000\mu m]}$ = \mbox{10$^{11}$-10$^{12}$ \lsun}) and ultraluminous (\ulirgs\twospace; \mbox{\lir = L$_{[8-1000\mu m]}$ = 10$^{12}$-10$^{13}$ \lsun\twospace}) infrared galaxies show the most extreme cases of star formation. 

It is not surprising that the studies carried out so far on (U)\lirgs have found young compact star forming regions.  However, only a few studies have been carried out to date for these kind of systems. This thesis work is devoted to the analysis of compact star forming regions (knots) in a representative sample of 32 (U)\lirgs\twospace, the largest sample used for this kind of study in these systems. The project is based mainly on optical high angular resolution images taken with the \acs and \wfpc cameras on board the \hst telescope, data from a high spatial resolution simulation of a major galaxy encounter, and with the combination of optical integral field spectroscopy (\ifs\twospace) taken with the INTEGRAL (WHT) and VIMOS (VLT) instruments. This is the first time that such combination of different types (photometric, spectroscopic and numerical) of a large amount of data, and such detailed study is performed on these systems. A few thousand knots --a factor of more than one order of magnitude higher than in previous studies-- are identified and their photometric properties are characterized as a function of the infrared luminosity of the system and of the interaction phase. These properties are compared with those of compact objects identified in simulations of galaxy encounters. Finally, and with the additional use of \ifs data, we search for suitable candidates to tidal dwarf galaxies, setting up constraints on the formation of these objects for the (U)\lirg class. The main findings and conclusions are summarized as follows:

\begin{itemize}
 
 \item With a typical size of tens of pc, the knots are in general compact. Most of them are likely to contain sub-structure, thus to constitute complexes or aggregates of star clusters. 
    
 \item Even though (U)\lirgs are known to have most of their star formation hidden by dust and re-emitted in the infrared, we have observed a fraction of 15\% of blue, almost free of extinction and luminous knots, with masses similar to or even higher than the super star clusters observed in other less luminous interacting systems.
 
 \item An extinction correction, characterized by an exponential probability density function, has to be applied to the colors of the compact stellar regions identified in simulations of major mergers so as to reproduce the rather broad range of colors sampled in knots in (U)\lirgs\twospace.
 
  \item Knots in \ulirgs\twospace, with higher star formation rate per unit area and gas content than less luminous interacting galaxies, are intrinsically more luminous (likely most massive) due to size-of-sample effects.

  \item Knots in \ulirgs can have both sizes and masses characteristic of stellar complexes or clumps detected in galaxies at high redshifts (\mbox{z $\gtrsim$ 1}), intrinsically more massive than stellar complexes in less luminous interacting galaxies.

 \item The aging of the knots rules the evolution of the color distribution during the interaction. Theoretical and observational evidence shows that, as a consequence of the interaction process, only the most massive knots remain when the system relaxes.
   
 \item The slope of the luminosity function (\lf\twospace) of the knots, compatible with \mbox{$\alpha$ $\backsimeq$ 2}, is independent of the luminosity of the galaxy. There are, however, slight indications that it varies with the interaction phase, becoming steeper from early to late phases of the interaction process. Supported by the simulation of a major galaxy encounter, higher knot formation rates at early phases of the interaction with respect to late phases, would explain this evolution of the \lf\twospace.
  
 \item Among extranuclear star-forming \ha clumps identified in 11 (U)\lirgs\twospace, with typical size up to several hundreds of pc, we identify 9 as candidates to tidal dwarf galaxies. They fulfill certain criteria of mass, self-gravitation and stability. With a production rate of 0.1 candidates per (U)\lirg systems, only a few fraction (\mbox{$<$ 10 \%}) of the general dwarf satellite population could be of tidal origin.

\end{itemize}

\cdp
\chapter*{Resumen}
\refstepcounter{chapter}
\pagestyle{plain} 
\markboth{{\sc Resumen}}{}
\phantomsection
\addcontentsline{toc}{section}{Resumen}

Como resultado de las interacciones gal\'acticas, las galaxias que intervienen pueden sufrir importantes transformaciones y, durante el proceso, se pueden dar fen\'omenos de una envergadura extraordinaria si se compara con los procesos que tienen lugar en ga\-laxias con poca formaci\'on estelar. En concreto, tanto las simulaciones como las observaciones han demostrado que estas interacciones desencadenan eventos de formaci\'on estelar en donde se pueden formar objetos j\'ovenes tan o m\'as masivos que los c\'umulos globulares. Las galaxias luminosas (\lirgs\twospace; L$_{\rm{bol}}$ $\sim$ \lir = L$_{[8-1000\mu m]}$ = \mbox{10$^{11}$-10$^{12}$ \lsun}) y ultraluminosas (\ulirgs\twospace; \mbox{\lir = L$_{[8-1000\mu m]}$ = 10$^{12}$-10$^{13}$ \lsun\twospace}) en el infrarrojo muestran los casos m\'as extremos de formaci\'on estelar dentro del entorno de las interacciones. 

Por ello, no es de extra\~nar que se hayan detectado regiones compactas de formaci\'on estelar reciente en todas las investigaciones llevadas a cabo en muestras de (U)\lirgs\twospace. No obstante, s\'olo unos pocos estudios se han preocupado por este tipo de sistemas. La presente tesis se centra en analizar regiones de formaci\'on estelar compacta (nodos) en una muestra representativa de 32 (U)\lirgs\twospace, la muestra m\'as cuantiosa que se ha utilizado para este tipo de estudio en estos sistemas. El proyecto se basa principalmente en el an\'alisis de im\'agenes de alta resoluci\'on angular tomadas en el visible con las c\'amaras \acs y \wfpc\twospace, instaladas a bordo del telescopio espacial Hubble (\hst\twospace), su combinaci\'on con espectroscop\'ia de campo integral (\ifs\twospace) obtenida con los espectr\'ografos INTEGRAL (WHT) y VIMOS (VLT), y con datos obtenidos con una simulaci\'on num\'erica de alta resoluci\'on espacial de una interacci\'on mayor de galaxias. Por primera vez, se ha realizado un estudio tan detallado en este tipo de sistemas y con tal combinaci\'on de diversos tipos (fotometr\'ia, espectroscop\'ia y simulaciones num\'ericas) y cantidad de datos. Se han identificado unos pocos miles de nodos, m\'as de un orden de magnitud que en estudios anteriores. Se ha procedido a realizar una  caracterizaci\'on completa de sus propiedades fotom\'etricas en funci\'on de la luminosidad en infrarrojo del sistema y seg\'un la fase de interacci\'on. Se han comparado dichas propiedades con las de objetos compactos que se han identificado en simulaciones de interacciones de galaxias. Por \'ultimo, y con el uso combinado de datos de espectrosco\'ia integral, hemos llevado a cabo una b\'usqueda de candidatos a galaxias enanas de marea, y se han podido establecer restricciones a la formaci\'on de este tipo de objectos para las galaxias tipo (U)\lirg\twospace. Los resultados m\'as relevantes se resumen a continuaci\'on:

\begin{itemize}

 \item Con un tama\~no t\'ipico de decenas de pc, los nodos son generalmente compactos. La mayor\'ia de los mismos contienen probablemente subestructura y constituyen, por lo tanto, complejos o agregados de c\'umulos estelares.\\

 \item Pese a que en los sistemas (U)\lirgs la mayor parte de la formaci\'on estelar est\'a escondida por el polvo, el 15\% de los nodos que se han detectado son muy azules, con poca extinci\'on y muy luminosos. Estos nodos azules poseen una masa semejante e incluso mayor los a superc\'umulos que se han observado en interacciones de galaxias menos luminosas. \\
 
  \item Para reproducir el ancho rango de colores de los nodos detectados en (U)\lirgs se tiene que aplicar cierta extinci\'on a los colores obtenidos de regiones estelares compactas que se han identificado en simulaciones de interacciones mayores. En concreto, la extinci\'on aplicada est\'a caracterizada por una funci\'on de densidad de probabilidad exponencial. \\
  
  \item Los nodos en \ulirgs\twospace, sistemas que poseen una mayor tasa de formaci\'on estelar por unidad de \'area y contenido en gas respecto a sistemas en interacci\'on menos luminosos, son intr\'insecamente m\'as luminosos (seguramente m\'as masivos) debido a efectos del tama\~no de la muestra.\\
  
  \item Los nodos en \ulirgs pueden ser de un tama\~no y tener una masa caracter\'istica similar a la de los complejos o estructuras estelares que se han detectado a altos desplazamientos al rojo (\mbox{z $\gtrsim$ 1}), intr\'insecamente m\'as massivos que los complejos estelares observados en interaciones de galaxias menos luminosas.\\
  
  \item El envejecimiento de los nodos gobierna la evoluci\'on de su distribuci\'on de colores a lo largo del proceso de interacci\'on. Hay evidencias te\'oricas y observacionales de que, debido a la interacci\'on, \'unicamente los nodos m\'as masivos sobreviven cuando el sistema se encuentra finalmente relajado. \\
  
  \item La pendiente de la funci\'on de luminosidad de los nodos, compatible con \mbox{$\alpha$ $\backsimeq$ 2}, no depende de la luminosidad de la galaxia. No obstante, hay indicaciones de una ligera variaci\'on con la fase de interacci\'on, de tal modo que dicha pendiente se vuelve m\'as pronunciada desde \'epocas tempranas a fases tard\'ias de la interacci\'on. Una simulaci\'on num\'erica de una interacci\'on mayor apoya este resultado, una variaci\'on de la pendiente que se puede explicar si la tasa de formaci\'on de nodos estelares es mayor en fases tempranas de la interacci\'on.\\

 \item De entre las estructuras de formaci\'on estelar extranucleares que se han identificado en 11 (U)\lirgs y que poseen un tama\~no t\'ipico de hasta varios cientos de pc, nueve de ellas se han reconocido como candidatas a galaxias enanas de marea. Estas candidatas cumplen ciertos criterios de masa, auto-gravitaci\'on y estabilidad. Con una tasa de producci\'on estimada de 0.1 candidatas por sistema (U)\lirg\twospace, s\'olo una peque\~na fracci\'on (\mbox{$<$ 10 \%}) de la polaci\'on de galaxias sat\'elites enanas se puede haber formado como una \tdg\twospace.
 
\end{itemize}

\vfill

\clearpage{\pagestyle{empty}\cleardoublepage}
\tableofcontents
\addtocounter{autotoc}{4} 

\clearpage{\pagestyle{empty}\cleardoublepage}
\addcontentsline{toc}{section}{List of figures}
\listoffigures\cdp
\addtocounter{autotoc}{5} 

\addcontentsline{toc}{section}{List of tables}
\listoftables
\addtocounter{autotoc}{6} 

\clearpage{\pagestyle{empty}\cleardoublepage}
\cdp

\chapter*{Acronyms}
\refstepcounter{section}
\markboth{{\sc Acronyms}}{}
\phantomsection
\addcontentsline{toc}{section}{Acronyms}

\vspace{-0.5cm}
\begin{longtable}{l@{\hspace{2cm}}l}


\textbf{	A\&AS	} & \textit{	Astronomy and Astrophysics			} \\[0.1cm]
\textbf{	A\&A	} & \textit{	Astronomy and Astrophysics Supplement			} \\[0.1cm]
\textbf{	ACS	} & \textit{	Advanced Camera for Surveys			} \\[0.1cm]
\textbf{	AGN	} & \textit{	Active Galactic Nucleus			} \\[0.1cm]
\textbf{	ALMA	} & \textit{	Atacama Large Millimeter/submillimeter Array			} \\[0.1cm]
\textbf{	AJ	} & \textit{	Astronomical Journal			} \\[0.1cm]
\textbf{	ApJ	} & \textit{	Astrophysical Journal			} \\[0.1cm]
\textbf{	Ap\&SS	} & \textit{	Astrophysics and Space Science			} \\[0.1cm]
\textbf{	ApJS	} & \textit{	Astrophysical Journal Supplement			} \\[0.1cm]
\textbf{	ApSSS	} & \textit{	Astrophysics and Space Science Supplement			} \\[0.1cm]
\textbf{	ARA\&A	} & \textit{	Annual Review of Astronomy and Astrophysics			} \\[0.1cm]
\textbf{	CCD	} & \textit{	Charged Couple Device			} \\[0.1cm]
\textbf{	ELT	} & \textit{	Extremely Large Telescope			} \\[0.1cm]
\textbf{	ESO	} & \textit{	European Southern Observatory			} \\[0.1cm]
\textbf{	EW	} & \textit{	Equivalent  Width			} \\[0.1cm]
\textbf{	FIR	} & \textit{	Far Infrared			} \\[0.1cm]
\textbf{	FoV	} & \textit{	Field of View			} \\[0.1cm]
\textbf{	FWHM	} & \textit{	Full Width at Half Maximum			} \\[0.1cm]
\textbf{	GC	} & \textit{	Globular Cluster			} \\[0.1cm]
\textbf{	GMC	} & \textit{	Giant Molecular Cloud			} \\[0.1cm]
\textbf{	GEMS	} & \textit{	Galaxy Evolution From Morphology and SEDs			} \\[0.1cm]
\textbf{	GOODS	} & \textit{	Great Observatory Origins Deep Survey			} \\[0.1cm]
\textbf{	GTC	} & \textit{	Gran Telescopio de Canarias			} \\[0.1cm]
\textbf{	\textit{HST}	} & \textit{	Hubble Space Telescope			} \\[0.1cm]
\textbf{	ICMF	} & \textit{	Initial Cluser Mass Function			} \\[0.1cm]
\textbf{	IDL	} & \textit{	Interface Definition Language			} \\[0.1cm]
\textbf{	IFS	} & \textit{	Integral Field Spectroscopy			} \\[0.1cm]
\textbf{	IFU	} & \textit{	Itegral Field Unit			} \\[0.1cm]
\textbf{	IMF	} & \textit{	Initial Mass Function			} \\[0.1cm]
\textbf{	IR	} & \textit{	Infrared			} \\[0.1cm]
\textbf{	IRAF	} & \textit{	Image Reduction and Analysis Facility			} \\[0.1cm]
\textbf{	\textit{IRAS}	} & \textit{	Infrared Astronomical Satellite			} \\[0.1cm]
\textbf{	ISM	} & \textit{	Interestellar Medium			} \\[0.1cm]
\textbf{	\textit{ISO}	} & \textit{	Infrared Space Observatory			} \\[0.1cm]
\textbf{	KW	} & \textit{	Kolmogorov-Smirnov			} \\[0.1cm]
\textbf{	LINER	} & \textit{	Low-Ionization Nuclear Emission-line Region			} \\[0.1cm]
\textbf{	LIRG	} & \textit{	Luminous Infrared Galaxy			} \\[0.1cm]
\textbf{	LF	} & \textit{	Luminosity Function			} \\[0.1cm]
\textbf{	MF	} & \textit{	Mass Function			} \\[0.1cm]
\textbf{	MNRAS	} & \textit{	Monthly Notices of the Royal Astronomical Society			} \\[0.1cm]
\textbf{	NASA	} & \textit{	National Aeronautics and Space Administration			} \\[0.1cm]
\textbf{	NICMOS	} & \textit{	Near Infrared Camera and Multi-Object  Spectrometer			} \\[0.1cm]
\textbf{	NIR	} & \textit{	Near-Infrared			} \\[0.1cm]
\textbf{	PAH	} & \textit{	Polycyclic Aromatic Hydrocarbon			} \\[0.1cm]
\textbf{	PASP	} & \textit{	Publications of the Astronomical Society of the Pacific			} \\[0.1cm]
\textbf{	PMAS	} & \textit{	Potsdam MultiAperture Spectrophotometer			} \\[0.1cm]
\textbf{	PSF	} & \textit{	Point Spread Function			} \\[0.1cm]
\textbf{	QSO	} & \textit{	Quasi Stellar Object			} \\[0.1cm]
\textbf{	SB99	} & \textit{	Starburst 99			} \\[0.1cm]
\textbf{	SED	} & \textit{	Spectral Energy Distribution			} \\[0.1cm]
\textbf{	SINFONI	} & \textit{	Spectrograph for INtegral Field Observations in the Near Infrared			} \\[0.1cm]
\textbf{	SFH	} & \textit{	Star Formation History			} \\[0.1cm]
\textbf{	SFR	} & \textit{	Star Formation Rate			} \\[0.1cm]
\textbf{	SMG	} & \textit{	Submillimeter Galaxies			} \\[0.1cm]
\textbf{	SSP	} & \textit{	Single Stellar Population			} \\[0.1cm]
\textbf{	SSC	} & \textit{	Super Star Cluster			} \\[0.1cm]
\textbf{	TDG	} & \textit{	Tidal Dwarf Galaxy			} \\[0.1cm]
\textbf{	UDF	} & \textit{	Ultra Deep Field			} \\[0.1cm]
\textbf{	ULIRG	} & \textit{	Ultraluminous Infrared Galaxy			} \\[0.1cm]
\textbf{	UV	} & \textit{	Ultraviolet			} \\[0.1cm]
\textbf{	VIMOS	} & \textit{	VIsible MultiObject Spectrograph			} \\[0.1cm]
\textbf{	VLT	} & \textit{	Very Large Telescope			} \\[0.1cm]
\textbf{	WFPC2	} & \textit{	Wide-Field Planetary Camera 2			} \\[0.1cm]
\textbf{	WR	} & \textit{	Wolf-Rayet			} \\[0.1cm]
\textbf{	WYFFOS	} & \textit{	Wide Field Fibre Optical Spectrograph			} \\[0.1cm]
\textbf{	YMC	} & \textit{	Young Massive Cluster			} \\[0.1cm]

\end{longtable}
\cdp


\setcounter{page}{1}\pagenumbering{arabic}

\chanonumber{General Introduction}

\chaphead{This Chapter gives a short review of the star formation processes in interacting galaxies and presents the main properties of the most extreme star-forming systems in the local Universe, upon which the project is based on. It also contains a short description of the formation and dynamical evolution processes of the compact star-forming regions that will be studied throughout this thesis.}


\sect[sf_interactions]{Star Formation in Interactions} 
\subsect[general_concept]{General Concept}

A large fraction, if not all, of stars are thought to be born within some type of cluster environment (e.g., \mycitealt{Lada03};~\mycitealt{Porras03}). Thus, the formation of star clusters is intimately linked to the process of star formation. The use of star clusters as a tool to study the (extra)galactic star formation processes is therefore of major importance. The most massive are long lived and therefore potentially carry information about the entire \acr[s]{}{SFR}{star formation histories} of their host galaxies and they are also bright enough to be observed at much greater distances than individual stars (\mycitealt{Larsen10}). 

In the last two decades young, massive, compact clusters have been observed in many external galaxies and different environments: dwarf galaxies (\mycitealt{Billett02}), spiral disks (\mycitealt{Larsen99a}), barred galaxies, starburst galaxies (\mycitealt{Meurer95a};~\mycitealt{Barth95}), interacting galaxies (\mycitealt{Whitmore93},~\myciteyear{Whitmore99},~\myciteyear{Whitmore07};~\mycitealt{Bik03};~\mycitealt{Weilbacher00};~\mycitealt{Knierman03};~\mycitealt{Peterson09}) and compact groups of galaxies (\mycitealt{Iglesias-Paramo01};~\mycitealt{Temporin03}). It is commonly assumed that at least some of these are young counterparts of the ancient \acr[s]{}{GC}{globular clusters}, which are ubiquitous in all major galaxies. However, the link between star formation, the formation of massive clusters and ulterior evolution is not completely clear. Much less clear is the understanding of the role played by the environment in establishing this link. This leads us to pose the following questions: How do star clusters form? Where do they form more efficiently? Are those clusters formed in very active star-forming systems similar to those formed in less efficient environments? How do they evolve, especially when they are formed during major interactions of galaxies? Under which conditions can they be, if possible, as massive and large as dwarf galaxies?

All these questions can be addressed from many directions, from the detailed study of individual star clusters to a more general and statistical study of a sample of thousands of clusters. Furthermore, some studies of star cluster populations have shown that young star clusters do not form in isolation, but tend to be clustered themselves (\mycitealt{Zhang01};~\mycitealt{Larsen04}). Hence, the study of associations of individual clusters can also help us understand the star formation occurring in active star-forming systems. 

The dominant force that governs the cluster formation and all the processes that affect them is gravity. However, the implementation of all the ingredients in the models (e.g., gravity, stellar evolution, binary interactions, the influence of the interestelar medium, the influence of the host galaxy and/or other galaxies in interactions) makes the task of simulating the evolution of the host galaxy and its cluster population very complex. Nevertheless, during the last few years the combination of detailed observations with more realistic models of cluster formation (i.e., via interactions of galaxies), is starting to set important constrains on the processes of cluster formation and evolution.

\subsect[inter_gal]{Star Formation in Violent Environments}

Clusters tend to form where strong star formation occurs and, especially, in starbursts triggered by violent environments such as galaxy interactions and mergers (\mycitealt{Schweizer98}). In the following we will overview the role of this extreme environment. 

\subsubsection{The Environment: Interacting Galaxies}

During the second half of the past century our understanding of galaxy formation and evolution, and the role of interactions between galaxies underwent considerable change. From the initial belief that galaxies were quiescent systems, they passed to be understood as dynamic, evolving entities that could produce cataclysmic interactions. Early n-body simulations (\mycitealt{Toomre72}) showed how colliding spiral galaxies could be turned into the way seen in the Arp Atlas catalogue (``Atlas of Peculiar Galaxies'';~\mycitealt{Arp66}), most of them interacting systems. Toomre (\myciteyear{Toomre77}) farther suggested that mergers may convert one galaxy type (spiral) into another (elliptical) along the Hubble sequence. In the 90s it was clear that mergers drive
galaxy evolution by forming elliptical galaxies (e.g., \mycitealt{Kormendy92};~\mycitealt{Kauffmann93}).


Mergers not only transform one galaxy type into another but, during their occurrence, drive phenomena that are extraordinary compared to the processes that take place in quiescent galaxies. Simulations by Toomre \& Toomre (\myciteyear{Toomre72}) also showed that large amount of gas would be funneled into the galaxy centers. Shortly after, evidence for recent burst of star formation was reported in the Arp Atlas (\mycitealt{Larson78}), which was suggested to occur due to galaxy interactions like those described by Toomre.

\subsubsection{A Typical Merger. End Product}

The sequential major events in a merger of gas-rich galaxies can be summarized as follows: as the two galaxies approach each other, their stellar disks collide and distort and their gas starts flowing toward the center of the system (phase called ``first pericenter passage''). As the distance between the galaxies decreases, the gravitational attraction increases and so does the gas infall toward the center of the merger as a result of dissipation. The stars follow the gas in its infall due to changes in the potential well of the merging system. Due to gravitational torques and tidal forces caused by angular momentum conservation constraints, long tails of stars and gas are formed. The two galaxies continue to approach one another before they finally merge and reach relaxation. After the nuclei coalesce, the gas infall to the center of the merger decreases and finally stops a few hundreds of Myr later due to outflowing supernova winds related to the ongoing star formation and to feedback to the \acr{}{ISM}{interestellar medium} from an \acr{}{AGN}{active galactic nucleus}. The final remnant is supported mainly by random motions instead of systematic rotation. The whole baryoninc matter-merger process usually takes about one Gyr. An example of the merger sequence is shown in \reffig{merger_bournaud}.

\begin{figure}[!tp]
\centering
 \hypertarget{fig:merger_bournaud}{}\hypertarget{autolof:\theautolof}{}\addtocounter{autolof}{1}
 \includegraphics[trim = 0cm 0cm 0cm 0cm,clip=true,width=1\textwidth]{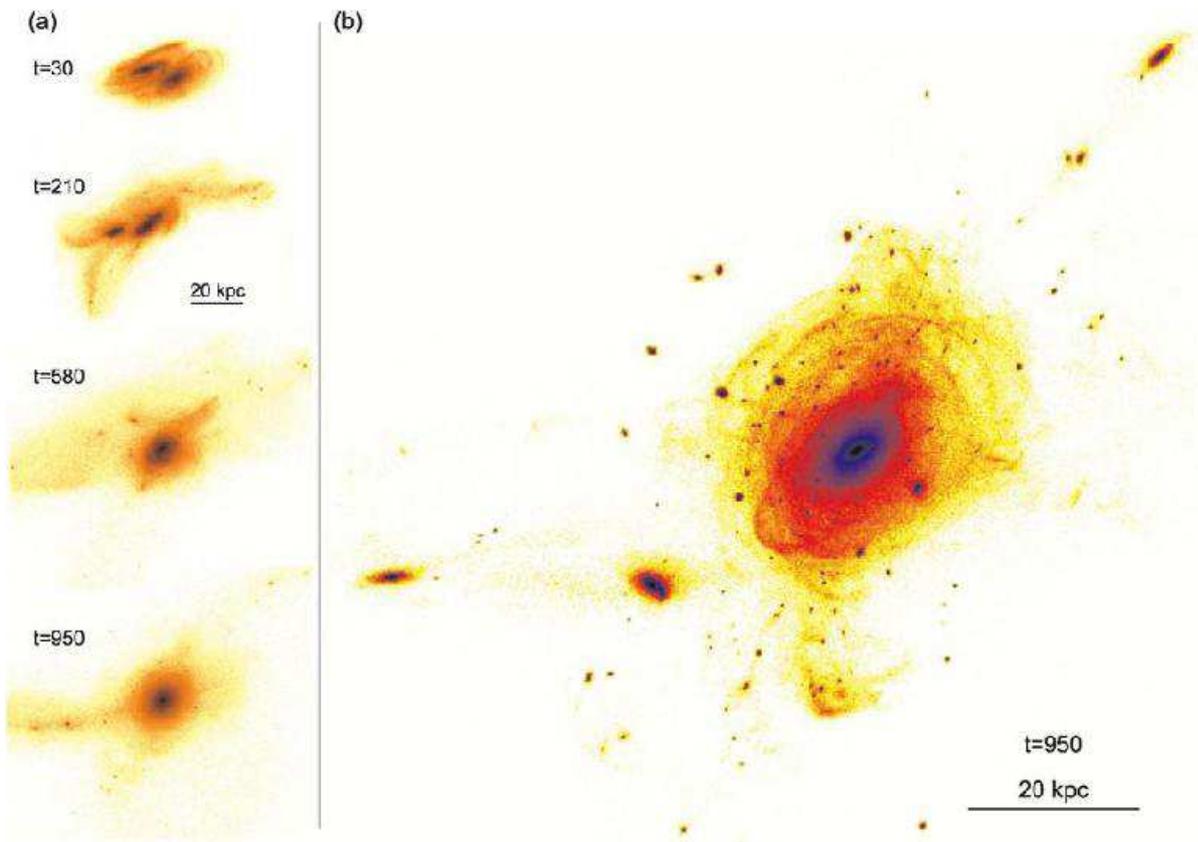}
 \caption[Sequence of four snapshots showing the merger evolution]{(a) Sequence of four snapshots showing the merger evolution. Time is indicated in Myr after the first pericenter passage. The total stellar mass
density is shown. (b) Surface density of ``young'' stars (defined as those formed after the first pericenter passage) at the end of the wet merger simulation, 950
Myr after the first pericenter passage. Figure taken from Bournaud et al. (\myciteyear{Bournaud08a}).}
 \label{fig:merger_bournaud}
\end{figure}

The main parameters that determine the end product of such mergers are well understood; the outcome
mainly depends upon the initial amount of gas and the mass ratio of the merging spiral galaxies (e.g., \mycitealt{Mihos96};~\mycitealt{Mihos98};~\mycitealt{Naab03}). The gas provides a dissipative component that assists the collapse of the mass to the center of the system by overcoming the angular momentum which tends to drive the mass toward the exterior of the system (\mycitealt{Hernquist93}); this way it enables the formation of systems with elliptical-like intensity profiles. The initial mass ratio of the colliding galaxies, connected to the depth of the potential well of the system, is therefore proportional to the amount of gas of the least massive component that inflows to the center of the system. The so-called major mergers  --mergers of roughly equal mass, typically 10\% of the stellar mass-- lead to the formation of elliptical galaxies (\mycitealt{Barnes92a};~\mycitealt{Naab03};~\mycitealt{Bournaud05}). In contrast, mergers of galaxies with a larger mass ratio ($>$ 4:1), which are called minor, can preserve or even enhance the pre-existing angular momentum depending on the orientation of the merging system (e.g.,~\mycitealt{Mihos94a},~\myciteyear{Mihos96};~\mycitealt{Bendo00}). 

The end product of major mergers at high redshifts spiral galaxies are diverse, since their disks contain larger amounts of gas (\mycitealt{Tacconi06}). The outcome of \mbox{high-z} gas-rich mergers may either be an elliptical or even a spiral galaxy that possesses a massive
bulge (\mycitealt{Springel05a};~\mycitealt{Springel05b}). The result again
depends on the amount of gas in the initial galaxies  and on the presence of a bulge (it delays the star formation events) and an AGN (it suppresses the star formation events, producing a \textit{negative feedback}). 

\subsubsection{Star Formation Bursts in Mergers}

While the picture we have for the end product of gas-rich mergers is well understood, that for the intermediate merger phases (i.e., between first pericenter passage and final relaxation) is rather unclear. The uncertainties mainly originate from the simplified treatment of the \ism (i.e., the gas and the dust). The role played by interactions and mergers in enhancing star formation has also been subject of intense debate in the past few decades by means of numerical simulations (e.g.,~\mycitealt{Barnes92a};~\mycitealt{Mihos94a};~\mycitealt{Mihos96};~\mycitealt{Springel00};~\mycitealt{Barnes04};~\mycitealt{Springel05a};~\mycitealt{Bournaud06};~\mycitealt{Cox06};~\mycitealt{diMatteo07};~\mycitealt{diMatteo08};~\mycitealt{Teyssier10}). Although it has become clear that the star formation efficiency is highly dependent upon the numerical recipe adopted for star formation (density or shock dependent) and feedback assumptions, all the studies have supported the fact that galaxy interactions induce star formation.

It is not well understood either at what point in time the gas infall to the center of the merger is maximum and hence, it is unclear at what point the strongest star formation or starburst events occur. Statistically, it is thought that the \acr{}{SFR}{star formation rate} peaks roughly between the first encounter and shortly after the nuclear coalescence (\mycitealt{Mihos96};~\mycitealt{Springel05a};~\mycitealt{diMatteo08}). The result mainly depends on the size of the galaxy bulges (large bulges stabilize the gas and delay its fall to the center of the system), the presence of a black hole (it acts as a strong gravitational point at the center of the system), and the strength of the winds from the supernovae and the \agn (they produce a negative feedback).

\sect[compact]{Compact Star-Forming Regions in Galaxy Interactions}
\subsect[ssclusters]{Super Star Clusters and Associations}

Among the diverse consequences of the violent processes that galaxy encounters undergo, the enhancement of the star formation in some epochs of the interaction leads to the formation of large number of star clusters. Most intriguingly was the discovery of a new class of young and massive (as globular clusters) star clusters, very much common  in this environment of galaxy mergers.

\subsubsection{Discovery and Characterization of Super Star Clusters}

We commonly accept that our Galaxy is ``normal'' and a good representative of spiral galaxies. The
Milky Way has two distinct populations of star clusters, open and globular clusters (\gcs\twospace). The open clusters are typically young (a few Myr - a few Gyr), low mass (\mbox{M $<$ 5$\times$10$^{4}$~\msun\twospace}), and low density objects (\mbox{$\rho$ $\sim$ 20~\msun\twospace/pc$^3$} in the central regions), and are clearly associated with the galactic disk. \gcs\twospace, on the other hand, are all old (\mbox{$\tau$ $>$ 10 Gyr}), high mass (\mbox{M = 10$^4$-10$^6$~\msun\twospace}) and high density objects (\mbox{$\rho$ $\sim$400~\msun\twospace/pc$^3$}), associated with the Galactic bulge/halo. 

The \acr[s]{}{MF}{mass functions} of open and globular clusters are also strikingly different. The \mf of young open clusters can be fitted by a power law, \mbox{dN/dM $\propto$ M$^{-2}$}, down to a few hundred~\msun (\mycitealt{Elmegreen97};~\mycitealt{Piskunov08}). In contrast, the globular cluster \mf is rather flat at low masses (\mbox{dN/dM $\propto$ constant} for \mbox{M $<$ 10$^5$~\msun\twospace};~\mycitealt{McLaughlin96}), whereas the high-mass end can be fitted by a power law with slope of -2, as for the open clusters. Given the clear dichotomy between these two systems, traditionally it has been established that open and globular clusters formed by two separate mechanisms.

However, for the past few decades a significant population of extragalactic massive (\mbox{M $>$ 5$\times$10$^{4}$~\msun\twospace}) young (\mbox{$\tau$ $<$ 1 Gyr}) clusters has been detected in many environments: in the LMC (\mycitealt{Elson85}), in normal spirals (e.g.,~\mycitealt{Larsen00};~\mycitealt{Larsen04}, in starburst galaxies (e.g., in M82;~\mycitealt{OConnell95};~\mycitealt{DeGrijs05}), in ongoing mergers (e.g., in The Antennae;~\mycitealt{Whitmore99}) and in merger remnants (e.g., in NGC 7252;~\mycitealt{Schweizer98};~\mycitealt{Maraston01}). Nowadays, these objects are known as \acr[s]{}{YMC}{young massive clusters} or \acr[s]{}{SSC}{super star clusters}.

The \hst caused a revolution in the field of extragalactic clusters. Its high angular resolution first allowed to establish an upper limit to the size on young extragalactic clusters. Shortly after, with improved optics, the \wfpc camera on board \hst was able to resolve individual star clusters in external galaxies, showing that they did indeed have sizes (1-20 pc) comparable to globular clusters in our Galaxy (e.g., \mycitealt{Whitmore99}).

The first environments that were discovered to contain copious amounts of massive star clusters were those of galaxy mergers. Holtzman et al. (\myciteyear{Holtzman92}) were the first in a long list of studies which used the \hst to discover massive young clusters in mergers. They studied a recent galaxy merger (NGC 1275), and found a population of blue point-like sources, with ages of less than 300 Myr old and masses between 10$^5$-10$^8$~\msun\twospace. These young massive clusters were also found in the prototypical merger remnant NGC 7252, with ages ranging from 34 to 500 Myr, consistent with having been formed during the interaction process (\mycitealt{Whitmore93}). Many more examples soon followed, like those in NGC 3597 (\mycitealt{Holtzman96}), in NGC 3921 (\mycitealt{Schweizer96}), in NGC 3256 (\mycitealt{Zepf99}), in NGC 4038/4039 (\mycitealt{Whitmore93},~\myciteyear{Whitmore95},~\myciteyear{Whitmore99},~\myciteyear{Whitmore10};~\mycitealt{Bastian06}), in M51 (\mycitealt{Bastian05a},~\myciteyear{Bastian05b}) and in Arp284 (\mycitealt{Peterson09}), among others. An example of many \sscs candidates found in The Antennae, the youngest and nearest example of a pair of merging disk galaxies in the Toomre (\myciteyear{Toomre77}) sequence, is shown in \reffig{whit99}.

\begin{figure}[!htp]
\centering
 \hypertarget{fig:whit99}{}\hypertarget{autolof:\theautolof}{}\addtocounter{autolof}{1}
 \includegraphics[trim = 0cm 0cm 0cm 0cm,clip=true,width=0.95\textwidth]{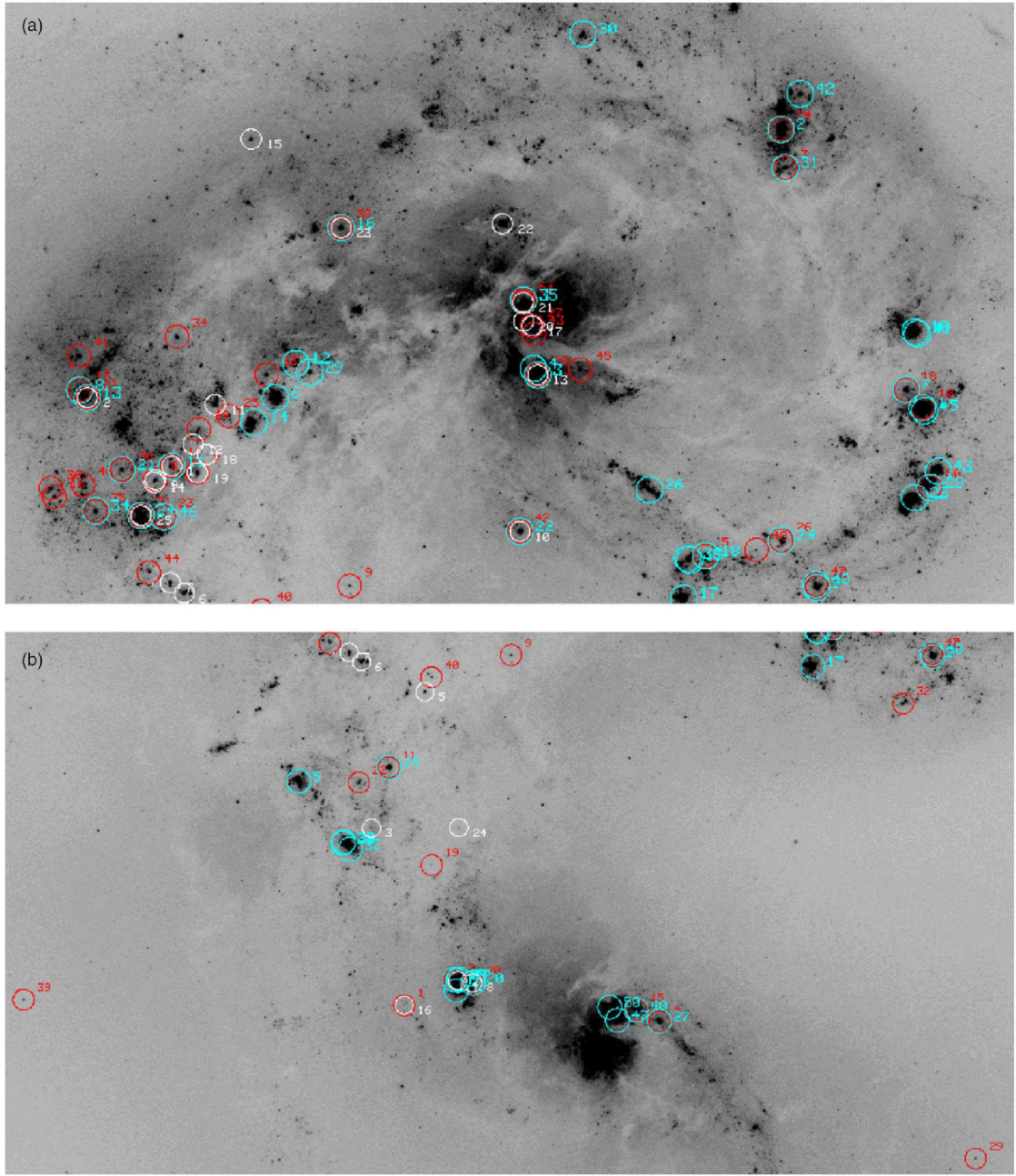}
 \caption[The main bodies of The Antennae viewed with the ACS/HST]{ The main bodies of The Antennae viewed with the Advanced Camera for Surveys (\acs\twospace/\hst\twospace). \textbf{Top:} Top half of The Antennae (NGC 4038) with the 50 most luminous clusters (blue circles) and the 50 most massive clusters (all with a mass higher than 5$\times$10$^5$~\msun\twospace; red circles) marked. Also marked are the 25 IR-brightest clusters (white circles). \textbf{Bottom:} Same information  but for the bottom half of The Antennae (NGC 4039). Figure taken from Whitmore et al. (\myciteyear{Whitmore10}).}
 \label{fig:whit99}
\end{figure}

The \sscs found in these systems have many of the same  properties as the galactic globular clusters, namely size and mass (and hence stellar density). On the other hand, their ages, metallicities, and mass functions are much more similar to those in galactic open clusters. This mixture of properties led to their designation as young globular clusters, implying that they are the same as the old globular clusters in our Galaxy (only younger) and that any differences between these populations are simply due to evolutionary effects.

\subsubsection{Formation of \sscs}

The problem of cluster formation is intimately linked to that of star formation. Star formation is closely  associated with dense molecular gas and proto-clusters are observed to form deeply embedded within \acr{}{GMC}{giant molecular clouds}. The \gmcs are themselves part of a larger hierarchy of structure in the interstellar medium (\mycitealt{Elmegreen96};~\mycitealt{Elmegreen07b}), and tend to be organized into giant molecular complexes (\mycitealt{Wilson03}). The average particle density of molecular gas in a \gmc is low (\mbox{n$_{H} \sim $10$^2$-10$^3$ cm$^{-3}$}), but stars only form in the densest regions (\mbox{i.e., n$_{H}>$ 10$^5$ cm$^{-3}$}). Therefore, in general star formation is an inefficient process and only a few percent of the mass of a given \gmc is converted into stars before the cloud is dispersed. 

The \gmc mass function in the Milky Way has a characteristic upper-mass of about 6$\times$10$^6$~\msun and the mass spectrum of clumps within \gmcs appears to follow a similar power law as that of the \gmcs\twospace, \mbox{dN/dM $\propto$ M$^{\alpha}$}, with \mbox{$\alpha \sim$ -2} (\mycitealt{MacLow04}). From there an upper-limit to a cluster mass of few 10$^5$~\msun is derived, consistent with 
other constraints on the initial cluster mass function in spiral galaxies (\mycitealt{Larsen09}).

It is hard to understand how the most massive clusters observed in some external galaxies, with masses of 10$^6$~\msun or higher, could form from Milky Way-like \gmcs\twospace. In order for a \gmc to reach the average mean density required to form a massive cluster (i.e., n$_{H}>$10$^5$ cm$^{-3}$), it should collect the amount of gas typical of a massive Galactic \gmc within a volume only a few pc across, essentially turning such a cloud into one big clump (\mycitealt{Larsen10}). Such high dense and massive clumps exist in starbursts and interacting galaxies, which may allow denser and more massive \gmcs to condense (\mycitealt{Escala08}). The presence of shocks and compressive tides in mergers also help these massive \gmcs condense (\mycitealt{Ashman01};~\mycitealt{Renaud08a},~\myciteyear{Renaud09}). In even more extreme environments, such as in (U)\lirgs (see \refsec{ulirgprop}), \gmcs may be both denser and more massive (\mycitealt{Murray10}), consistent with the presence of clusters with \mbox{M $>$ 10$^7$~\msun} in some merger remnants (W3 in NGC 7252; \mycitealt{Maraston04};~\mycitealt{Bastian06}).

\subsubsection{Dynamical Evolution of \sscs}

All clusters undergo at birth an embedded phase in the progenitor molecular cloud that lasts at least a few Myr (i.e., \mbox{$\tau \lesssim$ 3 Myr}), with high gas column densities and relatively large amounts of dust extinction of \mbox{\av $\sim$ 1-8 mag} (\mycitealt{Larsen09}). Stellar winds, ionizing flux from massive stars, and stellar feedback expel the remaining gas eventually (see~\mycitealt{Goodwin09} for a recent review). If all stars are formed in clusters, then only a small percentage remain still within clusters at the end of the embedded phase (\mycitealt{Lada03},~\myciteyear{Lada10}). In fact, while expelling the gas many clusters become unbound and subsequently disperse into the surrounding medium. 

An embedded cluster will not necessarily remain bound after gas expulsion (\mycitealt{Lada10}). If the stars and gas are initially in virial equilibrium, the velocity dispersion of the stars will be too high to match the shallower potential once the gas is expelled. In practise, cluster expansion and gas expulsion does not occur instantly so the stars have some time to adjust to the new potential. Simulations suggest that a small fraction of the stars may remain bound for \acr{s}{SFE}{star formation efficiencies}\footnote{SFE = $\frac{M_{cl}}{M_{cl}+M_{gas}}$, where M$_{cl}$ is the total mass of stars formed in the embedded cluster and M$_{gas}$ is the mass of the gas not converted into stars} of less than 20-30\% (\mycitealt{Boily03};~\mycitealt{Baumgardt07}). This process, known as 'infant mortality' or 'mass-independent disruption mechanism' can last for several tens of Myr until the cluster settles into a new equilibrium (\mycitealt{Goodwin06};~\mycitealt{Whitmore07}).

This picture is consistent with observations. Lada \& Lada (\myciteyear{Lada03}) estimated that about 95\% of clusters formed in the Milky Way dissolve in less than 100 Myr. Although the universality of this process is poorly quantified, some efforts have been made by observing the cluster demographics in extragalactic clusters, especially in interacting galaxies. For instance, the age distribution of mass-limited cluster samples in The Antennae galaxies, the well-known archetype of a major merger, is approximately dN/d$\tau\backsimeq\tau^{-1}$ for ages ($\tau$) of up to 10$^8$-10$^9$ yr (\mycitealt{Fall05}), suggesting that 80-90\% of the clusters disappear per decade in age, independent of mass (\mycitealt{Whitmore07}). However, over such a large age range, it is unlikely that disruption can still be attributed to gas expulsion. 

During the last few years the has been a controversy about this issue, since estimating disruption parameters for interacting systems is not straightforward. Bastian et al. (\myciteyear{Bastian09}) fitted the age distribution of clusters in The Antennae with a model in which the cluster formation rate has increased over the past few hundreds of Myr (\mycitealt{Mihos93}), as suggested by simulations of the ongoing interaction (e.g.,~\mycitealt{Cox08}). In this model the mass-independent disruption mechanism only lasts for $\lesssim$ 20 Myr. A high disruption fraction ($\sim$ 70\%) was also claimed in M51 (\mycitealt{Bastian05b}), but this can be partly due to age-dating artifacts around 10 Myr (\mycitealt{Gieles09}).

Clusters that survive the infant mortality process will continue to evolve dynamically on longer timescales as a result of internal two-body relaxation, external shocks and mass loss due to stellar evolution (\mycitealt{Vesperini10}). This evolution (known as 'secular evolution') will lead to the gradual evaporation of any star cluster and, eventually, its total dissolution on timescales dependent on the mass of the cluster (\mycitealt{Spitzer87};~\mycitealt{Baumgardt03}). Two-body relaxation causes the velocities of the stars in the cluster to approach a Maxwellian distribution, and stars with velocities above the escape velocity will gradually evaporate from the cluster. Shocks may be due to encounters with spiral arms, \gmcs or, for \gcs on eccentric orbits, passages through the galactic disc or near the bulge. Finally, stellar evolution causes mass loss as stars are turned into much less massive remnants. Over a 10 Gyr time span, about one third of the initial cluster stellar mass is lost this way (e.g.,~\mycitealt{Bruzual03}).

\subsubsection{Cluster Mass and Luminosity Functions}

The basic stellar dynamical mechanisms responsible for the evolution of the \mf are the same for all clusters, although the importance of one mechanism or another (i.e., two-body relaxation vs. shocks) differs. However, the  \acr[s]{}{ICMF}{initial cluster mass function} on which these mechanisms operate might still vary with environment. Nowadays, there are authors who support these dependent mechanisms (e.g.,~\mycitealt{Larsen09}) and others who claim that the physical processes responsible for the formation of the clusters are the same, thus the \icmf is universal (e.g.,~\mycitealt{Whitmore07};~\mycitealt{Fall09}).

\begin{figure}[!tp]

 \hypertarget{fig:mf_lf_intro}{}\hypertarget{autolof:\theautolof}{}\addtocounter{autolof}{1}
 \includegraphics[width=1.\textwidth]{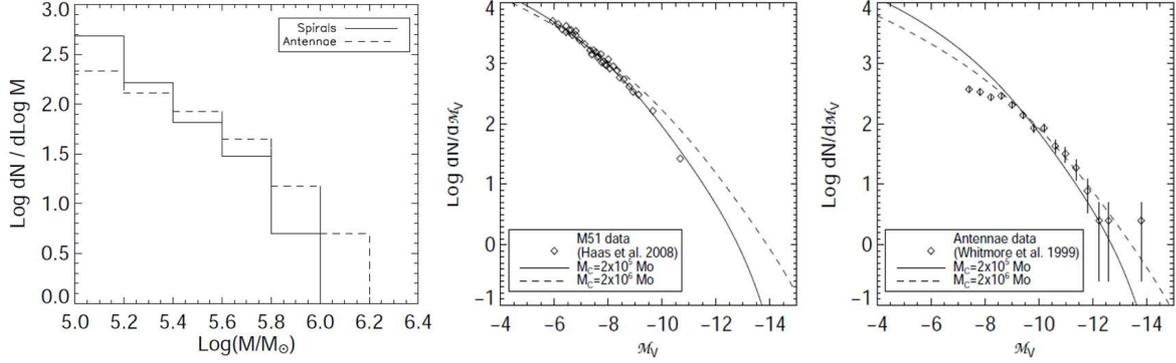}
 \caption[Mass and luminosity functions in The Antennae and M51]{ \textbf{Left:} Mass functions for young ($\tau <$200 Myr) clusters in spiral galaxies (\mycitealt{Larsen09} and The Antennae (\mycitealt{Whitmore99}). \textbf{Center} and \textbf{right:} Luminosity functions of clusters in M51 (center) and The Antennae galaxies (right). Also shown are modeled \lfs assuming Schechter initial cluster \mfs with \mbox{M$_c$ = 2$\times$10$^5$} and 2$\times$10$^6$~\msun (solid and dashed lines, respectively), scaled to match the data. Figure taken from Larsen (\myciteyear{Larsen10}). }
 \label{fig:mf_lf_intro}
\end{figure}

Determinations of the \mf are only available for a few young cluster systems. Generally, they are well-represented by power laws, \mbox{dN/dM $\propto$ M$^{-\alpha}$}, with slopes of \mbox{$\alpha \sim$ 2}, although the mass ranges over which these slopes are derived vary considerably (from \mbox{few $\times$ 10$^3$} to \mbox{few $\times$ 10$^6$~\msun\twospace};~\mycitealt{Gieles06};~\mycitealt{Larsen09}). However, when comparing the \mfs of young massive star clusters (i.e., M $>$ 10$^5$~\msun and $\tau < $200 Myr) in spirals and more active star-forming galaxies (like The Antennae), small differences may show up (see \reffig{mf_lf_intro}, left).  Although a uniform power law with no truncation is consistent with the data, a fit using a Schechter (\myciteyear{Schechter76}) function may be also possible,

\begin{equation}
 \frac{dN}{dM}\propto\left(\frac{M}{M_c}\right)^{-\alpha}exp\left(-\frac{M}{M_c}\right)
\end{equation} 

This function usually fits well the high-mass end of the \mf in old \gcs systems. A fit to the spiral data in \reffig{mf_lf_intro} for a fixed \mbox{$\alpha$ = 2} yields a truncation mass \mbox{M$_c$ = (2.1 $\pm$ 0.4) $\times$ 10$^5$~\msun} (\mycitealt{Larsen09}). A fit to The Antennae data gives \mbox{M$_c$ = (1.7 $\pm$ 0.7) $\times $10$^6$~\msun} (\mycitealt{Jordan07};~\mycitealt{Larsen09};~\mycitealt{Gieles09}), although a uniform power law with no truncation is also consistent with the data (\mycitealt{Whitmore07}). This suggests a lack of high-mass clusters in spirals (i.e., a different \icmf\twospace), compared to the more-active star-forming systems, like The Antennae. Whitmore et al. (\myciteyear{Whitmore07}) attributes the existence of such high mass clusters in The Antennae system to a size-of-sample effect (statistical effect) rather than a difference in the physical process responsible for the formation of the clusters: active mergers have the brightest and most massive young clusters  only because they contain more clusters. 

Once observational selection effects are accounted for, it is relatively straightforward to derive the \acr{}{LF}{luminosity function} of a cluster sample, assuming the distance is known. However, the interpretation of the \lf in terms of the physically more fundamental \mf is complicated by the fact that not all clusters have the same age and therefore mass-to-light ratio. 



If no disruption is present and the \mf is a uniform power law, \mbox{$\psi$(M)dM $\propto$ M$^{-\alpha}$dM}, and then the \lf is a power law with the same slope (\mbox{$\psi$(L)dL $\propto$ L$^{-\beta}$dL}, with \mbox{$\alpha$ = $\beta$}). However, the shape of the \lf can differ from that of the underlying \mf because of disruption and the age-dependent mass-to-light ratio. 

Again, the same opposite views outlined at the beginning of the section are currently being extensively discussed:

\begin{enumerate}[(i)]
 \item For Schechter-like \icmfs (i.e., it depends on the environment), the LF is not expected to be a single power law (\mycitealt{Gieles10}). An example of possible different fits to the \lf using the Schechter function is shown in the central and right panels in \reffig{mf_lf_intro}. The \lfs of clusters in M51 (\mycitealt{Haas08}) and The Antennae system (\mycitealt{Whitmore99}) are compared with modeled \lfs for Schechter \icmfs with different truncation mass, same disruption processes  and assuming a constant \sfr\twospace. It is clear that the M$_c$= 2$\times$10$^5$~\msun modeled \lf matches the M51 data quite well, while the M$_c$= 2$\times$10$^6$~\msun \lf is too shallow at the bright end. For The
Antennae, however, the M$_c$= 2$\times$10$^6$~\msun \lf provides a better fit. The different mass truncation suggests that the \mf in both systems are different, that in M51 being more similar to that in non-interacting spirals. 

\item There is no change in the \icmf because it is universal (\mycitealt{Whitmore07};~\mycitealt{Fall09}). The turnovers observed in the \mfs and \lfs in other studies are understood as selection effects. For instance, Anders et al. (\myciteyear{Anders07}) re-analyses the \wfpc data from Whitmore et al. (\myciteyear{Whitmore99}) on observations of The Antennae and found a truncation of the \lf at \mbox{M$_V \simeq$ -8.5}. However, Whitmore et al. (\myciteyear{Whitmore10}) argue that  the turnover appears to be caused by the introduction of a selection criterion that requires objects to have a \textit{U}-band photometric uncertainty of \mbox{$\sigma_U \leq$ 0.2} mag. They do not observe that truncation on either the \mf or the \lf in the latest survey of clusters in The Antennae down to magnitudes of \mbox{M$_V \simeq$ -6}, which is at the same time consistent with relaxation-driven cluster disruption models that predict the turnover should not be observed until \mbox{M$_V \simeq$ -4} (\mycitealt{Fall01}).
\end{enumerate}


\subsubsection{Star-forming Complexes}

Normally, clusters do not form in isolation but tend to be clustered themselves (\mycitealt{Zhang01};~\mycitealt{Larsen04}). Star-forming complexes represent the largest units of star formation in a galaxy (see the review by~\mycitealt{Efremov95}). They are typically kpc-scale regions encompassing several clusters, probably originated from the same parent giant molecular cloud. 

An interesting question is how the \mf of \gmcs is related to that of the clusters forming within them. The \gmc mass function in the MilkyWay can be approximated by a power law, \mbox{dN $\propto$ dM/M$^{-\alpha}$}, with \mbox{$\alpha$ = 1.6-1.8} (\mycitealt{Williams97}). The \icmf is unlikely to be a simple scaled-down version of the \gmc \mf, since a single \gmc may form more than one cluster (e.g.,,~\mycitealt{Kumar04}). The mass spectrum of clumps within \gmcs seems to be somewhat steeper, with \mbox{$\alpha~\backsimeq$ 2} (\mycitealt{MacLow04}), similar to that measured for individual star clusters.

Star-forming complexes have also been observed in some nearby interacting galaxies, such as in M51 (\mycitealt{Bastian05a}) and in NGC 4038/4039 (\mycitealt{Whitmore99}), with the aim of understanding the hierarchy of the star formation in embedded groupings. These works have confirmed that their properties resemble those of \gmcs from which they are formed. In particular, the mass-radius relation, namely \mbox{M$_{\rm{GMC}} \propto R_{\rm{GMC}}^2$}, reflects the original state of virial equilibrium in these clouds (\mycitealt{Solomon87}) and holds down to the scale of cloud clumps with a few parsecs in radius (\mycitealt{Williams95}). Complexes of young clusters in M51 have a similar correlation, showing the imprint from the parent \gmc (\mycitealt{Bastian05a}). Likewise, \ymcs with masses above 10$^6$~\msun follow the relation (\citealt{Kissler-Patig06}). However, young clusters with lower mass do not (\mycitealt{Larsen04};~\mycitealt{Bastian05b};~\mycitealt{Kissler-Patig06}). A star formation efficiency that depends on the binding energy (predicted by~\mycitealt{Elmegreen97}) of the progenitor \gmc could destroy such a relation during the formation of young clusters. Another possible explanation for this lack of a mass-radius relation in young clusters is that dynamical encounters between young clusters (and gas clouds) add energy into the forming clusters, thereby increasing their radii (\mycitealt{Bastian05a}).

\subsect[tdg_gal]{Tidal Dwarf Galaxies}

As a result of the enhancement of the star formation in galaxy mergers, \sscs are probably not the ultimate bound structures that can be created during the interaction. Condensations of large amounts of gas along the tidal tails and at the tips have been observed for the past few decades. In such condensations objects as massive and as large as the most extreme cases of \sscs and even more massive and larger, within the range of the dwarf galaxy population, can be formed. Some of these structures can be bound and long-lived. They are widely known as \acr[s]{}{TDG}{tidal dwarf galaxies}.

\subsubsection{Discovery \& Characterization of TDGs}

The scenario of low mass (i.e., dwarf) galaxy formation during giant galaxy collisions was first proposed by Zwicky (\myciteyear{Zwicky56}), though without strong observational evidence. During the following two decades the concept that galaxies were static entities was being diverted into an opposite direction. They started to be understood as dynamic, evolving objects  with a high possibility of close encounters that radically
change their morphology (see \refsec{inter_gal}).

The first thorough analysis of an apparent dwarf galaxy built in a tidal tail was presented by Schweizer (\myciteyear{Schweizer78}). He noted the young age of the stars in an area at the tip of the southern tail of The Antennae system and their higher than expected metal content for a location at such large radii. Questions about stable dwarf galaxies forming at this location arose. 

Some years later, Mirabel et al. (\myciteyear{Mirabel91},\myciteyear{Mirabel92}) went on with the investigation of this specific mode of galaxy formation by claiming to have found at least one dwarf galaxy in the long tails of The Superantennae merger (AM 1925-724). Hernquist (\myciteyear{Hernquist92a}) noted the possibility that more dwarf galaxies could be formed during the collisions of larger galaxies. 

The first numerical simulations of the dynamics of interacting galaxies with the aim of studying the formation of \tdgs were carried out by Barnes \& Hernquist (\myciteyear{Barnes92b}) and Elmegreen et al. (\myciteyear{Elmegreen93}), who showed that the formation of dwarf galaxies as a result of a major interaction is possible. Barnes \& Hernquist (\myciteyear{Barnes92b}) also noted that \tdgs are not likely to contain a large amount of dark matter, contrary to normal dwarf galaxies (in particular, dwarf spheroidals),  because their material is drawn from the spiral disk while the dark matter is thought to surround the galaxy in an extended halo.

\begin{figure}[!htp]
\centering
 \hypertarget{fig:ngc7252}{}\hypertarget{autolof:\theautolof}{}\addtocounter{autolof}{1}
 \includegraphics[width=0.9\textwidth]{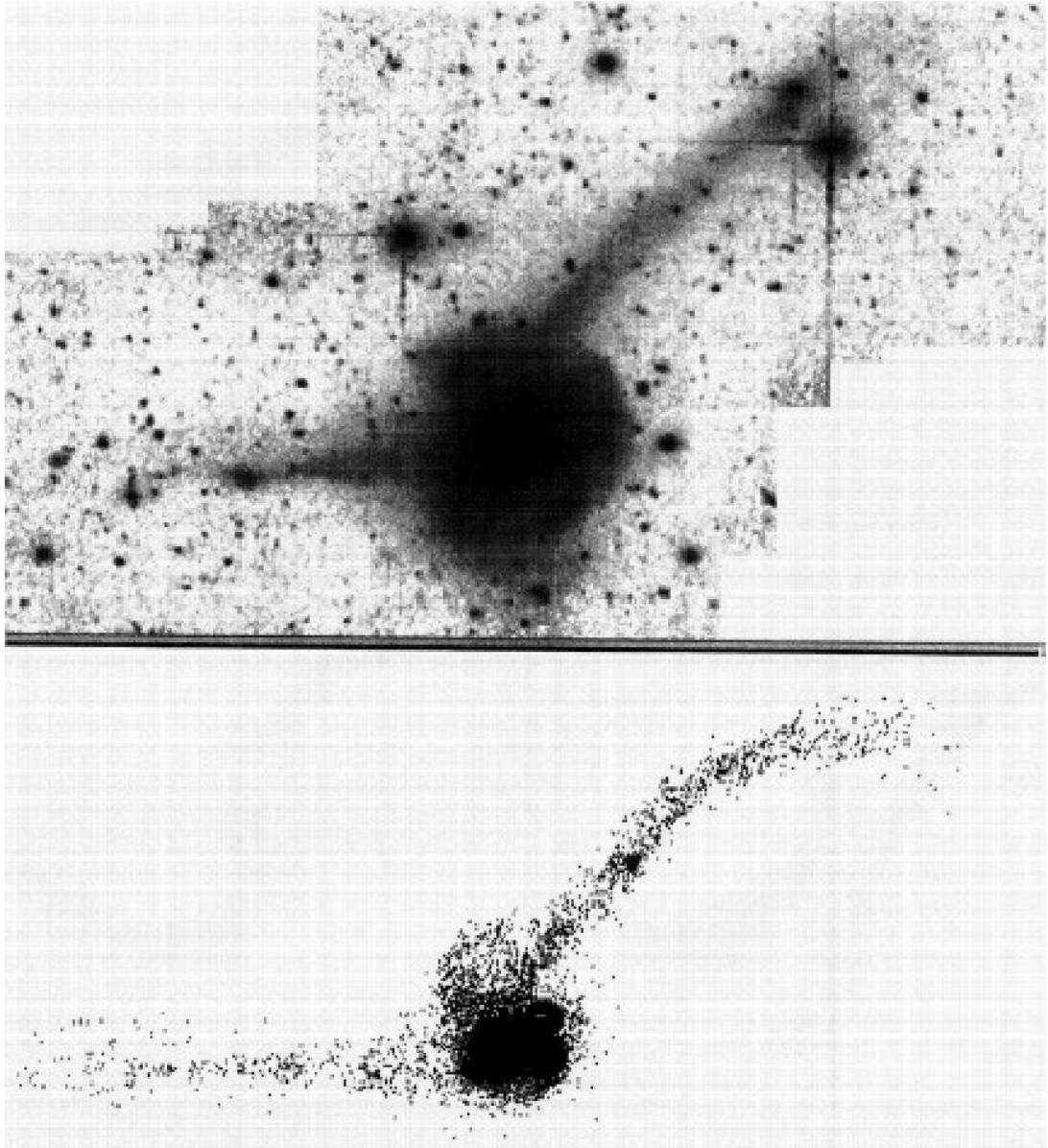}
 \caption[N-body model of NGC 7252]{\textbf{Top:} The CTIO 4m \textit{B}-band image of NGC 7252 from Hibbard et al. (\myciteyear{Hibbard94}). \textbf{Bottom:} The best-fit projection of the N-body model of NGC 7252 at 580 Myr since orbital periapse. A clump that develops in the NW tail is clearly seen. Figure taken from Hibbard \& Mihos (\myciteyear{Hibbard95}).}
 \label{fig:ngc7252}
\end{figure}

Motivated by the theoretical prediction of these models and in order to study the nature of these kind of objects, several campaigns were launched in interacting systems, such as in NGC 2782 (\mycitealt{Yoshida94}), Arp 105 (Duc \& Mirabel~\myciteyear{Duc94};~\mycitealt{Duc97}), NGC 7252 (\mycitealt{Hibbard94}), NGC 5291 (\mycitealt{Duc98};~\mycitealt{Higdon06}), Arp 245 (\mycitealt{Duc00}), small samples (\mycitealt{Weilbacher00},~\myciteyear{Weilbacher03};~\mycitealt{Knierman03};~\mycitealt{Monreal07}) and Arp 305 (\mycitealt{Hancock09}). Hibbard \& Mihos (\myciteyear{Hibbard95}) also developed a successful dynamical N-body model of NGC 7252 (see \reffig{ngc7252}). According to that model, most of the tidal material will remain bound to the central merger, but a significant amount of matter will not fall back within a Hubble time. It is expected that the two \tdg candidates in NGC 7252 --if they remain stable entities-- will attain long-lived orbits around the merger. Some other investigations have tried to find \tdgs in the specific environment of compact galaxy groups (e.g., \mycitealt{Hunsberger96};~\mycitealt{Iglesias-Paramo01};~\mycitealt{Temporin03};~\mycitealt{Nishiura02}).

Based on these observations, \tdg candidates are characterized by a luminosity higher than \mbox{\textit{B} = -10.65}, typically  \mbox{\textit{B} = -13}, and can be as luminous as  \textit{B} = -19. Their color is typically blue (\mbox{\textit{B-I}$=$0.7-2}), indicative of young stellar population. The presence of young population is firmly proved with the detection of strong \ha emission with and equivalent width higher than tens of \AA{}. The half-light radius of these kpc-sized structures ranges from few hundreds of pc to less than 2 kpc, comparable to that in observed dwarf galaxies in different galaxy groups and to the so-called Blue Compact Dwarf galaxies.  

All spectroscopic observations of \tdgs have showed that the luminosity-metallicity correlation found for normal dwarf galaxies does not hold up for \tdgs\twospace. While dwarf galaxies have lower oxygen abundances for lower luminosities, \tdgs have an approximately constant metallicity of around one-third of the solar value as determined from gaseous emission lines (see e.g.,~\mycitealt{Duc00};~\mycitealt{Weilbacher03}). On this basis, Hunter et al. (\myciteyear{Hunter00}) confirmed the metallicity of \tdgs as one of the best criteria to find old dwarf galaxies in merger remnants. Supplementing this criterion with estimates about rotational properties and stellar populations, they also presented a list of nearby dwarf irregulars which might be good candidates for old \tdgs\twospace.

For normal galaxies is easy to know if they are real galaxies and if they will survive long enough to deserve being
called galaxies. Most galaxies are very isolated, without much matter nearby to disturb their stability. Some starburst galaxies may lose much of their gas by strong stellar winds and supernova driven outflows, but in most cases there is no doubt that the stellar component will remain bound and stable, even without gas, for several billion years. The same is true for normal dwarf galaxies in less dense environments.

But is not straightforward the identification of the observed \tdg candidates as real galaxies. They are
embedded in a tidal tail and, most probably, do not have massive dark matter halos (\mycitealt{Barnes92b};~\mycitealt{Braine01};~\mycitealt{Wetzstein07}. Tidal forces of the parent galaxy disturb their gravitational field, strong star formation might blow away the recently accreted gas, and some of the \tdgs may even fall back into the central merger (\mycitealt{Hibbard95}). An accepted definition that tries to ensure that only those objects which, called Tidal Dwarf Galaxies, deserve to be named ``galaxy'' is: A Tidal Dwarf Galaxy is a self-gravitating entity of dwarf-galaxy mass built from the tidal material expelled during interactions  (\mycitealt{Duc00};~\mycitealt{Weilbacher01}). 

Taking into account evaporation and fragmentation processes plus tidal disruption, Duc et al. (\myciteyear{Duc04}) suggested that a total mass as high as 10$^9$~\msun may be necessary for a new formed \tdg to become a long-lived object. This is the typical mass of the giant HI accumulations observed near the tip of several long tidal tails. Less massive condensations may evolve, if they survive, into objects more similar to globular clusters. However this mass criterion is not well established, and less massive objects (a total mass of few 10$^7$-few 10$^9$~\msun\twospace) have been normally considered as \tdg candidates (\mycitealt{Duc07},~\mycitealt{Sheen09},~\mycitealt{Hancock09}).

\subsubsection{Formation Scenario for \tdgs}

From the observational evidence and especially from the theoretical modeling of \tdgs presented in the previous section, the formation sequence of Tidal Dwarf Galaxies can be summarized as follows:

An interaction occurs between two or more galaxies, of which at least one has to be a disk galaxy so that the interaction is able to produce tidal tails. One or more stellar or gaseous condensations appear within the tails, born from gravitational instabilities. If massive enough, the gravitational potential of these condensations will attract the surrounding matter. When a condensation has condensed enough, the neutral gas will be transformed into molecular gas and star formation can start on top of a possible old stellar population. If this proto dwarf galaxy withstands the gravitational forces of the surrounding matter and the supernova explosions in the course of its star formation episode, and does not fall back into the parent galaxy, it will survive as an independent dwarf galaxy as soon as the remaining matter in the tidal features falls back onto the remnant or thins out and disappears. Now, not all self-gravitating structures in tidal tails have the required initial mass to form a genuine galaxy. 

Parallel to this top-down scenario, Kroupa (\myciteyear{Kroupa98}) has suggested a bottom-up scenario of formation of \sscs\twospace, leading to the formation of spheroidal dwarf-type objects, like \tdgs\twospace. Kroupa studied a collisionless simulation of a \ssc without gas. After the formation of a central \ssc within a few tens of Myr, individual clusters continue to merge with the growing central system. Kroupa (\myciteyear{Kroupa98}) estimated between 40 and 60\% the fraction of merged \sscs that survive for 95 Myr and predicted that surviving spheroidal dwarf-type objects should exist and orbit around the main body. If the dwarf galaxy formed in the inner parts of a tidal arm, it could be ejected later on an eccentric orbit with a semi-major axis of tens of kpc. Even if the \ssc model that Kroupa focused on is gas poor (gas and dust are expelled from compact star clusters with radius of 10 pc in only about 0.1 Myr), he raised the possibility that the formation of massive \ssc may also be a mode of star formation in the outer regions of gas-rich tidal tails. If rich groups of globular clusters form in the \ssc\twospace, bound spheroidal dwarf galaxies may remain after gas expulsion. Bastian et al. (\myciteyear{Bastian05a}) provide observational evidence that support the formation of massive \sscs in M51 within this scenario.

\subsubsection{Cosmological Implications of TDG Formation}

The origin and properties of the satellites surrounding massive galaxies has recently been the subject of an active debate, triggered by its cosmological implications. In fact, a complete picture for the origin of dwarf galaxies is still missing. Generally speaking, a dwarf galaxy can be formed from either collapse of primordial gas cloud in the framework of cosmology (i.e., classical dwarf), or materials driven by tidal force away from massive galaxies in interactions and mergers (\tdg\twospace). The classical dwarf galaxies are characterized by small size and dominated by a dark matter halo (\mycitealt{Aaronson83};~\mycitealt{Simon07};~\mycitealt{Geha09}). They are metal poor because of inefficient chemical enrichment in shallow gravitational potential which keeps little metal against supernova winds (\mycitealt{Tremonti04}), and thus sensitive for testing physical mechanisms driving galaxy evolution (e.g. supernova feedback). The cosmologically-originated dwarf galaxies give rise to a well-known challenge to the theory of galaxy formation (i.e., the ``missing satellites problem'';~\mycitealt{Klypin99}) since, although the number of known satellites keeps increasing with time (e.g.,~Belokurov et al.~\myciteyear{Belokurov07}), it is still lower than the number of primordial satellites predicted by standard cosmological hierarchical scenarios. 

The situation could even be worse, as claimed by Bournaud \& Duc (\myciteyear{Bournaud06}), who pointed out that the cosmological models do not take into account the fact that second-generation (or recycled) galaxies may be formed during collisions. As mentioned in previous sections, \tdg galaxies are believed to contain no dark
matter halo and be metal rich. They often have episodic \acr[s]{}{SFHs}{star formation historie} (\mycitealt{Weisz08}) compared to the constant \sfh for classical dwarfs (\mycitealt{Marconi95};~\mycitealt{Tolstoy09}). Understanding the contribution of \tdgs to the local dwarf population is therefore an important issue in dwarf galaxy astrophysics. According to the hierarchical scenario, massive galaxies were gradually assembled through a number of merger events (\mycitealt{Eisenstein05}, and references therein), providing potential space for numerous \tdgs produced over cosmic time. However, \tdgs are exclusively associated with tidal tails driven by rotation-support systems. They are also affected by the tidal friction with their parent galaxies. It is likely that only a small fraction of \tdgs are able to live longer than 10 Gyr, depending on their mass and distance to the parent galaxies.

An early study by Okazaki \& Taniguchi (\myciteyear{Okazaki00}) showed that \tdgs can contribute significantly to the total population of dwarf satellites in addition to primordial dwarfs, even that the overall dwarf population could be of tidal origin. However, it has been recently claimed that only a marginal fraction (less than 10\%) of dwarf galaxies in the local universe could actually be of tidal origin (\mycitealt{Bournaud06};~\mycitealt{Wen11}). The issue is still controversial. Additionally, it is difficult to identify \tdgs once the tidal tail fades away. Only a few works have found \tdg candidates with an important fraction of old population in post-merger galaxies (\mycitealt{Duc07};~\mycitealt{Sheen09}).

\sect[ulirgprop]{(U)LIRGs, Extreme in Star Formation}

As outlined in previous sections, the observation of galactic mergers offers a special opportunity for learning more about the star formation in general and, specifically, the star cluster formation process. Interacting galaxies were common in the early Universe but they also continue today, producing the strongest known starbursts (\mycitealt{Sanders96}). In some systems, the extreme burst of star formation is hidden by large amounts of dust and its energy output re-emitted at longer wavelengths (i.e., the \acr{}{IR}{infrared}\footnote{The \ir spectral range covers from about 0.75 $\mu$m to 350 $\mu$m, and it is normally divided into three regions: the near-\ir (from 0.75 to 5 $\mu$m), the mid-\ir (from 5 to 25 $\mu$m), and the far-\ir (from 25 to 350$\mu$m) (\mycitealt{Glass99}).}). These systems, known as \acr{}{LIRG}{Luminous} and \acr[s]{}{ULIRG}{Ultraluminous Infrared Galaxies}, deserve special consideration in this context.

\subsect[history_ulirgs]{Discovery \& Characterization of (U)LIRGs}

The launch of the \acr{}{$IRAS$}{\textit{InfraRed Astronomical Satellite}} (\mycitealt{Neugebauer84}) caused a revolution for the infrared astronomy, since it increased the number of catalogued sources by about 70\%. It took images in four bands centered at 12, 25, 60 and 100 $\mu$m, and detected many extragalactic sources with an emission in the \ir higher than in any other spectral range.

\begin{table}
\hspace{0.8cm}
\hypertarget{table:definition}{}\hypertarget{autolot:\theautolot}{}\addtocounter{autolot}{1}
\begin{center}
\begin{minipage}{0.85\textwidth}
\renewcommand{\footnoterule}{}  
\caption{Acronyms and definitions used by the IRAS catalog}
\label{table:definition}
\begin{tabular}{ll}
\hline \hline
   \noalign{\smallskip}
$F_{\rm{FIR}}$ & 1.26 $\times$  10$^{-14}$ $\cdot$ (2.58$f_{60}$ + $f_{100}$)[Wm$^{-2}$] \\
$L_{\rm{FIR}}$ & $L$(40-500$\mu$m) = 4$\pi D_{\rm{L}}^{2}CF_{\rm{FIR}}$[\lsun\twospace] \\
$F_{\rm{IR}}$  & 1.8 $\times$ 10$^{-14}$ $\cdot$ (13.48$f_{12}$ + 5.16$f_{25}$ + 2.58$f_{60}$ + $f_{100}$)[W$^{-2}$] \\
$L_{\rm{IR}}$  & $L$(8-1000$\mu$m) = 4$\pi D_{\rm{L}}^{2}F_{\rm{IR}}$[\lsun\twospace] \\
LIRG      & Luminous IR Galaxy, $L_{\rm{IR}} > 10^{11}$ \lsun \\
ULIRG     & Ultra-Luminous IR Galaxy, $L_{\rm{IR}} > 10^{12}$ \lsun \\
HyLIRG    & Hyper-Luminous IR Galaxy, $L_{\rm{IR}} > 10^{13}$ \lsun \\
\hline
\noalign{\smallskip}
\multicolumn{2}{@{} p{\textwidth} @{}}{\textbf{\textsc{Notes.}} The quantities $f_{12}$, $f_{25}$, $f_{60}$, and $f_{100}$ are the IRAS flux densities in Jy at 12, 25, 60 and 100$\mu$m. The scale factor C (typically in the range \mbox{1.4-1.8}) is the correction factor required to account mainly for the extrapolated flux long-ward of the IRAS 100$\mu$m filter. \ld corresponds to the luminosity distance. Table adapted from Sanders \& Mirabel~\myciteyear{Sanders96}.}
\end{tabular}
\end{minipage}
\end{center}
\end{table}

Owing to the discovery of such luminous extragalactic objects in the \ir\twospace, the definition of Luminous and Ultraluminous Infrared Galaxies was made. \acr[s]{notarg}{LIRG}{} and \acr[s]{notarg}{ULIRG}{} are objects with infrared luminosities of \mbox{$10^{11} L_{\odot} \leq L_{\rm{bol}} \sim L_{\rm{IR}} < 10^{12} L_{\odot}$} and \mbox{$10^{12} L_{\odot} \leq L_{\rm{IR}}$}\footnote{For simplicity, we will identify the infrared luminosity as \mbox{\lir(\lsun) $\equiv$ log (\lir\twospace) }.}\mbox{ $< 10^{13} L_{\odot}$}, respectively (\mycitealt{Sanders96}). This classification is based on the definitions in~\reftab{definition}. Although they comprise the dominant population of extragalactic objects at $L_{\rm{bol}}~>$ 10$^{11}$ \lsun and are even twice as numerous as optically detected QSOs with same bolometric luminosity (see~\reffig{lf_class_ir}, left), they are relatively rare in the local Universe (\mycitealt{Soifer87},~\myciteyear{Soifer89}). In fact, their space density is several orders of magnitude
lower than that of the normal galaxies. 

Thanks to observations with \textit{Spitzer} (\mycitealt{Perez-Gonzalez05};~\mycitealt{leFloch05}) it is known that most of the galaxies selected at high-z are (U)\lirgs and they are major contributors to the star formation rate density at \mbox{z $\sim$ 1-2}.


\begin{figure}
 \hypertarget{fig:lf_class_ir}{}\hypertarget{autolof:\theautolof}{}\addtocounter{autolof}{1}
 \includegraphics[width=0.45\textwidth]{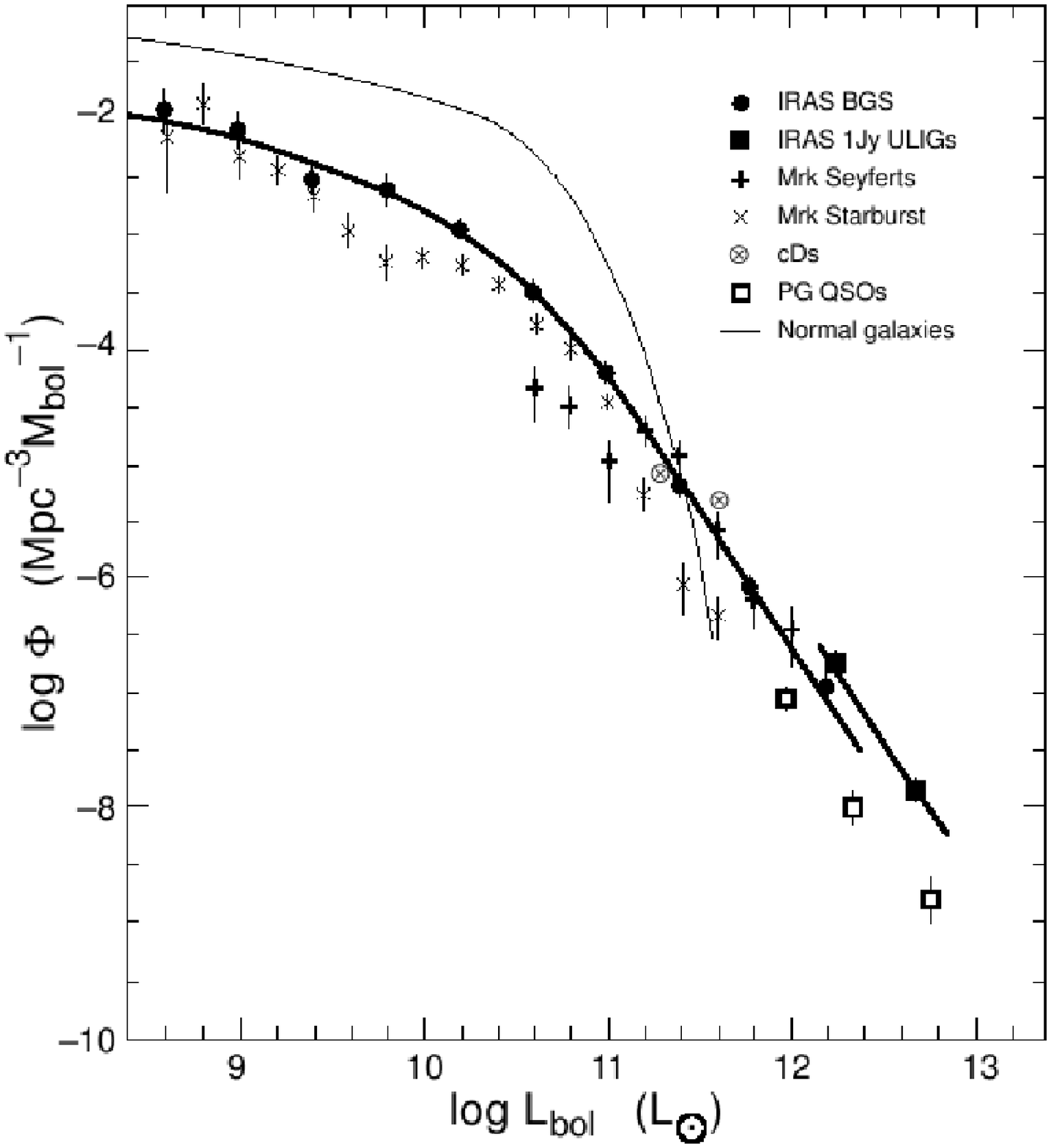}
 \includegraphics[width=0.53\textwidth,viewport=-25 -50 450 450]{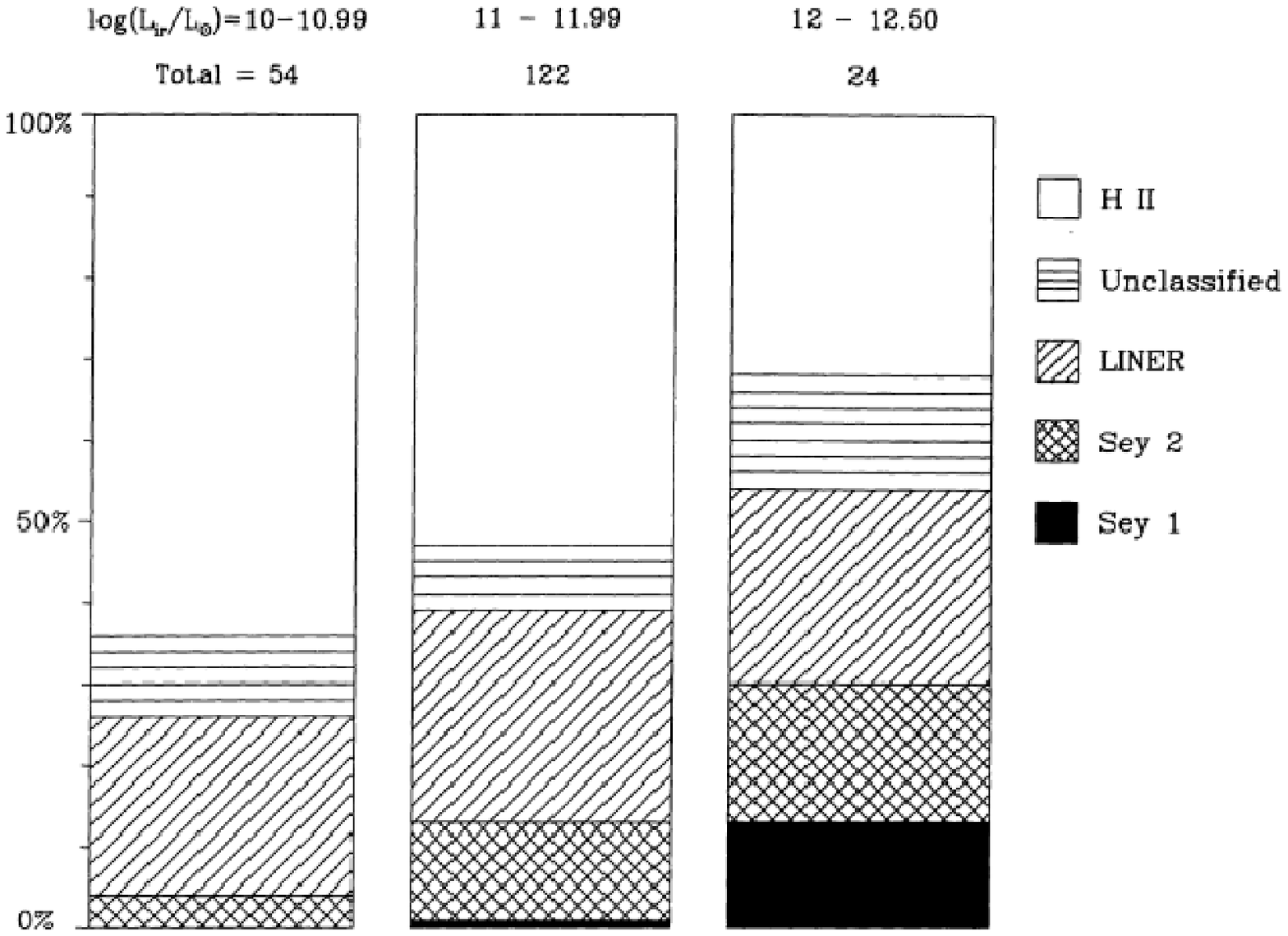}
 \caption[Luminosity function and SED for infrared galaxies]{\textbf{Left:} The luminosity function for infrared galaxies compared with other extragalactic sources. Figure taken from Sanders et al. (\myciteyear{Sanders96}). \textbf{Right:} Optical spectroscopic classification of starburst galaxies and (U)\lirgs as a function of the infrared luminosity. Figure taken from Veilleux et al. (\myciteyear{Veilleux95}). }
 \label{fig:lf_class_ir}
\end{figure}

The only way to explain these high infrared luminosities in (U)\lirgs is the absorption by dust of the available energy (photons) at shorter wavelengths and ulterior re-emission in the infrared. Since the first detections of \ir\twospace-bright galaxies, the nature of the dominant power source in (U)\lirgs has been one of the key questions and much effort has been devoted to resolving it. Today it is widely accepted that their main energy source corresponds to strong starburst activity, possibly triggered by a major merger (\mycitealt{Genzel98b}). However, an \agn may also be present, and even be the dominant energy source in a small percentage of the most luminous systems (\mycitealt{Genzel98a};~\mycitealt{Farrah03}). As the infrared luminosity increases the presence of a heavily obscured \agn is more likely (\mycitealt{kim95};~\mycitealt{Goldader97};~\mycitealt{Yuan10}).

Since (U)\lirgs are objects affected by the presence of large amounts of dust, extinction measurements are usually very high and patchy, up to \mbox{\av $\sim$ 10 mag} or even higher depending on the observation (\mycitealt{Lonsdale06};~\mycitealt{Alonso-Herrero06}).

LINER-like ionization has also been detected in this kind of systems and is thought to come out for the presence of galactic super-winds produced by strong starbursts or galaxy merging processes (\mycitealt{Heckman90};~\mycitealt{McDowell03};~\mycitealt{Colina04};~\mycitealt{Monreal10}).

\begin{figure}[!htp]
\centering
 \hypertarget{fig:morph_lirgs}{}\hypertarget{autolof:\theautolof}{}\addtocounter{autolof}{1}
 \includegraphics[width=0.95\textwidth]{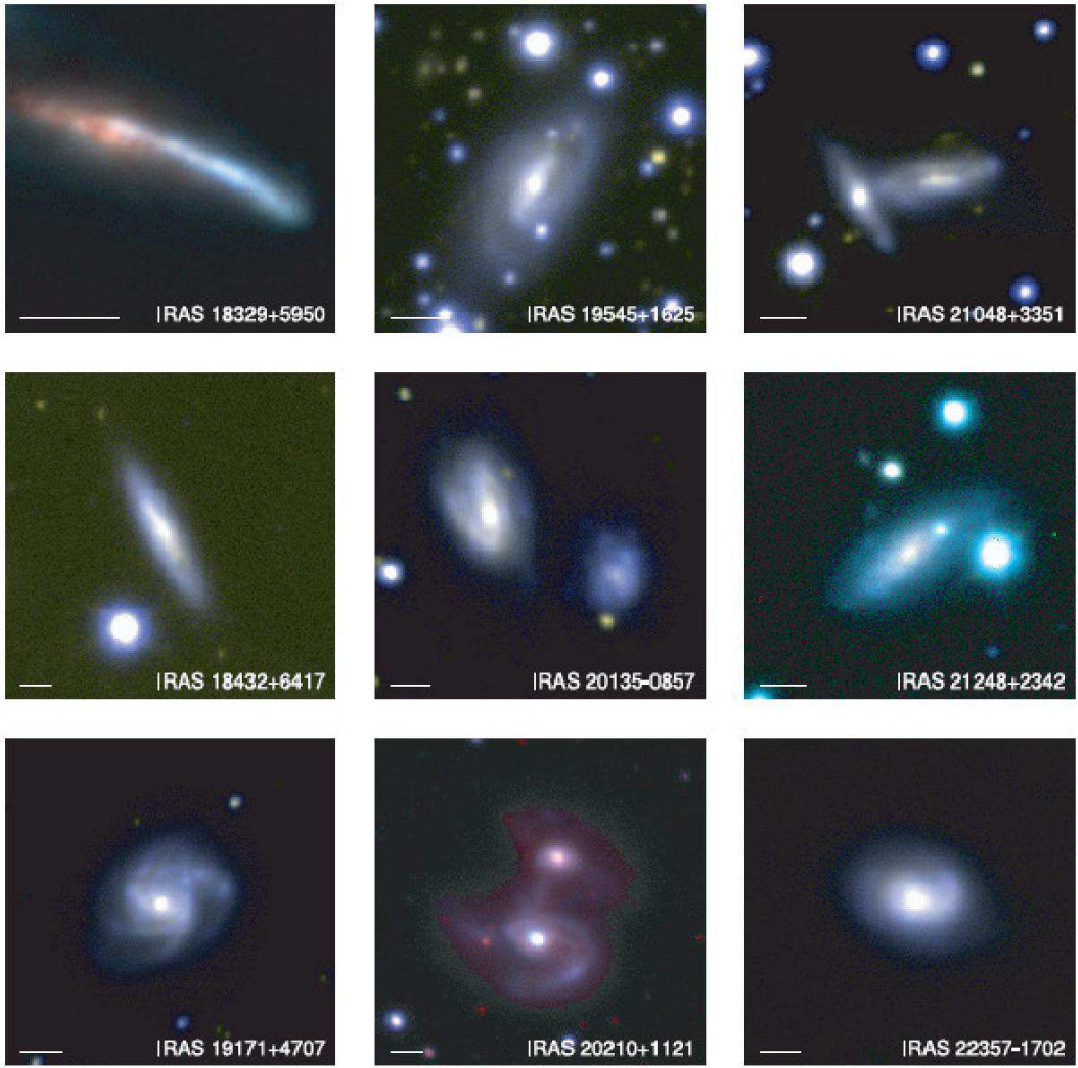}
 \caption[True-color images of some LIRGs taken with the NOT telescope]{True-color images of some local \lirgs taken with the NOT (Nordic Optical Telescope) from a sample of local \lirgs\twospace. Figure taken from Arribas et al. (\myciteyear{Arribas04}).}
 \label{fig:morph_lirgs}
\end{figure}

Most local \lirgs show the typical morphology of a spiral or a system undergoing an early stage of interaction, with the presence of star-forming regions and often tails with dust (\mycitealt{Surace00};~\mycitealt{Scoville00};~\mycitealt{Alonso-Herrero02},~\myciteyear{Alonso-Herrero06},~\myciteyear{Alonso-Herrero09};~\mycitealt{Arribas04};~\mycitealt{Rodriguez-Zaurin10}). Yet, an important fraction (at least 25\%) presents a perturbed morphology or is clearly involved in an interaction process. A few examples of the morphology of these systems are shown in \reffig{morph_lirgs}.

The vast majority of \ulirgs (at least 90\%), on the other hand, are interacting galaxies undergoing different phases of the merging process, according to optical and \acr{}{NIR}{near infrared} studies (\mycitealt{Melnick90};~\mycitealt{Clements96};~\mycitealt{Bushouse02};~\mycitealt{Evans02};~\mycitealt{Veilleux02},~\myciteyear{Veilleux06}). The images in these bands show \ulirgs with a peculiar morphology, tidal tails, bridges, shells and condensations of star formation, typical of systems undergoing a merging process. Not only are most or all the \ulirgs in the different samples studied involved in an interaction process, but their nuclei also show complex structures (\mycitealt{Surace98},~\myciteyear{Surace00};~\mycitealt{Farrah01},~\myciteyear{Farrah03}). Using high angular resolution images taken with the \acr[]{}{$HST$}{\textit{Hubble Space Telescope}}, some authors have suggested that a few \ulirgs are under a multiple merger process (e.g., \mycitealt{Borne00};~\mycitealt{Bushouse02}). A few examples of the morphology of these systems are shown in \reffig{morph_ulirgs}.

\begin{figure}[!t]
\centering
 \hypertarget{fig:morph_ulirgs}{}\hypertarget{autolof:\theautolof}{}\addtocounter{autolof}{1}
 \includegraphics[width=0.91\textwidth]{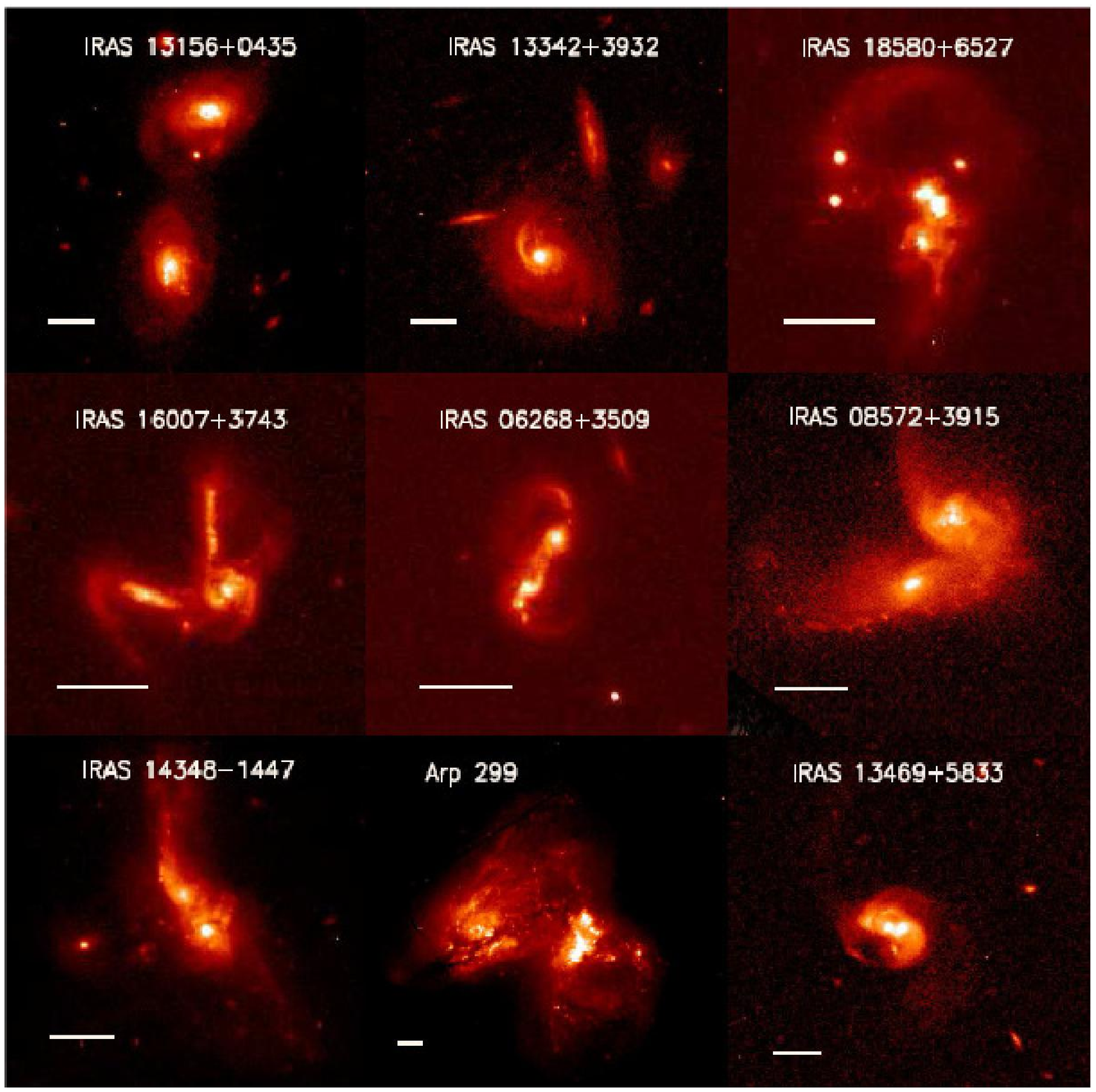}
 \caption[HST (WFPC2/F814W) images of some ULIRGs]{\hst (\wfpc\twospace/\textit{F814W}) images of some \ulirgs\twospace. Figure taken from Garc\'ia-Mar\'in et al. (\myciteyear{Garcia-Marin09a})}
 \label{fig:morph_ulirgs}
\end{figure}

\subsect[sf_dyn]{Star Formation \& Dynamical Processes in (U)LIRGs}

What causes a merger to behave as a (U)\lirg\onespace? Do all merger show at some epoch a \ulirg phase or special conditions are required to trigger the emission in the far-\ir\twospace? According to theoretical simulations, galaxy mergers enhance the star formation (\mycitealt{Mihos94a};~\myciteyear{Mihos96};~\mycitealt{Tissera02};~\mycitealt{Kapferer05};~\mycitealt{Cox06}). These models take into account the stellar and gas content in both galaxies that are interacting. The initial disk stability is the main parameter affecting the star formation sequence: late-type galaxies without bulge are more prone to violent bar instability during an encounter, which drives the internal gas toward the galaxy center to trigger a nuclear starburst (\mycitealt{Mihos96}). However, the availability of gas in the interacting galaxies is also one of the more constraining parameters, as well as the gas physics adopted (\mycitealt{Cox04}). 

\begin{figure}[!tp]
\centering
 \hypertarget{fig:sfh_sim}{}\hypertarget{autolof:\theautolof}{}\addtocounter{autolof}{1}
 \includegraphics[width=1\textwidth]{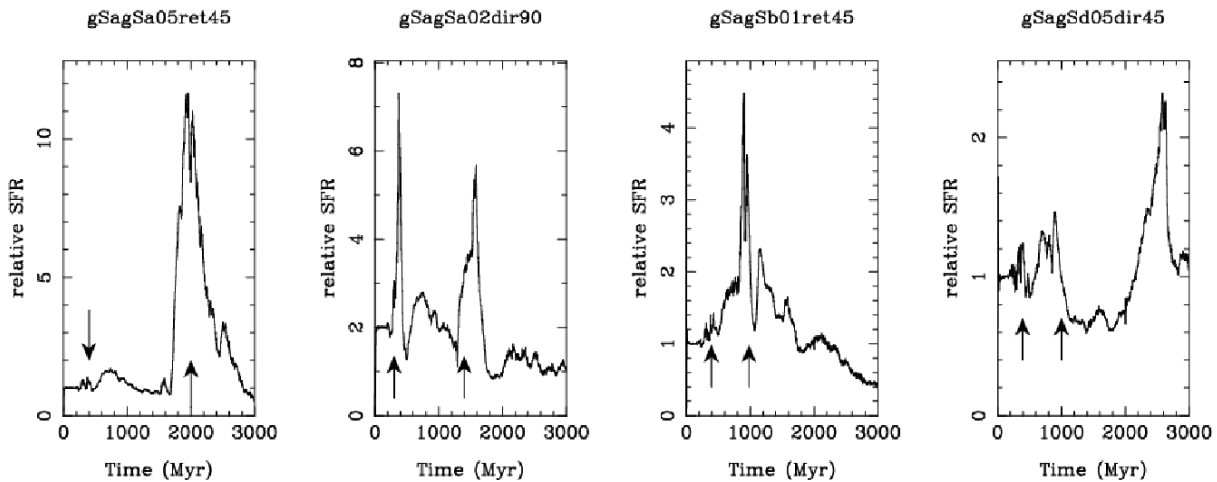}
 \caption[Star formation rate versus time for some galaxy mergers]{Star formation rate evolution with time during major mergers involving galaxies of different morphologies (gSa, gSb and gsd for giant-like Sa, Sb and Sd spirals respectively) and different orbital parameters: 05ret, 02dir, 01ret and 05dir correspond to different initial orbitals parameters, ``ret'' being retrograde and ``dir'' prograde ; the numbers 45 and 90 indicate that the second galaxy is inclined with respect to the orbital plane by this amount. The \sfr is normalized to that of the corresponding isolated galaxies. In each panel, the black arrows indicate, respectively, the first pericenter passage between the two galaxies and the time of nuclear coalescence. Figure taken from di Matteo et al. (\myciteyear{diMatteo08}).}
 \label{fig:sfh_sim}
\end{figure}

The observed massive concentrations of gas in some systems in interaction can be reproduced (\mycitealt{Barnes92a},~\myciteyear{Barnes96}). However, for the last decade there has been an intense debate on the star formation efficiencies, and hence \sfrs\twospace, predicted by the numerical simulations of galaxy encounters. There are models able to reproduce the observed \sfrs in many (U)\lirgs, though they claim that only interactions between two gas-rich spirals are able to reach such high luminosities to explain the emission in \ulirgs, and during short periods in time,~$\sim$50-150 Myr (\mycitealt{Mihos96};~\mycitealt{Bekki01},~\myciteyear{Bekki06}). Recent simulations show that a significant enhancement of the \sfr during a major merger at low redshifts (given by the linear resolution achieved and the initial gas content of the parent galaxies) is quite rare (\mycitealt{Bournaud06};~\mycitealt{Cox08};~\mycitealt{diMatteo08}). It seems to be highly dependent on the mass ratio of the galaxies involved in the interaction, an equal-mass ratio of two gas-rich spirals being the most favored. The orbital geometry represent a secondary factor, though the enhancement is more efficient for galaxies in retrograde orbits. Some examples of the \sfh of simulations of mergers are shown in \reffig{sfh_sim}.

There is both theoretical (\mycitealt{Toomre72};~\mycitealt{Hernquist92b};~\myciteyear{Hernquist93}) and observational (\mycitealt{Stanford91};~\mycitealt{Genzel01};~\mycitealt{Colina01};~\mycitealt{Tacconi02};~\mycitealt{Dasyra06}) evidence that in (U)\lirgs the end product of the encounter between two disk galaxies can be an early-type galaxy with a moderate mass of 0.1-1 m$_\ast$ (m$_\ast$ = 1.4 $\times$ 10$^{11}$~\msun\twospace) and a de Vaucoulers surface brightness profile (r$^{1/4}$), although this view was initially refused for two reasons:


\begin{itemize}
 \item How do the observed \gcs in elliptical form? The discovery of \acr[s]{}{SSC}{super star clusters} (see \refsect{ssclusters}) hidden by dust in interacting systems was proposed by some authors as precursors of \gcs (\mycitealt{Whitmore95};~\mycitealt{Kassin03}). 
 \item As ellipticals barely contain gas, how can such large amounts of gas involved in the merging process disappear? Galactic superwinds and supernovae explosions are a likely explanation. Additionally, the jets coming from the \agns also expel significant amount of gas toward the extra-galactic medium.
\end{itemize}

Another aspect of the dynamical evolution of \ulirgs which is worth mentioning is the presence of compact star-forming regions at all galactocentric scales, especially within the tidal tails, in which knots or condensations of star formation have been observed (e.g.,~\mycitealt{Surace98}). The next section will present in more detail the compact star-forming structures that have been observed in (U)\lirgs (i.e., \sscs and \tdg candidates).

\subsect[structures_ulirgs]{Young Compact Star-forming Structures in (U)LIRGs}

As already mentioned in previous sections, the high infrared luminosity in (U)\lirgs is believed to be mainly caused by the re-emission of hidden starburst activity, which is triggered by a merger process in a significant percentage of \lirgs and practically in all U\lirgs\twospace. It is not surprising that the studies carried out on these systems have found young compact star-forming regions at all galactocentric distances. Copious studies on The Antennae --on the border line between the non-\lirg and the \lirg class, with a luminosity of \lir\twospace=10.99-- have been carried out (e.g.,~\mycitealt{Whitmore95},~\myciteyear{Whitmore99},~\myciteyear{Whitmore05},~\myciteyear{Whitmore10};~\mycitealt{Fall05};~\mycitealt{Bastian06};~\mycitealt{Mengel08}). Given the proximity of the system and the use of high angular resolution images from the \hst\twospace, these works have reported the properties of thousands of clusters with typical half-light radii of a few pc, the most luminous ($M_V > -14.6$) being young (typically $\tau \lesssim $ 100 Myr) and spanning the mass range 10$^5$-$10^6$~\msun\twospace, typical properties of \sscs\twospace.

For more luminous systems only a few studies of compact, massive (i.e., 10$^5$-10$^7$~\msun\twospace) and extremely young (i.e., $\tau < 10$ Myr) circumnuclear star clusters have been reported (\mycitealt{Scoville00};~\mycitealt{Alonso-Herrero02};~\mycitealt{Diaz-Santos07}), using \nir images. Moreover, compact condensations compatible with massive young star formation in the outermost regions have been detected with near-\uv and optical images (\mycitealt{Surace98};~\mycitealt{Surace00}). The systems sampled in these works are located at such distances (between 40 to hundreds of Mpc) that the observation of intrinsically faint clusters was not possible. Additionally, blending can occur and thus, although several systems were sampled in each study, only a total of dozens or few hundreds of cluster-like objects were detected. 

With half-light radii between about ten to few hundreds of pc, most of the cluster-like ojects detected in (U)\lirgs may actually be complexes of star clusters. However, the size of the largest objects is similar to that of a typical dwarf galaxy. Given the interacting nature of (U)\lirgs (especially \ulirgs\twospace) there have also been searches of \tdg candidates there. Despite the fact that a priori they have the appropriate nature to harbour these dwarf galaxies, only few studies have been performed so far. The \tdg candidate in The Superantennae (\mycitealt{Mirabel91}) turned out to be a background object (\mycitealt{Weilbacher03}). Mihos \& Bothun (\myciteyear{Mihos98}) identified only three more candidates in a sample of one \lirg and three \ulirgs\twospace. More recently, Monreal-Ibero et al. (\myciteyear{Monreal07}) have performed a systematic search of \tdgs in a sample of nine \ulirgs\twospace. A total of twelve candidates were identified and characterized. They share similar properties to observed \tdg candidates in other samples, such as high metallicty (i.e., $Z_ \odot$ or $Z_ \odot$/3), effective radii of a few hundreds of pc and total masses higher than \mbox{10$^8$~\msun\twospace}. That work also studied the stability of their candidates against internal motions and forces from the parent galaxy and found that five out of the twelve show high- medium or high likelihood of survival as future \tdg\twospace.

\subsect[ulirgs_highz_intro]{(U)LIRGs in the High-z Universe}

With the advent of satellites and ground-based instruments working on the \ir\twospace, millimeter and radio wavelenghts, many high-z luminous infrared galaxies have been discovered. Although these galaxies do not dominate the luminosity function in the local Universe, they become more important as we observe them farther. Surveys carried on with the \acr{}{$ISO$}{InfraRed Space Observatory} showed that the evolution of the luminosity density in these galaxies is very rapid until \mbox{z = 1}, and no much farther increase is left for the higher redshifts (\mycitealt{Franceschini01}). 

\begin{figure}[!tp]
\centering
 \hypertarget{fig:sfr_density}{}\hypertarget{autolof:\theautolof}{}\addtocounter{autolof}{1}
 \includegraphics[width=0.9\textwidth]{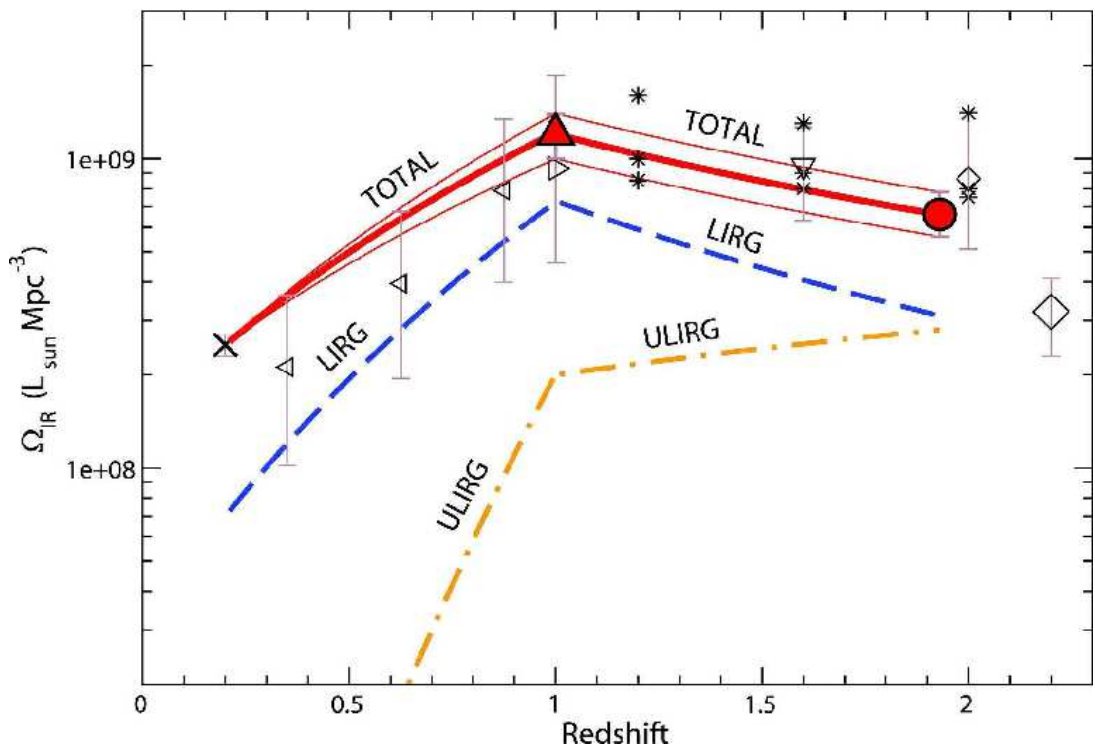}
 \caption[Evolution of the co-moving bolometric IR luminosity density with z]{Evolution of the co-moving IR luminosity density with redshift (red solid line). The blue dashed and the orange dotted-dashed lines show the contributions of \lirgs and \ulirgs, respectively, at different redshifts. Figure taken from Caputi et al. (\myciteyear{Caputi07}).}
 \label{fig:sfr_density}
\end{figure}

The \textit{Spitzer} telescope, working in the 3-180 $\mu$m range, allowed to survey the (U)\lirg population up to redshift \mbox{z $\sim$ 3} (\mycitealt{leFloch05};~\mycitealt{Franceschini05};~\mycitealt{Perez-Gonzalez05};~\mycitealt{Caputi07}). From these studies we know that at \mbox{z $>$ 1} \lirgs are the major contributors of the co-moving star formation rate density of the Universe, and at \mbox{z $\simeq$ 2}, (U)\lirgs have similar contributions (see~\reffig{sfr_density}), and they account for at least half of the newly born stars by \mbox{z $\sim$ 1.5}.

It is also believed that at these redshifts the main source of energy in (U)\lirgs corresponds to star formation activity. Studies carried out with \textit{Spitzer} data confirm the importance of the \acr[s]{}{PAH}{polycyclic aromatic hydrocarbons} up to \mbox{z $\sim$ 2.5} (\mycitealt{Yan05}), and works on the millimeter and radio wavelenghts have estimated that only about 25\% of the (U)\lirg population harbour an \agn (\mycitealt{Ivison04};~\mycitealt{Egami04}).

Sources with \sfrs up to ten times higher than in \ulirgs\twospace, known as \acr[s]{}{SMG}{sub-millimeter galaxies}, were detected with the sub-millimeter instrument SCUBA, operating at 450 and 850 $\mu$m (\mycitealt{Smail97}). Given their peculiar morphologies, high luminosities and energy spectral distributions, it is very likely that they are the counterparts of (U)\lirgs at high redshifts  (\mbox{z $\gtrsim$ 2};~\mycitealt{Smail97};~\mycitealt{Blain02};~\mycitealt{Frayer04}). Follow-up multi-wavelength observations of \smgs showed that some of these galaxies exhibit multiple components, a clear indication of ongoing interaction processes (\mycitealt{Ivison02};~\mycitealt{Smail02}). 

(U)\lirgs can also become relevant in the context of clumpy galaxies at high redshifts. Star-forming galaxies become increasingly irregular at higher redshifts, with a blue clumpy structure, asymmetry and lack of central concentration (\mycitealt{Abraham96};~\mycitealt{Vandenbergh96};~\mycitealt{Im99}). It has been claimed that clumps in clumpy galaxies represent star-forming complexes intrinsically more massive by one or two orders of magnitudes than similar-size complexes in local galaxies (\mycitealt{Efremov95};~\mycitealt{Elmegreen09};~\mycitealt{Gallagher10}). Although most of the clumpy galaxies, observed at \mbox{z = 0.7-2}, do not show signs of interactions, they might share the extreme star formation properties with local (U)\lirgs.

Since all these galaxies at high redshift seem to be equivalent to the (U)\lirgs observed in the local Universe, in order to understand them and the role they have played on the star and galaxy formation in the early Universe it is important to know in detail the physical properties and the dynamical processes that take place in local (U)\lirgs.

\sect[thesis_outline]{Thesis Project}

In general, the investigation of compact star-forming objects in galaxies can be done using two different approaches: (i) surveys of a significative number of galaxies, useful to study the global integrated properties and to make statistical studies; (ii) and the analysis of a particular galaxy to study in detail the star formation and disruption processes occurring there. For the past few decades much effort has been applied to study how the interaction of galaxies affects the star and galaxy formation, using the state-of-the-art instrumentation (e.g., \hst studies in M51, NGC 4038/4039 or NGC 7252; \acr{}{IFU}{integral field unit} spectroscopic observations of \tdg candidates).  Numerical modeling of galaxy encounters, with a continuous improvement of the spatial resolution and the enlargement of the number of simulations that allows us to explore each time more realistic scenarios, has also become an important complementation to observations to advance in the field. All these theoretical and observational studies have provided significant results and observational restrictions to the physical processes occurring in galaxy mergers (e.g., the role of the infant mortality, physical constrains of the galaxies under interaction for the star formation to be enhanced more effectively, etc.). 

But, is the star formation in (U)\lirgs ruled by the same processes as in less luminous galaxy interactions? Can a different mode of star formation, in which more massive objects can be formed as a result of the interaction, take place there? Does the star formation evolve as the interaction proceeds? Are the formation and disruption processes universal? These are some questions that remain unanswered because, apart from very few and limited studies, the most energetic systems in the local Universe have remained barely explored in this field.  This is in part because (U)\lirgs normally are located at such distances (i.e., normally at \acr[s]{}{D$_L$}{luminosity distances} farther than 60 Mpc and there is only one \ulirg closer than 100 Mpc) that the detection of large number of individual compact star-forming regions is not possible with the best instrumentation available today. The studies with the largest samples correspond to those of Surace et al. (\myciteyear{Surace98},~\myciteyear{Surace00}), that encompass less than a dozen of \ulirgs each and detect a few hundred compact regions. 

This thesis presents the study of compact star-forming regions in the most luminous low-z (U)\lirgs\twospace. With a sample of 32 systems, larger than in previous works, it increases considerably the statistics with the detection of a few thousand compact star-forming regions. The galaxies in the sample represent different interaction phases (first-contact, pre-merger, merger and post-merger) and cover a wide luminosity range (\mbox{11.46 $\leq$ \lir $\leq$ 12.54}). The project is based mainly on optical high resolution images taken with the \acs on board the \hst telescope, data from a high linear resolution simulation of a major galaxy encounter, and with the combination of optical \acr{}{IFS}{Integral Field Spectroscopy} from the INTEGRAL (WHT) and VIMOS (VLT) instruments. The main topics covered in this project are: (i) the identificantion and empirical characterization of the compact star-forming regions identified in our sample of low-z (U)\lirgs\twospace; (ii) the comparison of their properties with those of compact objects identified in a high linear resolution of a galaxy encounter; (iii) the identification and characterization of \ha-emitting complexes, and the study if their nature is compatible with being progenitors to \tdgs\twospace. The sample selection, the different sets of data analyzed and the general analysis techniques are presented in~\refcha{data_tech}. Then, this thesis is divided into the three main topics, each one covered in one Chapter:

\begin{itemize}
 \item \textbf{Characterization of compact star-forming regions}\\
 In \refcha{knots} we present an empirical characterization of optically-selected compact star-forming regions (knots) in a representative sample of 32 low-z (\mbox{i.e., z $<$ 0.1}) (U)\lirgs\twospace. It shows the first attempt at obtaining an homogeneous and statistically significant study of the photometric properties (luminosities, colors and sizes) of optically-selected knots of star formation found in these systems as a function of the infrared luminosity, morphology (i.e., interaction phase) and radial distance to the nucleus of the galaxy. The luminosity function is also evaluated and compared with that obtained in other studies. Finally, their properties are compared to clumpy structures identified in high-z star-forming galaxies.
 
 \item \textbf{Comparison with a high linear resolution simulation of a major merger} \\
 The previous chapter represents the starting point in \refcha{sim}, where the first direct comparison of the properties of stellar regions (simulated knots) identified in the high spatial resolution galaxy simulation of Bournaud et al. (\myciteyear{Bournaud08a}) with the knots detected in our sample is performed. Using stellar population synthesis techniques, the luminosities, colors and \lfs of the simulated knots are evaluated at different interaction phases. A direct comparison with the properties of the observed knots makes possible a physical interpretation of the results from observations. How close are the predictions of the simulation from real encounters? How can we understand some of the properties of the observed knots? These are some questions we investigate throughout the chapter.
 
 \item \textbf{Search and characterization of \tdg candidates}\\
 The third part of the thesis is presented in \refcha{tdgs}. We characterize the extranuclear star-forming
 regions of a sample of (U)\lirgs using the combined information of Integral Field Spectroscopy (IFS) data together with high resolution images from the \hst\twospace. This chapter is devoted to the search of potential \tdg candidates in local (U)\lirgs\twospace. Properties such as the metallicity, extinction, age and mass of the young stellar population, \ha luminosity, dynamical mass, velocity dispersion, relative velocity etc., are derived and compared with those of \tdg candidates in the literature.  We use these parameters to estimate the likelihood of survival of these regions as future \tdgs\twospace. We also estimate the implications of \tdg formation to the overall dwarf population and primordial dwarf formation at high-z.

\end{itemize}

\def\mychapname{Chap. } 
\cha{data_tech}{Sample Selection, Dataset and Data Treatment}
\chaphead{This Chapter presents the sample upon which this thesis is based. The selection criteria and the main morphological properties of the galaxies are described. The different sets of data presented and the general analysis techniques applied to perform this study are also discussed in detail.}

\sect[sample_prop]{Sample Selection and General Properties}

\subsect[sample_ini]{The Sample}

The galaxies for this thesis were selected from the flux-limited \iras Revisited Galaxy Sample (RBGS,~\mycitealt{Sanders03}) and from the sample of Sanders et al. (\myciteyear{Sanders88b}), using the following additional criteria: 

\begin{itemize}
 \item To sample the wide range of \ir luminosities in (U)\lirgs\twospace.
 \item To cover all types of nuclear activity (i.e., different ionization mechanisms) to avoid biases.
 \item To span different phases of interaction, from first approach to final merger phases. 
 \item To optimize the linear scales by selecting low-z galaxies, which would allow us to have a relatively high linear resolution and to detect intrinsically faint compact regions.
\end{itemize}

For all the selected systems, archival high angular resolution \hst blue and red optical images are available. The \textit{I}-band, less affected by extinction, can indicate the extent of the system and the tidal features emerging from it, whereas the \textit{B}-band can give us and idea of the young population. Two of the systems were not considered for this study, since they are so close (\mbox{\ld $\lesssim$ 40 Mpc}) that individual stars belonging to them start to contaminate, and selection effects can be encountered for these galaxies. For instance, the faintest objects detected in IRAS 10257-4338 by Knierman et al. (\myciteyear{Knierman03}) with the \wfpc can be individual stars (though they expect them to be few). Since the \acs images are more sensitive at similar exposure times, more individual stars can be detected. In addition, the galaxies in the sample were observed or planned to be observed with \ifs (VIMOS, INTEGRAL and PMAS).

\begin{figure}[!t]
\hspace{2cm}
 \hypertarget{fig:LirVsz}{}\hypertarget{autolof:\theautolof}{}\addtocounter{autolof}{1}
\includegraphics[angle=90,width=0.80\textwidth]{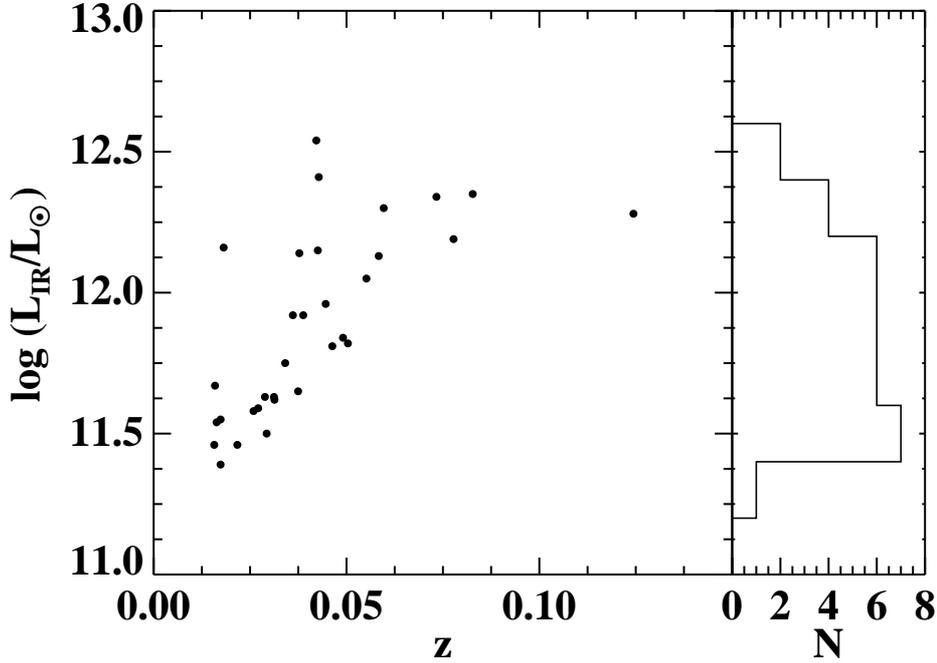}
      \caption[Distribution of the systems in the luminosity-redshift plane]{Distribution of the 32 systems of the sample in the luminosity-redshift plane. The sample consists of low-z (\mbox{z $\lesssim$ 0.1}) systems and covers a factor higher than 10 in infrared luminosity.}     
      \label{fig:LirVsz}
\end{figure}

Our sample comprises 32 low-z systems, 20 \lirgs and 12 \ulirgs\twospace. Individual properties of the systems are presented in \reftab{sample}. The sample is certainly not complete in either volume, flux or luminosity. However, it is essentially representative of the local (U)\lirg population (see also \reffig{LirVsz}): (i) it covers the luminosity range \mbox{11.39 $\leq$ \lir}\footnote{As defined in the previous chapter, for simplicity we adopt \lir to be \mbox{\lir(\lsun) $\equiv$ log ($L_{\rm{IR}}$)}}\mbox{ $\leq $ 12.54} and the redshift range \mbox{0.016 $\leq$ z $\leq$ 0.124} (from  65 to about 550 Mpc), with a median value of 0.037; (ii) it spans all types of nuclear activity, with different excitation mechanisms such as LINER (i.e., shocks, strong winds, weak \agn), \hii (star formation) and Seyfert-like (presence of an \agn); (iii) and the different morphologies usually identified in these systems are also sampled (see \refsec{morphologies} and \reffig{composite34}).  Therefore, the sample is appropriate for the purpose of the present work.

\begin{table}
\hypertarget{table:sample}{}\hypertarget{autolot:\theautolot}{}\addtocounter{autolot}{1}
\begin{minipage}{\textwidth}
\renewcommand{\footnoterule}{}  
\begin{scriptsize}
\caption[Sample of (U)LIRGs]{Sample of (U)\lirgs}
\label{table:sample}
\begin{center}
\begin{tabular}{l@{\hspace{0.22cm}}c@{\hspace{0.22cm}}c@{\hspace{0.22cm}}c@{\hspace{0.22cm}}c@{\hspace{0.22cm}}c@{\hspace{0.22cm}}c@{\hspace{0.22cm}}c@{\hspace{0.22cm}}c@{}c}
\hline \hline
   \noalign{\smallskip}
   
IRAS 	&	Other 	&	R. A.	&	DEC 	&	$D_{L}$	&	$L_{\rm{IR}}$	&	Morphology	&	Interaction	&	Spectral Class	&	Ref	\\
name &	name&	 (J2000)	&	(J2000)	&	(Mpc)	&		&		&	phase	&	(Optical)	&		\\
(1)	&	(2)	&	(3)	&	(4)	&	(5)	&	(6)	&	(7)	&	(8)	&	(9)	&	(10)	\\

\hline
   \noalign{\smallskip}
00506+7248	&		&	00:54:04.91 	&	+73:05:05.5 	&	65.3	&	11.46	&	IP	&	III	&	\hiiminus	&	[1]	\\
02512+1446	&		&	02:54:01.99 	&	+14:58:26.2 	&	131.3	&	11.63	&	IP	&	III	&	\hiiminus	&	[2]	\\
04315-0840	&	NGC 1614 	&	04:33:59.99 	&	-08:34:42.8 	&	66.1	&	11.67	&	SN	&	IV (2)	&	\hiiminus	&	[3]	\\
05189-2524	&		&	05:21:01.42 	&	-25:21:48.3	&	180.6	&	12.15	&	SN	&	V (2)	&	Sy2	&	[4],[5]	\\
06076-2139	&		&	06:09:45.98 	&	-21:40:25.6 	&	158.3	&	11.65	&	IP	&	III (1)	&	Sy/L-\hiiminus	&	[6]	\\
06259-4708	&		&	06:27:22.53 	&	-47:10:45.4 	&	164.2	&	11.92	&	MI	&	I-II (1)	&	\hiiminus\twospace-\hiiminus	&	[7]	\\
07027-6011	&		&	07:03:27.46 	&	-60:16:00.2 	&	131.8	&	11.62	&	IP	&	I-II (0)	&	Sy2	&	[7]	\\
08355-4944	&	NGC 2623 	&	08:37:01.98 	&	-49:54:27.5 	&	108.5	&	11.58	&	SN	&	IV (2)	&	\hiiminus	&	[8]	\\
08520-6850	&		&	08:52:28.86 	&	-69:02:00.3 	&	197.1	&	11.81	&	IP	&	I-II (1)	&	-	&	-	\\
08572+3915	&		&	09:00:25.62 	&	+39:03:54.1 	&	250.6	&	12.13	&	IP	&	III	&	L/\hiiminus	&	 [9], [10]	\\
09022-3615	&		&	09:04:12.26 	&	-36:26:58.1 	&	256.4	&	12.30	&	SN	&	IV (2)	&	-	&	-	\\
F10038-3338	&		&	10:06:05.04 	&	-33:53:14.8 	&	143.8	&	11.75	&	DN	&	IV (2)	&	-	&	-	\\
10173+0828	&		&	10:20:00.10 	&	+08:13:33.3 	&	209.4	&	11.84	&	SN	&	V	&	-	&	-	\\
12112+0305	&		&	12:13:45.82 	&	+02:48:38.0 	&	318.3	&	12.34	&	IP	&	III	&	L-L	&	[4], [5], [9]	\\
12116-5615	&		&	12:14:21.77 	&	-56:32:27.6 	&	113.7	&	11.59	&	SN	&	V (2)	&	\hiiminus	&	[7]	\\
12540+5708	&	Mrk 231	&	12:56:13.73 	&	+56:52:29.6 	&	178.9	&	12.54	&	SN	&	V (2)	&	Sy1	&	[2], [5]	\\
13001-2339	&		&	13:02:52.49 	&	-23:55:19.6 	&	90.7	&	11.46	&	DN	&	IV (2)	&	L	&	[11]	\\
13428+5608	&	Mrk 273	&	13:44:41.93 	&	+55:53:12.3 	&	159.8	&	12.14	&	DN	&	IV	&	L-Sy2	&	[2]	\\
13536+1836	&	Mrk 463	&	13:56:02.80	&	18:22:17.20	&	215.0	&	11.82	&	DN	&	III	&	Sy1-Sy2	&	[12]	\\
14348-1447	&		&	14:37:40.36 	&	-15:00:29.4 	&	361.6	&	12.35	&	DN	&	III	&	L	&	 [5], [9]	\\
15206+3342	&		&	15:22:38.00	&	+33:31:35.9	&	559.6	&	12.28	&	SN	&	V	&	\hiiminus	&	[5], [13]	\\
15250+3609	&		&	15:26:59.05 	&	+35:58:38.0	&	236.3	&	12.05	&	SN	&	IV	&	L	&	 [5]	\\
15327+2340	&	Arp 220	&	15:34:54.63 	&	+23:29:40.5 	&	75.5	&	12.16	&	DN	&	IV 	&	L	&	[13]	\\
16104+5235	&	NGC 6090	&	16:11:40.10 	&	+52:27:21.5 	&	123.1	&	11.50	&	IP	&	III	&	\hiiminus	&	[2]	\\
F17138-1017	&		&	17:16:35.64 	&	-10:20:34.9 	&	72.2	&	11.39	&	SN	&	V (2)	&	\hiiminus	&	[14]	\\
17208-0014	&		&	17:23:19.14 	&	-00:17:22.5 	&	181.7	&	12.41	&	SN	&	V	&	L	&	[5], [15]	\\
F18093-5744	&	IC 4686	&	18:13:38.66 	&	-57:43:53.7 	&	72.2	&	11.55	&	MI	&	III (1)	&	\hiiminus\twospace-\hiiminus	&	[7]	\\
18329+5950	&	NGC 6670	&	18:33:37.06 	&	+59:53:19.3	&	121.2	&	11.63	&	IP	&	III	&	-	&	-	\\
20550+1656	&		&	20:57:23.51 	&	+17:07:34.6 	&	152.5	&	11.92	&	IP	&	III	&	\hiiminus	&	[2]	\\
22491-1808	&		&	22:51:45.79 	&	-17:52:22.5 	&	338.6	&	12.19	&	DN	&	III (1)	&	\hiiminus	&	[5]	\\
23007+0836	&		&	23:03:16.92 	&	+08:53:06.4 	&	67.9	&	11.54	&	IP	&	I-II	&	Sy1-Sy2/L	&	 [2]	\\
23128-5919	&		&	23:15:46.46 	&	-59:04:01.9 	&	189.6	&	11.96	&	DN	&	III (1)	&	Sy/L-\hiiminus	&	 [5], [9]	\\
\hline
\noalign{\smallskip}
\multicolumn{10}{@{} p{\textwidth} @{}}{\footnotesize{\textbf{Notes.} Col (1): object designation in the IRAS Point and Faint Source catalogs. Col (2): other name. Cols (3) and (4): right ascension (hours, minutes and seconds) and declination (degrees, arcminutes, arcseconds) taken from the from the NASA Extragalactic Database (NED). Col (5): luminosity distances derived assuming a $\Lambda$CDM cosmology with H$_{0}$ = 73 $km s^{-1} Mpc^{-1}$, $\Omega_{\Lambda}$=0.73 and $\Omega_{\rm{M}}$=0.27, and using the Eduard L. Wright Cosmology calculator, which is based in the prescription given by Wright (\myciteyear{Wright06}). Col (6): logarithm of the infrared luminosity  in units of solar bolometric luminosity, computed using the fluxes in the four \iras bands, as prescribed in \reftab{definition}. Col (7): MI stands for Multiple Interacting galaxies, IP interacting pair, DN double nucleus and SN single nucleus. Col (8):  Morphological classification used in this study, where I-II stands for first approach, III for pre-merger, IV for post-merger and V for relaxed system (see text). Between brackets the classification by Rodriguez-Zaurin et al. (\myciteyear{Rodriguez-Zaurin10}) is given, when available. In  their study, class 0 corresponds to classes I-II, class 1 to class III and class 2 to classes IV-V in this study, respectively. Note a disagreement in IRAS 06259-4708 and IRAS 08520-6850 due to the slight different of the definitions of the early interaction phases in both studies (see text). Col (9): Spectral class in the optical. A slash between two different classes indicates that no clear classification is known between both, whereas a dash separates the class of each nucleus in the system when known. L stands for LINER and Sy refers to a Seyfert activity. Col (10): References to the spectral class-- [1]~Alonso-Herrero et al. (\myciteyear{Alonso-Herrero09}), [2]~Wu et al. (\myciteyear{Wu98}), [3]~Kotilainen et al. (\myciteyear{Kotilainen01}), [4]~Risaliti et al. (\myciteyear{Risaliti06}), [5]~Nardini et al. (\myciteyear{Nardini08}), [6]~Arribas et al. (\myciteyear{Arribas08}), [7]~Kewley et al. (\myciteyear{Kewley01b}), [8]~Cohen (\mycitealt{Cohen92}), [9]~Evans et al. (\myciteyear{Evans02}), [10]~Arribas et al. (\myciteyear{Arribas00}), [11]~Zenner \& Lenzen (\myciteyear{Zenner93}), [12]~Garc\'ia-Mar\'in et al. (\myciteyear{Garcia-Marin07}), [13]~Kim et al. (\myciteyear{Kim98}), [14]~Corbett et al. (\myciteyear{Corbett03}), [15]~Arribas \& Colina (\myciteyear{Arribas03})}. }

\end{tabular}
\end{center}
\end{scriptsize}
\end{minipage}
\end{table}

\subsect[morphologies]{Morphologies Sampled}

The galaxies exhibit a variety of morphologies, from a wide pair with a projected nuclear distance of 54 kpc, to close nuclei with a projected radial distance of d $\lesssim$ 5 kpc, and to a single nucleus with a distorted stellar envelope. We identified a morphology class for each system of the sample by using the red (\textit{F814W} filter in \acs and \wfpc instruments) \hst images (see \refsec{phot_data}). A merging classification scheme similar to that given by Veilleux et al. (\myciteyear{Veilleux02}) was considered. Yet, due to our limited number of objects some grouping was made, so as to have enough statistics in all phases. For instance, phases I and II in Veilleux et al. (\myciteyear{Veilleux02}) were grouped for this study. The sample was divided in four groups according to their projected morphology and trying to match well identified temporal snapshots in the evolution of the interaction according to the models (see images in~\reffig{composite34}):

\begin{enumerate}[]
 \item  ~I-II. \textit{First Approach}.-- In this early phase, the first passage has not been completed. It is prior to the first close passage or during that first passage. The galaxy disks remain relatively unperturbed and bars and tidal tails have not yet formed. In some cases the disks can overlap, but no projected morphological disruption is seen on either galaxy. Some examples of systems in this phase are IRAS 06259-4708 and IRAS 07027-6011. \\
\item  ~III. \textit{Pre-Merger}. --  Two identifiable nuclei, with very well identified tidal tails and bridges, characterizes this phase.  Both disks are still recognizable. This occurs 200-400 Myr after the first approach, according to the similar morphologies observed in Bournaud et al. (\myciteyear{Bournaud08a}) models (see \refsec{sim_data}). Some examples representing this phase are IRAS 08572+3915 and IRAS 12112+0305. \\
\item  ~IV. \textit{Merger}. -- We only included phase IVa of Veilleux et al. (\myciteyear{Veilleux02}), since the systems in phase IVb have morphologies more similar to those in phase V than in IVa. In this phase both nuclei have apparently coalesced. We consider that the nuclei have coalesced as soon as their separation is less than 1.5 kpc, since numerical simulations have shown that by the time the two nuclei have reached a separation \mbox{of $\leq$ 1 kpc}, the stellar system has basically achieved equilibrium (\mycitealt{Mihos99};~\mycitealt{Naab06}). Long tails and shell structures are seen in this phase. Both disks are no longer recognizable, but a common internal structure has formed. This phase is seen 500-700 Myr after the first approach, according to the models mentioned. IRAS 04315-0840 and IRAS F10038-3338 are representative galaxies of this phase. It has to be taken into account that we are considering projected distances. Hence, those systems with double nuclei allocated to this phase could well belong to the previous phase if the real separation of both nuclei is considerably greater (i.e., orbital position along our line of sight).  \\
\item  ~V. \textit{Post-Merger}. -- We also included phase IVb of Veilleux et al. (\myciteyear{Veilleux02}). In this phase there are no prominent and bright tidal tails or bridges, since their surface brightness starts to fall below the detection limit, though some shell structures around the nucleus can still be detected. These isolated nuclei can have disturbed stellar envelopes, similar to those in the previous phase, indirectly indicating the past interaction. Typical ages of \mbox{$\tau$ $\gtrsim$ 1 Gyr} after the first passage characterizes the timescale of this phase, according to the models mentioned. IRAS 05189-2524 and IRAS 12116-5615 are good examples of systems in this phase. 
\end{enumerate}

Since this classification was performed only by evaluating the apparent morphology of the systems under study, it is always subject to improvements based on data with a higher angular resolution and sensitivity, and deeper imaging (i.e., to observe envelopes in post-mergers). Therefore, misclassification could occur in a few systems. Moreover, for certain studies, slightly different and simpler classifications have been used previously (\mycitealt{Rodriguez-Zaurin10}; see also \reftab{sample}). Although they have the advantage of reducing subjectivity and therefore uncertainty, their classification is too coarse for
detail comparison with model predictions. While the classification of a few objects can differ, as is the case for IRAS 06259-4807 and IRAS 08520-6850 in common with Rodr\'{i}guez-Zaur\'{i}n's sample (the definitions of the early phases in both studies are slightly different), the agreement between both classifications is excellent. Using this classification, 4 systems were assigned to category I-II, 13 to category III, 8 to category IV and 7 to category V (see \reftab{sample}).

\begin{figure*}[!htp]
  \hypertarget{fig:composite34}{}\hypertarget{autolof:\theautolof}{}\addtocounter{autolof}{1}
  \centering
   \includegraphics[trim = 0cm 0cm 0cm 3cm,clip=true,width=0.95\textwidth]{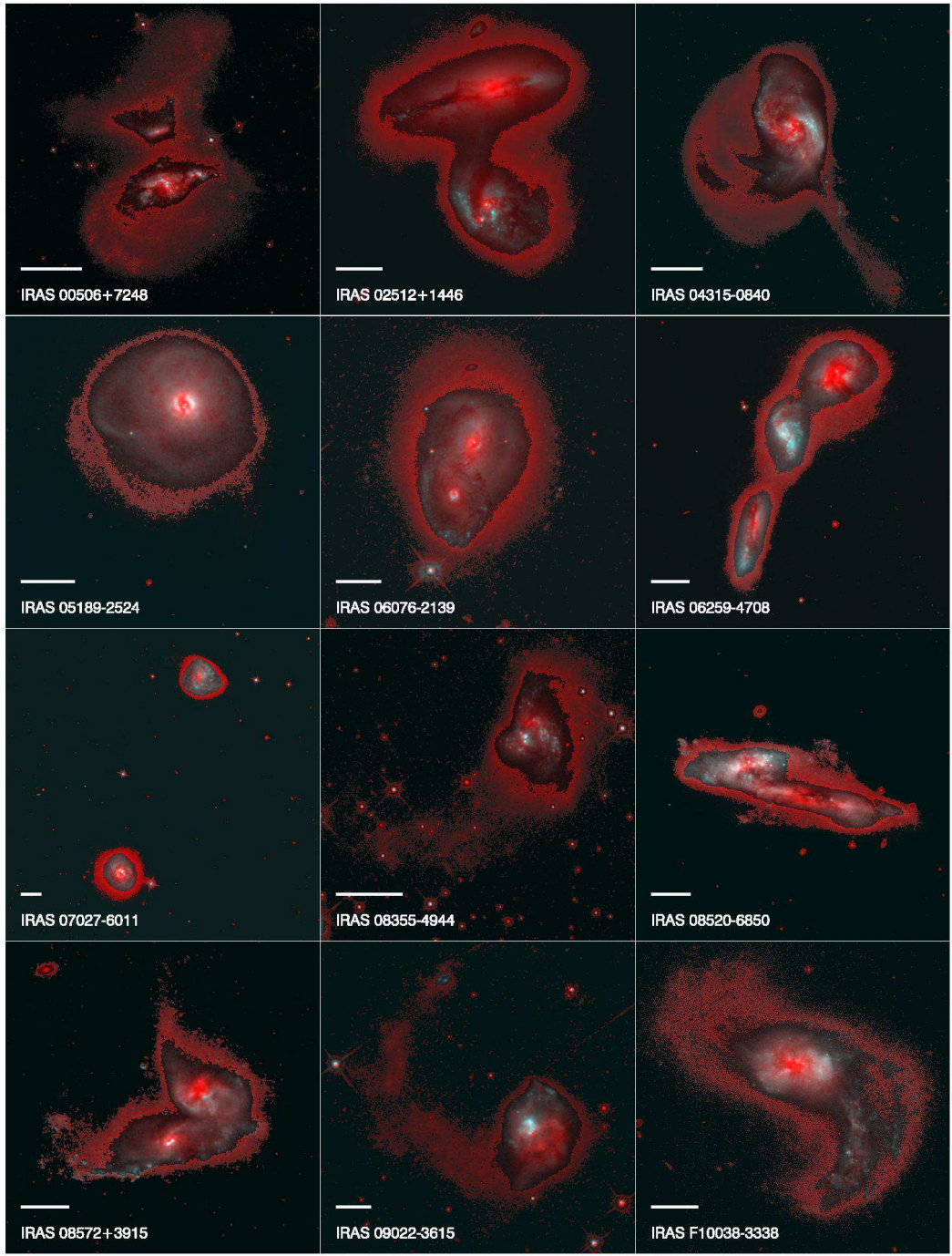} 
\caption[False color composite images of the (U)LIRGs under study]{Systems under study. False color composite image of the complete sample using \textit{F814W} (red) and \textit{F435W} (blue) \acs\twospace-\hst images. We have saturated the images from a given surface brightness (typically 5-10 times the global background deviation) faintward. The result of this, the diffuse red light, shows the lowest surface brightness features (tails, plums and shells).  The white horizontal line indicates a scale of 5 kpc. Blue knots along the tail are clearly visible in some systems. North points up and East to the left.}
\label{fig:composite34}
    \end{figure*}
  \begin{figure*}[!htp]
   \captionsetup{list=no}
   \centering
   \includegraphics[trim = 0cm 0cm 0cm 3cm,clip=true,width=\textwidth]{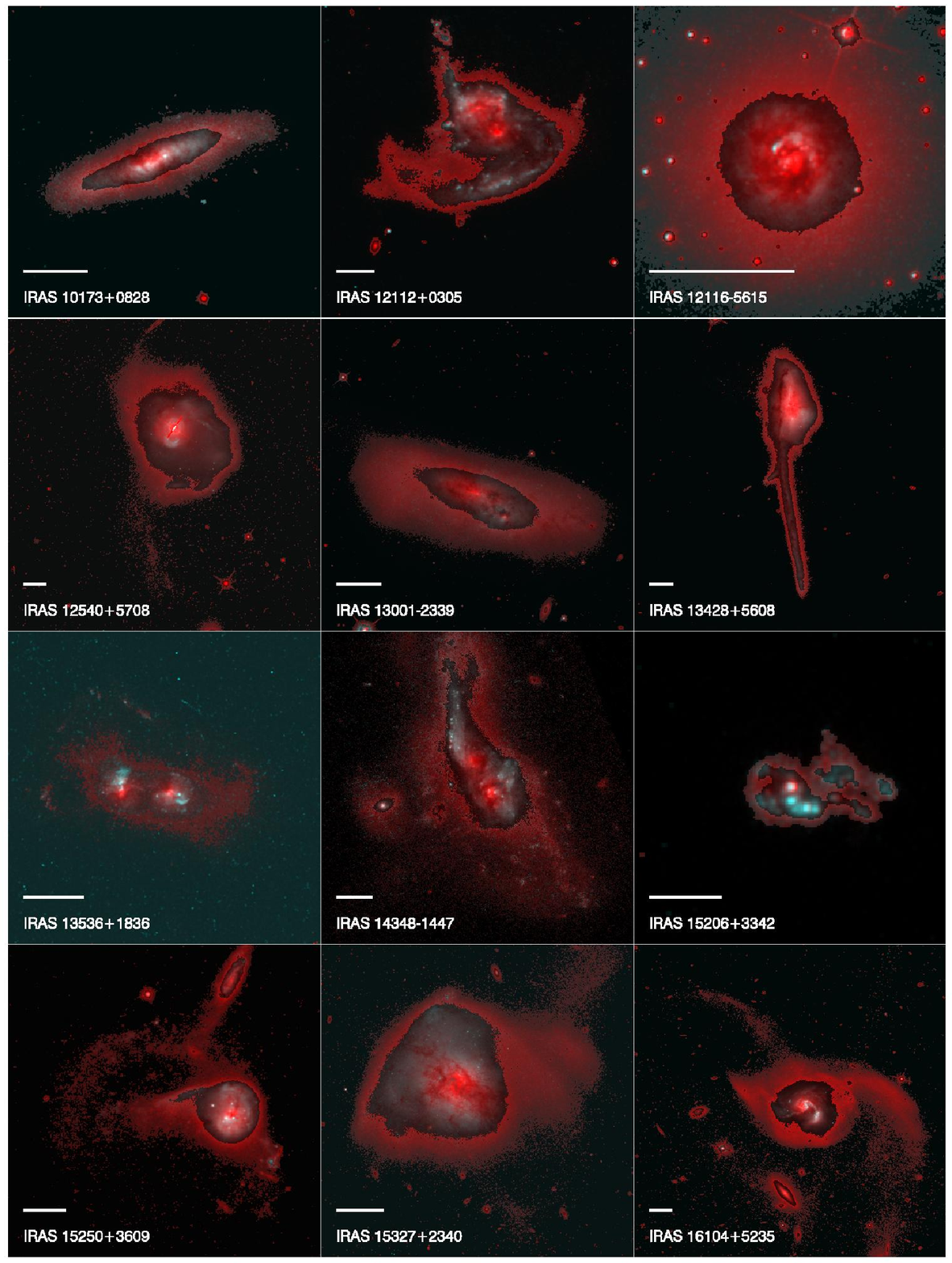} 
\addtocounter{figure}{-1}   
   \caption[]{- Continued}
    \end{figure*}
  \begin{figure*}[!htp]
   \centering
   \includegraphics[trim = 0cm 6cm 0cm 3cm,clip=true,width=\textwidth]{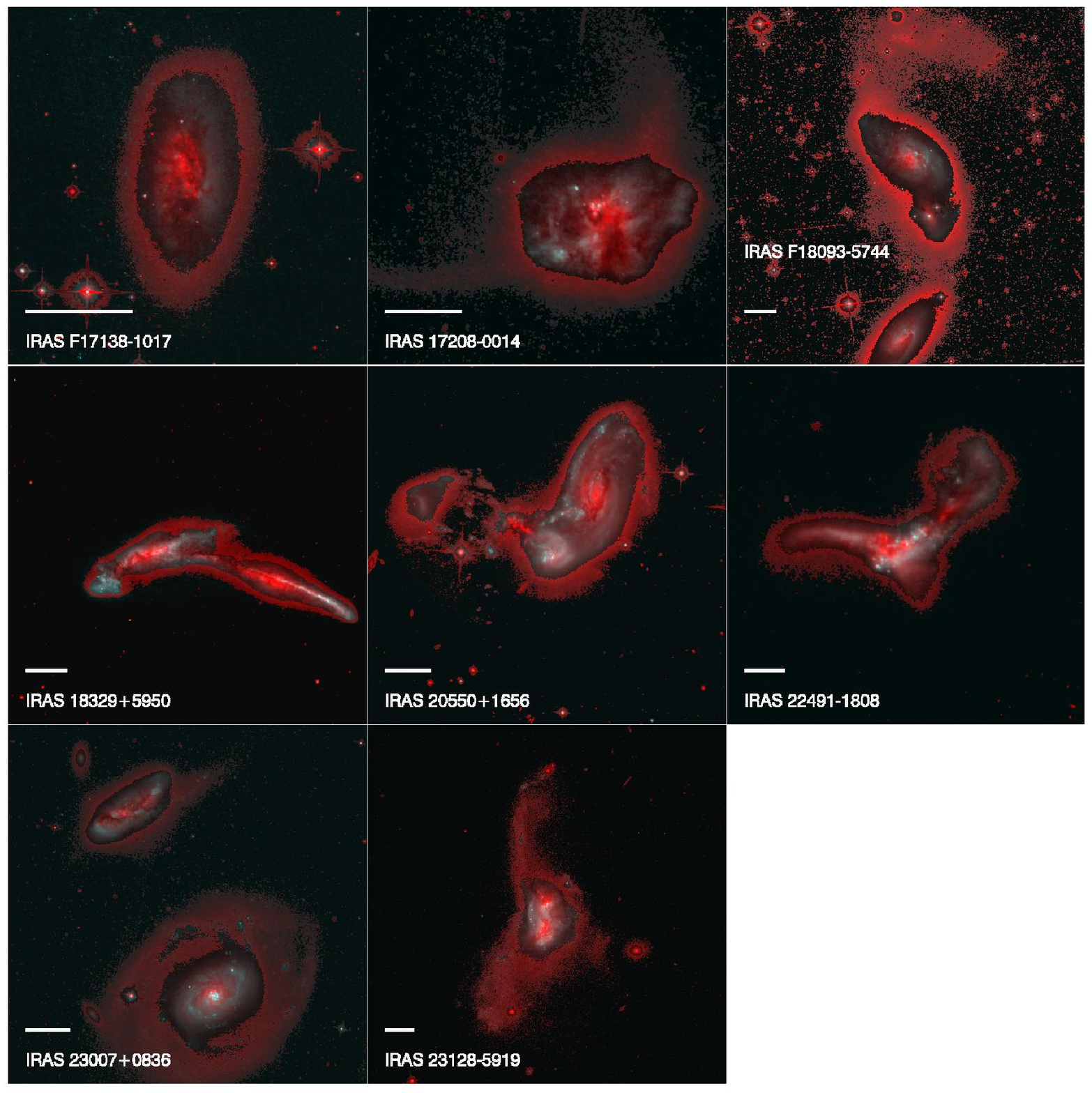} 
\vspace{-1cm}
\addtocounter{figure}{-1} 
   \caption[]{- Continued}
   \vspace{-0.5cm}
  
    \end{figure*}

\sect[data_ac]{Data Acquisition}
   
For this thesis we used a variety of different but complementary data sets: high angular resolution imaging as observed with the \textit{Hubble Space Telescope}, \ifs data from the fiber-based optical \ifu instruments and data from a high linear resolution simulation of a wet galaxy merger. They are all based on completely different concepts, and each one requires a different treatment, in terms of data reduction and analysis techniques. The available dataset from observations for each galaxy is shown in \reftab{data}.

\subsect[phot_data]{Photometric Data: Images from the $HST$}

Observations with the \hst\twospace, a 2.4m reflector telescope carried into orbit by a space shuttle in 1990,  have great advantages compared to ground-based facilities. Thanks to the absence of the atmospheric veil the observations are not limited by seeing conditions (i.e., it can provide an angular resolution in the optical of 0.05\arcsec~ vs. the 1.0\arcsec~ ground-based one), and the stability of the observing conditions makes the calibrations quite stable. The main photometric datasets used in this thesis cover the optical spectral range, though complementary infrared images are also used. 

\subsubsection{Optical Images}

\relax
\hypertarget{sub:optical_images}{}
\label{sub:optical_images}

\begin{figure}[!htp]
\centering
 \hypertarget{fig:filters}{}\hypertarget{autolof:\theautolof}{}\addtocounter{autolof}{1}
 \includegraphics[trim = 2.5cm 12cm 1.5cm 4cm,clip=true,width=0.98\textwidth]{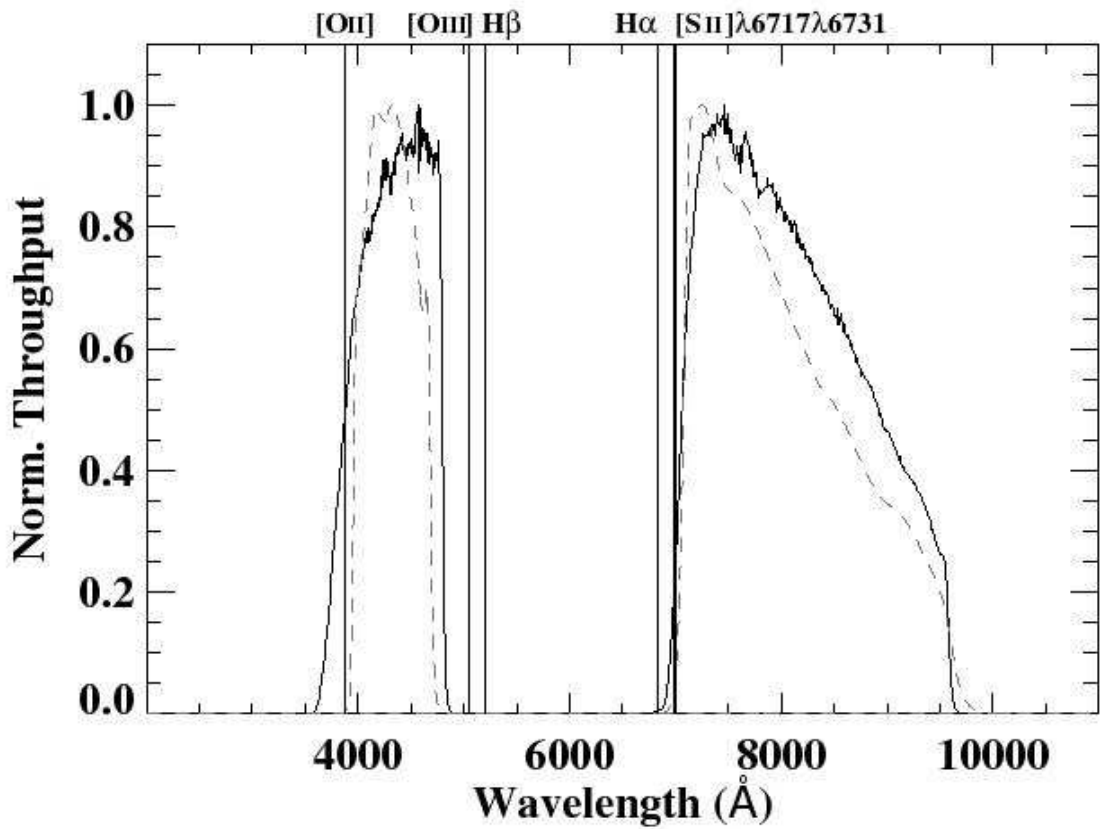}
 \caption[Normalized response curves for the optical filters]{Normalized response curve for the optical filters used in this thesis. The black solid lines represent the response curves for the \acs filters, while the dashed lines show the response curves for the \wfpc filters. Very well-known ionizing lines are also plotted (vertical lines) for \mbox{z = 0.4}, approximately the median redshift of the sample.}
 \label{fig:filters}
\end{figure}

In this thesis, we mainly use archive data from the \acr{}{ACS}{Advanced Camera for Surveys}, a third generation axial instrument aboard the \hst\twospace. It includes three independent channels that cover the ultraviolet to the near-infrared spectral range: the Wide Field Channel (WFC), the High-Resolution Channel (HRC) and the Solar Blind Channel. Most of the images in this study were taken with the WFC, which consists of two butted 2048$\times$4096 pixel CCD with a plate scale of 0.05\arcsec\onespace/pixel and an effective \acr{}{FoV}{field of view} of 202\arcsec$\times$202\arcsec. 

\acs broad-band images were retrieved for thirty systems from the Hubble Legacy Archive, with the filters \textit{F814W} and \textit{F435W} (see response curves in~\reffig{filters}). The former is equivalent to the ground-based Johnson-Cousins \textit{I}, whereas the latter differs from the ground-based Johnson-Cousins \textit{B} between 7 and 12\% in flux (\mycitealt{Sirianni05}). Total integration times were taken from 720 to 870s with filter \textit{F814W} and from 1260 to 1500s with filter \textit{F435W}.\\

For IRAS 13536+1836 and IRAS 15206+3342, \acr{}{WFPC2}{Wide Field Planetary Camera 2} images were taken. This instrument, which covers the spectral range from about 1150 to 10500 \AA{}, is composed of four cameras that operate simultaneously. Three of them (chips 2, 3, 4) have a plate scale of 0.1\arcsec\onespace/pixel and a \fov of 80\arcsec$\times$80\arcsec, and form the Wide Field Camera. The fourth (chip 1), with a plate scale of 0.046\arcsec\onespace/pixel and a \fov of 36\arcsec$\times$36\arcsec, is the so-called Planetary Camera (PC). Images were taken with the filters \textit{F814W} and \textit{F439W} (see response curves in~\reffig{filters}), being equivalent to ground-based Johnsons \textit{I} and \textit{B}, respectively (\mycitealt{Origlia00}). Total integration times of about 350s were taken with filter \textit{F814W} and of 1000s with filter \textit{F439W}. Here, the systems fall in the PC chip. 

Whenever both \acs and \wfpc images were available in both filters, \acs images were preferred for its superior sensitivity and larger \fov\twospace. From now on, we will refer to the blue filter as \textit{F435W} for \acs and \wfpc for simplicity. 

Images were reduced on the fly (\hst pipeline), with the highest quality available reference files at the time of retrieval. The calibrated \acs \textit{F435W} images had only few cosmic rays remaining, since 3 images had been observed and then combined. However, all the \acs \textit{F814W} images were severely contaminated with cosmic rays in a 6\arcsec-width band across the entire \fov\twospace. Two exposures had been taken and when they did not overlap the \hst pipeline was not able to discriminate the cosmic rays. In most cases that band went through part of the system, covering about 10-20\% of its extension. Cosmic ray rejection was then carried out using the IRAF\footnote{IRAF software is distributed by the National Optical Astronomy Observatory (NOAO), which is operated by the Association of Universities for Research in Astronomy (AURA), Inc., in cooperation with the National Science Foundation.} task \textit{credit} with the aid of the blue images since they were barely contaminated. Most of the systems have observations that cover the entire \fov with only this instrument and the two filters, so the aid of multi-band images was limited to few systems. 

Once the photometric data is processed, the derived counts can be transformed into the Vega system magnitudes
using: 

\eqn[phot_cal]{
 m_{\lambda}=-2.5\times log \left(DN\right) + Zeropoint_{\lambda}
}

where $\lambda$ refers to photometric band (\textit{I} or \textit{B}), DN corresponds to the integrated counts per second in the defined aperture, and Zeropoint a constant of value 25.536 and 25.759 for the \acs \textit{F814W} and \textit{F435W} broadbands, respectively (21.678 and 20.877 for the \wfpc filters). The Zeropoints were derived using the calibration reference papers by Sirianni et al. (\myciteyear{Sirianni05}) and Holtzman et al. (\myciteyear{Holtzman95}), for the \acs and \wfpc data respectively.

%
%
%
%
%
\begin{sidewaystable}
\hypertarget{table:data}{}\hypertarget{autolot:\theautolot}{}\addtocounter{autolot}{1}
\begin{minipage}{0.97\textwidth}
\renewcommand{\footnoterule}{}  
\begin{tiny}
\caption[Available data from observations for this thesis]{Available data: \hst archive and \ifs data}
\label{table:data}
\begin{tabular}{l@{\hspace{0.1cm}}c@{\hspace{0.1cm}}ccccccccc@{\hspace{0.3cm}}c@{\hspace{0.3cm}}c}
\hline \hline
   \noalign{\smallskip}
   IRAS name	&	\hst optical	&	PI and	&	Obs.	&	Exp.	&	\hst \nir	&	PI and	&	Obs.	&	Exp.	&	\ifs data	&	PI and	&	Obs.	&	Exp.	\\
	&	data	&	proposal ID	&	date	&	time (s)	&	data	&	proposal ID	&	date	&	time (s)	&		&	project ID	&	date	&	time (s)	\\
(1)	&	(2)	&	(3)	&	(4)	&	(5)	&	(6)	&	(7)	&	(8)	&	(9)	&	(10)	&	(11)	&	(12)	&	(13)	\\ 
\hline
    \noalign{\smallskip}
00506+7248	&	ACS	&	Evans, 10592	&	09/05	&	800, 1500	&	NICMOS	&	Alonso-Herrero, 10169	&	10/04	&	311	&	-	&	-	&	-	&	-	\\
02512+1446	&	ACS	&	Evans, 10592	&	07/06	&	720, 1260	&	-	&	-	&	-	&	-	&	-	&	-	&	-	&	-	\\
04315-0840	&	ACS	&	Evans, 10592	&	08/06	&	720, 1260	&	NICMOS	&	Rieke, 7218	&	02/98	&	191	&	VIMOS	&	Arribas, 076.B-0479(A)	&	03/08	&	2880	\\
05189-2524	&	ACS	&	Evans, 10592	&	08/06	&	730, 1275	&	NICMOS	&	Veilleux, 9875	&	08/04	&	2560	&	VIMOS	&	Arribas, 076.B-0479(A)	&	12/06	&	3000	\\
06076-2139	&	ACS	&	Evans, 10592	&	11/05	&	720, 1260	&	WFC3	&	Surace, 11235	&	06/09	&	2395	&	VIMOS	&	Arribas, 076.B-0479(A)	&	12/06	&	3000	\\
06259-4708	&	ACS	&	Evans, 10592	&	04/06	&	780, 1350	&	NICMOS	&	Surace, 11235	&	06/08	&	4990	&	VIMOS	&	Arribas, 076.B-0479(A)	&	12/06	&	3000	\\
07027-6011	&	ACS	&	Evans, 10592	&	09/05	&	830, 1425	&	NICMOS	&	Surace, 11235	&	06/08	&	4990	&	VIMOS	&	Arribas, 076.B-0479(A)	&	12/06	&	6000	\\
08355-4944	&	ACS	&	Evans, 10592	&	09/05	&	780, 1350	&	NICMOS	&	Surace, 11235	&	06/08	&	2495	&	VIMOS	&	Arribas, 076.B-0479(A)	&	03/07	&	3000	\\
08520-6850	&	ACS	&	Evans, 10592	&	04/06	&	870, 1485	&	NICMOS	&	Surace, 11235	&	05/08	&	2300	&	VIMOS	&	Arribas, 076.B-0479(A)	&	03/07	&	3000	\\
08572+3915	&	ACS	&	Evans, 10592	&	12/05	&	750, 1305	&	NICMOS	&	Maiolino, 9726	&	03/04	&	160	&	INTEGRAL	&	Colina, 98ACAT26 \& C4	&	04/98	&	16200	\\
	&	-	&	-	&	-	&	-	&	-	&	-	&	-	&	-	&	INTEGRAL	&	Colina, 98ACAT26 \& C4	&	01/11	&	9000	\\
09022-3615	&	ACS	&	Evans, 10592	&	06/06	&	750, 1305	&	-	&	-	&	-	&	-	&	VIMOS	&	Arribas, 076.B-0479(A)	&	03/07	&	3000	\\
F10038-3338	&	ACS	&	Evans, 10592	&	11/05	&	740, 1290	&	-	&	-	&	-	&	-	&	VIMOS	&	Arribas, 076.B-0479(A)	&	03/07	&	3000	\\
10173+0828	&	ACS	&	Evans, 10592	&	12/05	&	720, 1260	&	-	&	-	&	-	&	-	&	VIMOS	&	Arribas, 076.B-0479(A)	&	03/08	&	2880	\\
12112+0305	&	ACS	&	Evans, 10592	&	02/06	&	720, 1260	&	NICMOS	&	Scoville, 7219	&	11/97	&	192	&	INTEGRAL	&	Colina, 98ACAT26 \& C4	&	04/98	&	9000	\\
	&	-	&	-	&	-	&	-	&	-	&	-	&	-	&	-	&	INTEGRAL	&	Colina, 98ACAT26 \& C4	&	01/11	&	7500	\\
12116-5615	&	ACS	&	Evans, 10592	&	09/05	&	830, 1425	&	NICMOS	&	Surace, 11235	&	07/07	&	2495	&	VIMOS	&	Arribas, 076.B-0479(A)	&	03/07	&	2250	\\
12540+5708	&	ACS	&	Evans, 10592	&	05/06	&	830, 1425	&	NICMOS	&	Veilleux, 9875	&	09/04	&	2560	&	INTEGRAL	&	Colina, D	&	03/02	&	3600	\\
13001-2339	&	ACS	&	Evans, 10592	&	02/06	&	720, 1260	&	-	&	-	&	-	&	-	&	VIMOS	&	Arribas, 076.B-0479(A)	&	03/07	&	6000	\\
13428+5608	&	ACS	&	Evans, 10592	&	11/05	&	820, 1425	&	NICMOS	&	Maiolino, 9726	&	05/04	&	599	&	INTEGRAL	&	Colina, 98ACAT26 	&	04/98	&	4500	\\
13536+1836	&	WFPC2	&	Sanders, 5982	&	11/95	&	360, 1030	&	NICMOS	&	Low, 7213	&	12/97	&	480	&	INTEGRAL	&	Colina, C4	&	04/01	&	4500	\\
14348-1447	&	ACS	&	Evans, 10592	&	04/06	&	720, 1260	&	NICMOS	&	Scoville, 7219	&	12/97	&	480	&	INTEGRAL	&	Colina, 98ACAT26 	&	04/98	&	7200	\\
15206+3342	&	WFPC2	&	Sanders, 5982	&	09/95	&	343, 750	&	-	&	-	&	-	&	-	&	INTEGRAL	&	Colina, 98ACAT26 	&	04/98	&	7200	\\
15250+3609	&	ACS	&	Evans, 10592	&	01/06	&	750, 1305	&	NICMOS	&	Scoville, 7219	&	11/97	&	223	&	INTEGRAL	&	Colina, 98ACAT26 	&	04/98	&	9000	\\
15327+2340	&	ACS	&	Evans, 10592	&	01/06	&	720, 1260	&	NICMOS	&	Maiolino, 9726	&	01/04	&	599	&	INTEGRAL	&	Colina, W/200A/27	&	05/01	&	9000	\\
16104+5235	&	ACS	&	Evans, 10592	&	09/05	&	800, 1380	&	NICMOS	&	Maiolino, 9726	&	12/03	&	599	&	-	&	-	&	-	&	-	\\
F17138-1017	&	ACS	&	Evans, 10592	&	03/06	&	720, 1260	&	NICMOS	&	Alonso-Herrero, 10169	&	09/04	&	240	&	VIMOS	&	Arribas, 081.B-0108(A)	&	06/09	&	3400	\\
17208-0014	&	ACS	&	Evans, 10592	&	04/06	&	720, 1260	&	NICMOS	&	Scoville, 7219	&	10/97	&	223	&	INTEGRAL	&	Colina, 98ACAT26 	&	04/98	&	7800	\\
F18093-5744	&	ACS	&	Evans, 10592	&	04/06	&	830, 1425	&	NICMOS	&	Alonso-Herrero, 10169	&	09/04	&	240	&	VIMOS	&	Arribas, 081.B-0108(A)	&	06/09	&	6800	\\
18329+5950	&	ACS	&	Evans, 10592	&	10/05	&	830, 1425	&	NICMOS	&	Surace, 11235	&	03/09	&	2495	&	-	&	-	&	-	&	-	\\
20550+1656	&	ACS	&	Evans, 10592	&	04/06	&	720, 1260	&	-	&	-	&	-	&	-	&	INTEGRAL	&	Arribas, P15	&	07/06	&	4500	\\
22491-1808	&	ACS	&	Evans, 10592	&	05/06	&	720, 1260	&	NICMOS	&	Scoville, 7219	&	11/97	&	480	&	VIMOS	&	Arribas, 081.B-0108(A)	&	07/09	&	6800	\\
23007+0836	&	ACS	&	Evans, 10592	&	06/06	&	720, 1260	&	NICMOS	&	Scoville, 7219	&	11/97	&	351	&	-	&	-	&	-	&	-	\\
23128-5919	&	ACS	&	Evans, 10592	&	03/06	&	830, 1425	&	NICMOS	&	Maiolino, 9726	&	10/03	&	599	&	VIMOS	&	Arribas, 081.B-0108(A)	&	07/09	&	3400	\\
\hline
   \noalign{\smallskip}
   \multicolumn{13}{@{} p{\textwidth} @{}}{\footnotesize{\textbf{Notes.} Col (1): object designation in the IRAS Point and Source Catalogs. Col (2): optical instruments used on board \hst to observe the retrieved \filteri and \filterb images. For observations with the WFPC2 the blue filter corresponds to \textit{F439W}. Col (3): last name of the principal investigator, followed by the proposal ID of the optical observation. Col (4): day when the optical observation was taken. Col (5): total exposure times for each galaxy, with filter \filteri (left) and filter \filterb (right). Col (6): \nir instruments used on board \hst to observe the retrieved \filterh images. For all these observations the camera used was NIC2. Col (7): same as (3) for the infrared data. Col (8): same as (4) for the infrared data. Col (9): same as (5) for the infrared data. Col (10): optical ground-based instruments used to observe the \ifs data. Since two observations were taken for two sources at different epochs, two rows are provided for each of them. Col (11): same as (3) for the \ifs data. Col (12): sames as (4) for the \ifs data. Col (13): same as (5) for the \ifs data.}}
  \end{tabular}
\end{tiny}
\end{minipage}
\end{sidewaystable}
 

\newpage
\subsubsection{Complementary \nir Images}
\relax
\hypertarget{sub:complementary_nir}{}
\label{sub:complementary_nir}

The \acr{}{NICMOS}{Near Infrared Object and Spectrometer} consists of three individual cameras (NIC1, NIC2, NIC3) located in one side of the \hst \fov and designed to operate independently. These cameras operate at different magnification scales (0.043\arcsec, 0.075\arcsec~ and 0.2\arcsec~ for NIC1, NIC2 and NIC3, respectively). Its spectral coverage ranges from 0.8 to 2.5 $\mu$m. Since it works in the \nir\twospace, the detector is cooled to cryogenic temperatures to avoid the thermal contamination coming from the environment and the instrument electronics itself.

Complementary \nicmos broad-band images with filter F160W ($\sim$\textit{H}) were also available for the majority of galaxies, which were helpful to locate the nuclei of the systems. However, they were not included in the main photometric study since they have small \fov (20\arcsec$\times$20\arcsec), not covering the outskirts of the systems or interacting tail-bridge structures. 

These data were also calibrated on the fly (\hst pipeline), with the highest quality available reference files at the time of retrieval. There is one effect that appears in the \nicmos data and that is not included in the standard pipeline: the so-called pedestal effect. This is an additive signal that appears in the \nicmos images when the amplifiers are switched on, and that have different values depending on the camera quadrant. This effect is detected as flat-field residuals in the calibrated images. In the present case no relevant residuals related to the pedestal effects were detected, and thus no additional correction was applied to the images. The only aspect that was improved with respect to the pipeline was the re-combination of the dithered individual exposures to remove the values of the column 128, which is thought to be affected by uncertainties in detector shading corrections (it is considered as a bad column).

For IRAS 06076-2139 we also retrieved recently released data (end of 2010) from the \acr{}{WFC3}{Wide Field Camera 3}, a fourth-generation UVIS/\ir imager aboard the \hst\twospace. With a \fov of 135\arcsec$\times$127\arcsec~ and a pixel scale of 0.13\arcsec\onespace/pixel, the near infrared channel covers a much higher field than its predecessor, the \nicmos instrument. We retrieved the processed WFC3/IR \textit{H}-band (F160W) image. 

\subsect[spec_data]{Spectroscopic Data}

\subsubsection{Integral Field Spectroscopy}

The general term 3D-spectroscopy refers to those techniques that allow to obtain spatially-resolved spectra over a two dimensional field. In an extragalactic context, each point of the galaxy ($\alpha$, $\delta$) has associated a spectrum ($\lambda$), and all the information has to be stored in a two-dimensional detector. The Integral Field Spectroscopy (\ifs\twospace) is one of the most widely used 3D-spectroscopy methods. This technique simultaneously obtains the spectral and spatial information for a given telescope pointing (\mycitealt{Allington-Smith07}). One of the fundamental advantages with respect to other 3D-spectroscopy methods is the data homogeneity. This technique presents a good compromise between the exposure time, resolution, and homogeneity of the data.

\begin{figure}[!htp]
 \hypertarget{fig:ifs_tech}{}\hypertarget{autolof:\theautolof}{}\addtocounter{autolof}{1}
\hspace{0.3cm}
\includegraphics[width=0.95\textwidth]{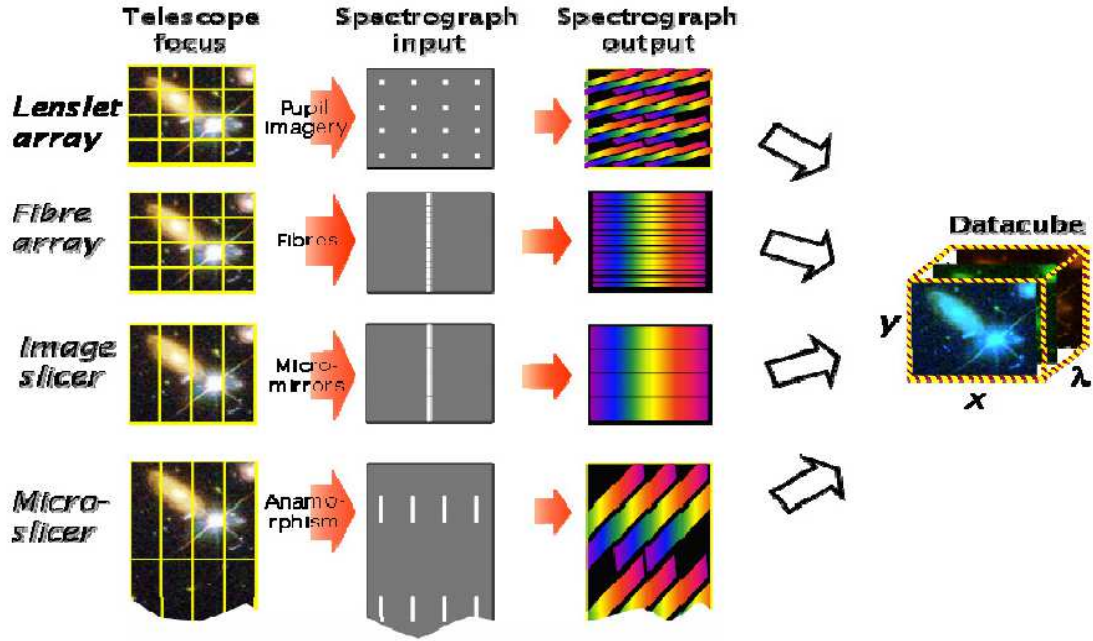}
      \caption[The four main techniques of integral field spectroscopy]{A summary of the four main techniques of integral field spectroscopy. Figure taken from Allington-Smith (\myciteyear{Allington-Smith07}).}     
      \label{fig:ifs_tech}
\end{figure}

Since the '80s several instruments have been designed using different \ifs techniques. Allington-Smith (\myciteyear{Allington-Smith07}) gives a recent review of the most popular present and future \ifus (see also \reffig{ifs_tech}). All these techniques are based on dividing the light coming from the source and dispersing each of the subdivisions to obtain its spectrum. The data for this thesis were taken using fiber arrays or their combination with lenslets as integral field unit (\ifu\twospace):

\begin{itemize}
 \item \textbf{Fiber Array:} In this case, a fiber array is used to subdivide the \fov in the focal plane of the telescope. The fibers are then reordered forming a pseudo-slit and the information dispersed by a conventional spectrograph (\reffig{ifs_tech}, second panel). The low cross-talk derived from the reordering of the fibers and the large spectral coverage represent the best advantages of this configuration. However, this configuration has two main disadvantages: round fibers cannot fill a two-dimensional region completely and the light beam is widen as it travels across the fiber (focal degradation). The INTEGRAL instrument (\mycitealt{Arribas98}) uses this \ifu\twospace.
\item \textbf{Lenslets+Fiber:} Some disadvantages associated to the pure fiber-based \ifu (e.g., flux losses and focal degradation) can be avoided by coupling them to a lens array at the focal plane. This way, the spatial coverage is optimal, and the small pupil created at the entrance of the fiber core avoid light losses and minimize the focal ratio degradation. However, in general these systems need for additional optical mechanisms that can lead to a decrease of the system efficiency. Among others, PMAS (\mycitealt{Roth05}) and VIMOS (\mycitealt{leFevre03}) instruments make use of this technique.
\end{itemize}

\subsubsection{Observations and Spectral Maps}

Most of the \lirgs and a few \ulirgs in our sample were observed with VIMOS (mounted on the Nasmyth focus B on the VLT). The observations were carried out with the high-resolution mode ``HR-Orange'' (grating GG435), whose spectral range covers from about 5200 to 7400 \AA{}. The \fov and the spatial scale in this mode are 27\arcsec$\times$27\arcsec~ and 0.67\arcsec~ per fiber, respectively (i.e., 40$\times$40 fibers, 1600 spectra). A square four-pointing dithering pattern was used, with a relative offset of 2.7\arcsec. The exposure time per pointing was between 720 and 850 seconds and therefore, the total integration time per galaxy is in general 2880-3400 seconds. Details on the data reduction and calibration can be found in Arribas et al. (\myciteyear{Arribas08}) and in Rodr\'iguez-Zaur\'in et al. (\myciteyear{Rodriguez-Zaurin10}). To summarize, the data were reduced with a combinations of the pipeline recipe Esorex (included in the pipeline provided by ESO), and a series of IDL and IRAF customized scripts. Esorex performed the basic data reduction (bias subtraction, flat field correction, spectra tracing and extraction, correction of fiber and pixel transmission and relative flux calibration). Then, the four quadrants per pointing were reduced individually and combined into a single data-cube associated to each pointing. The final data-cube per object was generated combining the four independent dithered pointings, containing a total of 1936 spectra.

The rest of the \ulirgs in our sample were observed with INTEGRAL, connected to the camera WYFFOS (\mycitealt{Bingham94}) and mounted on the ~\mbox{4.2 m} William Herschel Telescope. The observations were carried out with the grating R600B, with a spectral range that covers from about 5000 to 8000 \AA{}. The \fov varies depending on the INTEGRAL dithering and pointings used, though in general it is around \mbox{23.0\arcsec $\times$ 16.3\arcsec}. The angular sampling is 0.90\arcsec~ per fiber. In general, between 4 and 6 pointings of 1500 seconds were taken and therefore, the total integration time per galaxy is 6000-9000 seconds. Details on the data reduction and calibration can be found in Garc\'ia-Mar\'in et al. (\myciteyear{Garcia-Marin09a}) and references therein. To summarize, the data reduction of the two-dimensional fiber spectra was performed using the IRAF environment, following the standard procedures applied for this type of data: bias subtraction, scattered light subtraction and cross-talk corrections, spectra tracing and extraction, wavelength calibration, flat-field correction, sky subtraction, relative flux calibration, image combination and absolute flux calibration.

The absolute flux uncertainty for the VIMOS and INTEGRAL data corresponds to about \mbox{10-20\%}. After the data reduction, every galaxy has a set of spectra, each one associated with the region observed with the individual fibers. The emission lines from each galaxy were analyzed by fitting them to Gaussian profiles. In both sets of data, the \ha line lays within the spectral range and, once calibrated, \ha line maps were then generated. Each pixel on the map (commonly called spaxel) has the information of the region observed with the individual fibers. Maps with the \ha equivalent width, the \nii lines, and the oxygen  (\oiii$\lambda\lambda$4959,5007) and \hb lines for galaxies observed with INTEGRAL were also produced. In this thesis, we make use of the maps published in Garc\'ia-Mar\'in et al. (\myciteyear{Garcia-Marin09a}) and in Rodr\'iguez-Zaur\'in et al. (\myciteyear{Rodriguez-Zaurin10}) for the systems in our sample. 

\subsect[sim_data]{Data from numerical Simulations}

We analyzed the wet galaxy merger simulation of Bournaud et al. (\myciteyear{Bournaud08a}) at different times. The simulation is performed with a particle-mesh code based on a FFT Poisson solver for the gravity. The gas dynamics is modeled with a sticky-particle scheme, and a Schmidt law is used for star formation. Supernovae feedback is taken into account in the simulation with the scheme proposed by Mihos and Hernquist (\myciteyear{Mihos94b}). The initial setup was chosen so as to be representative of an equal-mass wet merger of two spirals at \mbox{z $<$ 1} with an initial gas fraction of 17\%. The encounter corresponds to a prograde orbit for one galaxy and a retrograde orbit for the other, with an inclination angle of the encounter of 30\textdegree~ and 70\textdegree, respectively. A total number of 36 million particles are used in the simulation, corresponding to 6 million particles per galaxy per component (gas, stars and dark matter), having each stellar disk an initial mass of 2$\times10^{11}$~\msun\onespace.

The dataset corresponds to 81 datacubes with the spatial distribution of young particles of equal mass (8333~\msun\twospace) regularly spaced in time, 13 Myr between two time-steps, covering the temporal evolution of the galaxy encounter up to 1053 Myr. Thus, each datacube corresponds to one snapshot of the simulation. All the components of the simulation (gas, young stars, old stars and dark matter) were also simulated but the study was carried out only on young stars. We extracted a total of 78 snapshots, since  the star formation is triggered at t=39 Myr, where our analysis starts. Example of total stellar mass density maps at different interaction times, showing how the morphologies evolve with the merging process, are shown in Bournaud et al. (\myciteyear{Bournaud08a}), while the whole series  is presented in Belles. et al. (in prep). \reffig{panel_sim} shows similar maps, but referring to the mass density of the stellar particles created in the simulation (excluding the old component prior to the merger epoch). This way we give an idea on how the new formed stars distribute throughout the merging process. 

\begin{figure}[!tp]
 \hypertarget{fig:panel_sim}{}\hypertarget{autolof:\theautolof}{}\addtocounter{autolof}{1}
 \hspace{-1.2cm}
\includegraphics[angle=90,trim = 0cm 0cm 6cm 0cm,clip=true,width=1.2\textwidth]{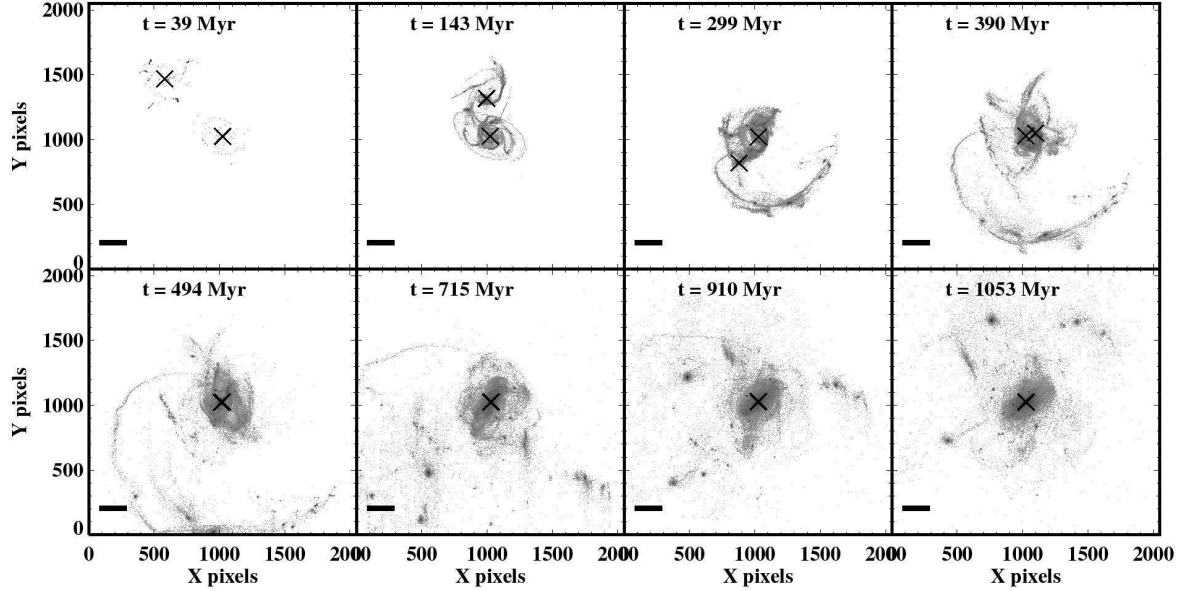}
      \caption[XY projection of the 3D particle distributions at different snapshots]{XY projection of the 3D particle distributions at different snapshots, where t indicates the time elapsed from the beginning of the simulation. The position of both nuclei is marked with crosses. The horizontal line indicates a scale of 10 kpc. From top to bottom and left to right, every 2 snapshots could be identified with the interaction phases defined in previous sections.}     
      \label{fig:panel_sim} 
\end{figure}

To identify the stellar regions (simulated knots), the datacube with the 3D position of the particles was projected on three 2D-planes. Knots were then identified using an algorithm based on particle counts and extracted above a S/N  with respect to the local background. Finally, they were cross-correlated between the planes to recover their 3D positions. Thus, the simulated knots represent a group of bound sticky particles. A thorough description of the analysis and the methodology applied to the detection, measurement of the mass and tracking of a knot is explained in Belles et al. (in preparation). The linear resolution achieved (the softening length of the gravitational potential) in this model corresponds to 32 pc. With a stellar mass resolution of about \mbox{10$^5$~\msun\onespace,} the simulated knots are typically more massive than \mbox{10$^6$~\msun} and up to 10$^9$~\msun\twospace.


\sect[tech]{General Photometric and Spectroscopic Analysis Techniques}


In this section, we describe the general methodology and tools used to derive the different properties of the objects under study in this thesis (i.e., compact regions of star formation in our sample). Some of these techniques are based only on the use of photometric data (e.g., analysis of the \lfs\twospace), others on spectroscopic data (e.g., metallicity derivations) and a few of them combine both datasets (e.g., stellar population models, dynamical mass derivations). The subsequent chapters in this thesis include additional analysis procedures (e.g., detection of the compact regions of interest) that have been kept there for a better comprehension of the text. 

\subsect[sizes]{Determination of Sizes}

The knowlege of the size of a stellar object is important to set up constraints on its nature and properties. The derivation of its size is not straightforward since some considerations have to be taken into account, such as the shape of the light profile, the contamination of the \acr{}{PSF}{point spread function}, the usage of different photometric filters, ect. 

In this section, we explain how the sizes of the stellar objects under study in this thesis were derived. Specifically, the effective radius --identified as an approximation to the half-light radius-- and the total size.

\subsubsection{Effective Radius}

\begin{figure}[!htp]
 \hypertarget{fig:sizes}{}\hypertarget{autolof:\theautolof}{}\addtocounter{autolof}{1}
\includegraphics[trim = -1.8cm 0cm 0cm 0cm,clip=true,angle=90,width=0.45\columnwidth]{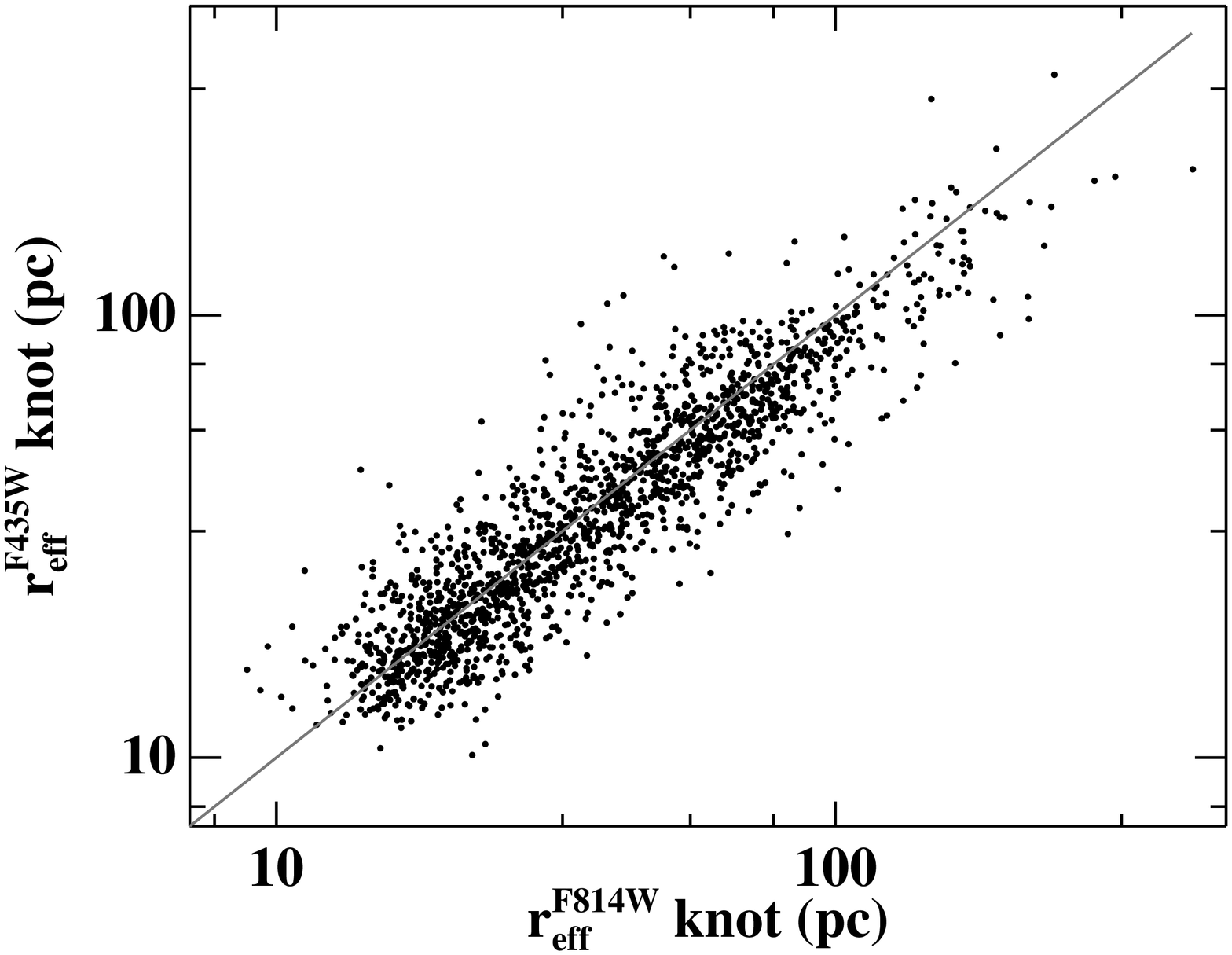} 
\hspace{0.2cm}
\includegraphics[angle=90,width=0.52\columnwidth]{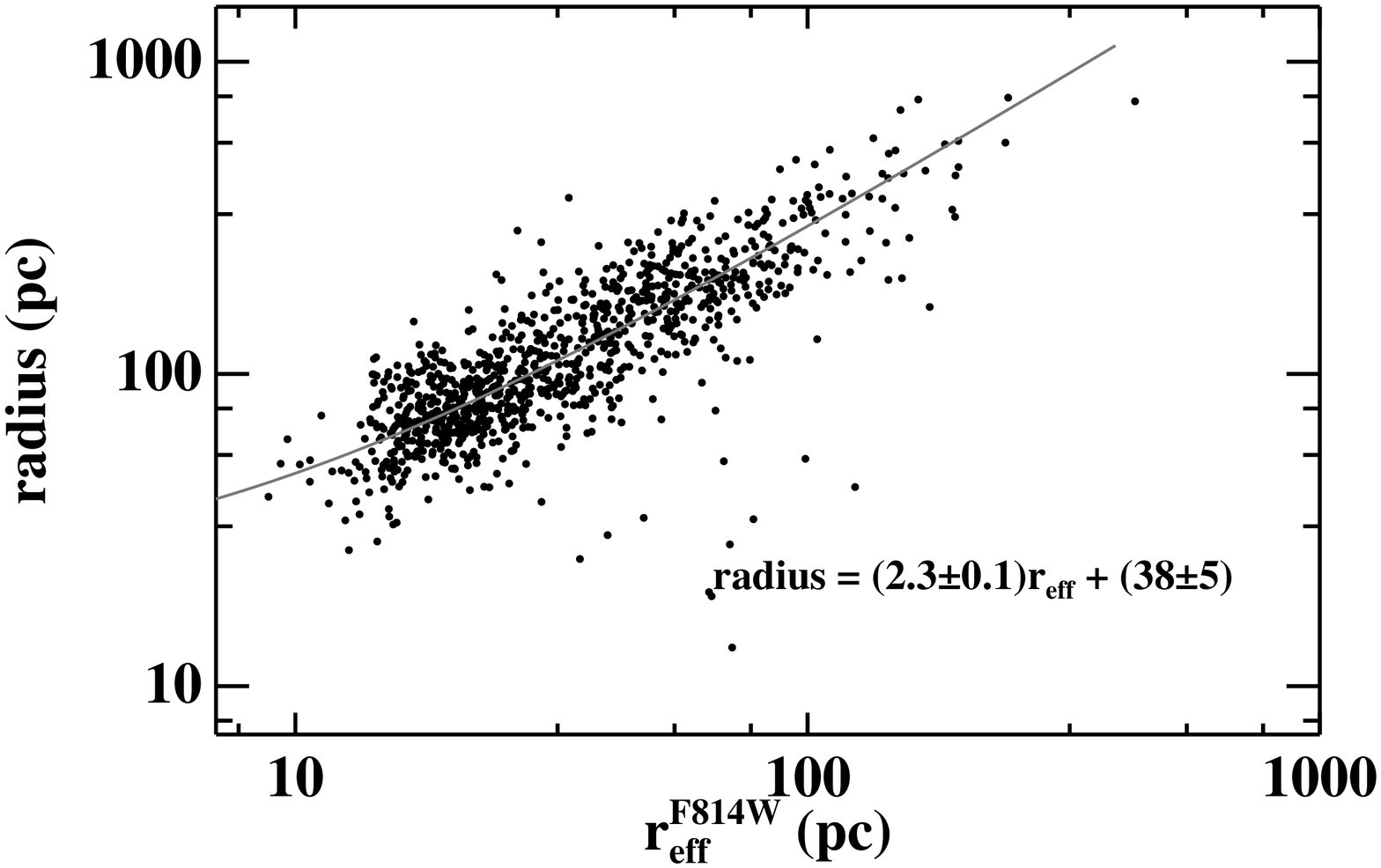} 
   \caption[Sizes of the knots]{\textbf{Left:} Relation between effective radii of all the knots measured in images with filter \textit{F814W} and \textit{F435W}. The line indicates a ratio of unity. \textbf{Right:} Relation between effective radii and total radii of the knots for which we could estimate the total size value with an uncertainty of less than 50\%. The fit for the data is shown at the bottom-right corner. Both axes are in log scale.}
   \label{fig:sizes}
              
 \end{figure}
 
Size measurements were performed by fitting radial Gaussians, technique that has been successfully used in other works (e.g.,~\citealt{Surace98,Whitmore93}). The measured radial profile of a knot (compact region of star formation)  corresponds to the convolved profile of the \psf with the intrinsic profile of the knot.  We then fitted \mbox{2D-Gaussians} to each knot and to foreground stars in the image (to compute the \psf profile). We can estimate the $\sigma$ for the intrinsic profile of the knot as \mbox{$\sigma$ = $\sqrt{\sigma_{m}^2 - \sigma_{PSF}^2}$}, where $\sigma_{m}$ corresponds to the $\sigma$ of the measured profile (average of $\sigma$ in the x and y direction) and $\sigma_{PSF}$ to that of the \psf profile.  An effective radius, \reff\twospace, was derived as the half value of the \acr{}{FWHM}{full width half maximum} derived from the computed $\sigma$ (\fwhm\twospace=2.354$\sigma$). 

Note that \reff defined in this study does not strictly correspond to the half-light radius (which is the common definition of \reff\twospace), since the light profile of the globular clusters and young massive clusters detected in the Milky way and in other galaxies is not necessarily Gaussian. However, given the fact that the distance of 80\% of the systems is larger than 100 Mpc (distance at which the linear resolution of our data is about 20 pc per pixel) we do not expect to measure resolved sizes as small as the typical size of \ymcs (typically, \mbox{\reff $<$ 20 pc};~\mycitealt{Whitmore99}). Sizes of globular clusters in the Milky Way are even slightly smaller. Therefore, it is not worth using more robust methods (e.g.,~\mycitealt{Larsen99b}) that better characterize the light profile of the source, and where the derived \reff is closer to the half-light radius. The main interest here is the relative changes in the sizes of the knots among the different systems.

The rms scatter around the 1:1 line in \reffig{sizes} (left) is about 30\%, from 3 pc to 90 pc for knots with sizes of 10 and 300 pc, respectively. This should give a reasonable estimate of the uncertainty in the measurement of \reff\twospace. Differential extinction, which is wavelength dependent, is also responsible for part of the scatter. Then, under similar local background conditions, knots look a bit more extended when observed with the \filteri filter. Another source of uncertainty comes from the young population. In fact, faint young blue knots easily detectable in the \filterb image but with low S/N ratio in the \filteri image will look more extended in the blue ones. In any case, this measurement is unlikely to be affected by the nebular emission-lines (i.e., \ha\twospace, \hb or oxygen lines), since in the vast majority of the systems these lines do not significantly contaminate the \hst filters used in this study (see~\reffig{filters}).

To assess the minimum size we can achieve by using this technique we estimated which would be the Gaussian that, convolved with that of the \psf profile, would give a total \mbox{\fwhm = \fwhm\twospace$_{PSF}$ + 3{\scriptsize{STD}}}, where  {\scriptsize{STD}} gives the standard deviation of the \fwhm distribution measured for all the stars that were used to compute the \psf profile. For the closest systems we are typically limited to about 10 pc and for the farthest systems, 40 pc.

The method described is appropriate to estimate \reff of cluster-like objects with a size similar to the \psf\twospace. If we want to compute \reff of a galaxy, whose light profile is normally very different, other techniques should be applied. \refapp{galfit} describes the procedure followed to estimate \reff of the galaxies by fitting their 2D profile. 

\subsubsection{Total Size}

The total size of the knots was obtained from the \textit{F814W} images, since this filter is less affected by extinction than the other photometric filter in this study and therefore is more convenient to measure the radial extent of the light of the knots. We derived the edge of the knots at the point where the surface brightness ($\Sigma$), as a function of radius, equals that of the local background. To determine this point, we assumed that the knots are circular and measured the surface brightness in concentric rings, as done in Bastian et al. (\myciteyear{Bastian05a}).

The size of the knots was then derived by fitting the radial profile of the light with a power-law of the form $\Sigma$ \mbox{$\propto$ r$^{-\beta}$}. This function follows the data quite well and corresponds to the projected density profile of a knot with \mbox{$\rho \propto$ r$^{-(\beta+1)}$}, where $\rho$ is in units of~\msun pc$^{-3}$ (\mycitealt{Bastian05a}). For slightly-resolved knots the \psf may be dominating the size measured. We then estimated the half-light radius of the \psf and applied the ratio \mbox{\reffknot/ \reffpsf} to the derived size.

Owing to the complex stellar structure of (U)\lirgs\twospace, a large fraction of them involved in an interaction process, the light profile of the local background of the galaxies is usually very steep and irregular. Thus, the application of this method is somewhat limited. Furthermore, the size of the unresolved knots cannot be derived. We could measure reliable sizes, with an uncertainty of less than 50\%, for a third of the detected knots. However, we could estimate the total size of the remaining knots (as long as they are resolved) by using the relation found between \reff and the total size (see \reffig{sizes}, right). 

\subsect[ssp]{Stellar Population Models}

In this thesis we make use of single stellar population synthesis models in order to derive the age, mass and extinction of the stellar populations under some circumstances (e.g., when the colors are so blue that the degeneracy of the tracks is minimized significantly; or when we know the population is young because \ha emission is detected). If we know the age and the mass of the stellar population we can also use these models the other way round and derive the emission we would measure with the broad-band filters, as done in \refcha{sim}.

\begin{figure}
\hspace{1.2cm}
 \hypertarget{fig:ssp_models}{}\hypertarget{autolof:\theautolof}{}\addtocounter{autolof}{1}
\includegraphics[angle=90,width=0.85\textwidth]{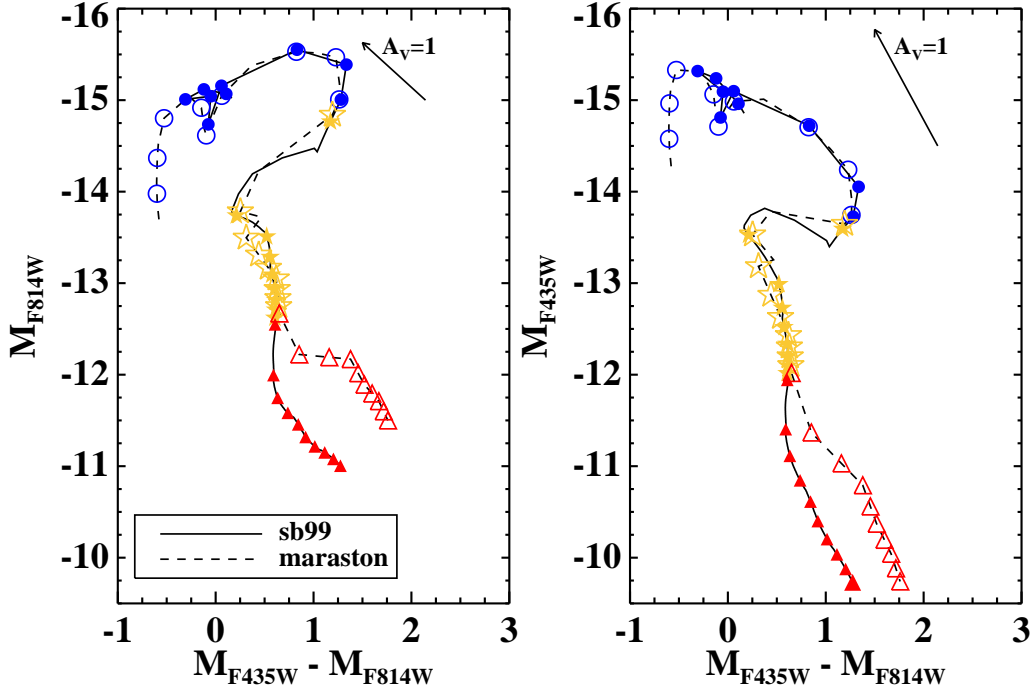}
      \caption[Color-magnitude diagrams of simple stellar population models]{Color-magnitude diagrams according to the \sbnn and the Maraston single stelar population models. The curves are normalized to \mbox{10$^6$~\msun\onespace}. Filled circles, stars and triangles indicate the age in the \sbnn track in steps of 1 Myr (from 1 to 9 Myr), 10 Myr (from 10 to 90 Myr) and 100 Myr (from 100 to 1000 Myr), respectively. Open symbols show the same for the Maraston tracks. The de-reddening vector (\mbox{\av= 1 mag}), computed using the curve of Calzetti et al.(\myciteyear{Calzetti00}), is shown in both plots.}
      \label{fig:ssp_models}
\end{figure}

To perform this kind of analysis, two independent stellar population mo\-dels were considered: the evolutionary synthesis code \acr{}{SB99}{Starburst99} v5.1 (\mycitealt{Leitherer99};~\mycitealt{Vazquez05}) and the code by Maraston (\myciteyear{Maraston05}). The former is optimized for young populations and also models the contribution due to ionized gas whereas the latter gives a rigorous treatment of the thermally pulsing asymptotic giant branch (TP-AGB) phase, which is re\-le\-vant within the interval 0.1-1 Gyr. In both cases we assumed instantaneous burst models with a Kroupa \acr{}{IMF}{initial mass function} (\mycitealt{Kroupa02}) over the range 0.1-120~\msun and solar metallicity. These models are normalized to 10$^6$~\msun\onespace. Mass estimates must be multiplied by a factor of 1.56 if the Salpeter \imf is considered.

The evolutionary tracks (up to 1 Gyr) using both \sbnn and Maraston SEDs are presented in \reffig{ssp_models}. At very young ages (\mbox{log $\tau$ $\lesssim$ 6.5}) the nebular continuum, which is considered in \sbnn\twospace, reddens the colors by up to 0.6 mag. From \mbox{$\tau$ $\gtrsim$ }50 Myr (log $\tau$ = 7.7) up to 1 Gyr the differences between both models also become significant. However, regardless of the model used, the de\-ge\-ne\-ra\-cy in color within the range \mbox{\mbi= 0.6-1.4} is considerable for two rea\-sons: stellar populations with ages within the interval 50-500 Myr have the same \mbi color in the range 0.6-0.8; the Red Super Giant population reddens the tracks at \mbox{ages $\tau \sim$ 7-14 Myr}, when this population dominates the lu\-mi\-no\-si\-ty, hence making it indistinguishable from a population of up to \mbox{$\tau \sim$ 1 Gyr}. Considering this, there is a clear age degeneracy for the color interval \mbox{\mbi= 0.6-1.4}. Since the mass-to-light ratio (M/L) depends on the age of the population, this age degeneracy translates into an uncertainty of about a factor of 100 in the mass estimates. Hence, although the ignorance of the extinction contributes as well, the age degeneracy represents the primary source of uncertainty.

\begin{figure}
\centering
 \hypertarget{fig:ew_plot}{}\hypertarget{autolof:\theautolof}{}\addtocounter{autolof}{1}
\includegraphics[trim = 0cm 0cm 0cm 0cm,clip=true,width=0.95\textwidth]{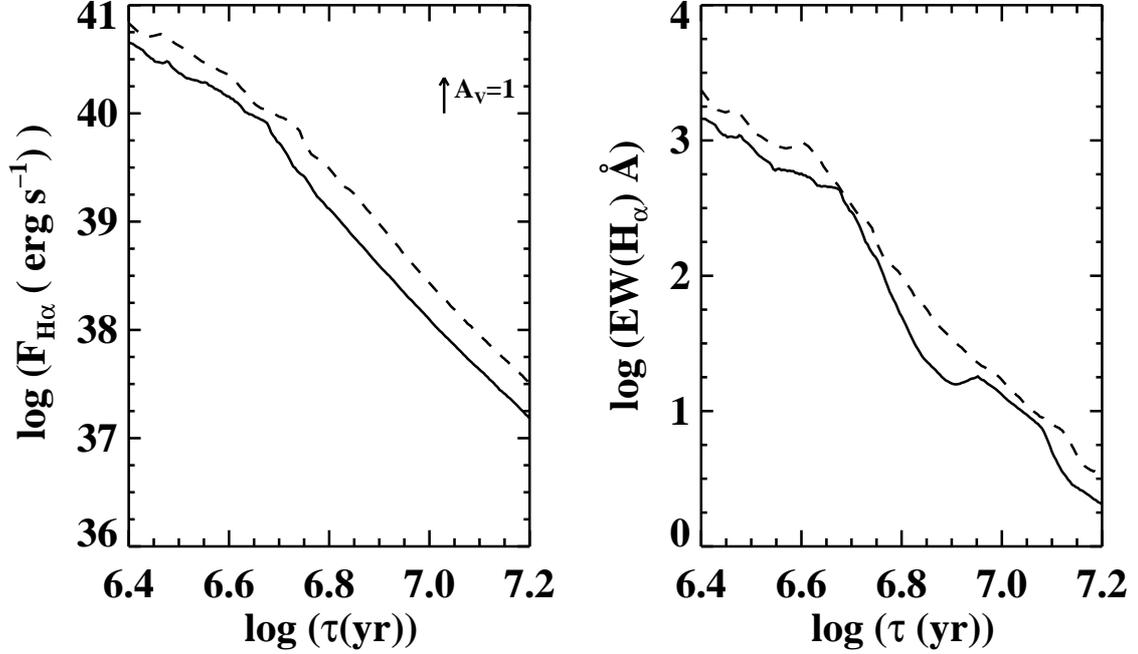}
      \caption[H$\alpha$ and EW evolution of young population, according to two SB99 models]{\textbf{Left:} \ha evolution of young population, according to two \sbnn instantaneous bursts of Z = 0.020 (solid line) and Z = 0.008 (dashed line), and the same IMF and mass as in~\reffig{ssp_models}. The de-reddening vector (\av\twospace=1 mag), computed using the curve of Calzetti et al.(\myciteyear{Calzetti00}), is shown in the top right corner. \textbf{Right:} Equivalent width evolution according to the same models. Note that this plot does not depend on the mass of the stellar population.}
      \label{fig:ew_plot}
\end{figure}

If we have access to spectroscopic measurements, we can use the \sbnn models to constrain the properties of recent episodes of star formation. It is widely known that \ha emission can be used to constrain the properties of recent episodes of star formation. In fact, if the population for which we want to estimate the mass is young (i.e., \mbox{$\tau$ $<$ 10 Myr}), the \ha emission line and its \acr[]{}{EW}{equivalent width} are generally used to constrain the age of such population (see \reffig{ew_plot}). 

\subsect[lf]{Luminosity and Mass Functions}

The luminosity function, which quantifies the number of objects per luminosity bin, can give us indirect information about the underlying mass function of the knots. The initial mass function and the processes that undergo any object once it is formed defines the shape of the \lf\twospace. As mentioned in the previous Chapter, for the young knots and clusters observed in merging systems and other environments, the \lf is well described as a power-law distribution: $dN \propto L^{-\alpha}_{\lambda} dL_{\lambda}$ or the equivalent form using magnitudes, $dN \propto 10^{\beta  M_{\lambda}} dM_{\lambda}$, $L_{\lambda}$ and $M_{\lambda}$ \mbox{being} the luminosity and magnitude respectively and N being the number of clusters. The slopes in both equations are related as $\alpha = 2.5\times  \beta + 1$. 

The shape of the LF has usually been  determined by fitting an equal sized bin distribution. Instead, we used a method described and tested in Ma\'iz-Apell\'aniz \& \'Ubeda (\myciteyear{Maiz05}), based on D'Agostino \& Stephens (\myciteyear{Dagostino86}). It is based on using bins variable in width, such that every bin contains approximately an equal number of knots. Hence, the same statistical weight is assigned to each bin and biases are minimized. Haas et al. (\myciteyear{Haas08}) showed that this method is more accurate than those based on fitting an equal sized bin distribution. The number of bins ($N_{\rm{bins}}$) is different for all samples and is related to the total number of objects in the samples. On the basis of the prescription in Ma\'iz-Apell\'aniz \& \'Ubeda (\myciteyear{Maiz05}), we computed the number of bins using the expression: $N_{\rm{bins}} = 2  N_{C}^{2/5} + 15$, where $N_{C}$ corresponds to the total number of knots. 

To properly fit the slope of the \lf a completeness limit must be set and the fit must be done until this limit. When possible, complenetness limits were derived for each galaxy. \refapp{completeness} describes in detail the method used to compute the completeness limits. The fit was performed on the whole range from the 90\% completeness limit to the brightest knot in the sample, thus the slope of the \lf might be sensitive to an incompleteness of about 10\%. Using a higher completeness limit cuts off a large number of the knots. By correcting for incompleteness at the 90 and 95\% limits the \lf would become steeper (we underestimate the slope of the \lf if we do not correct for incompleteness). The typical value of the underestimate of the slope of the \lf in Haas et al. (\myciteyear{Haas08}) is \mbox{$\Delta \alpha \lesssim$ 0.1} dex for the ACS \textit{F435W} and \textit{F814W} filters. We computed a similar value, so we adopt it for this study. 

The uncertainties associated with the fit of the LF (shown in Tables and Figures throughout the thesis) should be considered lower limits since they do not take into account other sources of error. We performed several tests to establish the robustness of the fits. When the brightest bins were dropped in the fit the values obtained for the slope in all cases agreed within 0.08 dex. When moving the magnitude limit higher by 0.5 mag changes the slope generally agreed within 0.05 dex. Finally, we also explored how the slope varies when performing a linear regression fit to  the cumulative luminosity distribution (see the third case in the Appendix in \mycitealt{Haas08}). The variations here generally agreed within 0.06 dex. Hence, combining all these effects, a more realistic uncertainty for the slopes of the LF that we obtained would be about 0.1 dex.

The determination of the slope of the mass function was done following the same procedure. In this case, the completeness limit was assumed to be the peak of the mass distribution. 

\subsect[rel_astrometry]{Relative $HST$-IFS Astrometry}

In \refcha{tdgs} the photometric images and spectral maps are combined, different datasets with different angular resolution. For an adequate comparison between both data sets, it is necessary to align the images to the same reference system. To that end, the high spatial resolution \hst images were degraded and compared with the VIMOS and INTEGRAL red continuum. Taking advantage of their morphological resemblance, we could establish the position of the spectral-line maps with respect to the \hst images with an uncertainty of about half a spaxel (i.e., about 0.3\arcsec~ for VIMOS maps and about 0.4\arcsec~ for INTEGRAL maps). An example is shown in \reffig{rel_astro}.

\begin{figure}
\centering
 \hypertarget{fig:rel_astro}{}\hypertarget{autolof:\theautolof}{}\addtocounter{autolof}{1}
\includegraphics[trim = 1.5cm 13cm 1cm 6cm,clip=true,width=0.95\textwidth]{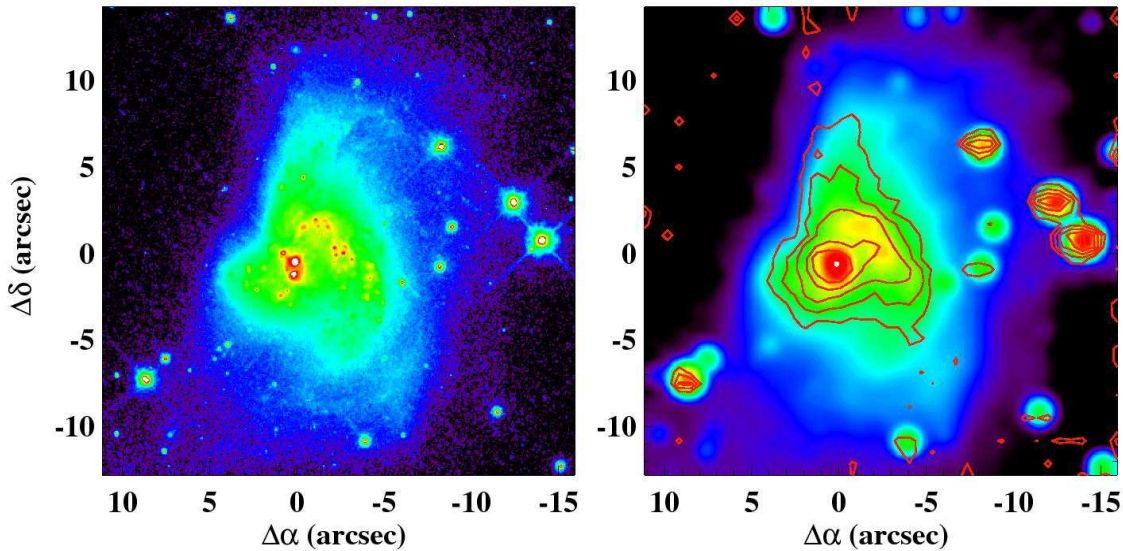}
      \caption[Relative HST-IFS Astrometry for IRAS 08355-4944]{\textbf{Left:} Iras 08355-4944 as observed with the \hst \acs \textit{F814W} filter, where the structural details are clearly visible, provided by the high spatial resolution. \textbf{Right:} The previous \hst image convolved with a Gaussian to simulate the atmosphere + telescope effect is shown in color. The contours represent the VIMOS red continuum close to the \ha line.}
      \label{fig:rel_astro}
\end{figure}

\subsect[metal]{Metallicity Calibrators}

Abundances are usually estimated using empirical methods based on the intensities of several optical lines. The most popular methods are the widely used R23 (\mycitealt{Torres-Peimbert89}) and the S23 (\mycitealt{Vilchez96}) calibrators. However, both methods make use of emission lines that our spectral range does not cover. We used instead the N2 calibrator proposed by Denicolo et al. (\myciteyear{Denicolo02}) and the empirical diagrams of Edmuns \& Pagel (\mycitealt{Edmunds84}). The former is based on the ratio between the \nii\twospace$\lambda$6584 and the \ha emission lines and the latter relates the ratio \oiii\twospace$\lambda$4959 + \oiii\twospace$\lambda$5007 to \hb to the oxygen abundance, parametrized as in Duc \& Mirabel (\myciteyear{Duc98}):

\eqn[n2_cal]{
[N2]:~~~~~~~~~~~ 12 + log (O/H) = 9.12(\pm 0.05) + 0.73(\pm 0.10)\times log \left( \frac{[\mbox{N\scriptsize{II}}] \lambda6584}{H_{\alpha}} \right)
}

\eqn[edmuns_cal]{
[Edmunds]:~~12 + log (O/H) = -0.68 \times log \left(\frac{[\mbox{O\scriptsize{III}}]\lambda4959 + [\mbox{O\scriptsize{III}}]\lambda5007}{H_{\beta}}\right) + 8.74
}

Given the spectral range of the VIMOS data, metallicity determinations were only possible using the N2 calibrator for systems observed with this instrument. The typical uncertainties of the determined metallicities are about 0.2 dex.  For the galaxies observed with INTEGRAL, given these uncertainties, the differences between both indicators (usually within 0.2 dex) are not relevant, and thus we adopted the average value of both determinations.

\subsect[dyn]{Mass Determinations}

\subsubsection{Stellar Masses}

The flux measured with broad-band filters is usually used for mass determinations of stellar populations. In this study, however, only two filters are used and degeneracy issues can make these estimates very much uncertain. However, under some circumstances (i.e., \mbox{\mbi$<$ 0.5}) a first-order estimate of the age of an instantaneous burst can be achieved, according to the \sbnn models (see \reffig{ssp_models} in \refsec{ssp}). Assuming a single burst, once the age of the stellar population is known, the mass-to-light ratio (M/L) given by the models is used to derive the stellar mass of the population that dominates the broad-band emission (in our case under the assumption that only one population dominates such emission). 

As mentioned in \refsec{ssp}, the \ha emission can be used to constrain the properties of recent episodes of star formation (see \reffig{ew_plot}). Alghough the EW does not depend on the mass of the population and is not affected by internal extinction, an underlying old population which does not contribute significantly to the broad-band filters (i.e., the \acs \filterb and \filteri\twospace), if considerably more massive than the young population (i.e., more massive than one order of magnitude), can affect the \ha continuum and consequently the value of the measured \ew\twospace. Therefore, the age determined by using the \ew is normally assumed to be an upper-limit to the real age of the young component in a hypothetical case of a composite stellar population (young and old). As in the case of using only broad-band filters, once the age is known, the mass-to-light ratio given by the models is applied to derive the young stellar mass of the population.

\vspace{2cm}
\subsubsection{Dynamical Masses}
\relax
\hypertarget{sub:eq_dyn_mass}{}
\label{sub:eq_dyn_mass}

The difference between dynamical (i.e., stars + gas + dark matter) and stellar masses relies on the fact that the first is derived taking into account the gravitational field, whereas the latter is obtained based on the integrated light assuming a mass-to-light ratio. The dynamical mass of a bound system can be derived using the virial theorem, under the following assumptions: (i) the system is spherically symmetric; (ii) it is gravitationally bound; and (iii) it has an isotropic velocity distribution [$\sigma^2$(total)=3$\times\sigma_{LOS}^2$], where $\sigma_{LOS}$ is the line-of-sight velocity dispersion.

Assuming virialization, the dynamical of a stellar system is given by two parameters: its velocity dispersion ($\sigma$) and its half mass radius (r$_{hm}$). Depending on the object under study (cluster- or galaxy-like) different relations are typically used:

\eqn[dyn_mass1]{
M_{dyn}/M_{\odot}=m10^6r_{hm}\times \sigma^2 
}

\eqn[dyn_mass2]{
M_{dyn}/M_{\odot}=\eta \frac{r_{eff}\times \sigma_{LOS}^2}{G}
}

\refeqn{dyn_mass1} is generally used for galaxy-/ellipsoidal-like objects. The parameter r$_{hm}$ is given in kpc, $\sigma$ in km s$^{-1}$ and the factor m is a dimensionless function of the assumed mass distribution and ranges from 1.4 for a King stellar mass distribution that adequately represents ellipticals (\mycitealt{Bender92};~\mycitealt{Tacconi02}) to 1.75 for a polytropic sphere with a density index covering a range of values (\mycitealt{Spitzer87}) and 2.09 for a de Vaucoluleurs mass distribution (\mycitealt{Combes95}). 

\refeqn{dyn_mass2} is generally used for cluster-like objects. Here, \reff is given in pc, $\sigma_{LOS}$ in km s$^{-1}$ and the gravitational constant G=4.3$\times10^{-3}$ pc~\msun\onespace$^{-1}$ (km s$^{-1}$)$^2$ (\mycitealt{Spitzer87}) and \mbox{$\eta$ = 9.75} for a wide range of light profiles. However, some studies have shown that $\eta$ is not a constant, and can vary with time (\mbox{$\eta \sim 3-10$}), depending on, for example, the degree of mass segregation and the binary fraction of the cluster (\mycitealt{Fleck06};~\mycitealt{Kouwenhoven08}). Since we do not have that information we use \mbox{$\eta$ = 9.75}.

The half-mass radius is not an observable and cannot be measured directly. It has to be inferred indirectly by measuring the half-light radius (\reff\twospace). The broadening of the emission lines is produced by a few fundamental effects, as described by Melnick et al. (\myciteyear{Melnick99}): mainly thermal broadening, virial broadening, natural broadening and local dynamical effects as expanding shells, filaments and outflows. Once the widths of the principal components ($\sigma _{obs}$) were measured and corrected for instrumental broadening ($\sigma _{ins}$), additional corrections had to be applied for natural ($\sigma _n$) and thermal ($\sigma _{th}$) broadening to obtain what we denote as the virial width according to the following relation:
\eqn[sigmas]{
\sigma = \sqrt{\sigma _{obs}^2- \sigma _{n}^2- \sigma _{th}^2- \sigma _{ins}^2}
}

The natural broadening is a constant for hydrogen and corresponds to 3 $km s^{-1}$ (\mycitealt{Odell88};~\mycitealt{Rozas06}). For \hii regions with temperatures of 5000 K, 10000 K, and 15000 K, Osterbrock (\myciteyear{Osterbrock89}) provides Doppler widths of $\sigma _{th}$ = 6.4, 9.1, and 11.1 $km s^{-1}$, respectively. In the absence of the temperature information for our sample, a uniform value of $\sigma _{th}$ = 9.1 $km s^{-1}$ was adopted. Given the uncertainties of $\sigma _{obs}$ (higher than 5-10 $km s^{-1}$ in our case), the ignorance of the temperature is not worrisome.

\cha{knots}{Optically-selected Compact Star-Forming Regions in (U)LIRGs}
\chaphead{This Chapter presents a comprehensive characterization of the general properties (luminosity functions, mass, size, ages, etc) of optically-selected compact stellar objects (knots) in a representative sample of 32 low-z Luminous and Ultraluminouos Infrared Galaxies. These properties are important to understand their formation and evolution in these systems, which represent the most extreme cases of starbursts in the low-z Universe. Specifically, we investigate how they depend on the infrared luminosity of the system and on the interaction phase it is undergoing. We also investigate how similar these knots are to clumpy structures in galaxies at high-z.}

\sect[intro_knots]{Introduction}

The observation of young and massive compact clusters in different environments (e.g.,~\mycitealt{Meurer95a};~\mycitealt{Larsen99a};~\mycitealt{Whitmore99},~\myciteyear{Whitmore07};~\mycitealt{Weilbacher00};~\mycitealt{Gallagher10};~\mycitealt{Mullan11}) has provided important clues to understand the formation and evolutionary processes in interacting systems. With masses of $10^{4}-10^{6}$~\msun and a median size in terms of effective radius around 3.5 pc, their light follow the same luminosity function (\lf\twospace) in the optical in a wide range of galactic environments, a power-law (\mbox{$\psi$(L)dL $\propto$ L$^{-\alpha}$dL}) with an index of $\alpha$ = 2. Yet, there is some controversy, since the range is rather large depending on the study (see General Introduction,~\refsec{ssclusters}). These clusters associate in star-forming complexes, which represent the largest units of star formation in a galaxy (\mycitealt{Efremov95}). The observation of these objects (e.g.,~\mycitealt{Whitmore99};~\mycitealt{Bastian05a}) has provided a deeper knowledge on the hierarchy of the star formation in embedded groupings.

Although much effort has been made to understand the star formation and evolutionary processes in interacting systems, little is known for Luminous (\lirgs\twospace) and  Ultraluminous (\ulirgs\twospace) Infrared Galaxies, which represent the most extreme cases of starbursts and interactions in our nearby Universe. Given the significant contribution of these galaxies to the star formation rate density at \mbox{z $\sim$ 1-2} (\mycitealt{Perez-Gonzalez05};~\mycitealt{Caputi07}) and their resemblance with sub-millimeter galaxies at higher redshifts (\mycitealt{Smail97};~\mycitealt{Blain02};~\mycitealt{Frayer04}), their properties can also have important cosmological implications. They are natural laboratories for probing how star formation is affected by major rearrangements in the structure and kinematics of galactic disks. Establishing the general properties (luminosity functions, mass, size, ages, etc) of the compact stellar objects in (U)\lirgs as a function of luminosity and interaction phase can provide relevant information in order to understand the mechanisms that govern the star formation and evolution in these systems. It is not known either whether the processes that (U)\lirgs undergo produce stellar objects similar or very different in terms of size or luminosity than in less luminous \mbox{(non-)} interacting galaxies.

In fact, previous studies of compact stellar structures in (U)\lirgs were focused only on small samples, mostly with low angular resolution ground-based imaging (\mycitealt{Surace98};~\mycitealt{Surace00}), or on detailed multi-frequency studies of compact stellar objects in individual galaxies (\mycitealt{Surace00};~\mycitealt{Diaz-Santos07}). Thus, no study has been done so far on a representative sample of luminous infrared galaxies covering the different phases of the interaction process as well as the entire \lirg and \ulirg luminosity range. 

This Chapter presents the first attempt at obtaining an homogeneous and statistically significant study of the photometric properties (magnitudes, colors, sizes and luminosity function) of optically-selected compact stellar objects found in these systems as a function of infrared luminosity, morphology (i.e., interaction phase) and radial distance to the nucleus of the galaxy. For this study we make use of the photometric \textit{B}- and \textit{I}-band \acs images (Chapter~\ref{cha:data_tech},~\refsec{phot_data}) of the representative 32 (U)\lirgs that comprise our sample (Chapter~\ref{cha:data_tech},~\refsec{sample_ini}).  The detected stellar objects, some of them identified as Super Star Cluster candidates for the closest galaxies (i.e., at \mbox{\ld$\lesssim$ 60 Mpc}) and most of them as cluster complexes further away, will be referred to as ``knots''. 

Specifically, the Chapter is organized as follows: we first describe the procedure followed to detected the knots under study in~\refsec{det_phot}; we then characterize the bias that we will encounter throughout the Chapter caused by the distribution of distances of the galaxies in our sample, distance effects, in~\refsec{distance_effect}; in ~\refsec{general_prop_knots} the main photometric properties of the identified knots are presented; next, we characterize these properties as a function of the infrared luminosity of the system in~\refsec{prop_lir}; the characterization as a function of the interaction phase is developed in~\refsec{prop_is};~\refsec{clumps_highz} compares the properties of the detected knots with star-forming clumps in high-z galaxies; finally,~\refsec{summary_knots} gives a brief summary of the most relevant results and the main conclusions.

\sect[det_phot]{Source Detection and Photometry}

Compact regions with high surface brightness (knots hereafter) that are usually bluer than the underlying galaxy  were identified for each galaxy. They were detected inside a box that encloses each system including the diffuse emission observed in the red band. They appear similar to the bright blue knots found in interacting galaxies like in NCG 7252 (\mycitealt{Whitmore93}), NGC 4038/39 (\mycitealt{Whitmore95}), some \ulirgs (\mycitealt{Surace98}), M51 (\mycitealt{Bik03};~\mycitealt{Lee05} and Arp284 (\mycitealt{Peterson09}). These knots were identified on the basis of having a flux above the 5$\sigma$ detection level of the local background. Owing to the irregular and distorted morphologies for most of the systems, we performed a two step analysis to identify the compact regions: we first smoothed the image, averaging in a box of 30x30 pixels, and subtracted this smoothed image from the original one in order to make a first rough local background subtraction; we then ran SExtractor (\mycitealt{Bertin96}) on the resulting image to detect the knots. Given the steepness of the light profile of the local background in the inner regions and its larger deviation with respect to outer regions, an appropriate background subtraction was not always possible and some inner knots were never detected. Thus, visual inspection was needed afterward to include some inner knots and also to eliminate some spurious detections. 

Once the knots were detected, all photometric measurements were done using the tasks PHOT and POLYPHOT within the IRAF (\mycitealt{Tody93}) environment. We first identified the point-like sources against the extended ones, since the  photometry of the former requires aperture corrections. We fitted the knots with a Moffat profile, and those with a \mbox{FWHM $\lesssim$ 1.5$\times$FHWM} the profile of the stars in the field were considered as point-like knots. Note that although they actually consist of point-like and slightly resolved objects, they are considered as point-like objects here just to assign an aperture radius, since both require aperture corrections when the photometry is done. Once their size was derived via psf fitting (see Chapter~\ref{cha:data_tech},~\refsec{sizes}) we differentiated between resolved and unresolved knots.

Aperture photometry with a 3 pixel aperture radius (0.15\arcsec) was performed for the point-like knots. For the resolved knots, we used aperture radii between 5 and 7 pixels (0.25-0.35\arcsec), depending on when the radial profile reached background levels. The \mbox{7-pixel} apertures were large enough to include practically all the light from the knot but small enough to avoid confusion with other sources and to avoid adding more noise to the measurements.  In some cases, the irregular shape of the knots demanded the use of polygonal apertures. In some cases, especially for knots belonging to the most distant galaxies, the polygonal apertures were larger than 7 pixels. Estimates of the underlying background flux were made by using the mode of flux values measured in an 8-13 pixel annulus centered on the computed centroid with POLYPHOT, and eliminating the 10\% of the extreme values at each side of the flux background distribution (e.g., if there is a nearby knot it adds spurious flux at the bright end of the distribution). Most of the knots are apart from each other by more than 7 pixels (0.35\arcsec), hence confusion does not seem to be a serious problem, though in the inner regions of few systems (IRAS 04315-0840, IRAS 15206+3342 and IRAS 20550+1656) there is some overcrowding. 

Photometric calibrations were performed following Sirianni et al. (\myciteyear{Sirianni05}) for the \acs images and Holtzman et al. (\myciteyear{Holtzman95}) for the \wfpc images (hereafter calibration reference papers), and magnitudes computed in VEGA system, as explained in Chapter~\ref{cha:data_tech},~\refsub{optical_images}. Correction for charge transfer efficiency (CTE) was needed in the \wfpc photometry but was not applied to the \acs photometry, since tests in some images showed that the effect was less than 1\% for the faintest regions identified.  Aperture corrections were considered for the point-like knots. To that end, we performed the photometry for isolated stars in the field (from a few to about a hundred, depending on the image) in the same way as we did for the point-like knots, then a second time but using a 10 pixel radius. The ratio of both values gives us an aperture correction up to 10 pixels. From 10 pixels to an \textit{infinite} aperture radius the values in the calibration reference papers were taken, since the \psf profile from 10 pixels on is stable enough regardless of the observing conditions. Typical aperture corrections were about 0.3 and 0.2 mag for the \textit{F814W} and \textit{F435W} filters, respectively. Using the deviation of the different aperture corrections computed with stars in the field we estimated the uncertainty due to the aperture corrections to be of the order of 0.05-0.1 mag in both filters. Typical photometric uncertainties for all the knots lie between 0.05 and 0.15 mag, depending on the brightness of the knot. According to the completeness test performed in~\refapp{completeness}, the systematics affect both filters nearly equally, thus the typical uncertainty in color is between 0.10 and 0.20 mag. 

We corrected all the magnitudes for reddening due to our galaxy taking the values directly from the NASA/IPAC Extragalactic Database (NED), computed following Schlegel et al. (\myciteyear{Schlegel98}). Apart from this, no internal reddening corrections have been applied to the magnitudes reported in this chapter. 

Initially, we probably detected old globular clusters, super star clusters, star complexes, \hii regions, foreground stars, background objects, etc. Some of them were rejected after the photometry was made. Foreground stars were mostly easily identified by their high brightness and red colors and their status as point-like objects. The fields in systems at low galactic latitudes (IRAS 08355-4944, IRAS 09022-3615 and IRAS 12116-5615) are so crowded with foreground stars that some may still  contaminate the photometric sample knots after rejection, especially if a faint foreground star lies within the inner regions of the galaxy. No cluster/knot with color \mbox{\mbi$\gtrsim$ 2.2} fits in stellar population models for an extinction-free starburst at redshift \mbox{z $\sim$ 0}. Since we do not expect much extinction in the outer parts of the galaxies (usually where we do not see diffuse emission from the galaxy in the red band), all the red knots outside the diffuse emission of the systems are likely to be background objects and were also rejected. After these rejections a total of 2961 knots were considered under study. 

The nuclei of the systems were identified using the \textit{H}-band images and with the aid of other works that have detected them in radio (e.g.,~\mycitealt{Dinshaw99}). The identification of the true nucleus is of relevance in the derivation of the projected distances of the knots with respect to the closest nucleus, to properly study the spatial distribution of these knots. 

\sect[distance_effect]{Distance Dependence of the Photometric Properties}

\begin{figure}
  \hypertarget{fig:simulation}{}\hypertarget{autolof:\theautolof}{}\addtocounter{autolof}{1}
  \hspace{0.6cm}
   \includegraphics[angle=90,width=0.95\textwidth]{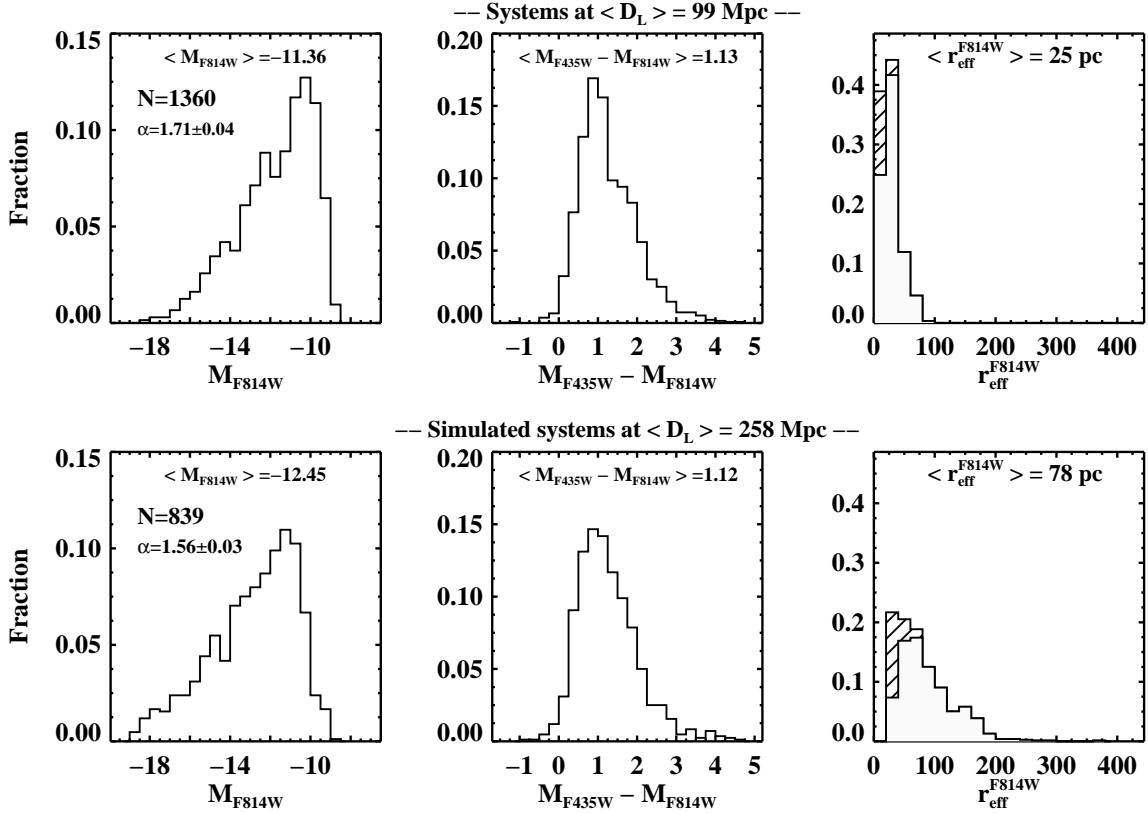}
   \caption[Photometric properties for real and simulated data]{Photometric properties for real (top) and simulated (bottom) data. The median values (between brackets $<~>$ throughout the chapter) of the magnitude, color and \reff distributions are shown. The slope of the \lf for the magnitude distributions is also presented ($\alpha$). Throughout the Chapter N represents the number of detected knots in any given distribution.}
   \label{fig:simulation}
\end{figure}

Before going any further, we first consider a selection effect that we can encounter when studying the photometric properties of the knots in (U)\lirgs located at different distances. The classical Malmquist bias is caused by the fact that systematically brighter objects are observed as distance (and volume) increases, as a result of a combination of the selection and the intrinsic scatter of absolute magnitudes. 

Our sample is affected by this bias, since the systems are located at different distances (from 65 to 560 Mpc). As the system stands further, the faintest knots become undetected and we can only see the bright ones. Furthermore, for the same angular resolution the spatial resolution decreases linearly with distance, and therefore the knots that we detect at large distances can consist of associations of knots instead of individual knots as those observed in less distant systems. Thus, larger sizes are computed. Though this is not a complete sample,~\reffig{LirVsz} in Chapter~\ref{cha:data_tech} shows that in general more distant \mbox{galaxies} have higher infrared luminosity. The same tendency is seen if the galaxies from the IRAS Revisited Galaxy Sample (RBGS,~\mycitealt{Sanders03}) are taken. This effect has to be subtracted in order to observe reliable intrinsic differences between systems at different distances.

We divided the sample in three \lir intervals, so as to study the dependence of the photometric properties with the total infrared luminosity of the system (see~\refsec{prop_lir}). To asses how the photometric properties depend on the distance of the system a simulation was performed. The median distance of the low luminosity interval (\mbox{\lir$<$ 11.65}) corresponds to 99 Mpc, whereas that of \ulirgs (the high luminosity interval) is \mbox{258 Mpc}. Hence, for this sample \ulirgs are located a factor of 2.6 further away than the low luminous \lirgs\twospace. We simulated the light emission of galaxies within the first \lir interval as if they were a factor of 2.6 further, obtaining the same median distance as in \ulirgs\twospace. To that end, the \hst images were convolved with a Gaussian at the resolution of the \psf at the new distance of each galaxy and re-binned in order to have the same ``pixel size'' at that distance. Using foreground stars we computed a multiplication factor for each image in order to preserve the flux, since the flux of the point-like objects must be preserved. Then, photometry was carried out as in the original images, with circular and in some cases polygonal apertures. 

The results of the simulation for the \textit{F814W} filter are presented in~\reffig{simulation}. The magnitudes and the sizes increase \mbox{by about 1.1} mag (a factor of 2.8 in flux) and by a factor of 3, respectively. The number of detected objects drops by about a factor of 1.6. Therefore, in the simulated image, individual knots are associations of knots from the original one. As a result, not only do the sizes become larger, but its distribution also flattens. The median values of the \reff distribution in the simulation are 25 and 78 pc for the real and simulated knots, respectively. 

The flattening of the luminosity function (from $\alpha$ = 1.71 to 1.56) is also expected, since as the knots are artificially grouped due to angular resolution effects the bright tail of the \lf starts to get more populated. Another expected result of the simulation is that the color distribution does not change significantly. 

If we consider the spatial distribution of the knots (inner and outer knots, with projected distances of less and more than 2.5 kpc to the nucleus, respectively), some changes are also observed; the ratio of inner to outer knots shifts from 0.7 to 0.5. Owing to the higher surface brightness and steeper background profile in the inner regions, when convolving the images to get lower linear resolution, a large fraction of inner knots are diluted by the local background, in contrast to a lower dilution degree in the outermost regions. Furthermore, more grouping is expected in the innermost regions, where there is more crowding, hence also losing a larger number of inner knots when compared with those located in the outer regions.

\sect[general_prop_knots]{General Properties of the Knots}

Owing to the high angular resolution and sensitivity of the \acs camera, and to the definition of our large sample, covering a wider \lir  range than previous studies focused only on \ulirgs (\mycitealt{Surace98};~\mycitealt{Surace00}), we have detected close to 3000 knots. This is more than a factor of ten larger than in previous investigations and allows us to carry out a statistical study exploring the different physical properties of the knots over the entire \lirg and \ulirg luminosity range.  We have then performed the photometry in both \textit{F435W} and \textit{F814W} filters for these knots. For this study we consider the knots with positive detection in both filters. The objects identified as the nuclei were not, since many of them are believed to be contaminated by an AGN (see Chapter~\ref{cha:data_tech},~\reftab{sample}, and the references regarding the spectral class).

\begin{figure}
  \hypertarget{fig:mag_col_gen_knots}{}\hypertarget{autolof:\theautolof}{}\addtocounter{autolof}{1}
   \hspace{1.7cm}\includegraphics[width=0.85\columnwidth]{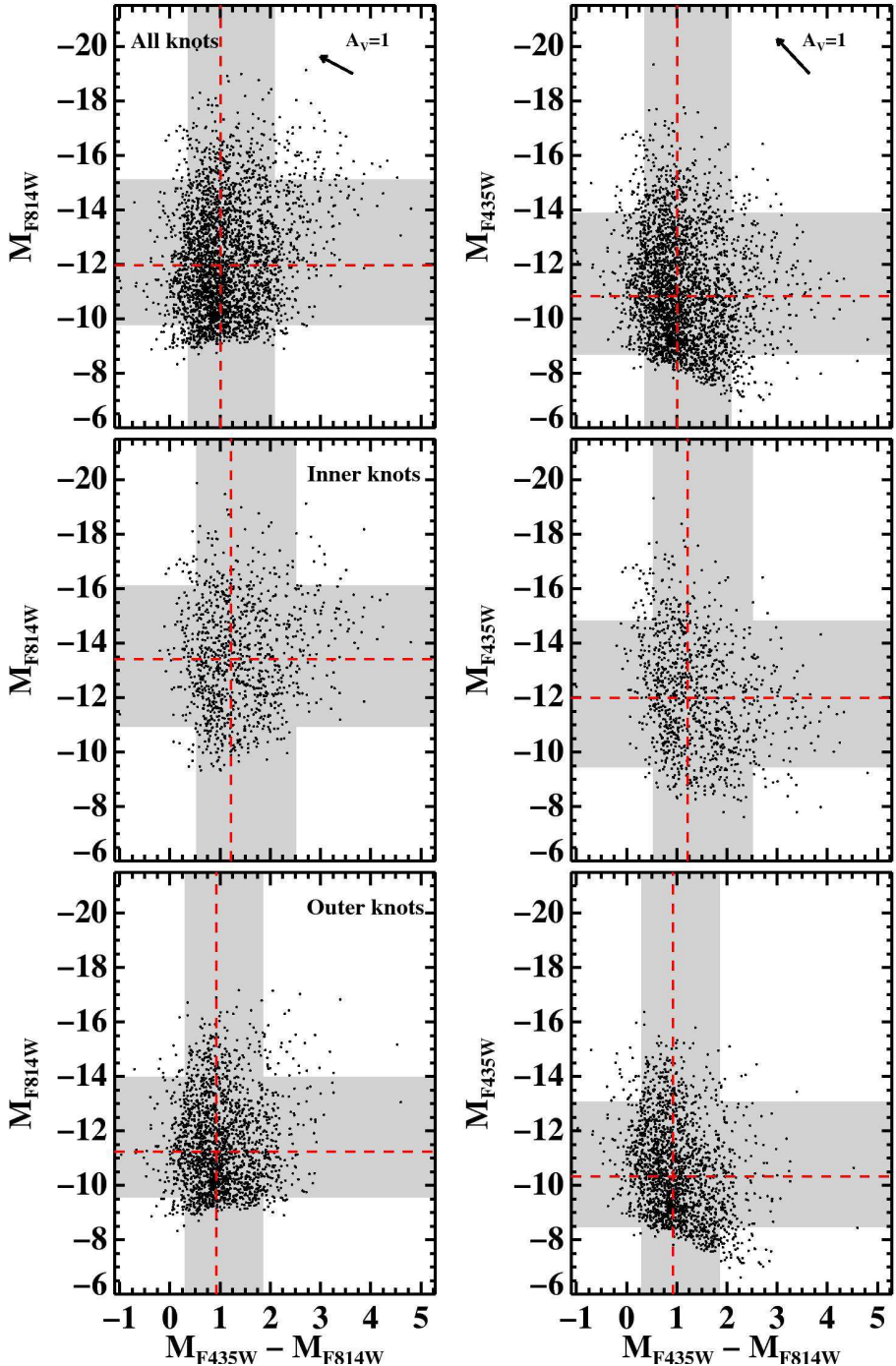}
   \caption[Color-magnitude diagrams for the identified knots]{Color-magnitude diagrams for the identified knots. \textbf{Top:} All knots. The red dashed line indicates the median of the distribution and the gray band covers 80 \% of the total number of knots in each plot, rejecting the 10 \% at either side of the distribution. No reddening corrections have been applied to the magnitudes given. A de-reddening vector of \mbox{A$_{V}$ = 1 mag} is shown in the top right corner. \textbf{Middle:} the same diagrams for the knots in this study within 2.5 kpc. Bottom: same as before but for knots further than 2.5 kpc from the closest nucleus.}
   \label{fig:mag_col_gen_knots}
 \end{figure}

\subsect[mag_col_gen_knots]{Magnitudes and Colors. Average Values}

The knots detected in our sample of luminous infrared galaxies have observed \textit{F814W} and \textit{F435W} absolute magnitudes (uncorrected for internal extinction) in the  \mbox{-20 $\lesssim$ \mi $\lesssim$ -9} and \mbox{-19.5 $\lesssim$ \mb $\lesssim$ -7} range, respectively, while colors cover the -1 $\lesssim$ \mbi $\lesssim$ 5 range (see~\reffig{mag_col_gen_knots}).  The corresponding median magnitudes are \mbox{$<$ \mi $>$=-11.96} and \mbox{$<$ \mb $>$=-10.84}. The median color of the knots  corresponds to \mbox{$<$ \mbi $>$ = 1.0}, which clearly indicates the presence of a young stellar population (about half of the knots have colors bluer than that value). By contrast, the mean color of the diffuse local background light, which traces the old stellar population in the host galaxies, is 1.72, consistent with the typical colors measured in spirals (\mbox{B-I=1.80};~\mycitealt{Lu93}). 

A small fraction of the knots (2\% of the total) show extreme red colors (\mbox{\mbi\twospace = 3} or redder), likely tracing regions with very high internal obscuration, usually located in the more central regions (\mycitealt{Alonso-Herrero06};~\mycitealt{Garcia-Marin09b}). If the stellar population in these knots was dominated by young stars (i.e., \mbox{$\tau$ $\sim$ 10 Myr}), the extinction could be as high as \mbox{A$_V$ = 5-7} mag. Heavily obscured young clusters with \mbox{\av = 5-10 mag} have also been detected in the optical in less luminous interacting systems like the Antennae (e.g., clusters S1\_1 and 2000\_1, whose extinction  was computed with the aid of infrared spectroscopy;~\mycitealt{Mengel08}). 

\subsect[mag_col_rad_knots]{Magnitudes and Colors. Radial Distribution} 

\begin{figure*}
  \hypertarget{fig:spatial_dist_knots}{}\hypertarget{autolof:\theautolof}{}\addtocounter{autolof}{1}
   \hspace{0.5cm}\includegraphics[trim = 0cm 1cm 1cm 4cm,clip=true,width=0.95\textwidth]{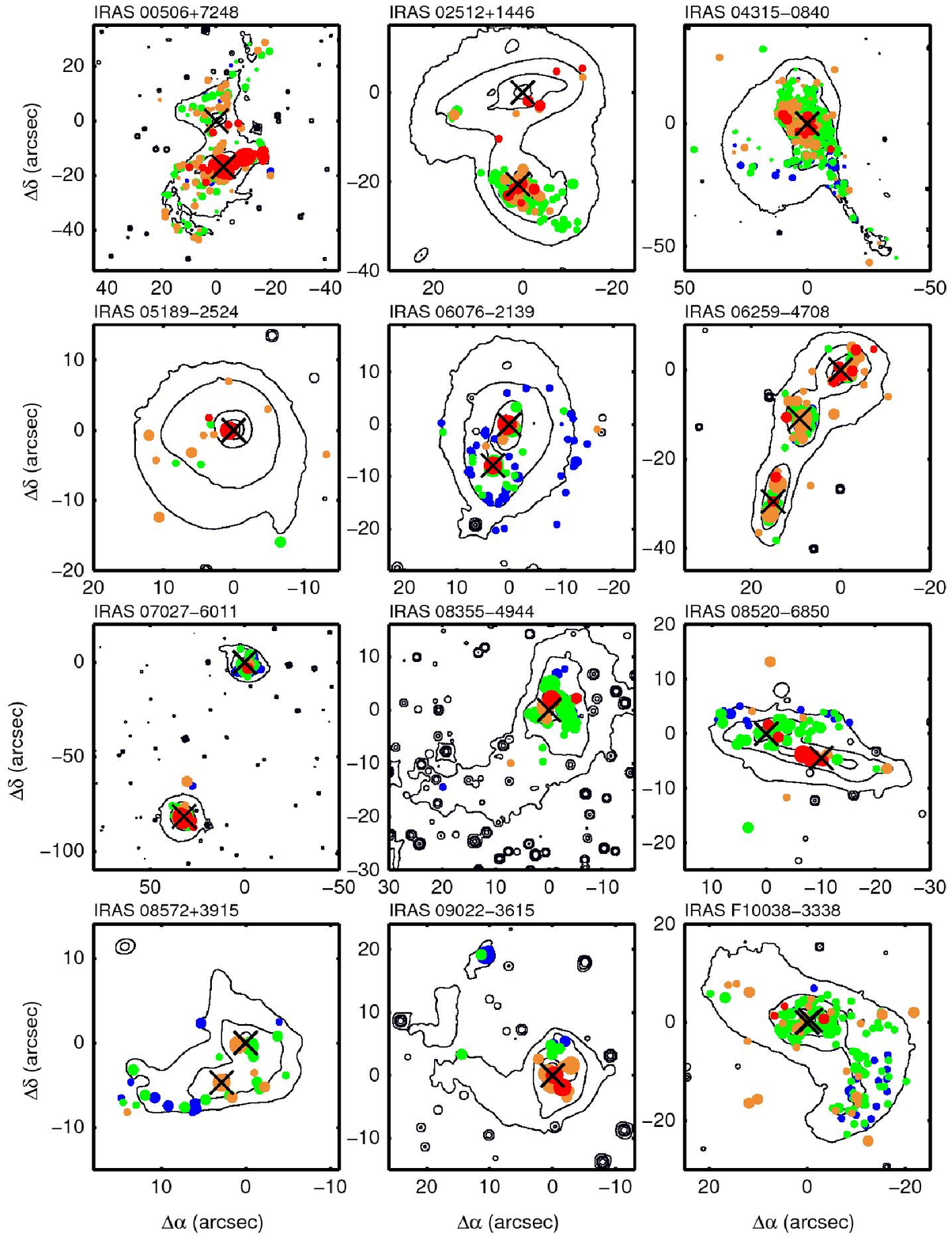}
   \vspace{-0.5cm}
   \caption[Spatial distribution of the knots]{Spatial distribution of the knots measured on the isophotal map of each system. Each \textit{F814W} image has been smoothed to diminish pixel-to-pixel noise and then avoid spurious contours. The field of view is the same as in~\reffig{composite34} in Chapter~\ref{cha:data_tech}. The symbol 'X' marks the knot identified as the nucleus . From bluer to redder colors, knots are blue, green, orange or red, and the sizes plotted depend on their \mi~(see the legend at the end of the figure). North points up and East to the left.}
   \label{fig:spatial_dist_knots}
              
    \end{figure*}

  \begin{figure*}
   \hspace{0.5cm}\includegraphics[trim = 0cm 0cm 0cm 4cm,clip=true,width=1.\textwidth]{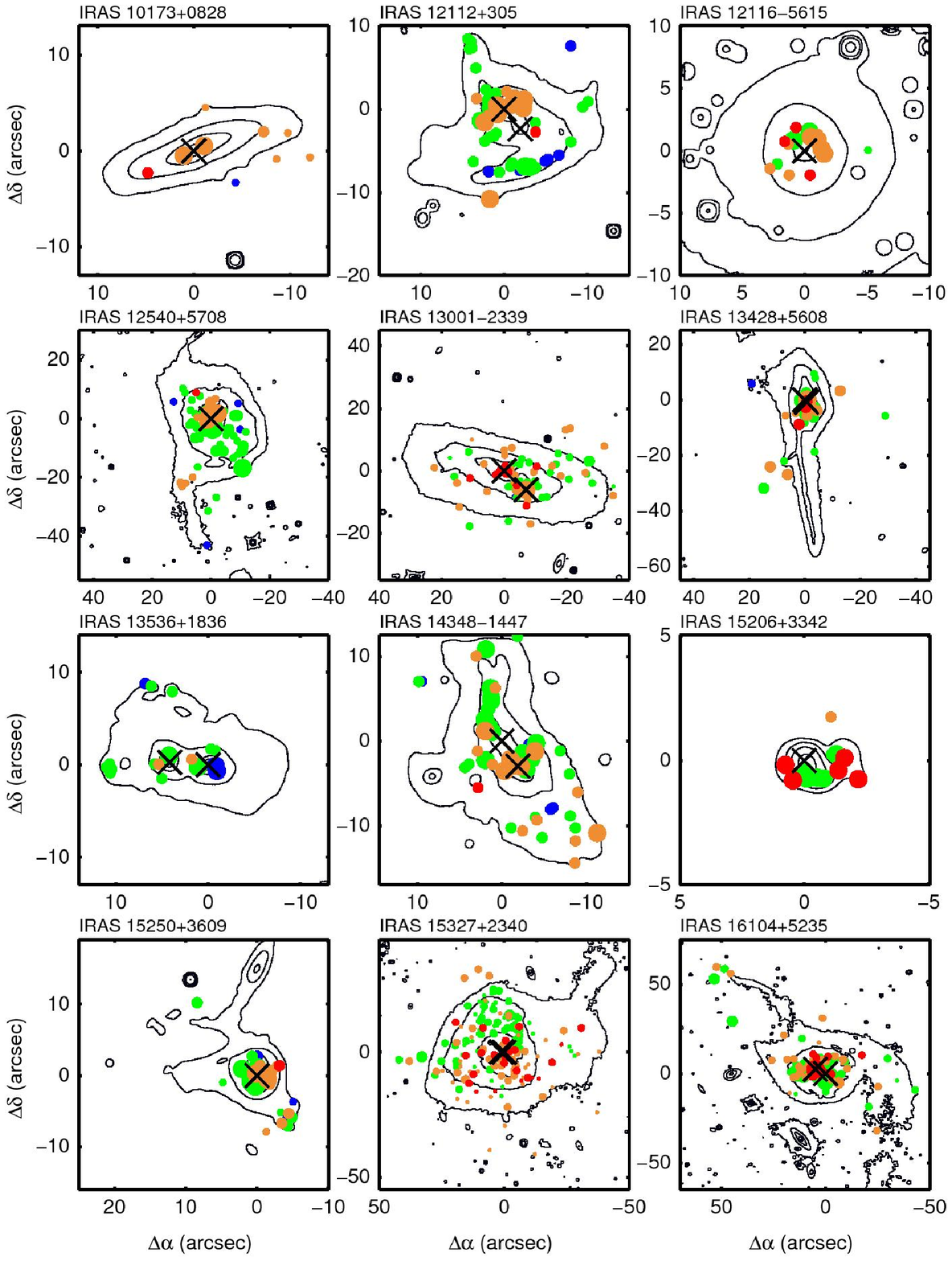}
\addtocounter{figure}{-1} 
 \vspace{-0.5cm}
   \caption[]{- Continued}
    \end{figure*}
  \begin{figure*}
   \includegraphics[trim = 0cm 3cm 0cm 4cm,clip=true,width=1.\textwidth]{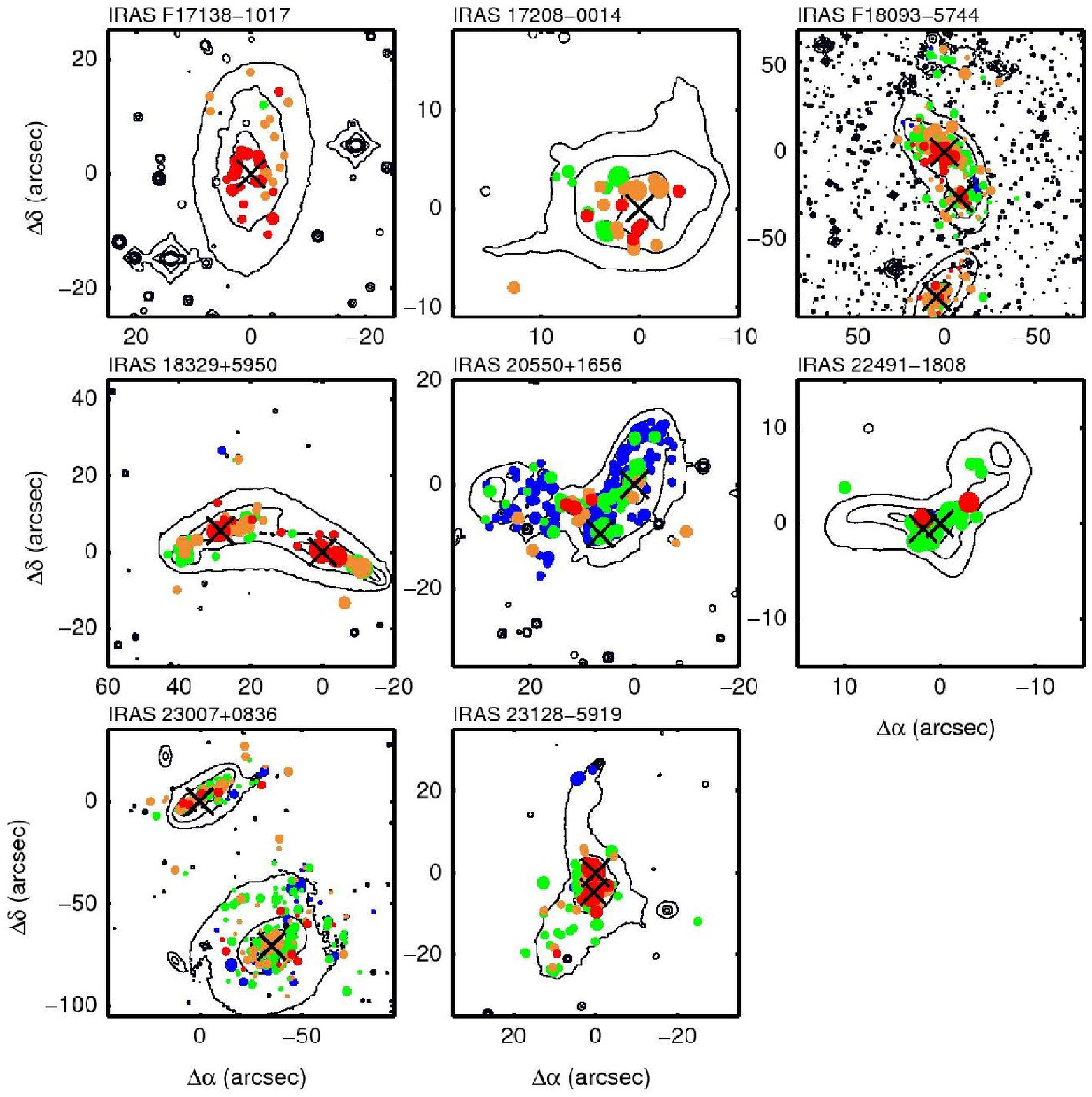}
	\vspace{-3cm} 
\addtocounter{figure}{-1}   
   \caption[]{- Continued}
   \vspace{-6.5cm}
  \hspace{9.4cm}
   \includegraphics[trim = 0cm 0cm 0cm 0cm,clip=true,angle=90,width=0.35\textwidth]{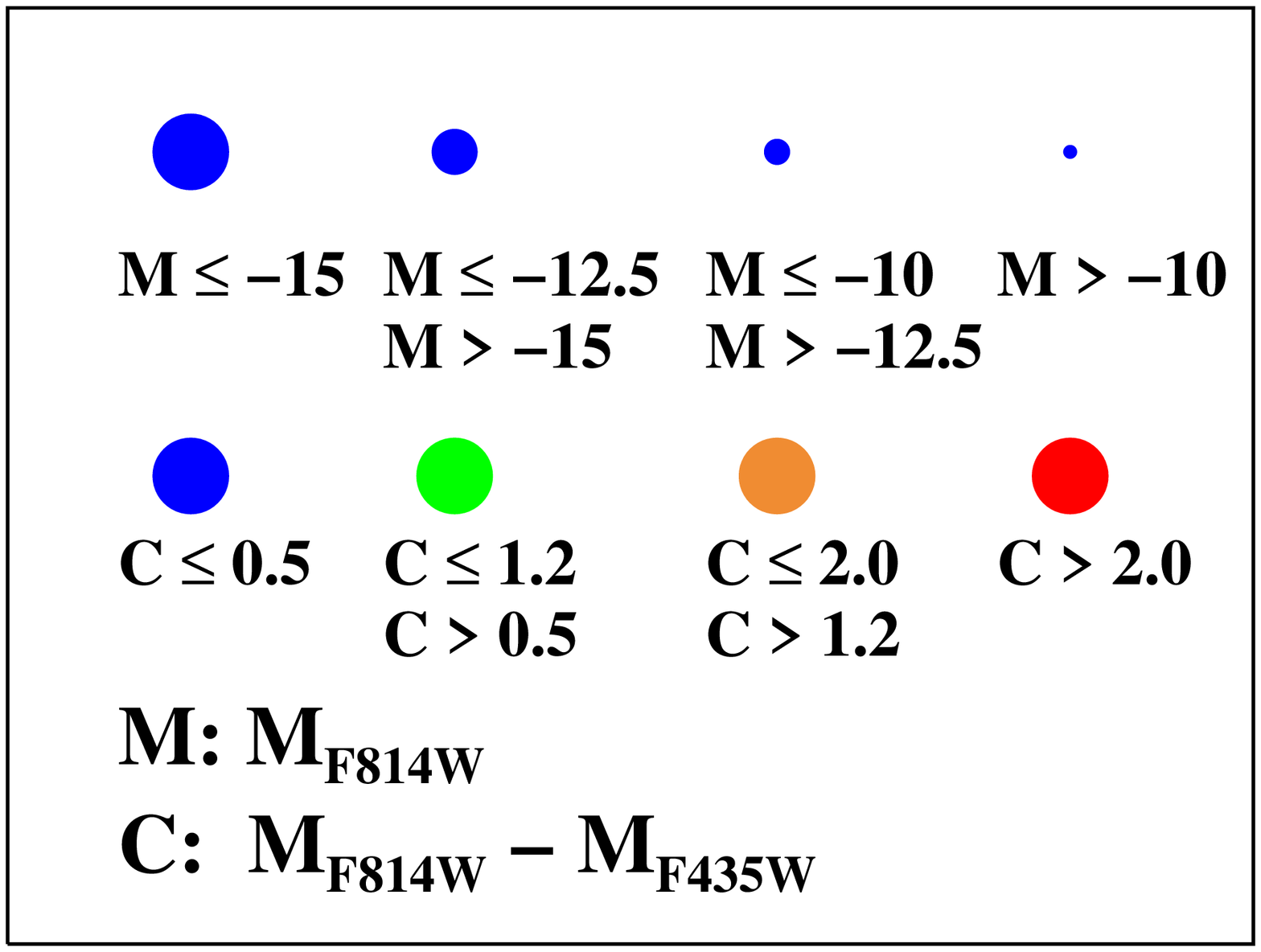}
	\vspace{2cm}
\end{figure*}

To understand the star formation in luminous infrared galaxies, it is important to investigate whether the observed properties of the knots (magnitudes, colors and sizes) do show dependence on the galactocentric distance. The spatial distribution of the knots is shown in~\reffig{spatial_dist_knots}, where their sizes are related to their absolute magnitude and their color to the photometric color \mbox{\mbi\twospace}. About one-third of the knots lie within a projected galactocentric radius of 2.5 kpc while this fraction increases up to about two-thirds within a radius of 5 kpc. Thus, a substantial fraction of the knots are located close to the nucleus at distances of less than 5 kpc. Also, as can be seen in the figure, they mainly represent the population with the reddest colors, likely indicative of higher internal extinction, although an older population can not be excluded. Yet, there is also an external population of knots (10.8\% of the total) at distances of more than 10 kpc and up to 40 kpc, located along the tidal tails and at the tip of these tails. Most of these external knots are blue (\mbox{\mbi$<$ 1}), and some (5.7\%) are very luminous (\mbox{\mb$<$ -12.5}), suggesting the presence of young massive (\mbox{M $>$ 10$^5$~\msun} in young stars) objects in the outer parts of these systems. Examples of these knots can be seen at the \mbox{northern} tip in IRAS 23128-5919 and in IRAS 09022-3675, along the northern tail in IRAS 14348-1447, etc. 

Throughout the chapter, we identify the inner sample of knots as those with projected distances of less than 2.5 kpc to the closest nucleus, and the outer sample as those with distances of at least 2.5 kpc. Both distributions have a significantly different magnitude and color distribution (see~\reffig{mag_col_gen_knots}), the inner knots being 2 mag brighter (in the \textit{F814W} filter) and 0.3 mag redder than the outer knots. These two distributions are different, and do not come from the same parent distribution according to the \acr{}{KW}{Advanced Kolmogorov-Smirnov} test (using the idl routine KSTWO;~\mycitealt{Press92}). Finally, the inner knots also have a larger color dispersion with a tail toward red colors. This is likely the effect of having higher and patchier internal extinction in the innermost regions, in agreement with recent spectroscopic studies of these systems (\mycitealt{Alonso-Herrero06};~\mycitealt{Garcia-Marin09b}). Additionally, the brighter blue magnitudes in the inner regions suggest that there must be more young star formation there than in the outer field.

\subsect[size_gen_knots]{Effective Radius of the Knots}

The angular resolution limit (and therefore upper sizes for unresolved knots) varies from around 10 to 40 pc, depending on the distance to the galaxy. The detected knots do show a wide range of sizes ranging from unresolved (12\%) to a few with very extended sizes (a radius of 200 pc, and up to 400 pc). The median size of the resolved knots is 32 pc, with about 65\% of them smaller than 40 pc. This is significantly larger than the largest knots found in other systems, and several times larger than their mean knot radii (e.g., 10 pc in~\mycitealt{Whitmore99}). Given the distances of the galaxies in the sample (from about 65 Myr to more than 500 Myr) and the resolution limit, in general  the knots identified in our galaxies are likely aggregates of individual clusters, which would increase their (apparent) size. According to the simulations (see~\refsec{distance_effect}), the apparent size of the knots changes from an average of 25 pc to 78 pc when galaxies at an average distance of 100 Mpc are moved at distances 2.6 times further away.

If we consider the sample of inner and outer knots, their distributions are similar (\reffig{reff_gen_inner_outer}), so there is no evidence of a radial dependence of the sizes at the angular resolution of our data.

\begin{figure}[!t]
 \hypertarget{fig:reff_gen_inner_outer}{}\hypertarget{autolof:\theautolof}{}\addtocounter{autolof}{1}
\hspace{0.5cm}
   \includegraphics[trim = -1cm -1cm 0cm 11cm,clip=true,width=0.85\columnwidth]{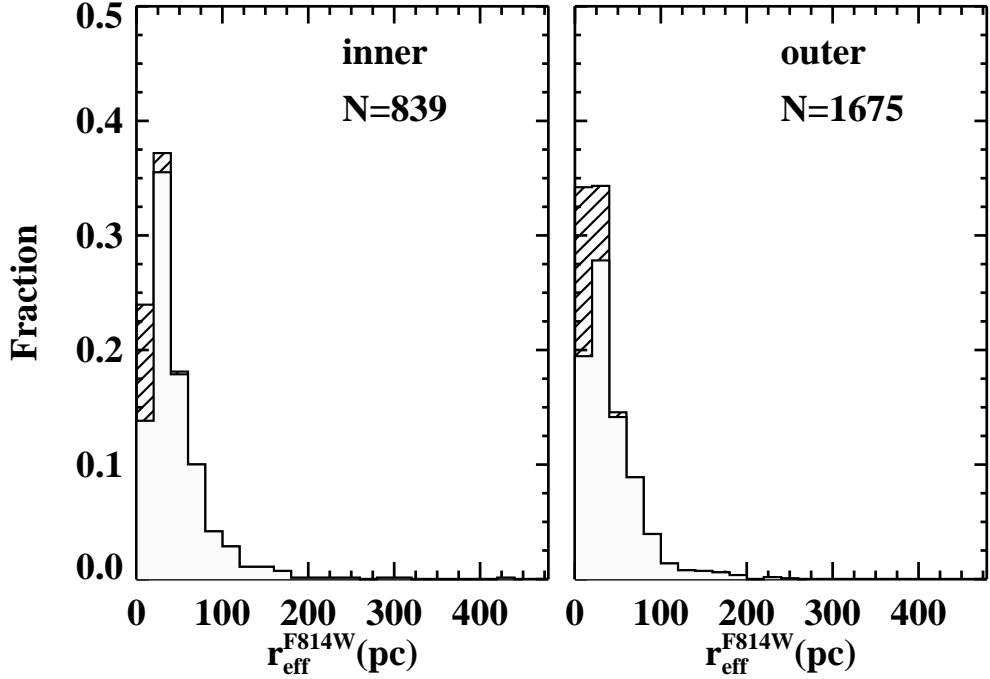}
   \caption[Effective radius for inner and outer knots]{Effective radius for inner and outer knots. Throughout the thesis the resolved sizes are grouped into blank bins. The knots with unresolved sizes are grouped into hatched bins, above the blank bins.}
   \label{fig:reff_gen_inner_outer}           
 \end{figure}
 
\subsect[ages_masses_knots]{The Bluest Knots. Age and Mass Estimates}

The determination of ages and masses of stellar populations is highly degenerate when using a single color index. Moreover, uncertainty (or lack of knowledge) in the internal extinction in heavily obscured systems like the luminous infrared galaxies considered here, adds more degeneracy (see Chapter~\ref{cha:data_tech},~\refsec{ssp}). However, a first-order estimate of the age and mass can be obtained for a specific range in magnitude and colors where degeneracy can be minimized. In particular, knots with colors \mbi $\lesssim$ 0.5 (hereafter blue knots) must be young and should not be very much affected by degeneracy and  extinction, according to the stellar population models considered in this study. They are bright (\mbox{$<$\mb\onespace$>$ = -11.5}) and in general found in the outer regions of the systems, with some exceptions such as IRAS 13536+1836 and IRAS 20550+1656 where they are also detected very close to the nuclei (see~\reffig{spatial_dist_knots}). A first-order estimation of the age and mass can be achieved for the blue knots with colors \mbox{\mbi $\lesssim$ 0.5}. In that color interval the mass uncertainty reduces considerably down to a factor of 2-4 if a single stellar population is considered (see tracks in Chapter~\ref{cha:data_tech},~\reffig{ssp_models}). That color can only indicate young population (if \mbox{\mbi$ \lesssim$ 0.5}, then \mbox{$\tau$ $\lesssim$ 30 Myr}), hence the \sbnn tracks were used for this estimation. In case of having some degeneracy (contribution from the RSG branch), the average age was taken, in order to minimize the uncertainty of the estimation.

Note that the embedded phase of the young population in any environment can last few Myr and that (U)\lirgs are known to have young population highly enshrouded in dust. Previous works have measured very high extinction values for clusters younger than about 3 Myr in different environments (\mycitealt{Larsen10}). They have measured extinction values as low as \av= 0.5 mag for some clusters older than 3 Myr. Finally, it is thought that the embedded phase does not last more than 5 Myr, since they have observed clusters older than that age which are already extinction-free. Hence, the knots for which we tried to make age and mass estimates, though young, are likely to be older than 3 Myr, probably even older than 5 Myr.

\begin{figure}[!t]
 \hypertarget{fig:ages_mass_knots}{}\hypertarget{autolof:\theautolof}{}\addtocounter{autolof}{1}
\hspace{0.8cm}\includegraphics[angle=90,width=0.9\columnwidth]{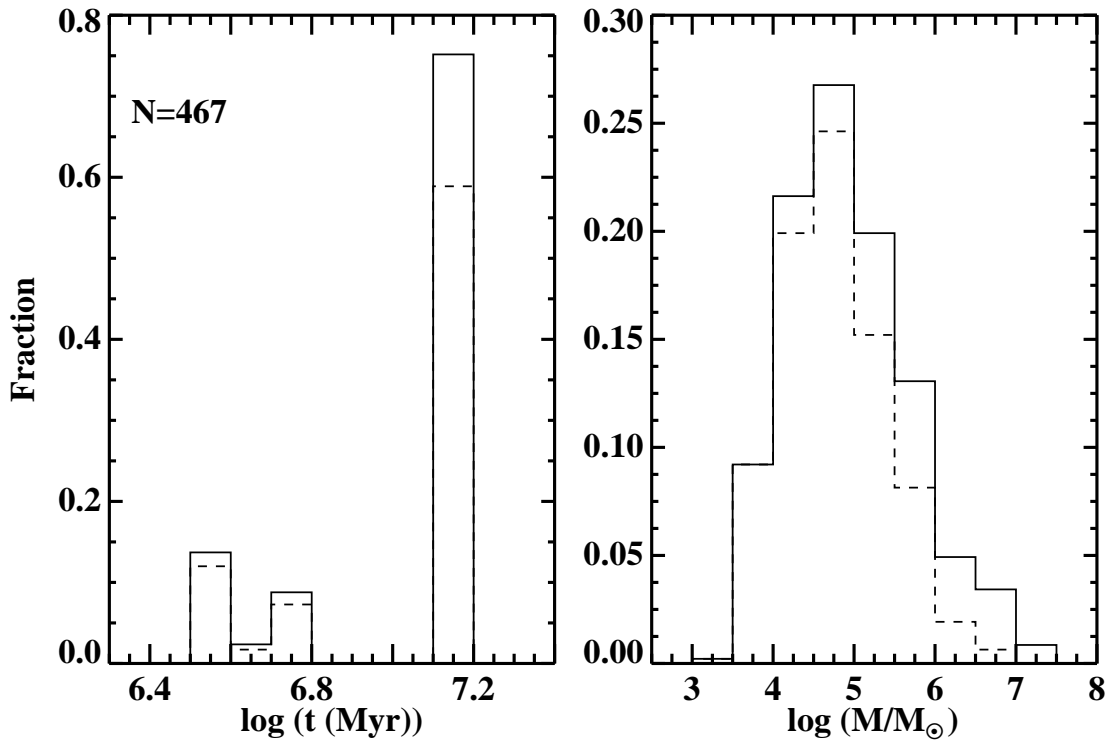}
\caption[Ages and masses estimated for the bluest knots]{Ages and masses estimated for the knots with color \mbi $\lesssim$ 0.5, where the degeneracy in the evolutionary tracks is less severe. The dashed line histograms refer to the external knots (projected distance $>$ 2.5 kpc).}
\label{fig:ages_mass_knots}
\vspace{-0.5cm}
\end{figure}

Under these assumptions and considering all caveats above, an estimate of the age and mass of the blue knots was performed, assuming a single stellar population (see~\reffig{ages_mass_knots}). If small extinction was present in these regions, the estimated mass would be increased by a small amount (up to a factor of 1.8 for A$_V$=1 mag). Some blue knots seem to be more massive than the most massive old globular clusters in the Milky Way, which have masses up to 10$^6$~\msun (\mycitealt{Harris01}). However, most of the blue knots have masses in the 10$^4$ to few 10$^6$~\msun range, similar to those of giant \hii regions and brightest Young Massive Clusters in other less luminous interacting galaxies (e.g., ~\mycitealt{Bastian05b};~\mycitealt{Whitmore99};~\mycitealt{Mengel08};~\mycitealt{Konstantopoulos09}). Complexes of less massive star clusters can also constitute these blue knots.

It is interesting to mention that the blue knots in the outermost regions (i.e., \mbox{d $>$ 2.5 kpc}) represent the 80\% of the total number, while most of the star formation is likely to be occurring within the central regions of the galaxies, as mentioned in~\refsec{mag_col_rad_knots}. This suggests that in general in the outermost regions the dust extinction is very low in (U)\lirgs\twospace. Hence, the most obscured knots must be located close to the nuclei, as mentioned in the previous section and in agreement with recent spectroscopic studies of these systems (\mycitealt{Alonso-Herrero06};~\mycitealt{Garcia-Marin09b}). Part of the difference between the number of blue knots in the inner and outer regions may also be caused by a stronger disruption process in the central regions, as claimed by Haas et al. (\myciteyear{Haas08}), as well as by different histories of star formation of the knots in both fields.

The number of blue knots with estimated masses higher than 10$^6$~\msun ($\sim$ 10\%) drops rapidly in the external areas (the dashed line histogram in~\reffig{ages_mass_knots}). Hence, the most massive young knots are also located close to the nuclei. Nevertheless, there are still some few knots in the outskirts of the galaxies as massive as \mbox{10$^6$-10$^7$~\msun} in young stars. Some contribution in mass of older population and gas would  easily boost their mass to the typical dynamical mass of a dwarf galaxy (10$^7$-10$^9$~\msun\twospace). Hence, the presence of such knots is very promising in the field of Tidal Dwarf Galaxies and the main motivation for searching and characterizing these kind of objects in (U)\lirgs\twospace, as done in \refcha{tdgs}.

\subsect[mass_radius_gen_knots]{Mass-radius Relation of the Bluest Knots}

\begin{figure}[!t]
  \hypertarget{fig:size_mass_all_knots}{}\hypertarget{autolof:\theautolof}{}\addtocounter{autolof}{1}
  \hspace{0.5cm}\includegraphics[angle=90,width=0.9\columnwidth]{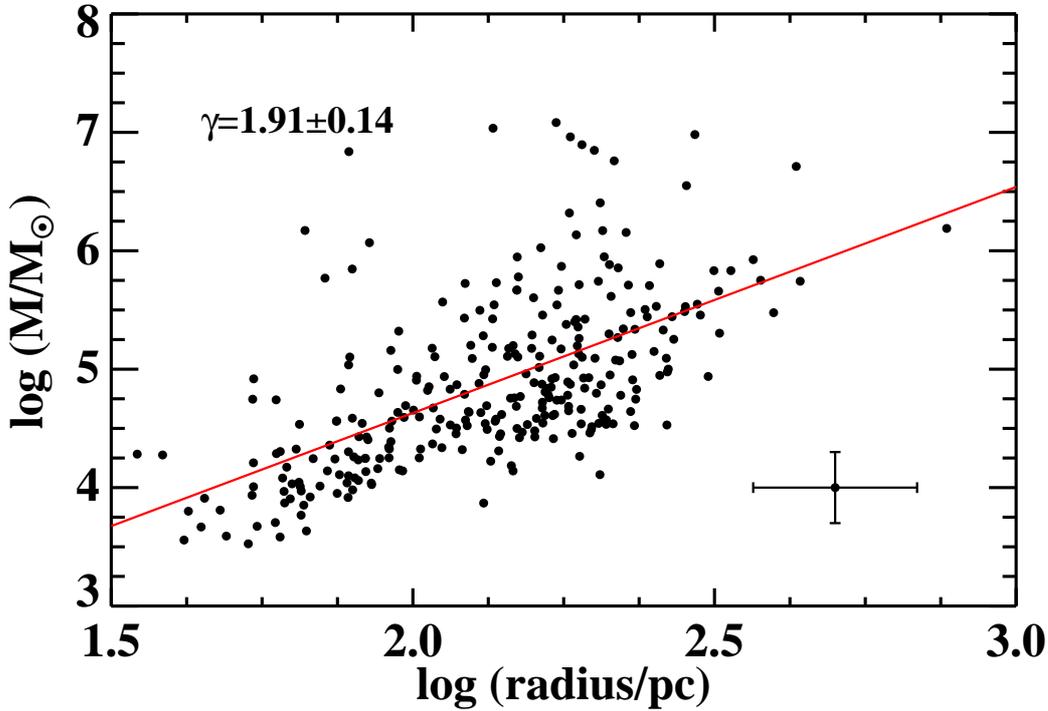}
   \caption[Mass vs. radius relation for knots in (U)LIRGs]{The mass vs. radius relation for knots in (U)\lirgs\twospace. The solid red line is a power-law fit to the data, with index $\gamma$. The typical uncertainty associated with the data is shown in the right bottom corner. }
   \label{fig:size_mass_all_knots}
              
 \end{figure}

In general, the knots represent associations of star clusters created in clumps. A single giant molecular cloud (\gmc) does not usually produce a single star cluster, but a complex of star clusters (\mycitealt{Elmegreen83};~\mycitealt{Bastian05a}). Thus, it is interesting to see how these complexes fit into the hierarchy of star formation by searching for a relation between the mass and the radius of the knots. This can be done for the bluest knots, for which an estimation of the mass with an uncertainty of less than a factor of 4 was achieved (\refsec{ages_masses_knots}). Their radius was derived as explained in Chapter~\ref{cha:data_tech},~\refsec{sizes}. We could obtain reliable measurements of both values for a total of 294 knots, which is about a 10\% of the total number of knots or about two-thirds of the total number of blue knots.

The surface brightness profile of the bluest knots is well fitted by a power-law with index \mbox{$\beta$ = 0.37 $\pm$ 0.33}, which corresponds to (assuming sphericity) a three dimensional density profile of \mbox{$\rho \propto r^{\beta}$}, with \mbox{$\beta$ = 1.37}. This power-law profile is similar to that observed for \gmcs associated with high-mass star-forming regions (\mbox{$\beta$ = 1.6 $\pm$ 0.3};~\mycitealt{Pirogov09}). The similarity between the profiles in \gmcs and the subsample of bluest knots suggests that the amount of luminous material formed is proportional to the gas density. Obviously, measurements of the gas in molecular clouds in (U)\lirgs are needed to prove this.

Although with high dispersion, the bluest knots follow a mass-radius relation (see~\reffig{size_mass_all_knots}). We fitted a function of the form \mbox{M $\propto$ $R^{\gamma}$}, with index \mbox{$\gamma$ = 1.91 $\pm$ 0.14}. The size and the mass of \gmcs in the Milky Way are also related, \mbox{M$_{\rm{cloud}} \propto$ $R_{\rm{cloud}}^{2}$}, as a consequence of virial equilibrium (\mycitealt{Solomon87}). This relation has also been found for extragalactic \gmcs (e.g.,~\mycitealt{Ashman01};~\mycitealt{Bastian05a}). Our result is therefore consistent with the bluest knots having the same mass-radius relation as \gmcs in general, contrary to what is measured for individual young star clusters (e.g., \mycitealt{Bastian05b}). An offset between both relations can be used to estimate the star formation efficiency within the cloud, as done in Bastian et al. (\myciteyear{Bastian05a}). Data from the ALMA observatory, which will be able to reach sub-arcsecond resolution, will permit the achivement of this estimation for (U)\lirg systems.

\subsect[lf_closest_knots]{Luminosity Function of the Knots in the Closest Systems}

The large number of detected knots allows us to study their luminosity function and compare it with that for other less luminous and/or interacting galaxies. However, this comparison should be restricted to the closest (U)\lirgs\twospace, since given the sizes measured complexes of star clusters are detected only for galaxies located at distances larger than 100 Mpc with the resolution of the \acs\twospace. Although many knots detected in our closest systems may still consist of complexes of star clusters, in order to minimize distance effects (see~\refsec{distance_effect}) we focus here on a subsample that comprises the seven systems closer to this distance (see specific distances in Chapter~\ref{cha:data_tech},~\reftab{sample}).

To compute the slope of the \lf for the knots in both the blue and red filters (as explained in
Chapter~\ref{cha:data_tech},~\refsec{lf}), we assigned a lower limit cutoff to the brightest value of the 90\% completeness limit in order to ensure the same level of completeness for the whole subsample. The single power-law fits to our blue and red luminosity functions are not significantly different, with slopes of $\alpha = 1.95$ and $\alpha = 1.89$ for the \textit{F435W} (blue) and the \textit{F814W} (red) filters, respectively (see~\reffig{lf_close}). 

The slopes in this study are in disagreement with the much flatter slopes previously measured in \ulirgs ~\citep{Surace98}. Surace and coworkers argued that the flattening of the slope is the artificial consequence of not resolving individual clusters, but associations of clusters (the bright end is overpopulated), the same bias that we encounter but more severe. In addition, other factors could contribute to the flattening of the \lf in that study, as well as in other works (see~\reftab{lf_knots}): (i) the slopes are not corrected for incompleteness (like in NGC 7252 in~\mycitealt{Miller97} and NGC 4038/39 in~\mycitealt{Whitmore99}), (ii) the statistics are more limited, since the distribution of less than 90 knots is fitted within their completeness limit interval, and (iii) \acs imaging is more sensitive than \wfpc\twospace, and since we are observing closer systems, the \lf can be measured to intrinsically fainter magnitudes. 

\begin{figure}
  \hypertarget{fig:lf_close}{}\hypertarget{autolof:\theautolof}{}\addtocounter{autolof}{1}
   \hspace{1.3cm}\includegraphics[angle=90,width=0.85\columnwidth]{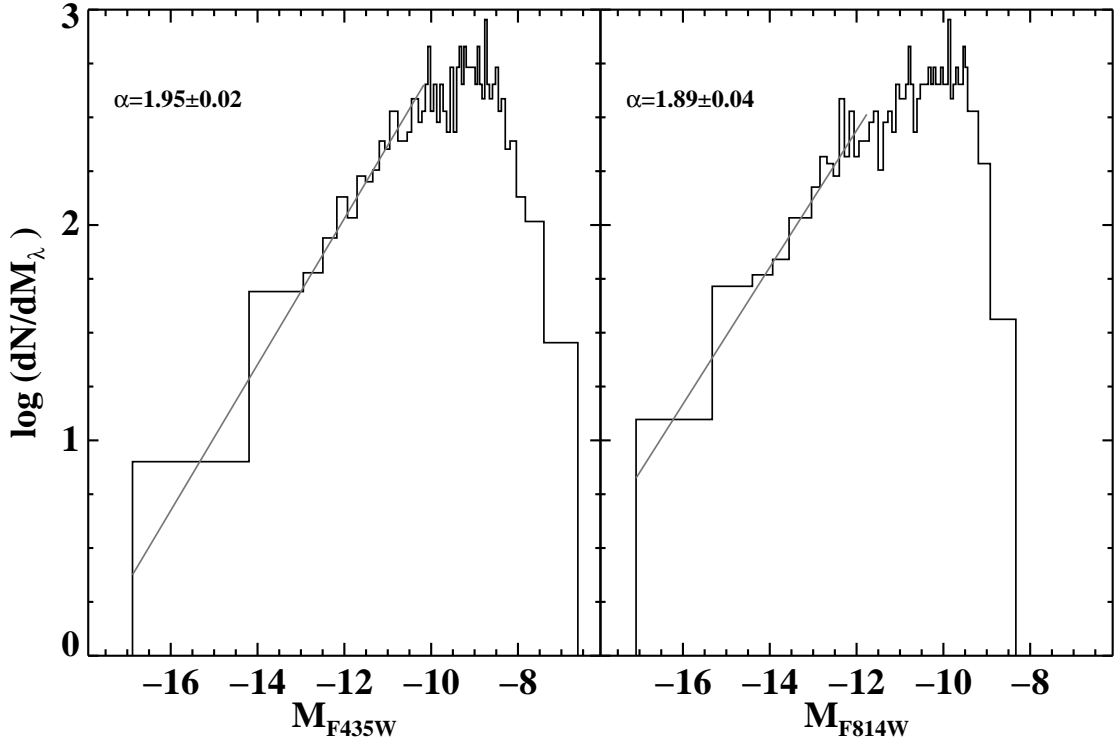}
   \vspace{0.5cm}
   \caption[Luminosity functions for the knots identified in the closest systems]{Luminosity functions for the knots identified in both the red and the blue filters for the subsample of (U)\lirgs closer than 100 Mpc. The line shows the fit up to the completeness limit computed for each filter.}
    \label{fig:lf_close}          
 \end{figure}

The slopes measured for our subsample are however similar to those obtained for other less luminous interacting galaxies such as in NGC 4038/39, NGC 3921, NGC 7252 and Arp 284 (see~\reftab{lf_knots}). As the studies start to resolve individual clusters and corrections for incompleteness are applied the slopes tend to consolidate around a value of 2-2.2. For instance, \reff (between 5 or unresolved, and 50 pc) derived in early studies (\mycitealt{Whitmore95}) for clusters in NGC 4038/39, indicate that some blending existed and thus the slope of the \lf is somewhat flatter than in more recent studies which detect only individual clusters and corrections for incompleteness are applied as well (\mycitealt{Whitmore99},\myciteyear{Whitmore10}); see values in~\reftab{lf_knots}. In any case, we did not expected noticeable flattening of the slope, since the mass function of molecular clouds in the Local Group (\mycitealt{Blitz07}) and in different \ism models (\mycitealt{Fleck96};~\mycitealt{Wada00};~\mycitealt{Elmegreen02}) is consistent with a slope of index 2, like the mass function for clusters in interacting systems (e.g.,~\mycitealt{Bik03};~\mycitealt{Gieles09}). This similarity can be reflected in the \lf\twospace. Given the fact that several clusters can be formed from a single molecular cloud, if there is not much blending (like in our case of nearby systems), a similar slope of the \lf is expected.

Therefore, the result obtained in this study extends the universality of the slope of the luminosity function regardless of the intensity of the star formation in interacting galaxies, to extreme star-forming systems like the luminous and ultraluminous infrared galaxies, at least for systems closer than 100 Mpc.

\begin{table}
\hypertarget{table:lf_knots}{}\hypertarget{autolot:\theautolot}{}\addtocounter{autolot}{1}
\hspace{1cm}
\begin{minipage}{0.85\textwidth}
\renewcommand{\footnoterule}{}  
\begin{small}
\caption[Slopes of the LF for nearby systems]{Slopes of the \lf computed for this work and for other interacting systems }
\label{table:lf_knots}
\begin{center}
\begin{tabular}{lcccccc}
\hline \hline
   \noalign{\smallskip}
Galaxy & \ld (Mpc) & Band & $\alpha$ & Uncertainty & Ref. \\
(1) & (2) & (3) & (4) & (5) & (6) \\
\hline
   \noalign{\smallskip}
ULIRGs Surace	&	422	&	I	&	1.39	&	0.08	&	[1]	\\
NGC 3921	&	84.5	&	V	&	2.10	&	0.30	&	[2]	\\
Nearby (U)LIRGs	&	72.8	&	B,I	&	1.95,1.89	&	0.02,0.04	&	This work*	\\
NCG 7252	&	62.2	&	V	&	1.84	&	0.06	&	[3]	\\
Arp 284	&	33.6	&	I	&	2.30	&	0.30	&	[4]	\\
NGC 4038/39	&	27.5	&	V	&	1.78	&	0.05	&	[5]	\\
NGC 4038/39\footnote{Computed for cluster-rich regions on the PC chip.}	&	27.5	&	V	&	2.01	&	0.11	&	[6]	\\
NGC 4038/39	&	27.5	&	V	&	2.13	&	0.07	&	[7]*	\\
\hline
\noalign{\smallskip}
\multicolumn{7}{@{} p{\textwidth} @{}}{\textbf{Notes.} Col (1): Galaxy name or designation for a sample of galaxies. Col (2): Luminosity distance, taken from NED. For the (U)\lirg samples (this work and that by \mycitealt{Surace98}) the mean distance is given. Col (3): Photometric band. Col (4): Derived slope of the luminosity function. In some works they also fit a double power-law, but the slopes given in this table correspond to a single power-law fitting. Col (5): Uncertainty of $\alpha$. In our case only uncertainties associated to the fit. Col (6): References-- [1] Surace et al. (\myciteyear{Surace98}), [2]Schweizer et al. (\myciteyear{Schweizer96}), [3] Miller et al. (\myciteyear{Miller97}), [4] Peterson et al. (\myciteyear{Peterson09}), [5] Whitmore \& Schweizer (\myciteyear{Whitmore95}), [6] Whitmore et al. (\myciteyear{Whitmore99}) and [7] Whitmore et al. (\myciteyear{Whitmore10}). In asterisk, studies that have used the prescription in Ma{\'{\i}}z-Apell{\'a}niz \& {\'U}beda (\myciteyear{Maiz05}).}
\end{tabular}
\end{center}
\end{small}
\end{minipage}
\end{table}



\sect[prop_lir]{Properties of the Knots as a Function of Infrared Luminosity}

\begin{figure}
   \hypertarget{fig:prop_lir}{}\hypertarget{autolof:\theautolof}{}\addtocounter{autolof}{1}
	\centering
  \hspace{-0.5cm}\includegraphics[trim = -2cm 12cm -1cm -4cm,clip=true,width=1.\textwidth]{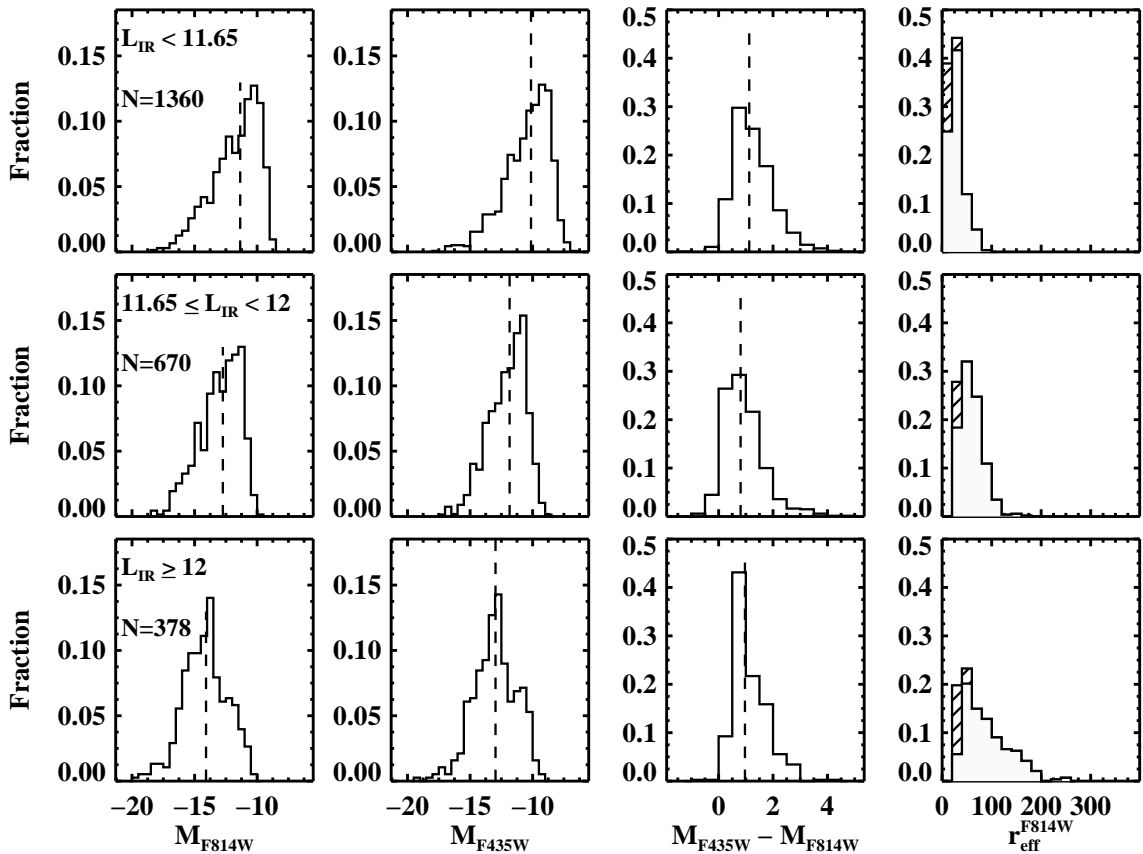}
   \caption[Photometric properties of the sample in three \lirnorm intervals]{Photometric properties of the sample in three \lir intervals, excluding 3 systems in the sample, as explained in the text. The dashed vertical line indicates the median value of the distributions of magnitudes \mb and \mi\onespace, and color \mbi\onespace.\label{fig:prop_lir}}
 \end{figure}

\begin{sidewaystable}
\hypertarget{table:prop_lir}{}\hypertarget{autolot:\theautolot}{}\addtocounter{autolot}{1}
\begin{minipage}{1.\textwidth}
\renewcommand{\footnoterule}{}  
\begin{footnotesize}
\caption[Photometric properties of the sample as a function of \lirnorm]{Photometric properties of the sample as a function of \lir}
\label{table:prop_lir}
\begin{tabular}{@{}l@{\hspace{-0.45cm}}c@{}c@{\hspace{0.15cm}}c@{\hspace{0.15cm}}c@{}c@{}c@{}c@{}c@{}c@{\hspace{0.15cm}}c@{\hspace{0.15cm}}c@{}}
\hline \hline
   \noalign{\smallskip}
System/s & \lir & N$_{sys}$ & knots per &  $<$ D$_{L}$ $>$ & $<$ $M_{F814W}$ $>$ & $<$ $M_{F435W}$ $>$ & $<$ C $>$ & $<$ $r_{\rm{eff}}^{F814W}$ $>$ & $\alpha$ LF & $\alpha$ LF \\
& (or interval) & & system & (Mpc) & & & & (pc) & \textit{F814W} & \textit{F435W} \\
(1) & (2) & (3) & (4) & (5) & (6) & (7) & (8) & (9) & (10) & (11) \\
 \hline
   \noalign{\smallskip}
 
(U)LIRGs& \lir $<$ 11.65 (all)& 11& 113 $\pm$ 102& 99 $\pm$ 26& -11.36& -10.13& 1.13& 25& 1.80 $\pm$ 0.02& 1.77 $\pm$ 0.02\\
(this work)& 11.65 $\le$ \lir $<$ 12.0& 8& 84 $\pm$ 59& 178 $\pm$ 27& -12.74& -11.84& 0.81& 55& 1.73 $\pm$ 0.04& 1.79 $\pm$ 0.04\\
& \lir $\ge$ 12.0& 10& 38 $\pm$ 15& 246 $\pm$ 72& -14.09& -12.97& 0.97& 81& 1.83 $\pm$ 0.03& 1.78 $\pm$ 0.04\\
& 11.65 $\le$ \lir $<$ 12.0 (all)& 9& 115 $\pm$ 109& 166 $\pm$ 45& -12.11& -11.18& 0.82& 45& 1.78 $\pm$ 0.04& 1.78 $\pm$ 0.04\\
& \lir $\ge$ 12.0 (all)& 12& 47 $\pm$ 41& 258 $\pm$ 125& -13.39& -12.20& 1.10& 59& 1.78 $\pm$ 0.04& 1.78 $\pm$ 0.04\\
 \hline
   \noalign{\smallskip}
M51& 10& 1& 881& 10.6& -8.4& -7.6& 0.7& -& - & - \\
NGC 4038/4039& 10.7& 1& $\sim$ 10$^4$& 27.5 & -11.7& -11.0& 0.9& 16.8& - & - \\
ULIRGs (Surace98)& \lir $>$ 12.0& 9& 12 $\pm$ 10& 520 $\pm$ 210& -14.7& -13.7& 0.8& 65& - & - \\
 \hline
   \noalign{\smallskip}
   \multicolumn{11}{@{} p{\textwidth} @{}}{\textbf{Notes.} The first three rows show the statistics for the systems considered in the text (some few have been excluded in the intermediate and high infrared luminosity intervals; see beginning of~\refsec{prop_lir}). The following two rows implement the statistics for all the systems that define the \lir interval. Below the horizontal middle line the values for clusters and knots from other works are shown (M51,~\mycitealt{Bik03}; NGC 4038/4039,~\mycitealt{Whitmore99}; and ULIRGs,~\mycitealt{Surace98}). Col (1): Galaxy name or designation for a sample of galaxies. Col (2): Infrared luminosity (or interval) of the galaxy/sample. Col (3): Number of systems that comprises each sample. Col (4): Average number of knots per system. Note that even though more than 10$^4$ clusters were detected, the values for NGC 4038/4039 refer to the 86 brightest clusters. Col (5): Luminosity distance or average value. Col (6): Median value of the \mi absolute magnitude distribution. Col (7): Median value of the \mb absolute magnitude distribution. Col (8): Median value of the \mbi color distribution. Col (9): Median value of the effective radius distribution. Col (10): Slope of the \textit{F814W} luminosity function distribution. The slopes of the \lfs for interacting systems in other works are already showed in~\reftab{lf_knots}. Col (11): Slope of the \textit{F435W} luminosity function distribution.}
\end{tabular}
\end{footnotesize}
\end{minipage}
\end{sidewaystable}

For most (U)\lirgs the infrared luminosity comes mainly from the re-emission by dust of light emitted in intense episodes of star formation. In this study we span  a range of a factor of 15 in infrared luminosities from \mbox{\lir= 11.39} to \mbox{\lir = 12.54}. To investigate whether the properties of the star-forming knots do show dependence on the infrared luminosity (i.e.,  star formation rate), we divided the initial sample into three \lir intervals. We aim at covering the low, intermediate and high luminosity regimes while preserving a similar number of systems per luminosity bin for the subsequent statistical analysis. We therefore chose the following luminosity intervals: \mbox{\lir $<$ 11.65} (low), \mbox{11.65 $\leq$ \lir $<$ 12.0} (intermediate) and \mbox{\lir $\geq$ 12.0} (high). The number of systems that lie in each interval is 11, 9 and 12, being at an average distance of 99, 166 and 258 Mpc, respectively. A summary of the properties of the knots per luminosity bin is given in~\reftab{prop_lir} and will be discussed in detail in subsequent subsections. 

As shown in Chapter~\ref{cha:data_tech},~\reffig{LirVsz}, systems with higher infrared luminosity are in general intrinsically more distant. We tried to get rid of any distance effects by selecting a subsample of systems located at a similar distance (those at 65-75 Mpc) and sampling all the \lir bins. This subsample, not homogeneously distributed, includes only one system in the intermediate and high luminosity bins (4,1 and 1 systems, respectively per luminosity bin). Furthermore, the results based only on the \ulirg Arp 220 (IRAS 15327+2340) are not reliable since it is known to suffer extreme obscuration from its silicate features (\mycitealt{Spoon07}). We need to include a larger number of systems in the defined luminosity bins in order to achieve more reliable statistics,  thus we inevitably have to deal with distance effects.

Finally, for each \lir bin, we excluded the knots of systems located at the two extremes of the distance scale, in order to diminish distance effects within the intervals. Thus, the statistics that are shown first in~\reftab{prop_lir} for the intermediate luminosity interval do not take into account IRAS 04315-0840, and for the high luminosity interval the knots in IRAS 15206+3342 and in IRAS 15327+2340 are not considered either. However, for completeness, the statistics is also provided considering all the sources in each \lir bin.

\subsect[mc_lir]{Magnitudes and Colors of the Knots as a Function of \lirnorm} 

\begin{figure}
  \hypertarget{fig:mag_col_lira}{}\hypertarget{autolof:\theautolof}{}\addtocounter{autolof}{1}
   
    \hspace{-0.5cm}\includegraphics[angle=90,trim = 4cm 0cm 4cm 0cm,clip=true,width=1.05\textwidth]{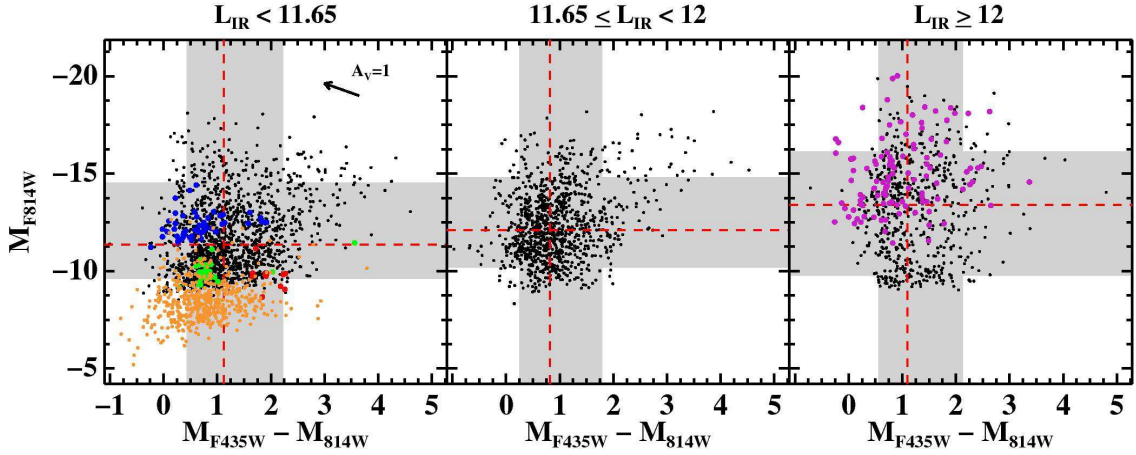}
   \caption[Color-magnitude diagrams of the knots at different \lirnorm intervals]{Color-magnitude diagrams of all the identified knots (black) at different \lir intervals, compared with other interacting systems from the literature. The gray band covers the 80\% of the total number of knots in each plot, as in~\reffig{mag_col_gen_knots}. The colors refer to clusters/knots measured in M51 (orange) taken from Bik et al. (\myciteyear{Bik03}), in the \ulirg sample of Surace et al. (\myciteyear{Surace98}) (pink) and in NGC 4038/4039 (blue, those younger than 10 Myr; green, those between 250 and 1000 Myr; and red, those older than 1 Gyr), taken from Whitmore et al. (\myciteyear{Whitmore99}).}
   \label{fig:mag_col_lira}           
\end{figure}

Knots show a dramatic increase ($\times$ 12) in their blue and red luminosities \mbox{(i.e., 2.7 mag)} with the infrared luminosity of the entire galaxy while basically preserving their colors (see histograms in~\reffig{prop_lir}, and average properties listed in~\reftab{prop_lir}). Our simulations (Section \ref{sec:distance_effect},~\reffig{simulation}) suggest that distance effects can explain only part of the differences revealed by the data. If the same type of galaxies are located at the average distance of the low, intermediate and high luminosity subsamples, the distance effect does not change significantly the colors while increasing the luminosity (absolute magnitude) by only a factor of 2.8 (1.1 mag). Therefore, we conclude that knots in highly luminous galaxies are at least a factor of four more luminous than knots located in systems with lower infrared luminosity.

It is interesting to mention that when comparing these results with those independently obtained for 
nearby, less luminous systems (the weakly and strongly interacting galaxies M51 and NGC 4038/4039, respectively), and more distant \ulirgs (\mycitealt{Surace98}), the same trend with infrared luminosity appears (see ~\reffig{mag_col_lira} and~\reftab{prop_lir}). In nearby systems like M51 and NGC 4038/4039, the angular resolution is such that individual clusters, instead of aggregates as in more distant galaxies, are detected, and therefore the difference in the luminosity of clusters and knots can be partly due to a distance effect.
Knots in distant \ulirgs tend to be as luminous as knots detected in our closer high-luminosity galaxies (also \ulirgs\twospace) while having colors similar to those knots detected in our sample and clusters in M51 and NGC 4038/4039 (The Antennae). The most luminous clusters in the Antennae cover a luminosity range close  to the median value for knots detected in our low-luminosity systems. Finally, clusters in the very low luminosity, weakly interacting M51 galaxy are on average about 3 mag fainter (i.e., x15 less luminous) than knots in our low-luminosity galaxies. 

If we assume a statistically similar extinction and age for the star-forming knots regardless of the \lir of the system, the most plausible explanation for the luminosity excess measured in the intermediate and high luminosity systems has to invoke a mass and/or density effects:

\begin{itemize}
\item \textit{Mass effect}. More high-mass knots are sampled in the most luminous systems.

\item \textit{Density effect}. The knots are aggregates of an intrinsically larger number of clusters as the infrared luminosity of the system increases.
\end{itemize}

The star formation rate as well as the gas content in \ulirgs is higher than in less luminous systems and as a consequence, these systems form more clusters. Therefore, it is matter of simple statistics rather than a difference in physical formation mechanisms that in \ulirgs we find the brightest knots. Given the stability of the colors among the different infrared luminosity sample, they likely represent the most massive knots. This stochastic process, known as size-of-sample effect (\mycitealt{Whitmore07}), can explain the demographics of star cluster systems in the merging environment, where the star formation is enhanced. Therefore, although density effects cannot be discarded our result is compatible with a mass effect (which in turn corresponds to a size-of-sample effect), explained by statistics.
  
\subsect[spa_knots_lir]{Spatial Distribution of the Knots as a Function of \lirnorm} 

\begin{figure}
  \hypertarget{fig:reff_lir}{}\hypertarget{autolof:\theautolof}{}\addtocounter{autolof}{1}
\centering
\includegraphics[angle=90,trim =0cm 0cm 1.5cm 0cm,clip=true,width=0.95\textwidth]{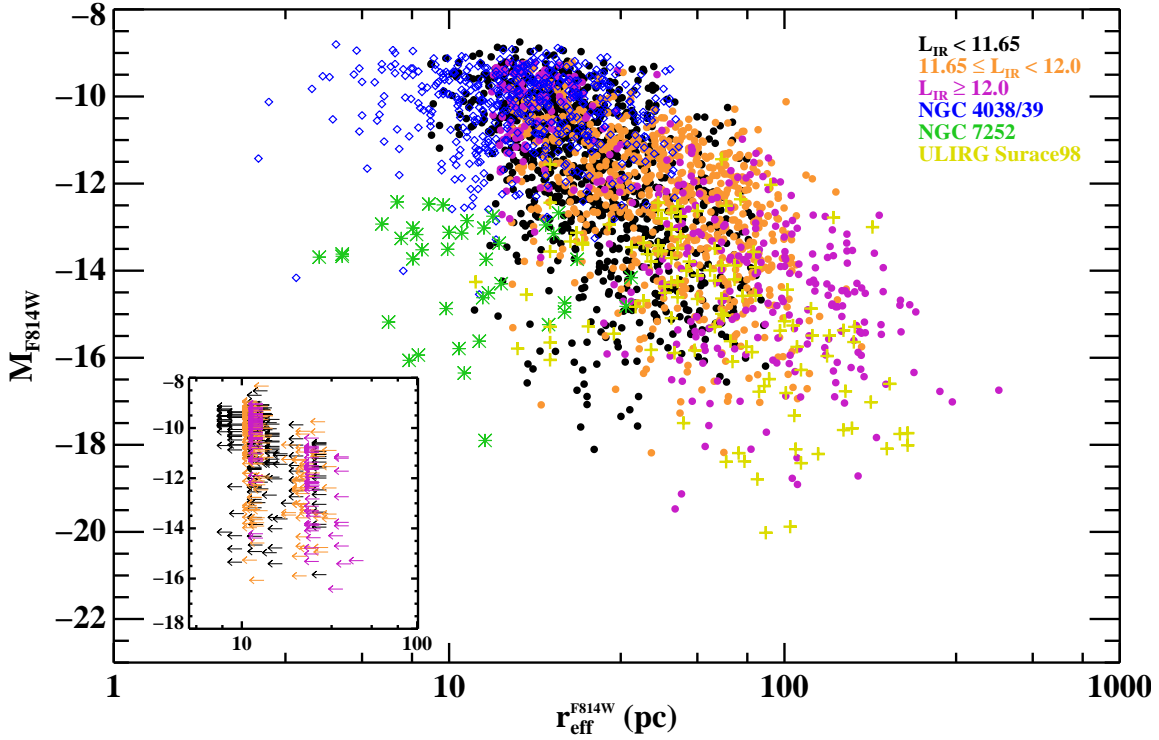}
\caption[Absolute \textit{F814W} magnitude as a function of size for clusters]{Absolute \textit{F814W} magnitude (uncorrected for internal extinction) as a function of size for clusters in interacting systems. Note that the x-axis is in log scale. We have divided the knots in our sample in three groups depending on their infrared luminosity. Values for clusters in the Antennae were taken from Whitmore \& Schweizer (\myciteyear{Whitmore95}), in NGC 7252 (with \mbox{\lir= 10.70}) from Whitmore et al. (\myciteyear{Whitmore93}) and in the \ulirg sample from Surace et al. (\myciteyear{Surace98}). The inset plot shows the same for the unresolved knots in this study.}
\label{fig:reff_lir}
 \end{figure}

One important aspect to investigate is whether the spatial distribution of the knots shows dependence on the luminosity of the system. We measure a ratio of inner to outer knots ($N_{\rm{inner}}$/$N_{\rm{outer}}$) that changes with \lir from 0.68 to 0.49 and 0.31 for the three (low-, intermediate- and high-) luminosity intervals, respectively). According to our simulations, these differences are consistent with being purely due to a distance effect (see~\reftab{prop_lir} for the average distance of the different luminosity bins), and therefore no evidence for the dependence of the spatial distribution of the knots on the infrared luminosity of the systems is found.

\subsect[reff_lir]{Effective Radius of the Knots as a Function of \lirnorm}

If there is a change in the luminosity of the knots with the bolometric luminosity of the system, their size could show a variation as well. At first glance, this seems to be the case (see~\reffig{prop_lir} and~\reffig{reff_lir}). As the luminosity of the system increases, the distribution of \reff shows a broadening with a non negligible fraction of the knots having sizes larger than 100 pc. This is more clear for the high-luminosity systems (\ulirgs). However, as our simulations have shown (see~\reffig{simulation}), this changes can be explained purely as a distance effect. On the one hand, the distribution of \reff of the knots in \ulirgs is similar to the simulated \reff distribution. Moreover, the median value of \reff in \ulirgs is 81 pc (see~\reftab{prop_lir}), slightly larger, but consistent within the uncertainties,  with the 75 pc derived from the simulations. In fact, the sizes of the knots in the closest systems (i.e., \mbox{\ld$<$ 100 Mpc}), where individual clusters are almost resolved, are closer to the sizes of clusters measured in less luminous interacting galaxies (e.g., see clusters of NGC 4038/4039 and NGC 7252 in ~\reffig{reff_lir}). Therefore, with the present angular resolution there is no evidence for a variation in size with luminosity. 

\subsect[mr_lir]{Mass-radius Relation of the Bluest Knots as a Function of \lirnorm}

\begin{figure}
  \hypertarget{fig:size_mass_lir}{}\hypertarget{autolof:\theautolof}{}\addtocounter{autolof}{1}
\centering
\includegraphics[trim = 0cm -1cm 0 -2.5cm,clip=true,width=0.8\columnwidth]{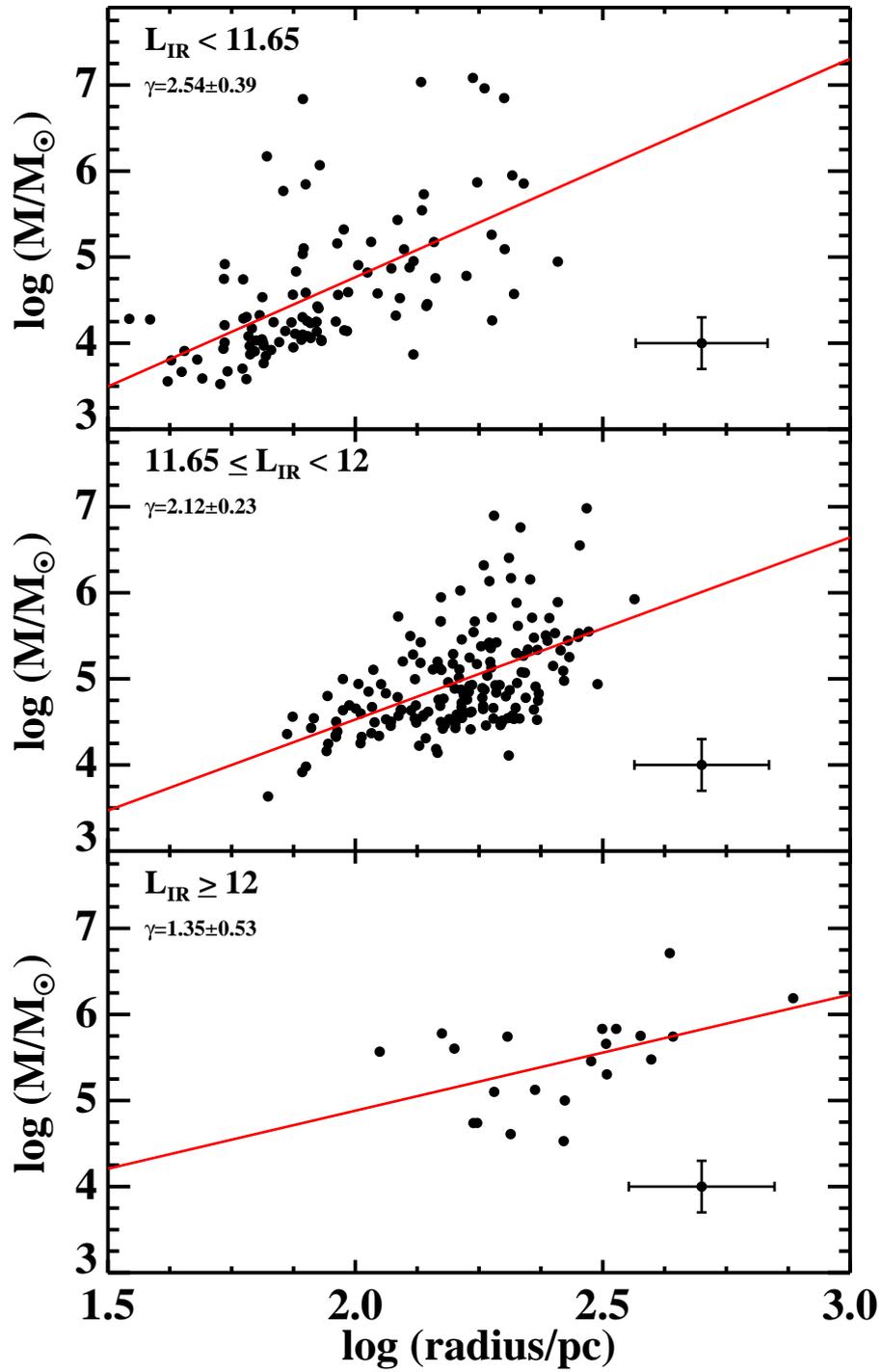} 
   \caption[Mass vs. radius relation for knots in (U)LIRGs with \lirnorm\twospace]{The mass vs. radius relation for knots in (U)\lirgs as a function of \lir\twospace. The solid red line is a power-law fit to the data, with index $\gamma$.} 
   \label{fig:size_mass_lir}
 \end{figure}

We have seen that in general terms the bluest knots in our sample follow a mass-radius relation with index $\gamma$$\sim$2, similar to that in \gmcs and complexes of clusters in less luminous interacting systems (see ~\refsec{mass_radius_gen_knots}). Here we have the opportunity to characterize this relation as a function of the infrared luminosity of the system.

Though with high dispersion, the correlation holds for the low and intermediate infrared luminosity intervals (see~\reffig{size_mass_lir}). The fit for the bluest knots in \ulirgs gives a shallower index, though it could still be consistent with a value of $\gamma$=1.9 given the large dispersion on the index values. The fit is affected in any case by the small number of blue knots in \ulirgs and probably by distance effects. Therefore, according to our data, there is no clear variation of the mass-radius relation with the infrared luminosity of the system.

\subsect[lf_lir]{Luminosity Function of the Knots as a Function of \lirnorm}

The derived slopes of the luminosity functions with values between 1.7 and 1.8 (see~\reftab{prop_lir})  do not show any clear trend of variation with the luminosity of the system. However, our simulations of the distance effect show that the slope of the \lf becomes shallower by about 0.15 dex  when moving the low-luminosity systems to the average distance of the high-luminosity subsample  (see~\reffig{simulation}). This change in the slope is due to the reduction in the linear resolution such that knots that appear resolved in the closer, low-luminosity systems appear to be more luminous aggregates in the more distant luminous systems. Therefore, the distance-corrected slopes for the intermediate and high-luminosity subsamples would be about 1.9-2, in the range of those measured in other low-z interacting galaxies (see~\reftab{lf_knots}).

We know that these slopes were derived using the luminosity of complexes of clusters, especially in the case of systems in the intermediate and high infrared luminosity intervals. Thus, the slopes here might not reflect the physics beneath the formation and dissolution of star clusters, as normally do for nearby galaxies. However, we have investigated how distance effects can distort these slopes. Therefore, we conclude that, in absence of better angular resolution images, our data suggest that the \lf remains universal with slopes about \mbox{$\alpha$ = 2} within the (U)\lirg luminosity range. Based on the universality of the \lf\twospace, Whitmore et al. (\myciteyear{Whitmore07}) claim that active mergers have the brightest clusters only because they contain more clusters (i.e., size-of-sample effect). Thus, if the discrepancy between the luminosities of knots in \ulirgs and in \lirgs is explained by this effect (which is likely to be the case), then the physics underlying the disruption of clusters in the highest luminous infrared galaxies can be the same as in less luminous systems.

\sect[prop_is]{Properties of the Knots as a Function of Interaction Phase}

\begin{figure}
   \hypertarget{fig:propIS_sameD}{}\hypertarget{autolof:\theautolof}{}\addtocounter{autolof}{1}
  \includegraphics[trim = -2cm 2cm -1cm 0cm,clip=true,width=0.95\textwidth]{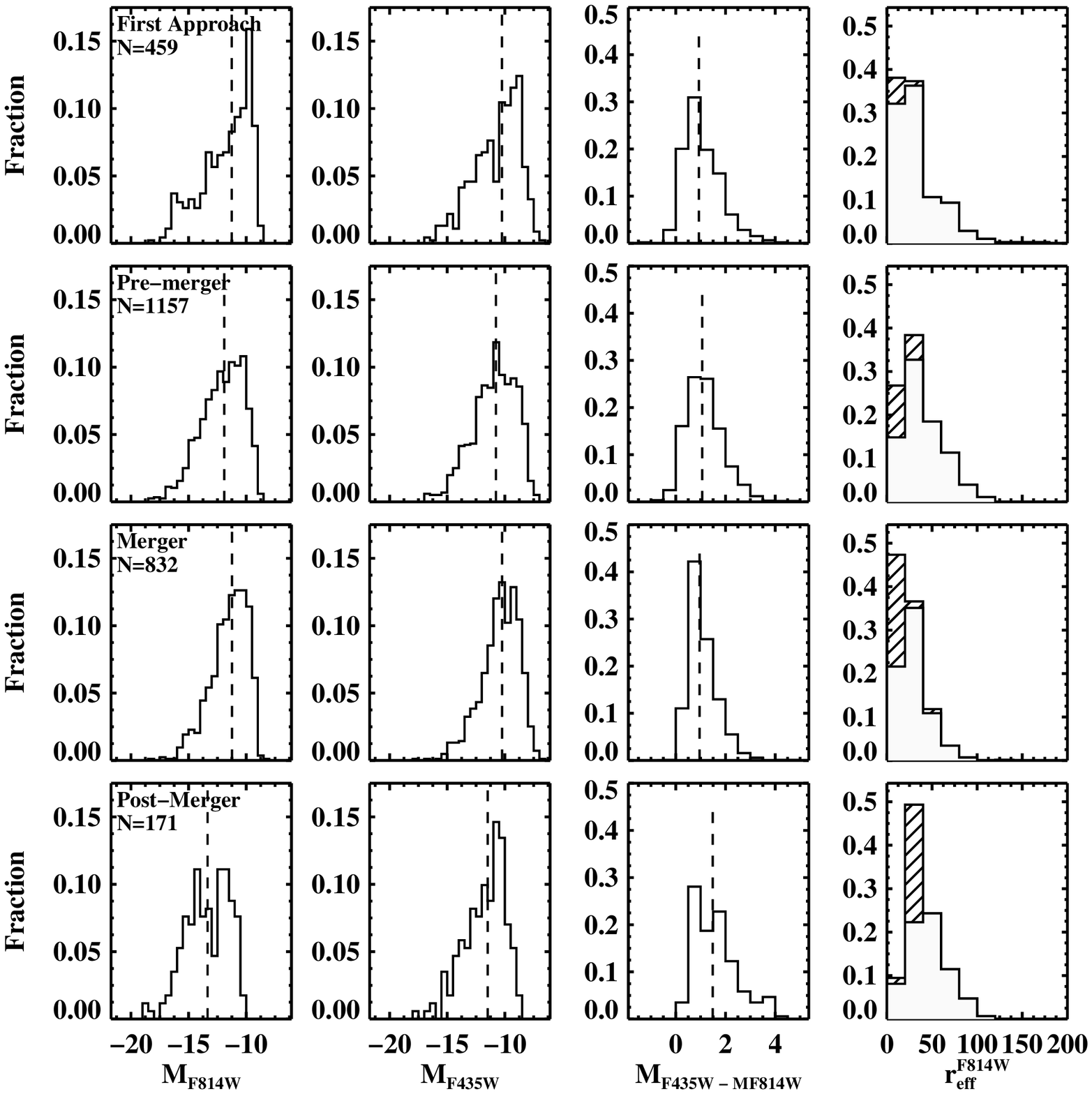}
   \vspace{-1cm}
   \caption[Photometric properties as a function of merging phase]{Photometric properties as a function of merging phase for systems with similar D$_{L}$. Dashed vertical lines indicate the median value of the distributions of magnitudes \mi and \mb, and color \mbi, as in~\reffig{prop_lir}. The relevant parameters of the plots are shown in the first four rows in~\reftab{prop_is}.}
   \label{fig:propIS_sameD}             
 \end{figure}

\begin{sidewaystable}
\hypertarget{table:prop_is}{}\hypertarget{autolot:\theautolot}{}\addtocounter{autolot}{1}
\begin{minipage}{1.\textwidth}
\renewcommand{\footnoterule}{}  
\begin{footnotesize}
\caption[Photometric properties of the sample as a function of interaction phase]{Photometric properties of the sample as a function of interaction phase }
\label{table:prop_is}
\begin{tabular}{lccccccccc}
\hline \hline
   \noalign{\smallskip}
Interaction phase & N$_{sys}$ & knots per & $<$ D$_{L}$ $>$ & $<$ $M_{F814W}$ $>$ & $<$ $M_{F435W}$ $>$ & $<$ C $>$ & $<$ $r_{\rm{eff}}^{F814W}$ $>$ & $\alpha$ LF \textit{F814W} & $\alpha$ LF \textit{F435W} \\
& & system & (Mpc) & & & & (pc) & & \\
\hline
   \noalign{\smallskip}
First Approach& 3& 153 $\pm$ 127& 121 $\pm$ 49& -11.25& -10.26& 0.93& 24& 1.50 $\pm$ 0.09& 1.63 $\pm$ 0.06\\
Pre-merger& 8& 145 $\pm$ 81& 126 $\pm$ 42& -11.91& -10.77& 1.06& 35& 1.84 $\pm$ 0.03& 1.86 $\pm$ 0.04\\
Merger& 6& 139 $\pm$ 123& 107 $\pm$ 37& -11.23& -10.24& 0.96& 25& 2.00 $\pm$ 0.04& 1.86 $\pm$ 0.03\\
Post-Merger& 5& 34 $\pm$ 19& 145 $\pm$ 50& -13.35& -11.49& 1.49& 44& 1.76 $\pm$ 0.12& 1.58 $\pm$ 0.17\\
 \hline
   \noalign{\smallskip}
First Approach& 4& 131 $\pm$ 113& 140 $\pm$ 55& -11.54& -10.66& 0.92& 27& 1.53 $\pm$ 0.10& 1.67 $\pm$ 0.02\\ 
Pre-merger& 13& 103 $\pm$ 83& 192 $\pm$ 98& -12.29& -11.12& 1.04& 40& 1.93 $\pm$ 0.03& 1.86 $\pm$ 0.03\\ 
Merger& 8& 112 $\pm$ 115& 142 $\pm$ 71& -11.41& -10.38& 0.96& 26& 1.83 $\pm$ 0.02& 1.78 $\pm$ 0.03\\ 
Post-Merger& 7& 29 $\pm$ 18& 213 $\pm$ 159& -13.51& -11.73& 1.55& 49& - & - \\
 \hline
   \noalign{\smallskip}
 \multicolumn{10}{@{} p{\textwidth} @{}}{\textbf{Notes.}  The interaction groups above the horizontal middle line includes only systems having similar distance. Below the horizontal middle line all the systems have been considered, regardless their distance  (note the higher dispersion in the distance of the systems here). Apart from the first column, which refers to the different interaction groups defined in Chapter~\ref{cha:data_tech} (\refsec{morphologies}), the rest indicate the same as shown in~\reftab{prop_lir}.}
\end{tabular}
\end{footnotesize}
\end{minipage}
\end{sidewaystable}

Strong interactions and mergers provide a natural laboratory for probing how star formation is affected by tidal forces and major rearrangements of the gaseous component during the different phases of the interaction. Most of the systems in our sample are interacting systems, hence we divided the sample into four groups, depending on the morphology seen in the \textit{F814W} \hst images. Each group represents snapshots of the  different phases of the interaction/merger: first approach, pre-merger, merger and  post-merger. The morphology class is explained in Chapter~\ref{cha:data_tech} (\refsec{morphologies}) and the selection is shown in the same chapter (\refsec{sample_ini}). In order to know how the star formation evolves with the interaction process, in this section we study the photometric properties of the knots for each of the four morphology groups.

\subsect[mc_is]{Magnitudes and Colors as a Function of Interaction Phase}

\begin{table}
\hypertarget{table:dust}{}\hypertarget{autolot:\theautolot}{}\addtocounter{autolot}{1}
\hspace{0.5cm}
\begin{minipage}{0.9\textwidth}
\begin{small}
\renewcommand{\footnoterule}{}  
\caption[Infrared to ultraviolet luminosity ratio as a function of interaction phase]{Infrared to ultraviolet luminosity ratio as a function of interaction phase}
\label{table:dust}
\begin{tabular}{lccccc}
\hline \hline
   \noalign{\smallskip}
   Interaction phase	&	N$_{sys}$	&	$<$ IR/UV $>$	&	IR /UV	& $<$ IR$_{25}$/UV $>$ & IR$_{25}$ /UV \\
	&	&		&	range & & range \\
(1) & (2) & (3) & (4) & (5) & (6)\\
\hline 
   \noalign{\smallskip}
First Approach	&	2	&	-	&	[15,130]	&  -	&	[4,11]	\\
Pre-merger	&	6	&	48	&	[14,189]	&   7	&	[2,45]	\\
Merger	&	6	&	248	&	[114,1235]	& 32	&	[14,117]	\\
Post-merger	&	2	&	-	&	[385,895]	&  -	&	[86,93]	\\
\hline 
   \noalign{\smallskip}
\multicolumn{6}{@{} p{\textwidth} @{}}{\textbf{Notes.} Since the IR/UV ratio is not affected by distance effects (we compute it for the whole system) we give the statistics for all the systems per interaction phase with available UV image. Col (1): interaction phase. Col (2): number of systems with IR and UV data available. Col (3): median of the IR/UV ratio when more than two values are derived. Col (4): range covered of the ratio IR/UV for each interaction phase. Col (5): same as (3), but in this case only the contribution of the flux at 25$\mu$m is taken to compute \lir\twospace. Col (6): same as (4), but only the contribution of the flux at 25$\mu$m is taken to compute \lir\twospace.}
\end{tabular}
\end{small}
\end{minipage}
\end{table}

The magnitudes and colors of the knots in the first three interaction phases differ significantly from those of the knots in the post-merger phase (see results in~\reftab{prop_is}). 
Even if in a strictly statistical sense (Kolmogoroff-Smirnov test at a confidence level of at least 98\%) the population of knots in the first three phases are different, their median values and distributions are very similar suggesting the lack of strong differences. This is true in particular if, in order to mi\-ni\-mi\-ze the distance effect,  only galaxies at similar distances in each phase are considered (see~\reftab{prop_is} and~\reffig{propIS_sameD}). However, the population of knots in the post-merger phase are significantly more luminous (median \textit{I}-band magnitude  difference of up to 2 mag), and significantly redder than in any of the three earlier phases. Furthermore, the population of knots in the post-merger phase shows in its color distribution an extension toward redder colors (\mbox{\mbi $>$ 3}). The fact that knots in the post-merger phase have redder colors means that they likely represent either a more obscured or an older population. In the following we explore both alternatives:

\begin{enumerate}
\item

To investigate whether the knots in the post-merger phase are intrinsically more obscured than in any of the  previous phases,  we define the infrared to \acr{}{UV}{ultraviolet} luminosity ratio (IR/UV) as:

\eqn[ir_uv]{
{\large \rm{IR/UV} =  \frac{\textit{L}_{\rm{IR}}}{\lambda \textit{L}_{\lambda}(\rm{FUV})+\lambda \textit{L}_{\lambda}(\rm{NUV})}}
}

where \lir stands for the infrared luminosity, $\lambda$ to the pivot wavelength of a given filter and L$_{\lambda}$ to the flux in the near (NUV) and far (FUV) ultraviolet emission (\mycitealt{Howell10}). This ratio is a useful empirical measurement of the dust absorption and emission in these heavily obscured systems. We searched for \uv images in the Galaxy Evolution Explorer (\textit{GALEX}) catalog, using the Multimission Archive at STScI (MAST\footnote{http://archive.stsci.edu/}) facility. With a typical angular resolution of about 4-5\arcsec in the FUV-NUV ($\lambda$ = 1528-2271$\AA{}$) range, we can estimate an average value of IR/UV for each system assuming that the far infrared and ultraviolet flux is emitted in the same regions. Even if this is not necessarily true, taking into account that most of the nuclear star formation is obscured in (U)\lirgs\twospace, we can have a first-order estimate on which systems have a higher average extinction.

A total of 16 systems were detected in the GR5 data set, most of them in the All-Sky Imaging survey (ASI\footnote{Visit the ~\textit{GALEX} website, http://www.galex.caltech.edu/, for more information}). The other systems were either not detected or not observed. In this dataset, the images are already reduced, the sources detected and the photometry performed. In general the photometry was performed using an aperture that covers the entire galaxy (if two galaxies form the system we add up the flux values). We then take the value of the flux provided by the data set facility.

With the flux in the far and near \uv and the infrared luminosity we estimated an average value of the IR/UV ratio for each system. Although the assumption by Kennicutt (\myciteyear{Kennicutt98}) that the great majority of the luminosity from young stars would be absorbed by dust and re-radiated in the far-\ir is likely to be correct, observationally there may be other contributions to the total IR luminosity from older populations of stars or other luminosity sources (e.g., see the review by Tuffs \& Popescu~\myciteyear{Tuffs05}). For that reason, we also computed the IR/UV ratio taking only the contribution at 25$\mu$m in IRAS (IR$_{25}$), based on the good relation between the mid-IR 24$\mu$m-25$\mu$m (MIPS-IRAS) and the number of ionizing photons as derived from extinction corrected Pa$ \alpha$ and \ha luminosities (\mycitealt{Wu05};~\mycitealt{Alonso-Herrero06};~\mycitealt{Calzetti07}). This way we select only hot dust, heated by the radiation of young population, and avoid any contribution from colder dust, heated by older stars.

Bearing in mind that we are dealing with a small number of statistics, there is a trend toward higher values of the IR/UV ratio in not only the post-merger but also the merger phase, with respect to early phases represented by the first approach and pre-merger (see values in~\reftab{dust}). Since this ratio is related to an average dust absorption, this result suggests that knots in the late phases of an interaction could be suffering on average intrinsically more extinction than in early phases.  For instance, an additional average extinction of \av\onespace=1 can already redden the colors by about 0.6, according to the \sbnn models (see Chapter~\ref{cha:data_tech},~\reffig{ssp_models}). If this were the case, the population of clusters in the late phases of a merger would be more luminous than in previous phases. Therefore, we would need to invoke a physical process responsible for this significant increase in luminosity. Note however that dust absorption shows significant structure on scales of a kiloparsec or less (\mycitealt{Garcia-Marin09b}), and therefore the measured IR/UV ratios do not necessarily represent the intrinsic absorption toward the individual optically-selected clusters. A larger set of  high angular resolution \uv and mid-\ir imaging is required before any firm conclusion is derived. 

\item

An alternative scenario to explain the differences in color and luminosity of the knots identified in the post-merger phase with respect to early phases can invoke evolutionary effects. As proposed by Kroupa (\myciteyear{Kroupa98}), many dozens or even hundreds of massive individual star clusters with masses of 10$^5$-10$^7$~\msun  could merge into extremely massive super star clusters (\sscs\twospace) after a few hundred Myr. This hierarchical star formation evolution was also suggested to be happening in complexes of clusters in M51 (\mycitealt{Bastian05a}). As the interaction proceeds, the smaller \sscs would be destroyed by disruption effects (e.g., by the mechanisms proposed in~\mycitealt{Whitmore07}) and a few massive individual clusters would still be able to form. Note that, although with a large dispersion, while the number of knots per system stays about the same (140-150) in the first three phases, this number drops by a factor of 4 to 5 in the post-merger phase (34 knots per system). Thus, in the post-merger phase systems we might be detecting only the evolved, extremely massive \sscs and a few massive individual clusters. The knots with the reddest colors could be evolved \sscs as massive as 10$^{7-9}$~\msun with an optical extinction (\av\twospace) of 3 mag, or massive young clusters with higher extinction. 

Kroupa (\myciteyear{Kroupa98}) estimated that after 100 Myr or so an object with these characteristics can be formed. If the merging of clusters started during the early phases of the interaction we would have already detected some of them in the merger phase. Given the degeneracy of the \mbi color we could only set an upper-limit to the number of merged superclusters in this phase. Thus, with a color higher than \mbox{\mbi= 0.6} and a luminosity higher than \mbox{\mi= -15}, an object could be more massive than 10$^7$~\msun\twospace, were it around or older than 100 Myr. About three such knots per system are detected in our sample of (U)LIRGs. 

Obviously, photometry at other wavelengths is necessary to better constrain the ages and masses of these knots and thus to check the validity of this scenario for (U)\lirg systems. Likewise, simulations of galactic encounters help us know if this scenario can occur in this environment. A detailed comparison of the present observational results with state-of-the-art high-resolution numerical simulations (\mycitealt{Bournaud08a}) is presented in~\refcha{sim}, to further investigate this scenario.

\end{enumerate}

\subsect[spa_dist_is]{Spatial Distribution as a Function of Interaction Phase}

We investigated the relative fraction of inner (inside 2.5 kpc radius) and outer (outside 2.5 kpc radius) knots  in the four interaction phases defined above in order to know if major mergers have an impact on the spatial distribution of the knots. Although the number of systems in each phase is small (3, 8, 6 and 5 for the first approach, pre-merger, merger, and post-merger) the ratio $N_{\rm{inner}}$/$N_{\rm{outer}}$ is 0.50, 0.65, 0.52, 0.68 from the first approach to the post-merger. 

We do not see any correlation toward a more dispersed or concentrated spatial distribution of the knots with  interaction phase. However, in the pre- and post-merger phases, where the magnitudes are higher compared to the other phases, there is some evidence of greater concentration toward the nuclear regions. The greater the number of knots in the central regions, the harder it is to detect them, and therefore, only the brightest ones are identified given the local background conditions (bright and steep light profiles). Hence, this effect can contribute to a small increase in the luminosity distribution of the knots in these phases. Assuming that the knots in the pre-merger phase have a luminosity similar to that in the first contact and merger phases, this effect would increase the median luminosity by a factor of almost 2. Since the ratio $N_{\rm{inner}}$/$N_{\rm{outer}}$ in this phase is similar to the ratio in the post-merger phase, an additional increase of a factor of about 5 would still be unexplained by the first scenario explored in the previous section. A larger sample of galaxies would be required to establish a firmer result.

\subsect[reff_is]{Effective Radius as a Function of Interaction Phase}

All size distributions are statistically different with a high degree of confidence (higher than 99\%), yet the median value of \reff is similar during the three first interaction phases indicating that these differences must be rather small. However, knots in the post-merger phase tend to be larger by factors of 1.3 to 2 than in any of the previous phases (see~\reftab{prop_is}). This result is not very robust, given the uncertainties in the determination of sizes, but if confirmed, it would support the hierarchical star formation scenario proposed by Kroupa (\myciteyear{Kroupa98}). According to this scenario, the merged super clusters would have a half-mass radius of about \mbox{$r_{\rm{h}}$ = 45-95 pc}. Under the assumption of virial equilibrium and isotropic velocity distribution the effective and half-mass radius can be related as \mbox{\reff= 3/4$\times r_{\rm{h}}$}. Then, the merged extremely massive super clusters would have \mbox{\reff= 30-70 pc}, similar to the values measured for knots in the post-merger sample.

\subsect[mr_is]{Mass-radius Relation of the Bluest Knots with Interaction Phase}

\begin{figure}
 \hypertarget{fig:size_mass_is}{}\hypertarget{autolof:\theautolof}{}\addtocounter{autolof}{1}
\includegraphics[trim = 0cm 3.5cm 1cm 0cm,angle=90,width=0.9\textwidth]{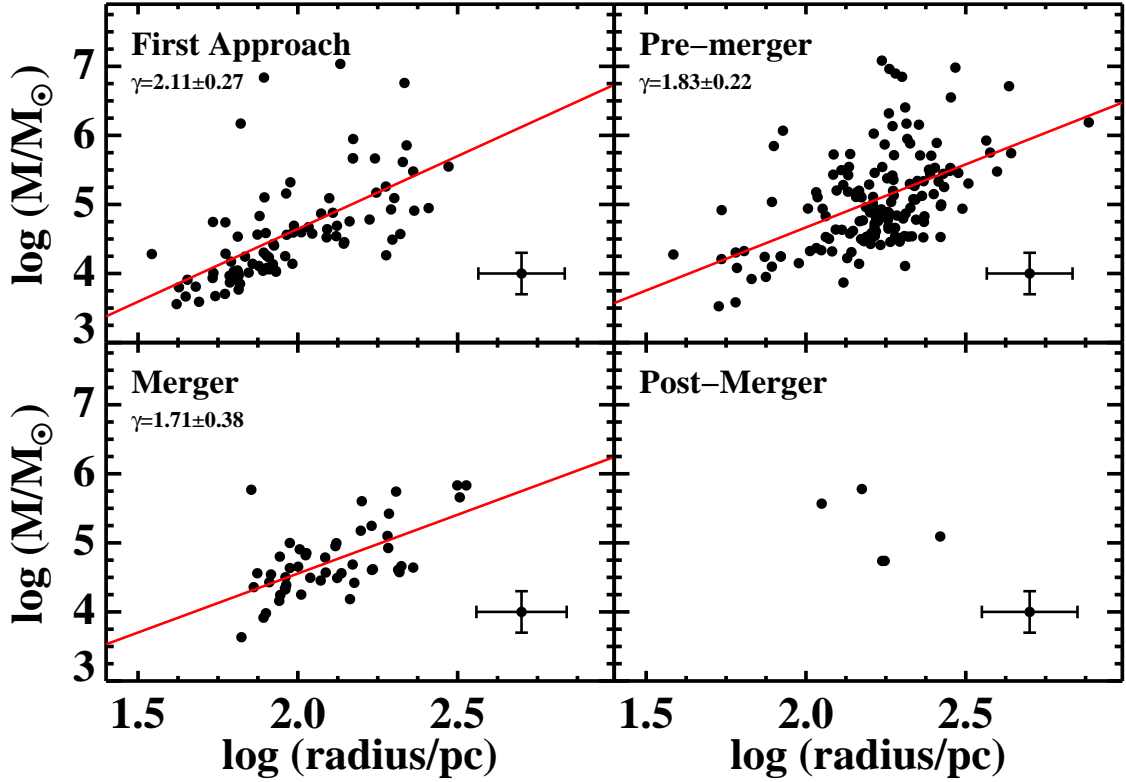}
  \caption[Mass vs. radius relation for knots in (U)LIRGs with interaction phase]{The mass vs. radius relation for knots in (U)\lirgs as a function of the interaction phase. The solid red line is a power-law fit to the data, with index $\gamma$.} 
  \label{fig:size_mass_is}            
 \end{figure}
 
The broad-band emission of the knots for which we have measured the mass is very much likely to come from young population (i.e., \mbox{$\tau$ $<$ 30 Myr}; see~\refsec{ages_masses_knots}). Based on equivalent width measurements, Bastian et al. (\myciteyear{Bastian05a}) determined the age of the complexes in M51 to be in the range 5-8 Myr. On the other hand, the phases of interaction are separated by hundreds of Myr. Thus, although we cannot study how the mass-radius relation of the knots behaves as they evolve, we can see if they all have a similar relation at a short time after they are born during the interaction process.

The mass-size relation of the bluest knots in the three interaction  phases for which a fit could be achieved is consistent with an index of \mbox{$\gamma$ = 2} (\reffig{size_mass_is}). Although the line-fit seems to become shallower as the interaction proceeds from the first approach up to the merger phase (from \mbox{$\gamma$ = 2.11} to 1.71), the dispersion is too high to accept that it varies. Actually, the fit that we obtained using all the knots gives a slope inbetween both values (1.91; see~\refsec{mass_radius_gen_knots}).

\subsect[lf_is]{Luminosity Function as a Function of Interaction Phase}

There appears to be a trend in the slopes of the \lfs of the knots  with interaction phase such that they become steeper as the interaction evolves from first contact to pre-merger and up to merger and 
post-merger (see~\reftab{prop_is}). This is particularly more visible when only systems at similar distances and \textit{I}-band LFs, i.e., less affected by extinction effects, are considered (slopes of 1.50, 1.84, 2.0 and 1.76, respectively). Note, however, that the slope of the \lf for the post-merger phase is computed with less than 100 knots (brighter than the completeness limit), hence it is statistically less significant than the other slopes. Also note that the systematic uncertainties of the slope of the LF can be as high as about 0.1 dex (the tables give the uncertainty related only to the fit), as mentioned in Chapter~\ref{cha:data_tech} (\refsec{lf}). Thus, the observed trend is very weak.

If we select only systems closer than 100 Mpc (see~\refsec{lf_closest_knots}), the phases are also sampled. We can try to further minimize distance effects by computing the \lf only for these systems. While 400-600 measurements can be used in each of the other phases, on the post-merger phase we are instead restricted to only about 30 knots and thus, we will not include this phase in this analysis. The slopes derived for the closest systems are 1.68-1.68-1.92 (1.75-1.77-1.99)  for the \textit{I}- (\textit{B}-) band \lf for the first contact, pre-merger and merger phases, respectively. It is interesting to note that in this case the LF still becomes steeper when the systems evolve from the pre-merger to the merger phases.

(U)\lirgs are known to have high \sfrs at all phases of interaction. However, this can be concentrated at small scales or more dispersed and, at the same time, less or more hidden by dust. For instance, 80\% of the total infrared luminosity in IRAS 20550+1656 comes from an extremely compact, red source not associated with the nuclei of the merging galaxies but with a buried starburst (\mycitealt{Inami10}). If, on average, we detect more young knots in the optical at early phases of interaction than in more evolved phases, they can overpopulate the bright end of the luminosity function, and thus the measured slope would be shallower. In~\refcha{sim} we compare the observed \lfs with those computed from the mass functions in numerical simulations by Bournaud et al. (\mycitealt{Bournaud08a}) and stellar population synthesis models (i.e., mass to light ratios). This comparision  can help us learn if the observed trend (change of the slope as the interaction evolves) is real and allow us to investigate its origin.

\sect[clumps_highz]{Knots in (U)LIRGs vs. Star-forming Clumps at high-z}

(U)\lirgs represent the extreme cases of low-redshift star-forming galaxies  associated with spirals, interacting galaxies and mergers. Deep cosmological imaging and spectroscopic surveys at redshifts \mbox{z $\sim$ 1-4} have identified massive star-forming galaxies with irregular structures where the star formation appears to proceed in large clumps,with sizes in the 0.1-1.5 kpc range. The most recent are the surveys conducted in the  GEMS and GOODS fields (\mycitealt{Elmegreen06},~\myciteyear{Elmegreen07a}), in the Ultra Deep Field (UDF;~\mycitealt{Elmegreen05},~\myciteyear{Elmegreen07b},~\myciteyear{Elmegreen09}) and in the Spectroscopic Imaging survey in the Near-infrared with SINFONI (SINS;~\mycitealt{Schreiber11b}). While some of these galaxies show structural characteristics of being involved in mergers (\mycitealt{Elmegreen07a};~\mycitealt{Schreiber11a}), most systems appear as thick turbulent disks (\mycitealt{Elmegreen09}). Faint tidal features are difficult to discern at higher redshifts because of cosmological dimming. It is important therefore to make a direct comparison between the star-forming knots in (U)\lirgs and in clumpy high-z galaxies in order to understand the similarities and differences of the star formation in the present Universe and at cosmological distances.

\begin{table}
\hypertarget{table:ULIRG_highz}{}\hypertarget{autolot:\theautolot}{}\addtocounter{autolot}{1}
\begin{minipage}{\textwidth}
\renewcommand{\footnoterule}{}  
\begin{scriptsize}
\caption[Comparison of knots in ULIGRs with high-z clumps]{Comparison of knots in ULIGRs with clumps in the high-z Universe}
\label{table:ULIRG_highz}
\begin{center}
\begin{tabular}{l@{\hspace{0.3cm}}c@{\hspace{0.3cm}}c@{\hspace{0.3cm}}c@{\hspace{0.3cm}}c@{\hspace{0.3cm}}c@{\hspace{0.3cm}}c}
\hline \hline
   \noalign{\smallskip}
 System /field & rest-frame & redshift & N clumps & Mass range & Total size & Ref. \\
 & band / line & per galaxy & (\msun\twospace) & (kpc) & \\
 (1) & (2) & (3) & (4) & (5) & (6) & (7) \\
\hline 
   \noalign{\smallskip} 
UDF field	&	B	&	0.5 - 4	&	$\sim$ 5	&	10$^7$ - 10$^9$	&	1 - 2.5	&	[1], [2] 	\\
Lensed galaxies	&	\ha	&	1.7 - 3.1 	&	-	&	6$\times$10$^8$ - 3$\times$10$^9$	&	0.3 - 1 	&	[3]	\\
star-forming galaxies at z$\sim$2	&	optical g, UV, \ha	&	2.2 - 2.4	&	4.5	&	9$\times$10$^7$ - 9$\times$10$^9$	&	0.3 - 1.5	&	[4]	\\
GEMS \& GOODS fields	&	B	&	0.1 - 1.4	&	$\sim$ 4	&	5$\times$10$^6$ - 10$^8$	&	0.2 - 7	&	[5]	\\
Simulated ULIRGs at z=1	&	I	&	1	&	6.7	&	 2$\times$10$^5$ - 7$\times$10$^7$	&	0.4 - 3.5	&	This work	\\
TDG candidates in (U)LIRGs	&	I	&	0 - 0.1	&	$\sim$ 1	&	10$^6$ - 4$\times$10$^7$	& 	1 - 2 	&	[6] \\
\hline 
   \noalign{\smallskip} 
   \hline
\noalign{\smallskip}
\multicolumn{7}{@{} p{\textwidth} @{}}{\footnotesize{\textbf{Notes.} Col (1): Type of systems and deep fields. Col (2): Rest-frame photometric band. Col (3): Redshift interval of the galaxies. Col (4): Average number of clumps per galaxy. Col (5): Total stellar mass range of the clumps, except for the lensed galaxies, where the dynamical mass of the clumps is provided. Col (6): Total diameter of the clumps. For star-forming galaxies at \mbox{z = 2} the full-width at half-maximum from the clump radial light profiles after subtraction of the local background is given. Col (7): References--[1] Elmegreen \& Elmegreen (\myciteyear{Elmegreen05});~[2] Elmegreen (\myciteyear{Elmegreen07b});~[3] Jones et al. (\myciteyear{Jones10});~[4] Schreiber et al. (\myciteyear{Schreiber11b});~[5] Elmegreen \& Elmegreen (\myciteyear{Elmegreen06});~[6]~\refcha{tdgs} in this thesis.} }
\end{tabular}
\end{center}
\end{scriptsize}
\end{minipage}
\end{table}

The clumps in high-z galaxies are less numerous and intrinsically brighter and more massive (by a factor of 10-1000) than the largest star-forming complexes in local galaxies (\mycitealt{Efremov95};~\mycitealt{Elmegreen05};~\mycitealt{Elmegreen07a},~\myciteyear{Elmegreen09};~\mycitealt{Schreiber11b}). The integrated luminosity of these clumps  usually accounts for about 10-25\% of the total luminosity. The larger masses in the high-redshift galaxies are thought to be the result of high turbulence in the disks. In fact, velocity dispersions of 40 km s$^{-1}$ have been derived from spectroscopic observations of UDF6462, a \mbox{z = 1.6} galaxy (\mycitealt{Bournaud08b}). This large-scale star formation can also be reproduced by means of numerical simulations which require velocity dispersions several times higher than in local quiescent galaxies (\mycitealt{Bournaud07}). Finally, these high velocity dispersions have already been observed in clumps in local \ulirgs (\mycitealt{Monreal07}), since gravitational torques in major mergers can also induce star formation in large condensations of gas. Yet, the last mention has to be taken with care since strong winds and shocks in mergers can also induce large gradients in the velocity fields (i.e., high velocity dispersions).

In the standard $\Lambda$CDM cosmology (\mycitealt{Spergel07}) the angular size is nearly constant over the redshift range \mbox{0.7 $<$ z $<$ 4}. At closer redshifts, the angular size decreases approximately linearly with increasing redshift. A kpc-sized clump  corresponds to 3-4 pixels in the \acs images at high redshift values. Thus, clumps much smaller than this could not be measured at high redshifts. On the other hand, if there were clumps significantly larger than this in nearby galaxies, they would be easily discernible. The lack of spatial resolution at high redshifts can make a complex of star formation look like a clump since the light from the different clusters combines, blurs and some contrast with the local background is lost. Whether the high-z clumps correspond to a single entity or to complexes of star formation, the largest complexes can be identified and compared in both local and high-redshift galaxies. The \mbox{high-z} galaxy clumps have masses typically between 10$^7$ and 10$^9$~\msun from \mbox{z = 1 to 4}, decreasing to about 10$^6$~\msun at \mbox{z = 0.1-0.2}. The star-forming regions in the few interacting clumpy galaxies observed have sizes similar to star complexes in local interacting systems (i.e., the Antennae or M51), and their mass is 10-1000 times higher. On the other hand, complexes in the Hickson Compact Group 31 (HCG 31), stellar complexes, which are sensitive to the magnitude of disk turbulence, have both sizes and masses more characteristic of z = 1-2 galaxies (\mycitealt{Gallagher10}).

In~\refsec{mc_lir} we saw that knots in \ulirgs are intrinsically more luminous than knots in less luminous systems. This means that for complexes of similar sizes and similar population (\reftab{prop_lir} shows that the colors of the knots in \lirgs and \ulirgs are similar) the knots in \ulirgs (they could also be understood as clumps or complexes of star clusters) can be more massive than in our low luminous \lirgs by at least a factor of 4. Although the knots in our \ulirgs only account for 0.5-2\% of the total luminosity (much less than clumps at high redshift), we can compare them with the complexes in high-redshift galaxies in order to determine if the star formation is similar or not to that in clumpy galaxies . Already at \mbox{z $\simeq$ 0.05-0.1} a significant percentage of knots in \ulirgs are more luminous than \mbox{\mi\twospace= -15 mag} (see~\reffig{mag_col_lira}). According to the models used in this study, if we assume that these knots are formed by a single young (e.g., \mbox{ $\tau \sim$ 6 Myr}) population  they can be as massive as \mbox{few 10$^4$ - few 10$^7$~\msun\twospace}, with a median value of \mbox{1.4$\times$10$^5$~\msun\onespace}. An older population (e.g., \mbox{$\tau$ $\sim$ 50 Myr}) implies a knot with higher mass by one order of magnitude. Correction for internal extinction makes the population even more massive. Thus, these values have to be understood as a lower limit. To compare the knots in these \ulirgs with the clumps at high-redshift galaxies, we applied a Gaussian blur to our images in such a way that we simulate their appearance at \mbox{z = 1} (the scale at this redshift corresponds to 7.69 kpc/\arcsec), using the same technique as explained in~\refsec{distance_effect}.

\begin{figure}
  \hypertarget{fig:deg_highz}{}\hypertarget{autolof:\theautolof}{}\addtocounter{autolof}{1}
\hspace{-0.8cm}
\includegraphics[angle=90,width=1.\columnwidth]{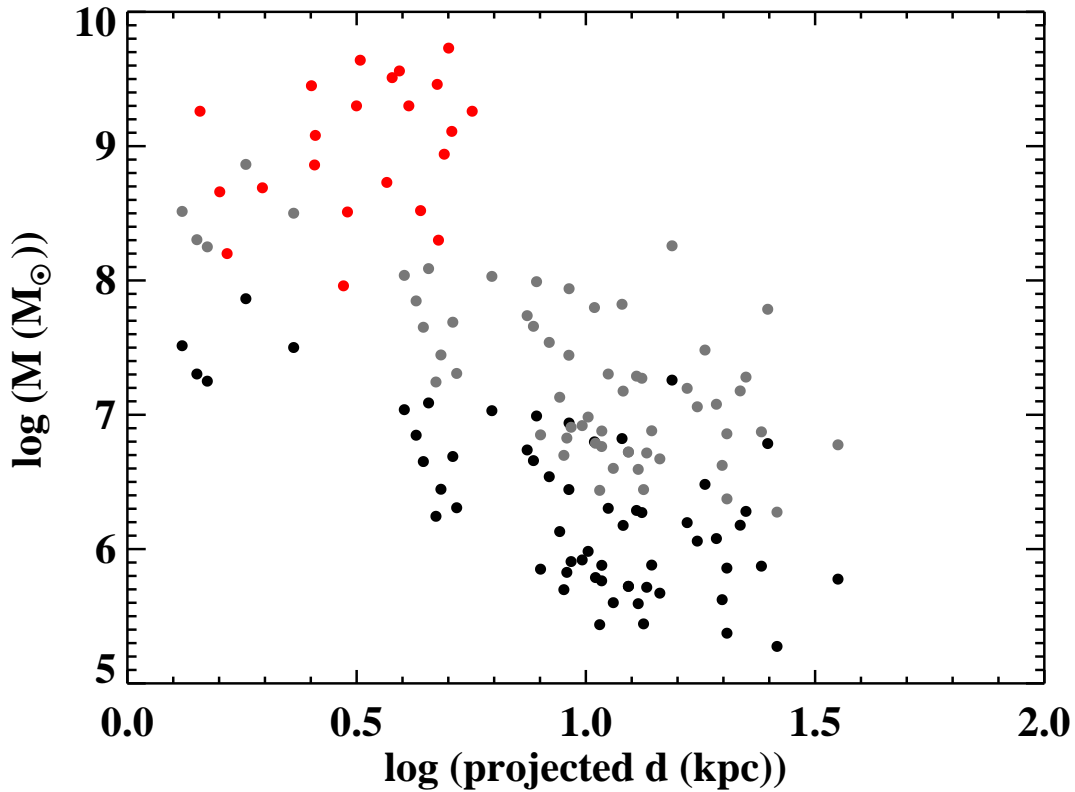}
      \caption[Mass vs. projected distance for complexes in simulated ULIRGs at z=1]{Mass vs projected distance for complexes in simulated \lirgs at z=1 (in black for a population of 6 Myr and in gray for a population of 50 Myr) and clumps in z=2 star-forming galaxies (red; values taken from \mycitealt{Schreiber11b}).}
      \label{fig:deg_highz}    
\end{figure}


The knots in the blurred images (hereafter high-z knots) were measured.~\reftab{ULIRG_highz} show the results together with the values for clumps in galaxies at high redshifts and \tdgs candidates found in our sample of (U)\lirgs (\refcha{tdgs}). Some of the high-z knots were not recognizable in the blurred images because they were too faint or too small; a few were blended with other knots that were too close or with the nucleus. Most of the lowest mass knots and those close to the nucleus disappear, although the remaining high-z knots range from 2$\times$10$^5$~\msun to 7$\times$10$^7$~\msun\twospace, with a median value of \mbox{2.5$\times$10$^6$~\msun\onespace}. The effective radius averages about 700 pc and the total size is typically \mbox{$r \gtrsim$ 1.5 kpc}. Taking into account that the derived masses correspond to lower limits, they are typically one or two orders of magnitude less massive than clumps in high-z galaxies, having similar sizes. If the population that dominates the light of the high-z knots were older (e.g., $\tau$ $>$ 50 Myr), the masses of the high-z knots would significantly approach those of clumps in high-z galaxies.

Having set a lower limit for the mass of knots in \ulirgs\twospace, the mass of kpc-complexes in these systems is not likely to lie far from clumps in high-z galaxies with a similar size. Therefore, the extreme cases of low-redshift star-forming galaxies (\ulirgs\twospace) are likely to induce large-scale star formation similar to that observed in \mbox{high-z} disks (and some few mergers) with a high gas fraction.

{We have also investigated how the projected distance of the high-z knots relates to their mass (see~\reffig{deg_highz}. While other works do not find any clear relation, we measure lower mass high-z knots with increasing distance. Although, according to our results it is clear that the mass is mainly concentrated at small distances in \ulirgs\twospace, we can only compare with one study of high-z galaxies and we explore larger distances by a factor of 10.

\sect[summary_knots]{Summary and Conclusions}

A comprehensive study of bright compact knots (most of them apparent associations of clusters) in a representative sample of 32 (U)\lirgs as a function of infrared luminosity and interaction phase has been performed using \hst\twospace/\acs \mbox{\textit{B}- and} \textit{I}-band imaging, providing linear resolutions of 10 to 40 pc, depending on the distance of the galaxy. The sample has been divided into three infrared luminosity intervals (low, \mbox{\lir $<$ 11.65}; intermediate, \mbox{11.65 $\leq$ \lir $<$ 12.0}; and high, \mbox{\lir $\geq$ 12.0}) and in the 4 interaction phases defined in Chapter~\ref{cha:data_tech} (\refsec{morphologies}), namely  first approach, pre-merger, merger and post-merger. We have extended toward lower infrared luminosities previous stu\-dies focused only on \ulirgs (e.g.,~\mycitealt{Surace98},~\myciteyear{Surace00}), and have detected close to 3000 knots, larger than a factor of ten more than in the previous studies of \ulirgs\twospace. The main conclusions derived from this study are the following:

\begin{enumerate}

\item The knots span a wide range in magnitudes  (\mbox{-20 $\lesssim$ \mi $\lesssim$ -9} and \mbox{-19.5 $\lesssim$ \mb $\lesssim$ -7.5}), and color \mbox{(-1 $\lesssim$ \mbi $\lesssim$ 5)}. The median values are \mbox{\mb= -10.84}, \mbox{\mi= -11.96} and \mbox{\mbi = 1.0}. \\

\item The knots are in general compact, with a median effective radius of 32 pc, being 12\% unresolved and a few very extended, up to 200-400 pc. With these sizes, in general (and particularly for galaxies located at 100 Mpc or more) the knots constitute complexes or aggregates of star clusters.  Within the resolution limits there is no evidence of a luminosity dependence on the size. A slight dependence on the interaction phase, in particular in the post-merger phase, needs confirmation with larger samples and better angular resolution.\\

\item A non-negligible fraction (15\%) of knots are blue and luminous. Given their color (\mbox{\mbi$<$ 0.5}) and magnitude (\mbox{$<$\mb\twospace$>$= -11.5}) they appear to be young (ages of about 5 to 30 Myr), almost free of extinction, and have masses similar to and up to one order of magnitude higher than the young massive clusters detected in other less luminous interacting systems. In addition, some (5.7\%) very luminous (\mbox{\mb $<$ -12.5}),  and relatively blue (\mbox{\mbi $<$ 1.0}) knots appear along the tidal tails and at their tips. These objects could be candidates to Tidal Dwarf galaxies. \\

\item Unlike for isolated young star clusters in other systems, we find a clear correlation between the knot mass and radius, M$\propto R^{2}$, similar to that found for complexes of star clusters in less luminous interacting galaxies (e.g., M51 and the Antennae) and Galactic and extragalactic giant molecular clouds. This relation does not seem to be dependent on the infrared luminosity of the system or on the interaction phase.

\item The star formation is characterized by \textit{B}- and \textit{I}-band luminosity functions (\lf\twospace) with slopes close to 2, extending therefore the universality of the \lf measured in interacting galaxies at least for nearby systems (i.e., \mbox{\ld$<$ 100 Mpc}), regardless of the total luminosity (i.e., the strength of the global star formation). This result has to be taken with caution because the \lf of our knots (which can be mainly formed of complexes of star clusters) may not reflect the same physics as the \lf of individual star clusters. Nevertheless, there are slight indications that the \lf evolves as the interaction progresses becoming steeper (from about 1.5 to 2 for the \textit{I}-band) from first approach to merger and post-merger phases.\\

\item Taking into account distance effects, knots in high luminosity systems (\ulirgs\twospace) are intrinsically more luminous (a factor of about 4) than knots in less luminous systems. Given the star formation rate in \ulirgs is higher than in less luminous systems, size-of-sample effects are likely to be the natural explanation for this. \\

\item Knots in \ulirgs would have both sizes and masses characteristic of stellar complexes or clumps detected in galaxies at high redshifts (\mbox{z $\gtrsim$ 1}), were their population about 50 Myr or older. Thus, there is evidence that the larger-scale star formation structures are reminiscent of those seen during the epoch of morphological galaxy transformations in dense environments.

\item Knots in systems undergoing the post-merger phase are on average redder and more luminous than those in systems in earlier phases. Scenarios involving either a different spatial distribution of the obscuration or a different evolutionary phase in which small knots do not survive for long are considered. The evolutionary scenario could also explain the trend in size toward higher values in post-merger systems. A comparison with numerical simulations to further investigate these scenarios are presented in~\refcha{sim}. \\

\end{enumerate}

\clearpage{\pagestyle{empty}\cleardoublepage}

\cha{sim}{Comparative Study with High Spatial Resolution Simulations}
\chaphead{This Chapter complements the previous one by including the theoretical frame in the analysis and characterization of the data. We present the first direct comparison of the properties of stellar regions (simulated knots) identified in a high spatial resolution simulation of a major merger with knots detected in our sample of (U)\lirgs\twospace. The main purpose of the comparison is to be able to interpret some of the properties measured in the previous Chapter and to constrain the results of the simulation.}

\sect[intro_sim]{Introduction}

The role played by interactions and mergers in enhancing star formation has been widely studied in the past few decades by means of numerical simulations (e.g., ~\mycitealt{Barnes92a};~\mycitealt{Mihos94a};~\mycitealt{Mihos96};~\mycitealt{Springel00};~\mycitealt{Barnes04};~\mycitealt{Springel05b};~\mycitealt{Bournaud06};~\mycitealt{Cox06};~\mycitealt{diMatteo07};~\mycitealt{diMatteo08};~\mycitealt{Teyssier10}). These studies have supported the fact that major mergers induce star formation, though it has become clear that the star formation efficiency is highly dependent upon the numerical recipe adopted for star formation (density or shock dependent) and feedback assumptions. The spatial resolution achieved by the model is also decisive (\mycitealt{Teyssier10}). The higher the resolution the better the models can handle the inner phenomena affecting the star formation.  

Even though numerical simulations represent a powerful tool to study galaxy mergers, none of the previous studies has been able to monitor the physical evolution of the star-forming regions at typical scales of Super Star Clusters (\sscs\twospace) or associations of few clusters. Bournaud et al. (\myciteyear{Bournaud08a}) presented a simulation of a wet galaxy merger with the highest resolution in mass and spatial sampling reached until then. With a spatial resolution of 32 pc, they were able to directly identify dense structures with typical masses of 10$^6$-10$^8$~\msun\onespace, similar to the most massive \sscs and aggregates of few clusters. They argued that these dense structures are tightly bound and likely progenitors of long-lived globular clusters. Models capable of resolving compact star-forming regions at similar spatial scales to observations at \mbox{z $\leq$ 0.1} offer the opportunity to make a one-to-one comparison. 

This Chapter presents the first direct comparison of the properties of stellar regions (simulated knots) identified in the galaxy simulation by Bournaud et al. (\myciteyear{Bournaud08a}) with knots detected in (U)\lirgs\twospace. Belles et al. (in preparation) has performed the physical characterization of the most massive simulated knots (i.e., \mbox{log M (\msun\onespace) $\ge$ 7.5}). In this thesis we also extend the statistics to all mass ranges, down to about \mbox{log M (\msun\onespace) $\sim$ 6}. With the available data from the simulation (see Chapter~\ref{cha:data_tech},~\refsec{sim_data}}) and using stellar population synthesis techniques, the magnitudes, colors, and \lfs of the simulated knots are evaluated at different interaction phases. This enables us to perform a direct comparison with the properties of the observed knots studied in the previous Chapter, a study of stellar regions in an observational and theoretical frame simultaneously.  

To avoid distance effects in this study, we only consider the knots detected in the previous Chapter that belong to systems located at similar distances in each phase (see Chapter~\ref{cha:knots},~\refsec{prop_is}). That is,  a total of about 2200 knots in 22 (U)\lirgs\twospace. The galaxies are located at a median distance of 125 Mpc, with a dispersion of 40 Mpc.

The layout of the Chapter is as follows:~\refsec{def_add} provides the description of additional analysis treatment of the data from the simulations and the definition of the interaction phases under comparison; the comparison of the properties of the knots detected in observations and simulations is performed in~\refsec{comparison}, along with the temporal evolution of their physical and spatial distributions with the interaction process; finally, we summarize the main conclusions in~\refsec{summary_sim}.

\sect[def_add]{Additional Analysis Treatment and Definitions}
\subsect[phot_sim]{Derivation of the Photometric Properties of the Simulated Knots}

The output of the simulation corresponds to the mass and the position of each particle at any given snapshot. However, the observational data consist of photometric (light) measurements. To compare both sets of data, we have chosen to convert the mass of the simulated knots into light and compare the modeled luminosities because we know their age with an uncertainty of a few Myr. These estimates are more reliable than the determination of masses from the observed luminosity, which can be uncertain by large factors (10-100 or even larger) if we only have the knowledge of the luminosities in the \filterb and \filteri bands (see~\reffig{ssp_models} in Chapter~\ref{cha:data_tech}, Section~\ref{sec:ssp}). 

To perform such analysis, we first have to know the star formation history of the simulated knots and then apply stellar population synthesis techniques to obtain the observables (luminosities) we need to compare with the observed knots.

\subsubsection{Initial Mass and Growth Phase in the Simulation}

Reliable estimates of the magnitudes of the simulated knots require to know as well as possible their mass and age, and if they are formed by one or more populations. Belles et al. (in prep) argue that the knots undergo a rapid growth phase in which their mass increases by up to a factor of about 100 during their first 100 Myr or so of existence. During this phase, the number of new stars within each knot increases, while the number of old stars does not vary significantly. This suggests a scenario in which clumps of gas form first and then collapse to form stars. However, the model does not have enough resolution for us to know if the clump fragments and form little clusters which then merge together, or if the collapse is monolithic.

Once this phase ends, the mass of the knots stabilizes to what it is defined as the initial mass (\mini\twospace). Later on, they lose mass (due to dynamical friction, two-body relaxation, etc.), generally in a much moderate way. Since the simulated knots are tracked until they disappear or until the end of the simulation, we assumed their initial mass to be the mass value 80-100 Myr (i.e., 6-8 snapshots) after it is detected for the first time. Then, if a knot is massive enough for us to follow all the growth phase in the simulation, after 6-8 snapshots it will have reached the initial mass already. If, on the other hand, the knot is less massive, we will have missed this phase partially or completely. However, after 6-8 snapshots its mass will be very similar to the initial mass because statistically it will lose mass in a much moderate way after having reached \mini\twospace. Thus, the initial mass was taken as the peak value 80-100 Myr after it is detected for the first time.

Once the initial mass of the knots was known, we studied this phase of rapid growth for the most massive knots. If a knot has an initial mass higher than \mbox{\mini = 10$^8$~\msun\twospace}, at the time when it is identified for the first time (at the beginning of this phase) it is still massive enough (i.e., \mbox{M $>$ 10$^{6.2}$~\msun\twospace}) to be detected against the local background. Hence, the rapid growth phase is completely tracked for the most massive knots. In contrast, this phase is missed partially or completely for a knot with lower initial mass since, assuming that the rapid growth phase is similar, its mass at the beginning of the phase would be too low for the knot to be detected. In the extreme case where the initial mass of a knot is lower than \mbox{M $>$ 10$^{6.5}$~\msun}\twospace, we identify it for the first time when it has already reached the initial mass. This mass (\mbox{10$^{6.5}$~\msun\twospace}) represents the \textit{global mass completition limit} defined in Belles et al. (in preparation).

We decomposed the growth phase in successive bursts of star formation. Since this phase lasts approximately six snapshots, the knots are formed by a combination of six different stellar population, with different masses and ages. For each population, we derived the mass fraction and dragged it throughout the whole evolution of the simulation. Thus, for instance, when a massive knot appears for the first time, it has only one population. One additional population is created per snapshot until the knot reaches its initial mass. Then, the mass fraction of each population is frozen and used as such until the end of the simulation or its dissolution.

\subsubsection{Modeling the Stellar Populations}

Characterizing the growth phase of the simulated knots involves identifying each population present in a knot and, especially, its age.

To characterize the growth phase of the simulated knots we need to identify as well as possible each population present in a knot. With that information and the use of stellar population synthesis models we were able to derive observables, such as the luminosities of the simulated knots.

The tracks of the \sbnn models with solar metallicity, optimized for young ages, are used up to \mbox{$\tau$ = 100 Myr}. At older ages, the tracks by Maraston et al. (\myciteyear{Maraston05}) are considered, where a rigorous treatment of the thermally pulsing asymptotic giant branch (TP-AGB) phase has been applied (relevant within the interval 0.1-1 Gyr). We consider this rather high metallicity even at large radii because global mixing of the ISM in mergers, due to gas inflows and outflows, flattens any metallicity gradient of the parent galaxies (\mycitealt{Rupke10};~\mycitealt{diMatteo11};~\mycitealt{Perez11}). Given the relatively small gas mass at large radii, mixing of small amounts of more enriched gas easily raises the gas metallicity in these regions. 
  
The growth phase is not completely tracked in the simulation for knots less massive than 10$^8$~\msun\twospace, thus we assumed that they are characterized in the same way as the massive ones. When computing the mass fraction of the six stellar populations, we applied a gradual correction:
\begin{enumerate}[(i)]
 \item For knots in the interval 10$^7$ $<$ \mini (\msun\twospace) $<$ 10$^8$ (\textit{first interval}) we assumed we had missed the growth phase for 2 snapshots.
 \item For knots in the interval 10$^{6.5}$ $<$ \mini (\msun\twospace) $\leq$ 10$^7$ (\textit{second interval}), we had missed it for 4 snapshots.
 \item Finally, for knots less massive than the detection threshold (\textit{third interval}, i.e., \mbox{\mini $<$ 10$^{6.5}$~\msun\twospace}), we had missed it for 6 snapshots (i.e., the whole phase).
\end{enumerate}

Thus, when a knot in the first interval was detected for the first time we assumed it was already formed by three stellar populations, with ages \mbox{$\tau$ = 6.5 Myr} (detected), 6.5+13 (19.5) Myr (missed) and 6.5+13+13 (32.5) Myr (missed), respectively. Then, in order to estimate the luminosity we used the averaged mass fractions computed for all the massive knots when they are formed by three populations. In the following snapshot, the knot was assumed to be formed by four stellar populations (the older two we missed and the younger two in which the knot was detected). The new mass fractions of the non-detected with respect to the detected populations were computed using the average mass fractions that we had derived for all the massive knots. Once the knot reached the six stellar populations (at the fourth snapshot in this case), the final mass fraction values were frozen and dragged until the knot disappeared or until the end of the simulation. A similar analysis procedure was followed for the knots in the second interval. The knots in the third interval were assumed to have already reached their initial mass when they were first detected (i.e., they are already formed by the six stellar populations). Hence, we dragged all the populations from the beginning and the mass fractions applied correspond to the average values that we had derived for the most massive knots.

This process introduce a typical uncertainty of 5-25 \% in the knowledge of the mass for each of the six populations for the least massive knots. Another major source of uncertainty comes from the age of the young population. When a knot appeared for the first time in the simulation we assumed the youngest out of the six populations is 6.5 Myr old, since the age elapsed between two consecutive snapshots is 13 Myr. To know how this is translated into an uncertainty in magnitude, we assumed that the ages distribute following a Gaussian function in the interval 0-13 Myr, with a mean value of 6.5 Myr and a one-sigma (1$\sigma$) deviation of 6.5/3 (i.e., 2.17) Myr. According to the \sbnn models, this introduces a 1$\sigma$ uncertainty of up to 0.5 mag in the \filteri band and up to 0.3 mag in the \filterb band. These uncertainties decrease considerably due to the aging of the knots while they evolve throughout the interaction. After about 50 Myr, they become negligible (less than 0.1 mag) compared to the uncertainties related to the mass fraction for each of the six populations. Hence, considering both major sources of uncertainties, the error in the estimated magnitudes for the new knots (at the snapshot where they appear for the first time)  can be as high as 0.75 mag. About three snapshots later this error drops below 0.2-0.3 mag, since at that time the new population with an age of 6.5 Myr represents a very small fraction in mass. About seven snapshots later the main source of uncertainty comes from the mass fraction for each of the six populations, and the magnitudes computed are reliable within less than 0.2 mag. 

\subsect[phases_def]{Interaction Phases Under Comparison}

Although a single 1:1 simulation was carried out, it is rather typical and characteristic of many observed real systems. For instance, NGC 7252, the Mice and The Antennae can be modeled by roughly equal-mass mergers with high gas fractions, from 12.5 to 20\% (\mycitealt{Hibbard95};~\mycitealt{Barnes04}). Three out of the four double system \ulirgs in Colina et al. (\myciteyear{Colina05}) also have a mass ratio of 1:1, and most \ulirgs in Garc{\'i}a-Mar{\'i}n et al. (\myciteyear{Garcia-Marin07}) are compatible with a mass ratio of 1:1 and 1:2. However, other parameters such as the orbit inclination differs significantly from system to system. On the other hand, the observational data come from a sample of dozens of systems, surely with galaxies having different initial configurations. Therefore, we only expect to find general trends between both observed and simulated data.

In Chapter~\ref{cha:data_tech} (\refsec{morphologies}), we classified the (U)\lirg sample according to four interaction phases, depending on the optical morphology of the systems. A similar classification scheme to that in Veilleux et al. (\myciteyear{Veilleux02}) was followed, though with a few simplifications.  An age interval for each morphological class was assigned. Given the uncertainties involved when performing morphological classifications and in order to make a more realistic comparison, we assigned in the simulation a typical age to each morphological class defined in Chapter~\ref{cha:data_tech} (\refsec{morphologies}) and considered a window of about 150 Myr to characterize the properties of each class. We then identify in the simulation typical ages (central ages in the window) for the pre-merger, merger and post-merger interaction phases to be 299, 585 and 988 Myr. We thus considered snapshots covering $\pm$ 75 Myr to define the interaction phase. Owing to computing time issues, the simulation in this study extends only up to \mbox{1 Gyr}. However, the duration of a merger can be more extended in time until the system relaxes (e.g., simulations by~\mycitealt{diMatteo08} extend to longer than 2 Gyr). As a consequence, the comparison of this simulation with very relaxed galaxies (i.e., a very old elliptical galaxy) is not appropriate. Nevertheless, a comparison with a (U)\lirg remnant, with past signs of interaction (e.g., outer asymmetries), no fully relaxed, is not worrysome. 

Some galaxies were identified to be in a first-contact phase (before the first pericenter passage). This phase, corresponding to the first snapshots in the simulation, was not considered for this study because only a few simulated knots were detected (e.g., only 34 knots in the snapshot number 8, at t=104 Myr). This unrealistic early scenario is the consequence of how the star formation was implemented in the code, in such a way that no star formation existed before the encounter. Therefore, the knots in the three interaction phases pre-merger, merger and post-merger were compared for both sets of data. 

\sect[comparison]{Observed Knots vs. Simulated Knots}

\subsect[mag_col_sim]{Luminosities and Colors}

\begin{figure}
 \hypertarget{fig:number_time}{}\hypertarget{autolof:\theautolof}{}\addtocounter{autolof}{1}
\hspace{1cm}
\includegraphics[angle=90,width=0.85\textwidth]{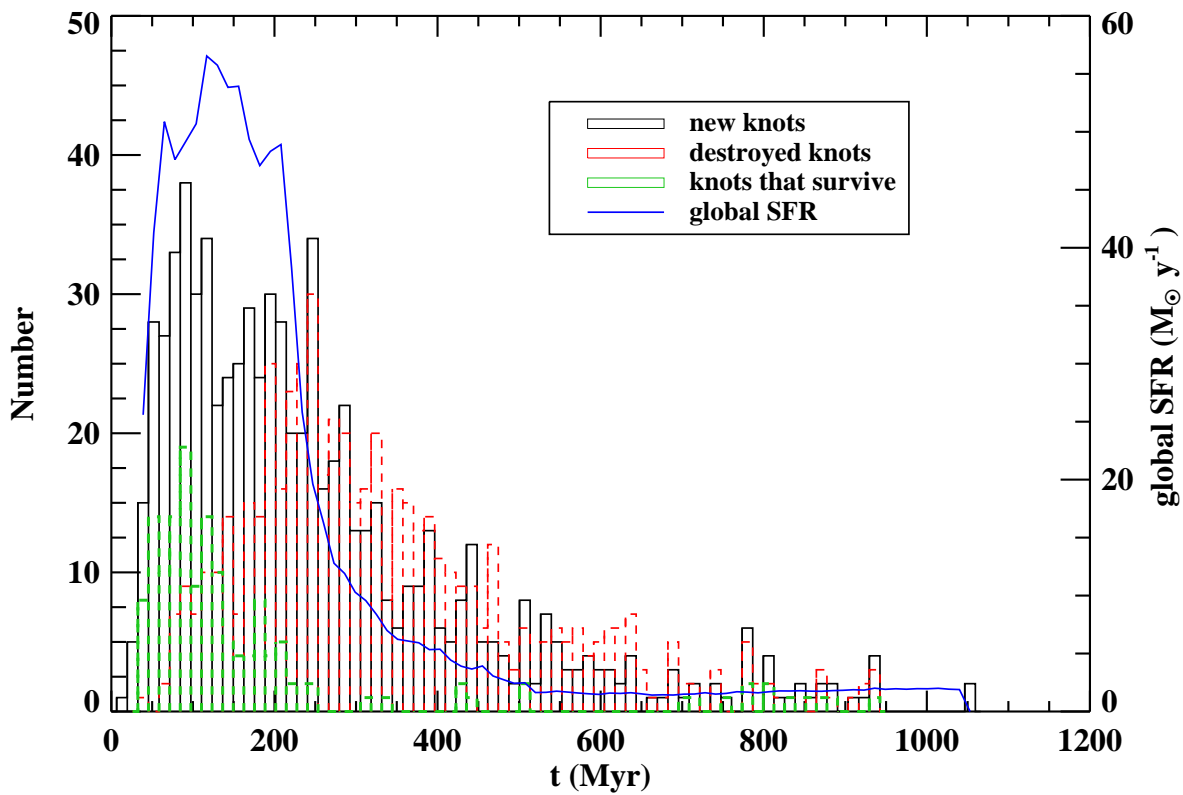}
\vspace{1cm}
      \caption[Number of simulated knots newly formed and destroyed with time]{Number of simulated knots newly formed (black) and destroyed (dashed red) per snapshot. The green dashed histogram shows how many of the formed knots at a given snapshot survive until the end of the simulation. The blue line shows the global \sfr\twospace, which refers to the total mass of new particles formed into stars divided by the time extent of the snapshot. The peak of the global \sfr\twospace, at about \mbox{t = 150 Myr}, occurs during the first pericenter passage. The detection delay of the knots less massive than \mbox{\mini = 10$^8$~\msun} has been corrected in the plot.}
      \label{fig:number_time}
\end{figure}

\subsubsection{Preliminary Considerations}

In this section we aim at comparing the luminosities and color distributions of both the simulated and observed knots during the interaction process. In order to understand differences and similarities we first make the following considerations:

\begin{itemize}
 \item The time resolution of the simulation is 13 Myr, which means that any variation in luminosity and color during that time cannot be known. \\
 \item According to the \sbnn models and considering the previous limitation, during the first 100 Myr of the life of any knot, the color does not change significantly while the luminosity does. We define this period as a ``constant-color'' phase. 
 \item After 100 Myr simulated knots become redder and fainter.
 \item The knots that survive until the end of the simulation (long-lived knots) are formed early on (see ~\reffig{number_time}), and it turns out tha they are also the most massive ones (see Belles et al. in preparation). Most of the other knots are destroyed before reaching 100 Myr. Thus, we can assume that the color distribution is actually driven by the most massive knots in the simulation. 
 \item The global star formation rate (\sfr\twospace) in the simulation is peaked at about 150-200 Myr after the first pericenter passage and then it drops quickly, such that it holds very low from about 400 Myr after the first pericenter passage until the end of the simulation (see ~\reffig{number_time}). In contrast, (U)\lirgs are systems with high \sfrs\twospace, independently of the interaction phase.
\end{itemize}

With the knowledge of these facts, we are now able to investigate what occurs in each interaction phase. The distributions of all the observables that we discuss in this section are shown in~\reffig{mass_mag_sim}. The averages values are given in~\reftab{sim_res}.

\begin{figure}
 \hypertarget{fig:mass_mag_sim}{}\hypertarget{autolof:\theautolof}{}\addtocounter{autolof}{1}
\includegraphics[trim = -2.5cm -2cm 0cm 6cm,clip=true,width=0.95\textwidth]{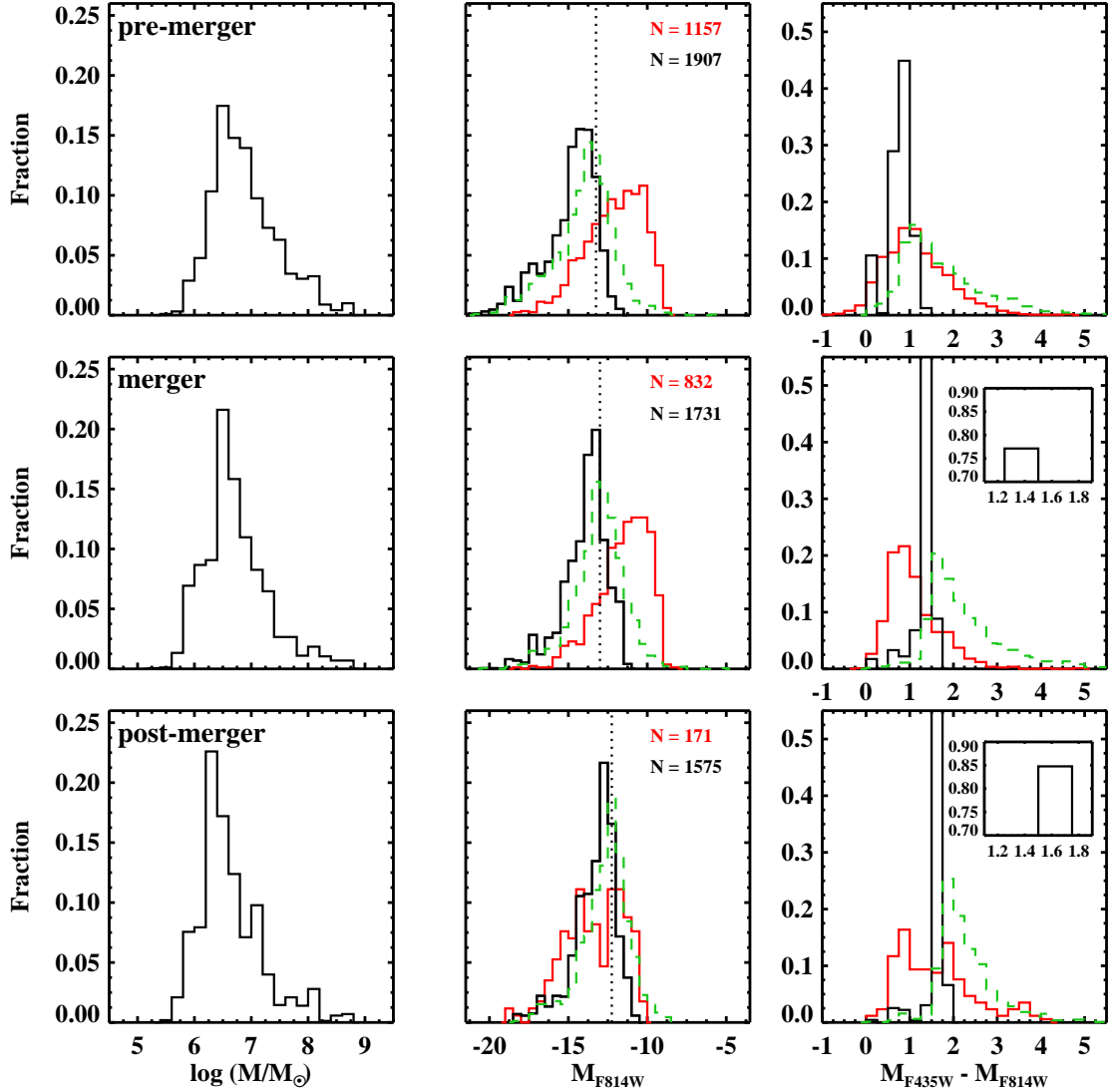}
      \caption[Mass, \textit{I}-band magnitude and color distributions for the simulated knots]{Mass, \textit{I}-band magnitude and color distributions for the simulated knots (in black) adding up the knots identified in each of the 11 snapshots in which we defined an interaction phase. The vertical dashed line indicates the luminosity of a single stellar population with a mass of \mbox{10$^{6.2-6.5}$~\msun} (the mass completeness, as inferred from the mass distributions showed) and the same age as the oldest simulated knot in each phase, in order to give an idea of the magnitude threshold of the distribution. Red line histograms show the same distributions for the observed data (see~\refcha{knots}). N gives the number of detected knots in each case. Green line histograms show the distribution of the simulated knots applying an extinction following an exponential probability density function and that best matches the red wing (i.e., \mbox{\mbi $>$ 1-1.5}) of the color distribution of the observed knots (see text). The inset plots in the third column extend the y-axis to better visualize the peak of the color distribution of the simulated data.}
      \label{fig:mass_mag_sim}
\end{figure}

\subsubsection{Results of the Simulation}

\begin{table}
 \hypertarget{table:sim_res}{}\hypertarget{autolot:\theautolot}{}\addtocounter{autolot}{1}
\begin{minipage}{\textwidth}
\renewcommand{\footnoterule}{}  
\begin{scriptsize}
\caption[Mean properties of the simulated and observed knots]{Mean properties of the mass and luminosity distributions of the simulated and observed knots at different interaction phases}
\label{table:sim_res}
\begin{center}
\begin{tabular}{l@{\hspace{0.20cm}}c@{\hspace{0.20cm}}c@{\hspace{0.20cm}}c@{\hspace{0.20cm}}c@{\hspace{0.20cm}}c@{\hspace{0.20cm}}c@{\hspace{0.20cm}}c@{\hspace{0.20cm}}c@{\hspace{0.20cm}}c}
\hline \hline
   \noalign{\smallskip}
Interaction & N & N & Mass$_{\rm{sim}}$ & \mi & \mi & \mb & \mb & C & C \\
phase & sim & obs & log (M/\msun$\!$) & sim & obs & sim & obs &sim & obs\\
(1) & (2) & (3) & (4) & (5) & (6) & (7) & (8) & (9) & (10)\\
\hline
   \noalign{\smallskip}
pre-merger& 178$\pm$11& 145$\pm$ 81& 6.8$\pm$ 0.6& -14.5 $\pm$ 1.6& -11.9 $\pm$ 0.9& -13.7 $\pm$ 1.8& -10.8 $\pm$ 0.9& 0.8 $\pm$ 0.3& 1.1 $\pm$ 0.3\\
merger& 160$\pm$6& 139$\pm$ 123& 6.6$\pm$ 0.6& -13.7 $\pm$ 1.4& -11.2 $\pm$ 0.7& -12.3 $\pm$ 1.4& -10.2 $\pm$ 0.7& 1.4 $\pm$ 0.3& 1.0 $\pm$ 0.2\\
post-merger& 145$\pm$2& 34$\pm$ 19& 6.5$\pm$ 0.6& -13.0 $\pm$ 1.4& -13.4 $\pm$ 1.1& -11.4 $\pm$ 1.4& -11.5 $\pm$ 0.8& 1.7 $\pm$ 0.2& 1.5 $\pm$ 0.4\\
\hline
\noalign{\smallskip}
\multicolumn{10}{@{} p{\textwidth} @{}}{\footnotesize{\textbf{Notes.} Col (1): defined interaction phase. Col (2): average number of simulated knots per snapshot during a given phase. Col (3): average number of detected knots per (U)\lirg system. Col (4): median value of the mass of the simulated knots in the 11 snapshots in which we have defined the interaction phase. Col (5): median value of the magnitude \mi of the simulated knots. Col (6): idem for knots in observed (U)\lirgs\twospace. Col (7): median value of the magnitude \mb of the simulated knots. Col (8): idem for knots in observed (U)\lirgs\twospace. Col (9): median value of the  \mbi color of the simulated knots. Col (10): idem for knots in observed (U)\lirgs\twospace.}}
\end{tabular}
\end{center}
\end{scriptsize}
\end{minipage}
\end{table}

In the simulation we are sampling knots typically more massive than 10$^6$~\msun\twospace. As observed in ~\reffig{mass_mag_sim}, the mass completeness in the simulation is about \mbox{log (M/\msun\twospace)=6.2-6.5} depending on the phase. According to our magnitude estimates, this translates into a luminosity threshold of \mbox{\mi = [-12,-13]}.  

The median value of the luminosity distributions of the simulated knots becomes fainter as the interaction proceeds (see~\reftab{sim_res}). Most of the short-lived knots are formed during the pre-merger phase (see~\reffig{number_time}). Since they are rather young when they disappear, a higher median value of the luminosity distribution is expected in more evolved phases. Moreover, at later phases (i.e., merger phase), the system relaxes and starts to forget the initial conditions. As a consequence of stellar evolution (aging), the knots also become fainter. In the post-merger phase the system is practically dead and aging effects dominate the fate of the system.

The colors become redder by about 1 mag (see~\reftab{sim_res}) and their distribution becomes more peaked (see~\reffig{mass_mag_sim}) as the interaction proceeds. In the simulation the \sfr peaks just after the pre-merger phase starts, then it drops very rapidly in a period which already exceeds the ``constant color'' phase (see~\reffig{number_time}). All the long-lived knots, which are likely to affect most significantly the color distribution, are formed. As a consequence, basically blue populations (those which formed last), intermediate-color populations and rather minor red populations are present in this early phase. In the merger phase, as the system starts to relax and to forget the initial conditions, the color distribution is basically driven by the most massive knots (formed early on), and it starts to redden. Since the presence of a young population is hugely diminished, the color becomes less spread and peaks toward redder values. In the post-merger phase, the knots become even redder, since only aging effects play the role. 

\subsubsection{Observations vs. Simulations. General Comparison}

In the previous section we estimated that the faint end of the luminosity distribution of the simulated knots is affected by the limited mass resolution of the simulation. This can explain why in the pre-merger and merger phases the luminosity of the observed knots is on average up to a factor of 15 (3 mag) lower than the luminosity of the simulated knots (see~\reftab{sim_res}). In fact, the luminosity threshold in the simulation is about this amount higher than the luminosities at the peak of the distributions for the observed knots (see~\reffig{mass_mag_sim}), more or less the completeness limits.

The numerical simulation was designed to reproduce correctly the most massive structures, while it misses the low mass ones. Thus, if this less massive population were not missed the median value of the luminosity distributions of observed and simulated knots would approach. Once this bias understood, we see that the evolutionary trend due to aging is also observed in (U)\lirgs during these two early phases (knots become fainter by about 0.7 mag). 

During the post-merger phase we see the following changes:

\begin{itemize}
 \item  On the one hand, there is a good agreement between the observed and simulated luminosity distributions. In this case, the faintest knots in observed systems seem to have disappeared. This might not be just a coincidence since, according to studies on long-term evolution of cluster populations (e.g.,~\mycitealt{Whitmore07};~\mycitealt{Fall09}), low mass clusters (populating the faint end of the luminosity distribution) have more probabilities than the most massive ones to be disrupted. Thus, during the pre-merger and merger phases this faint end of the luminosity distribution is not sampled in the simulation by resolution issues. And in the post-merger phase it is probably not sampled because there are no low-mass knots to sample, since they have dissolved in the interestelar medium.
  
 \item On the other hand, the observed knots in this phase are more luminous (by about 2 mag) than in previous phases, (see~\reffig{mass_mag_sim} and~\reftab{sim_res}). We would expect that, by aging effects (predicted by the simulations), the population became fainter.   The merging of smaller clusters into massive superclusters, as suggested by Kroupa (\myciteyear{Kroupa98}) and argued in~\refcha{knots}, could be responsible for the lack of fainter knots in post-mergers (U)\lirgs\twospace. As mentioned before, this process can not resolved in the simulation given its spatial resolution.

\end{itemize}

In summary, both theoretical and observational data support that after the dynamical process of interaction in galaxy mergers the population is concentrated in the most massive structures.

\subsubsection{Broadening of the Color Distribution of Knots in (U)LIRGs}

The color distributions of the observed knots become redder as the interaction proceeds, a behavior that the simulation predicts and it is explained by aging effects (see specific values in~\reftab{sim_res}). However, the rather peaked color distributions of the simulated knots are quite different from the broader color distributions of the observed knots, as~\reffig{mass_mag_sim} shows. These differences are actually expected, given the intrinsic characteristics of observations and simulations. In the following, we investigate the broadening of the color distribution of knots in (U)\lirgs\twospace.

Despite the fact that no extinction is implemented in the simulations, a significant percentage of the observed knots are still bluer than the simulated knots at any interaction phase (blue wing; typically \mbox{\mbi$<$ 0.5-1}). As mentioned before, (U)\lirgs are known to preserve high SFRs independently of the interaction phase, while in the simulation the \sfr is high only during the pre-merger phase (see~\reffig{number_time}). Thus, the star formation history of (U)\lirgs implies that the young knots, with blue colors, are continuously being formed in all phases. In contrast, the formation of knots is quickly suppressed in the simulations after the pre-merger phase. Moreover, given the time resolution on the simulation, we are not able to sample colors as blue as \mbox{\mbi $\lesssim$ 0.4}, whereas in (U)\lirgs younger population with bluer colors can exist.

Since the effect of obscuration by dust is not implemented in the simulation, it is worth investigating the difference of the color distributions of simulated and observed knot to understand how the extinction is distributed in (U)\lirgs\twospace. In this case, we focus on the red wing (i.e., \mbox{\mbi$>$ 1.5}) of the color distribution of the observed knots.

We investigated how the colors of the simulated knots would change if they were suffering some extinction, so as to recover the red wing of the color distribution of the observed knots. To that end, a test was performed by applying an extinction distribution with a probability density function of an exponential decay with the form P(\av\twospace)=$\lambda e^{-\lambda A_V}$ to the simulated colors, where 1/$\lambda$ (ln2/$\lambda$) represents the mean (median) value of the distribution. The test was performed from 1/$\lambda$ values of 0.01 to 2 mag and the best match by means of a Kolmogorov-Smirnov test (KS) of the simulated (with extinction) and observed color distribution was selected. The comparison between both distributions was done from the peak of the initial distribution of the simulated colors toward redder colors. 

We were able to obtain similar shapes for the red wing of the observed and simulated knots (see green histograms in~\reffig{mass_mag_sim}). We derived the mean values of \mbox{1/$\lambda$ = [1.3,1.4,1.1] mag} for the pre-merger, merger and post-merger phase, respectively. With this mean values, the simulated knots would suffer extinctions typically in the range \mbox{\av = 0-5 mag}, with a probability distribution that peaks around 0-1 mag and decays exponentially. Therefore, based on this result we conclude that the obscuration for optically-selected knots in (U)\lirgs is likely to be distributed with a probability density function of an exponential decay. In fact, this range of extinctions is only slightly smaller than that reported in (U)\lirgs in the optical and near infrared (\mycitealt{Alonso-Herrero06};~\mycitealt{Garcia-Marin09b}). 

We estimated the corrected magnitude distribution of the simulated knots as if they were affected by dust extinction (green histograms in~\reffig{mass_mag_sim}),  using the obtained values of 1/$\lambda$. Interestingly, the extinction values derived, similar to those observed in (U)\lirgs\twospace, do not contribute significantly to dimming the luminosity of the simulated knots (less than 1 mag on average). 

\subsect[mf_lf_sim]{Mass and Luminosity Functions}

\subsubsection{MF and LF of the Simulated Knots}

We have the opportunity to study the evolution of the \mf and the \lf of the simulated knots with the merging process. In general, the \mf and \lf of young stellar objects (clusters, \sscs\twospace, \ymcs\twospace, complexes of clusters) are well described as power-law distributions: $dN \propto m^{-\alpha_{MF}}$ dm (MF) and $dN \propto L^{-\alpha_{LF}}_{\lambda} dL_{\lambda}$ (LF). We applied the same methodology as in~\refcha{knots}, explained in Chapter~\ref{cha:data_tech} (\refsec{lf}), based on fitting equal-sized bin distributions, to fit both the MF and LF. The lower mass and brightness for which the simulated knots sample is complete was determined assuming a simple power law form for the knot mass distribution. The uncertainties associated to the fit of the LF (shown in tables and figures) do not include other sources of systematic error (e.g., dropping the brightest bins, adopting another method of binning, moving the magnitude limit). Therefore, like in~\refcha{knots}, we consider a systematic uncertainty of 0.1 dex.

\begin{figure}
 \hypertarget{fig:lm_mf_all}{}\hypertarget{autolof:\theautolof}{}\addtocounter{autolof}{1}
 \hspace{-0.3cm}
\includegraphics[trim = -2cm -3cm 1cm -3cm,clip=true,width=0.5\textwidth]{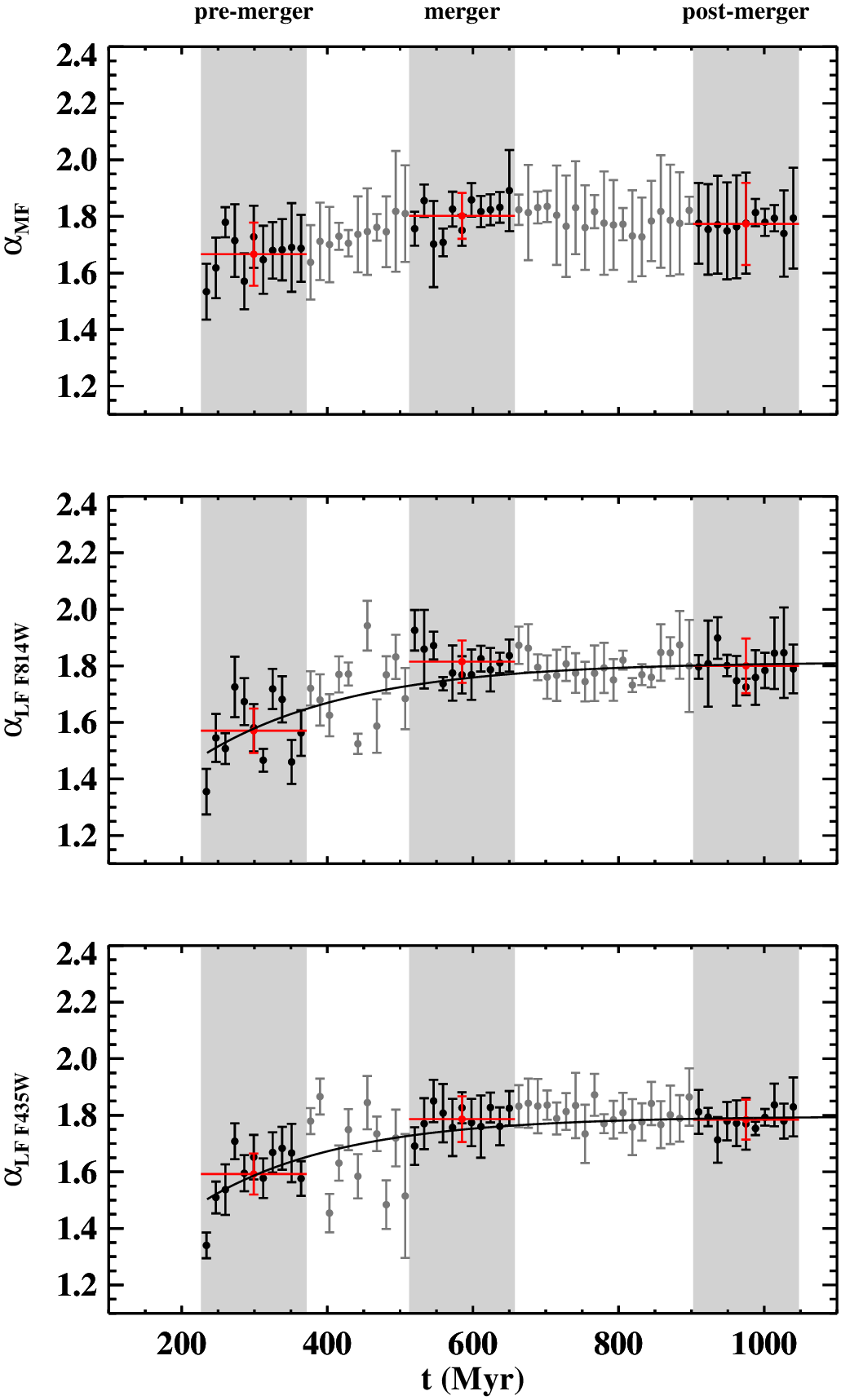}
\includegraphics[trim = -2cm -3cm 1cm -3cm,clip=true,width=0.5\textwidth]{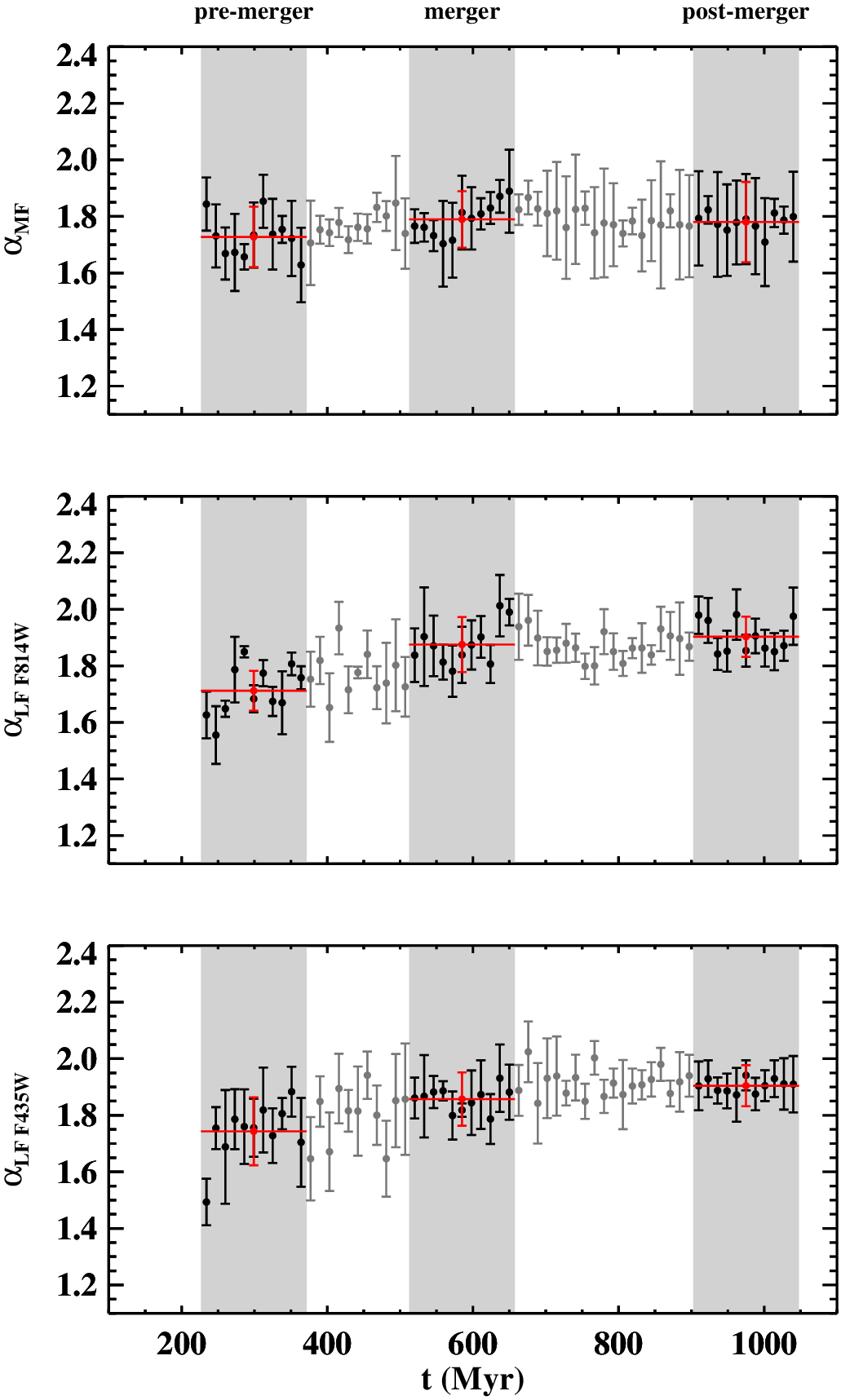}
\caption[Temporal evolution of the  slope of the MF LF of the simulated knots]{Computed slope of the \mf (top) and the \lf for filter \filteri (middle) and \filterb (bottom) of the simulated knots with the interaction. The shaded areas mark the time period, identified with the different interaction phases, for which the median value of the slopes enclosed were computed in each case (in red). \textbf{Left:} For all the simulated knots at any given snapshot. The line shows the best fit of the function \mbox{$\alpha_{LF}$=-10$^{\beta t/t_{0}}$+c} (see text). \textbf{Right:} Excluding the knots that appear in the current and in the previous snapshot (i.e., the younger knots).}
\label{fig:lm_mf_all}
\end{figure}

The slope of the mass function of the simulated knots ($\alpha_{MF}$) is typically 1.8 (see~\reffig{lm_mf_all}). Although a visual trend may be observed toward steeper slopes from the pre-merger to the merger phase, given the uncertainties involved there is no clear evidence of an evolution of the slope with the phase of the merger. Even though in general the simulated knots do not represent individual clusters, the slope computed here is compatible with the slope of the mass function of young star cluster populations in other interacting and galaxy mergers: \mbox{$\alpha_{MF}$ = 1.95 $\pm$ 0.03} in the Antennae (\mycitealt{Zhang99}); \mbox{$\alpha_{MF}$ = 1.5 $\pm$ 0.4} in Arp 284 (\mycitealt{Peterson09}); \mbox{$\alpha_{MF}$ = 1.85 $\pm$ 0.11} in NGC 6872 (\mycitealt{Bastian05c}); and \mbox{$\alpha_{MF}$ = 1.85 $\pm$ 0.12} in NGC 3256 (\mycitealt{Goddard10}). Goddard and co-workers used the same methodology as in this study, whereas the others fitted equal bin distributions of the \mf\twospace. The mass of the clusters was computed by means of broad-band photometry in all the studies mentioned.

We were able to derive the slope of the \lf ($\alpha_{LF}$) because we computed the absolute magnitude distribution of the simulated knots and \mbox{$\alpha_{LF} = 2.5  \beta + 1$}, where $\beta$ is the slope of the fit for the magnitude distribution. In this case, there is a slight trend toward steeper $\alpha_{LF}$ for the simulated knots as the merger proceeds up to about 500 Myr (see~\reffig{lm_mf_all}, left). This trend is less significant for the filter \filterb (see~\reftab{sim_slopes}). We tried to fit empirically the values of $\alpha_{LF}$ at each snapshot with a function of the form \mbox{$\alpha_{LF}$ = -10$^{\gamma t/t_{0}}$ + c}, a kind of an exponential decay (with index $\gamma$). However, as shown in~\reffig{lm_mf_all} (left), the fit is not very well constrained since the dispersion is very high. Hence, even though the trend can be real, within the uncertainties we cannot characterize it given the large dispersions in the different slopes.

We suspect the young population can be responsible for the variation of the slope of the \lf\twospace, since it can be up to a factor of 10-100 more luminous than an old population (i.e., 1 Gyr) with the same mass. (see~\reffig{ssp_models} in Chapter~\ref{cha:data_tech}). For instance, a population of 100 Myr can be up to 5 mag fainter than a population of less than 10 Myr having the same mass. If there is a significant fraction of young population, it can overpopulate the bright end of the \lf\twospace, making it flat.   

To prove the validity of this hypothesis we explore the same plot ($\alpha_{LF}$ vs. time) but excluding those knots that appear for the first time in the simulation in the current snapshot and in the previous one (i.e., normally the youngest knots). We refer to them as the \textit{young simulated knots}. When these young knots are excluded from the fit, the trend tends to disappear in both filters (\reffig{lm_mf_all}, right). Since in the simulation the knot formation rate is very low during the merger and the post-merger phases, the number of young simulated knots is negligible then compared to the pre-merger phase (see~\reftab{sim_slopes}). Therefore, these youngest knots are the responsible for overpopulating the bright end of the \lf\twospace, which makes its slope flatten.

\begin{table}
\hypertarget{table:sim_slopes}{}\hypertarget{autolot:\theautolot}{}\addtocounter{autolot}{1}
\begin{minipage}{\textwidth}
\renewcommand{\footnoterule}{}  
\begin{small}
\caption[Slopes of the MF and LF of the simulated and observed knots with time]{Slopes of the mass and luminosity distributions of the simulated and observed knots at different interaction phases}
\label{table:sim_slopes}
\begin{center}
\begin{tabular}{lcccccc}
\hline \hline
   \noalign{\smallskip}

Interaction & N$_Y$ (\%) & $\alpha_{MF}$ & $\alpha_{LF\ F814W}$ & $\alpha_{LF\ F814W}$ & $\alpha_{LF\ F435W}$ & $\alpha_{LF\ F435W}$ \\
phase & & sim & obs & sim & obs & sim \\
\hline 
   \noalign{\smallskip}
pre-merger& 12$\pm$7& 1.67 $\pm$ 0.11& 1.84 $\pm$ 0.03& 1.57 $\pm$ 0.08& 1.86 $\pm$ 0.05& 1.59 $\pm$ 0.07\\
merger& 2$\pm$1& 1.80 $\pm$ 0.08& 2.00 $\pm$ 0.04& 1.81 $\pm$ 0.07& 1.86 $\pm$ 0.02& 1.79 $\pm$ 0.08\\
post-merger& 1$\pm$1& 1.77 $\pm$ 0.15& 1.76 $\pm$ 0.12& 1.80 $\pm$ 0.10& 1.58 $\pm$ 0.17& 1.78 $\pm$ 0.07\\
  
\hline
\noalign{\smallskip}
\multicolumn{7}{@{} p{\textwidth} @{}}{\footnotesize{\textbf{Notes.} Col (1): defined interaction phase. Col (2): number of \textit{young simulated knots} (see text) with respect to the total number of simulated knots. Col (3): 
average of the median values of the slopes of the \mf computed for the 11 distributions per interaction phase. Col (4): slope of the luminosity function (filter \textit{F814W}) at a given interaction phase. Col (5): average of the median values of the slopes of the \lf (filter \textit{F814W}) computed for the 11 distributions per interaction phase. Col (6): same as column (4), but with data observed with the filter \textit{F435W}. Col (7) same as column (5), but with filter \textit{F435W}.}}
\end{tabular}
\end{center}
\end{small}
\end{minipage}
\end{table}

\subsubsection{LF: Models vs. Observations}

In~\refcha{knots} we observed (including the first contact phase in the analysis) a marginally significant indication that for the red filter the \lf evolves during the interaction process (see also~\reftab{sim_slopes}). For the blue filter the slope does not change perceptibly in the observed sample of (U)\lirgs\twospace. This variation of the slope could have the same origin in both observations and simulations. Then, even if there is substantial active star formation (i.e., young population) in all interacting phases in (U)\lirgs\twospace, this may imply that in early phases the young population represents more percentage (\mbox{N$_{Y}\sim$ 5-20\%}; see~\reftab{sim_slopes}) with respect to the total population than in most evolved phases (\mbox{N$_{Y}\sim$ 1-2\%}) in the optical band. The contribution of this overpopulation of young knots would flatten the slope of the \lf in early interaction phases.

In absolute values, the \lfs in the observed sample of (U)\lirgs are systematically about 0.1 dex steeper than in the simulation, as shown in~\reftab{sim_slopes}. Given the uncertainties involved (specially the systematics), this is not very relevant. In any case, the reasons why the slopes are systematically different can be: (1) the \lfs of the simulated knots were not been corrected for incompleteness (if corrected, the slopes would be steeper); (2) the spatial resolution is a bit lower in the simulation (if improved, the slopes would be steeper); (3) the knots do not suffer internal extinction (if they did, the slopes would flatten).

Is this variation of the slope with the interaction phase observed in other mergers? In other interacting galaxies the slope is steeper, its value being around 2 (\mycitealt{Schweizer96};~\mycitealt{Miller97};~\mycitealt{Whitmore99};~\mycitealt{Gieles06};~\mycitealt{Haas08};\mycitealt{Peterson09};~\mycitealt{Santiago-Cortes10}). If we take individual snapshots in the simulation, the slopes can vary considerably within the same interaction phase (see~\reffig{lm_mf_all}). The trend is detected only when we consider the whole range from 200 Myr to about 500 Myr. Thus, measuring steeper slopes than the predicted by the simulation at early phases when studying individual systems would be possible. Caution must be taken with these comparisons because the slopes computed in other interacting galaxies correspond in general to the \lf of individual clusters. Since in this study we do not have enough spatial resolution to resolve individual clusters, the slopes of the \lf of the simulated knots can be distorted. In fact, this happens as well in the sample of observed (U)\lirgs\twospace. In Chapter~\ref{cha:knots} (\refsec{distance_effect}) we made a simulation to understand how the slope behaves under the situation of not having enough angular resolution, and concluded that it can be systematically flattened by about 0.1 dex.

\subsubsection{What Traces the Variation of the Slope of the LF?}

We have seen that the slight variation of the slope of the \lf can be explained by the presence of a larger fraction of young population in early phases of the interaction compared to more evolved phases. But, does that mean that the \sfr of the system traces the slope of the \lf\twospace? This is a key issue, since (U)\lirgs are known to have high \sfrs\twospace, but in advanced phases (merger and post-merger) the slope of the \lf of the knots is the same.

At t = 250 Myr the global \sfr in the simulation drops dramatically (see~\reffig{number_time}). The shape of the histogram that represents the knot formation rate (black histogram in~\reffig{number_time}) is similar, though wider. It is a difference not related to the detection threshold of the knot, because the black histogram was corrected for that. At \mbox{t = 300 Myr} the global \sfr is already very low (less than \mbox{10~\msun yr$^{-1}$}) compared to the peak (\mbox{Peak$_{SFR}$ $\sim$ 50~\msun yr$^{-1}$}), and the slope of the \lf can still be quite flat (\mbox{$\alpha_{LF}$ = 1.5-1.6}). Yet, the knot formation rate remains still high. Therefore, the variation of the slope of the \lf is likely to be related only to the knot formation rate, and not necessarily to the global \sfr\twospace. This implies that if a system has a high \sfr (i.e., a \lirg or a \ulirg\twospace), but the current star formation is concentrated on very few knots or mainly obscured (i.e., not observable in the optical), the young population does surely not affect the slope of the \lf\twospace. Might this be occurring in advanced phases (merger and post-merger) in (U)\lirgs\twospace?

\subsect[spa_evol]{Evolution of the Spatial Distributions of the Properties of the Knots}

The comparison of the evolution of the spatial distributions of the simulated and observed knots as a function of the interaction phase can provide additional information on the physical processes that govern the production of star clusters. For this study, we defined 6 bins of galactocentric distance, each separated by a factor of 2 from d=2.5 kpc:  0-2.5, 2.5-5, 5-10, 10-20, 20-40, and 40-80 kpc. The distances for the observed knots in~\reffig{spatial_num} correspond to projected galactocentric distances. However, in the simulation we have access to the real galactocentric distance of the knots at any given snapshot (the 3D-distance). We thus study the spatial distribution of the photometric properties using the 3D-distance and the 3 projected distances on each of the three planes in the simulation (XY, XZ and YZ). In general, either if we are dealing with the 3D or with any projected distance the trends in the simulation do not change significantly. In any case, we decided to make the comparison with the observations taking a projected distance because we have more statistics at smaller galactocentric distances and the distances measured in the sample of (U)\lirgs are also projected. 

Finally, if there is not enough statistics in a given distance bin, then the mean value (red and green in~\reffig{spatial_sim} and~\reffig{spatial_obs}) is not shown in the plot.

\subsubsection{Spatial Distribution of the Knots}

\begin{figure}
 \hypertarget{fig:spatial_num}{}\hypertarget{autolof:\theautolof}{}\addtocounter{autolof}{1}
\hspace{0.7cm}
\includegraphics[angle=90,width=0.95\columnwidth]{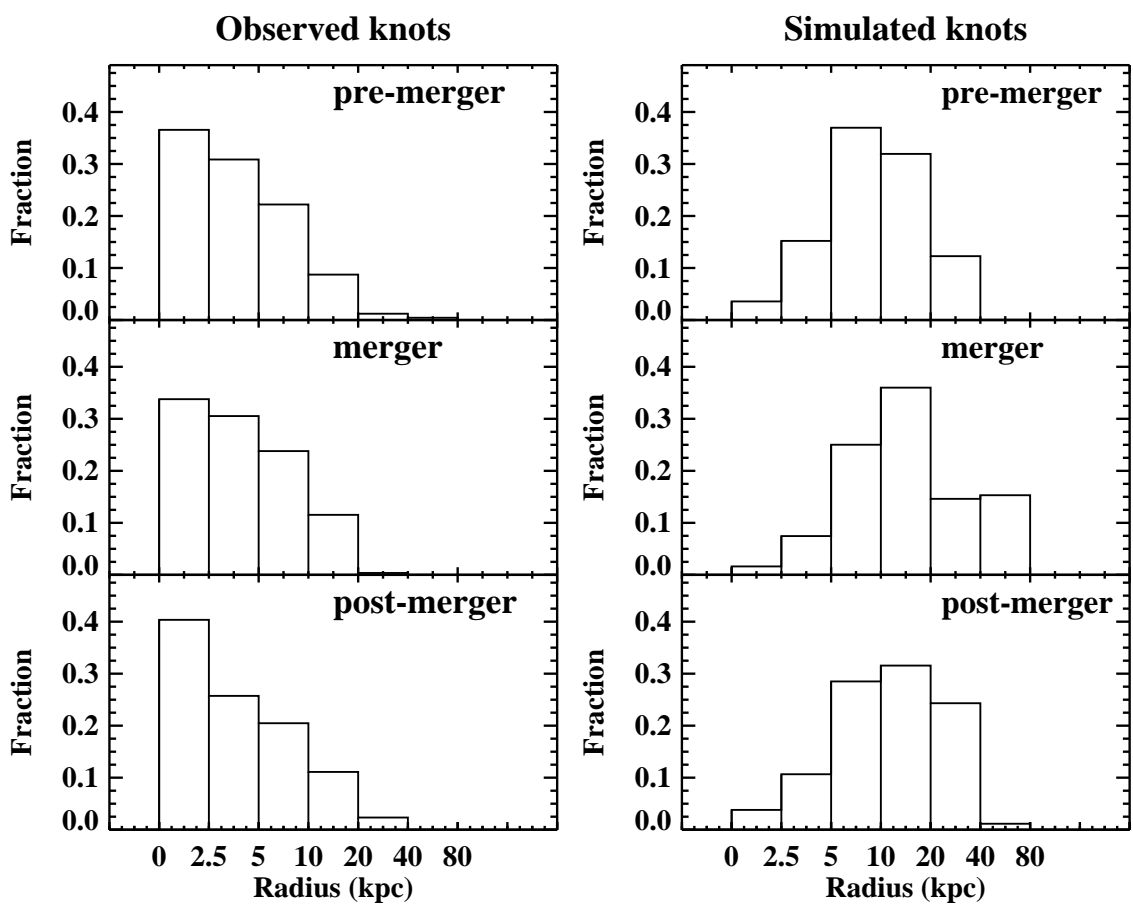}
\vspace{0.2cm}
\caption[Spatial distribution of the observed and simulated knots]{Spatial distribution of the observed (left) and simulated (right) knots. For the simulated knots the distance correspond to the projected galactocentric distance in the XY plane.}
  \label{fig:spatial_num}
\end{figure}

The spatial distribution of the observed and simulated knots is significantly different at any given interaction phase. While more than 60\% of the total number of knots are within a projected distance of 5 kpc in the sample of (U)\lirgs\twospace, only about 20\% are located below that distance in the simulation. Recent hydrodynamic simulations of a major merger (\mycitealt{Teyssier10}) showed that models limited in spatial and mass resolution, and/or the absence of a model for gas cooling below $\sim$10$^4$ K, cannot resolve properly the process of gas fragmentation into massive clouds and rapid star formation. Teyssier and co-workers ran their simulation in a low and high resolution mode, using a softening of $\Delta x$=96 and 12 pc, respectively. The \ism becomes much more clumpy in the high resolution mode, especially in the inner regions and then more clusters/knots can be formed. Furthermore, gas is more extended, allowing formation of \sscs at higher radii.

In the simulation by Bournaud et al. (\mycitealt{Bournaud08a}), even if the spatial resolution is still very high, it is three times lower than the high resolution mode in the hydrodynamic simulations mentioned and neglect the thermal pressure of the gas\footnote{The sticky particle scheme used in this simulation assumes that the turbulent pressure dominates: the turbulent pressure and turbulence dissipation are modeled by the interactions of sticky particles, and the thermal pressure is neglected. This may be a realistic sub-grid treatment for molecular and cool/cold atomic phases (which are dominated by supersonic turbulence, e.g.,~\mycitealt{Burkert96}) but the more important thermal pressure in warmer phases (\mbox{$\geq$ 10$^4$ K}) is neglected.}. Additionally, it may be more difficult to detect low-mass knots in the central regions of the simulated galaxies given the large diffuse background there. Therefore, we have to be cautious when comparing the evolution of spatial distributions of the magnitudes and colors of the observed and simulated knots at galactocentric distances smaller than \mbox{d $<$ 5 kpc}.

\subsubsection{Evolution of the Spatial Distribution of Luminosities.}

\begin{sidewaysfigure}
 \hypertarget{fig:spatial_sim}{}\hypertarget{autolof:\theautolof}{}\addtocounter{autolof}{1}
\includegraphics[angle=90,trim = 3.5cm 1cm 2cm 3cm,clip=true,width=1.\textwidth]{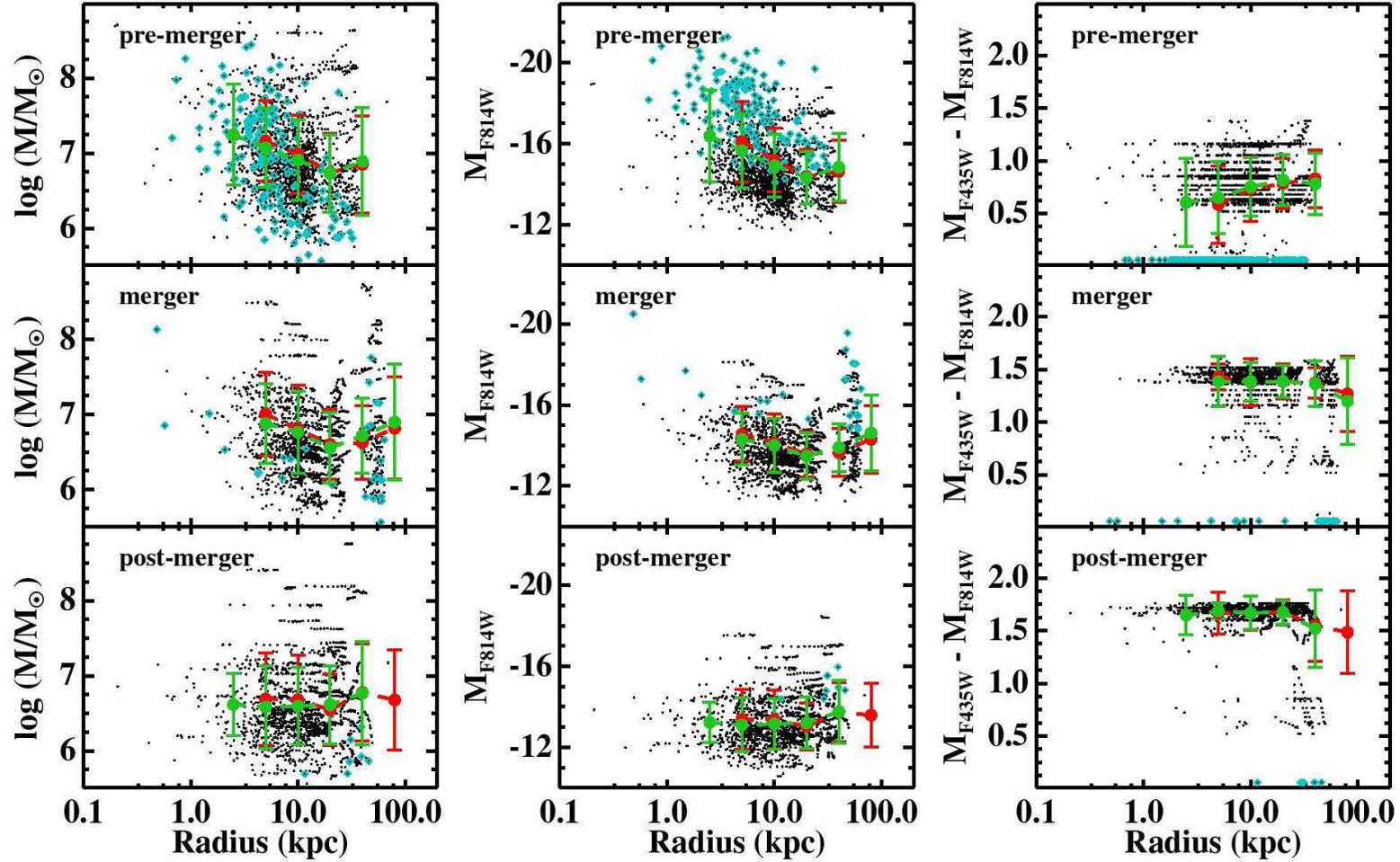}
\caption[Properties of the simulated knots and galactocentric distance]{Spatial distribution of the mass, luminosity and color for the simulated knots at the three interaction phases defined in this study. In each case we plot the values in the 11 snapshots for which the interaction phases have been defined, and taking the projected galactocentric distance in the XY plane (black dots). Those dots circled in light blue refer to the knots that appear for the first time in one of the 11 snapshots that define an interaction phase. Mean and standard deviation of each 0.3 dex (log 2) bin in distance is shown in red (green) taking the 3D-distance (projected distance in the XY plane), starting from the bin [0,2.5] kpc. The red line is not a fit and marks the mean values within each bin.}
\label{fig:spatial_sim}  
\end{sidewaysfigure}

\begin{sidewaysfigure}
 \hypertarget{fig:spatial_obs}{}\hypertarget{autolof:\theautolof}{}\addtocounter{autolof}{1}
\includegraphics[angle=90,trim = 3.5cm 1cm 2cm 3cm,clip=true,width=1.\textwidth]{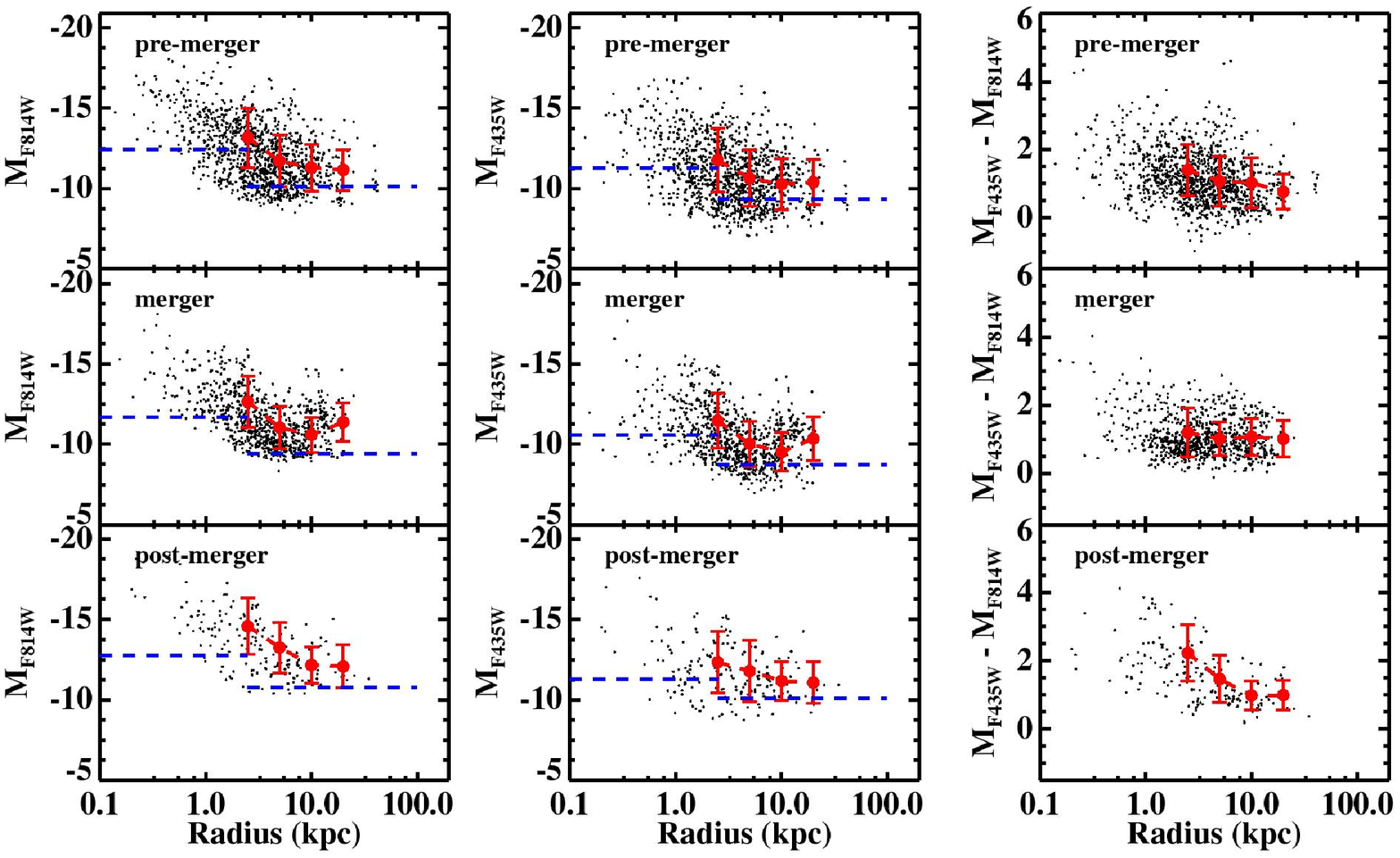}
\caption[Properties of the observed knots and galactocentric distance]{Relation between the photometric properties (luminosity, color and \reff\twospace) of the observed knots and the galactocentric distance, as a function of interaction phase. Mean and standard deviation of each 0.3 dex (log 2) bin in distance is shown in red, starting from the bin [0,2.5] kpc. The blue horizontal dashed lines on the magnitude distribution plots give the average 90\% completeness for the inner (i.e., \mbox{d $<$ 2.5 kpc}) and outer (i.e., \mbox{d $>$ 2.5 kpc}) knots.}
\label{fig:spatial_obs} 
\end{sidewaysfigure}

During the pre-merger phase, we see a trend of the luminosities of the simulated knots with the radial distance, in such a way that they become fainter as the distance increases. As the interaction proceeds, in the merger phase this trend (if real) is very weak. By the time the post-merger phase is reached the spatial distribution of luminosities is completely flat (see~\reffig{spatial_sim}).  Both the spatial distribution of magnitudes and masses follow the same trends at any given interaction phase.  This could indicate that the star formation in the simulation is not centrally concentrated. However, if we select the youngest simulated knots (\textit{new knots}, circled light blue dots in~\reffig{spatial_sim}) we realize that during the pre-merger phase, close to the peak of the \sfr\twospace, 45\% are formed within the central 5 kpc and the median value of their mass distribution corresponds to \mbox{log (M/\msun\twospace) = 7.25} against the value of 6.57 for knots located at larger distances. Later on, this percentage drops to 30\% and 0\% in the merger and post-merger phase, respectively. In the merger phase the new knots in the central 5 kpc are still significantly more massive than the more external ones (log (M/\msun\twospace) of 6.55 against 6.18). Hence, the trend can be due to two contributions: 

\begin{enumerate}[(1)]
 \item the star formation in the simulation is more concentrated in the nuclear regions during the first phases of the interaction. The knots are more luminous there, since they are on average younger than knots in the outermost regions. 
 
 \item the most massive knots tend to form in the central regions. Thus, they are also more massive, i.e., more luminous there. 
\end{enumerate}

The trend tends to disappear as the interaction proceeds since at later phases the knot formation process (though not very strong given that \sfr is very low in the simulation) is more efficient in the external regions (i.e., in tidal tails). Younger population will be more luminous being even less massive there. Note that given the limitation of the simulation on forming knots in the innermost regions, discussed in the previous section, the percentages given should be considered as a lower limits. Thus, in the pre-merger phase, we expect that actually more than half of the knots are formed within the central 5 kpc instead of 45\%. However, given this limitation we cannot quantify by how much. 

In the sample of (U)\lirgs\twospace, a similar variation to that in the pre-merger phase of the simulation of luminosities with distance persists up to the merger phase (see~\reffig{spatial_obs}). Actually, part of this trend can be due to incompleteness in detecting the knots in the central regions of the galaxies, where the surface brightness of the underlying background is high. For that reason, knots fainter than the 90\% completeness limit are not detected. Nevertheless, there are not very bright knots in the outermost regions, independently of the completeness limit. Therefore, this variation might remain, meaning that on average the brightest objects are located within the first few kpc of the galaxies at any given interaction phase, as the simulation predicts only for the pre-merger phase. Then, why does the trend remain up to more advanced phases in (U)\lirgs\twospace? These systems can sustain high rates of star formation at any interaction phase. If we assume that this variation has the same origin as in the simulation, then at least half of the current star formation is always concentrated within 5 kpc, no matter which phase of the interaction the systems are undergoing (the variation is kept during the whole interaction). This has been observed in galaxies at different phases of interaction, like in the central regions of the Antennae (\mycitealt{Mengel08};~\mycitealt{Whitmore10} and in NGC 7252 (\mycitealt{Miller97}). As a main conclusion, our theoretical and empirical data suggest that whenever the global star formation rate is high in interacting environments, it is likely to be mainly concentrated in the central regions.

\subsubsection{Evolution of the Spatial Distribution of Colors.}

As we have seen before, the young knots in the simulation are more concentrated during the pre-merger phase. The spatial distribution of colors confirm this, since younger knots are bluer, and knots with bluer color are located in central regions during this phase (\reffig{spatial_sim}). In contrast, the knots in (U)\lirgs show a trend toward redder colors in the inner regions, in the pre-merger and post-merger phase (\reffig{spatial_obs}). This is an indication of higher internal extinction in the central regions of this systems. Based on the previous conclusion that most of the current star formation appears to be concentrated toward the central regions in (U)\lirgs\twospace, then the extinction must be even more severe, as it has been measured in these kind of systems (e.g.,~\mycitealt{Alonso-Herrero06};~\mycitealt{Garcia-Marin09a}).

\sect[summary_sim]{Summary and Conclusions}

This Chapter shows the results of the first direct comparison of the star formation in interacting galaxies obtained with high spatial resolution numerical simulations with observations of galaxy interactions on scales of 30-100 pc. We have compared the evolution of the observed photometric properties of the stellar regions (knots) in a sample of 22 (U)\lirgs with those predicted in a simulation of a major merger of two galaxies with roughly equal initial mass ratio and large fraction of gas, similar to some observed systems. Three different interaction phases have been defined: pre-merger, merger and post-merger, occurring in the simulation at about \mbox{t = 200-400 Myr}, \mbox{500-700 Myr} and \mbox{1 Gyr}, respectively. The high resolution model reproduces and/or explains in a coherent way some observed properties of the star-forming regions in major mergers. We draw the following conclusions:

\begin{enumerate}

\item As a consequence of the aging of the knots during the merging process, observations and simulations show an evolutionary trend in which knots become fainter and have redder \mbi colors as the interaction proceeds from the pre-merger to the merger phase. Later on, during the post-merger phase, there is theoretical and observational evidence that only the most massive structures remain, although it is still not clear how they form: (i) as individual massive entities; (ii) or after a merging process of clusters into superclusters. \\

\item The color distribution of the simulated knots span a very narrow range (\mbox{\mbi\onespace = 0.5-2}) compared to the observed color distribution of the knots in (U)\lirgs\twospace. To reproduce the observed color distribution an extinction correction has to be applied to the colors of the simulated knots, characterized by an exponential probability density function of median value \mbox{\av = 1.1-1.4 mag}. The derived typical obscuration peaks at \mbox{\av = 0-1 mag}, and ranges from 0 to 5 mag.\\

\item The shape of the mass function (\mf\twospace) of the simulated knots is practically constant along the interaction process. The derived slope of the \mf\twospace, typically \mbox{$\alpha_{MF} \sim$ 1.8}, is similar to that obtained for observed individual star clusters in less luminous interacting environments.\\

\item The slope of the luminosity function (\lf\twospace) may evolve throughout the interaction process. A weak trend toward steeper slopes of the \lfs of the simulated knots is observed during the pre-merger phase up to the merger phase, when it stabilizes around \mbox{$\alpha_{LF}$ = 1.8-1.9}. According to the simulation, this slight variation is likely to be due to high knot formation rates at early phases of the interaction with respect to the knot formation rates at late phases. However, there is not clear evidence that it is correlated with the global star formation rate of the system. The same origin could explain the slight variation of the \lf of the observed knots in (U)\lirgs provided that the global star formation is concentrated in few knots during intermediate and advanced interaction phases.   \\

\item Numerical issues considered on the formation and detection of knots at small distances (galactocentric distance \mbox{d $<$ 5 kpc}), the simulation confirms that when the global star formation is high (pre-merger) the star formation is mainly concentrated in the central regions. The spatial distribution of luminosities in (U)\lirgs suggests that the star formation in these systems is mainly concentrated in the central regions in general (no matter the interaction phase), in agreement with previous studies of these systems.\\

\end{enumerate}

\clearpage{\pagestyle{empty}\cleardoublepage}

\cha{tdgs}{\tdg Candidates in Low-z (U)LIRGs}
\chaphead{In this Chapter we use the combination of the photometric data from previous chapters, and spectroscopic data to investigate the nature of the extreme extranuclear young population, detected in our sample of (U)\lirgs in the context of \tdg formation. We make use of several criteria (e.g., \ha luminosity, dynamical mass content, signs of self-gravitation, etc.) to investigate their nature as \tdgs and the chances of survival as long-lived entities. The knowledge of the production rate of long-lived \tdgs for the (U)\lirg class also provides clues to constrain the dwarf galaxy formation at high redshifts.}

\sect[intro_tdgs]{Introduction}

Interacting systems (and remnants) constitute the optimal environment for finding the so-called tidal dwarf galaxies (\tdg\twospace): a self-gravitating entity of dwarf-galaxy mass built from the tidal material expelled during interactions (General Introduction,~\refsec{tdg_gal}). Although many studies devoted to searching and characterizing \tdgs in nearby interacting systems have been carried out in the past few decades, only a few searches for \tdg candidates have been performed to date in \lirgs and \ulirgs\twospace.

In previous chapters we have disentangled the photometric properties of knots of star formation in a representative sample of (U)\lirgs\twospace. Interestingly, very bright (i.e.,  \mbox{\mi $<$ -12.5}) and blue (i.e.,  \mbox{\mbi $<$ 0.5}) knots along the tidal tails and at the tip, indicative of massive (i.e.,  \mbox{M = 10$^4$-10$^6$ \msun\twospace}) young populations, have been detected in those systems. On the basis of previous studies of nearby interacting galaxies (e.g.,~\mycitealt{Duc97},~\myciteyear{Duc98};~\mycitealt{Weilbacher00};~\mycitealt{Temporin03};~\mycitealt{Monreal07}), \tdg candidates are normally identified as extreme star forming regions in tidal tails. With the additional aid of optical \acr{}{IFS}{Integral Field Spectroscopy} data, this chapter is devoted to studying more in detail extranuclear \ha\twospace-emitting clumps linked to the knots and associations investigated in previous chapters. With dynamical masses up to \mbox{10$^8$-10$^{10}$ \msun\twospace}, these knots and associations represent the most massive case of young star formation in (U)\lirgs\twospace, toward the typical range of dwarf galaxies.

This Chapter extends the previous study of Monreal-Ibero et al. (\myciteyear{Monreal07}) by including galaxies in the LIRG class and with additional (i.e.,  \textit{B}-band) and more sensitive photometric data.~\refsec{sample_tdgs} presents the subsample with available photometric and \ifs data simultaneously. We then identify and characterize the luminosities, colors, metallicities, etc., of extranuclear \ha\twospace-emitting complexes in~\refsec{ha_complexes} and~\refsec{characterization_tdgs}, respectively. In section~\refsec{discussion_tdgs} we select the complexes that are most likely to constitute \tdg candidates and discuss their total mass content and the likeliness of the survival of these regions as future \tdgs\twospace, and the implications for the formation of \tdgs at high-z. Finally, we draw our conclusions in~\refsec{summary_tdgs}.

\sect[sample_tdgs]{The Subsample}

The initial sample for this study is presented in Chapter~\ref{cha:data_tech} (\refsec{sample_ini}). Since this study combines our photometric (see Chapter~\ref{cha:data_tech},~\refsec{phot_data}) and spectroscopic (see Chapter~\ref{cha:data_tech},~\refsec{spec_data}) data, we have selected those systems with available data in both sets simultaneously at the moment of the analysis. That is a total of 27 systems.

As explained in section~\refsec{id_tdgs}, regions of interest were selected as any high surface brightness compact region (\ha clumps) in the emission line maps (obtained from the \ifs data) at a projected distance from the nucleus of the galaxy greater than 2.5 kpc, and associated with one or several of the star-forming regions (knots) in the \acs images. On the basis of these criteria, \ha\twospace-emitting complexes have been identified in 11 systems. We did not identify other complexes in the remaining sample basically for two reasons: (1) based on the above criteria, in some systems knots are detected but no \ha emission is found in the outskirts in the \ha line map (e.g., in IRAS 22491-1808); (2) the \fov covered by the \ifs data is not large enough to encompass the full extent of some systems. Hence, the nature of blue knots along the tidal tail and at its tip, where \tdgs are normally found, cannot be assessed using the \ifs data.

Therefore, the final subsample consists of 11 systems with at least one region of interest (see table~\reftab{phot_prop_tdgs}), or more precisely 7 \lirgs and 4 \ulirgs\twospace. For completeness we also include in our tables the results of Monreal-Ibero et al. (\myciteyear{Monreal07}) for IRAS 16007+3743, for which an image in the \textit{B} photometric band was unavailable. According to the classification scheme defined in  Chapter~\ref{cha:data_tech} (\refsec{morphologies}), IRAS 04315-0840, IRAS 08355-4944, IRAS F10038-3338 and IRAS 15250+3609 are in an advanced phase of the merger (merger phase) where the nuclei have apparently coalesced. However, the tails are still recognizable in the images, in some cases very prominently. The rest of the systems are in earlier stages of the merging process, first contact or pre-merger phases, according to the classification scheme we use. 

\sect[ha_complexes]{\ha\onespace-Emitting Complexes in (U)LIRGs}

\subsect[id_tdgs]{Identification}

\begin{sidewaysfigure}
\hypertarget{fig:hst_halfa}{}\hypertarget{autolof:\theautolof}{}\addtocounter{autolof}{1}
\hspace{1.5cm}
\includegraphics[trim = 0cm 0cm 0cm 0cm,width=0.82\textheight]{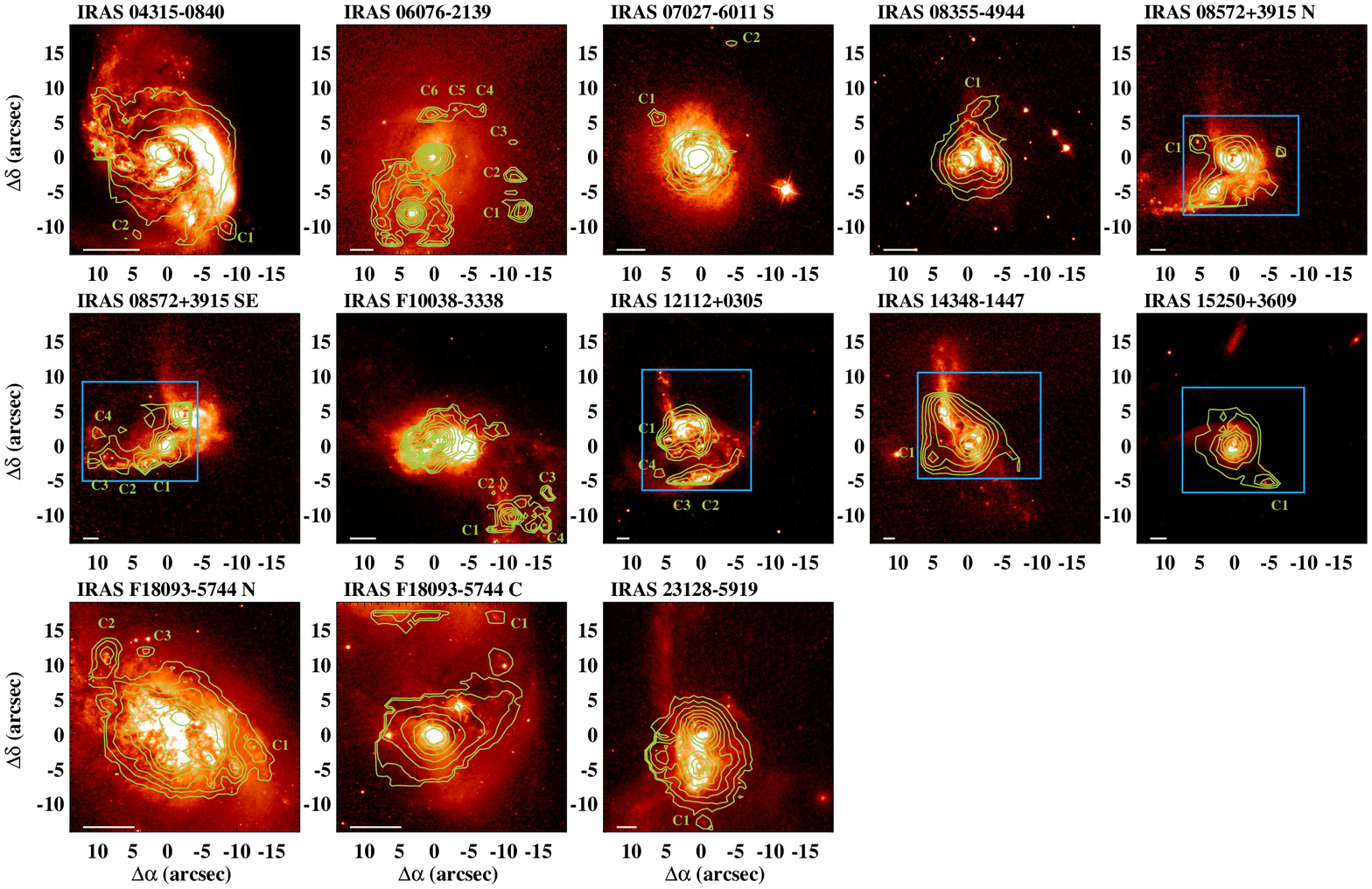} 
\caption[Systems under study with the \ha maps obtained from IFS]{Systems under study with the \ha maps obtained from integral field spectroscopy. The \textit{F435W} \acs images are shown and contours corresponding to the \ha are over-plotted. The contours are in arbitrary units, optimized to show and identify the structures of the \ha complexes. In some cases, there are two different pointings for the same system. The labels show the location of the complexes under study. The horizontal line at the bottom left corner shows a scale of 2.5 kpc. The VIMOS \fov is shown approximately. For INTEGRAL sources, the INTEGRAL \fov is over-plotted in the blue boxes. Most of the \ha peaks match up with several knots identified in the blue images.  Note that the brightest knot on the \ha clump below C1 corresponds to a red foreground star, not a knot of star formation.}
\label{fig:hst_halfa} 
 \end{sidewaysfigure}

\begin{sidewaysfigure}
 \hypertarget{fig:hst_halfa_zoom}{}\hypertarget{autolof:\theautolof}{}\addtocounter{autolof}{1}
\hspace{1.5cm}\includegraphics[trim = 0cm 0cm 0cm 0cm,width=0.82\textheight]{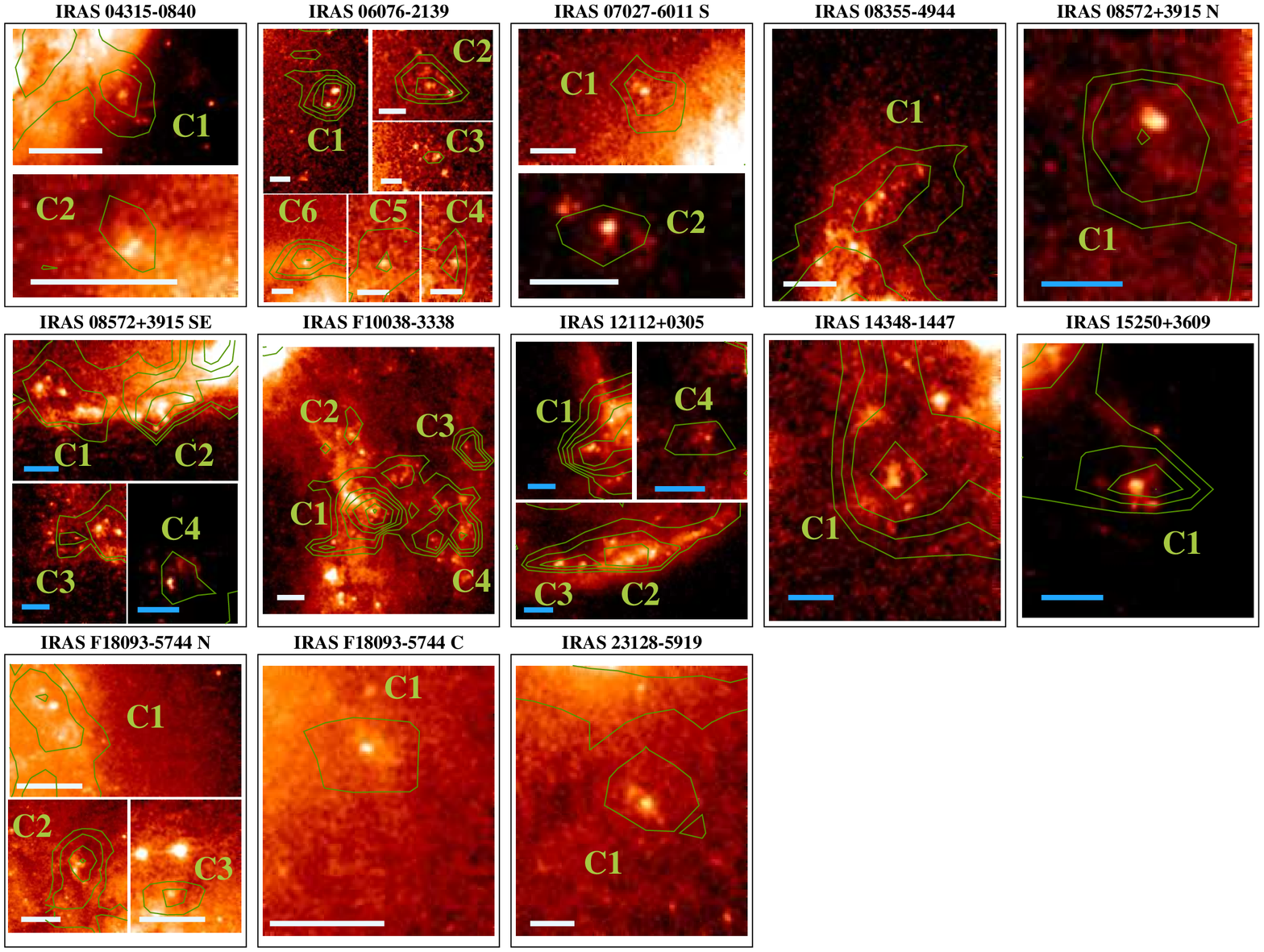} 
\caption[Zoomed view of all the identified \ha complexes]{Zoomed view of all the identified \ha complexes. In each case the FoV has been selected to visualize the complex more clearly. The horizontal line at the bottom left corner shows a scale of 2 kpc (blue) for complexes detected on INTEGRAL maps and 1 kpc (white) for complexes detected on VIMOS maps.}
\label{fig:hst_halfa_zoom}
 \end{sidewaysfigure}
 
The identification of the stellar population that ionizes the surrounding interstellar medium, and subsequently produces the \ha emission, is not necessarily straightforward. Depending on the resolution of the data and the contamination of the field by other ionizing sources, the establishment of this matching is not always possible. We first selected bright condensations in the \ha maps that were coincident with at least one blue knot in the \filterb \acs images. We chose to use these high resolution images instead of the \filteri ones because the most likely responsible for the ionization were detected most efficiently with the blue filter.

We defined a complex of young star formation as an \ha clump (a set of spaxels\footnote{As explained in Chapter~\ref{cha:data_tech} (\refsec{spec_data}), a spaxel corresponds to the spatial element from which we have one independent spectrum, and it is used to reconstruct the map.}) that had coincident position with one or several knots identified in the \hst images. In general, around 12 spaxels \mbox{(e.g., 3 $\times$ 4} spaxels) are needed to define a complex, depending on how isolated it is and the extent of the \ha emission before it is affected by another source or diffuse nuclear emission. In one case (IRAS 08572N) a prominent \ha peak is observed at around 7 kpc west of the northern nucleus, but with no corresponding region in the \hst images. 

We visually inspected each \ha map and found 11 systems with prominent peaks apart from the nuclei (see~\reffig{hst_halfa}). We defined the inner complexes as those located within a projected distance of 2.5 kpc to the closest nucleus, and the outer complexes as those further out than 2.5 kpc. Given the resolution of the spectral maps, approximately 18$^2$ pixels on the \acs images correspond to 1$^2$ spaxel on the INTEGRAL maps, and 13$^2$ pixels to 1$^2$ spaxel on the VIMOS maps. Thus, the identification of the ionizing stellar knots is highly uncertain in inner complexes, since the \ha flux there is very much likely to be contaminated by flux from nearby knots. Together with the motivation of searching for \tdg candidates, they constitute the main reason why we exclude these complexes from the study. Therefore, we focus on the outer complexes. Moreover, we do not select complexes in rings of star formation (i.e., those at the external ring in IRAS 06076-2139), since they are not normally associated with tidal forces, hence with \tdgs\twospace. All considered, we identified for our study 31 outer young, \ha\onespace-emitting complexes of star formation, shown in~\reffig{hst_halfa} and~\reffig{hst_halfa_zoom}. 

\subsect[str_complex]{Structure and Location}

Owing to the higher angular resolution of the \hst images, we identified the \ha clumps (observed in the spectral maps) with a complex of star-forming knots (observed in the photometric images). The number of knots in the ACS image per \ha complex is on average 2.2. However, the structure of these clumps is in general quite simple. As can be seen in~\reffig{hst_halfa_zoom} and~\reftab{phot_prop_tdgs}, the position of the \ha\twospace-emission peak is coincident or quite close to a prominent blue knot. In most cases, only one bright knot is observed within the area defined for the \ha complex (e.g., C1 in IRAS 08572+3915 N) or a centered bright knot plus a few fainter knots located at the border of the \ha complex (e.g., C1 in IRAS 04315-0840). In a few cases, the structure is quite complex, with several bright blue knots spread inside the area of the clump (e.g., C2 in IRAS 12112+0305).

When characterizing the \ha complexes, given the rich structure of some of them, we assume that every knot detected in the \hst images within the area of the complex is an \ha emitter. With the current data, the estimate of the contribution of each knot to the total \ha emission or if a knot is not an \ha emitter (i.e., it is older than 10 Myr, at which the \ha emission declines abruptly) for all the complexes, is not possible. We also assume that only the detected knots are responsible for the whole \ha emission of the complex. 

The \ha complexes are located at distances in the range 3.3-13.7 kpc from the nucleus of the parent galaxy, with a median value of 9.3 kpc (see~\reftab{phot_prop_tdgs}). These distances, owing to projection effects, represent lower limits to the real distances. They are relatively close distances to the parent galaxy compared to those of the \tdg candidates (35-100 kpc; e.g.,~\mycitealt{Duc98};~\mycitealt{Sheen09};~\mycitealt{Hancock09}), although some \tdg candidates have been reported at such distances (e.g.,~\mycitealt{Iglesias-Paramo01};~\mycitealt{Weilbacher03}). At these distances, some of the \ha complexes lie along the tidal tails (see~\reffig{hst_halfa}), and the diffuse low surface brightness \ha emission (which is not subtracted in our measurements of the \ha flux) is either  undetected or considerably diminished, thus does not contaminate the emission from the knots. Others also appear to be simply a non-nuclear giant \hii region.

\sect[characterization_tdgs]{Characterization of the \ha\onespace-Emitting Complexes}

We now investigate the physical properties of the bright \ha complexes identified. This characterization includes the analysis of the stellar continuum magnitudes (\mb and color),  spectral features (\ha luminosity and equivalent width), as well as physical sizes and an estimate of the metallicity. All these observables are compared to those measured in extragalactic \hii regions, dwarf galaxies, and other \tdg candidates.

\subsect[phot_complex]{Photometric Properties of the Complexes}

\subsubsection{Broad-band Luminosities and Colors}

The integrated blue absolute magnitude \mb and color (\mbi\twospace) of the young stellar population within the \ha complexes were estimated by adding all the flux of the knots inside a complex. We cover a magnitude range \mbox{\mb= [-9.32, -15.77]}, and a color range  \mbox{\mbi= [-0.23, 2.34]} (see~\reftab{phot_prop_tdgs}). The median values correspond to -12.06 and 0.63 mag, respectively. The typical uncertainties are between 0.05 and 0.1 mag. 

The integrated luminosities are in general higher than in embedded clusters in nearby extragalactic \hii regions in spirals (\mycitealt{Bresolin97}) and more typical of those found in extragalactic giant \hii regions (\mycitealt{Mayya94};~\mycitealt{Ferreiro08}). Although early studies in extragalactic \hii regions only focus on nearby galaxies (\mbox{\ld $<$ 20 Mpc}), Ferreiro and co-workers observed minor mergers at distances ({\ld = 40-170 Mpc}) similar to our sample, detecting also only the most luminous giant \hii regions, the external ones covering the range \mbox{$M_B~\simeq$ [-11.7, -17.4]}, typically 2 magnitudes more luminous than the range sampled in this study.

Nearby dwarf galaxies are typically more luminous than \textit{B} $\lesssim$ -13.0, with colors in the range \mbox{\mbi = 0.7-2} (\mycitealt{Hunter86};~\mycitealt{Marlowe97};~\mycitealt{Cairos01}). In addition, the luminosities observed for nearby (\mycitealt{Duc97},~\myciteyear{Duc98},~\myciteyear{Duc07}) and more distant (\mycitealt{Weilbacher00};~\mycitealt{Temporin03}) \tdg candidates are generally more luminous than \mbox{\textit{B} = -10.65}, as for most of our \ha complexes.

The integrated broad-band luminosities of the embedded stellar population in our \ha complexes is, therefore, compatible with the luminosities measured in giant \hii regions and \tdg candidates.

\begin{table}
\hypertarget{table:phot_prop_tdgs}{}\hypertarget{autolot:\theautolot}{}\addtocounter{autolot}{1}
\begin{minipage}{1\textwidth}
\renewcommand{\footnoterule}{}  
\begin{scriptsize}
\caption[Photometric properties of the identified star-forming complexes]{Photometric properties of the identified star-forming complexes: distances, magnitudes, and sizes \label{table:phot_prop_tdgs}}
\begin{center}
\begin{tabular}{l@{\hspace{0.3cm}}c@{\hspace{0.3cm}}c@{\hspace{0.3cm}}c@{\hspace{0.3cm}}ccccccc}
\hline \hline
   \noalign{\smallskip}
   IRAS & Nuclei &Complex & Number & d$_{near}$ & d$_{CM}$ & M$_{F435}$ & M$_{F435W}$ - M$_{F814W}$ & \reff & $r$ & $r_{\rm{H} \alpha}$ \\
 & \& IP & number  & of knots &(kpc)  & (kpc) & &  & (pc) & (pc) & (pc) \\
 (1) & (2)  & (3)  & (4) & (5) & (6) & (7) & (8) & (9) & (10) & (11)  \\
 \hline
   \noalign{\smallskip}
04315-0840& 1,IV& 1& 4& 4.6& - & -11.37& 1.02& 38& 166& 386\\
& & 2& 1& 4.2& - & -9.34& 0.85& 21& 93& 260\\
06076-2139& 2,III& 1& 2& 11.5& - & -12.86& 0.34& 77& 306& 1216\\
& & 2& 2& 11.3& - & -12.02& 0.21& 59& 283& 902\\
& & 3& 1& 12.6& - & -10.53& 0.29& 53& 121& 384\\
& & 4& 1& 12.3& - & -10.27& 0.26& 51& 116& 544\\
& & 5& 1& 11.1& - & -10.77& -0.23& 41& 89& 608\\
& & 6& 1& 9.9& - & -11.88& 0.35& 66& 137& 942\\
07027-6011 S& 2,I-II& 1& 1& 4.6& - & -11.20& 0.62& 31& 127& 740\\
& & 2& 1& 9.5& - & -10.73& 0.07& 61& 186& 670\\
08355-4944& 1,IV& 1& 2& 3.3& - & -12.94& 0.26& 52& 149& 705\\
08572+3915 N& 2,III& 1& 1& 6.5& 6.8& -12.96& 0.20& 105& 405& 1716\\
08572+3915 SE& 2,III& 1& 5& 4.4& 6.8& -14.54& 0.44& 176& 636& 1514\\
& & 2& 3& 9.3& 11.2& -14.05& 0.39& 199& 560& 1809\\
& & 3& 2& 9.3& 11.2& -12.10& 0.56& 76& 322& 809\\
& & 4& 1& 11.0& 12.1& -11.83& 0.90& 191& 312& 1144\\
F10038-3338& 1,IV& 1& 8& 10.9& - & -14.33& 0.67& 235& 630& 1016\\
& & 2& 4& 8.3& - & -12.47& 0.98& 58& 414& 551\\
& & 3& 2& 12.4& - & -11.34& 0.76& 43& 269& 603\\
& & 4& 2& 13.4& - & -11.72& 0.50& 88& 252& 603\\
12112+0305& 2,III& 1& 3& 3.4& 3.8& -15.56& 1.26& 200& 887& 1721\\
& & 2& 3& 9.4& 9.6& -15.77& 0.56& 212& 506& 1405\\
& & 3& 2& 10.4& 10.8& -14.82& 0.68& 200& 583& 1217\\
& & 4& 1& 11.1& 11.6& -12.34& 0.56& 82& 288& 1405\\
14348-1447& 2,III& 1& 2& 7.8& 7.8& -14.26& 2.34& 280& 909& 2167\\
15250+3609& 1,IV& 1& 4& 8.8& - & -14.13& 1.09& 165& 627& 1176\\
F18093-5744 N& 3,III& 1& 3& 3.9& 3.1& -11.91& 0.56& 48& 170& 497\\
& & 2& 3& 4.8& 6.8& -12.06& 1.11& 46& 178& 602\\
& & 3& 1& 3.8& 6.0& -10.11& 0.86& 20& 74& 363\\
F18093-5744 C& 3,III& 1& 1& 6.5& 4.3& -9.92& 0.84& 37& 78& 482\\
23128-5919& 2,III& 1& 2& 6.8& 8.3& -12.30& 0.87& 83& 376& 1098\\
\hline																							
\noalign{\smallskip}																							
16007+3743	&	3, III	&	R1	&	1	&	16.9	&	17.1	&	-	&	-	&	828	&-	&		1151	\\
	&		&	R2	&	1	&	8.9	&	9.4	&	-	&	-	&	884	&-	&		1375	\\
	&		&	R3	&	1	&	9	&	5.8	&	-	&	-	&	851	&-	&		1562	\\
\hline
\noalign{\smallskip}
\multicolumn{11}{@{} p{\textwidth} @{}}{\footnotesize{\textbf{Notes.} Col (1): object designation in the IRAS Point and Faint Source catalogues. In case of several pointings taken with VIMOS data, the orientation (N: North, S: South, SE: South-East, C: Center) of the centered nucleus in that pointing is given. Col(2): number of nuclei in the system (1-3) and interaction phase (IP) that the system is undergoing according to the classification scheme defined in Chapter~\ref{cha:data_tech},~\refsec{morphologies}, where I-II indicates first approach, III pre-merger and IV merger phases, respectively. Col (3): identified complex. Col (4): number of knots per complex. Col (5): projected distance to the nearest galaxy. Col (6): projected distance to the mass center of the system. Col (7): absolute magnitude,  \mb\twospace, not corrected for internal extinction. Col (8): derived photometric color of the complex. Col (9): \reff of the brightest knot inside the complex. Col (10): equivalent radius, derived using the size of the knots (see text). Col (11): radius of the complex measured on the \ha map.}}
\end{tabular}
\end{center}
\end{scriptsize}
\end{minipage}
\end{table}

\subsubsection{Sizes and Compactnesses}


Owing to the higher angular resolution of the \hst images, estimates of the sizes and compactnesses of the \ha emitting clumps were derived from the \acs images. All the knots within a complex were assumed to represent the ionizing stellar population. We then added the area of each knot (derived as explained in Chapter~\ref{cha:data_tech},~\refsec{sizes}) within an \ha complex, computing a total area $A_{\rm{T}}$. With this area, we derived an equivalent total radius, given by \mbox{$r$ = $\sqrt{A_{\rm{T}}/\pi}$} (see~\reftab{phot_prop_tdgs}). Had we used the spectral maps to measure the radius of the complexes ($r_{\rm{H} \alpha}$ in~\reftab{phot_prop_tdgs}), we would have overestimated it in general by more than a factor of two. The radii $r$ range from somewhat less than 100 pc to about 900 pc, the median value being 280 pc. These sizes are similar to those for the largest giant extragalactic \hii regions observed in nearby (\mbox{\ld $<$ 20 Mpc}) spirals (\mycitealt{Mayya94};~\mycitealt{Rozas06};~\mycitealt{Diaz07}), measured using ground-based instrumentation.

To evaluate the compactness of the \ha complexes, we also derived an effective radius (\reff\onespace) for each, as an approximation of the half-light radius. We again used the \hst images for this determination. In this case, since in most cases one knot dominates the total broad-band luminosity within the complex, we identified the effective radius of an \ha complex  with the effective radius of the most luminous knot (derived as explained in Chapter~\ref{cha:data_tech},~\refsec{sizes}). In complexes where more than one knot dominates the luminosity, this approach leads to an underestimate of the effective radius.

The effective radii of complexes observed with VIMOS are smaller than 100 pc (see~\reftab{phot_prop_tdgs}), from 20 pc to about 80 pc. For those observed with INTEGRAL, the values range from  100 pc to about 300 pc. These sizes are in general larger than the effective radii of the so-called ultra compact dwarf galaxies (e.g.,~\mycitealt{Dabringhausen08}), similar to those in giant \hii regions (e.g.,\mycitealt{Relanyo05};~\mycitealt{Rozas06}), and within the range of  the smallest blue compact dwarf galaxies (with \mbox{\reff = 0.2-1.8 kpc}; e.g.,~\mycitealt{Cairos03};~\mycitealt{Papaderos06};~\mycitealt{Amorin09}) and Local, NGC 1407, and Leo groups (with \mbox{\reff $\sim$ 0.3-1} kpc;~\mycitealt{Mateo98};~\mycitealt{Forbes11}). 

The ratio of the effective to equivalent radii gives an idea of the compactness of a certain complex. They are very compact, with a ratio of typically \mbox{\reff\twospace/$r$ = 0.2}, ranging from 0.15 to 0.6.

\subsect[ha_ew_tdgs]{\ha Luminosities and Equivalent Widths}

\begin{table}
\hspace{-0.4cm}
\hypertarget{table:spec_prop_tdgs}{}\hypertarget{autolot:\theautolot}{}\addtocounter{autolot}{1}
\begin{minipage}{1.03\textwidth}
\renewcommand{\footnoterule}{}  
\begin{scriptsize}
\caption[Spectral observables and dynamical parameters of the star-forming complexes]{Spectral observables, metallicity, and dynamical parameters of the identified star-forming complexes \label{table:spec_prop_tdgs}}
\begin{center}
\begin{tabular}{l@{}c@{\hspace{0.2cm}}c@{\hspace{0.2cm}}c@{\hspace{0.2cm}}c@{\hspace{0.2cm}}c@{\hspace{0.2cm}}c@{\hspace{0.2cm}}c@{\hspace{0.2cm}}c@{\hspace{0.2cm}}c}
\hline \hline
   \noalign{\smallskip}
IRAS & \tiny{Complex} & \tiny{$F_{\rm{obs}}$ (\ha\twospace)} & \tiny{$L_{\rm{obs}}$ (\ha\twospace)} & \tiny{EW (\ha\twospace)} & \tiny{Peak   EW} & $\frac{\rm{Peak~EW}}{\rm{EW}~(\rm{H}_\alpha)}$ & \tiny{$\sigma$ (\ha\onespace)} & $v_{\rm{rel}}$ &Adopted \\
 & \tiny{number} & \tiny{(10$^{-16}$ erg s$^{-1}$ cm$^{-2}$)} & \tiny{(10$^{39}$ erg s$^{-1}$)} & (\AA{}) & (\AA{}) &  & \tiny{(km s$^{-1}$)} & \tiny{(km s$^{-1}$)} & \tiny{12+log(O/H)} \\
 (1) & (2)  & (3)  & (4) & (5) & (6) & (7) & (8) & (9)  & 10\\
 \hline
   \noalign{\smallskip}
04315-0840& 1& 19.3& 1.4& 41.7& 78& 1.9& 16& -26& 8.72\\
& 2& 2.2& 0.2& 8.6& 12& 1.5& 34& -103& 8.99\\
06076-2139& 1& 15.1& 5.8& 48.8& 173& 3.6& 22& 421& 8.50\\
& 2& 5.8& 2.2& 41.3& 80& 2.0& 26& 442& 8.68\\
& 3& 0.6& 0.2& 26.3& 44& 1.7& 23& 442& 8.75\\
& 4& 1.1& 0.4& 27.0& 41& 1.6& 26& 432& 8.53\\
& 5& 1.6& 0.6& 17.8& 28& 1.6& 18& 430& 8.73\\
& 6& 6.8& 2.6& 11.0& 24& 2.2& - & 426& 8.66\\
07027-6011 S& 1& 10.4& 2.9& 68.9& 303& 4.4& 31& 24& 8.65\\
& 2& 2.2& 0.6& - & - & - & - & 73& - \\
08355-4944& 1& 40.8& 40.5& 83.0& 190& 2.3& 30& -25& 8.64\\
08572+3915 N& 1& 7.0& 6.0& 56.6& 242& 4.3& 46& 196& 8.53\\
08572+3915 SE& 1& 15.8& 13.6& 52.2& 183& 3.5& 74& 148& 8.54\\
& 2& 5.8& 5.0& 4.1& 85& 21.0& 62& 207& 8.50\\
& 3& 1.0& 0.8& - & - & - & - & 228& 8.53\\
& 4& 1.0& 0.8& 8.8& 28& 3.2& - & - & 8.56\\
F10038-3338& 1& 12.6& 4.1& 86.0& 374& 4.4& 19& 19& 8.68\\
& 2& 0.6& 0.2& 34.3& 74& 2.2& - & -11& 8.32\\
& 3& 2.0& 0.6& - & - & - & 17& -32& - \\
& 4& 2.2& 0.7& - & - & - & 27& 4& 8.24\\
12112+0305& 1& 44.8& 60.6& 71.1& 252& 3.6& 72& 96& 8.59\\
& 2& 11.9& 16.1& 26.3& 41& 1.6& 81& -204& 8.81\\
& 3& 8.7& 11.8& 11.5& 36& 3.1& - & -349& 8.59\\
& 4& 4.9& 6.7& 6.9& 53& 7.7& - & -266& 8.62\\
14348-1447& 1& 29.5& 65.7& 88.9& 149& 1.7& 54& 106& 8.63\\
15250+3609& 1& 8.1& 5.7& 168.9& 258& 1.5& 69& 124& 8.61\\
F18093-5744 N& 1& 32.8& 2.6& 48.1& 213& 4.4& 23& -76& 8.75\\
& 2& 71.8& 5.6& 67.0& 263& 3.9& 24& 60& 8.77\\
& 3& 9.5& 0.7& 15.8& 21& 1.4& 39& 48& 8.90\\
F18093-5744 C& 1& 3.8& 0.3& 26.0& 57& 2.2& 18& 119& 8.62\\
23128-5919& 1& 5.8& 2.8& 32.0& 51& 1.6& 69& 2& - \\
\hline																			
\noalign{\smallskip}																			
16007+3743	&	R1	&	12	&	57.1	&	40	&	-	&	-	&	61	&	-41	&	8.7	\\
	&	R2	&	35.5	&	189.5	&	234	&	-	&	-	&	76	&	92	&	8.7	\\
	&	R3	&	5.85	&	8.8	&	80	&	-	&	-	&	85	&	338	&	8.7	\\

\hline
\noalign{\smallskip}
\multicolumn{10}{@{} p{\textwidth} @{}}{\footnotesize{\textbf{Notes.} Col (1): object designation as in~\reftab{phot_prop_tdgs}. Col (2): identified complex. Col (3): observed \ha flux. As stated in Garc\'ia-Mar\'in et al. (\myciteyear{Garcia-Marin09a}) and in Rodr\'iguez-Zaur\'in et al. (\myciteyear{Rodriguez-Zaurin10}), typical uncertainties in the absolute flux calibration are between 15 and 20\%.  Col (4): derived \ha luminosity with no correction for internal extinction, except for IRAS 16007+3743. Col (5): computed equivalent width (EW) from the \ha and \ew maps (see text). Col (6): spaxel with the largest value of the EW. Col (7): ratio of the peak of the \ew to the derived \ew for the whole complex. Col (8): central velocity dispersion. Col (9): relative velocity with respect to the  mass center of the system. Col (10): adopted metallicity for the complex, derived using two different line-ratio calibrators (see text).}}
\end{tabular}
\end{center}
\end{scriptsize}
\end{minipage}
\end{table}

\begin{figure}
 \hypertarget{fig:halfa_ew_lit}{}\hypertarget{autolof:\theautolof}{}\addtocounter{autolof}{1}

\includegraphics[angle=90,width=0.49\columnwidth]{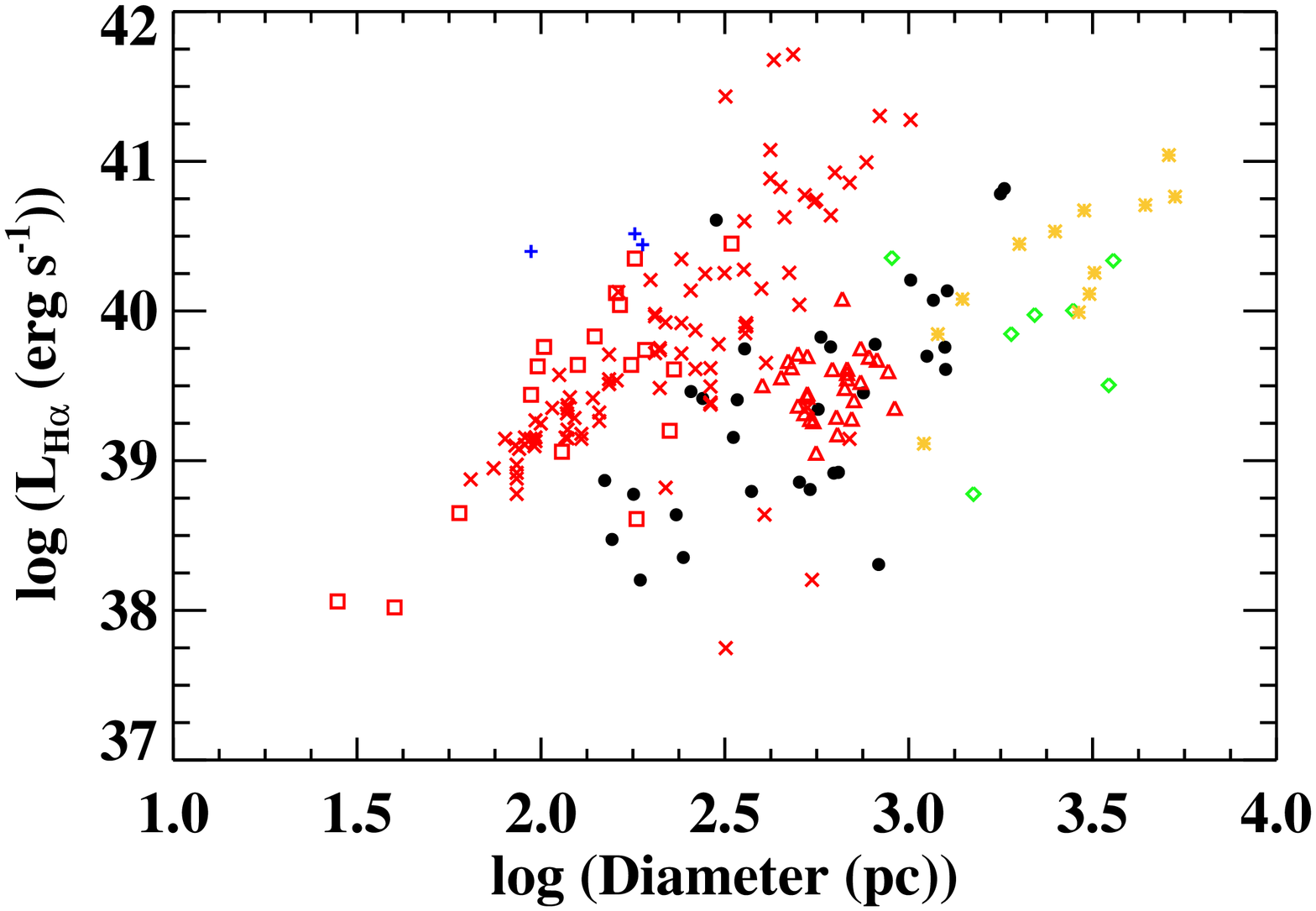}
\includegraphics[angle=90,width=0.49\columnwidth]{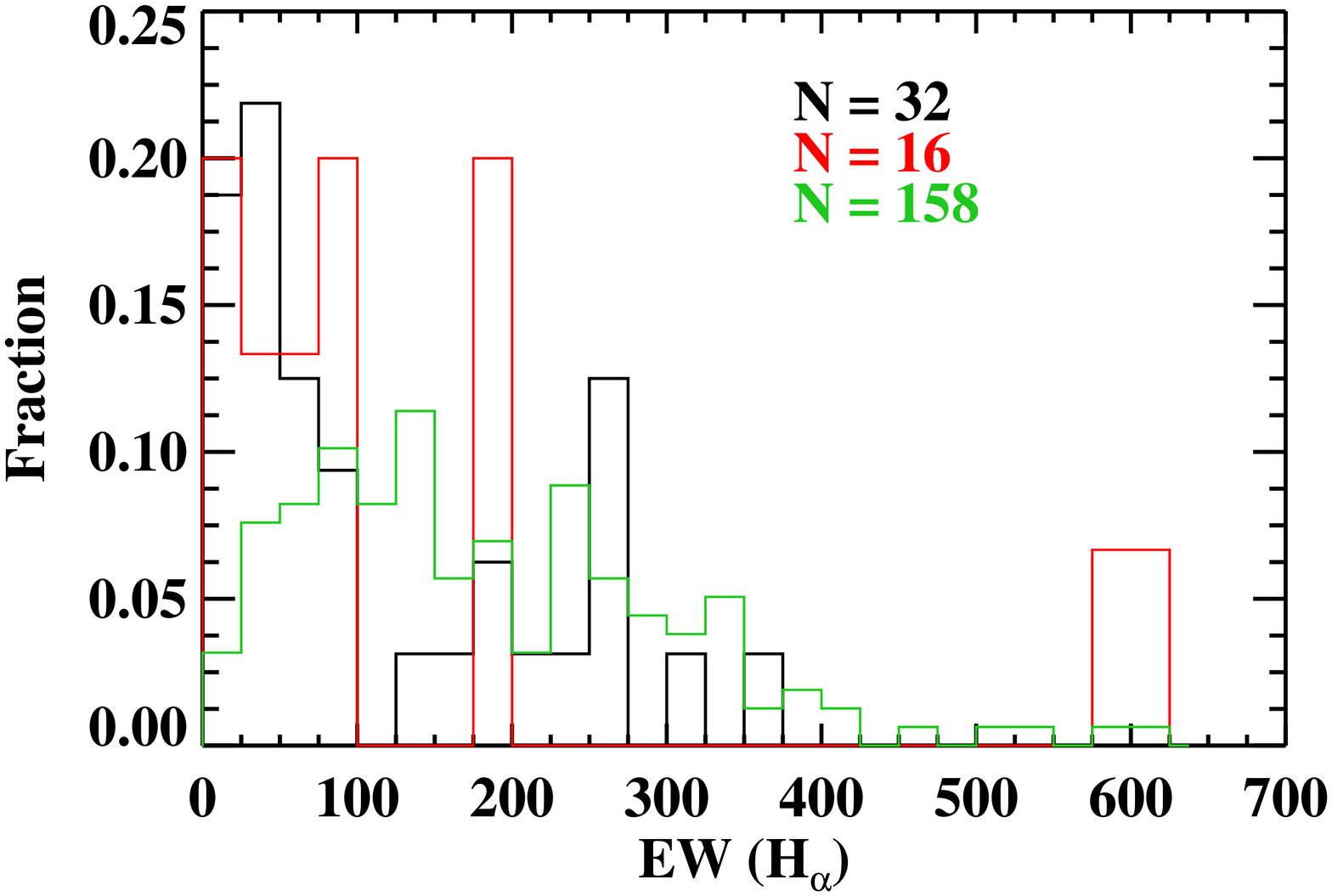}

\vspace{-0.3cm}
\caption[\ha luminosity and EW distribution of the identified complexes]{\textbf{Left:} Relation between the \ha luminosity and the size of the identified \ha complexes in our sample (black dots), extremely luminous \ha complexes in The Antennae (blue plusses;~\mycitealt{Bastian06}), different samples of isolated giant \hii regions in spirals and minor mergers (red symbols;~\mycitealt{Terlevich81};~\mycitealt{Ferreiro04};~\mycitealt{Rozas06}), \tdg candidates in CG J1720-67.8 (green diamonds;~\mycitealt{Temporin03}) and nearby dwarf amorphous (yellow crosses;~\mycitealt{Marlowe97}). None of the  luminosities on the plot are corrected for internal extinction. \textbf{Right:} Equivalent widths measured for the complexes  in this study (black), for extragalactic \hii regions (green;~\mycitealt{Mayya94}) and for \tdg candidates (red;~\mycitealt{Iglesias-Paramo01};~\mycitealt{Temporin03}).The number of objects is indicated for each case.}
\label{fig:halfa_ew_lit}
\end{figure}

Both \ha flux and equivalent width measurements (\ew\twospace) were obtained for the 31 \ha complexes. Their observed \ha luminosity is typically higher than \mbox{10$^{39}$ \ergs} (see~\reftab{spec_prop_tdgs}). The luminosity of about 20\% of them is even higher than \mbox{10$^{40}$ \ergs\twospace}. The median value of the \ha flux is \mbox{2.9$\times$10$^{39}$ \ergs\twospace}. Monreal-Ibero et al. (\myciteyear{Monreal07}) also measured the \ha fluxes for four of our systems with INTEGRAL data (in IRAS 08572+3915 only for the northern pointing) and most of them are systematically lower by up to a half. The discrepancy occurs because they used apertures of 0.5\arcsec~ or 1.0\arcsec, whereas we have integrated the fluxes of several spaxels, considering a greater effective aperture (one spaxel on INTEGRAL maps already corresponds to an aperture of 0.45\arcsec~in radius). In the case of IRAS 12112+0305, we grouped their kc and k1 regions and did not consider R2 because it is too close to the northern nucleus, which makes its \ha flux determination very uncertain.

\reffig{halfa_ew_lit} (left) compares the \ha luminosities and sizes of our complexes with those of extragalactic \hii regions, \ha complexes in nearby galaxies, \tdg candidates, and dwarf galaxies. This plot shows that using the \ha luminosity alone to establish the nature of a given object might be misleading. For instance, the complexes of star formation in The Antennae have an \ha luminosity comparable to the brightest extragalactic \hii or even to dwarf galaxies, while these complexes are generally much smaller. Our complexes are typically located where the giant extragalactic \hii regions and the dwarf-type objects lie on the plot. 

We also estimated the total equivalent width of a given complex from the \ew and the \ha spectroscopic line maps. Since in a spaxel the \ew $\propto$ {\small{Flux}} \ha$\!\!$ / {\small{Flux continuum}}, we first determined the continuum per spaxel. We then integrated all the continuum within a complex to compute the total continuum of the complex. Finally, we divided the total \ha flux by the total continuum flux for each complex. Their integrated \ew span a range of two orders of magnitudes, from 4.1\AA{} to about 170\AA{} (see~\reftab{spec_prop_tdgs}). For each complex, we also give the value of the spaxel with the highest \ew (EW$_{\rm{peak}}$). The highest value corresponds to \mbox{EW$_{\rm{peak}}$ = 374\AA{}}. Its median value (\mbox{$<\rm{EW}_{\rm{peak}}>$ = 85\AA{}}) doubles the median value of the integrated \ew computed for the complexes. The \ha equivalent widths derived, which is indicative of a very young stellar population, are comparable to those measured for extragalactic \hii regions and  \tdg candidates  (see~\reffig{halfa_ew_lit}, right).

The ratio $\frac{\rm{Peak}~\rm{EW}}{\rm{EW}~(\rm{H}\alpha)}$ was defined for each complex, where \mbox{\textit{Peak EW}} refers to EW$_{\rm{peak}}$. This ratio helps us understand how the ionizing population is distributed. If two complexes have a similar underlying old population, the ionizing population in the one with the larger ratio is probably more concentrated (i.e., less knots) than the other with the smaller ratio. And if it is less concentrated (i.e., has more knots),  most of the knots are probably somewhat older (low \ew\twospace) and only a few are very young (high \ew\twospace), leading to an older average population. In fact, the complexes with large ratios (\mbox{$\frac{\rm{Peak}~\rm{EW}}{\rm{EW}~(\rm{H}\alpha)}$ $\gtrsim$ 3}) have typically a few knots (1-3), such as C1 in IRAS 06076-2139, C1 in IRAS 08572+3915 N, and C4 in IRAS 12112+0305. 

\subsect[metal_tdgs]{Metallicities}

\begin{figure}
 \hypertarget{fig:met_relations}{}\hypertarget{autolof:\theautolof}{}\addtocounter{autolof}{1}
\hspace{1cm}
\includegraphics[angle=90,width=0.80\columnwidth]{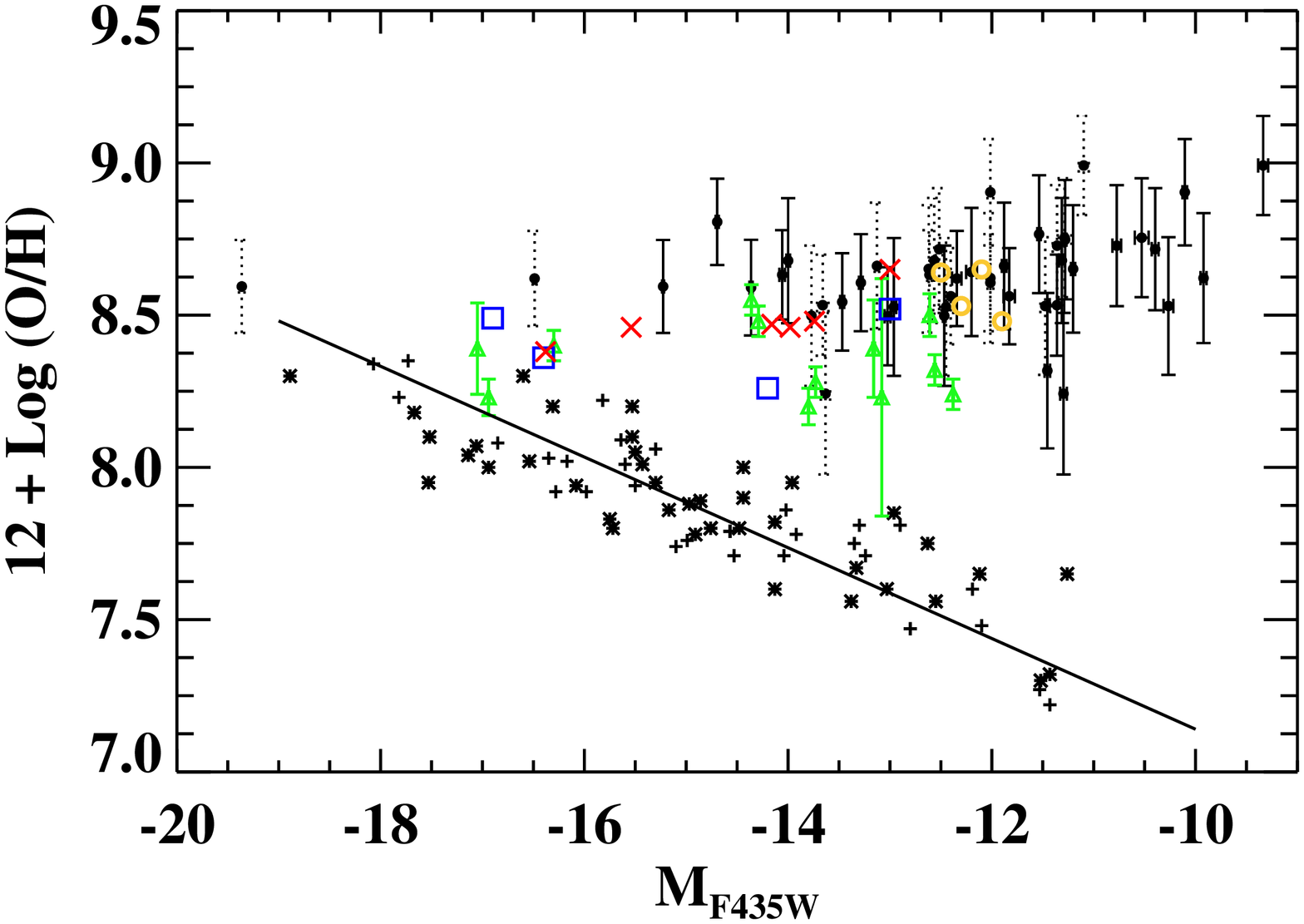}
\caption[Metallicity-luminosity relation of the complexes]{Metallicity-luminosity relation for external \hii regions in Stephan'{}s Quintet (yellow open circles;~\mycitealt{Oliveira04}), nearby irregular galaxies (black plusses,~\mycitealt{Pilyugin04}; black asterisks,~\mycitealt{VanZee06}), \tdg candidates (green filled triangles,~\mycitealt{Weilbacher00},~\myciteyear{Weilbacher03}; red crosses,~\mycitealt{Duc98}; blue open squares,~\mycitealt{Temporin03}), and our complexes (black dots with errors). The line shows the correlation found by van Zee. Only the luminosities given by Pilyugin are corrected from internal extinction. For those \ha complexes for which an estimate of the extinction has been made (see~\refsub{age_mass_complex}), the metallicity-luminosity relation is also shown (dots with dashed line errors).}
\label{fig:met_relations}
\end{figure}

Numerical studies have claimed that during an interaction large amounts of gas flow toward the central regions, carrying less enriched gas from the outskirts of the galaxy into the central regions (\mycitealt{Rupke10};~\mycitealt{Montuori10}). Indications of this mixing process, which usually disrupts any metallicity gradient and dilutes the central metallicity, have been observed in merging (U)\lirgs (\mycitealt{Rupke08}), in galaxy pairs (\mycitealt{Kewley06};~\mycitealt{Ellison08};~\mycitealt{Michel-Dansac08}), and in star-forming galaxies at high-z (\mycitealt{Cresci10}). As a consequence of the metal enrichment of the external regions of the mergers, additional star formation can occur from reprocessed material there. We investigated whether the interaction process affects the metallicity of our \ha complexes.

The adopted metallicities (see Chapter~\ref{cha:data_tech},~\refsec{metal}) for all the candidates are shown in the last column in~\reftab{spec_prop_tdgs}. With values compatible to a solar metallicity, they generally range from 12 + logO/H = 8.5 to 8.8 (with the exception of two candidates with lower abundances). The values for complexes observed with VIMOS have to be considered with care and must be checked with other metallicity indicators, since the N2 calibrator might be affected by ionization from strong shocks in the external regions of these galaxies (\mycitealt{Monreal10}). Moreover, spatially resolved studies of star-forming regions indicate that the assumption of spherical geometry is unrealistic in most cases, which has a direct impact on the derivation of metallicities. Ercolano et al. (\myciteyear{Ercolano07}) estimated the systematic errors in the metallicity determinations when assuming spherical geometry using different calibrations. In the worst-case scenario, the derived oxygen abundances might be overestimated by 0.2-0.3 dex by the use of the calibrators considered in this study, in which case the derived metallicites for the \ha complexes would then approach Z$_{\odot}$/3.

The \ha complexes in this study do not follow the well-known metallicity-luminosity relation for nearby isolated dwarf galaxies (see~\reffig{met_relations}). Most galaxies also follow this relation (\mycitealt{Weilbacher03}). On the other hand, \hii regions in compact groups of galaxies and \tdgs in general deviate significantly from the relation and have a metallicity that is independent of their luminosity, an indication that all these objects consist of recycled material.

\sect[discussion_tdgs]{From H$\alpha$ Complexes to TDG Candidates in (U)LIRGs}

Most \ha\twospace-emitting complexes have similar observational properties (i.e., \ha equivalent widths and luminosities, \mb magnitudes, metallicities, radii) to the most luminous extragalactic giant \hii regions in spirals and mergers, as well as more massive objects such as \tdgs or \tdg progenitors. Associations of young star-forming regions with a large HI reservoir have been found in \tdgs\twospace, for instance by Duc et al. (\myciteyear{Duc07}). We may then consider whether the \ha\twospace-emitting complexes in our sample are dynamically unbound associations of objects with masses similar to observed Super Star Clusters (\sscs\twospace) or larger, more massive, and self-gravitating objects such as dwarf galaxies.

\tdgs are self-gravitating objects with masses and sizes typical of dwarf galaxies (i.e., a total mass of 10$^7$-10$^9$~\msun\twospace), formed with recycled material from the parent galaxies involved in an interaction/merger (\mycitealt{Duc00}; see also General Introduction,~\refsec{tdg_gal}). Hence, they are stable entities with their own established dynamical structure. To evaluate whether our complexes constitute real \tdgs or \tdg progenitors and assess their chance of survival, we need to answer a few basic questions: Is the complex massive enough to be considered as a dwarf galaxy? Is it stable enough to be unaffected by its internal motion? Are the gravitational forces too strong to disrupt it? We disscuss here these questions.

\subsect[selection_tdgs]{Selection of \ha\twospace-emitting Likely to Represent TDG Candidates}

Although old \tdgs have been observed (e.g.,~\mycitealt{Duc07}), their identification is not straightforward. When an old \tdg is detected, the tail from its place of origin will probably have completely disappeared and the \tdg may be classified as another type of dwarf galaxy. A determination of the metallicity and, especially, the total mass of the object is needed to establish its tidal origin and mass. With spectroscopic data, we are able to study (though with some limitations and biases) the metal content, the dynamical mass, and different methods to establish whether a given candidate can withstand both internal and external forces. It is awkward to identify luminous enough condensations of old populations (i.e., very massive) and obtain kinematic information because the stellar absorption lines that would have to be used often have too low a signal-to-noise ratio (S/N). Although \tdgs containing only old populations may be present in our sample, we are unable to detect them. Another way of searching for \tdg candidates is based on either analyzing the \ha emission clumps (for that we have a sufficiently high S/N) and/or combining spectral and photometric data of blue \ha\twospace-emitting objects in interacting systems (e.g.,~\mycitealt{Iglesias-Paramo01};~\mycitealt{Weilbacher03};~\mycitealt{Temporin03};~\mycitealt{Lopez-Sanchez04}). Therefore, the selection of \tdg candidates in our sample of (U)\lirgs is based on the detection of \ha emission clumps with stellar counterparts, which is similar to our identification of \ha complexes.

However, we are limited by the angular resolution of the spectral data. To make our estimates as accurate as possible, we do not consider in our discussion here complexes with multiple luminous star-forming knots. Different types of motions within the components of a given complex can strongly affect the derivations mentioned before. The most characteristic examples of complicated complexes are C1 in IRAS F10038-3338 and C2 in IRAS 12112+0305. This does not mean that the possibility of these complexes being \tdgs is ruled out, just that we  do not have enough resolution to study the kinematics of each knot, thus determine whether they are kinematically bound or not. 

We consider only complexes with simple structures, that is, mainly where only one knot is detected. We also include complexes with several knots, one of which (normally centered on the \ha peak emission) dominates the broad-band luminosity. The complex C1 in IRAS 04315-0840 would be a good example. On the basis of these criteria, the kinematic and dynamical properties of 22 complexes are derived and compared with those expected for a \tdg\twospace.

If we detect more than one knot within the selected complexes, we consider only the broad-band luminosity of the brightest knot. In these cases, the \ha luminosity of the complex is scaled in the same way as the \textit{F435W} flux is scaled to the total \textit{F435W} broad-band flux of the complex in~\reftab{phot_prop_tdgs}. In practice, we multiply the \ha luminosity by a factor that in general is higher than 0.5 (a factor of 1 means that there is only one knot emitting the whole broad-band flux, thus responsible for the full \ha emission). The kinematic measurements, such as the velocity dispersion, are not performed for the whole complex, but only for the spaxels close to and located at the peak of the \ha emission. This permits us to avoid any contamination by a faint knot in the border of the complex (e.g., faint knots within C1 in IRAS 04315-0840).

Using the kinematic information provided by the IFU data as well as the photometric measurements, we assess the origin and likelihood of survival of the 22 selected \ha complexes, and evaluate whether they can be the progenitors of Tidal Dwarf Galaxies.

\subsect[massive_tdgs]{Are They Massive Enough to Constitute a TDG?}

\subsubsection{Age and Mass Estimates of the Young Population}
\relax
\hypertarget{sub:age_mass_complex}{}
\label{sub:age_mass_complex}

The \ha emission can be used to constrain the properties of any recent episodes of star formation. According to the starburst99 models (\sbnn,~\mycitealt{Leitherer99};~\mycitealt{Vazquez05}), the detection of an \ha flux and EW larger than about 13 \AA{} (see~\reftab{spec_prop_tdgs} for specific values) implies that the stellar population is younger than 10 Myr (see~\reffig{ew_plot} in Chapter~\ref{cha:data_tech}). With the broad-band luminosities of the brightest knot within the complex and the scaled \ha luminosity, we can estimate to first order the age and the mass of the young population of the selected complexes. We note that, owing to the distance of the systems (\mbox{\ld$<$ 270} Mpc) and the shape of the \hst filters, the \ha emission-line does not contaminate these broad-band filters. We have made use of the synthetic stellar population model \sbnn\twospace, which is optimized for young population and also provides information on the ionized gas. The parameters used in the model (\imf, metallicity, etc.) are defined in Chapter~\ref{cha:data_tech},~\refsec{ssp}.

\begin{figure}
 \hypertarget{fig:halfa_com}{}\hypertarget{autolof:\theautolof}{}\addtocounter{autolof}{1}
 \hspace{1.5cm}
\includegraphics[angle=90,width=0.85\columnwidth]{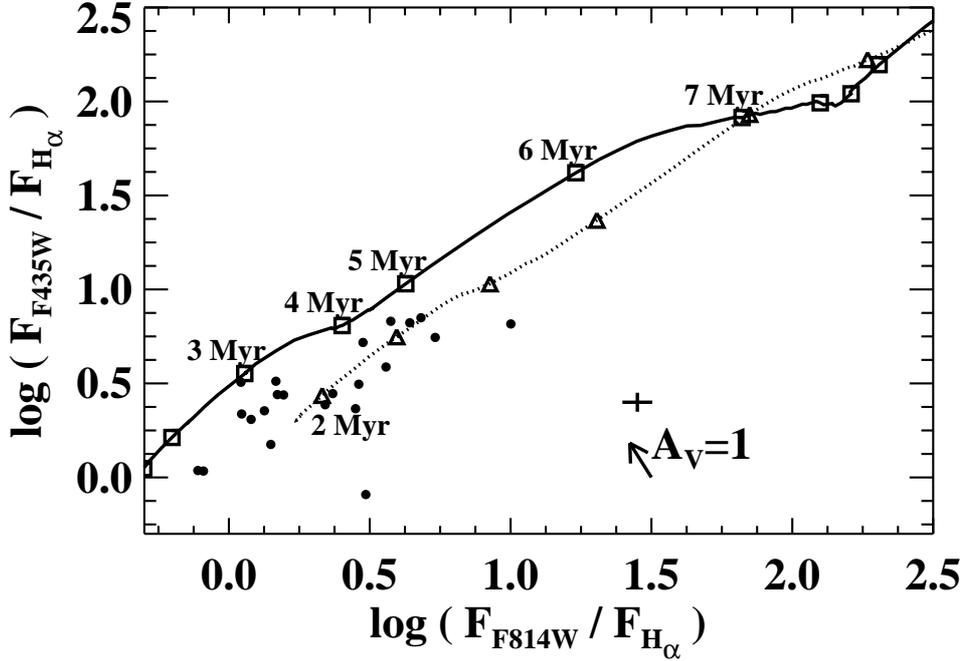}
\caption[Flux ratios of the blue and red filters with H$\alpha$]{Flux ratios of the blue and red filters when the \ha emission can be measured in a complex. The broad-band filter fluxes have been multiplied by the band-width to obtain the values with the same units as the \ha flux. Solid line shows the evolutionary track of the same ratios for a single burst using \sbnn models (see text). Dotted line shows the track for a composite population where 99\% of the mass corresponds to a 1Gyr population. The small triangle almost at the beginning (zero-age young population) of the second track indicates that the young population is 2Myr old and we add 1 Myr to the subsequent triangles. The typical uncertainty associated with the data is shown above the extinction vector, which is drawn at the bottom right-hand corner.}
\label{fig:halfa_com}
\end{figure}

We computed the mass-independent evolutionary track shown in~\reffig{halfa_com} using information about the ratio of the flux of the modeled population in the broad-band filters to the flux of the ionized gas in the \ha line. Placing these ratios on a plot for all observed complexes, we now estimate the age of the young population. Once the age is known (hence a M/L ratio), the mass can be derived. However, before going any further and based on the assumptions made, we first assess the sources of uncertainty that a priori could affect our results:

\begin{itemize}

\item  The internal extinction is unknown. In general, the extinction in the inner regions is patchy with high peaks in (U)\lirgs\twospace, up to \av $\simeq$ 8 mags (see~\mycitealt{Alonso-Herrero06} and~\mycitealt{Garcia-Marin09b}), though it decreases considerably with the galactocentric distance. Were complexes to have low extinction values (\av $\lesssim$ 1.5mag) their flux quotients would tend to follow the evolutionary track (solid line in~\reffig{halfa_com}), and the age would remain practically unchanged. Only one complex probably has a higher internal extinction (C1 in IRAS 14349-1447). A high extinction value of \av $\sim$ 3.5 mags was indeed measured for this complex in Moneral-Ibero et al. (\myciteyear{Monreal07}). The typical extinctions of the \tdg candidates studied in Monreal-Ibero are between \mbox{\av= 0.7 and 2 mag}, which is consistent with the extinctions derived for our selected complexes (see~\reffig{halfa_com}).\\

\item The local background flux (from the underlying parent galaxy) assumed to be associated with either older populations or non-\ha line-emitters, was subtracted when the photometry was performed for the knots. However, a knot itself can be formed by a composite of young and old population. If we were to assume that a significant fraction of the red and blue fluxes measured for the knots originated from an older population, the evolutionary track would change (see the composite populations in~\reffig{halfa_com}); the young population would then be even younger than initially predicted and consequently less massive. Almost half of the complexes are incompatible with the composite track, since their flux ratios would be too small, even smaller than for a zero-age young population. The other complexes, for which the values are compatible with the composite track, would typically have a young population of about or younger than 4 Myr. Were this to be the case, the extinction would be negligible and the equivalent widths would be about 200 \AA{}. \\

\item It can also happen that the measured \ha flux is emitted by possible embedded (and undetected) star-forming knots. The existence of undetected \ha\twospace-emitters is justified with the \ha\twospace-emitter in IRAS 08572+3915 N (clump on the North-West). We estimated how many undetected knots there can be with an age derived from the peak values of the \ew (typically between 5 and 10 Myr, according to the stellar population models used in~\refsub{age_mass_complex}) that can contribute significantly to the total broad-band luminosity of the knots. Since they would have a similar age, an embedded population of knots with a combined broad-band flux similar to that of the detected knots would emit half of the \ha flux. We then estimated how many knots at the detection limit would be needed to double either the \textit{F435W} or the \textit{F814W} flux. A total of  9 and 75 undetected knots per complex would be needed and we detected an average of 2.2 knots per complex. Thus, between 4 and 37 times more knots fainter by 2.4-4.7 magnitudes would be needed only in one complex. Assuming a luminosity function with slope of 2 we would need practically one-third of the total knots predicted for the whole galaxy inside a complex, which is unlikely. \\

\item Had we included some \ha contribution from the neighboring zones then the \ha flux measured would have been overestimated. The \ha flux drops considerably as age increases across the age interval we consider (by up to more than one order of magnitude), but if the \ha flux is weaker the mass of the young population decreases accordingly. The mass estimated using the red filter can vary by up to a factor of four within the age interval 1-10 Myr, according to the stellar population models used in this study. The age estimate would be older by no more than 1 Myr, if we had overestimated the flux by a half. Thus, this correction would in our case be negligible.

\end{itemize}

Taking into account these sources of uncertainty, the first-order estimates of the age were performed for the young population inside a complex, assuming a single burst. We estimated the age by considering the extinction value that we would need to shift the location of each dot (flux quotients) along the evolutionary tracks. The estimated ages range from 2 to 5 Myr (Age$_{ph}$; see~\reftab{derived_prop_tdgs}), the typical age range in which an instantaneous burst of star formation displays \acr{}{WR}{Wolf-Rayet} features in its spectra (\mycitealt{Leitherer99}). Given the spectral range covered in this study, we would expect to see the well-known red \wolfr bump, which is the result of the blend of the \ciii $\lambda 5698$ and  \civ $\lambda 5808$ broad emission lines (\mycitealt{Kunth86}). We tried to find these features in the spectra of the complexes, but could find no clear evidence of these bumps. This bump is much more difficult to detect than the other bump (such as the blue bump) characteristic of \wolfrs because it is always weaker (e.g.,~\mycitealt{Fernandes04};~\mycitealt{Lopez-Sanchez09}). Sometimes it is not even detected again in Wolf-Rayet galaxies where it was previously seen (\mycitealt{Lopez-Sanchez09}). Therefore, the non-detection of \wolfr features does not definitely exclude the hypothesis that the young population in our complexes spans the age range 2-5 Myr.

An upper limit to the age can be estimated by using the \ew (\ha\twospace), as it decreases strongly with time. Using the \ew peaks (see~\reftab{spec_prop_tdgs}), we estimate that the age is in the range \mbox{5-10} Myr (Age$_{EW}$; see~\reftab{derived_prop_tdgs}). In many cases, the \ew peaks are not strong enough to explain population as young as our former age estimates. This is unsurprising because within a complex and a spaxel itself the underlying older population contributes significantly to the \ha continuum but not to the emission line. The contribution of this continuum minimizes the total \ew within a complex. The broad-band flux of the knots typically represents between 1\% and 40\% of the total flux within the area of the complex. We consider for instance a 3 Myr-old population whose broad-band flux represents 10\% of its overall broad-band flux within the complex. According to the models we use, the equivalent width of the young population without contamination would be $\sim$1000 \AA{}. However, the contamination of a 1 Gyr-old population would diminish the equivalent width to a measured value of 70 \AA{} for the whole complex, the same value as a single population of 6 Myr without contamination. Therefore, with the observed \ew we can only set an upper limit to the age in each case.

As a conservative approach, instead of directly using the youngest estimates to derive the mass of the young population, we use the average of the two estimates, Age$_{ph}$ and Age$_{EW}$. Thus, once the age and extinction were estimated, the mass was directly obtained via the extinction-corrected \textit{F814W} magnitude.

Under these assumptions, the derived mass of the young population of the selected \ha complexes is between 10$^{4.5}$~\msun and 10$^{5.5}$~\msun\twospace, with the exception of three complexes for which the derived mass is about or close to 10$^{7}$~\msun (M$_{[I]}$;~\reftab{derived_prop_tdgs}). These complexes have extinctions of \mbox{\av $\sim$ 1-2} mag, with the exception of C1 in IRAS 14348-1447 (\mbox{\av = 4.2 mag}). The uncertainty in this mass is typically smaller than a factor of two. This rather small uncertainty is expected, since during the first 1-7 Myr of the starburst the broad-band luminosities do not change significantly. However, if we compute the mass using the \ha luminosity (M$_{[\rm{H}\alpha]}$;~\reftab{derived_prop_tdgs}) the uncertainties increase considerably. This is also expected because the \ha flux for a population of a given mass evolves significantly with the age of the starburst. The few cases in which both measurements are incompatible suggest that the age of the burst is closer to the youngest value (Age$_{\rm{ph}}$). The corresponding M$_{[\rm{H}\alpha]}$ would be lower toward a value similar to M$_{[I]}$.

An old population of a few Gyr (prior to the interaction, and in the parent galaxy) in \tdg candidates that contributes to most of the stellar mass has been reported (e.g.,\mycitealt{Sheen09}). Large \hi reservoirs have also been found in a few \tdgs (\mycitealt{Duc07}) that can sustain star formation on a Gyr scale. We have seen previously that the colors of some complexes can be compatible with a composite population where a 1 Gyr-old population is 99\% more massive than a young burst. Thus, we consider it worth investigating this possibility also here.~\reftab{derived_prop_tdgs} shows the estimates of the age of the young (Age$_{\rm{Cyoung}}$) and the mass of both populations (M$_{\rm{Cyoung}}$ and M$_{\rm{Cold}}$). The stellar mass of the complexes generally ranges from 10$^{5.5}$~\msun to 10$^{6.5}$~\msun, an order of magnitude higher than the previous estimates.

\begin{sidewaystable}
\hypertarget{table:derived_prop_tdgs}{}\hypertarget{autolot:\theautolot}{}\addtocounter{autolot}{1}
\begin{minipage}{1.02\textwidth}
\renewcommand{\footnoterule}{}  
\begin{scriptsize}
\caption[Derived characteristics and dynamics of the star-forming complexes]{Derived characteristics of the stellar populations and dynamics of the identified star-forming complexes \label{table:derived_prop_tdgs}}
\begin{center}
\begin{tabular}{@{}l@{\hspace{-0.3cm}}c@{\hspace{0.1cm}}c@{\hspace{0.1cm}}c@{\hspace{0.1cm}}c@{\hspace{0.3cm}}c@{\hspace{0.3cm}}c@{\hspace{0.1cm}}c@{\hspace{0.1cm}}c@{\hspace{0.3cm}}c@{\hspace{0.3cm}}c@{\hspace{0.3cm}}c@{\hspace{0.3cm}}c@{\hspace{0.3cm}}c@{\hspace{0.1cm}}c@{\hspace{0.1cm}}c@{\hspace{0.1cm}}c@{}}
\hline \hline
   \noalign{\smallskip}
IRAS & Complex & Age$_{\rm{ph}}$ & Age$_{\rm{EW}}$ &  \av & M$_{[I]}$ & M$_{[H_{\alpha}]}$ & Age$_{\rm{Cyoung}}$ & \av\twospace$_C$ & M$_{\rm{Cold}}$ & M$_{\rm{Cyoung}}$  & M$_{\rm{dyn}}$ & M$_{\rm{tid}}^{\rm{near}}$ & M$_{\rm{tid}}^{\rm{CM}}$ & M$_{\rm{dyn}}/\rm{M}_{\rm{tidal}}$ & $v_{\rm{esc}}$ & $\frac{|v_{\rm{esc}}|}{v_{\rm{rel}}}$ \\
\noalign{\smallskip}
 & number  & (Myr)  & (Myr) & (mags) &  & & (Myr) & (mags) & &  & &  & & (km s$^{-1})$ &  \\
 (1) & (2)  & (3)  & (4) & (5) & (6) & (7) & (8) & (9) & (10) & (11) & (12) & (13) & (14) & (15) & (16) & (17) \\

 \hline
   \noalign{\smallskip}
04315-0840& 1& 3.4& 6.0& 1.7 $\pm$ 0.3& 4.9 $\pm$ 0.1& 5.4 $\pm$ 0.3& 2.1 & 0.2 $\pm$ 0.4& 5.8 $\pm$ 0.1& 3.8 $\pm$ 0.1& 7.5 $\pm$ 0.3 & 6.6 $\pm$ 0.3 & - & 8.1 & 224& 8.6\\
& 2& 4.0& 10.0& 1.5 $\pm$ 0.4& 4.2 $\pm$ 0.2& 5.8 $\pm$ 0.5& 2.8 & 0.7 $\pm$ 0.5& 5.5 $\pm$ 0.2& 3.5 $\pm$ 0.2& 7.8 $\pm$ 0.3 & 6.0 $\pm$ 0.3 & - & $>$ 10 & 235& 2.3\\
06076-2139& 1& 2.9& 5.4& 1.1 $\pm$ 0.3& 5.4 $\pm$ 0.2& 5.9 $\pm$ 0.2& -& -& -& -& 8.0 $\pm$ 0.5 & 5.8 $\pm$ 0.2 & - & $>$ 10 & 96& 0.2\\
& 2& 3.0& 6.0& 1.0 $\pm$ 0.4& 4.9 $\pm$ 0.2& 5.4 $\pm$ 0.3& -& -& -& -& 8.0 $\pm$ 0.3 & 5.7 $\pm$ 0.2 & - & $>$ 10 & 96& 0.2\\
& 3& 4.4& 6.4& 0.6 $\pm$ 0.3& 4.6 $\pm$ 0.1& 5.1 $\pm$ 0.2& 2.9 & 0.0 $\pm$ 0.0& 5.6 $\pm$ 0.1& 3.6 $\pm$ 0.1& 7.9 $\pm$ 0.3 & 4.5 $\pm$ 0.2 & - & $>$ 10 & 92& 0.2\\
& 4& 3.0& 6.5& 0.8 $\pm$ 0.4& 4.5 $\pm$ 0.2& 5.0 $\pm$ 0.5& -& -& -& -& 8.0 $\pm$ 0.3 & 4.5 $\pm$ 0.2 & - & $>$ 10 & 93& 0.2\\
& 5& 2.9& 6.8& 0.4 $\pm$ 0.2& 4.4 $\pm$ 0.2& 5.1 $\pm$ 0.4& -& -& -& -& 7.6 $\pm$ 0.4 & 4.3 $\pm$ 0.2 & - & $>$ 10 & 97& 0.2\\
& 6& 2.8& 7.0& 1.0 $\pm$ 0.3& 5.2 $\pm$ 0.1& 5.9 $\pm$ 0.8& -& -& -& -& - & 5.0 $\pm$ 0.2 & - & - & 103& 0.2\\
07027-6011 S& 1& 2.1& 5.0& 1.2 $\pm$ 0.3& 4.9 $\pm$ 0.2& 5.6 $\pm$ 0.3& -& -& -& -& 7.9 $\pm$ 0.4 & 6.1 $\pm$ 0.3 & - & $>$ 10 & 180& 7.5\\
& 2& 3.2& -& 0.6 $\pm$ 0.3& 4.5 $\pm$ 0.1& 4.6 $\pm$ 0.2& -& -& -& -& - & 5.6 $\pm$ 0.3 & - & - & 125& 1.7\\
08572+3915 N& 1& 2.8& 5.2& 0.7 $\pm$ 0.3& 5.4 $\pm$ 0.2& 5.9 $\pm$ 0.2& -& -& -& -& 8.7 $\pm$ 0.3 & 7.2 $\pm$ 0.4 & 7.4 $\pm$ 0.3 & $>$ 10 & 233& 1.2\\
08572+3915 SE& 3& 4.9& -& 0.9 $\pm$ 0.3& 5.1 $\pm$ 0.1& 5.1 $\pm$ 0.2& 3.1 & 0.0 $\pm$ 0.6& 6.1 $\pm$ 0.2& 4.1 $\pm$ 0.2& - & 6.6 $\pm$ 0.4 & 6.6 $\pm$ 0.3 & - & 225& 1.0\\
& 4& 4.9& 6.9& 1.4 $\pm$ 0.4& 5.5 $\pm$ 0.2& 6.1 $\pm$ 0.2& 3.2 & 0.6 $\pm$ 0.5& 6.5 $\pm$ 0.2& 4.5 $\pm$ 0.2& - & 6.3 $\pm$ 0.4 & 6.5 $\pm$ 0.3 & - & 207& - \\
F10038-3338& 3& 3.6& -& 0.7 $\pm$ 0.3& 4.4 $\pm$ 0.1& 4.5 $\pm$ 0.1& 2.7 & 0.0 $\pm$ 0.0& 5.7 $\pm$ 0.1& 3.7 $\pm$ 0.1& 7.6 $\pm$ 0.3 & 5.8 $\pm$ 0.3 & - & $>$ 10 & 117& 3.6\\
& 4& 4.9& -& 0.9 $\pm$ 0.3& 5.0 $\pm$ 0.1& 5.2 $\pm$ 0.1& 3.0 & 0.0 $\pm$ 0.6& 6.0 $\pm$ 0.2& 4.0 $\pm$ 0.2& 8.2 $\pm$ 0.3 & 5.6 $\pm$ 0.3 & - & $>$ 10 & 113& 26.2\\
12112+0305& 1& 3.4& 5.2& 2.2 $\pm$ 0.3& 7.1 $\pm$ 0.2& 7.3 $\pm$ 0.2& 2.2 & 0.8 $\pm$ 0.4& 8.0 $\pm$ 0.1& 6.0 $\pm$ 0.1& 9.4 $\pm$ 0.2 & 9.2 $\pm$ 0.3 & 9.6 $\pm$ 0.2 & 0.7 & 496& 5.1\\
& 4& 2.1& 6.3& 1.1 $\pm$ 0.3& 5.4 $\pm$ 0.2& 6.1 $\pm$ 0.3& -& -& -& -& - & 6.2 $\pm$ 0.3 & 6.7 $\pm$ 0.2 & - & 275& 1.0\\
14348-1447& 1& 3.3& 5.5& 4.2 $\pm$ 0.3& 7.7 $\pm$ 0.1& 8.1 $\pm$ 0.2& 1.5 & 2.5 $\pm$ 0.3& 8.5 $\pm$ 0.1& 6.5 $\pm$ 0.1& 9.3 $\pm$ 0.2 & 9.1 $\pm$ 1.1 & 9.2 $\pm$ 0.3 & 1.3 & 575& 5.4\\
15250+3609& 1& 5.4& 5.2& 2.4 $\pm$ 0.3& 6.6 $\pm$ 0.1& 6.7 $\pm$ 0.2& 3.9 & 1.1 $\pm$ 0.4& 7.5 $\pm$ 0.1& 5.5 $\pm$ 0.1& 9.3 $\pm$ 0.3 & 7.7 $\pm$ 0.2 & - & $>$ 10 & 214& 1.7\\
F18093-5744 N& 3& 2.9& 7.2& 1.9 $\pm$ 0.3& 5.0 $\pm$ 0.2& 5.8 $\pm$ 0.5& -& -& -& -& 7.9 $\pm$ 0.4 & 4.1 $\pm$ 0.1 & 5.1 $\pm$ 0.2 & $>$ 10 & 220& 4.6\\
F18093-5744 C& 1& 3.5& 6.2& 1.6 $\pm$ 0.3& 4.8 $\pm$ 0.1& 5.2 $\pm$ 0.3& 2.4 & 0.4 $\pm$ 0.4& 5.7 $\pm$ 0.1& 3.7 $\pm$ 0.1& 7.6 $\pm$ 0.5 & 3.6 $\pm$ 0.3 & 5.6 $\pm$ 0.2 & $>$ 10 & 175& 1.5\\
23128-5919& 1& 3.3& 6.3& 1.8 $\pm$ 0.3& 5.7 $\pm$ 0.2& 6.1 $\pm$ 0.5& 2.0 & 0.3 $\pm$ 0.4& 6.5 $\pm$ 0.2& 4.5 $\pm$ 0.2& 9.0 $\pm$ 0.2 & 7.1 $\pm$ 0.2 & 7.0 $\pm$ 0.3 & $>$ 10 & 222& 85.7\\
\hline																																	
\noalign{\smallskip}																																	
16007+3743	&	R1	&	-	&	7.1	&	1.7	&	8.8	&	-	&	-	&	-	&	-	&	-	&	9.8	&	8.2	&	8	&	$>$ 10	&	302	&	7.4	\\
	&	R2	&	-	&	5.4	&	1.2	&	8.8	&	-	&	-	&	-	&	-	&	-	&	10	&	9.2	&	10.1	&	0.9	&	408	&	4.4	\\
	&	R3	&	-	&	6.4	&	2.3	&	7.8	&	-	&	-	&	-	&	-	&	-	&	10.1	&	10	&	9.1	&	1.2	&	519	&	1.5	\\

\hline
\noalign{\smallskip}

\multicolumn{17}{@{} p{\textwidth} @{}}{\footnotesize{\textbf{Notes.}  Col (1): object designation as in~\reftab {phot_prop_tdgs}. Col (2): identified complex. Col (3): age of the young population, derived using the photometric information (\reffig{halfa_com}). Col (4): age of the young population using the \ew (\ha\twospace). Col (5): internal extinction. Col (6): derived mass of the young population using the \textit{I} magnitude.  All masses in the table are given in log (M (\msun\twospace)). Col (7): derived mass of the young population  using the \ha flux. Col (8): derived age of the young component, assuming that the candidate consists of a composite population. Col(9): derived internal extinction of the composite population. Col (10): derived mass of the old population (i.e., 1Gyr), assuming a composite population (see text). Col (11): derived mass of the estimated young population, assuming a composite population.  Col (12): dynamical mass. Col (13): tidal mass assuming the potential is created by the nearest galaxy. Col (14): tidal mass assuming the potential is created by a point mass at the mass center of the system. Col (15): ratio of the dynamical to the tidal mass. Col (16): escape velocity. Col (17): ratio of the escape to the relative velocity. }}
\end{tabular}
\end{center}
\end{scriptsize}
\end{minipage}
\end{sidewaystable}

\subsubsection{Conditions for Self-gravitation}
\relax
\hypertarget{sub:internal_motions}{}
\label{sub:internal_motions}

\begin{figure}
\hypertarget{fig:autograv}{}\hypertarget{autolof:\theautolof}{}\addtocounter{autolof}{1}
\hspace{-0.5cm}\includegraphics[angle=90,width=0.53\columnwidth]{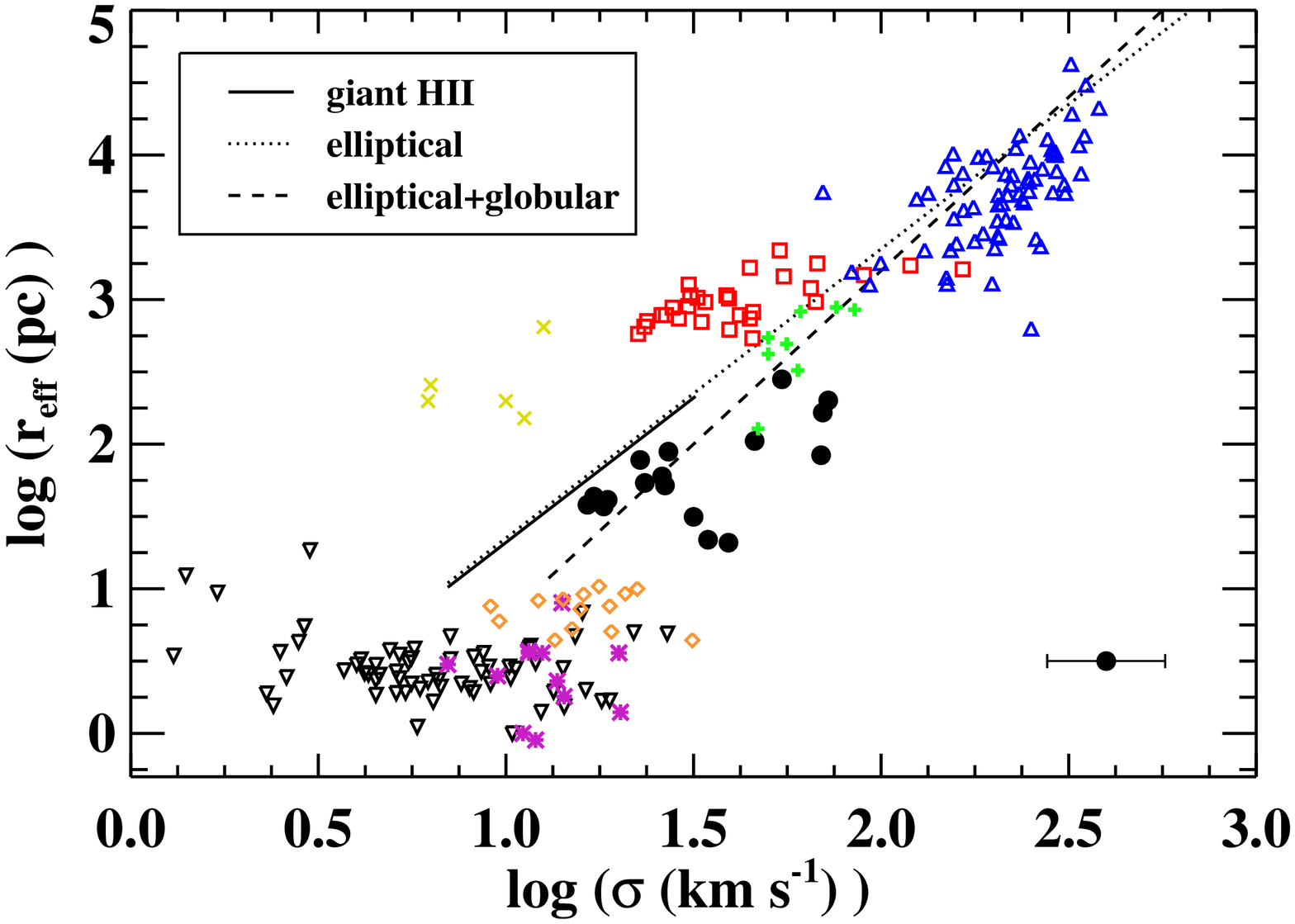}
\hspace{-0.5cm}\includegraphics[angle=90,width=0.53\columnwidth]{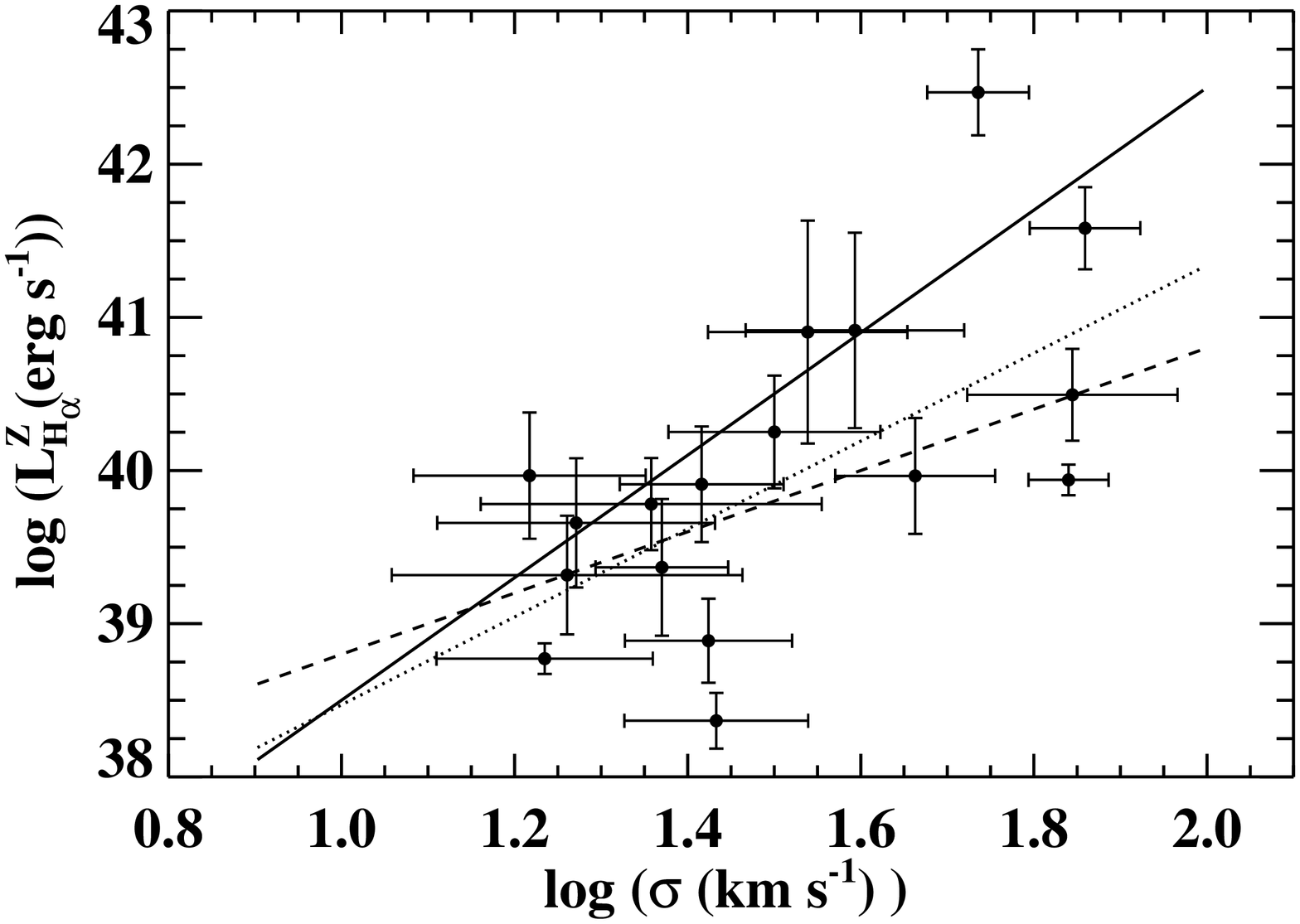}
\caption[Velocity dispersion vs. estimated effective radius and vs. \lha]{\textbf{Left:} elocity dispersion vs. estimated effective radius. Dots correspond to the relation for our selected complexes. The typical size for errors is shown in the bottom right corner. The lines show the fit for extragalactic \hii regions (continuous), elliptical galaxies (dashed) and globular + elliptical galaxies (dotted) obtained by Terlevich \& Melnick (\myciteyear{Terlevich81}). The other symbols represent different samples of dynamically hot systems: open triangles in blue, intermediate and giant ellipticals (\mycitealt{Bender92}); open squares in red, dwarf ellipticals (\mycitealt{Bender92};~\mycitealt{Geha03}); crosses in yellow, dwarf spheroidals (\mycitealt{Bender92}); plusses in green, the \tdg candidates in Monreal-Ibero et al. (\myciteyear{Monreal07}); diamonds in orange, massive globular clusters in NGC 5128 (\mycitealt{Martini04}); inverse triangles, globular clusters in the Galaxy (\mycitealt{Trager93}) and in M31 (\mycitealt{Dubath97}); and asterisks in pink, clusters in the Antennae (\mycitealt{Mengel08}). \textbf{Right:} Velocity dispersion vs. \ha luminosity corrected for internal extinction and metallicity effects. The lines show different fits for extragalactic giant \hii regions which are approximately in virial equilibrium, given by:~Terlevich \& Melnick (\myciteyear{Terlevich81}) (solid line),~Rela{\~n}o et al. (\myciteyear{Relanyo05})  (dotted line) and~Rozas et al. (\myciteyear{Rozas06}) (dashed line).}
\label{fig:autograv}
\end{figure} 

Measuring velocity gradients helps us assess whether there is either any independent rotation (\mycitealt{Weilbacher02};~\mycitealt{Bournaud04};~\mycitealt{Bournaud08a}) or outflows. However, given the spatial resolution achieved with the \ifs instruments, we cannot in general resolve any velocity field across the extranuclear condensations. In this case, other indirect methods must be used to assess whether the internal motions of our complexes can affect their stability.

Iglesias-P{\'a}ramo \& V{\'{\i}}lchez (\myciteyear{Iglesias-Paramo01}) established an empirical luminosity criterion \mbox{(L(H$\alpha$) $>$ 10$^{39}$ erg s$^{-1}$)} that must be reached by their \hii complexes to ensure self-gravitation. This criterion, however, has not been supported by any kinematic study. More than 50\% of our own complexes fulfill this criterion, as shown in~\reftab{spec_prop_tdgs}. If we corrected for internal extinction, only three would be less luminous than 10$^{39}$ erg s$^{-1}$. 

Another method that can be used to establish whether the selected complexes are stable is to study their location in well-known empirical correlations followed by other self-gravitating entities such as elliptical galaxies, bulges of spiral galaxies, globular clusters, and/or giant \hii regions: the radius-velocity dispersion and the luminosity-velocity dispersion relations (\mycitealt{Terlevich81}). Here, we consider these relations because they provide more reliable constraints than the \ha luminosity of the complexes. We plot these relations for our complexes in~\reffig{autograv}.

In the radius-velocity dispersion diagram, we superimpose the data of samples of dynamically hot systems such as massive globular clusters, dwarf elliptical and spheroidal galaxies, intermediate and giant ellipticals, and \tdg candidates. Within all of the uncertainties, the velocity dispersions of six of the complexes selected here are too high to ensure self-gravitation.


For a giant \hii region, the lower envelope in the log \lha - log$\sigma$ plane is closely represented by a straight line of the form \lha = c  + $\Gamma \times log{\sigma}$, where $\sigma$ refers to the velocity dispersion, the constant c ranges from 34.5 to 36.8, and $\Gamma$ is between 2 and 4 (see line fits and references in~\reffig{autograv}, right). This is known as the luminosity-velocity (L-$\sigma$) dispersion relation, and implies that the giant extragalactic \hii regions above the envelope are gravitationally bound complexes of stars and gas and that the widths of the nebular emission lines reflect the motions of discrete ionized gas clouds in the gravitational field of the underlying stellar and gaseous mass (\mycitealt{Terlevich81}). The kinematics of the \hii regions below the envelope may be dominated by processes other than gravitational motions, such as stellar winds and expanding shells (\mycitealt{Relanyo05};~\mycitealt{Relanyo05b};~\mycitealt{Rozas06};~\mycitealt{Rozas06b}). That is, if the relation for giant \hii regions lies above the envolope, then the non-gravitational processes are identified with broad, low-intensity components that do not considerably affect the physical properties of the main spectral component (i.e., gravitational motions). Thus, although the velocity of these fenomena could be similar or even larger than the $\sigma$ measured in this paper, if the relation for our complexes is above the envelope, the spectral component corresponding to the gravitational motions will probably not be affected. 

In any case, we estimated the expansion speeds of bubbles in the interstellar medium, so as to quantify how broad these non-gravitational components are. We evaluated the equivalent number of O3(V) stars using the \ha luminosity of a given complex and the value given by Vacca et al. (\myciteyear{Vacca96}) for the ionizing photon output of an O3(V) star. We then derived the kinetic energy input from the stellar winds, using the estimate by Leitherer (\myciteyear{Leitherer98}) of the wind luminosity for an O3(V) star, integrated over the time the star is on the main sequence. Assuming a typical average electron density of \mbox{100 cm$^{-3}$}, obseved in (U)\lirgs (\mycitealt{Veilleux95},~\myciteyear{Veilleux99}), the expansion speeds are expected to be (\mycitealt{Lamers99}) between several km s$^{-1}$ for the least luminous complexes and less than 30 km s$^{-1}$ for the most luminous ones.

Given the metallicity range of our selected complexes, we applied a metallicity correction to the \ha luminosity, in such a way that \mbox{$\Delta$log (F(\ha\twospace)) = 127z - 1.17} (\mycitealt{Terlevich81}), where z denotes the metallicity value (z=0.02 for solar metallicity). Within the uncertainties, all the selected complexes but two are consistent with the line fits or above the envelope (\reffig{autograv}, right). Interestingly, these two outlying complexes are consistent with the radius-velocity dispersion correlation. Although the values for another complex are below the envelope, we note that its \ha luminosity was not corrected because its metallicity is unknown. Thus, no complex fails both criteria at the same time. According to these results, self-gravitation should be ensured for more than a half of the complexes for which a determination of the velocity dispersion was achieved. For the remainder, self-gravitation is neither guaranteed nor excluded because at least they generally satisfy one of the two criteria used.

\subsubsection{Dynamical Mass Estimates}
\relax
\hypertarget{sub:dyn_mass}{}
\label{sub:dyn_mass}

Upper limits to the dynamical masses (\mdyn\twospace) of the selected complexes were derived under the following assumptions: (i) the systems are spherically symmetric; (ii) they are gravitationally bound; and (iii) they have isotropic velocity distributions [\mbox{$\sigma^2$(total) = 3$\times\sigma_{LOS}^2$}], where $\sigma_{LOS}$ is the line-of-sight velocity dispersion. In the previous section, we have seen that more than a half of these complexes are likely to be gravitationally bound.

The equations to obtain the dynamical mass of a virialized stellar (cluster- or galaxy-like) system are presented in Chapter~\ref{cha:data_tech},~\refsub{eq_dyn_mass}. The observables needed to derived it are the effective radius and the velocity dispersion. Since the star-forming knots have a cluster-like light profile (similar to a Gaussian) the use of~\refeqn{dyn_mass2} to derive the dynamics of the selected complexes seems appropriate. The corresponding masses, given in~\reftab{derived_prop_tdgs}, are between a few 10$^7$ and a few 10$^9$~\msun\twospace, with an uncertainty of a factor of 2-3.

\subsubsection{Total Mass: Discrepancy Between the Stellar and the Dynamical Masses}

It is unclear where to draw the line between a high-mass young super cluster and a low-mass dwarf galaxy in terms of stellar mass. We derive masses only for young population, and those complexes less massive than 10$^6$~\msun  are comparable to the brightest young clusters or complexes hosting recent star formation in other less luminous interacting galaxies (e.g.,~\mycitealt{Mengel08};~\mycitealt{Bastian09}). However, does the estimated mass of the young stellar population represent the total mass of the candidates? Only a few studies have reported stellar masses for \tdgs\twospace. The dynamical mass of our complexes is in the range of the typical total baryonic mass of a \tdg (\mbox{M = 10$^7$-10$^9$~\msun\twospace}; e.g.,~\mycitealt{Duc04};~\mycitealt{Higdon06};~\mycitealt{Hancock09}). However, the ratio of the dynamical to the most massive photometric mass in our complexes is suspiciously high (\mbox{\mdyn\twospace/m$_{[M_I]}\sim$ 50-1000)}. Several factors can explain this ratio.

If we assume that the stellar mass represents the total mass of a given complex, this mass might be underestimated, because:

\begin{itemize}
\item Although the most recent bursts of stars are responsible for the bulk of the ionization of the gas - hence of the emitted flux- the kinematics of the ionizing gas is affected by those of previous stellar generations. In regions of mixed populations, the light is generally dominated by the young stars and the mass by the older population. As outlined before, Sheen et al. (\myciteyear{Sheen09}) derived a total stellar mass of 3.1$\times$10$^7$~\msun for one of their \tdg candidates, a fraction of no more than 2\% being contained in populations younger than 6 Myr. The mass-to-light (M/L) ratio of young (e.g., 5 Myr) and old (e.g., 1 Gyr) populations differs by about a factor of 100. The \ha to the broad-band luminosity colors of some of our candidates are also consistent with those of composite populations. Our complexes might indeed consist of young star-forming knots and evolved population from the complexes themselves and the parent galaxy that has been accreted by the complex. We estimated the total stellar mass of the complex if practically all the \textit{F814W} flux measured inside the whole complex (minus the flux from the knots) belonged to it and came from a 1 Gyr-old population. This total stellar mass approaches significantly to the dynamical mass measured for most of the complexes.\\
 
\item The total stellar population does not normally represent the total baryonic mass (gas + stellar), and we do not have access to the mass of the gas. Normally, the efficiency of star formation in molecular clouds is very low (typically about 1\%). In (U)\lirgs the efficiency can be higher, but the mass of a molecular cloud undergoing a starburst episode may still be higher than the mass of the starburst by more than one order of magnitude.  The candidates in Hancock et al. (\myciteyear{Hancock09}) have total stellar masses of \mbox{1-7$\times$10$^6$~\msun\twospace}, the HI mass being \mbox{6$\times$10$^7$~\msun\onespace}. Studies based on radio \hi observations have found that the gas mass of a \tdg with a typical diameter of a few kpcs lies within the range between a few 10$^7$ and several 10$^{9}$~\msun (e.g.,~\mycitealt{Duc00};~\mycitealt{Bournaud04};~\mycitealt{Duc07}). \\
 
 \end{itemize}

On the other hand, if we assume that the dynamical mass represents the total mass of a given complex, we might be overestimating it, because:

\begin{itemize}

\item In each case, \mdyn represents normally an upper limit to the dynamical mass. For instance, given the complex dynamics in interacting systems, high velocity dispersions might not necessarily indicate high masses, but might alternatively represent tidal-flows induced from the merging process and strong winds from LINER-like regions (\mycitealt{Colina05};~\mycitealt{Monreal10}). However, at least for the systems that we previously observed with VIMOS, these strong winds may not affect significantly the determination of the velocity dispersions (\mycitealt{Monreal10}).\\ 
 
 \item Dark matter might contribute significantly to the total \tdg mass. However, \tdgs are unlikely to contain a large amount of dark matter (\mycitealt{Barnes92b}) because their material is drawn from the spiral disk, while the dark matter is thought to surround the galaxy in an extended halo.
 
\end{itemize}

On the basis of all these arguments, it is clear that the total masses of these complexes is somewhere between the photometric and the dynamical masses. Other independent criteria also show that more than a half of the selected complexes are in virial equilibrium, thus the velocity dispersion traces more likely the dynamical mass rather than other processes (see~\refsub{internal_motions}). Given all the caveats and uncertainties previously considered when determining the dynamical and stellar masses, we assume as a mass criterion that the complexes with photometric mass estimates using the broad-band luminosities (single or composite population) that are compatible with 10$^6$~\msun are likely to have a sufficient total mass to become a tidal dwarf galaxy. This assumption should obviously be verified using multi-wavelength photometric and spectroscopic data. Only six of the selected complexes (plus the three candidates in IRAS 16007+3743) fulfill this mass criterion. Had we assumed that a complex with a dynamical mass compatible with \mbox{10$^8$~\msun} or higher has sufficient mass to become a \tdg\twospace, then 15 of the selected complexes would fulfill the mass criterion.

\subsect[forces_parent_gal]{Are They Unaffected by the Forces from the Parent Galaxy?}
\subsubsection{Tidal Forces}

If the complex is massive enough to constitute a dwarf galaxy and gravitationally bound, its fate basically depends on the ratio of its mass to the so-called tidal mass (\mycitealt{Binney87};~\mycitealt{MendesO01}). The tidal mass condition will tell us whether it is massive enough to survive the tidal forces exerted by the parent galaxy. The tidal mass is defined as

\eqn[tidal_mass]{
 \rm{M}_{\rm{tid}} = 3 \rm{M} \times \left(\frac{\rm{R}}{\rm{D}}\right)^3
}

where M refers to the mass of the parent galaxy, R is the radius of the complex (here the size of the most luminous  knot in a complex), and D is the distance to the parent galaxy. This equation is valid when the size (R) of a certain region  is small in comparison with the distance (D) to the parent galaxy, which is the case for all of the complexes (the ratio ``size/projected distance'' is typically below 0.04). If the tidal mass of an object is lower than the total mass, then the object is unaffected by the forces applied by the parent galaxy.

In general, the gravitational potential of interacting galaxies, as in \ulirgs\twospace, is a complex function of the mass distribution of the system, which evolves with time. Nevertheless, as a first approximation, we assumed that the distribution is dominated by the masses of the main bodies of the system. Two different approaches were applied in this study: the gravitational potential depends either on:  i) the nearest galaxy (M$_{\rm{tid}}^{\rm{near}}$) or ii) a point mass in the mass center and the total mass of the system (M$_{\rm{tid}}^{\rm{CM}}$). The highest value obtained using the different approaches was taken to compare with the dynamical mass of the complex. 

In each case, we used the measured projected distances $d_{\rm{near}}$ and $d_{\rm{CM}}$, respectively (see~\reftab{phot_prop_tdgs}). Since the projected distance is always smaller than the real one, we actually estimated upper limits to the real tidal masses. To derive the tidal mass, we need to know the mass of the parent galaxy, which in each case is identified with its dynamical mass. Under the same hypothesis as in~\refsub{dyn_mass} of a virialized system, the dynamical mass of a galaxy-like object can be obtained using ~\refeqn{dyn_mass1} (Chapter~\ref{cha:data_tech},~\refsub{eq_dyn_mass}), but assuming this time a constant value of \mbox{m = 1.75} (a scaling factor for a polytropic sphere), like in Colina et al. (\myciteyear{Colina05}). In each case, we used the average value of the velocity dispersion at the peak and nearby surrounding spaxels, covering about a radius of 1\arcsec. Finally, the effective radii of the parent galaxies were derived using the \textit{H}-band \nicmos and WFC3 images of the galaxies and fitting their two-dimensional light profiles with GALFIT (\mycitealt{Peng02};~\mycitealt{Peng10}). In many cases, the \nicmos images do not cover the full extent of the galaxies, which meant that we had to use the values given in Arribas et al. (in prep.).

In summary, two values (M$_{\rm{tid}}^{\rm{near}}$ and M$_{\rm{tid}}^{\rm{CM}}$) for the tidal mass were obtained using~\refeqn{tidal_mass}, and the maximum value was compared to the dynamical mass of the selected complexes derived in~\refsub{dyn_mass}. The ratio M$_{\rm{dyn}}$/M$_{\rm{tid}}$ is in general \mbox{$>$ 10} (see~\reftab{derived_prop_tdgs}), which ensures that the complexes are unaffected by the tidal forces exerted by the parent galaxy. Only one complex fails this condition and another one is close to failing because of large uncertainties in its dynamical mass estimate.

\subsubsection{Escape Velocity}

If the complex does not have a large enough relative velocity ($v_{\rm{rel}}$) with respect to the galaxy to escape, it might still fall back towards the center of the system because of the gravitational force exerted by the parent galaxy. It is interesting to consider whether a complex does indeed exceed the effective escape velocity ($v_{\rm{esc}}$). This criterion, however, should not have the same weight as the others since it is subject to many uncertainties: (i) the real distance of the complex is always larger than the projected one; (ii) only one component of the velocity is measured and there is no information about the movements in the plane of the sky; and (iii) two possibilities for the relative movements between the region and the system are always possible, since for a given configuration it is impossible to determine whether the complex is closer or further away from the observer than the mass center. We include this criterion in this study for completeness.

For simplicity, we assume that the gravitational potential is created by a point mass representing the total mass of the system, that is located at the mass center. The ratio $|v_{\rm{esc}}|/v_{\rm{rel}}$ (last column in~\reftab{derived_prop_tdgs}) is smaller than 1 for 7 of the selected complexes out of the 22, that is, they could escape. If we take into account projection effects statistically, a complex escapes if \mbox{$|v_{\rm{rel}}| - |v_{\rm{esc}}~\times ~cos(\pi/4)|~>$ 0}. In this case, an additional complex satisfies the criterion.

\subsect[tdg_common]{How Common is TDG Formation in (U)LIRGs?}
\subsubsection{\small{\tdg Candidates in (U)LIRGs}}

\begin{table}
\hypertarget{table:tests}{}\hypertarget{autolot:\theautolot}{}\addtocounter{autolot}{1}
\begin{minipage}{0.9\textwidth}
\renewcommand{\footnoterule}{}  
\begin{scriptsize}
\caption[Summary of the criteria used to investigate the nature of the complexes]{Summary of the different criteria used to investigate the nature of the complexes}
\label{table:tests}
\begin{center}
\begin{tabular}{l@{\hspace{0.2cm}}c@{\hspace{0.2cm}}c@{\hspace{0.2cm}}c@{\hspace{0.2cm}}c@{\hspace{0.2cm}}c@{\hspace{0.2cm}}c@{\hspace{0.2cm}}c@{}c}
\hline \hline
   \noalign{\smallskip}
IRAS	&	Complex	&	Mass	&	L (\ha\twospace)	&	$\sigma$ vs. \reff	&	$\sigma$ vs. \lha	&	M$_{\rm{tid}}$ vs. M$_{\rm{dyn}}$	&	$v_{\rm{esc}}$ vs. $v_{\rm{rel}}$	&	Prob	\\
	&	number	&		&		&		&		&		&		&		\\
 \hline
   \noalign{\smallskip}
 04315-0840	&	1	&	N	&	Y	&	Y	&	Y	&	Y	&	N	&	Medium	\\
	&	2	&	N	&	N	&	N	&	Y	&	Y	&	N	&	Low	\\
06076-2139	&	1	&	N	&	Y	&	Y	&	Y	&	Y	&	Y	&	Medium-high	\\
	&	2	&	N	&	Y	&	Y	&	Y	&	Y	&	Y	&	Medium-high	\\
	&	3	&	N	&	N	&	Y	&	Y	&	Y	&	Y	&	Medium	\\
	&	4	&	N	&	N	&	Y	&	N	&	Y	&	Y	&	Medium	\\
	&	5	&	N	&	Y	&	Y	&	Y	&	Y	&	Y	&	Medium-high	\\
	&	6	&	?	&	Y	&	?	&	?	&	?	&	Y	&	Low	\\
07027-6011 S	&	1	&	N	&	Y	&	N	&	Y	&	Y	&	N	&	Medium	\\
	&	2	&	N	&	Y	&	?	&	?	&	?	&	Y	&	Low	\\
08572+3915 N	&	1	&	N	&	Y	&	Y	&	Y	&	Y	&	N	&	Medium	\\
08572+3915 SE	&	3	&	?	&	Y	&	?	&	?	&	Y	&	Y	&	Low	\\
	&	4	&	Y	&	Y	&	?	&	?	&	?	&	?	&	Low	\\
F10038-3338	&	3	&	N	&	Y	&	Y	&	?	&	Y	&	N	&	Medium	\\
	&	4	&	Y	&	Y	&	Y	&	N	&	Y	&	N	&	Medium-high	\\
12112+0305	&	1	&	Y	&	Y	&	N	&	Y	&	N	&	N	&	Medium	\\
	&	4	&	?	&	Y	&	?	&	?	&	?	&	Y	&	Low	\\
14348-1447	&	1	&	Y	&	Y	&	Y	&	Y	&	Y	&	N	&	High	\\
15250+3609	&	1	&	Y	&	Y	&	N	&	Y	&	Y	&	N	&	Medium-high	\\
F18093-5744 N	&	3	&	N	&	Y	&	N	&	Y	&	Y	&	N	&	Medium	\\
F18093-5744 C	&	2	&	N	&	Y	&	Y	&	Y	&	Y	&	N	&	Medium	\\
23128-5919 	&	1	&	Y	&	Y	&	N	&	?	&	N	&	N	&	Low	\\

\hline
\noalign{\smallskip}																	
16007+3743	&	R1	&	Y	&	Y	&	Y	&	Y	&	Y	&	N	&	High	\\
	&	R2	&	Y	&	Y	&	Y	&	Y	&	N	&	N	&	High-medium	\\
	&	R3	&	Y	&	Y	&	Y	&	Y	&	Y	&	N	&	High	\\
\hline
\noalign{\smallskip}
\multicolumn{9}{@{} p{1.1\textwidth} @{}}{\footnotesize{\textbf{Notes.} The different criteria are explained throughout the text. The letter Y indicates that the given complex satisfies the criterion, while the letter N indicates that it fails. Symbols with a question mark are either doubtful or indicate that we do not have the data to study the criterion.  The last column indicates the probability that a given complex constitutes a TDG candidate based on these criteria.}}
\end{tabular}
\end{center}
\end{scriptsize}
\end{minipage}
\end{table}

None of the developed criteria can help us determine wheter a given complex will survive the merging process. Projection effects and observational constraints, especially the need for higher angular resolution spectroscopic data (which would allow us to search for velocity gradients), ensure that it is difficult to assess the fate of the selected complexes. In any case, we can investigate which have the higher probabilities of surviving as a \tdg\twospace, based on the fulfillment of a few or most of the criteria viewed in this study. The results are summarized in~\reftab{tests}. 

We assigned different weights to the diverse criteria we studied and derived a probability that a certain complex could survive as a \tdg by adding the weights for the criteria that it achieves. The \textit{mass} criterion is considered to be the most important since, even if it is self-gravitating, if it does not have enough mass the complex could be either a bound super cluster or any other entity rather than a low-mass dwarf galaxy. For this reason, it is assigned a weight of 30\%. The criteria with the least importance (each having a weight of 10\%) correspond to the \ha luminosity and the escape velocity ones, since the former comes from empirical considerations and the latter is the least reliable of all. We assigned a weight of 20\% to the M$_{\rm{tid}}$ versus (vs.) M$_{\rm{dyn}}$ criterion. Finally, we considered a somewhat lower weight (15\%) for each of the self-gravitating criterion ($\sigma$ vs. \reff and $\sigma$ vs. \lha), since they prove the same condition. In practice, the complexes that satisfy most of the criteria have the highest probabilities of being \tdg progenitors.

Once the percentage was computed based on which criteria the complexes fulfill, different probabilities that a complex constitutes a \tdgs were defined: \textit{low} (\mbox{prob $\leq$ 40\%}), \textit{medium} (\mbox{40 $<$ prob $\leq$ 60}), \textit{medium-high} (\mbox{60 $<$ prob $\leq$ 80}), and \textit{high} (\mbox{prob $>$ 80}). Only three complexes (counting IRAS 16007+3743 have a high probability. A total of 6 of the 22 selected complexes (9 of 25 if we include those from IRAS 16007+3743) have \textit{medium-high} or \textit{high} probabilities of being \tdg progenitors, and from now on we consider these as our \tdg candidates.

We detect candidates in both \lirgs and \ulirgs\twospace, although the three candidates with a high probability are only found in \lirgs\twospace. This suggests that \tdg production may be more efficient in systems with higher infrared luminosities. However, this statistic is of very limited robustness because of the low number of candidates found. If we consider only systems for which \ifs data are available and located at a distance (\mbox{i.e.,~\ld $>$ 130 Mpc}) where at least a galactocentric radius of 10-15 kpc is covered with the spectroscopic data, we obtain a production rate of about 0.3 candidates per system for the (U)\lirg class. Other \tdgs have been observed at larger distances (\mbox{d $>$ 30 kpc};~\mycitealt{Sheen09};~\mycitealt{Hancock09}) not covered by the \fov of our \ifs data. The detection of bright and blue knots (i.e., \mbox{\mi $>$ -12.5} and \mbox{\mbi $<$ 0.5}) in the tidal tails and their tips in the \acs images of some (U)\lirgs (regions not covered with the \ifs data; see~\reffig{spatial_dist_knots} in Chapter~\ref{cha:knots}), indicative of a young population more massive than 10$^5$~\msun\twospace, may increase the number of candidates for the (U)\lirg class.

\subsubsection{Dynamical Evolution of the TDG Candidates}

The number of candidates in systems undergoing the early phases of the interaction process (i.e., phases I-II and II) is 7, significantly larger than the 2 detected in more evolved systems. Normalized to the total number of systems in early and advanced phases, we obtain 0.5 and 0.13 candidates per system, respectively. Therefore, \tdgs are more likely to be formed during the first phases of the interaction in (U)\lirgs.

To analyze the meaning of this trend we need to consider both the life expectancy of these candidates as well as their detectability during their lifetime. Our result does not mean that the total number of candidates in each phase decreases at all. Although large \hi reservoirs have been found in a few \tdgs (\mycitealt{Duc07}) that would be able of sustaining star formation on a Gyr scale, it is unknown whether these are common. Our candidates are indeed \ha\twospace-selected. We might have been unable to identify candidate objects as old as 20 Myr and older, for which their \ha emission is undetectable. There is a score of knots in the initial sample of 32 galaxies at a projected galactocentric distance larger than 2.5 kpc, with colors (\mbox{\mbi $>$ 1.5}) and luminosities (\mbox{\mi $<$ -15}), such that if the population were about 100-1000 Myr old the stellar mass would be higher than 10$^7$-10$^8$~\msun\twospace, close to or similar to the total stellar mass of observed \tdg candidates.

Bournaud \& Duc (\myciteyear{Bournaud06}) ivestigated a set of 96 N-body simulations of colliding galaxies with various mass ratios and encounter geometries, including gas dynamics and star formation. They investigated the dynamical evolution of the \tdg candidates found in the various simulations up to \mbox{t = 2 Gyr} after the first pericenter of the relative orbit of the two galaxies. On the basis of the comparison with a higher resolution simulation of a major merger by Bournaud et al. (\myciteyear{Bournaud08a}), we can roughly assign different dynamical times after the first pericenter for the different phases of interaction defined in this thesis. In this way, the first phases of the interaction (I-II and III) are likely to occurr during the first 400 Myr of the interaction process. If we consider only the observed candidates in (U)\lirgs during these early phases and these systems for which we have available \ifs data that cover at least 10-15 kpc of galactocentric radius we derive a production rate of about 0.6 candidates per system. Although the simulations of Bournaud \& Duc (\myciteyear{Bournaud06}) continue up to \mbox{t = 2 Gyr}, they fit the number of their \tdg candidates versus time with an exponential decay and a lifetime of 2.5 Gyr. They estimate the number of candidates that survive (long-lived candidates that neither fall back nor lose a large fraction of their mass) after 10 Gyr corresponds to 20\% of the \tdg candidates formed during the early phases of the interaction (i.e., \mbox{t $<$ 500 Myr} after the first pericenter). According to this percentage, we estimate a production rate of about 0.1 (20\% of 0.6) long-lived \tdg candidates per system for the (U)\lirg class. 

The average production rate of 0.1 long-lived \tdgs per system can either become lower or higher depending on follow-up studies with \ifs capable of providing higher spatial resolution (i.e., adaptive optics assisted systems) and covering a \fov that will even allow us to study galactocentric distances greater than 30 kpc. Modeling and observations claim that the most prominent \tdgs are found along the tidal tails and in particular at their tips (e.g~\mycitealt{Hibbard95};~\mycitealt{Duc98};~\mycitealt{Weilbacher00};~\mycitealt{Higdon06};~\mycitealt{Duc07};~\mycitealt{Bournaud08a};~\mycitealt{Hancock09}). Thus, it would be unsurprising that when covering the whole field of the local (U)\lirg systems with \ifs facilities, more \tdg candidates will be found. Moreover, kinematic data of higher spatial resolution would allow us to determine more reliably whether the candidates are bound. 

\subsubsection{Implications for TDG Formation at High-z}

After estimating the production rate of long-lived \tdg candidates per system for the (U)\lirg class, we can now roughly estimate whether satellites of tidal origin are common, and more specifically, the contribution of \tdgs in the early Universe to the dwarf population we see in the local universe. Local (U)\lirgs represent an appropriate class of objects to study the importance of \tdgs at high redshift because: (i) they are major contributors to the star formation rate density at \mbox{z $\sim$ 1-2} (\mycitealt{Perez-Gonzalez05}); and (ii) \ulirgs resemble the sub-millimeter galaxies detected at higher redshifts in the sense that they are merging systems with extremely high rates of star formation (\mycitealt{Chapman03};~\mycitealt{Frayer03};~\mycitealt{Engel10}).

Okazaki \& Taniguchi (\myciteyear{Okazaki00}) studied a scenario in which galaxy interactions and/or merger events act as the dominant formation mechanism of dwarf galaxies in any environment and these interactions occur in the context of the hierarchical structure formation in the Universe. They claimed that all dwarfs in the Universe might have a tidal origin. On the basis of this scenario, their statement assumed a production rate of 1 to 2 tidal dwarfs per merger, which has a lifetime of at least 10 Gyr. The production rate we estimate for long-lived candidates in (U)\lirgs is a factor of 10-20 lower, such that their contribution to the overall dwarf population is only 10-5\%. Therefore, we do not find strong evidence that all the local dwarf galaxy population have a tidal origin. This result is consistent with recent estimates implying that the contribution of \tdgs to the overall dwarf population is insignificant (\mycitealt{Bournaud06};~\mycitealt{Wen11};~\mycitealt{Kaviraj11}).

The study by Okazaki \& Taniguchi (\myciteyear{Okazaki00}) covers a wide range of mass ratios for the interacting systems. In our sample, the systems with more than one nucleus span mass ratios from 1:1.7 to 1:3.6, which is the most favorable range for long-lived \tdgs\twospace, according to the simulations of Bournaud \& Duc (\myciteyear{Bournaud06}). Were high-z interactions to have similar mass ratios, the production rate estimated by Okazaki \& Taniguchi would be higher than the predicted 1-2 tidal dwarf per merger. Thus, our production rate would be a factor of more than 10-20 higher, and the contribution of \tdgs to the overall dwarf population would become even lower than 10-5\%. Rodighiero et al. (\myciteyear{Rodighiero11}) claimed that mergers at \mbox{z $\sim$ 2} have a very low contribution (\mbox{$\sim$ 10\%}) to the cosmic star formation rate density. Were this result to be confirmed, the contribution of \tdgs to the overall dwarf population would still be lower, to a negligible value.

\sect[summary_tdgs]{Summary and Conclusions}

In this Chapter, we have combined high angular resolution \hst images with spectroscopic data from \ifs facilities to characterize \mbox{\ha\twospace-emitting} clumps in an initial sample of 27 (U)\lirgs\twospace. Our study has extended the search of \tdg candidates by Monreal-Ibero et al. (\mycitealt{Monreal07}) by considering both \lirg systems and additional and more sensitive images. In particular, we have detected a total of 31 extranuclear star-forming complexes in 11 (U)\lirgs and characterized the main physical and kinematic properties of these complexes. Using structural, physical, and kinematic information, we have also estimated the stellar and dynamical mass content of the complexes and studied their likelihood of resisting the effects of internal as well as external forces based on different dynamical tracers. With all these parameters, we have identified which complexes have the highest probabilities of being long-lived \tdg candidates in (U)\lirgs\twospace.  We draw the following conclusions:

\begin{enumerate}

\item Located at an average projected distance of 9.3 kpc from the nucleus, within the range 3.3-13.4 kpc, the structures of the complexes are generally simple, consisting of one or a few compact star-forming regions (knots) in the \hst images, though a few of them reside in a richer cluster environment. The complexes have typical  \textit{B} luminosities (\mbox{\mb $<$ -10.65}), sizes (from few hundreds of pc to about 2 kpc), and \textit{B - I} colors (\mbox{\mbi $\lesssim$ 1.0}) that are similar to those observed in giant \hii regions and dwarf-like objects. The relatively high metallicities derived, of normally Z$_{\odot}$-Z$_{\odot}$/3 (independent on the luminosity of the complex), reflect the mixing of the metal content in interacting environments, as observed in a few extragalactic \hii regions and \tdg candidates. \\

\item The measured \ha luminosities of the complexes are comparable to those of extremely luminous \ha complexes in nearby systems, giant extragalactic \hii regions, \tdg candidates, and normal dwarfs. Their sizes seem to be more related to the nature of the system. In our case, many complexes have sizes typical of dwarf-like objects.\\

\item Twenty-two complexes with very simple structures were selected to study their nature as \tdgs\twospace. 
The stellar masses derived for these complexes using the \ha and broad-band luminosities and equivalent widths of a single burst or a composite population (young burst + 1 Gyr) range from a few 10$^4$ to \mbox{10$^8$~\msun\onespace}. Nevertheless, it is typically below \mbox{10$^{6.5}$~\msun\onespace}, which is the lower limit to the total mass. In contrast, the complexes have dynamical masses that are a factor of 50-1000 higher, which provides an upper limit to the total mass. \\

\item A total of 9 complexes, namely \tdg candidates, have the highest probabilities of becoming dwarf galaxies, at least up to 1-2 Gyr. They are probably massive enough and satisfy most of our criteria for self-gravitation (i.e., the position in the radius-$\sigma$ and the luminosity-$\sigma$ planes) and resistance to the forces from the parent galaxy. We have found evidence that their formation takes place more often in early phases of the interaction. \\

\item When we consider only systems for which the \ifs data cover a significant fraction of the whole system, the production rate of candidates averages about 0.3 per system. This rate is expected to decrease to 0.1 for a long-lived 10 Gyr dwarf, according to recent galaxy merger simulations. Were this to be the case, fewer than 5-10\% of the general dwarf satellite population could be of tidal origin. \\

\end{enumerate}

\clearpage{\pagestyle{empty}\cleardoublepage}

\chanonumber{General Conclusions \& Perspectives}
\chaphead{This Chapter gives a general overview of this thesis. We summarize the most relevant conclusions achieved throughout this study and propose new perspectives that this work opens. The most interesting lines of research upon the work carried out in this thesis, together with the possibilities that telescopes and facilities available currently and in the near future will offer, are also outlined.}


This thesis work has explored the physical and dynamical properties of optically-selected compact stellar regions in low-z galaxies with extreme episodes of star formation, Luminous and Ultraluminous Infrared Galaxies. The cornerstone of this project is the use of high angular resolution \hst imaging of a representative sample of (U)\lirgs\twospace. To that end, archival data from the optical cameras \acs and \wfpc were retrieved. For the first time, this kind of analysis has been carried out in an extensive (32) sample of these  systems, covering the entire infrared luminosity range and morphological types (indicative of different phases in the dynamical evolution of the interaction process they are undergoing). These data were combined with optical integral field spectroscopic (\ifs\twospace) data obtained with INTEGRAL and VIMOS spectrographs, mounted on the William Herschel Telescope and on the VLT, respectively. Finally, with the use of a dataset from a numerical simulation of a galaxy encounter, we have also been able to investigate the dynamical evolution of the properties of the compact stellar regions all along the interaction process. This thesis covers three main aspects:

\begin{enumerate}[-]
\item With a sample of 32 representative (U)\lirgs we are able to detect close to 3000 compact star-forming regions (knots), more than one order of magnitude higher than in previous studies in these kind of sytems. The present work thus takes a step forward in charaterizing their properties. We study the dependence of the physical properties of the knots (e.g., luminosities, colors, sizes, etc.) on different parameters such as the infrared luminosity of the system (i.e., a proxy of its global star formation rate) or the interaction phase (i.e., dynamical evolution). Given the importance of (U)\lirgs at \mbox{z $\sim$ 1-2}, we also evaluate the resemblance of these knots with clumpy structures observed at intermediate and high redshifts. \\

\item We also performed a direct comparison with the predicted properties of compact star-forming regions in a high spatial resolution simulation of a galaxy encounter. Since simulated and observed systems cover different phases of the interaction with similar spatial resolutions (30-100 pc), the sample used in this thesis is appropriate for such comparison. This study allows us to test the predictions of the simulations on real data and interpret empirical results obtained with the observations of (U)\lirgs\twospace. \\

\item The search for potential long-lived tidal dwarf galaxy (\tdg\twospace) candidates in local (U)\lirgs is the last issue we investigate in this thesis. We characterize the extranuclear \ha\twospace-emitting complexes and, with the derived properties (metallicity, \ha luminosity, dynamical mass, etc.), we select and evaluate the likelihood of survival of those complexes likely to represent long-lived \tdgs\twospace. This study is relevant to understand the formation of primordial dwarf galaxies at high-z and their contribution to the overall dwarf population.
\end{enumerate}

The most relevant results of this thesis are the following:

\begin{itemize}
 \item The knots in (U)\lirgs span a wide range in observed luminosities (\mbox{-20 $\lesssim$ \mi $\lesssim$ -9} and \mbox{-19.5 $\lesssim$ \mb $\lesssim$ -7.5}). They are in general more luminous than observed star-forming regions in less luminous interacting systems such as the Antennae or M51. In fact, distance effects considered, knots in \ulirgs are intrinsically a factor of 4 more luminous than in \lirgs\twospace. Size-of-sample effects are likely to be the natural explanations, since the star formation rate in \ulirgs is higher than in less luminous systems. \\
 
 \item The knots are in general compact, with a median effective radius of 32 pc, 12\% being unresolved and a few very extended, up to 200-400 pc. Most of them are likely to contain sub-structure and therefore to constitute complexes or aggregates of star clusters. In fact, we find a clear correlation between the knot mass and total radius, M$\propto$R$^{2}$, similar to that found for complexes of star clusters in less luminous interacting galaxies (e.g., M51 and the Antennae) and Galactic and extragalactic giant molecular clouds. \\
 
 \item  Even though (U)\lirgs are known to have most of their star formation hidden by dust, a fraction of 15\% of the optically-selected knots in these systems are extremely blue (\mbox{\mbi$<$ 0.5}) and luminous (\mbox{$<$\mb\twospace$>$ -11.5}). This clearly indicates the presence of a young stellar population (i.e., between a few to about few tens of Myr), almost free of extinction and with masses (\mbox{10$^4$-10$^7$~\msun\twospace}) similar to and up to one order of magnitude higher than the Young Massive Clusters detected in other less luminous interacting systems.\\
 
 \item Knots in \ulirgs can have both sizes and masses characteristic of stellar complexes or clumps detected in clumpy galaxies at high redshifts (\mbox{z $\gtrsim$ 1}), as long as their population is a few dozens of Myr or older. Thus, there is evidence that the larger-scale star formation structures are reminiscent of those seen during the epoch of mass assembly and galaxy formation. \\
 
\item The optically-selected knots span a wide range in color (\mbox{\mbi = [-1,5]}). In contrast, the color distribution of the simulated knots in simulations of major mergers span a narrower range (\mbox{\mbi\onespace= [0.5,2]}). Agreement between the reddish tail (i.e., \mbi$>1$) of the observed and simulated colors is obtained if an exponential probability density function of median value \av\onespace=1.1-1.4 mag (typical obscuration in the range \av\onespace=[0,5] mag) is applied to the simulated colors. On the other hand, in order to explain the bluish part of the broadening of the color distribution (bluish tail, \mbi$<0-1$), the combination of two effects must be taken into account: (1) the age uncertainty in the simulation, which affects significantly the color estimation of the youngest population; (2) the fact that (U)\lirgs constitute systems with high \sfrs (very young population, i.e., blue colors) at all interaction phases, whereas in the simulation the \sfr is only enhanced in the pre-merger phase.\\

\item The star formation in (U)\lirgs is characterized by \textit{B}- and \textit{I}-band luminosity functions (\lf\twospace) with slopes close to 2, extending therefore the universality of the \lf measured in interacting galaxies at least for nearby systems (\mbox{i.e., \ld$<$ 100 Mpc}), regardless of the bolometric luminosity of the systems (i.e., the strength of the global star formation). There are, however, slight indications that the \lf evolves with the interaction phase, becoming steeper (from about 1.5 to 2 for the \textit{I}-band) from early to late phases of interaction. A high linear resolution simulation of a major merger shows some evidence for the variation of the slope of the \lf\twospace with the interaction, which is likely due to higher knot formation rates during the early phases of the interaction with respect to late phases. However, there is no evidence that this variation is correlated with the global star formation rate of the system. The same origin could explain the slight variation of the \lf of the knots in (U)\lirgs provided that the global star formation is concentrated in few knots during intermediate and advanced interaction phases.  \\
  
\item We find theoretical and observational evidence that only the most massive structures remain after the interaction process is completed, although it is still not clear how they form: (i) as individual massive entities; (ii) or after a merging process of clusters into superclusters. \\
  
 \item A total of 31 luminous extranuclear star-forming \ha clumps have been identified in 11 (U)\lirgs\twospace. These clumps, associated to stellar complexes, are usually located along the tidal tails of the systems and do show in general simple structure (i.e., one or a few knots on the \hst images). They have integrated properties (luminosities, colors and sizes) similar to those observed in giant \hii regions and dwarf-like objects, and a relatively high metallicities (Z$_{\odot}$-Z$_{\odot}$/3), which reflects the mixing of the metal content in interacting environments. The measured \ha luminosities of the complexes are comparable with extremely luminous \ha complexes in nearby systems, giant extragalactic \hii regions, \tdg candidates and normal dwarfs.\\
 
 \item Twenty-two of the complexes with very simple structures were selected to study their nature as \tdg\twospace. Applying several different criteria related to the luminosity, mass and empirical evidence that suggests self-gravitation and stability against forces from the parent galaxy, only 9 complexes, namely \tdg candidates, have the highest probabilities to survive as dwarfs. If we consider only systems for which the \ifs covers a high fraction of its optical extent we obtain a rate production of about 0.1  of long-lived \tdg candidates per (U)\lirg system. This result suggests that only a small fraction (\mbox{$<$ 10 \%}, and compatible with a negligible fraction) of the general dwarf satellite population could be of tidal origin.
 
\end{itemize}

All this information has helped us draw a global picture of the unobscured or mildy obscured star formation and evolution of compact structures in systems with extreme star formation. Although some properties of these optically-selected star-forming regions can be similar to those in less luminous interacting environment (e.g., presence of young population in the optical despite the high extinction measured in the innermost regions of some systems, luminosity functions, presence of \tdg candidates), others reflect the peculiarities of (U)\lirg systems. The identification of intrinsically more luminous populations in systems with higher luminosity, the presence of young populations (i.e., blue colors) throughout all the phases of interaction, the evidence that clusters may merge in superclusters efficiently, and the similarity (in terms of mass and size) of the knots to observed structures in high-z clumpy galaxies, show that the extreme environment in (U)\lirgs have some effect on the the star formation and evolution at local scales.

However, this study has to be seen as a starting point, since there is a lot to be done in order to obtain a deeper understanding of all the peculiarities of the star formation and evolution of compact structures in (U)\lirgs\twospace. First, it is of paramount importance to observe these structures at other wavelengths to break the degeneracy between the extinction and the age of the stellar population, as well as to characterize knots with multiple stellar populations and to have a detailed knowledge of the star formation history of the systems. For instance, Lee et al. (\myciteyear{Lee05}) show how photometric observations in the \textit{U}-band breaks the degeneracy in young reddened clusters in M51. Information from high resolution near-infrared (\nir\twospace) photometry which covers the complete \fov of the systems is also important so as to make reliable estimations of the mass of the compact stellar structures. With the use of WFC3 on board the \hst\twospace, we can now deal with these challenges, since this imager covers the \uv\twospace, visible and near-infrared bands at resolutions similar to the optical images used in this study. 

Simulations of major mergers are catching up observations of nearby systems in terms of spatial resolution. The development of these numerical tools that let us understand the environment of interacting galaxies is nowadays growing up very rapidly. Specifically, new hydrodynamic simulations of major mergers by Teyssier et al. (\myciteyear{Teyssier10}), which achieve spatial resolutions up to 12 pc and implement gas cooling below $\sim$10$^4$ K, resolve properly the process of gas fragmentation into massive clouds and rapid star formation. The model used in this thesis has some issues (i.e., gas fragmentation) at small galactocentric distances (i.e., \mbox{d $<$ 5 kpc}) that the new set of models are able to sort out, and with higher spatial resolution. A large number of compact stellar objects extending to lower masses can be identified, thus the comparison with the knots in our systems can be improved significantly. 

The formation, evolution and survival of tidal dwarf galaxies is subject to intense debate. In order to farther investigate the existence and true nature of the candidates we first need to observe in \ha the full extent of the systems, sampling the tidal tails and especially the tips. With a \fov larger than the \acs camera and the possibility to tune the narrow-band \ha filter throughout the redshift range our sample spans, the OSIRIS imager, currently operating on the \acr{}{GTC}{Gran Telescopio de Canarias}, is the ideal instrument to perform this task. In the near-infrared we can also find tracers of recent star formation (i.e., Br$\gamma$, at \mbox{$\sim$ 2.17 $\mu$m}). Thus, once the interesting \ha\twospace-emitting complexes are selected, their kinematic and dynamical properties can be explored with \ifs observations taken with the \acr{}{SINFONI}{Spectrograph for INtegral Field Observations in the Near Infrared}, mounted on the \acr{}{VLT}{Very Large Telescope}. The use of adaptive optics would allow us to achieve an angular resolution of $\sim$ 0.1\arcsec. Since the Br$\gamma$ line is relatively weak, with a flux about 100 times weaker than that of \ha (\mycitealt{Osterbrock89}), only the most luminous candidates (i.e., those with \mbox{\lha~$\gtrsim$ a few $\times$10$^{39}$ erg s$^{-1}$} for the typical distance of the local \lirgs sampled and \mbox{\lha~$\gtrsim$ a few $\times$10$^{40}$ erg s$^{-1}$} for \ulirgs) would be observed. A similar study but using the Pa$\alpha$ line (at 1.87 $\mu$ and less than a factor 10 fainter than the \ha line) can be done for less luminous candidates. However, it is not possible to observe this line from the ground and we will have to wait until the \acr{}{JWST}{\textit{James Web Space Telescope}} is launched. The integral field unit mounted on NIRSpec (a \nir spectrograph) will be able to perform such task. Once the spectroscopic data retrieved, the velocity fields of the candidates would also be resolved and a more detailed study than that presented in this thesis could be performed. 

Information at (sub-)millimetric wavelengths also represents a valuable tool to characterize the structure of the cold molecular gas in (U)\lirgs\twospace. With the \acr{}{ALMA}{Atacama larger millimeter$/$submillimeter array}, planned to be commissioned for observers at the end of 2011, we will be able to reach sub-arcsecond resolution while covering a relatively large \fov\twospace. With this interferometer, the field of search and characterization of \tdgs in (U)\lirgs will be surely revolutionized. Furthermore, the knowledge of the properties of a molecular cloud in (U)\lirgs will provide the necessary  information to estimate the star formation efficiency within the cloud, as done in Bastian et al. (\myciteyear{Bastian05a}) for the nearby M51 by computing an offset between mass-radius relation of \gmcs and complexes of star clusters. 

Obviously, higher spatial resolution is always needed to observe substructure (i.e., individual clusters instead of knots in our case). In fact, (U)\lirgs are yet to be placed in the clustered star formation framework since most of them lie farther away than 100 Mpc and at these distances individual star clusters cannot be distinguished (normally apparent associations are detected). Future large telescopes, as the \acr{}{ELT}{extremely large telescope}, with extremely improved adaptative optics will allow the observation on pc or even sub-pc scales at the distances where our local (U)\lirgs are. Then, individual clusters will be observed and some results obtained in this thesis will be tested (e.g., the universality of the slope of the luminosity function, if clusters in \ulirgs are intrinsically more luminous than in less luminous systems, etc.). However, we may still have to wait for long and of course new questions will arise ...

\chanonumber{Conclusiones y Perspectivas de Futuro}

\chapheadesp{Este Cap\'itulo trata de dar una visi\'on general del trabajo de tesis. Se resumen las conclusiones m\'as importantes de dicho estudio. Asmismo, se exponen las l\'ineas de investigaci\'on que podr\'ian resultar m\'as interesantes en base en lo realizado en esta tesis, junto con las posibilidades que ofrecen los telescopios y las instalaciones operativas actualmente y en el futuro pr\'oximo.}

En esta tesis se han explorado las propiedades f\'isicas y din\'amicas de regiones estelares compactas seleccionadas en el visible en galaxias a bajo z con episodios extremos de formaci\'on estelar, las Galaxias Luminosas y Ultraluminosas en el Infrarrojo, (U)\lirgs (por sus siglas en ingl\'es). La piedra angular de este proyecto radica en el uso de im\'agenes de alta resoluci\'on angular tomadas con el \hst de una muestra representativa de (U)\lirgs\twospace. Para ello se han empleado im\'agenes de archivo tomadas con las c\'amaras \'opticas \acs y \wfpc\twospace. Es la primera vez que este tipo de estudio se ha completado para una muestra extensa (32) de estos sistemas. La muestra cubre por completo el rango de luminosidades en el infrarrojo y diferentes morfolog\'ias, que representan diferentes fases en la evoluci\'on din\'amica del proceso de interacci\'on que se est\'a llevando a cabo. Los datos fotom\'etricos se han combinado con datos de espectroscop\'ia de campo integral (\ifs\twospace, por sus siglas en ingl\'es), tomados con los espectr\'ografos INTEGRAL y VIMOS, instalados en los telescopios William Herschel y VLT, respectivamente. Por \'ultimo, hemos sido capaces de investigar la evoluci\'on din\'amica de las propiedades de las regiones compactas estelares durante el proceso de interacci\'on gracias a los datos obtenidos en una simulaci\'on num\'erica de una colisi\'on de dos galaxias. Los tres objetivos fundamentales que se desarrollan en esta tesis son: 

\begin{enumerate}[-]
\item Hemos sido capaces de detectar cerca de 3000 regiones de formaci\'on estelar compacta (nodos) en una muestra representativa de 32 (U)\lirgs\twospace, m\'as de un orden de magnitud que en estudios previos llevados a cabo con muestras de estos sistemas. Se ha estudiado la dependencia de las propiedades f\'isicas de los nodos (luminosidades, colores, tama\~nos, etc.) con diferentes par\'ametros tales como la luminosidad en el infrarrojo (indicador de la tasa de formaci\'on estelar global) o la fase de interacci\'on (evoluci\'on din\'amica). Puesto que la presencia de los sistemas (U)\lirgs es importante a \mbox{z $\sim$ 1-2}, tambi\'en se ha valorado la semejanza de estos nodos con estructuras aglutinadas que se han observado a desplazamientos al rojo moderados y altos.\\

\item Asimismo, se ha llevado a cabo una comparaci\'on directa con las propiedades de regiones compactas de formaci\'on estelar que se predicen en una simulaci\'on de una interaci\'on de galaxias de alta resoluci\'on espacial. Dado que las observaciones y las simulaciones abarcan las diferentes fases de interacci\'on y las resoluciones espaciales son similares (30-100 pc), la muestra observacional que utilizamos en estas tesis es adecuada para este tipo de comparaciones. Este estudio nos permite comprobar las predicciones de las simulaci\'on con datos reales e interpretar los resultados emp\'iricos que se han obtenido con las observaciones de los sistemas (U)\lirgs\twospace.\\

\item La b\'usqueda de cadidatos a galaxias enanas de marea (\tdg\twospace, por sus siglas en ingl\'es) en (U)\lirgs locales constituye el \'ultimo punto a tratar en esta tesis. Se han caracterizado propiedades tales como la metalicidad, luminosidad en \ha\twospace, masa din\'amica, etc., de los complejos de emisi\'on de \ha extra-nucleares. Con ello se han seleccionado y evaluado la probabilidad de supervivencia de aquellos complejos propensos a constituir \tdgs de larga duraci\'on. Este estudio es relevante para entender la formaci\'on de las galaxias enanas primigenias a alto z y la contribuci\'on de enanas de marea en el conjunto de la poblaci\'on de galaxias enanas. 

\end{enumerate}

Los resultados m\'as notables que se han obtenido en esta tesis son los siguientes:

\begin{itemize}
 
 \item Los nodos detectados en la muestra de (U)\lirgs abarcan un amplio rango de luminosidades observadas (\mbox{-20 $\lesssim$ \mi $\lesssim$ -9} y \mbox{-19.5 $\lesssim$ \mb $\lesssim$ -7.5}). En general, son m\'as luminosos que las regiones de formaci\'on estelar que se han observado en sistemas con menor luminosidad en el infrarrojo, tales como en La Antenna o M51. De hecho, y teniendo en cuenta efectos de distancia, los nodos en \ulirgs son intr\'insicamente cuatro veces m\'as luminosos que los nodos en \lirgs. Efectos debidos al  ``tama\~no de la muestra'' suponen la explicaci\'on m\'as l\'ogica, ya que la tasa de formaci\'on estelar en \ulirgs es mayor que en sistemas menos luminosos.\\
 
 \item Los nodos son compactos, siendo 32 pc el valor mediano del radio efectivo. El tama\~no de un 12\% de los nodos no se puede resolver, mientras que unos pocos se extienden considerablemente, hasta los 200-400 pc. Es muy probable que la mayor\'ia de ellos contengan subescructura y, por tanto, constituyan complejos o agregados de c\'umulos estelares. De hecho, existe una clara correlaci\'on entre la mass de los nodos y su radio total, M$\propto$R$^{2}$, parecida a la que se ha observado para complejos de c\'umulos estelares en galaxias en interacci\'on con menor luminosidad (por ejemplo, en M51 o en La Antena) y para nubes moleculares gigantes gal\'acticas y extragal\'acticas.\\
 
 \item Pese a que se sabe que la mayor parte de la formaci\'on estelar se encuentra escondida por el polvo en los sistemas (U)\lirgs\twospace, aproximadamente un 15\% de los nodos --que se han detectado en el visible-- muestran unos colores azules extremos (\mbox{\mbi$<$ 0.5}) y son muy luminosos (\mbox{$<$\mb$>$ = -11.5}). Estas propiedades indican claramente la presencia de poblaci\'on estelar joven (entre unos pocos Myr y unas pocas decenas de Myr) y con muy poca extinci\'on. Con unas masas entre 10$^4$ y 10$^7$~\msun\twospace, estos nodos de poblaci\'on estelar joven son tan masivos (y en algunos casos hasta un orden de magnitud m\'as masivos) como los superc\'umulos j\'ovenes masivos que se han observado en galaxias en interacci\'on menos luminosas. \\
 
 \item El tama\~no y las masas t\'ipicas de los nodos de formaci\'on estelar en \ulirgs y en estructuras aglutinadas que se han detectado en galaxias a alto redshift (\mbox{z $\gtrsim$ 1}) son muy similares, siempre que dicha poblaci\'on en \ulirgs sea mayor a unas pocas decenas de Myr. Por lo tanto, se han encontrado evidencias de que la formaci\'on de estructuras a alta escala siempre ha sido muy semejante (tanto en el Universo Local como en \'epocas tempranas). \\

 \item Los nodos abarcan un amplio rango de colores (\mbox{\mbi = [-1,5]}). Sin embargo, la distribuci\'on de los colores de los nodos en las simulaciones de interacciones mayores se extiende en un rago muy estrecho (\mbox{\mbi\onespace= [0.5,2]}). Se puede lograr un acuerdo entre la parte roja (i.e., \mbi$>1$) de la distribuci\'on de colores de los nodos simulados y observados si se le aplica una extinci\'on a los colores simulados, tal que siga una funci\'on con una densidad de probabilidad exponencial con un valor mediano  \av\onespace=1.1-1.4 mag (recorre un extincci\'on t\'ipica entre 0 y 5 mag). Por otro lado, se puede explicar el ensanchamiento de la parte azul (cola azul, \mbi$<0-1$) de la distribuci\'on de colores de los nodos observados mediante la combinaci\'on de dos efectos: (1) la incertidumbre en edad en la simulaci\'on, lo cual afecta considerablemente la estimaci\'on del color de la poblaci\'on m\'as joven; (2) el hecho de que los sistemas (U)\lirgs tengan tasas de formaci\'on estelar tan grandes (poblaci\'on muy joven o, lo que es lo mismo, colores azules), mientras que en la simulaci\'on la tasa de formaci\'on estelar s\'olo aumenta significativamente la fase pre-fusi\'on. \\
 
 \item La formaci\'on estelar en (U)\lirgs se caracteriza por una funci\'on de luminosidad (\lf\twospace), obtenida en bandas \textit{B} e \textit{I}, con una pendiente pr\'oxima a 2. Este resultado extiende la universalidad de la \lf en sistemas en interacci\'on con independencia de la luminosidad bolom\'etrica de los sistemas (es decir, la intensidad de la formaci\'on estelar global), almenos para sistemas cercanos (\mbox{\ld$<$ 100 Mpc}). No obstante, se ha encontrado una ligera evidencia de que la funci\'on de luminosidad evoluciona con la fase de interacci\'on, de tal modo que durante el paso de fases tempranas a fases tard\'ias de la interacci\'on la pendiente se vuelve m\'as pronunciada (de 1.5 a 2 para el caso de la banda \textit{I}). El an\'alisis de una simulaci\'on de una interacci\'on mayor de galaxias arroja un resultado similar. Seg\'un la simulaci\'on, ello es debido a que la tasa de formaci\'on de nodos con respecto al total de la poblaci\'on es mayor en fases tempranas con relaci\'on a fases tard\'ias de la interacci\'on. Sin embargo, no hay evidencias claras de que este comportamiento de la pendiente est\'e relacionado con la tasa de formaci\'on global del sistema. Este mismo origen podr\'ia explicar la variaci\'on de la pendiente de la \lf de los nodos observados en los sistemas (U)\lirgs\twospace. Para ello la formaci\'on estelar global se deber\'ia concentrar en pocos nodos durante las fases intermedias y avanzadas de la interacci\'on.\\

 \item Se encuentran evidencias te\'oricas y observacionales de que \'unicamente las estructuras m\'as masivas sobreviven cuando el proceso de interacci\'on se ha completado, aunque todav\'ia no est\'a claro c\'omo se forman: (i) entidades masivas individuales; (ii) o tras un proceso de interacci\'on y fusi\'on de c\'umulos en superc\'umulos.\\

 \item Se han identificado 31 regiones de formaci\'on estelar, emisores en \ha y extranucleares en un total de 11 (U)\lirgs\twospace. Se pueden asociar complejos de formaci\'on estelar a dichas regiones, las cuales se detectan generalmente a lo largo de las colas de marea de los sistemas y cuya estructura es normalmente bastante simple (uno o dos nodos en las im\'agenes del \hst\twospace). Las propiedades integradas (luminosidades, colores y tama\~nos) de estas regiones/complejos  son semejantes a las que se han obtenido en regiones \hii gigantes y objetos tipo galaxia enana. Asimismo, su metalicidad es relativamente alta (Z$_{\odot}$-Z$_{\odot}$/3), lo cual refleja la mezcla de metales en interacciones. Las luminosidades en \ha de estas regiones son comparables con las de complejos extremadamente luminosos en sistemas cercanos, las de regiones \hii gigantes extragal\'acicas, candidatos a \tdg y galaxias enanas normales. \\
  
 \item Se ha realizado un estudio acerca de la naturaleza de 22 complejos con una estructura muy simple, como posibles candidatos a galaxias enanas de marea. Tras aplicar varios criterios relacionados con la luminosidad, masa y evidencia emp\'irica que sugiere autogravitaci\'on y estabilidad frente a las fuerzas de la galaxia progenitora, 9 complejos se han identificado como candidatos a \tdg\twospace. La probabilidad de que estos candidatos sobrevivan como galaxias enanas son mayores que el resto de complejos. Si nos basamos en aquellos sistemas cuyo campo de visi\'on de los datos de espectroscop\'ia integral cubre casi \'integramente, la tasa de producci\'on de candidatos a \tdg por sistema se situa alrededor de 0.1. Este resultado sugiere que \'unicamente una peque\~na fracci\'on (\mbox{$<$ 10 \%}, e incluso despreciable) de la poblaci\'on de galaxias enanas sat\'elite se puede haber formado como una \tdg\twospace.
 
\end{itemize}

Toda esta informaci\'on nos ha ayudado a obtener una idea global de la formaci\'on estelar y evoluci\'on de estructuras compactas poco oscurecidas o sin extinci\'on en sistemas con formaci\'on estelar extrema. Algunas de las propiedades de estas regiones de formaci\'on estelar son muy parecidas a las que se observan en interacciones con menor luminosidad: la presencia de poblaci\'on joven en el visible a pesar de las altas extinciones que se miden en las zonas internas de algunos sistemas, las funciones de luminosidad, la presencia de candidatos a \tdg\twospace, etc. Sin embargo, otras propiedades reflejan las peculiaridades de los sistemas (U)\lirgs\twospace, tales como: la identificaci\'on de una poblaci\'on intr\'insecamente m\'as luminosa en galaxias con mayor luminosidad en el infrarrojo, la presencia de poblaci\'on joven (colores azules) durante todas las fases de interaci\'on, la evidencia de que los c\'umulos puedan interaccionar entre ellos de un modo eficiente y fusionarse formando superc\'umulos, y la semejanza (en t\'erminos de masa y tama\~no) entre los nodos en (U)\lirgs y las estructuras que se han observado galaxias de formaci\'on estelar a alto z. Estas peculiaridades demuestran que el entorno extremo en (U)\lirgs se plasma de alg\'un modo en la formaci\'on y evoluci\'on estelar a escala local. 

No obstante, este trabajo se tiene que considerar como un punto de partida para lograr entender en profundidad todas las peculiaridades de la formaci\'on y evoluci\'on estelar de estructuras compactas en (U)\lirgs\twospace, pues todav\'ia queda mucho por hacer. En primer lugar, es imprescindible llevar a cabo observaciones de dichas estructuras en otras longitudes de onda, para as\'i ser capaces de romper la degeneraci\'on entre la extinci\'on y la edad de la poblaci\'on estelar. Estas observaciones tambi\'en nos permitir\'ian caracterizar poblaciones estelares m\'ultiples en los nodos y obtener un conocimiento m\'as detallado de la historia de formaci\'on estelar de los sistemas. A modo de ejemplo, Lee et al. (\myciteyear{Lee05}) demuestra que con observaciones en la banda \textit{U} es capaz de romper la degeneraci\'on de c\'umulos j\'ovenes oscurecidos en la galaxia M51. Asimismo, la fotometr\'ia de im\'agenes de alta resoluci\'on en el infrarrojo cercano (\nir\twospace, por sus siglas en ingl\'es) que abarquen toda la extensi\'on de los sistemas es necesaria a la hora de realizar estimaciones fidedignas de la masa de structuras estelares compactas. Hoy en d\'ia, todos estos retos se pueden llevar a cabo con la c\'amara WFC3, a bordo del \hst\twospace. Con una resoluci\'on similar a las im\'agenes utilizadas en esta tesis, la c\'amara WFC3 cubre los rango ultravioleta, visible y \nir\twospace.

Las resoluciones espaciales que alcanzan las simulaciones de interacciones mayores de galaxias se acercan cada vez m\'as a las resoluciones de las observaciones. El desarrollo de estas herramientas num\'ericas, que son de gran utilidad para entender el entorno de interacciones de galaxias, sigue un ritmo fren\'etico. En concreto, las nuevas simulaciones hidrodin\'amicas desarrolladas por Teyssier et al. (\myciteyear{Teyssier10}) ya alcanzan resoluciones espaciales alrededor de los 12 pc y, a su vez, implementan el enfriamento del gas por debajo de $\sim$10$^4$ K. Estas nuevas simulaciones resuelven de un modo m\'as fehaciente el proceso de fragmentaci\'on del gas en nubes masivas y subsiguiente formaci\'on estelar r\'apida. El modelo num\'erico que se ha utilizado en esta tesis tiene problemas para resolver correctamente la fragmentaci\'on del gas a distancias galactoc\'entricas peque\~nas (\mbox{d $<$ 5 kpc}). El nuevo set de modelos es capaz de solventar estos problemas, y a la vez, alcanzar mayor resoluci\'on espacial. Con ello se podr\'an identificar un n\'umero mayor de objetos estelares compactos y con mayor resoluci\'on en masa. Por lo tanto, se podr\'a mejorar el estudio comparativo con los nodos estelares observados en (U)\lirgs\twospace. 

En la actualidad persiste un debate muy intenso en torno a la formaci\'on, evoluci\'on y supervivencia de galaxias enanas de marea. Es imprescindible realizar observaciones que sean capaces de cubrir la extensi\'on completa de los sistemas en interacci\'on a la hora de investigar m\'as en detalle la existencia y naturaleza real de candidatos.  De este modo, se podr\'an muestrear las colas de marea y, en especial, los extremos. Con un campo de visi\'on mayor al de la c\'amara \acs y la posibilidad de sintonizar el filtro de banda estrecha \ha a lo largo de los diferentes desplazamientos al rojo que la muestra de esta tesis recorre, la c\'amara OSIRIS, operativa en el Gran Telescopio de Canarias, supone el instrumento ideal para realizar tal tarea. En el infrarrjo cercano tambi\'en podemos encontrar trazadores de formaci\'on estelar reciente, tales como la l\'inea de recombinaci\'on Br$\gamma$, a \mbox{$\sim$ 2.17 $\mu$m}. As\'i pues, una vez se seleccionen los complejos emisores en \ha de inter\'es utilizando OSIRIS, se pueden ivestigar sus propiedades cinem\'aticas y din\'amicas con observaciones de espectroscop\'ia de campo integral con SINFONI, que se encuentra instalado en el VLT. Con el uso de t\'ecnicas de \'optica adaptativa se pueden alacanzar resoluciones angulares del orden de 0.1\arcsec. Debido a que la l\'inea de emisi\'on Br$\gamma$ es relativamente d\'ebil en comparaci\'on con \ha (aproximadamente 100 m\'as d\'ebil;~\mycitealt{Osterbrock89}), \'unicamente se podr\'an detectar las candidatas m\'as luminosas con: \lha~$\gtrsim$ pocos $\times$10$^{39}$ erg s$^{-1}$ a las distancias t\'ipicas de las \lirgs locales muestreadas; y 
\lha~$\gtrsim$ pocos $\times$10$^{40}$ erg s$^{-1}$ a las distancias t\'ipicas de las \ulirgs locales muestreadas. Se puede realizar un estudio similar para candidatas menos luminosas utilizando la l\'inea de emisi\'on Pa$\alpha$ (a \mbox{1.87 $\mu$} y menos de un factor 10 m\'as d\'ebil que \ha\twospace). Sin embargo, esta l\'inea no se puede observar desde tierra y tendremos que esperar hasta el lanzamiento del telescopio espacial James Web  (JWST). Dicha tarea se podr\'a realizar con la unidad integral instalada en el espectr\'ografo en el infrarrojo cercano NIRSpec. Una vez se tengan los datos espectrosc\'opicos, se podr\'a resolver el campo de velocidades de las candidatas y se podr\'a realizar un estudio m\'as detallado que el que se presenta en esta tesis. 

Observaciones en longitudes de onda (sub-)milim\'etricas permiten caracterizar la estructura del gas molecular fr\'io en (U)\lirgs\twospace. El interfer\'ometro ALMA, cuyo servicio a la comunidad cient\'ifica se iniciar\'a a finales de 2011, ser\'a capaz de alcanzar resoluciones por debajo del segundo de arco y cubriendo al mismo tiempo un campo de visi\'on relativamente grande. Seguramente este instrumento supondr\'a un avance revolucionario en la investigaci\'on en la b\'usqueda y caracterizaci\'on de \tdgs en (U)\lirgs\twospace. Asimismo, el conocimiento de las propiedades de las nubes moleculares en (U)\lirgs proporcionar\'a la informaci\'on necesaria para estimar la eficiencia de la formaci\'on estelar, como en Bastian et al. (\myciteyear{Bastian05a}). Bastian y colaboradores estiman dicha eficiencia para complejos estelares en M51, calculado el offset entre la relaci\'on de masa-tama\~no para las nubes moleculares y para complejos de c\'umulos estelares.  

Obviamente, simpre se necesita mayor resoluci\'on espacial para observar subestructura (c\'umulos individuales en lugar de nodos en nuestro caso). De hecho, todav\'ia no se sabe qu\'e lugar ocupan los sistemas (U)\lirgs en el marco de formaci\'on estelar de c\'umulos, ya que la mayor\'ia de estos sistemas se encuentran a distancias m\'as lejanas a 100 Mpc. Hoy en d\'ia, a estas distancias no se pueden distinguir c\'umulos indiviuales, sino que m\'as bien lo que se suele detectar es agrupaciones aparentes de c\'umulos. Con telescopios de gran tama\~no, tales como el ELT, que posean un sistema de \'optica adaptativa considerablemente mejorada, se podr\'an llevar a cabo observaciones a escalas del pc (e incluso por debajo) de sistemas que se encuentren a la misma distancia que las (U)\lirgs de este estudio. En tal caso, se podr\'an observar c\'umulos individuales y algunos de los resultados que se han obtenido en este trabajo de tesis (la universalidad de la pendiente de la funci\'on de luminosidad, si los c\'umulos en \ulirgs son intr\'insicamente m\'as luminosos que en sistemas con menor luminosidad, etc. ) se pondr\'an a prueba. No obstante, puede que tengamos que esperar todav\'ia largo tiempo y nuevas preguntas surgir\'an ... 

\clearpage{\pagestyle{empty}\cleardoublepage}

\appendix\def\mychapname{Appendix } 
\cha{completeness}{Completeness Tests}
\chaphead{This Appendix describes the technique used to estimate the incompleteness of our photometric data.}

We have run a series of artificial star tests in order to assess the completeness and measurement quality of the photometric data set. Some considerations were taken before applying the tests:
\begin{enumerate}[-]
 \item Since each galaxy has a different local background level we had to run a test per \hst image. The same happens if we consider different filters. \\
 \item In the central regions of the galaxies the local background level is higher and steeper than in the outer regions. Hence, different completeness levels are expected as the galactocentric distance increases. To that end we defined two surface brightness regions based on the median value across the system: the inner region, corresponding to high surface brightness levels, and the outer region, corresponding to low surface brightness levels. Using this approach, we ran completeness tests separately for each region and afterwards we weighted each computed value by the number of knots within each region. Some works have used more varying levels of background (e.g., 7 in~\mycitealt{Whitmore99}, 3 in~\mycitealt{Haas08}), but given the angular size of most of the systems in this study it is not worth defining such number of levels in this study. \\
 \item  We have also taken into account the size of the knots, since most of them are actually slightly resolved. To include this effect we also ran two tests separately. First, we added artificial stars with a \psf computed using foreground stars. Additionally, we artificially broadened that \psf by convolving it with a Gaussian that gives a \mbox{\fwhm $\sim$ 1.5 $\times$ \fwhm} of the initial \psf\twospace. We have then taken the average value of the completeness limit in each case. 
\end{enumerate}

The tests were run by adding artificial stars from apparent magnitudes of 20 to 30, in bins of 0.2 mag. We followed several steps to run a completeness for each of the 50 magnitude bins defined. We first added 1000 artificial stars per bin in a random position inside a field that covers the entire system. Poisson noise was added and only stars which were more than 10 pixels apart were selected to avoid contamination between them. Next, we determined in which of the two defined regions with different surface brightness (inner and outer) each artificial star was located. Finally, artificial stars were recovered by per\-for\-ming the photometry in the same way as we did for the knots.  

\begin{figure}
 \hypertarget{fig:completeness}{}\hypertarget{autolof:\theautolof}{}\addtocounter{autolof}{1}
   \centering
   
   \vspace{2cm}\includegraphics[width=0.27\textwidth]{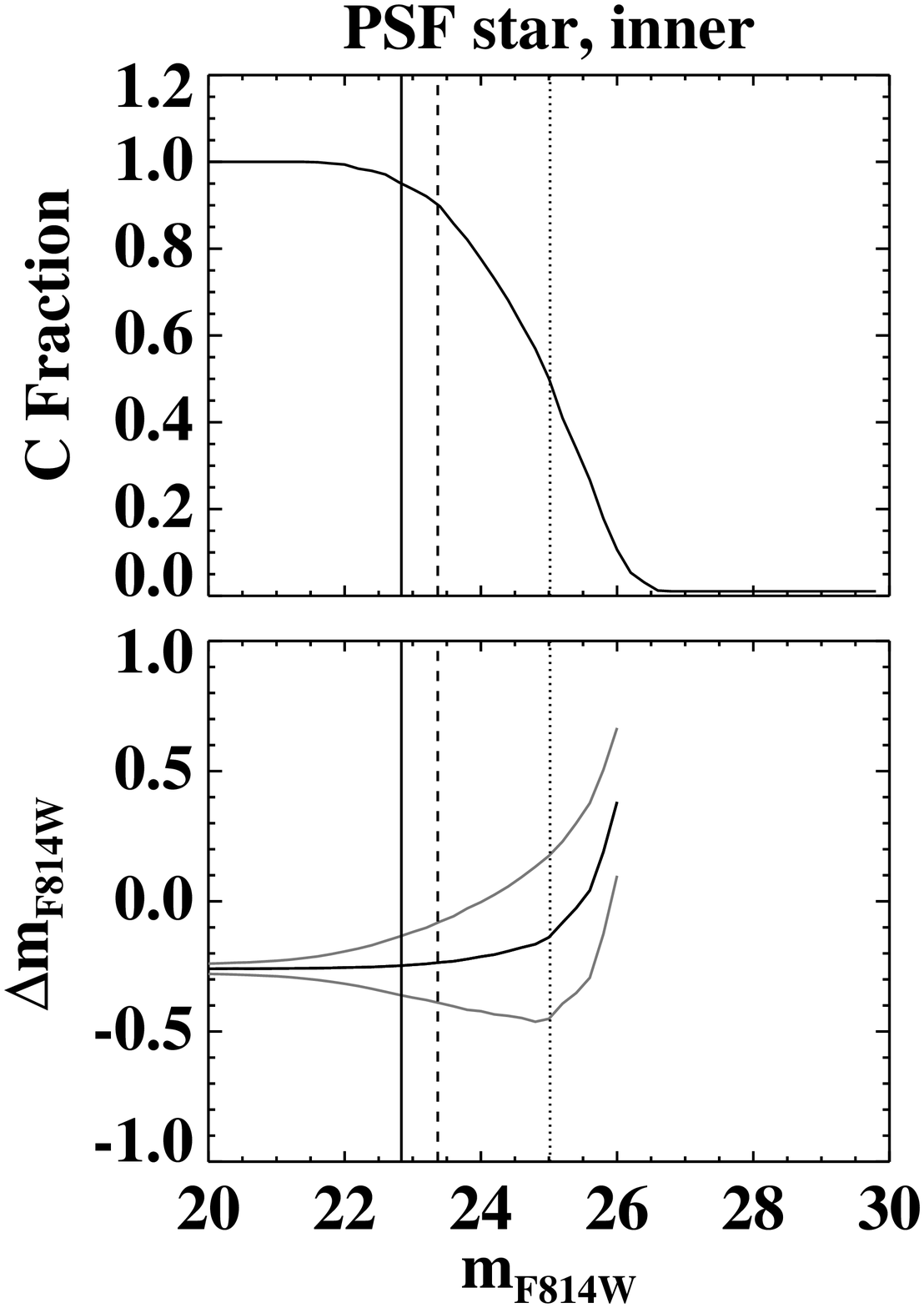}
    \hspace{-0.9cm}
   \includegraphics[width=0.27\textwidth]{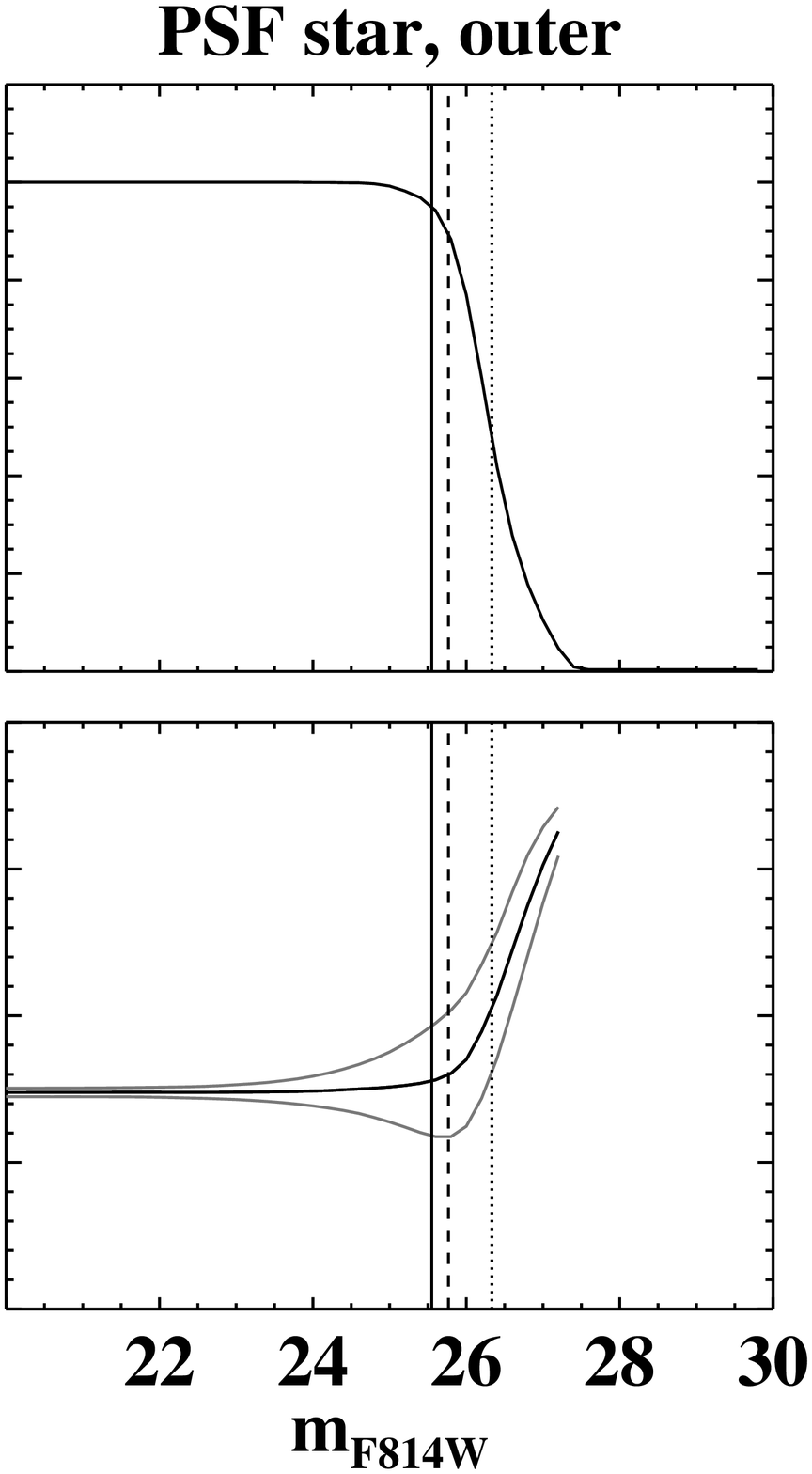}
    \hspace{-0.9cm}
   \includegraphics[width=0.27\textwidth]{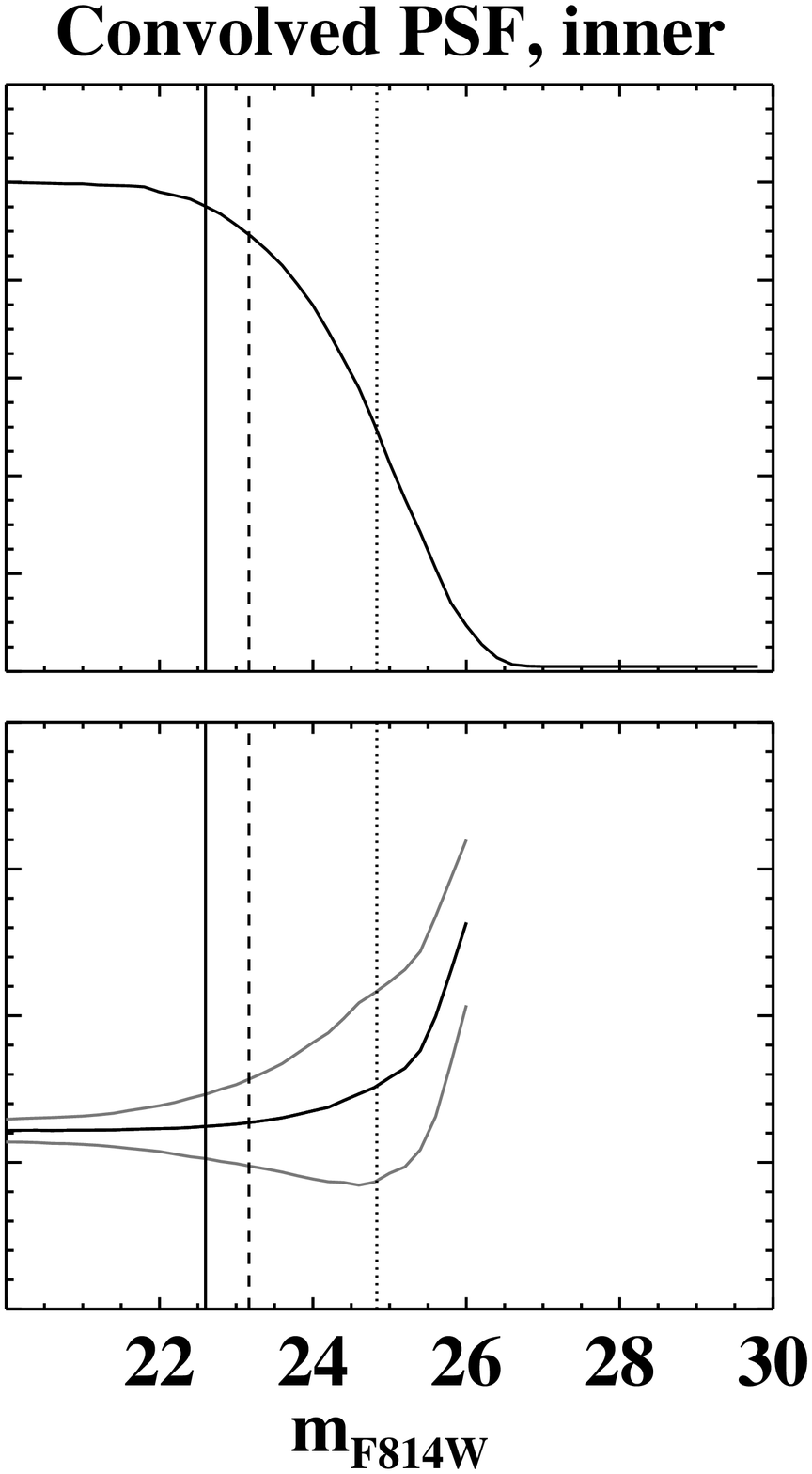}
    \hspace{-0.9cm}
   \includegraphics[width=0.27\textwidth]{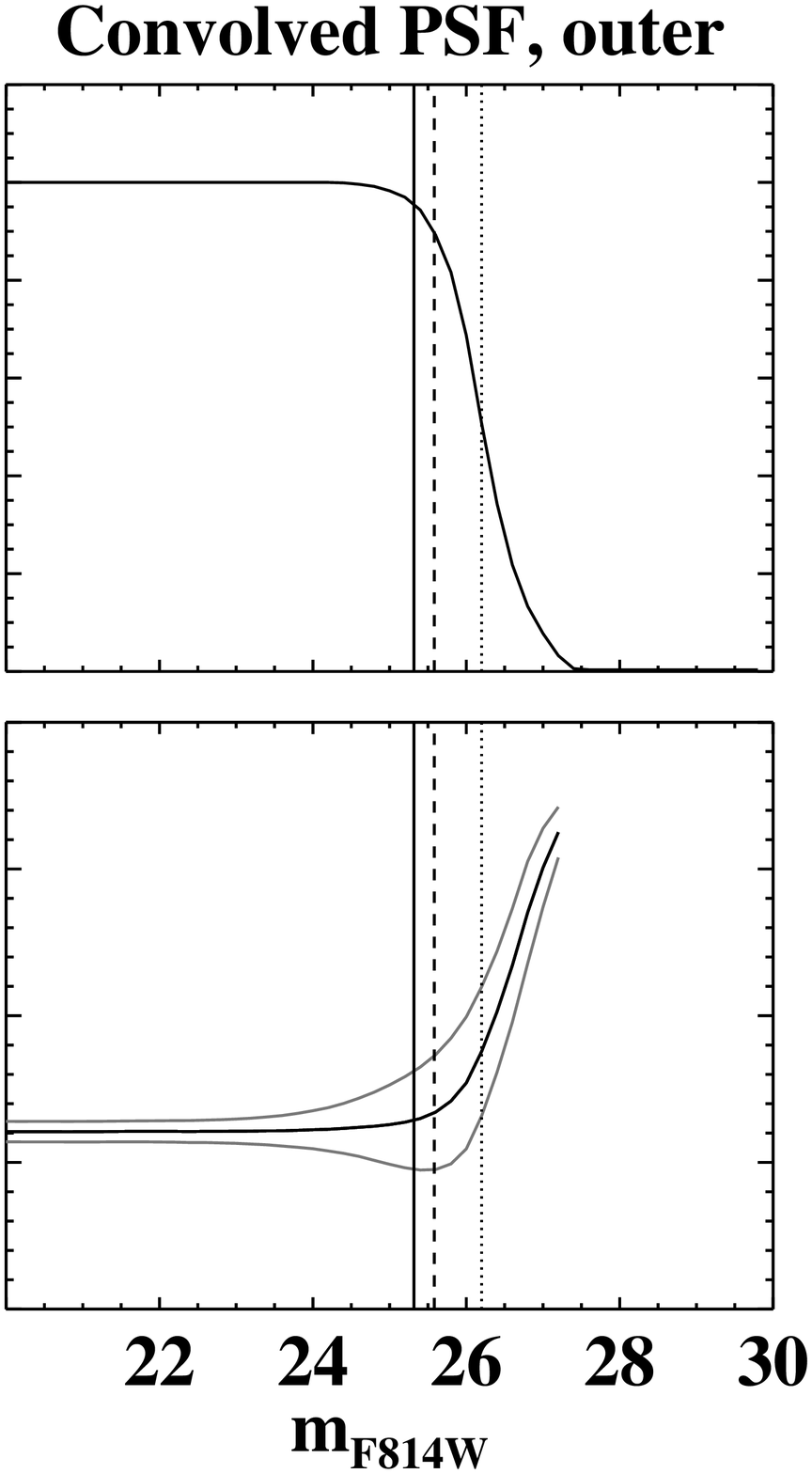}
 
   \vspace{1.5cm}\includegraphics[width=0.27\textwidth]{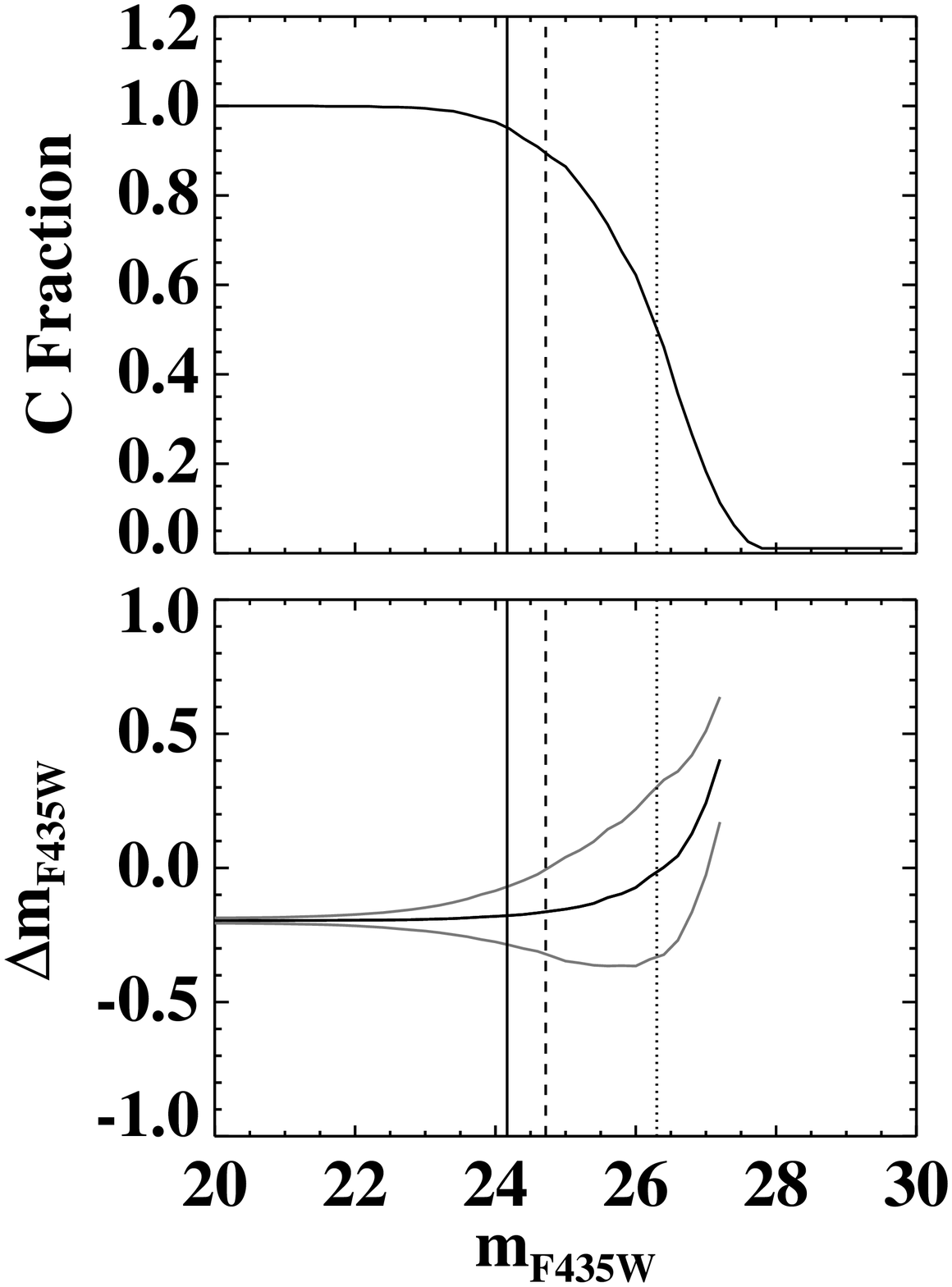}
    \hspace{-0.9cm}
   \includegraphics[width=0.27\textwidth]{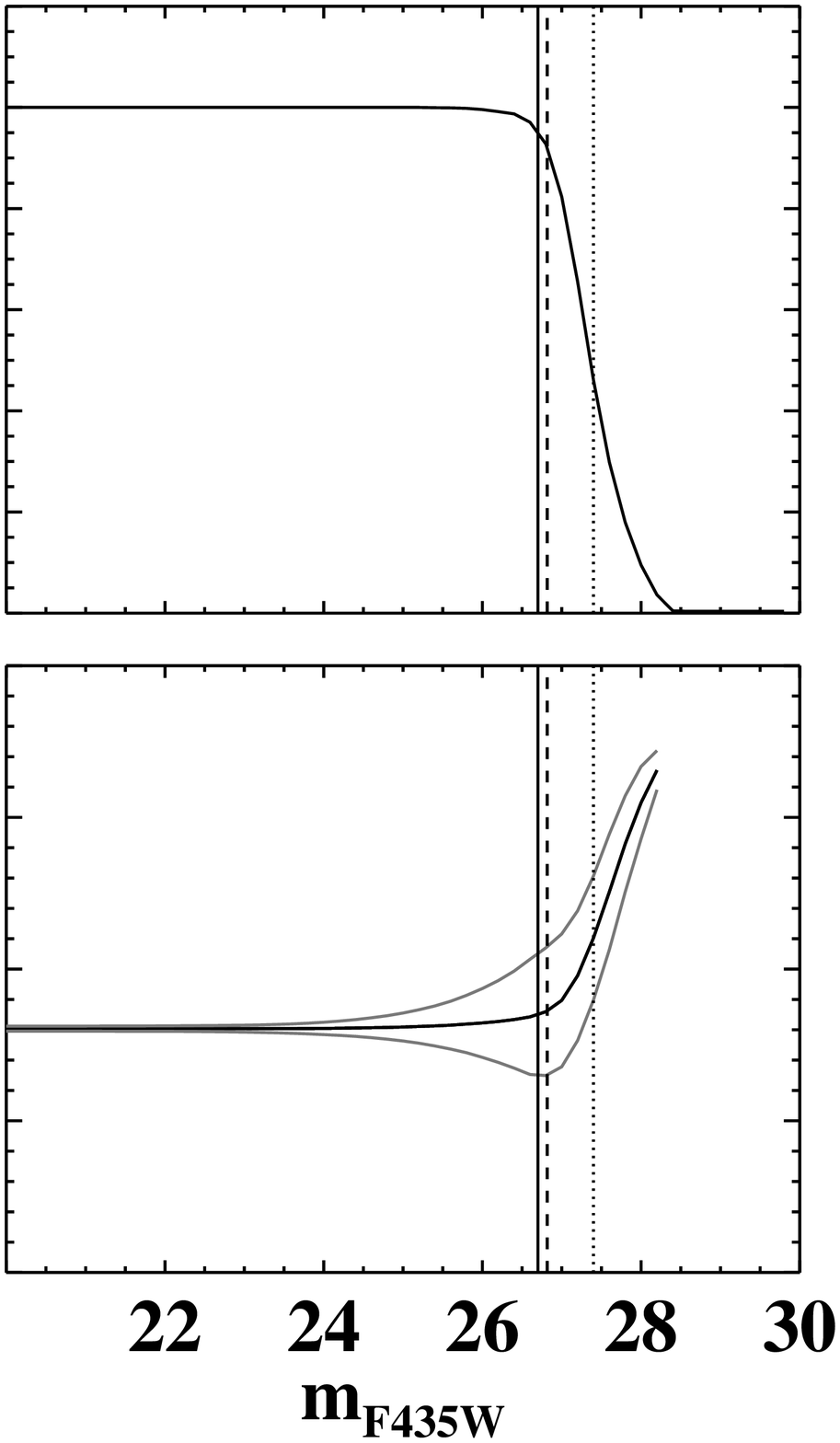}
    \hspace{-0.9cm}
   \includegraphics[width=0.27\textwidth]{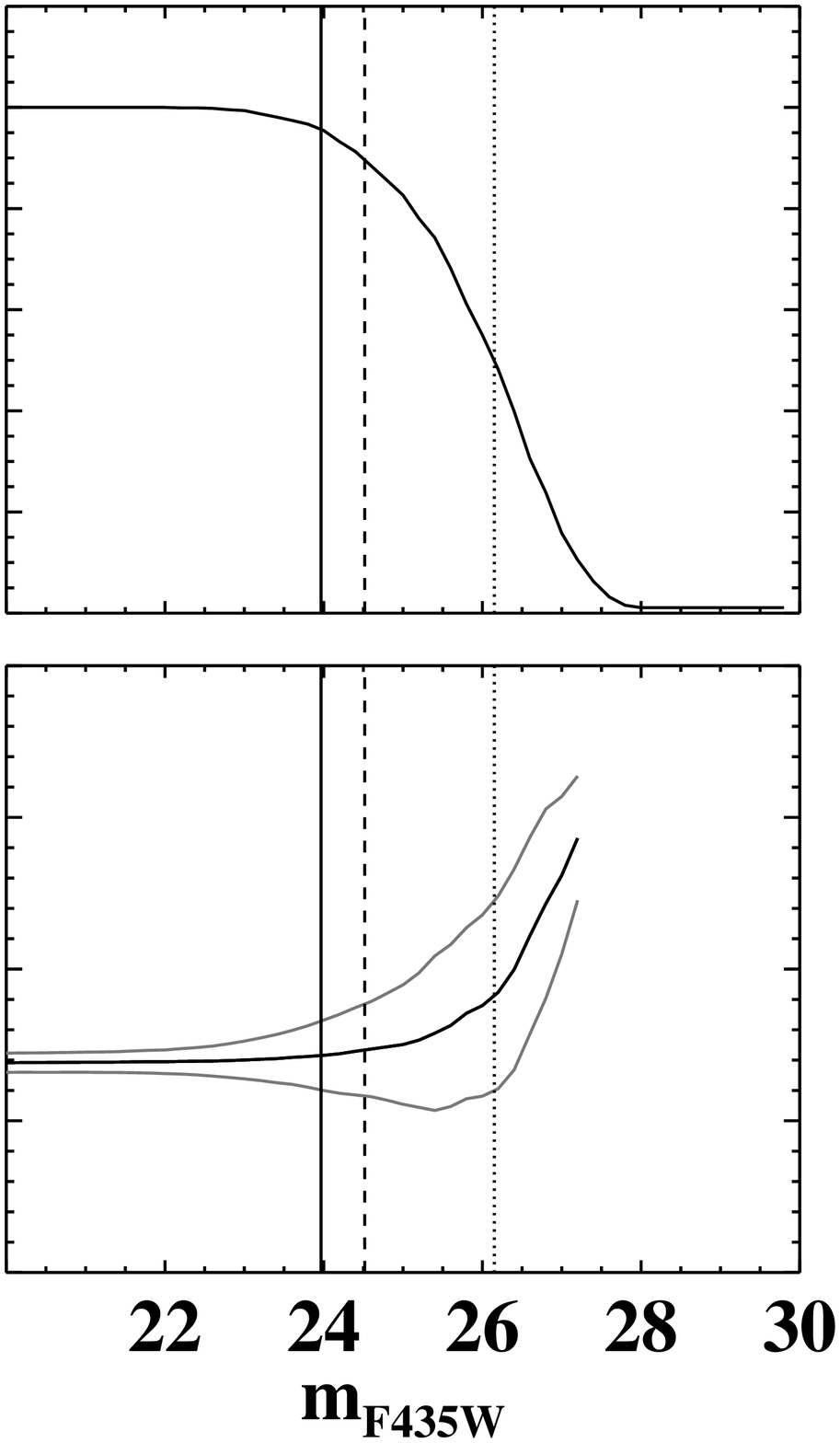}
    \hspace{-0.9cm}
   \includegraphics[width=0.27\textwidth]{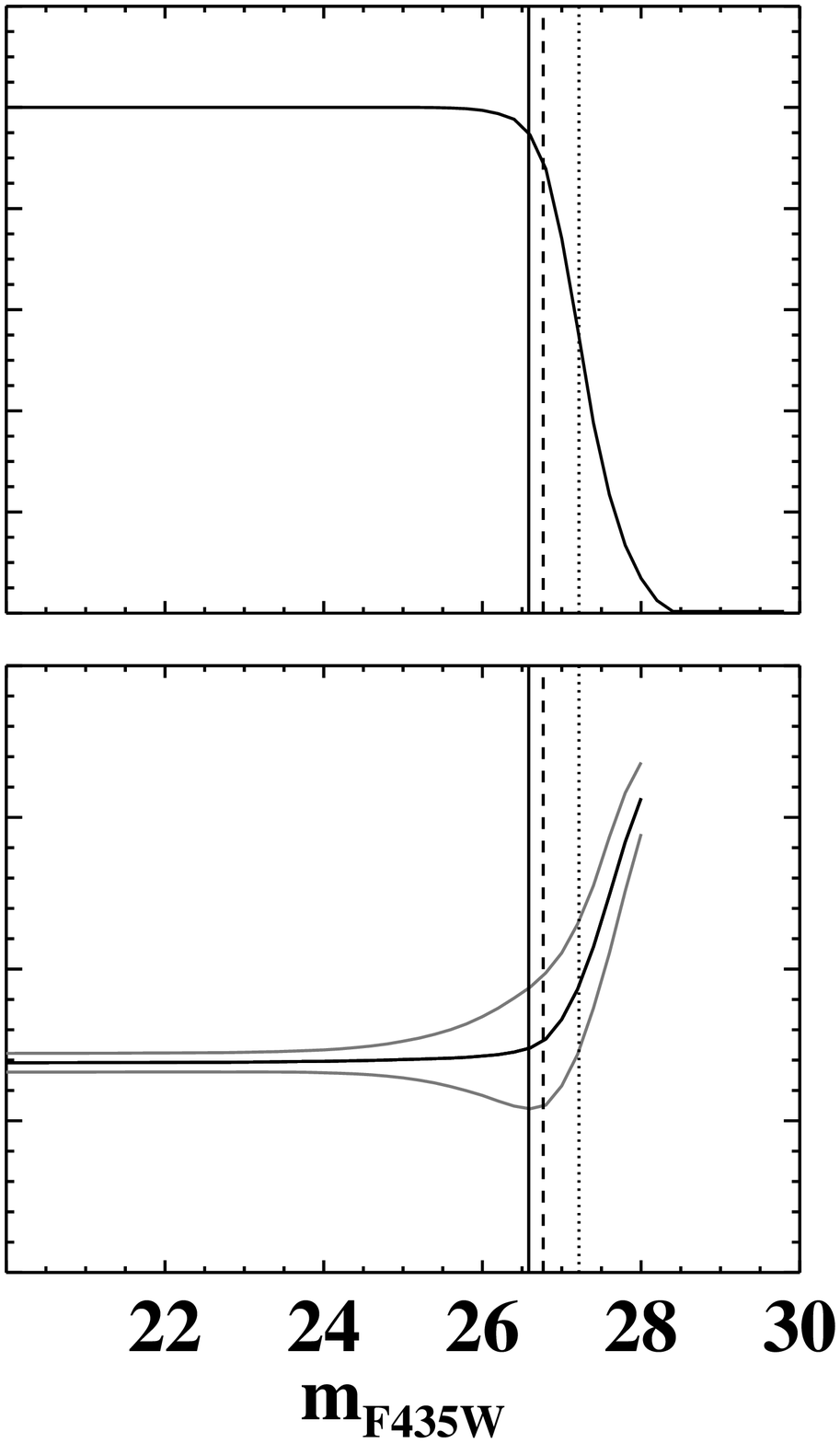}
 
   \caption[Result of the completeness test for IRAS 04315-0840]{Result of the completeness test for IRAS 04315-0840. In each case, from left to right, the curve is shown for the inner region using the \psf of foreground stars, the outer region with the same \psf, and then the same using the convolved \psf\twospace. From top to bottom, completeness fraction for filter \textit{F814W}, input-output of the recovered stars in black and 1$\sigma$ value added and subtracted in gray are shown, and the same for filter \textit{F435W}. The solid vertical line indicates a completeness level of 95\%, the dashed line a level of 90\% and the pointed one a level of 50\%.}
    \label{fig:completeness}          
 \end{figure}

\begin{figure}
 \hypertarget{fig:completeness2}{}\hypertarget{autolof:\theautolof}{}\addtocounter{autolof}{1}
   \centering
   
   \vspace{2cm}\includegraphics[width=0.27\textwidth]{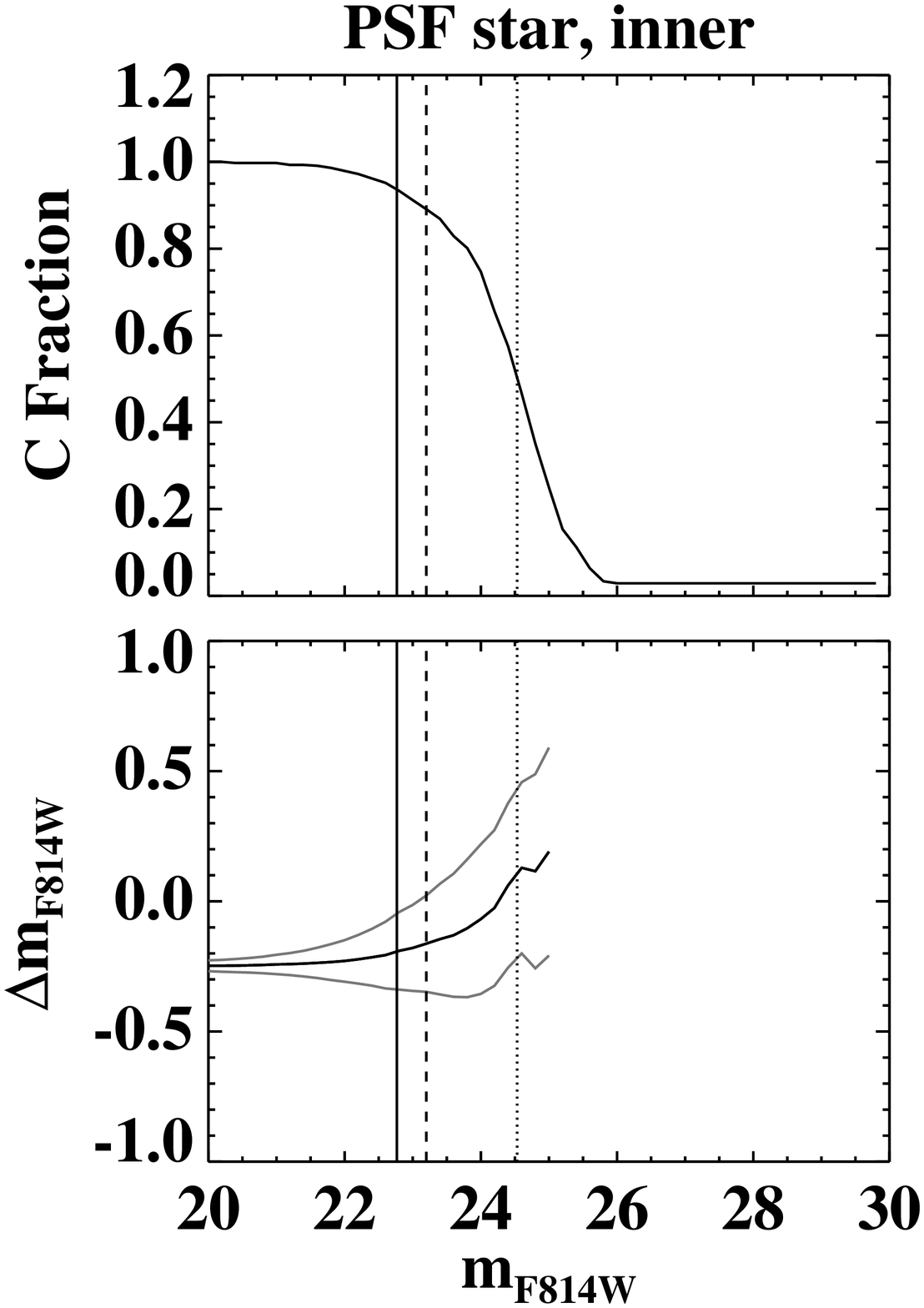}
    \hspace{-0.9cm}
   \includegraphics[width=0.27\textwidth]{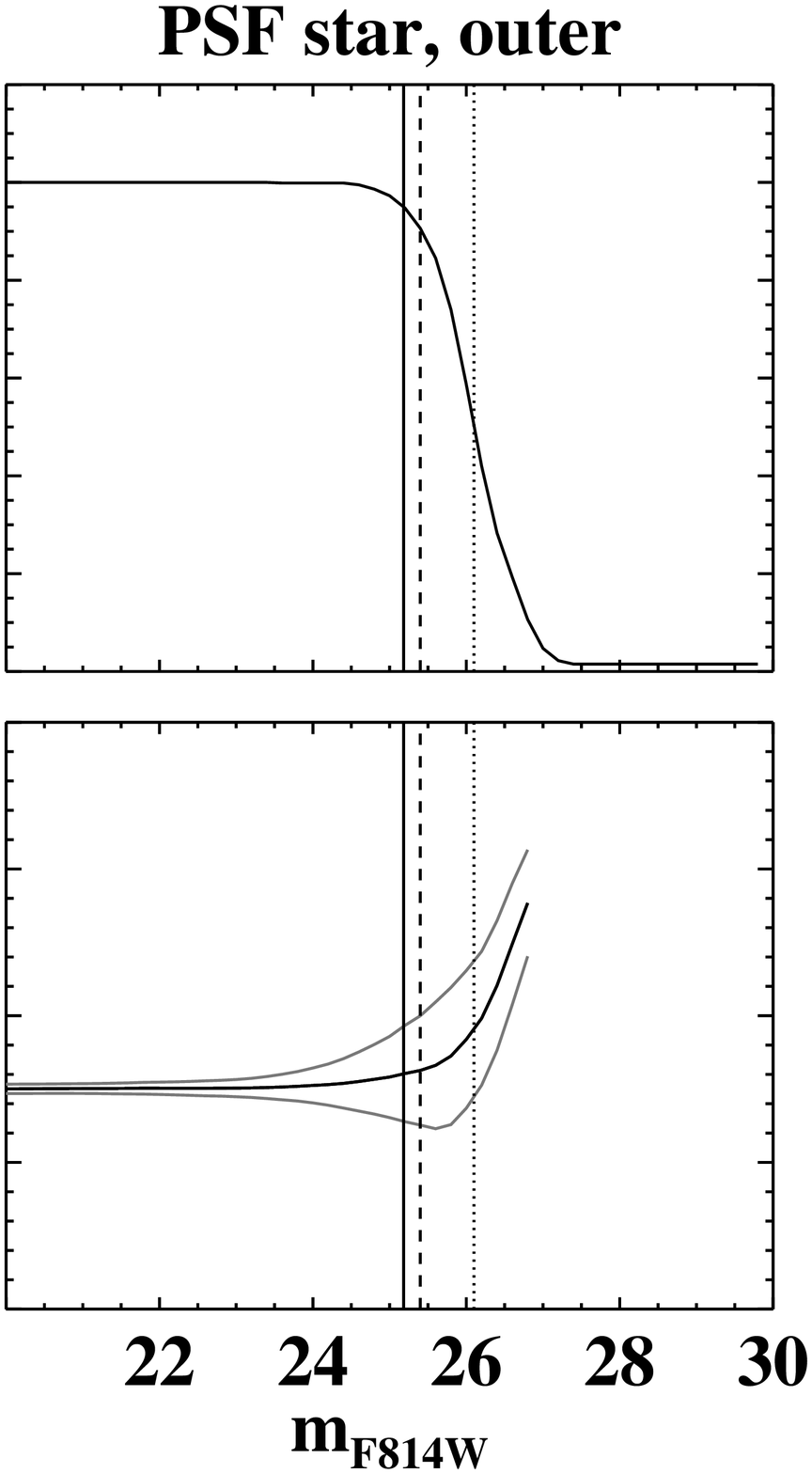}
    \hspace{-0.9cm}
   \includegraphics[width=0.27\textwidth]{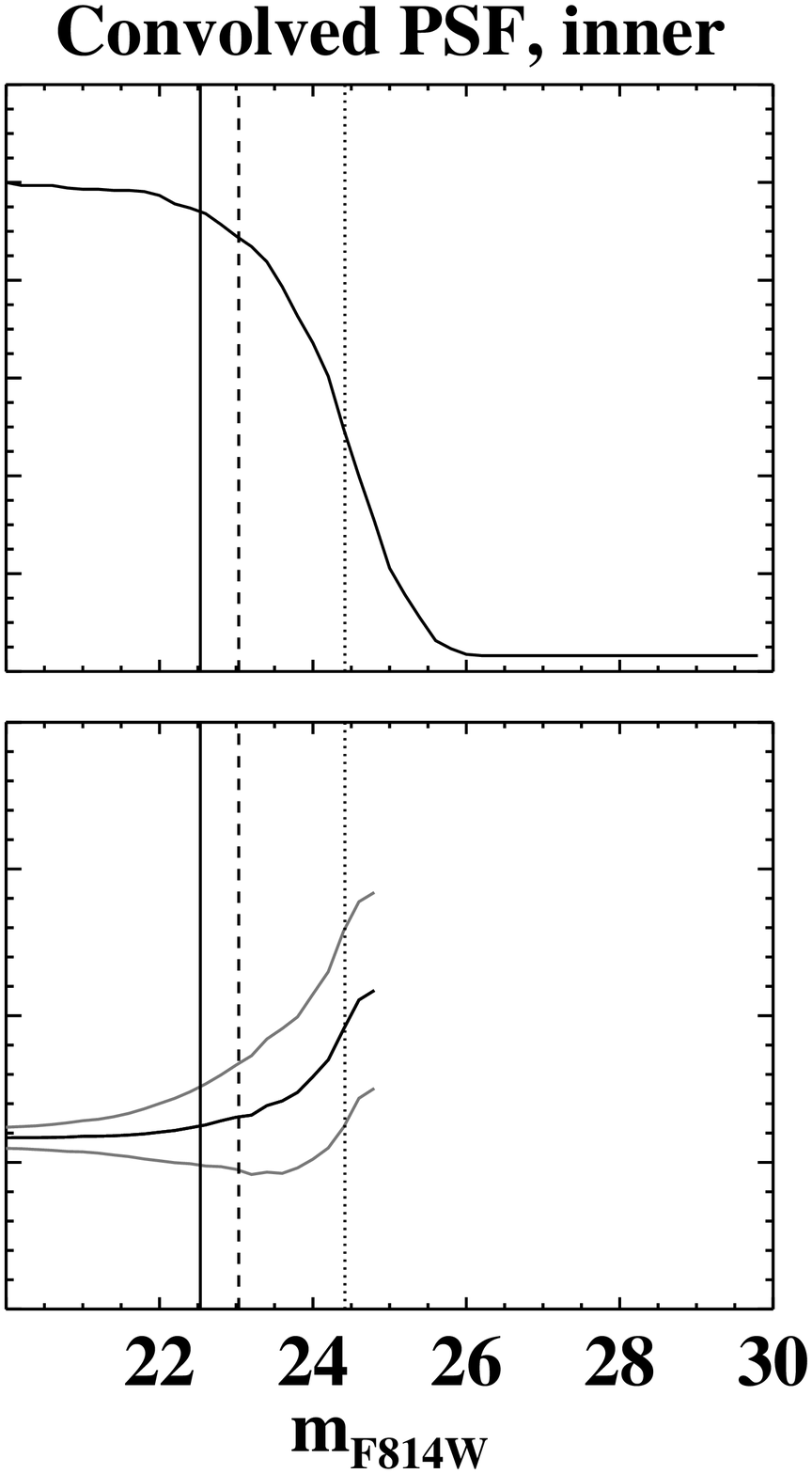}
    \hspace{-0.9cm}
   \includegraphics[width=0.27\textwidth]{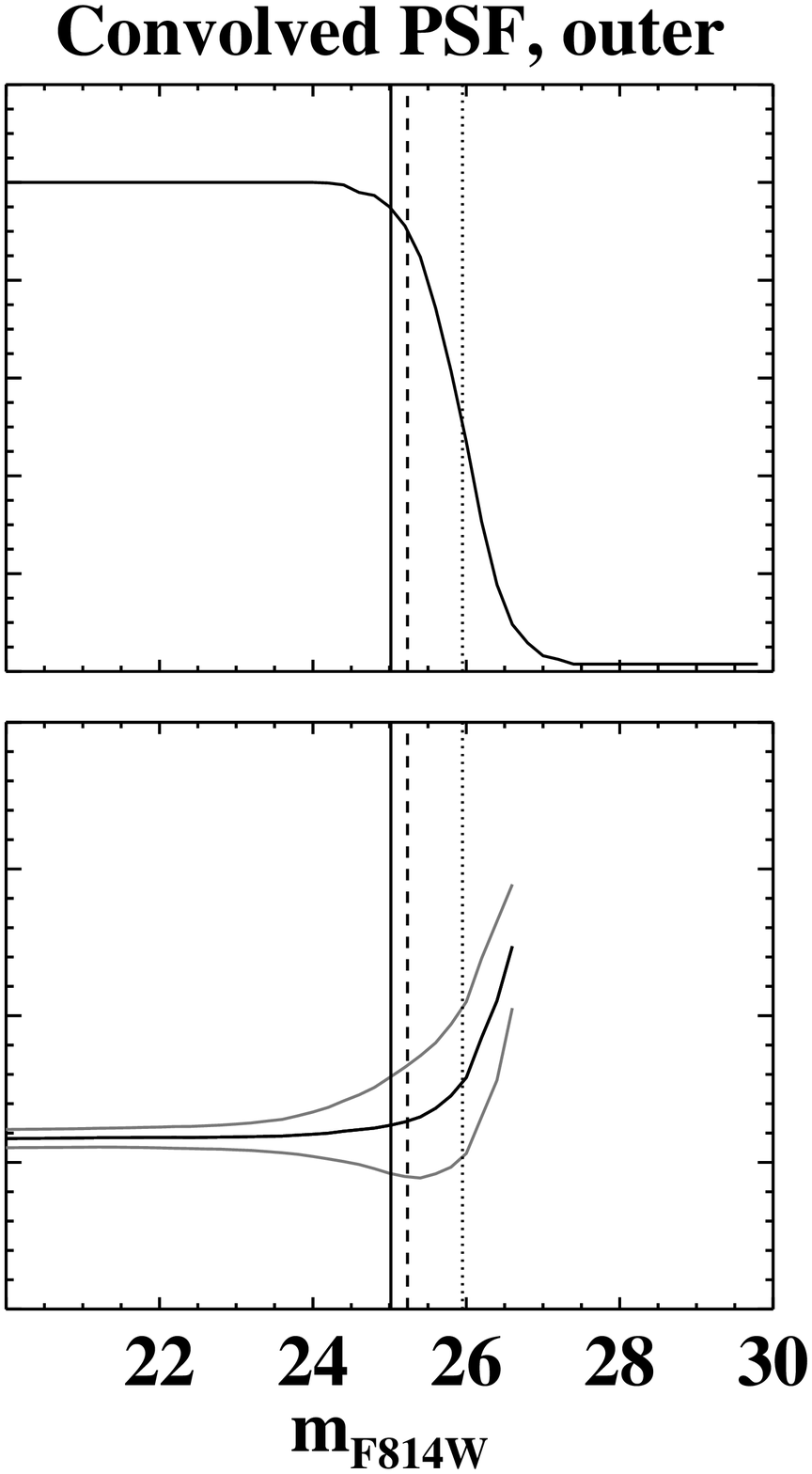}
 
   \vspace{1.5cm}\includegraphics[width=0.27\textwidth]{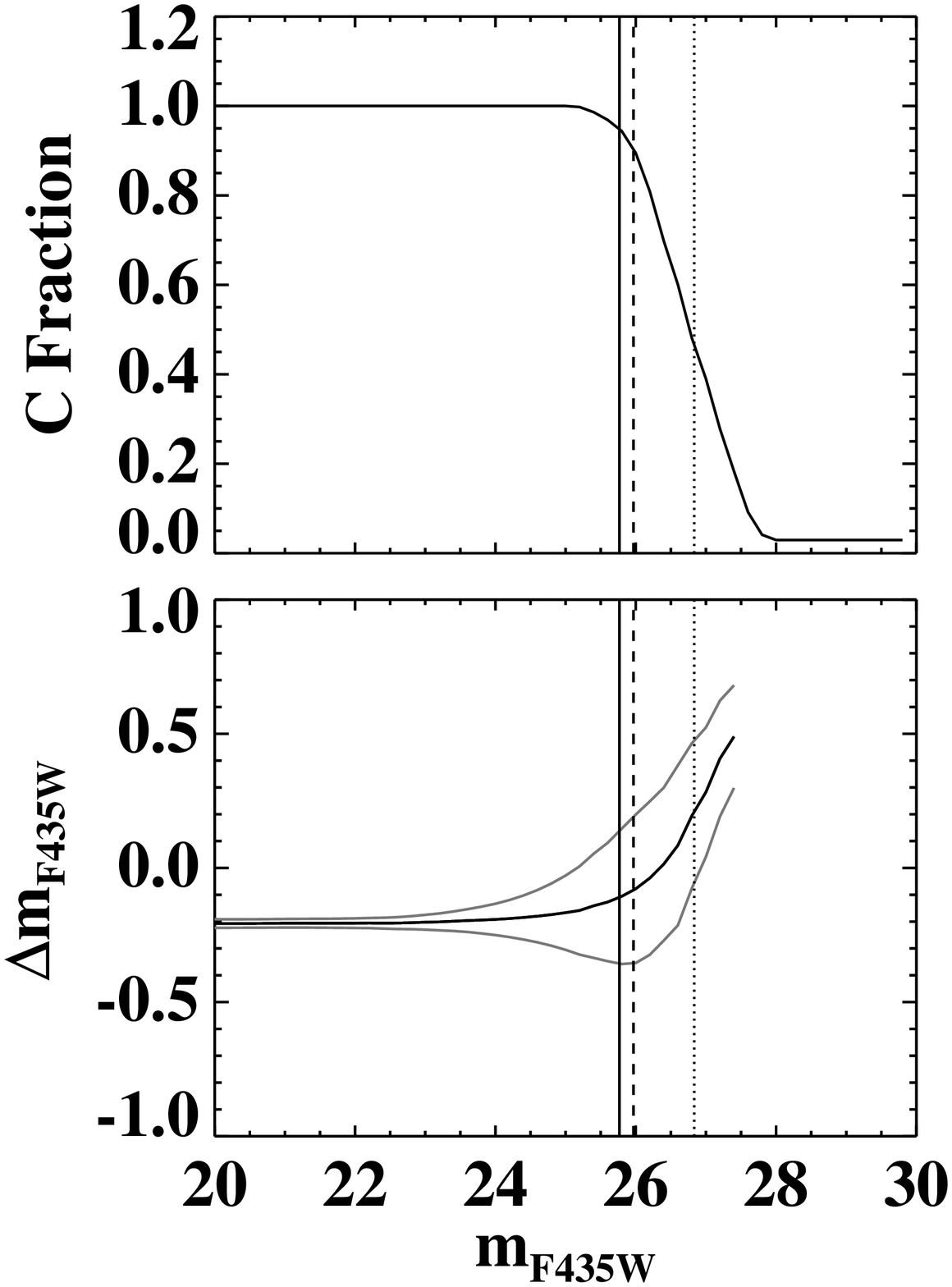}
    \hspace{-0.9cm}
   \includegraphics[width=0.27\textwidth]{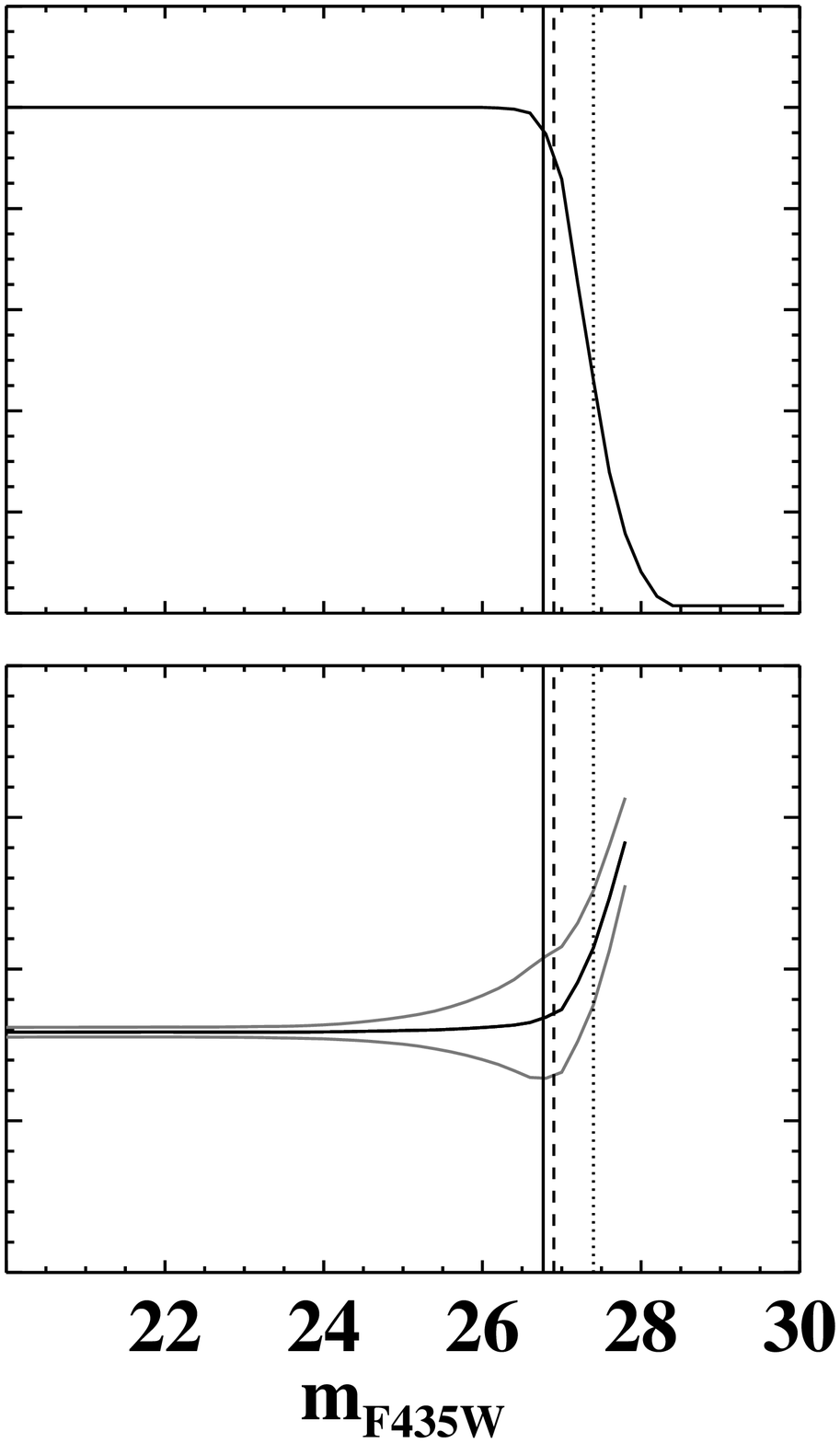}
    \hspace{-0.9cm}
   \includegraphics[width=0.27\textwidth]{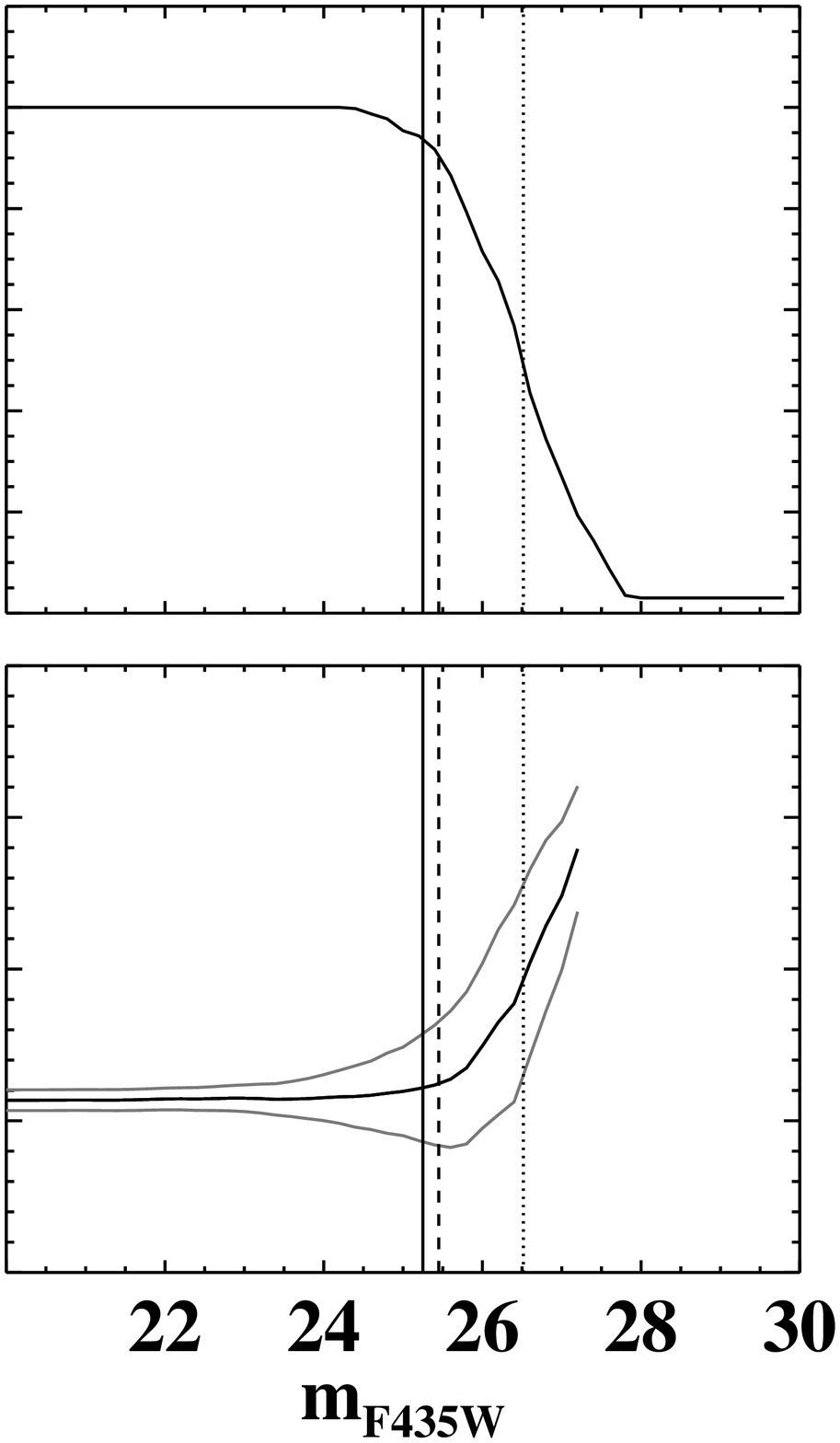}
    \hspace{-0.9cm}
   \includegraphics[width=0.27\textwidth]{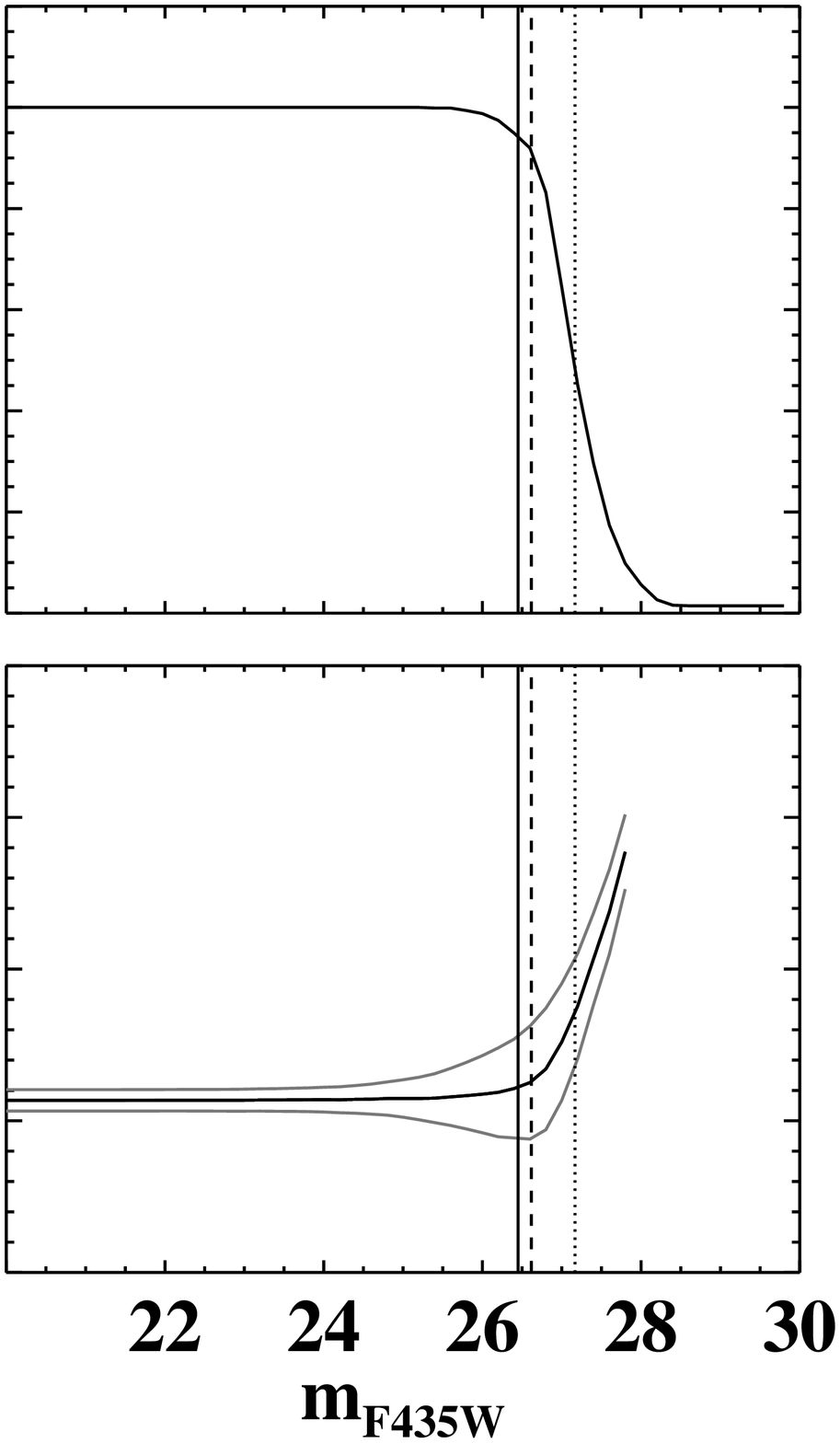}
 
   \caption[Result of the completeness test for IRAS 17208-0014]{Result of the completeness test for IRAS 17208-0014. Same labels and completeness levels are showed as in~\reffig{completeness}.}
    \label{fig:completeness2}          
 \end{figure}
 
Since the artificial stars were placed in random positions, some could reside in the same position as a real cluster/star/knot in the original image, increasing the measured flux. In order to avoid this, we rejected those cases where a recovered magnitude differed from the input magnitude by more than a certain value. We have taken the value of 0.75 mag because, as seen in the Figures, the systematic uncertainty for faint magnitudes is of the order of 0.4 mag, thus a 2$\sigma$ rejection level seems reasonable. False-positive detections with that filtering for the brightest artificial stars are not expected, since few real bright objects are encountered in the images. The completeness at each magnitude level is computed as the number ratio of recovered  over added artificial stars. All the process was done 10 times and an average completeness value at each magnitude level was computed. Two examples of the result, for the images of IRAS 04315-0840 and IRAS 17208-0014, are shown in~\reffig{completeness} and~\reffig{completeness2}.

The mean difference between input and output magnitudes of the artificial stars ($\Delta m_{F814W}$ and $\Delta m_{F435W}$ in the Figures) gives an estimate of the aperture correction, which is about -0.3 mag for the \textit{F814W} filter and a value close to -0.25 mag for the \textit{F435W} filter. This is consistent with the aperture corrections applied in the photometry of the sources (see \refcha{knots}, \refsec{det_phot}). Its standard deviation also represents a robust estimate of the actual photometric error to be associated with each magnitude bin.  This error is obviously larger than that computed by Poisson statistics only, especially towards the fainter magnitudes, since the former consider both the systematic and random uncertainties (see gray curve in Figures). Therefore, a more realistic uncertainty value for the knots with magnitudes near the 50\% completeness limit would move up to 0.3-0.4 mag. 

The systems are 90\% complete within the intervals $m_{F814W}$=[23,25] and $m_{F435W}$=[25,26.5], but due to the colors of the regions the \textit{F435W} completeness limit usually shifts to brighter magnitudes. We do not see a significant difference in the completeness limits when the \psf is convolved for the \textit{F435W} filter (less than 0.05 mag), but for the red filter the value of $\Delta m_{F814W}$ is increased by 0.1 mag when the \psf is convolved. Some difference is expected, since as the \psf widens we recover less flux by using the same aperture.

\cha{galfit}{Galfit Analysis}
\chaphead{This Appendix details the technique to perform 2D fits to the galaxy profile in order to derive the effective radius of the galaxies.}

To derive detailed information about the structural properties of our galaxies we performed a 2-dimensional decomposition of them using GALFIT (\mycitealt{Peng02};~\mycitealt{Peng10}) to fit the light profiles of the \textit{F160W} images (from \nicmos or the WFC3). Since merger systems consists of two or more galaxies, we have taken advantage of GALFITs ability
to fit simultaneously multiple components at different spatial regions.

Although most of the galaxies in our sample are ongoing mergers, and hence exhibit disturbed outer disks, in general we did not attempt to obtain a precise model for the entire merger system, since the global parameters (i.e., the total galaxy luminosity) are not affected (\mycitealt{Peng10}). Instead, we mainly focused on an accurate measurement of the bulge luminosity and central concentration indices, as these parameters provide important information about the build-up of stellar mass in the center of galaxies during the merger process. The main goal of the fit was to obtain an estimate of the \reff of each galaxy and we checked that in general if we try to fit the spiral arms the modeling becomes much more complicated and the derived \reff does not change significantly. Yet, we always tried to fit the asymmetric tidal features of the systems , since they can contribute to a considerable change of the total light of the system (i.e., the determination of the \reff\twospace). We performed a careful analysis of each galaxy, also identifying cases where a global bulge-disk decomposition would not work. 

Running GALFIT on an image requires initial preparation. The desired fitting region and sky background must be provided, like the \psf image. Images of \ulirgs cover the full extent of the merger, whereas in \lirgs only the central regions are sampled. Thus, the derived \reff for \lirgs must be understood as a lower limit to the real value. Sky backgrounds were measured separately using a region where no emission is present and designated as fixed parameters for all GALFIT procedures. 

After these preparation steps were carried out, we used an iterative process to perform a 1-3 component decomposition of the images. GALFIT requires initial guesses for each component it fits and uses a Levenberg-Marquardt downhill-gradient algorithm to determine the minimum $\chi^2$ based on the input guesses. The iterative process is described as follows: 
\begin{itemize}
 \item  We first fitted single S\'ersic profiles that are centered on the galaxies nuclei (typically 1-2 nuclei per image). The S\'ersic profile has the following form:
 \eqn[sersic]{
 \Sigma = \Sigma_e e^{-k[(r/r_{eff})^{1/n_s}-1]}
 }
 with the effective radius \reff\twospace, the surface brightness $\Sigma_e$ at \reff\twospace, the power law index $n_s$ and k coupled to $n_s$ such that half of the total flux is always within \reff\twospace. The advantage of the S\'ersic profile is that it is very useful for modeling different components of a galaxy such as bulges, flat disks, and bars, simply by varying the exponent. As the index $n_s$ becomes smaller, the the core flattens faster for \mbox{\textit{r} $<$ \reff\twospace}, and the intensity drops more steeply for \textit{r} $>$ \reff\twospace. Classical bulges typically have a S\'ersic index of \mbox{n = 4} (in general known as de Vaucouleurs profile), but can vary between \mbox{1 $<$ n $<$ 10}, while the outer disk follows an exponential profile \mbox{(n = 1)}. We then added another component (S\'ersic or exponential). When tidal features are present we also included in the model Fourier modes, which accounted for asymmetries of the light distribution. The GALFIT output parameters based on a single S\'ersic profile from the previous step were used as initial parameters for the multi-component fitting. The center coordinates were then fixed for the subsequent fitting steps. For each galaxy, a variety of different multi-component fits were performed (typically 1-3; S\'ersic \mbox{bulge + exponential} disk was normally a good solution) and compared afterwards. Fitted models were selected only as long as the model parameters were all well behaved, i.e., requiring that all parameters such as bulge radius and S\'ersic index converged within the range of reasonable values (e.g., \mbox{$n_{\rm{s}}$ $<$ 10}). Three examples of GALFIT models and residual (difference between the original and the modeled image) images are shown in \reffig{galfit}.
\end{itemize}

\begin{figure}[!htp]
 \hypertarget{fig:galfit}{}\hypertarget{autolof:\theautolof}{}\addtocounter{autolof}{1}
\hspace{0.3cm}\includegraphics[trim = 1cm 1cm 0cm 9.5cm,clip=true,width=1.\textwidth]{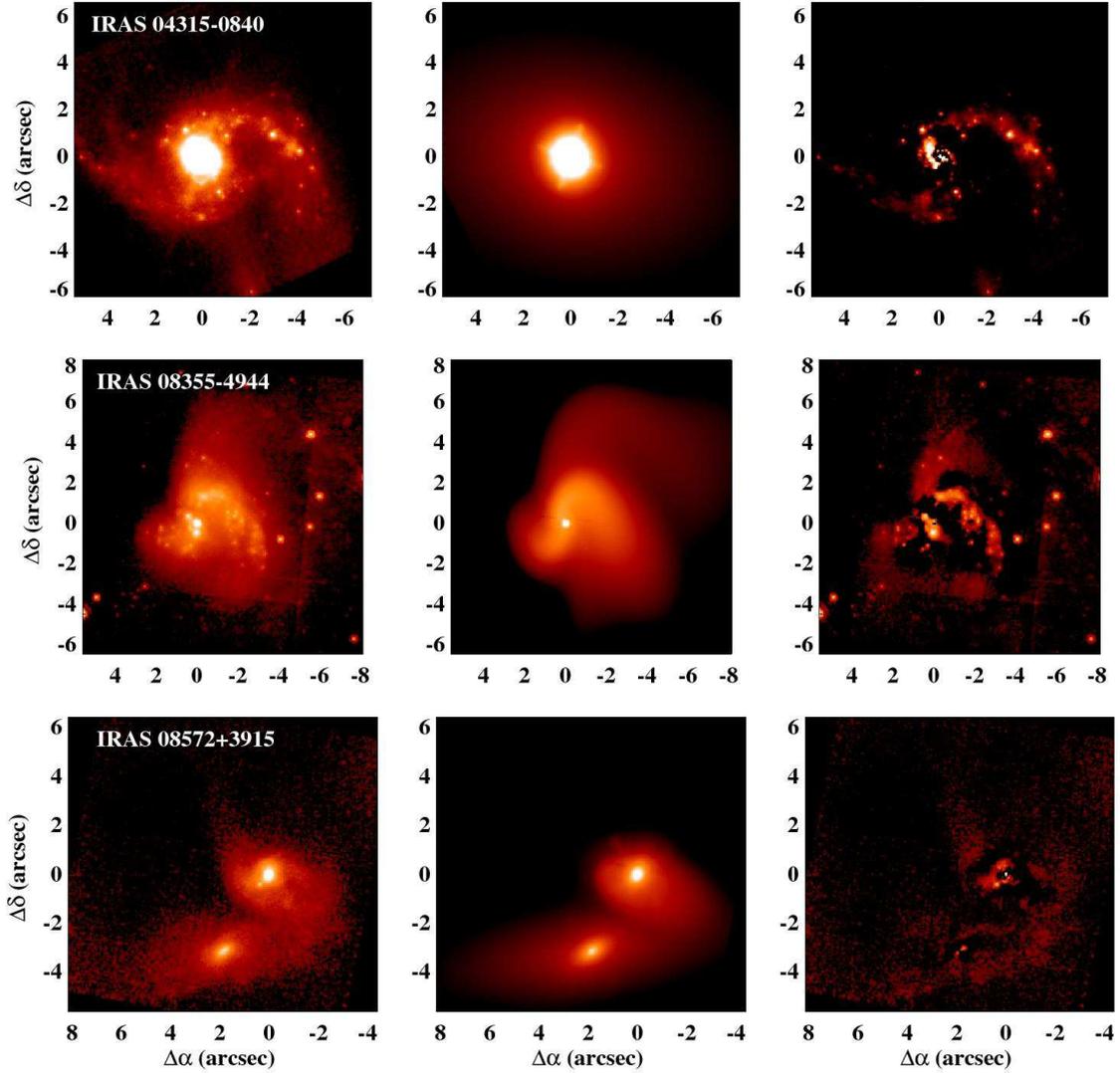}
      \caption[GALFIT output images]{GALFIT output images for IRAS 04315-0840 (top), IRAS 08355-4944 (middle) and IRAS 08572+3915 (bottom). In all cases, from left to right, the original, modeled and residual images are showed. For each system, the three images are drawn using the same scale. Faint diffuse emission, spiral structure, and stellar regions remain in the model-subtracted (residual) image.}     
      \label{fig:galfit}
\end{figure}

GALFIT calculates the $\chi^2$ for each model, which generally declines as more free parameters or S\'ersic components are given. However, a small $\chi^2$ does not necessarily mean that the solution with an extra component is more correct physically. Thus, we added extra components as long as the morphology of the modeled galaxy was improved and resembled significantly better that of the \textit{F160W} image.

For objects with central point sources such as an \agn\twospace, the bulge S\'ersic index can be overestimated unless an extra nuclear component (\psf component) is added to the model.  We used GALFIT in a first attempt without a central \psf component and calculated the central residual excess light from the residual image of the final fit. In cases where a central residual was present we re-ran GALFIT with an additional \psf component to take into account the extra central light originating from a central point source. In general the \psf component did not dominate the total emission of the galaxy, thus was not relevant in the computation of the \reff of the galaxy.

Once all the components were fitted, the \reff of a galaxy was derived by integrating its analytical form and computing at which \reff the total emission decreased by a half.  

\clearpage{\pagestyle{empty}\cleardoublepage}

\cleardoublepage

\newpage
\pagestyle{plain}

\begin{thebibliography}{351}
\expandafter\ifx\csname natexlab\endcsname\relax\def\natexlab#1{#1}\fi

\bibitem[{{Aaronson}(1983)}]{Aaronson83}	\linkup{autobib:Aaronson83}
{Aaronson}, M. 1983, \apjl, 266, L11,
  \adsurl{http://adsabs.harvard.edu/abs/1983ApJ...266L..11A}

\bibitem[{{Abraham} {et~al.}(1996){Abraham}, {Tanvir}, {Santiago}, {Ellis},
  {Glazebrook}, \& {van den Bergh}}]{Abraham96}	\linkup{autobib:Abraham96}
{Abraham}, R.~G., {Tanvir}, N.~R., {Santiago}, B.~X., {Ellis}, R.~S.,
  {Glazebrook}, K., \& {van den Bergh}, S. 1996, \mnras, 279, L47,
  \adsurl{http://adsabs.harvard.edu/abs/1996MNRAS.279L..47A},
  \eprint{arXiv:astro-ph/9602044}

\bibitem[{{Allington-Smith}(2007)}]{Allington-Smith07}	\linkup{autobib:Allington-Smith07}
{Allington-Smith}, J.~R. 2007, in Revista Mexicana de Astronomia y Astrofisica,
  vol. 27, Vol.~28, Revista Mexicana de Astronomia y Astrofisica Conference
  Series, ed. {S.~Kurtz}, 17--23,
  \adsurl{http://adsabs.harvard.edu/abs/2007RMxAC..28...17A}

\bibitem[{{Alonso-Herrero} {et~al.}(2009){Alonso-Herrero},
  {Garc{\'{\i}}a-Mar{\'{\i}}n}, {Monreal-Ibero}, {Colina}, {Arribas},
  {Alfonso-Garz{\'o}n}, \& {Labiano}}]{Alonso-Herrero09}	\linkup{autobib:Alonso-Herrero09}
{Alonso-Herrero}, A., {Garc{\'{\i}}a-Mar{\'{\i}}n}, M., {Monreal-Ibero}, A.,
  {Colina}, L., {Arribas}, S., {Alfonso-Garz{\'o}n}, J., \& {Labiano}, A. 2009,
  \aap, 506, 1541,
  \adsurl{http://adsabs.harvard.edu/abs/2009A\%26A...506.1541A},
  \eprint{0907.5105}

\bibitem[{{Alonso-Herrero} {et~al.}(2006){Alonso-Herrero}, {Rieke}, {Rieke},
  {Colina}, {P{\'e}rez-Gonz{\'a}lez}, \& {Ryder}}]{Alonso-Herrero06}	\linkup{autobib:Alonso-Herrero06}
{Alonso-Herrero}, A., {Rieke}, G.~H., {Rieke}, M.~J., {Colina}, L.,
  {P{\'e}rez-Gonz{\'a}lez}, P.~G., \& {Ryder}, S.~D. 2006, \apj, 650, 835,
  \adsurl{http://adsabs.harvard.edu/abs/2006ApJ...650..835A},
  \eprint{arXiv:astro-ph/0606186}

\bibitem[{{Alonso-Herrero} {et~al.}(2002){Alonso-Herrero}, {Rieke}, {Rieke}, \&
  {Scoville}}]{Alonso-Herrero02}	\linkup{autobib:Alonso-Herrero02}
{Alonso-Herrero}, A., {Rieke}, G.~H., {Rieke}, M.~J., \& {Scoville}, N.~Z.
  2002, \aj, 124, 166,
  \adsurl{http://adsabs.harvard.edu/abs/2002AJ....124..166A},
  \eprint{arXiv:astro-ph/0203494}

\bibitem[{{Amor{\'{\i}}n} {et~al.}(2009){Amor{\'{\i}}n}, {Aguerri},
  {Mu{\~n}oz-Tu{\~n}{\'o}n}, \& {Cair{\'o}s}}]{Amorin09}	\linkup{autobib:Amorin09}
{Amor{\'{\i}}n}, R., {Aguerri}, J.~A.~L., {Mu{\~n}oz-Tu{\~n}{\'o}n}, C., \&
  {Cair{\'o}s}, L.~M. 2009, \aap, 501, 75,
  \adsurl{http://adsabs.harvard.edu/abs/2009A\%26A...501...75A},
  \eprint{0903.2861}

\bibitem[{{Anders} {et~al.}(2007){Anders}, {Bissantz}, {Boysen}, {de Grijs}, \&
  {Fritze-v.~Alvensleben}}]{Anders07}	\linkup{autobib:Anders07}
{Anders}, P., {Bissantz}, N., {Boysen}, L., {de Grijs}, R., \&
  {Fritze-v.~Alvensleben}, U. 2007, \mnras, 377, 91,
  \adsurl{http://adsabs.harvard.edu/abs/2007MNRAS.377...91A},
  \eprint{arXiv:astro-ph/0702413}

\bibitem[{{Arp}(1966)}]{Arp66}	\linkup{autobib:Arp66}
{Arp}, H. 1966, \apjs, 14, 1,
  \adsurl{http://adsabs.harvard.edu/abs/1966ApJS...14....1A}

\bibitem[{{Arribas} {et~al.}(2004){Arribas}, {Bushouse}, {Lucas}, {Colina}, \&
  {Borne}}]{Arribas04}	\linkup{autobib:Arribas04}
{Arribas}, S., {Bushouse}, H., {Lucas}, R.~A., {Colina}, L., \& {Borne}, K.~D.
  2004, \aj, 127, 2522,
  \adsurl{http://adsabs.harvard.edu/abs/2004AJ....127.2522A},
  \eprint{arXiv:astro-ph/0402375}

\bibitem[{{Arribas} {et~al.}(1998){Arribas}, {Carter}, {Cavaller}, {del Burgo},
  {Edwards}, {Fuentes}, {Garcia}, {Herreros}, {Jones}, {Mediavilla}, {Pi},
  {Pollacco}, {Rasilla}, {Rees}, \& {Sosa}}]{Arribas98}	\linkup{autobib:Arribas98}
{Arribas}, S. {et~al.} 1998, in Society of Photo-Optical Instrumentation
  Engineers (SPIE) Conference Series, Vol. 3355, Society of Photo-Optical
  Instrumentation Engineers (SPIE) Conference Series, ed. S.~{D'Odorico},
  821--827, \adsurl{http://adsabs.harvard.edu/abs/1998SPIE.3355..821A}

\bibitem[{{Arribas} \& {Colina}(2003)}]{Arribas03}	\linkup{autobib:Arribas03}
{Arribas}, S., \& {Colina}, L. 2003, \apj, 591, 791,
  \adsurl{http://adsabs.harvard.edu/abs/2003ApJ...591..791A},
  \eprint{arXiv:astro-ph/0304138}

\bibitem[{{Arribas} {et~al.}(2000){Arribas}, {Colina}, \& {Borne}}]{Arribas00}	\linkup{autobib:Arribas00}
{Arribas}, S., {Colina}, L., \& {Borne}, K.~D. 2000, \apj, 545, 228,
  \adsurl{http://adsabs.harvard.edu/abs/2000ApJ...545..228A}

\bibitem[{{Arribas} {et~al.}(2008){Arribas}, {Colina}, {Monreal-Ibero},
  {Alfonso}, {Garc{\'{\i}}a-Mar{\'{\i}}n}, \& {Alonso-Herrero}}]{Arribas08}	\linkup{autobib:Arribas08}
{Arribas}, S., {Colina}, L., {Monreal-Ibero}, A., {Alfonso}, J.,
  {Garc{\'{\i}}a-Mar{\'{\i}}n}, M., \& {Alonso-Herrero}, A. 2008, \aap, 479,
  687, \adsurl{http://adsabs.harvard.edu/abs/2008A\%26A...479..687A},
  \eprint{0710.2761}

\bibitem[{{Ashman} \& {Zepf}(2001)}]{Ashman01}	\linkup{autobib:Ashman01}
{Ashman}, K.~M., \& {Zepf}, S.~E. 2001, \aj, 122, 1888,
  \adsurl{http://adsabs.harvard.edu/abs/2001AJ....122.1888A},
  \eprint{arXiv:astro-ph/0107146}

\bibitem[{{Barnes}(2004)}]{Barnes04}	\linkup{autobib:Barnes04}
{Barnes}, J.~E. 2004, \mnras, 350, 798,
  \adsurl{http://adsabs.harvard.edu/abs/2004MNRAS.350..798B},
  \eprint{arXiv:astro-ph/0402248}

\bibitem[{{Barnes} \& {Hernquist}(1992{\natexlab{a}})}]{Barnes92a}	\linkup{autobib:Barnes92a}
{Barnes}, J.~E., \& {Hernquist}, L. 1992{\natexlab{a}}, \araa, 30, 705,
  \adsurl{http://adsabs.harvard.edu/abs/1992ARA\%26A..30..705B}

\bibitem[{{Barnes} \& {Hernquist}(1992{\natexlab{b}})}]{Barnes92b}	\linkup{autobib:Barnes92b}
{Barnes}, J.~E., \& {Hernquist}, L. 1992{\natexlab{b}}, \nat, 360, 715,
  \adsurl{http://adsabs.harvard.edu/abs/1992Natur.360..715B}

\bibitem[{{Barnes} \& {Hernquist}(1996)}]{Barnes96}	\linkup{autobib:Barnes96}
{Barnes}, J.~E., \& {Hernquist}, L. 1996, \apj, 471, 115,
  \adsurl{http://adsabs.harvard.edu/abs/1996ApJ...471..115B}

\bibitem[{{Barth} {et~al.}(1995){Barth}, {Ho}, {Filippenko}, \&
  {Sargent}}]{Barth95}	\linkup{autobib:Barth95}
{Barth}, A.~J., {Ho}, L.~C., {Filippenko}, A.~V., \& {Sargent}, W.~L. 1995,
  \aj, 110, 1009, \adsurl{http://adsabs.harvard.edu/abs/1995AJ....110.1009B}

\bibitem[{{Bastian} {et~al.}(2006){Bastian}, {Emsellem}, {Kissler-Patig}, \&
  {Maraston}}]{Bastian06}	\linkup{autobib:Bastian06}
{Bastian}, N., {Emsellem}, E., {Kissler-Patig}, M., \& {Maraston}, C. 2006,
  \aap, 445, 471, \adsurl{http://adsabs.harvard.edu/abs/2006A\%26A...445..471B},
  \eprint{arXiv:astro-ph/0509249}

\bibitem[{{Bastian} {et~al.}(2005{\natexlab{a}}){Bastian}, {Gieles}, {Efremov},
  \& {Lamers}}]{Bastian05a}	\linkup{autobib:Bastian05a}
{Bastian}, N., {Gieles}, M., {Efremov}, Y.~N., \& {Lamers}, H.~J.~G.~L.~M.
  2005{\natexlab{a}}, \aap, 443, 79,
  \adsurl{http://adsabs.harvard.edu/abs/2005A\%26A...443...79B},
  \eprint{arXiv:astro-ph/0508110}

\bibitem[{{Bastian} {et~al.}(2005{\natexlab{b}}){Bastian}, {Gieles}, {Lamers},
  {Scheepmaker}, \& {de Grijs}}]{Bastian05b}	\linkup{autobib:Bastian05b}
{Bastian}, N., {Gieles}, M., {Lamers}, H.~J.~G.~L.~M., {Scheepmaker}, R.~A., \&
  {de Grijs}, R. 2005{\natexlab{b}}, \aap, 431, 905,
  \adsurl{http://adsabs.harvard.edu/abs/2005A\%26A...431..905B},
  \eprint{arXiv:astro-ph/0408043}

\bibitem[{{Bastian} {et~al.}(2005{\natexlab{c}}){Bastian}, {Hempel},
  {Kissler-Patig}, {Homeier}, \& {Trancho}}]{Bastian05c}	\linkup{autobib:Bastian05c}
{Bastian}, N., {Hempel}, M., {Kissler-Patig}, M., {Homeier}, N.~L., \&
  {Trancho}, G. 2005{\natexlab{c}}, \aap, 435, 65,
  \adsurl{http://adsabs.harvard.edu/abs/2005A\%26A...435...65B},
  \eprint{arXiv:astro-ph/0502561}

\bibitem[{{Bastian} {et~al.}(2009){Bastian}, {Trancho}, {Konstantopoulos}, \&
  {Miller}}]{Bastian09}	\linkup{autobib:Bastian09}
{Bastian}, N., {Trancho}, G., {Konstantopoulos}, I.~S., \& {Miller}, B.~W.
  2009, \apj, 701, 607,
  \adsurl{http://adsabs.harvard.edu/abs/2009ApJ...701..607B},
  \eprint{0906.2210}

\bibitem[{{Baumgardt} \& {Kroupa}(2007)}]{Baumgardt07}	\linkup{autobib:Baumgardt07}
{Baumgardt}, H., \& {Kroupa}, P. 2007, \mnras, 380, 1589,
  \adsurl{http://adsabs.harvard.edu/abs/2007MNRAS.380.1589B},
  \eprint{0707.1944}

\bibitem[{{Baumgardt} \& {Makino}(2003)}]{Baumgardt03}	\linkup{autobib:Baumgardt03}
{Baumgardt}, H., \& {Makino}, J. 2003, \mnras, 340, 227,
  \adsurl{http://adsabs.harvard.edu/abs/2003MNRAS.340..227B},
  \eprint{arXiv:astro-ph/0211471}

\bibitem[{{Bekki} \& {Shioya}(2001)}]{Bekki01}	\linkup{autobib:Bekki01}
{Bekki}, K., \& {Shioya}, Y. 2001, \apjs, 134, 241,
  \adsurl{http://adsabs.harvard.edu/abs/2001ApJS..134..241B},
  \eprint{arXiv:astro-ph/0012365}

\bibitem[{{Bekki} {et~al.}(2006){Bekki}, {Shioya}, \& {Whiting}}]{Bekki06}	\linkup{autobib:Bekki06}
{Bekki}, K., {Shioya}, Y., \& {Whiting}, M. 2006, \mnras, 371, 805,
  \adsurl{http://adsabs.harvard.edu/abs/2006MNRAS.371..805B},
  \eprint{arXiv:astro-ph/0607349}

\bibitem[{{Belokurov} {et~al.}(2007){Belokurov}, {Zucker}, {Evans}, {Kleyna},
  {Koposov}, {Hodgkin}, {Irwin}, {Gilmore}, {Wilkinson}, {Fellhauer},
  {Bramich}, {Hewett}, \& {Vidrih}}]{Belokurov07}	\linkup{autobib:Belokurov07}
{Belokurov}, V. {et~al.} 2007, \apj, 654, 897,
  \adsurl{http://adsabs.harvard.edu/abs/2007ApJ...654..897B},
  \eprint{arXiv:astro-ph/0608448}

\bibitem[{{Bender} {et~al.}(1992){Bender}, {Burstein}, \& {Faber}}]{Bender92}	\linkup{autobib:Bender92}
{Bender}, R., {Burstein}, D., \& {Faber}, S.~M. 1992, \apj, 399, 462,
  \adsurl{http://adsabs.harvard.edu/abs/1992ApJ...399..462B}

\bibitem[{{Bendo} \& {Barnes}(2000)}]{Bendo00}	\linkup{autobib:Bendo00}
{Bendo}, G.~J., \& {Barnes}, J.~E. 2000, \mnras, 316, 315,
  \adsurl{http://adsabs.harvard.edu/abs/2000MNRAS.316..315B},
  \eprint{arXiv:astro-ph/0003248}

\bibitem[{{Bertin} \& {Arnouts}(1996)}]{Bertin96}	\linkup{autobib:Bertin96}
{Bertin}, E., \& {Arnouts}, S. 1996, \aaps, 117, 393,
  \adsurl{http://adsabs.harvard.edu/abs/1996A\%26AS..117..393B}

\bibitem[{{Bik} {et~al.}(2003){Bik}, {Lamers}, {Bastian}, {Panagia}, \&
  {Romaniello}}]{Bik03}	\linkup{autobib:Bik03}
{Bik}, A., {Lamers}, H.~J.~G.~L.~M., {Bastian}, N., {Panagia}, N., \&
  {Romaniello}, M. 2003, \aap, 397, 473,
  \adsurl{http://adsabs.harvard.edu/abs/2003A\%26A...397..473B},
  \eprint{arXiv:astro-ph/0210594}

\bibitem[{{Billett} {et~al.}(2002){Billett}, {Hunter}, \&
  {Elmegreen}}]{Billett02}	\linkup{autobib:Billett02}
{Billett}, O.~H., {Hunter}, D.~A., \& {Elmegreen}, B.~G. 2002, \aj, 123, 1454,
  \adsurl{http://adsabs.harvard.edu/abs/2002AJ....123.1454B},
  \eprint{arXiv:astro-ph/0112260}

\bibitem[{{Bingham} {et~al.}(1994){Bingham}, {Gellatly}, {Jenkins}, \&
  {Worswick}}]{Bingham94}	\linkup{autobib:Bingham94}
{Bingham}, R.~G., {Gellatly}, D.~W., {Jenkins}, C.~R., \& {Worswick}, S.~P.
  1994, in Society of Photo-Optical Instrumentation Engineers (SPIE) Conference
  Series, Vol. 2198, Society of Photo-Optical Instrumentation Engineers (SPIE)
  Conference Series, ed. {D.~L.~Crawford \& E.~R.~Craine}, 56--64,
  \adsurl{http://adsabs.harvard.edu/abs/1994SPIE.2198...56B}

\bibitem[{{Binney} \& {Tremaine}(1987)}]{Binney87}	\linkup{autobib:Binney87}
{Binney}, J., \& {Tremaine}, S. 1987, {Galactic dynamics}, ed. {Binney, J.~\&
  Tremaine, S.}, \adsurl{http://adsabs.harvard.edu/abs/1987gady.book.....B}

\bibitem[{{Blain} {et~al.}(2002){Blain}, {Smail}, {Ivison}, {Kneib}, \&
  {Frayer}}]{Blain02}	\linkup{autobib:Blain02}
{Blain}, A.~W., {Smail}, I., {Ivison}, R.~J., {Kneib}, J., \& {Frayer}, D.~T.
  2002, \physrep, 369, 111,
  \adsurl{http://adsabs.harvard.edu/abs/2002PhR...369..111B},
  \eprint{arXiv:astro-ph/0202228}

\bibitem[{{Blitz} {et~al.}(2007){Blitz}, {Fukui}, {Kawamura}, {Leroy},
  {Mizuno}, \& {Rosolowsky}}]{Blitz07}	\linkup{autobib:Blitz07}
{Blitz}, L., {Fukui}, Y., {Kawamura}, A., {Leroy}, A., {Mizuno}, N., \&
  {Rosolowsky}, E. 2007, Protostars and Planets V, 81,
  \adsurl{http://adsabs.harvard.edu/abs/2007prpl.conf...81B},
  \eprint{arXiv:astro-ph/0602600}

\bibitem[{{Boily} \& {Kroupa}(2003)}]{Boily03}	\linkup{autobib:Boily03}
{Boily}, C.~M., \& {Kroupa}, P. 2003, \mnras, 338, 673,
  \adsurl{http://adsabs.harvard.edu/abs/2003MNRAS.338..673B},
  \eprint{arXiv:astro-ph/0211026}

\bibitem[{{Borne} {et~al.}(2000){Borne}, {Bushouse}, {Lucas}, \&
  {Colina}}]{Borne00}	\linkup{autobib:Borne00}
{Borne}, K.~D., {Bushouse}, H., {Lucas}, R.~A., \& {Colina}, L. 2000, \apjl,
  529, L77, \adsurl{http://adsabs.harvard.edu/abs/2000ApJ...529L..77B},
  \eprint{arXiv:astro-ph/9912151}

\bibitem[{{Bosch} {et~al.}(2009){Bosch}, {Terlevich}, \& {Terlevich}}]{Bosch09}	\linkup{autobib:Bosch09}
{Bosch}, G., {Terlevich}, E., \& {Terlevich}, R. 2009, \aj, 137, 3437,
  \adsurl{http://adsabs.harvard.edu/abs/2009AJ....137.3437B},
  \eprint{0811.4748}

\bibitem[{{Bournaud} {et~al.}(2008{\natexlab{a}}){Bournaud}, {Daddi},
  {Elmegreen}, {Elmegreen}, {Nesvadba}, {Vanzella}, {Di Matteo}, {Le Tiran},
  {Lehnert}, \& {Elbaz}}]{Bournaud08b}	\linkup{autobib:Bournaud08b}
{Bournaud}, F. {et~al.} 2008{\natexlab{a}}, \aap, 486, 741,
  \adsurl{http://adsabs.harvard.edu/abs/2008A\%26A...486..741B},
  \eprint{0803.3831}

\bibitem[{{Bournaud} {et~al.}(2004){Bournaud}, {Duc}, {Amram}, {Combes}, \&
  {Gach}}]{Bournaud04}	\linkup{autobib:Bournaud04}
{Bournaud}, F., {Duc}, P., {Amram}, P., {Combes}, F., \& {Gach}, J. 2004, \aap,
  425, 813, \adsurl{http://adsabs.harvard.edu/abs/2004A\%26A...425..813B},
  \eprint{arXiv:astro-ph/0406169}

\bibitem[{{Bournaud} {et~al.}(2008{\natexlab{b}}){Bournaud}, {Duc}, \&
  {Emsellem}}]{Bournaud08a}	\linkup{autobib:Bournaud08a}
{Bournaud}, F., {Duc}, P., \& {Emsellem}, E. 2008{\natexlab{b}}, \mnras, 389,
  L8, \adsurl{http://adsabs.harvard.edu/abs/2008MNRAS.389L...8B},
  \eprint{0806.1386}

\bibitem[{{Bournaud} \& {Duc}(2006)}]{Bournaud06}	\linkup{autobib:Bournaud06}
{Bournaud}, F., \& {Duc}, P.-A. 2006, \aap, 456, 481,
  \adsurl{http://adsabs.harvard.edu/abs/2006A\%26A...456..481B},
  \eprint{arXiv:astro-ph/0605350}

\bibitem[{{Bournaud} {et~al.}(2007){Bournaud}, {Elmegreen}, \&
  {Elmegreen}}]{Bournaud07}	\linkup{autobib:Bournaud07}
{Bournaud}, F., {Elmegreen}, B.~G., \& {Elmegreen}, D.~M. 2007, \apj, 670, 237,
  \adsurl{http://adsabs.harvard.edu/abs/2007ApJ...670..237B},
  \eprint{0708.0306}

\bibitem[{{Bournaud} {et~al.}(2005){Bournaud}, {Jog}, \& {Combes}}]{Bournaud05}	\linkup{autobib:Bournaud05}
{Bournaud}, F., {Jog}, C.~J., \& {Combes}, F. 2005, \aap, 437, 69,
  \adsurl{http://adsabs.harvard.edu/abs/2005A\%26A...437...69B},
  \eprint{arXiv:astro-ph/0503189}

\bibitem[{{Braine} {et~al.}(2001){Braine}, {Duc}, {Lisenfeld}, {Charmandaris},
  {Vallejo}, {Leon}, \& {Brinks}}]{Braine01}	\linkup{autobib:Braine01}
{Braine}, J., {Duc}, P.-A., {Lisenfeld}, U., {Charmandaris}, V., {Vallejo}, O.,
  {Leon}, S., \& {Brinks}, E. 2001, \aap, 378, 51,
  \adsurl{http://adsabs.harvard.edu/abs/2001A\%26A...378...51B},
  \eprint{arXiv:astro-ph/0108513}

\bibitem[{{Bresolin} \& {Kennicutt}(1997)}]{Bresolin97}	\linkup{autobib:Bresolin97}
{Bresolin}, F., \& {Kennicutt}, Jr., R.~C. 1997, \aj, 113, 975,
  \adsurl{http://adsabs.harvard.edu/abs/1997AJ....113..975B}

\bibitem[{{Bruzual} \& {Charlot}(2003)}]{Bruzual03}	\linkup{autobib:Bruzual03}
{Bruzual}, G., \& {Charlot}, S. 2003, \mnras, 344, 1000,
  \adsurl{http://adsabs.harvard.edu/abs/2003MNRAS.344.1000B},
  \eprint{arXiv:astro-ph/0309134}

\bibitem[{{Burkert} \& {Bodenheimer}(1996)}]{Burkert96}	\linkup{autobib:Burkert96}
{Burkert}, A., \& {Bodenheimer}, P. 1996, \mnras, 280, 1190,
  \adsurl{http://adsabs.harvard.edu/abs/1996MNRAS.280.1190B}

\bibitem[{{Bushouse} {et~al.}(2002){Bushouse}, {Borne}, {Colina}, {Lucas},
  {Rowan-Robinson}, {Baker}, {Clements}, {Lawrence}, \& {Oliver}}]{Bushouse02}	\linkup{autobib:Bushouse02}
{Bushouse}, H.~A. {et~al.} 2002, \apjs, 138, 1,
  \adsurl{http://adsabs.harvard.edu/abs/2002ApJS..138....1B},
  \eprint{arXiv:astro-ph/0108261}

\bibitem[{{Cair{\'o}s} {et~al.}(2003){Cair{\'o}s}, {Caon}, {Papaderos},
  {Noeske}, {V{\'{\i}}lchez}, {Garc{\'{\i}}a Lorenzo}, \&
  {Mu{\~n}oz-Tu{\~n}{\'o}n}}]{Cairos03}	\linkup{autobib:Cairos03}
{Cair{\'o}s}, L.~M., {Caon}, N., {Papaderos}, P., {Noeske}, K.,
  {V{\'{\i}}lchez}, J.~M., {Garc{\'{\i}}a Lorenzo}, B., \&
  {Mu{\~n}oz-Tu{\~n}{\'o}n}, C. 2003, \apj, 593, 312,
  \adsurl{http://adsabs.harvard.edu/abs/2003ApJ...593..312C},
  \eprint{arXiv:astro-ph/0306307}

\bibitem[{{Cair{\'o}s} {et~al.}(2001){Cair{\'o}s}, {Caon}, {V{\'{\i}}lchez},
  {Gonz{\'a}lez-P{\'e}rez}, \& {Mu{\~n}oz-Tu{\~n}{\'o}n}}]{Cairos01}	\linkup{autobib:Cairos01}
{Cair{\'o}s}, L.~M., {Caon}, N., {V{\'{\i}}lchez}, J.~M.,
  {Gonz{\'a}lez-P{\'e}rez}, J.~N., \& {Mu{\~n}oz-Tu{\~n}{\'o}n}, C. 2001,
  \apjs, 136, 393, \adsurl{http://adsabs.harvard.edu/abs/2001ApJS..136..393C}

\bibitem[{{Calzetti} {et~al.}(2000){Calzetti}, {Armus}, {Bohlin}, {Kinney},
  {Koornneef}, \& {Storchi-Bergmann}}]{Calzetti00}	\linkup{autobib:Calzetti00}
{Calzetti}, D., {Armus}, L., {Bohlin}, R.~C., {Kinney}, A.~L., {Koornneef}, J.,
  \& {Storchi-Bergmann}, T. 2000, \apj, 533, 682,
  \adsurl{http://adsabs.harvard.edu/abs/2000ApJ...533..682C},
  \eprint{arXiv:astro-ph/9911459}

\bibitem[{{Calzetti} {et~al.}(2007){Calzetti}, {Kennicutt}, {Engelbracht},
  {Leitherer}, {Draine}, {Kewley}, {Moustakas}, {Sosey}, {Dale}, {Gordon},
  {Helou}, {Hollenbach}, {Armus}, {Bendo}, {Bot}, {Buckalew}, {Jarrett}, {Li},
  {Meyer}, {Murphy}, {Prescott}, {Regan}, {Rieke}, {Roussel}, {Sheth}, {Smith},
  {Thornley}, \& {Walter}}]{Calzetti07}	\linkup{autobib:Calzetti07}
{Calzetti}, D. {et~al.} 2007, \apj, 666, 870,
  \adsurl{http://adsabs.harvard.edu/abs/2007ApJ...666..870C},
  \eprint{0705.3377}

\bibitem[{{Caputi} {et~al.}(2007){Caputi}, {Lagache}, {Yan}, {Dole},
  {Bavouzet}, {Le Floc'h}, {Choi}, {Helou}, \& {Reddy}}]{Caputi07}	\linkup{autobib:Caputi07}
{Caputi}, K.~I. {et~al.} 2007, \apj, 660, 97,
  \adsurl{http://adsabs.harvard.edu/abs/2007ApJ...660...97C},
  \eprint{arXiv:astro-ph/0701283}

\bibitem[{{Chapman} {et~al.}(2003){Chapman}, {Windhorst}, {Odewahn}, {Yan}, \&
  {Conselice}}]{Chapman03}	\linkup{autobib:Chapman03}
{Chapman}, S.~C., {Windhorst}, R., {Odewahn}, S., {Yan}, H., \& {Conselice}, C.
  2003, \apj, 599, 92,
  \adsurl{http://adsabs.harvard.edu/abs/2003ApJ...599...92C},
  \eprint{arXiv:astro-ph/0308197}

\bibitem[{{Clements} {et~al.}(1996){Clements}, {Sutherland}, {McMahon}, \&
  {Saunders}}]{Clements96}	\linkup{autobib:Clements96}
{Clements}, D.~L., {Sutherland}, W.~J., {McMahon}, R.~G., \& {Saunders}, W.
  1996, \mnras, 279, 477,
  \adsurl{http://adsabs.harvard.edu/abs/1996MNRAS.279..477C}

\bibitem[{{Cohen}(1992)}]{Cohen92}	\linkup{autobib:Cohen92}
{Cohen}, M. 1992, \aj, 103, 1734,
  \adsurl{http://adsabs.harvard.edu/abs/1992AJ....103.1734C}

\bibitem[{{Colina} {et~al.}(2004){Colina}, {Arribas}, \& {Clements}}]{Colina04}	\linkup{autobib:Colina04}
{Colina}, L., {Arribas}, S., \& {Clements}, D. 2004, \apj, 602, 181,
  \adsurl{http://adsabs.harvard.edu/abs/2004ApJ...602..181C},
  \eprint{arXiv:astro-ph/0310604}

\bibitem[{{Colina} {et~al.}(2005){Colina}, {Arribas}, \&
  {Monreal-Ibero}}]{Colina05}	\linkup{autobib:Colina05}
{Colina}, L., {Arribas}, S., \& {Monreal-Ibero}, A. 2005, \apj, 621, 725,
  \adsurl{http://adsabs.harvard.edu/abs/2005ApJ...621..725C}

\bibitem[{{Colina} {et~al.}(2001){Colina}, {Borne}, {Bushouse}, {Lucas},
  {Rowan-Robinson}, {Lawrence}, {Clements}, {Baker}, \& {Oliver}}]{Colina01}	\linkup{autobib:Colina01}
{Colina}, L. {et~al.} 2001, \apj, 563, 546,
  \adsurl{http://adsabs.harvard.edu/abs/2001ApJ...563..546C}

\bibitem[{{Combes} {et~al.}(1995){Combes}, {Boisse}, {Mazure}, {Blanchard}, \&
  {Seymour}}]{Combes95}	\linkup{autobib:Combes95}
{Combes}, F., {Boisse}, P., {Mazure}, A., {Blanchard}, A., \& {Seymour}, M.
  1995, {Galaxies and Cosmology}, ed. {Combes, F., Boisse, P., Mazure, A.,
  Blanchard, A., \& Seymour, M. },
  \adsurl{http://adsabs.harvard.edu/abs/1995gaco.book.....C}

\bibitem[{{Corbett} {et~al.}(2003){Corbett}, {Kewley}, {Appleton},
  {Charmandaris}, {Dopita}, {Heisler}, {Norris}, {Zezas}, \&
  {Marston}}]{Corbett03}	\linkup{autobib:Corbett03}
{Corbett}, E.~A. {et~al.} 2003, \apj, 583, 670,
  \adsurl{http://adsabs.harvard.edu/abs/2003ApJ...583..670C},
  \eprint{arXiv:astro-ph/0210197}

\bibitem[{{Cox} {et~al.}(2006){Cox}, {Jonsson}, {Primack}, \&
  {Somerville}}]{Cox06}	\linkup{autobib:Cox06}
{Cox}, T.~J., {Jonsson}, P., {Primack}, J.~R., \& {Somerville}, R.~S. 2006,
  \mnras, 373, 1013,
  \adsurl{http://adsabs.harvard.edu/abs/2006MNRAS.373.1013C},
  \eprint{arXiv:astro-ph/0503201}

\bibitem[{{Cox} {et~al.}(2008){Cox}, {Jonsson}, {Somerville}, {Primack}, \&
  {Dekel}}]{Cox08}	\linkup{autobib:Cox08}
{Cox}, T.~J., {Jonsson}, P., {Somerville}, R.~S., {Primack}, J.~R., \& {Dekel},
  A. 2008, \mnras, 384, 386,
  \adsurl{http://adsabs.harvard.edu/abs/2008MNRAS.384..386C},
  \eprint{0709.3511}

\bibitem[{{Cox} {et~al.}(2004){Cox}, {Primack}, {Jonsson}, \&
  {Somerville}}]{Cox04}	\linkup{autobib:Cox04}
{Cox}, T.~J., {Primack}, J., {Jonsson}, P., \& {Somerville}, R.~S. 2004, \apjl,
  607, L87, \adsurl{http://adsabs.harvard.edu/abs/2004ApJ...607L..87C},
  \eprint{arXiv:astro-ph/0402675}

\bibitem[{{Cresci} {et~al.}(2010){Cresci}, {Mannucci}, {Maiolino}, {Marconi},
  {Gnerucci}, \& {Magrini}}]{Cresci10}	\linkup{autobib:Cresci10}
{Cresci}, G., {Mannucci}, F., {Maiolino}, R., {Marconi}, A., {Gnerucci}, A., \&
  {Magrini}, L. 2010, \nat, 467, 811,
  \adsurl{http://adsabs.harvard.edu/abs/2010Natur.467..811C},
  \eprint{1010.2534}

\bibitem[{{Dabringhausen} {et~al.}(2008){Dabringhausen}, {Hilker}, \&
  {Kroupa}}]{Dabringhausen08}	\linkup{autobib:Dabringhausen08}
{Dabringhausen}, J., {Hilker}, M., \& {Kroupa}, P. 2008, \mnras, 386, 864,
  \adsurl{http://adsabs.harvard.edu/abs/2008MNRAS.386..864D},
  \eprint{0802.0703}

\bibitem[{{D'Agostino} \& {Stephens}(1986)}]{Dagostino86}	\linkup{autobib:Dagostino86}
{D'Agostino}, R.~B., \& {Stephens}, M.~A. 1986, {Goodness-of-fit techniques},
  ed. M.~A. D'Agostino, R. B. \&~Stephens,
  \adsurl{http://adsabs.harvard.edu/abs/1986gft..book.....D}

\bibitem[{{Dasyra} {et~al.}(2006){Dasyra}, {Tacconi}, {Davies}, {Naab},
  {Genzel}, {Lutz}, {Sturm}, {Baker}, {Veilleux}, {Sanders}, \&
  {Burkert}}]{Dasyra06}	\linkup{autobib:Dasyra06}
{Dasyra}, K.~M. {et~al.} 2006, \apj, 651, 835,
  \adsurl{http://adsabs.harvard.edu/abs/2006ApJ...651..835D},
  \eprint{arXiv:astro-ph/0607468}

\bibitem[{{de Grijs} {et~al.}(2005){de Grijs}, {Parmentier}, \&
  {Lamers}}]{DeGrijs05}	\linkup{autobib:DeGrijs05}
{de Grijs}, R., {Parmentier}, G., \& {Lamers}, H.~J.~G.~L.~M. 2005, \mnras,
  364, 1054, \adsurl{http://adsabs.harvard.edu/abs/2005MNRAS.364.1054D},
  \eprint{arXiv:astro-ph/0509721}

\bibitem[{{Denicol{\'o}} {et~al.}(2002){Denicol{\'o}}, {Terlevich}, \&
  {Terlevich}}]{Denicolo02}	\linkup{autobib:Denicolo02}
{Denicol{\'o}}, G., {Terlevich}, R., \& {Terlevich}, E. 2002, \mnras, 330, 69,
  \adsurl{http://adsabs.harvard.edu/abs/2002MNRAS.330...69D},
  \eprint{arXiv:astro-ph/0110356}

\bibitem[{{Di Matteo} {et~al.}(2008){Di Matteo}, {Bournaud}, {Martig},
  {Combes}, {Melchior}, \& {Semelin}}]{diMatteo08}	\linkup{autobib:diMatteo08}
{Di Matteo}, P., {Bournaud}, F., {Martig}, M., {Combes}, F., {Melchior}, A., \&
  {Semelin}, B. 2008, \aap, 492, 31,
  \adsurl{http://adsabs.harvard.edu/abs/2008A\%26A...492...31D},
  \eprint{0809.2592}

\bibitem[{{Di Matteo} {et~al.}(2007){Di Matteo}, {Combes}, {Melchior}, \&
  {Semelin}}]{diMatteo07}	\linkup{autobib:diMatteo07}
{Di Matteo}, P., {Combes}, F., {Melchior}, A., \& {Semelin}, B. 2007, \aap,
  468, 61, \adsurl{http://adsabs.harvard.edu/abs/2007A\%26A...468...61D},
  \eprint{arXiv:astro-ph/0703212}

\bibitem[{{Di Matteo} {et~al.}(2011){Di Matteo}, {Montuori}, {Lehnert},
  {Combes}, \& {Semelin}}]{diMatteo11}	\linkup{autobib:diMatteo11}
{Di Matteo}, P., {Montuori}, M., {Lehnert}, M.~D., {Combes}, F., \& {Semelin},
  B. 2011, ArXiv e-prints,
  \adsurl{http://adsabs.harvard.edu/abs/2011arXiv1102.2234D},
  \eprint{1102.2234}

\bibitem[{{D{\'{\i}}az} {et~al.}(2007){D{\'{\i}}az}, {Terlevich},
  {Castellanos}, \& {H{\"a}gele}}]{Diaz07}	\linkup{autobib:Diaz07}
{D{\'{\i}}az}, {\'A}.~I., {Terlevich}, E., {Castellanos}, M., \& {H{\"a}gele},
  G.~F. 2007, \mnras, 382, 251,
  \adsurl{http://adsabs.harvard.edu/abs/2007MNRAS.382..251D},
  \eprint{0709.1236}

\bibitem[{{D{\'{\i}}az-Santos} {et~al.}(2007){D{\'{\i}}az-Santos},
  {Alonso-Herrero}, {Colina}, {Ryder}, \& {Knapen}}]{Diaz-Santos07}	\linkup{autobib:Diaz-Santos07}
{D{\'{\i}}az-Santos}, T., {Alonso-Herrero}, A., {Colina}, L., {Ryder}, S.~D.,
  \& {Knapen}, J.~H. 2007, \apj, 661, 149,
  \adsurl{http://adsabs.harvard.edu/abs/2007ApJ...661..149D},
  \eprint{arXiv:astro-ph/0701557}

\bibitem[{{Dinshaw} {et~al.}(1999){Dinshaw}, {Evans}, {Epps}, {Scoville}, \&
  {Rieke}}]{Dinshaw99}	\linkup{autobib:Dinshaw99}
{Dinshaw}, N., {Evans}, A.~S., {Epps}, H., {Scoville}, N.~Z., \& {Rieke}, M.
  1999, \apj, 525, 702, \adsurl{http://ads.bao.ac.cn/abs/1999ApJ...525..702D},
  \eprint{arXiv:astro-ph/9907395}

\bibitem[{{Dubath} \& {Grillmair}(1997)}]{Dubath97}	\linkup{autobib:Dubath97}
{Dubath}, P., \& {Grillmair}, C.~J. 1997, \aap, 321, 379,
  \adsurl{http://adsabs.harvard.edu/abs/1997A\%26A...321..379D},
  \eprint{arXiv:astro-ph/9610194}

\bibitem[{{Duc} {et~al.}(2004){Duc}, {Bournaud}, \& {Masset}}]{Duc04}	\linkup{autobib:Duc04}
{Duc}, P., {Bournaud}, F., \& {Masset}, F. 2004, \aap, 427, 803,
  \adsurl{http://adsabs.harvard.edu/abs/2004A\%26A...427..803D},
  \eprint{arXiv:astro-ph/0408524}

\bibitem[{{Duc} {et~al.}(2007){Duc}, {Braine}, {Lisenfeld}, {Brinks}, \&
  {Boquien}}]{Duc07}	\linkup{autobib:Duc07}
{Duc}, P., {Braine}, J., {Lisenfeld}, U., {Brinks}, E., \& {Boquien}, M. 2007,
  \aap, 475, 187, \adsurl{http://adsabs.harvard.edu/abs/2007A\%26A...475..187D},
  \eprint{0709.2733}

\bibitem[{{Duc} {et~al.}(2000){Duc}, {Brinks}, {Springel}, {Pichardo},
  {Weilbacher}, \& {Mirabel}}]{Duc00}	\linkup{autobib:Duc00}
{Duc}, P., {Brinks}, E., {Springel}, V., {Pichardo}, B., {Weilbacher}, P., \&
  {Mirabel}, I.~F. 2000, \aj, 120, 1238,
  \adsurl{http://adsabs.harvard.edu/abs/2000AJ....120.1238D},
  \eprint{arXiv:astro-ph/0006038}

\bibitem[{{Duc} {et~al.}(1997){Duc}, {Brinks}, {Wink}, \& {Mirabel}}]{Duc97}	\linkup{autobib:Duc97}
{Duc}, P.-A., {Brinks}, E., {Wink}, J.~E., \& {Mirabel}, I.~F. 1997, \aap, 326,
  537, \adsurl{http://adsabs.harvard.edu/abs/1997A\%26A...326..537D}

\bibitem[{{Duc} \& {Mirabel}(1994)}]{Duc94}	\linkup{autobib:Duc94}
{Duc}, P.-A., \& {Mirabel}, I.~F. 1994, \aap, 289, 83,
  \adsurl{http://adsabs.harvard.edu/abs/1994A\%26A...289...83D}

\bibitem[{{Duc} \& {Mirabel}(1998)}]{Duc98}	\linkup{autobib:Duc98}
{Duc}, P.-A., \& {Mirabel}, I.~F. 1998, \aap, 333, 813,
  \adsurl{http://adsabs.harvard.edu/abs/1998A\%26A...333..813D}

\bibitem[{{Edmunds} \& {Pagel}(1984)}]{Edmunds84}	\linkup{autobib:Edmunds84}
{Edmunds}, M.~G., \& {Pagel}, B.~E.~J. 1984, \mnras, 211, 507,
  \adsurl{http://adsabs.harvard.edu/abs/1984MNRAS.211..507E}

\bibitem[{{Efremov}(1995)}]{Efremov95}	\linkup{autobib:Efremov95}
{Efremov}, Y.~N. 1995, \aj, 110, 2757,
  \adsurl{http://adsabs.harvard.edu/abs/1995AJ....110.2757E}

\bibitem[{{Egami} {et~al.}(2004){Egami}, {Dole}, {Huang}, {P{\'e}rez-Gonzalez},
  {Le Floc'h}, {Papovich}, {Barmby}, {Ivison}, {Serjeant}, {Mortier}, {Frayer},
  \& {Rigopoulou}}]{Egami04}	\linkup{autobib:Egami04}
{Egami}, E. {et~al.} 2004, \apjs, 154, 130,
  \adsurl{http://adsabs.harvard.edu/abs/2004ApJS..154..130E},
  \eprint{arXiv:astro-ph/0406359}

\bibitem[{{Eisenstein} {et~al.}(2005){Eisenstein}, {Zehavi}, {Hogg},
  {Scoccimarro}, {Blanton}, {Nichol}, {Scranton}, {Seo}, {Tegmark}, {Zheng},
  {Anderson}, {Annis}, {Bahcall}, \& {Brinkmann}}]{Eisenstein05}	\linkup{autobib:Eisenstein05}
{Eisenstein}, D.~J. {et~al.} 2005, \apj, 633, 560,
  \adsurl{http://adsabs.harvard.edu/abs/2005ApJ...633..560E},
  \eprint{arXiv:astro-ph/0501171}

\bibitem[{{Ellison} {et~al.}(2008){Ellison}, {Patton}, {Simard}, \&
  {McConnachie}}]{Ellison08}	\linkup{autobib:Ellison08}
{Ellison}, S.~L., {Patton}, D.~R., {Simard}, L., \& {McConnachie}, A.~W. 2008,
  \aj, 135, 1877, \adsurl{http://adsabs.harvard.edu/abs/2008AJ....135.1877E},
  \eprint{0803.0161}

\bibitem[{{Elmegreen}(2002)}]{Elmegreen02}	\linkup{autobib:Elmegreen02}
{Elmegreen}, B.~G. 2002, \apj, 564, 773,
  \adsurl{http://adsabs.harvard.edu/abs/2002ApJ...564..773E},
  \eprint{arXiv:astro-ph/0112528}

\bibitem[{{Elmegreen}(2007)}]{Elmegreen07b}	\linkup{autobib:Elmegreen07b}
{Elmegreen}, B.~G. 2007, \apj, 668, 1064,
  \adsurl{http://adsabs.harvard.edu/abs/2007ApJ...668.1064E},
  \eprint{0707.2252}

\bibitem[{{Elmegreen} \& {Efremov}(1997)}]{Elmegreen97}	\linkup{autobib:Elmegreen97}
{Elmegreen}, B.~G., \& {Efremov}, Y.~N. 1997, \apj, 480, 235,
  \adsurl{http://adsabs.harvard.edu/abs/1997ApJ...480..235E}

\bibitem[{{Elmegreen} \& {Elmegreen}(1983)}]{Elmegreen83}	\linkup{autobib:Elmegreen83}
{Elmegreen}, B.~G., \& {Elmegreen}, D.~M. 1983, \mnras, 203, 31,
  \adsurl{http://adsabs.harvard.edu/abs/1983MNRAS.203...31E}

\bibitem[{{Elmegreen} \& {Elmegreen}(2005)}]{Elmegreen05}	\linkup{autobib:Elmegreen05}
{Elmegreen}, B.~G., \& {Elmegreen}, D.~M. 2005, \apj, 627, 632,
  \adsurl{http://adsabs.harvard.edu/abs/2005ApJ...627..632E},
  \eprint{arXiv:astro-ph/0504032}

\bibitem[{{Elmegreen} \& {Falgarone}(1996)}]{Elmegreen96}	\linkup{autobib:Elmegreen96}
{Elmegreen}, B.~G., \& {Falgarone}, E. 1996, \apj, 471, 816,
  \adsurl{http://adsabs.harvard.edu/abs/1996ApJ...471..816E}

\bibitem[{{Elmegreen} {et~al.}(1993){Elmegreen}, {Kaufman}, \&
  {Thomasson}}]{Elmegreen93}	\linkup{autobib:Elmegreen93}
{Elmegreen}, B.~G., {Kaufman}, M., \& {Thomasson}, M. 1993, \apj, 412, 90,
  \adsurl{http://adsabs.harvard.edu/abs/1993ApJ...412...90E}

\bibitem[{{Elmegreen} \& {Elmegreen}(2006)}]{Elmegreen06}	\linkup{autobib:Elmegreen06}
{Elmegreen}, D.~M., \& {Elmegreen}, B.~G. 2006, \apj, 651, 676,
  \adsurl{http://adsabs.harvard.edu/abs/2006ApJ...651..676E},
  \eprint{arXiv:astro-ph/0607579}

\bibitem[{{Elmegreen} {et~al.}(2007){Elmegreen}, {Elmegreen}, {Ferguson}, \&
  {Mullan}}]{Elmegreen07a}	\linkup{autobib:Elmegreen07a}
{Elmegreen}, D.~M., {Elmegreen}, B.~G., {Ferguson}, T., \& {Mullan}, B. 2007,
  \apj, 663, 734, \adsurl{http://adsabs.harvard.edu/abs/2007ApJ...663..734E},
  \eprint{0704.0911}

\bibitem[{{Elmegreen} {et~al.}(2009){Elmegreen}, {Elmegreen}, {Marcus},
  {Shahinyan}, {Yau}, \& {Petersen}}]{Elmegreen09}	\linkup{autobib:Elmegreen09}
{Elmegreen}, D.~M., {Elmegreen}, B.~G., {Marcus}, M.~T., {Shahinyan}, K.,
  {Yau}, A., \& {Petersen}, M. 2009, \apj, 701, 306,
  \adsurl{http://adsabs.harvard.edu/abs/2009ApJ...701..306E},
  \eprint{0906.2660}

\bibitem[{{Elson} \& {Fall}(1985)}]{Elson85}	\linkup{autobib:Elson85}
{Elson}, R.~A.~W., \& {Fall}, S.~M. 1985, \pasp, 97, 692,
  \adsurl{http://adsabs.harvard.edu/abs/1985PASP...97..692E}

\bibitem[{{Engel} {et~al.}(2010){Engel}, {Tacconi}, {Davies}, {Neri}, {Smail},
  {Chapman}, {Genzel}, {Cox}, {Greve}, {Ivison}, {Blain}, {Bertoldi}, \&
  {Omont}}]{Engel10}	\linkup{autobib:Engel10}
{Engel}, H. {et~al.} 2010, \apj, 724, 233,
  \adsurl{http://adsabs.harvard.edu/abs/2010ApJ...724..233E}

\bibitem[{{Ercolano} {et~al.}(2007){Ercolano}, {Bastian}, \&
  {Stasi{\'n}ska}}]{Ercolano07}	\linkup{autobib:Ercolano07}
{Ercolano}, B., {Bastian}, N., \& {Stasi{\'n}ska}, G. 2007, \mnras, 379, 945,
  \adsurl{http://adsabs.harvard.edu/abs/2007MNRAS.379..945E},
  \eprint{0705.2726}

\bibitem[{{Escala} \& {Larson}(2008)}]{Escala08}	\linkup{autobib:Escala08}
{Escala}, A., \& {Larson}, R.~B. 2008, \apjl, 685, L31,
  \adsurl{http://adsabs.harvard.edu/abs/2008ApJ...685L..31E},
  \eprint{0806.0853}

\bibitem[{{Evans} {et~al.}(2002){Evans}, {Mazzarella}, {Surace}, \&
  {Sanders}}]{Evans02}	\linkup{autobib:Evans02}
{Evans}, A.~S., {Mazzarella}, J.~M., {Surace}, J.~A., \& {Sanders}, D.~B. 2002,
  \apj, 580, 749, \adsurl{http://adsabs.harvard.edu/abs/2002ApJ...580..749E},
  \eprint{arXiv:astro-ph/0208541}

\bibitem[{{Fall} {et~al.}(2005){Fall}, {Chandar}, \& {Whitmore}}]{Fall05}	\linkup{autobib:Fall05}
{Fall}, S.~M., {Chandar}, R., \& {Whitmore}, B.~C. 2005, \apjl, 631, L133,
  \adsurl{http://adsabs.harvard.edu/abs/2005ApJ...631L.133F},
  \eprint{arXiv:astro-ph/0509293}

\bibitem[{{Fall} {et~al.}(2009){Fall}, {Chandar}, \& {Whitmore}}]{Fall09}	\linkup{autobib:Fall09}
{Fall}, S.~M., {Chandar}, R., \& {Whitmore}, B.~C. 2009, \apj, 704, 453,
  \adsurl{http://adsabs.harvard.edu/abs/2009ApJ...704..453F},
  \eprint{0910.1044}

\bibitem[{{Fall} \& {Zhang}(2001)}]{Fall01}	\linkup{autobib:Fall01}
{Fall}, S.~M., \& {Zhang}, Q. 2001, \apj, 561, 751,
  \adsurl{http://adsabs.harvard.edu/abs/2001ApJ...561..751F},
  \eprint{arXiv:astro-ph/0107298}

\bibitem[{{Farrah} {et~al.}(2003){Farrah}, {Afonso}, {Efstathiou},
  {Rowan-Robinson}, {Fox}, \& {Clements}}]{Farrah03}	\linkup{autobib:Farrah03}
{Farrah}, D., {Afonso}, J., {Efstathiou}, A., {Rowan-Robinson}, M., {Fox}, M.,
  \& {Clements}, D. 2003, \mnras, 343, 585,
  \adsurl{http://adsabs.harvard.edu/abs/2003MNRAS.343..585F},
  \eprint{arXiv:astro-ph/0304154}

\bibitem[{{Farrah} {et~al.}(2001){Farrah}, {Rowan-Robinson}, {Oliver},
  {Serjeant}, {Borne}, {Lawrence}, {Lucas}, {Bushouse}, \& {Colina}}]{Farrah01}	\linkup{autobib:Farrah01}
{Farrah}, D. {et~al.} 2001, \mnras, 326, 1333,
  \adsurl{http://adsabs.harvard.edu/abs/2001MNRAS.326.1333F},
  \eprint{arXiv:astro-ph/0106275}

\bibitem[{{Fernandes} {et~al.}(2004){Fernandes}, {de Carvalho}, {Contini}, \&
  {Gal}}]{Fernandes04}	\linkup{autobib:Fernandes04}
{Fernandes}, I.~F., {de Carvalho}, R., {Contini}, T., \& {Gal}, R.~R. 2004,
  \mnras, 355, 728, \adsurl{http://adsabs.harvard.edu/abs/2004MNRAS.355..728F},
  \eprint{arXiv:astro-ph/0409114}

\bibitem[{{Ferreiro} \& {Pastoriza}(2004)}]{Ferreiro04}	\linkup{autobib:Ferreiro04}
{Ferreiro}, D.~L., \& {Pastoriza}, M.~G. 2004, \aap, 428, 837,
  \adsurl{http://adsabs.harvard.edu/abs/2004A\%26A...428..837F},
  \eprint{arXiv:astro-ph/0409150}

\bibitem[{{Ferreiro} {et~al.}(2008){Ferreiro}, {Pastoriza}, \&
  {Rickes}}]{Ferreiro08}	\linkup{autobib:Ferreiro08}
{Ferreiro}, D.~L., {Pastoriza}, M.~G., \& {Rickes}, M. 2008, \aap, 481, 645,
  \adsurl{http://adsabs.harvard.edu/abs/2008A\%26A...481..645F}

\bibitem[{{Fleck} {et~al.}(2006){Fleck}, {Boily}, {Lan{\c c}on}, \&
  {Deiters}}]{Fleck06}	\linkup{autobib:Fleck06}
{Fleck}, J.-J., {Boily}, C.~M., {Lan{\c c}on}, A., \& {Deiters}, S. 2006,
  \mnras, 369, 1392,
  \adsurl{http://adsabs.harvard.edu/abs/2006MNRAS.369.1392F},
  \eprint{arXiv:astro-ph/0604194}

\bibitem[{{Fleck}(1996)}]{Fleck96}	\linkup{autobib:Fleck96}
{Fleck}, Jr., R.~C. 1996, \apj, 458, 739,
  \adsurl{http://adsabs.harvard.edu/abs/1996ApJ...458..739F}

\bibitem[{{Forbes} {et~al.}(2011){Forbes}, {Spitler}, {Graham}, {Foster},
  {Hau}, \& {Benson}}]{Forbes11}	\linkup{autobib:Forbes11}
{Forbes}, D., {Spitler}, L., {Graham}, A., {Foster}, C., {Hau}, G., \&
  {Benson}, A. 2011, ArXiv e-prints,
  \adsurl{http://adsabs.harvard.edu/abs/2011arXiv1101.2460F},
  \eprint{1101.2460}

\bibitem[{{F{\"o}rster Schreiber} {et~al.}(2011{\natexlab{a}}){F{\"o}rster
  Schreiber}, {Shapley}, {Erb}, {Genzel}, {Steidel}, {Bouch{\'e}}, {Cresci}, \&
  {Davies}}]{Schreiber11a}	\linkup{autobib:Schreiber11a}
{F{\"o}rster Schreiber}, N.~M., {Shapley}, A.~E., {Erb}, D.~K., {Genzel}, R.,
  {Steidel}, C.~C., {Bouch{\'e}}, N., {Cresci}, G., \& {Davies}, R.
  2011{\natexlab{a}}, \apj, 731, 65,
  \adsurl{http://adsabs.harvard.edu/abs/2011ApJ...731...65F},
  \eprint{1011.1507}

\bibitem[{{F{\"o}rster Schreiber} {et~al.}(2011{\natexlab{b}}){F{\"o}rster
  Schreiber}, {Shapley}, {Genzel}, {Bouch{\'e}}, {Cresci}, {Davies}, {Erb},
  {Genel}, {Lutz}, {Newman}, {Shapiro}, {Steidel}, {Sternberg}, \&
  {Tacconi}}]{Schreiber11b}	\linkup{autobib:Schreiber11b}
{F{\"o}rster Schreiber}, N.~M. {et~al.} 2011{\natexlab{b}}, ArXiv e-prints,
  \adsurl{http://adsabs.harvard.edu/abs/2011arXiv1104.0248F},
  \eprint{1104.0248}

\bibitem[{{Franceschini} {et~al.}(2001){Franceschini}, {Aussel}, {Cesarsky},
  {Elbaz}, \& {Fadda}}]{Franceschini01}	\linkup{autobib:Franceschini01}
{Franceschini}, A., {Aussel}, H., {Cesarsky}, C.~J., {Elbaz}, D., \& {Fadda},
  D. 2001, \aap, 378, 1,
  \adsurl{http://adsabs.harvard.edu/abs/2001A\%26A...378....1F},
  \eprint{arXiv:astro-ph/0108292}

\bibitem[{{Franceschini} {et~al.}(2005){Franceschini}, {Manners}, {Polletta},
  {Lonsdale}, {Gonzalez-Solares}, {Surace}, {Shupe}, {Fang}, {Xu}, {Farrah},
  {Berta}, {Rodighiero}, \& {Perez-Fournon}}]{Franceschini05}	\linkup{autobib:Franceschini05}
{Franceschini}, A. {et~al.} 2005, \aj, 129, 2074,
  \adsurl{http://adsabs.harvard.edu/abs/2005AJ....129.2074F},
  \eprint{arXiv:astro-ph/0412476}

\bibitem[{{Frayer} {et~al.}(2003){Frayer}, {Armus}, {Scoville}, {Blain},
  {Reddy}, {Ivison}, \& {Smail}}]{Frayer03}	\linkup{autobib:Frayer03}
{Frayer}, D.~T., {Armus}, L., {Scoville}, N.~Z., {Blain}, A.~W., {Reddy},
  N.~A., {Ivison}, R.~J., \& {Smail}, I. 2003, \aj, 126, 73,
  \adsurl{http://adsabs.harvard.edu/abs/2003AJ....126...73F},
  \eprint{arXiv:astro-ph/0304043}

\bibitem[{{Frayer} {et~al.}(2004){Frayer}, {Chapman}, {Yan}, {Armus}, {Helou},
  {Fadda}, {Morganti}, {Garrett}, {Appleton}, {Choi}, {Fang}, {Heinrichsen},
  {Im}, {Lacy}, \& {Marleau}}]{Frayer04}	\linkup{autobib:Frayer04}
{Frayer}, D.~T. {et~al.} 2004, \apjs, 154, 137,
  \adsurl{http://adsabs.harvard.edu/abs/2004ApJS..154..137F},
  \eprint{arXiv:astro-ph/0406351}

\bibitem[{{Gallagher} {et~al.}(2010){Gallagher}, {Durrell}, {Elmegreen},
  {Chandar}, {English}, {Charlton}, {Gronwall}, {Young}, {Tzanavaris},
  {Johnson}, {Mendes de Oliveira}, {Whitmore}, {Hornschemeier}, {Maybhate}, \&
  {Zabludoff}}]{Gallagher10}	\linkup{autobib:Gallagher10}
{Gallagher}, S.~C. {et~al.} 2010, \aj, 139, 545,
  \adsurl{http://adsabs.harvard.edu/abs/2010AJ....139..545G},
  \eprint{1002.3323}

\bibitem[{{Garc{\'i}a-Mar{\'i}n}(2007)}]{Garcia-Marin07}	\linkup{autobib:Garcia-Marin07}
{Garc{\'i}a-Mar{\'i}n}, M. 2007, {2D structure and kinematics of a
  representative sample of low-z ultra-luminous infrared galaxies},
  \eprint{http://www.tesisenred.net/handle/10803/13774}

\bibitem[{{Garc{\'{\i}}a-Mar{\'{\i}}n}
  {et~al.}(2009{\natexlab{a}}){Garc{\'{\i}}a-Mar{\'{\i}}n}, {Colina}, \&
  {Arribas}}]{Garcia-Marin09b}	\linkup{autobib:Garcia-Marin09b}
{Garc{\'{\i}}a-Mar{\'{\i}}n}, M., {Colina}, L., \& {Arribas}, S.
  2009{\natexlab{a}}, \aap, 505, 1017,
  \adsurl{http://adsabs.harvard.edu/abs/2009A\%26A...505.1017G},
  \eprint{0907.2218}

\bibitem[{{Garc{\'{\i}}a-Mar{\'{\i}}n}
  {et~al.}(2009{\natexlab{b}}){Garc{\'{\i}}a-Mar{\'{\i}}n}, {Colina},
  {Arribas}, \& {Monreal-Ibero}}]{Garcia-Marin09a}	\linkup{autobib:Garcia-Marin09a}
{Garc{\'{\i}}a-Mar{\'{\i}}n}, M., {Colina}, L., {Arribas}, S., \&
  {Monreal-Ibero}, A. 2009{\natexlab{b}}, \aap, 505, 1319,
  \adsurl{http://adsabs.harvard.edu/abs/2009A\%26A...505.1319G},
  \eprint{0907.2408}

\bibitem[{{Geha} {et~al.}(2003){Geha}, {Guhathakurta}, \& {van der
  Marel}}]{Geha03}	\linkup{autobib:Geha03}
{Geha}, M., {Guhathakurta}, P., \& {van der Marel}, R.~P. 2003, \aj, 126, 1794,
  \adsurl{http://adsabs.harvard.edu/abs/2003AJ....126.1794G},
  \eprint{arXiv:astro-ph/0304537}

\bibitem[{{Geha} {et~al.}(2009){Geha}, {Willman}, {Simon}, {Strigari}, {Kirby},
  {Law}, \& {Strader}}]{Geha09}	\linkup{autobib:Geha09}
{Geha}, M., {Willman}, B., {Simon}, J.~D., {Strigari}, L.~E., {Kirby}, E.~N.,
  {Law}, D.~R., \& {Strader}, J. 2009, \apj, 692, 1464,
  \adsurl{http://adsabs.harvard.edu/abs/2009ApJ...692.1464G},
  \eprint{0809.2781}

\bibitem[{{Genzel} {et~al.}(1998{\natexlab{a}}){Genzel}, {Lutz}, {Sturm},
  {Egami}, {Kunze}, {Moorwood}, {Rigopoulou}, {Spoon}, {Sternberg},
  {Tacconi-Garman}, {Tacconi}, \& {Thatte}}]{Genzel98a}	\linkup{autobib:Genzel98a}
{Genzel}, R. {et~al.} 1998{\natexlab{a}}, \apj, 498, 579,
  \adsurl{http://adsabs.harvard.edu/abs/1998ApJ...498..579G},
  \eprint{arXiv:astro-ph/9711255}

\bibitem[{{Genzel} {et~al.}(1998{\natexlab{b}}){Genzel}, {Lutz}, \&
  {Tacconi}}]{Genzel98b}	\linkup{autobib:Genzel98b}
{Genzel}, R., {Lutz}, D., \& {Tacconi}, L. 1998{\natexlab{b}}, \nat, 395, 859,
  \adsurl{http://adsabs.harvard.edu/abs/1998Natur.395..859G}

\bibitem[{{Genzel} {et~al.}(2001){Genzel}, {Tacconi}, {Rigopoulou}, {Lutz}, \&
  {Tecza}}]{Genzel01}	\linkup{autobib:Genzel01}
{Genzel}, R., {Tacconi}, L.~J., {Rigopoulou}, D., {Lutz}, D., \& {Tecza}, M.
  2001, \apj, 563, 527,
  \adsurl{http://adsabs.harvard.edu/abs/2001ApJ...563..527G},
  \eprint{arXiv:astro-ph/0106032}

\bibitem[{{Gieles}(2009{\natexlab{a}})}]{Gieles09}	\linkup{autobib:Gieles09}
{Gieles}, M. 2009{\natexlab{a}}, \mnras, 394, 2113,
  \adsurl{http://adsabs.harvard.edu/abs/2009MNRAS.394.2113G},
  \eprint{0901.0830}

\bibitem[{{Gieles}(2009{\natexlab{b}})}]{Gieles09b}	\linkup{autobib:Gieles09b}
{Gieles}, M. 2009{\natexlab{b}}, \apss, 324, 299,
  \adsurl{http://adsabs.harvard.edu/abs/2009Ap\%26SS.324..299G},
  \eprint{0801.2676}

\bibitem[{{Gieles}(2010)}]{Gieles10}	\linkup{autobib:Gieles10}
{Gieles}, M. 2010, in Astronomical Society of the Pacific Conference Series,
  Vol. 423, Galaxy Wars: Stellar Populations and Star Formation in Interacting
  Galaxies, ed. {B.~Smith, J.~Higdon, S.~Higdon, \& N.~Bastian}, 123--+,
  \eprint{0908.2974},
  \adsurl{http://adsabs.harvard.edu/abs/2010ASPC..423..123G}

\bibitem[{{Gieles} {et~al.}(2006){Gieles}, {Larsen}, {Bastian}, \&
  {Stein}}]{Gieles06}	\linkup{autobib:Gieles06}
{Gieles}, M., {Larsen}, S.~S., {Bastian}, N., \& {Stein}, I.~T. 2006, \aap,
  450, 129, \adsurl{http://adsabs.harvard.edu/abs/2006A\%26A...450..129G},
  \eprint{arXiv:astro-ph/0512297}

\bibitem[{{Glass}(1999)}]{Glass99}	\linkup{autobib:Glass99}
{Glass}, I.~S. 1999, {Handbook of Infrared Astronomy}, ed. {Glass, I.~S.},
  \adsurl{http://adsabs.harvard.edu/abs/1999hia..book.....G}

\bibitem[{{Goddard} {et~al.}(2010){Goddard}, {Bastian}, \&
  {Kennicutt}}]{Goddard10}	\linkup{autobib:Goddard10}
{Goddard}, Q.~E., {Bastian}, N., \& {Kennicutt}, R.~C. 2010, \mnras, 491,
  \adsurl{http://adsabs.harvard.edu/abs/2010MNRAS.tmp..491G},
  \eprint{1002.2894}

\bibitem[{{Goldader} {et~al.}(1997){Goldader}, {Joseph}, {Doyon}, \&
  {Sanders}}]{Goldader97}	\linkup{autobib:Goldader97}
{Goldader}, J.~D., {Joseph}, R.~D., {Doyon}, R., \& {Sanders}, D.~B. 1997,
  \apjs, 108, 449, \adsurl{http://adsabs.harvard.edu/abs/1997ApJS..108..449G}

\bibitem[{{Goodwin}(2009)}]{Goodwin09}	\linkup{autobib:Goodwin09}
{Goodwin}, S.~P. 2009, \apss, 324, 259,
  \adsurl{http://adsabs.harvard.edu/abs/2009Ap\%26SS.324..259G},
  \eprint{0802.2207}

\bibitem[{{Goodwin} \& {Bastian}(2006)}]{Goodwin06}	\linkup{autobib:Goodwin06}
{Goodwin}, S.~P., \& {Bastian}, N. 2006, \mnras, 373, 752,
  \adsurl{http://adsabs.harvard.edu/abs/2006MNRAS.373..752G},
  \eprint{arXiv:astro-ph/0609477}

\bibitem[{{Haas} {et~al.}(2008){Haas}, {Gieles}, {Scheepmaker}, {Larsen}, \&
  {Lamers}}]{Haas08}	\linkup{autobib:Haas08}
{Haas}, M.~R., {Gieles}, M., {Scheepmaker}, R.~A., {Larsen}, S.~S., \&
  {Lamers}, H.~J.~G.~L.~M. 2008, \aap, 487, 937,
  \adsurl{http://adsabs.harvard.edu/abs/2008A\%26A...487..937H},
  \eprint{0806.3752}

\bibitem[{{Hancock} {et~al.}(2009){Hancock}, {Smith}, {Struck}, {Giroux}, \&
  {Hurlock}}]{Hancock09}	\linkup{autobib:Hancock09}
{Hancock}, M., {Smith}, B.~J., {Struck}, C., {Giroux}, M.~L., \& {Hurlock}, S.
  2009, \aj, 137, 4643,
  \adsurl{http://adsabs.harvard.edu/abs/2009AJ....137.4643H},
  \eprint{0904.0670}

\bibitem[{{Harris}(2001)}]{Harris01}	\linkup{autobib:Harris01}
{Harris}, W.~E. 2001, in Saas-Fee Advanced Course 28: Star Clusters, ed.
  {L.~Labhardt \& B.~Binggeli}, 223--+,
  \adsurl{http://adsabs.harvard.edu/abs/2001stcl.conf..223H}

\bibitem[{{Heckman} {et~al.}(1990){Heckman}, {Armus}, \& {Miley}}]{Heckman90}	\linkup{autobib:Heckman90}
{Heckman}, T.~M., {Armus}, L., \& {Miley}, G.~K. 1990, \apjs, 74, 833,
  \adsurl{http://adsabs.harvard.edu/abs/1990ApJS...74..833H}

\bibitem[{{Hernquist}(1992{\natexlab{a}})}]{Hernquist92a}	\linkup{autobib:Hernquist92a}
{Hernquist}, L. 1992{\natexlab{a}}, \nat, 360, 105,
  \adsurl{http://adsabs.harvard.edu/abs/1992Natur.360..105H}

\bibitem[{{Hernquist}(1992{\natexlab{b}})}]{Hernquist92b}	\linkup{autobib:Hernquist92b}
{Hernquist}, L. 1992{\natexlab{b}}, \apj, 400, 460,
  \adsurl{http://adsabs.harvard.edu/abs/1992ApJ...400..460H}

\bibitem[{{Hernquist}(1993)}]{Hernquist93}	\linkup{autobib:Hernquist93}
{Hernquist}, L. 1993, \apj, 409, 548,
  \adsurl{http://adsabs.harvard.edu/abs/1993ApJ...409..548H}

\bibitem[{{Hibbard} {et~al.}(1994){Hibbard}, {Guhathakurta}, {van Gorkom}, \&
  {Schweizer}}]{Hibbard94}	\linkup{autobib:Hibbard94}
{Hibbard}, J.~E., {Guhathakurta}, P., {van Gorkom}, J.~H., \& {Schweizer}, F.
  1994, \aj, 107, 67,
  \adsurl{http://adsabs.harvard.edu/abs/1994AJ....107...67H}

\bibitem[{{Hibbard} \& {Mihos}(1995)}]{Hibbard95}	\linkup{autobib:Hibbard95}
{Hibbard}, J.~E., \& {Mihos}, J.~C. 1995, \aj, 110, 140,
  \adsurl{http://adsabs.harvard.edu/abs/1995AJ....110..140H},
  \eprint{arXiv:astro-ph/9503030}

\bibitem[{{Higdon} {et~al.}(2006){Higdon}, {Higdon}, \& {Marshall}}]{Higdon06}	\linkup{autobib:Higdon06}
{Higdon}, S.~J., {Higdon}, J.~L., \& {Marshall}, J. 2006, \apj, 640, 768,
  \adsurl{http://adsabs.harvard.edu/abs/2006ApJ...640..768H},
  \eprint{arXiv:astro-ph/0511422}

\bibitem[{{Holtzman} {et~al.}(1992){Holtzman}, {Faber}, {Shaya}, {Lauer},
  {Groth}, {Hunter}, {Baum}, {Ewald}, {Hester}, \& {Light}}]{Holtzman92}	\linkup{autobib:Holtzman92}
{Holtzman}, J.~A. {et~al.} 1992, \aj, 103, 691,
  \adsurl{http://adsabs.harvard.edu/abs/1992AJ....103..691H}

\bibitem[{{Holtzman} {et~al.}(1995){Holtzman}, {Hester}, {Casertano},
  {Trauger}, {Watson}, {Ballester}, {Burrows}, {Clarke}, {Crisp}, {Evans},
  {Gallagher}, \& {Griffiths}}]{Holtzman95}	\linkup{autobib:Holtzman95}
{Holtzman}, J.~A. {et~al.} 1995, \pasp, 107, 156,
  \adsurl{http://adsabs.harvard.edu/abs/1995PASP..107..156H}

\bibitem[{{Holtzman} {et~al.}(1996){Holtzman}, {Watson}, {Mould}, {Gallagher},
  {Ballester}, {Burrows}, {Clarke}, {Crisp}, \& {Evans}}]{Holtzman96}	\linkup{autobib:Holtzman96}
{Holtzman}, J.~A. {et~al.} 1996, \aj, 112, 416,
  \adsurl{http://adsabs.harvard.edu/abs/1996AJ....112..416H}

\bibitem[{{Howell} {et~al.}(2010){Howell}, {Armus}, {Mazzarella}, {Evans},
  {Surace}, {Sanders}, {Petric}, {Appleton}, {Bothun}, {Bridge}, {Chan}, \&
  {Charmandaris}}]{Howell10}	\linkup{autobib:Howell10}
{Howell}, J.~H. {et~al.} 2010, \apj, 715, 572,
  \adsurl{http://adsabs.harvard.edu/abs/2010ApJ...715..572H},
  \eprint{1004.0985}

\bibitem[{{Hunsberger} {et~al.}(1996){Hunsberger}, {Charlton}, \&
  {Zaritsky}}]{Hunsberger96}	\linkup{autobib:Hunsberger96}
{Hunsberger}, S.~D., {Charlton}, J.~C., \& {Zaritsky}, D. 1996, \apj, 462, 50,
  \adsurl{http://adsabs.harvard.edu/abs/1996ApJ...462...50H},
  \eprint{arXiv:astro-ph/9510160}

\bibitem[{{Hunter} \& {Gallagher}(1986)}]{Hunter86}	\linkup{autobib:Hunter86}
{Hunter}, D.~A., \& {Gallagher}, III, J.~S. 1986, \pasp, 98, 5,
  \adsurl{http://adsabs.harvard.edu/abs/1986PASP...98....5H}

\bibitem[{{Hunter} {et~al.}(2000){Hunter}, {Hunsberger}, \& {Roye}}]{Hunter00}	\linkup{autobib:Hunter00}
{Hunter}, D.~A., {Hunsberger}, S.~D., \& {Roye}, E.~W. 2000, \apj, 542, 137,
  \adsurl{http://adsabs.harvard.edu/abs/2000ApJ...542..137H},
  \eprint{arXiv:astro-ph/0005257}

\bibitem[{{Iglesias-P{\'a}ramo} \& {V{\'{\i}}lchez}(2001)}]{Iglesias-Paramo01}	\linkup{autobib:Iglesias-Paramo01}
{Iglesias-P{\'a}ramo}, J., \& {V{\'{\i}}lchez}, J.~M. 2001, \apj, 550, 204,
  \adsurl{http://adsabs.harvard.edu/abs/2001ApJ...550..204I},
  \eprint{arXiv:astro-ph/0011115}

\bibitem[{{Im} {et~al.}(1999){Im}, {Griffiths}, {Naim}, {Ratnatunga}, {Roche},
  {Green}, \& {Sarajedini}}]{Im99}	\linkup{autobib:Im99}
{Im}, M., {Griffiths}, R.~E., {Naim}, A., {Ratnatunga}, K.~U., {Roche}, N.,
  {Green}, R.~F., \& {Sarajedini}, V.~L. 1999, \apj, 510, 82,
  \adsurl{http://adsabs.harvard.edu/abs/1999ApJ...510...82I},
  \eprint{arXiv:astro-ph/9901424}

\bibitem[{{Inami} {et~al.}(2010){Inami}, {Armus}, {Surace}, {Mazzarella},
  {Evans}, {Sanders}, {Howell}, {Petric}, {Vavilkin}, {Iwasawa}, {Haan},
  {Murphy}, {Stierwalt}, \& {Appleton}}]{Inami10}	\linkup{autobib:Inami10}
{Inami}, H. {et~al.} 2010, \aj, 140, 63,
  \adsurl{http://adsabs.harvard.edu/abs/2010AJ....140...63I},
  \eprint{1004.3543}

\bibitem[{{Ivison} {et~al.}(2004){Ivison}, {Greve}, {Serjeant}, {Bertoldi},
  {Egami}, {Mortier}, {Alonso-Herrero}, {Barmby}, {Bei}, {Dole}, {Engelbracht},
  {Fazio}, \& {Frayer}}]{Ivison04}	\linkup{autobib:Ivison04}
{Ivison}, R.~J. {et~al.} 2004, \apjs, 154, 124,
  \adsurl{http://adsabs.harvard.edu/abs/2004ApJS..154..124I},
  \eprint{arXiv:astro-ph/0406158}

\bibitem[{{Ivison} {et~al.}(2002){Ivison}, {Greve}, {Smail}, {Dunlop}, {Roche},
  {Scott}, {Page}, {Stevens}, {Almaini}, {Blain}, {Willott}, {Fox}, {Gilbank},
  {Serjeant}, \& {Hughes}}]{Ivison02}	\linkup{autobib:Ivison02}
{Ivison}, R.~J. {et~al.} 2002, \mnras, 337, 1,
  \adsurl{http://adsabs.harvard.edu/abs/2002MNRAS.337....1I},
  \eprint{arXiv:astro-ph/0206432}

\bibitem[{{Jones} {et~al.}(2010){Jones}, {Swinbank}, {Ellis}, {Richard}, \&
  {Stark}}]{Jones10}	\linkup{autobib:Jones10}
{Jones}, T.~A., {Swinbank}, A.~M., {Ellis}, R.~S., {Richard}, J., \& {Stark},
  D.~P. 2010, \mnras, 404, 1247,
  \adsurl{http://adsabs.harvard.edu/abs/2010MNRAS.404.1247J},
  \eprint{0910.4488}

\bibitem[{{Jord{\'a}n} {et~al.}(2007){Jord{\'a}n}, {McLaughlin},
  {C{\^o}t{\'e}}, {Ferrarese}, {Peng}, {Mei}, {Villegas}, {Merritt}, {Tonry},
  \& {West}}]{Jordan07}	\linkup{autobib:Jordan07}
{Jord{\'a}n}, A. {et~al.} 2007, \apjs, 171, 101,
  \adsurl{http://adsabs.harvard.edu/abs/2007ApJS..171..101J},
  \eprint{arXiv:astro-ph/0702496}

\bibitem[{{Kapferer} {et~al.}(2005){Kapferer}, {Knapp}, {Schindler},
  {Kimeswenger}, \& {van Kampen}}]{Kapferer05}	\linkup{autobib:Kapferer05}
{Kapferer}, W., {Knapp}, A., {Schindler}, S., {Kimeswenger}, S., \& {van
  Kampen}, E. 2005, \aap, 438, 87,
  \adsurl{http://adsabs.harvard.edu/abs/2005A\%26A...438...87K},
  \eprint{arXiv:astro-ph/0503559}

\bibitem[{{Kassin} {et~al.}(2003){Kassin}, {Frogel}, {Pogge}, {Tiede}, \&
  {Sellgren}}]{Kassin03}	\linkup{autobib:Kassin03}
{Kassin}, S.~A., {Frogel}, J.~A., {Pogge}, R.~W., {Tiede}, G.~P., \&
  {Sellgren}, K. 2003, \aj, 126, 1276,
  \adsurl{http://adsabs.harvard.edu/abs/2003AJ....126.1276K},
  \eprint{arXiv:astro-ph/0306325}

\bibitem[{{Kauffmann} \& {White}(1993)}]{Kauffmann93}	\linkup{autobib:Kauffmann93}
{Kauffmann}, G., \& {White}, S.~D.~M. 1993, \mnras, 261, 921,
  \adsurl{http://adsabs.harvard.edu/abs/1993MNRAS.261..921K}

\bibitem[{{Kaviraj} {et~al.}(2011){Kaviraj}, {Darg}, {Lintott}, {Schawinski},
  \& {Silk}}]{Kaviraj11}	\linkup{autobib:Kaviraj11}
{Kaviraj}, S., {Darg}, D., {Lintott}, C., {Schawinski}, K., \& {Silk}, J. 2011,
  ArXiv e-prints, \adsurl{http://adsabs.harvard.edu/abs/2011arXiv1108.4410K},
  \eprint{1108.4410}

\bibitem[{{Kennicutt}(1998)}]{Kennicutt98}	\linkup{autobib:Kennicutt98}
{Kennicutt}, Jr., R.~C. 1998, \araa, 36, 189,
  \adsurl{http://adsabs.harvard.edu/abs/1998ARA\%26A..36..189K},
  \eprint{arXiv:astro-ph/9807187}

\bibitem[{{Kewley} {et~al.}(2006){Kewley}, {Groves}, {Kauffmann}, \&
  {Heckman}}]{Kewley06}	\linkup{autobib:Kewley06}
{Kewley}, L.~J., {Groves}, B., {Kauffmann}, G., \& {Heckman}, T. 2006, \mnras,
  372, 961, \adsurl{http://adsabs.harvard.edu/abs/2006MNRAS.372..961K},
  \eprint{arXiv:astro-ph/0605681}

\bibitem[{{Kewley} {et~al.}(2001){Kewley}, {Heisler}, {Dopita}, \&
  {Lumsden}}]{Kewley01b}	\linkup{autobib:Kewley01b}
{Kewley}, L.~J., {Heisler}, C.~A., {Dopita}, M.~A., \& {Lumsden}, S. 2001,
  \apjs, 132, 37, \adsurl{http://adsabs.harvard.edu/abs/2001ApJS..132...37K}

\bibitem[{{Kim} {et~al.}(1995){Kim}, {Sanders}, {Veilleux}, {Mazzarella}, \&
  {Soifer}}]{kim95}	\linkup{autobib:kim95}
{Kim}, D., {Sanders}, D.~B., {Veilleux}, S., {Mazzarella}, J.~M., \& {Soifer},
  B.~T. 1995, \apjs, 98, 129,
  \adsurl{http://adsabs.harvard.edu/abs/1995ApJS...98..129K}

\bibitem[{{Kim} {et~al.}(1998){Kim}, {Veilleux}, \& {Sanders}}]{Kim98}	\linkup{autobib:Kim98}
{Kim}, D.-C., {Veilleux}, S., \& {Sanders}, D.~B. 1998, \apj, 508, 627,
  \adsurl{http://adsabs.harvard.edu/abs/1998ApJ...508..627K},
  \eprint{arXiv:astro-ph/9806149}

\bibitem[{{Kissler-Patig} {et~al.}(2006){Kissler-Patig}, {Jord{\'a}n}, \&
  {Bastian}}]{Kissler-Patig06}	\linkup{autobib:Kissler-Patig06}
{Kissler-Patig}, M., {Jord{\'a}n}, A., \& {Bastian}, N. 2006, \aap, 448, 1031,
  \adsurl{http://adsabs.harvard.edu/abs/2006A\%26A...448.1031K},
  \eprint{arXiv:astro-ph/0512360}

\bibitem[{{Klypin} {et~al.}(1999){Klypin}, {Kravtsov}, {Valenzuela}, \&
  {Prada}}]{Klypin99}	\linkup{autobib:Klypin99}
{Klypin}, A., {Kravtsov}, A.~V., {Valenzuela}, O., \& {Prada}, F. 1999, \apj,
  522, 82, \adsurl{http://adsabs.harvard.edu/abs/1999ApJ...522...82K},
  \eprint{arXiv:astro-ph/9901240}

\bibitem[{{Knierman} {et~al.}(2003){Knierman}, {Gallagher}, {Charlton},
  {Hunsberger}, {Whitmore}, {Kundu}, {Hibbard}, \& {Zaritsky}}]{Knierman03}	\linkup{autobib:Knierman03}
{Knierman}, K.~A., {Gallagher}, S.~C., {Charlton}, J.~C., {Hunsberger}, S.~D.,
  {Whitmore}, B., {Kundu}, A., {Hibbard}, J.~E., \& {Zaritsky}, D. 2003, \aj,
  126, 1227, \adsurl{http://adsabs.harvard.edu/abs/2003AJ....126.1227K},
  \eprint{arXiv:astro-ph/0307383}

\bibitem[{{Konstantopoulos} {et~al.}(2009){Konstantopoulos}, {Bastian},
  {Smith}, {Westmoquette}, {Trancho}, \& {Gallagher}}]{Konstantopoulos09}	\linkup{autobib:Konstantopoulos09}
{Konstantopoulos}, I.~S., {Bastian}, N., {Smith}, L.~J., {Westmoquette}, M.~S.,
  {Trancho}, G., \& {Gallagher}, J.~S. 2009, \apj, 701, 1015,
  \adsurl{http://adsabs.harvard.edu/abs/2009ApJ...701.1015K},
  \eprint{0906.2006}

\bibitem[{{Kormendy} \& {Sanders}(1992)}]{Kormendy92}	\linkup{autobib:Kormendy92}
{Kormendy}, J., \& {Sanders}, D.~B. 1992, \apjl, 390, L53,
  \adsurl{http://adsabs.harvard.edu/abs/1992ApJ...390L..53K}

\bibitem[{{Kotilainen} {et~al.}(2001){Kotilainen}, {Reunanen}, {Laine}, \&
  {Ryder}}]{Kotilainen01}	\linkup{autobib:Kotilainen01}
{Kotilainen}, J.~K., {Reunanen}, J., {Laine}, S., \& {Ryder}, S.~D. 2001, \aap,
  366, 439, \adsurl{http://adsabs.harvard.edu/abs/2001A\%26A...366..439K},
  \eprint{arXiv:astro-ph/0011428}

\bibitem[{{Kouwenhoven} \& {de Grijs}(2008)}]{Kouwenhoven08}	\linkup{autobib:Kouwenhoven08}
{Kouwenhoven}, M.~B.~N., \& {de Grijs}, R. 2008, \aap, 480, 103,
  \adsurl{http://adsabs.harvard.edu/abs/2008A\%26A...480..103K},
  \eprint{0712.1748}

\bibitem[{{Kroupa}(1998)}]{Kroupa98}	\linkup{autobib:Kroupa98}
{Kroupa}, P. 1998, \mnras, 300, 200,
  \adsurl{http://adsabs.harvard.edu/abs/1998MNRAS.300..200K},
  \eprint{arXiv:astro-ph/9806206}

\bibitem[{{Kroupa}(2002)}]{Kroupa02}	\linkup{autobib:Kroupa02}
{Kroupa}, P. 2002, Science, 295, 82,
  \adsurl{http://adsabs.harvard.edu/abs/2002Sci...295...82K},
  \eprint{arXiv:astro-ph/0201098}

\bibitem[{{Kumar} {et~al.}(2004){Kumar}, {Kamath}, \& {Davis}}]{Kumar04}	\linkup{autobib:Kumar04}
{Kumar}, M.~S.~N., {Kamath}, U.~S., \& {Davis}, C.~J. 2004, \mnras, 353, 1025,
  \adsurl{http://adsabs.harvard.edu/abs/2004MNRAS.353.1025K},
  \eprint{arXiv:astro-ph/0406639}

\bibitem[{{Kunth} \& {Schild}(1986)}]{Kunth86}	\linkup{autobib:Kunth86}
{Kunth}, D., \& {Schild}, H. 1986, \aap, 169, 71,
  \adsurl{http://adsabs.harvard.edu/abs/1986A\%26A...169...71K}

\bibitem[{{Lada} \& {Lada}(2003)}]{Lada03}	\linkup{autobib:Lada03}
{Lada}, C.~J., \& {Lada}, E.~A. 2003, \araa, 41, 57,
  \adsurl{http://adsabs.harvard.edu/abs/2003ARA\%26A..41...57L},
  \eprint{arXiv:astro-ph/0301540}

\bibitem[{{Lada} {et~al.}(2010){Lada}, {Lombardi}, \& {Alves}}]{Lada10}	\linkup{autobib:Lada10}
{Lada}, C.~J., {Lombardi}, M., \& {Alves}, J.~F. 2010, \apj, 724, 687,
  \adsurl{http://adsabs.harvard.edu/abs/2010ApJ...724..687L},
  \eprint{1009.2985}

\bibitem[{{Lamers} \& {Cassinelli}(1999)}]{Lamers99}	\linkup{autobib:Lamers99}
{Lamers}, H.~J.~G.~L.~M., \& {Cassinelli}, J.~P. 1999, {Introduction to Stellar
  Winds}, ed. {Lamers, H.~J.~G.~L.~M.~\& Cassinelli, J.~P.},
  \adsurl{http://adsabs.harvard.edu/abs/1999isw..book.....L}

\bibitem[{{Larsen}(1999)}]{Larsen99b}	\linkup{autobib:Larsen99b}
{Larsen}, S.~S. 1999, \aaps, 139, 393,
  \adsurl{http://adsabs.harvard.edu/abs/1999A\%26AS..139..393L},
  \eprint{arXiv:astro-ph/9907163}

\bibitem[{{Larsen}(2000)}]{Larsen00}	\linkup{autobib:Larsen00}
{Larsen}, S.~S. 2000, \mnras, 319, 893,
  \adsurl{http://adsabs.harvard.edu/abs/2000MNRAS.319..893L},
  \eprint{arXiv:astro-ph/0008191}

\bibitem[{{Larsen}(2004)}]{Larsen04}	\linkup{autobib:Larsen04}
{Larsen}, S.~S. 2004, \aap, 416, 537,
  \adsurl{http://adsabs.harvard.edu/abs/2004A\%26A...416..537L},
  \eprint{arXiv:astro-ph/0312338}

\bibitem[{{Larsen}(2009)}]{Larsen09}	\linkup{autobib:Larsen09}
{Larsen}, S.~S. 2009, \aap, 494, 539,
  \adsurl{http://adsabs.harvard.edu/abs/2009A\%26A...494..539L},
  \eprint{0812.1400}

\bibitem[{{Larsen}(2010)}]{Larsen10}	\linkup{autobib:Larsen10}
{Larsen}, S.~S. 2010, Royal Society of London Philosophical Transactions Series
  A, 368, 867, \adsurl{http://adsabs.harvard.edu/abs/2010RSPTA.368..867L},
  \eprint{0911.0796}

\bibitem[{{Larsen} \& {Richtler}(1999)}]{Larsen99a}	\linkup{autobib:Larsen99a}
{Larsen}, S.~S., \& {Richtler}, T. 1999, \aap, 345, 59,
  \adsurl{http://adsabs.harvard.edu/abs/1999A\%26A...345...59L},
  \eprint{arXiv:astro-ph/9902227}

\bibitem[{{Larson} \& {Tinsley}(1978)}]{Larson78}	\linkup{autobib:Larson78}
{Larson}, R.~B., \& {Tinsley}, B.~M. 1978, \apj, 219, 46,
  \adsurl{http://adsabs.harvard.edu/abs/1978ApJ...219...46L}

\bibitem[{{Le Floc'h} {et~al.}(2005){Le Floc'h}, {Papovich}, {Dole}, {Bell},
  {Lagache}, {Rieke}, {Egami}, {P{\'e}rez-Gonz{\'a}lez}, {Alonso-Herrero},
  {Rieke}, {Blaylock}, {Engelbracht}, {Gordon}, {Hines}, {Misselt}, \&
  {Morrison}}]{leFloch05}	\linkup{autobib:leFloch05}
{Le Floc'h}, E. {et~al.} 2005, \apj, 632, 169,
  \adsurl{http://adsabs.harvard.edu/abs/2005ApJ...632..169L},
  \eprint{arXiv:astro-ph/0506462}

\bibitem[{{Lee} {et~al.}(2005){Lee}, {Chandar}, \& {Whitmore}}]{Lee05}	\linkup{autobib:Lee05}
{Lee}, M.~G., {Chandar}, R., \& {Whitmore}, B.~C. 2005, \aj, 130, 2128,
  \adsurl{http://adsabs.harvard.edu/abs/2005AJ....130.2128L},
  \eprint{arXiv:astro-ph/0510144}

\bibitem[{{LeFevre} {et~al.}(2003){LeFevre}, {Saisse}, {Mancini}, {Brau-Nogue},
  {Caputi}, {Castinel}, {D'Odorico}, {Garilli}, {Kissler-Patig}, {Lucuix},
  {Mancini}, {Pauget}, {Sciarretta}, {Scodeggio}, {Tresse}, \&
  {Vettolani}}]{leFevre03}	\linkup{autobib:leFevre03}
{LeFevre}, O. {et~al.} 2003, in Society of Photo-Optical Instrumentation
  Engineers (SPIE) Conference Series, Vol. 4841, Society of Photo-Optical
  Instrumentation Engineers (SPIE) Conference Series, ed. M.~{Iye} \& A.~F.~M.
  {Moorwood}, 1670--1681,
  \adsurl{http://adsabs.harvard.edu/abs/2003SPIE.4841.1670L}

\bibitem[{{Leitherer}(1998)}]{Leitherer98}	\linkup{autobib:Leitherer98}
{Leitherer}, C., ed. 1998, {Stellar astrophysics for the local group : VIII
  Canary Islands Winter School of Astrophysics, eds., Aparicio, C., Herrero,
  A., S{\'a}nchez, F., p527},
  \adsurl{http://adsabs.harvard.edu/abs/1998salg.conf.....A}

\bibitem[{{Leitherer} {et~al.}(1999){Leitherer}, {Schaerer}, {Goldader},
  {Gonz{\'a}lez Delgado}, {Robert}, {Kune}, {de Mello}, {Devost}, \&
  {Heckman}}]{Leitherer99}	\linkup{autobib:Leitherer99}
{Leitherer}, C. {et~al.} 1999, \apjs, 123, 3,
  \adsurl{http://adsabs.harvard.edu/abs/1999ApJS..123....3L},
  \eprint{arXiv:astro-ph/9902334}

\bibitem[{{Lonsdale} {et~al.}(2006){Lonsdale}, {Farrah}, \&
  {Smith}}]{Lonsdale06}	\linkup{autobib:Lonsdale06}
{Lonsdale}, C.~J., {Farrah}, D., \& {Smith}, H.~E. 2006, {Ultraluminous
  Infrared Galaxies}, ed. {Mason, J.~W.} (Springer Verlag), 285--+,
  \adsurl{http://adsabs.harvard.edu/abs/2006asup.book..285L}

\bibitem[{{L{\'o}pez-S{\'a}nchez} \& {Esteban}(2009)}]{Lopez-Sanchez09}	\linkup{autobib:Lopez-Sanchez09}
{L{\'o}pez-S{\'a}nchez}, A.~R., \& {Esteban}, C. 2009, \aap, 508, 615,
  \adsurl{http://adsabs.harvard.edu/abs/2009A\%26A...508..615L},
  \eprint{0910.1578}

\bibitem[{{L{\'o}pez-S{\'a}nchez} {et~al.}(2004){L{\'o}pez-S{\'a}nchez},
  {Esteban}, \& {Rodr{\'{\i}}guez}}]{Lopez-Sanchez04}	\linkup{autobib:Lopez-Sanchez04}
{L{\'o}pez-S{\'a}nchez}, {\'A}.~R., {Esteban}, C., \& {Rodr{\'{\i}}guez}, M.
  2004, \aap, 428, 425,
  \adsurl{http://adsabs.harvard.edu/abs/2004A\%26A...428..425L},
  \eprint{arXiv:astro-ph/0409050}

\bibitem[{{Lu} {et~al.}(1993){Lu}, {Hoffman}, {Groff}, {Roos}, \&
  {Lamphier}}]{Lu93}	\linkup{autobib:Lu93}
{Lu}, N.~Y., {Hoffman}, G.~L., {Groff}, T., {Roos}, T., \& {Lamphier}, C. 1993,
  \apjs, 88, 383, \adsurl{http://adsabs.harvard.edu/abs/1993ApJS...88..383L}

\bibitem[{{Mac Low} \& {Klessen}(2004)}]{MacLow04}	\linkup{autobib:MacLow04}
{Mac Low}, M., \& {Klessen}, R.~S. 2004, Reviews of Modern Physics, 76, 125,
  \adsurl{http://adsabs.harvard.edu/abs/2004RvMP...76..125M},
  \eprint{arXiv:astro-ph/0301093}

\bibitem[{{Ma{\'{\i}}z Apell{\'a}niz} \& {{\'U}beda}(2005)}]{Maiz05}	\linkup{autobib:Maiz05}
{Ma{\'{\i}}z Apell{\'a}niz}, J., \& {{\'U}beda}, L. 2005, \apj, 629, 873,
  \adsurl{http://adsabs.harvard.edu/abs/2005ApJ...629..873M},
  \eprint{arXiv:astro-ph/0505012}

\bibitem[{{Maraston}(2005)}]{Maraston05}	\linkup{autobib:Maraston05}
{Maraston}, C. 2005, \mnras, 362, 799,
  \adsurl{http://adsabs.harvard.edu/abs/2005MNRAS.362..799M},
  \eprint{arXiv:astro-ph/0410207}

\bibitem[{{Maraston} {et~al.}(2004){Maraston}, {Bastian}, {Saglia},
  {Kissler-Patig}, {Schweizer}, \& {Goudfrooij}}]{Maraston04}	\linkup{autobib:Maraston04}
{Maraston}, C., {Bastian}, N., {Saglia}, R.~P., {Kissler-Patig}, M.,
  {Schweizer}, F., \& {Goudfrooij}, P. 2004, \aap, 416, 467,
  \adsurl{http://adsabs.harvard.edu/abs/2004A\%26A...416..467M},
  \eprint{arXiv:astro-ph/0311232}

\bibitem[{{Maraston} {et~al.}(2001){Maraston}, {Kissler-Patig}, {Brodie},
  {Barmby}, \& {Huchra}}]{Maraston01}	\linkup{autobib:Maraston01}
{Maraston}, C., {Kissler-Patig}, M., {Brodie}, J.~P., {Barmby}, P., \&
  {Huchra}, J.~P. 2001, \aap, 370, 176,
  \adsurl{http://adsabs.harvard.edu/abs/2001A\%26A...370..176M},
  \eprint{arXiv:astro-ph/0101556}

\bibitem[{{Marconi} {et~al.}(1995){Marconi}, {Tosi}, {Greggio}, \&
  {Focardi}}]{Marconi95}	\linkup{autobib:Marconi95}
{Marconi}, G., {Tosi}, M., {Greggio}, L., \& {Focardi}, P. 1995, \aj, 109, 173,
  \adsurl{http://adsabs.harvard.edu/abs/1995AJ....109..173M},
  \eprint{arXiv:astro-ph/9409013}

\bibitem[{{Marlowe} {et~al.}(1997){Marlowe}, {Meurer}, {Heckman}, \&
  {Schommer}}]{Marlowe97}	\linkup{autobib:Marlowe97}
{Marlowe}, A.~T., {Meurer}, G.~R., {Heckman}, T.~M., \& {Schommer}, R. 1997,
  \apjs, 112, 285, \adsurl{http://adsabs.harvard.edu/abs/1997ApJS..112..285M}

\bibitem[{{Martini} \& {Ho}(2004)}]{Martini04}	\linkup{autobib:Martini04}
{Martini}, P., \& {Ho}, L.~C. 2004, \apj, 610, 233,
  \adsurl{http://adsabs.harvard.edu/abs/2004ApJ...610..233M},
  \eprint{arXiv:astro-ph/0404003}

\bibitem[{{Mateo}(1998)}]{Mateo98}	\linkup{autobib:Mateo98}
{Mateo}, M.~L. 1998, \araa, 36, 435,
  \adsurl{http://adsabs.harvard.edu/abs/1998ARA\%26A..36..435M},
  \eprint{arXiv:astro-ph/9810070}

\bibitem[{{Mayya}(1994)}]{Mayya94}	\linkup{autobib:Mayya94}
{Mayya}, Y.~D. 1994, \aj, 108, 1276,
  \adsurl{http://adsabs.harvard.edu/abs/1994AJ....108.1276M}

\bibitem[{{McDowell} {et~al.}(2003){McDowell}, {Clements}, {Lamb}, {Shaked},
  {Hearn}, {Colina}, {Mundell}, {Borne}, {Baker}, \& {Arribas}}]{McDowell03}	\linkup{autobib:McDowell03}
{McDowell}, J.~C. {et~al.} 2003, \apj, 591, 154,
  \adsurl{http://adsabs.harvard.edu/abs/2003ApJ...591..154M},
  \eprint{arXiv:astro-ph/0303316}

\bibitem[{{McLaughlin} \& {Pudritz}(1996)}]{McLaughlin96}	\linkup{autobib:McLaughlin96}
{McLaughlin}, D.~E., \& {Pudritz}, R.~E. 1996, \apj, 457, 578,
  \adsurl{http://adsabs.harvard.edu/abs/1996ApJ...457..578M}

\bibitem[{{Melnick} \& {Mirabel}(1990)}]{Melnick90}	\linkup{autobib:Melnick90}
{Melnick}, J., \& {Mirabel}, I.~F. 1990, \aap, 231, L19,
  \adsurl{http://adsabs.harvard.edu/abs/1990A\%26A...231L..19M}

\bibitem[{{Melnick} {et~al.}(1999){Melnick}, {Tenorio-Tagle}, \&
  {Terlevich}}]{Melnick99}	\linkup{autobib:Melnick99}
{Melnick}, J., {Tenorio-Tagle}, G., \& {Terlevich}, R. 1999, \mnras, 302, 677,
  \adsurl{http://adsabs.harvard.edu/abs/1999MNRAS.302..677M}

\bibitem[{{Mendes de Oliveira} {et~al.}(2004){Mendes de Oliveira}, {Cypriano},
  {Sodr{\'e}}, \& {Balkowski}}]{Oliveira04}	\linkup{autobib:Oliveira04}
{Mendes de Oliveira}, C., {Cypriano}, E.~S., {Sodr{\'e}}, Jr., L., \&
  {Balkowski}, C. 2004, \apjl, 605, L17,
  \adsurl{http://adsabs.harvard.edu/abs/2004ApJ...605L..17M}

\bibitem[{{Mendes de Oliveira} {et~al.}(2001){Mendes de Oliveira}, {Plana},
  {Amram}, {Balkowski}, \& {Bolte}}]{MendesO01}	\linkup{autobib:MendesO01}
{Mendes de Oliveira}, C., {Plana}, H., {Amram}, P., {Balkowski}, C., \&
  {Bolte}, M. 2001, \aj, 121, 2524,
  \adsurl{http://adsabs.harvard.edu/abs/2001AJ....121.2524M},
  \eprint{arXiv:astro-ph/0101226}

\bibitem[{{Mengel} {et~al.}(2008){Mengel}, {Lehnert}, {Thatte}, {Vacca},
  {Whitmore}, \& {Chandar}}]{Mengel08}	\linkup{autobib:Mengel08}
{Mengel}, S., {Lehnert}, M.~D., {Thatte}, N.~A., {Vacca}, W.~D., {Whitmore},
  B., \& {Chandar}, R. 2008, \aap, 489, 1091,
  \adsurl{http://adsabs.harvard.edu/abs/2008A\%26A...489.1091M},
  \eprint{0805.2559}

\bibitem[{{Meurer} {et~al.}(1995){Meurer}, {Heckman}, {Leitherer}, {Kinney},
  {Robert}, \& {Garnett}}]{Meurer95a}	\linkup{autobib:Meurer95a}
{Meurer}, G.~R., {Heckman}, T.~M., {Leitherer}, C., {Kinney}, A., {Robert}, C.,
  \& {Garnett}, D.~R. 1995, \aj, 110, 2665,
  \adsurl{http://adsabs.harvard.edu/abs/1995AJ....110.2665M},
  \eprint{arXiv:astro-ph/9509038}

\bibitem[{{Michel-Dansac} {et~al.}(2008){Michel-Dansac}, {Lambas}, {Alonso}, \&
  {Tissera}}]{Michel-Dansac08}	\linkup{autobib:Michel-Dansac08}
{Michel-Dansac}, L., {Lambas}, D.~G., {Alonso}, M.~S., \& {Tissera}, P. 2008,
  \mnras, 386, L82, \adsurl{http://adsabs.harvard.edu/abs/2008MNRAS.386L..82M},
  \eprint{0802.3904}

\bibitem[{{Mihos}(1999)}]{Mihos99}	\linkup{autobib:Mihos99}
{Mihos}, C. 1999, \apss, 266, 195,
  \adsurl{http://adsabs.harvard.edu/abs/1999Ap\%26SS.266..195M},
  \eprint{arXiv:astro-ph/9903115}

\bibitem[{{Mihos} \& {Bothun}(1998)}]{Mihos98}	\linkup{autobib:Mihos98}
{Mihos}, J.~C., \& {Bothun}, G.~D. 1998, \apj, 500, 619,
  \adsurl{http://adsabs.harvard.edu/abs/1998ApJ...500..619M}

\bibitem[{{Mihos} {et~al.}(1993){Mihos}, {Bothun}, \& {Richstone}}]{Mihos93}	\linkup{autobib:Mihos93}
{Mihos}, J.~C., {Bothun}, G.~D., \& {Richstone}, D.~O. 1993, \apj, 418, 82,
  \adsurl{http://adsabs.harvard.edu/abs/1993ApJ...418...82M}

\bibitem[{{Mihos} \& {Hernquist}(1994{\natexlab{a}})}]{Mihos94b}	\linkup{autobib:Mihos94b}
{Mihos}, J.~C., \& {Hernquist}, L. 1994{\natexlab{a}}, \apj, 437, 611,
  \adsurl{http://adsabs.harvard.edu/abs/1994ApJ...437..611M}

\bibitem[{{Mihos} \& {Hernquist}(1994{\natexlab{b}})}]{Mihos94a}	\linkup{autobib:Mihos94a}
{Mihos}, J.~C., \& {Hernquist}, L. 1994{\natexlab{b}}, \apjl, 431, L9,
  \adsurl{http://adsabs.harvard.edu/abs/1994ApJ...431L...9M}

\bibitem[{{Mihos} \& {Hernquist}(1996)}]{Mihos96}	\linkup{autobib:Mihos96}
{Mihos}, J.~C., \& {Hernquist}, L. 1996, \apj, 464, 641,
  \adsurl{http://adsabs.harvard.edu/abs/1996ApJ...464..641M},
  \eprint{arXiv:astro-ph/9512099}

\bibitem[{{Miller} {et~al.}(1997){Miller}, {Whitmore}, {Schweizer}, \&
  {Fall}}]{Miller97}	\linkup{autobib:Miller97}
{Miller}, B.~W., {Whitmore}, B.~C., {Schweizer}, F., \& {Fall}, S.~M. 1997,
  \aj, 114, 2381, \adsurl{http://adsabs.harvard.edu/abs/1997AJ....114.2381M}

\bibitem[{{Mirabel} {et~al.}(1992){Mirabel}, {Dottori}, \& {Lutz}}]{Mirabel92}	\linkup{autobib:Mirabel92}
{Mirabel}, I.~F., {Dottori}, H., \& {Lutz}, D. 1992, \aap, 256, L19,
  \adsurl{http://adsabs.harvard.edu/abs/1992A\%26A...256L..19M}

\bibitem[{{Mirabel} {et~al.}(1991){Mirabel}, {Lutz}, \& {Maza}}]{Mirabel91}	\linkup{autobib:Mirabel91}
{Mirabel}, I.~F., {Lutz}, D., \& {Maza}, J. 1991, \aap, 243, 367,
  \adsurl{http://adsabs.harvard.edu/abs/1991A\%26A...243..367M}

\bibitem[{{Monreal-Ibero} {et~al.}(2007){Monreal-Ibero}, {Colina}, {Arribas},
  \& {Garc{\'{\i}}a-Mar{\'{\i}}n}}]{Monreal07}	\linkup{autobib:Monreal07}
{Monreal-Ibero}, A., {Colina}, L., {Arribas}, S., \&
  {Garc{\'{\i}}a-Mar{\'{\i}}n}, M. 2007, \aap, 472, 421,
  \adsurl{http://adsabs.harvard.edu/abs/2007A\%26A...472..421M},
  \eprint{0706.1145}

\bibitem[{{Monreal-Ibero} {et~al.}(2010){Monreal-Ibero}, {V{\'{\i}}lchez},
  {Walsh}, \& {Mu{\~n}oz-Tu{\~n}{\'o}n}}]{Monreal10}	\linkup{autobib:Monreal10}
{Monreal-Ibero}, A., {V{\'{\i}}lchez}, J.~M., {Walsh}, J.~R., \&
  {Mu{\~n}oz-Tu{\~n}{\'o}n}, C. 2010, \aap, 517, A27+,
  \adsurl{http://adsabs.harvard.edu/abs/2010A\%26A...517A..27M},
  \eprint{1003.5329}

\bibitem[{{Montuori} {et~al.}(2010){Montuori}, {Di Matteo}, {Lehnert},
  {Combes}, \& {Semelin}}]{Montuori10}	\linkup{autobib:Montuori10}
{Montuori}, M., {Di Matteo}, P., {Lehnert}, M.~D., {Combes}, F., \& {Semelin},
  B. 2010, \aap, 518, A56+,
  \adsurl{http://adsabs.harvard.edu/abs/2010A\%26A...518A..56M},
  \eprint{1003.1374}

\bibitem[{{Mullan} {et~al.}(2011){Mullan}, {Konstantopoulos}, {Kepley}, {Lee},
  {Charlton}, {Knierman}, {Bastian}, {Chandar}, {Durrell}, {Elmegreen},
  {English}, {Gallagher}, {Gronwall}, {Hibbard}, {Hunsberger}, {Johnson},
  {Maybhate}, {Palma}, {Trancho}, \& {Vacca}}]{Mullan11}	\linkup{autobib:Mullan11}
{Mullan}, B. {et~al.} 2011, \apj, 731, 93,
  \adsurl{http://adsabs.harvard.edu/abs/2011ApJ...731...93M},
  \eprint{1101.5393}

\bibitem[{{Murray} {et~al.}(2010){Murray}, {Quataert}, \&
  {Thompson}}]{Murray10}	\linkup{autobib:Murray10}
{Murray}, N., {Quataert}, E., \& {Thompson}, T.~A. 2010, \apj, 709, 191,
  \adsurl{http://adsabs.harvard.edu/abs/2010ApJ...709..191M},
  \eprint{0906.5358}

\bibitem[{{Naab} \& {Burkert}(2003)}]{Naab03}	\linkup{autobib:Naab03}
{Naab}, T., \& {Burkert}, A. 2003, \apj, 597, 893,
  \adsurl{http://adsabs.harvard.edu/abs/2003ApJ...597..893N},
  \eprint{arXiv:astro-ph/0110179}

\bibitem[{{Naab} {et~al.}(2006){Naab}, {Jesseit}, \& {Burkert}}]{Naab06}	\linkup{autobib:Naab06}
{Naab}, T., {Jesseit}, R., \& {Burkert}, A. 2006, \mnras, 372, 839,
  \adsurl{http://adsabs.harvard.edu/abs/2006MNRAS.372..839N},
  \eprint{arXiv:astro-ph/0605155}

\bibitem[{{Nardini} {et~al.}(2008){Nardini}, {Risaliti}, {Salvati}, {Sani},
  {Imanishi}, {Marconi}, \& {Maiolino}}]{Nardini08}	\linkup{autobib:Nardini08}
{Nardini}, E., {Risaliti}, G., {Salvati}, M., {Sani}, E., {Imanishi}, M.,
  {Marconi}, A., \& {Maiolino}, R. 2008, \mnras, 385, L130,
  \adsurl{http://adsabs.harvard.edu/abs/2008MNRAS.385L.130N}

\bibitem[{{Neugebauer} {et~al.}(1984){Neugebauer}, {Habing}, {van Duinen},
  {Aumann}, {Baud}, {Beichman}, {Beintema}, {Boggess}, {Clegg}, {de Jong},
  {Emerson}, {Gautier}, {Gillett}, \& {Harris}}]{Neugebauer84}	\linkup{autobib:Neugebauer84}
{Neugebauer}, G. {et~al.} 1984, \apjl, 278, L1,
  \adsurl{http://adsabs.harvard.edu/abs/1984ApJ...278L...1N}

\bibitem[{{Nishiura} {et~al.}(2002){Nishiura}, {Shioya}, {Murayama}, {Sato},
  {Nagao}, {Taniguchi}, \& {Sanders}}]{Nishiura02}	\linkup{autobib:Nishiura02}
{Nishiura}, S., {Shioya}, Y., {Murayama}, T., {Sato}, Y., {Nagao}, T.,
  {Taniguchi}, Y., \& {Sanders}, D.~B. 2002, \pasj, 54, 21,
  \adsurl{http://adsabs.harvard.edu/abs/2002PASJ...54...21N},
  \eprint{arXiv:astro-ph/0201178}

\bibitem[{{O'Connell} {et~al.}(1995){O'Connell}, {Gallagher}, {Hunter}, \&
  {Colley}}]{OConnell95}	\linkup{autobib:OConnell95}
{O'Connell}, R.~W., {Gallagher}, III, J.~S., {Hunter}, D.~A., \& {Colley},
  W.~N. 1995, \apjl, 446, L1+,
  \adsurl{http://adsabs.harvard.edu/abs/1995ApJ...446L...1O}

\bibitem[{{O'dell} \& {Townsley}(1988)}]{Odell88}	\linkup{autobib:Odell88}
{O'dell}, C.~R., \& {Townsley}, L.~K. 1988, \aap, 198, 283,
  \adsurl{http://adsabs.harvard.edu/abs/1988A\%26A...198..283O}

\bibitem[{{Okazaki} \& {Taniguchi}(2000)}]{Okazaki00}	\linkup{autobib:Okazaki00}
{Okazaki}, T., \& {Taniguchi}, Y. 2000, \apj, 543, 149,
  \adsurl{http://adsabs.harvard.edu/abs/2000ApJ...543..149O},
  \eprint{arXiv:astro-ph/0006006}

\bibitem[{{Origlia} \& {Leitherer}(2000)}]{Origlia00}	\linkup{autobib:Origlia00}
{Origlia}, L., \& {Leitherer}, C. 2000, \aj, 119, 2018,
  \adsurl{http://adsabs.harvard.edu/abs/2000AJ....119.2018O},
  \eprint{arXiv:astro-ph/0001408}

\bibitem[{{Osterbrock}(1989)}]{Osterbrock89}	\linkup{autobib:Osterbrock89}
{Osterbrock}, D.~E. 1989, {Astrophysics of gaseous nebulae and active galactic
  nuclei}, ed. {Osterbrock, D.~E.},
  \adsurl{http://adsabs.harvard.edu/abs/1989agna.book.....O}

\bibitem[{{Papaderos} {et~al.}(2006){Papaderos}, {Guseva}, {Izotov}, {Noeske},
  {Thuan}, \& {Fricke}}]{Papaderos06}	\linkup{autobib:Papaderos06}
{Papaderos}, P., {Guseva}, N.~G., {Izotov}, Y.~I., {Noeske}, K.~G., {Thuan},
  T.~X., \& {Fricke}, K.~J. 2006, \aap, 457, 45,
  \adsurl{http://adsabs.harvard.edu/abs/2006A\%26A...457...45P},
  \eprint{arXiv:astro-ph/0607443}

\bibitem[{{Peng} {et~al.}(2002){Peng}, {Ho}, {Impey}, \& {Rix}}]{Peng02}	\linkup{autobib:Peng02}
{Peng}, C.~Y., {Ho}, L.~C., {Impey}, C.~D., \& {Rix}, H.-W. 2002, \aj, 124,
  266, \adsurl{http://adsabs.harvard.edu/abs/2002AJ....124..266P},
  \eprint{arXiv:astro-ph/0204182}

\bibitem[{{Peng} {et~al.}(2010){Peng}, {Ho}, {Impey}, \& {Rix}}]{Peng10}	\linkup{autobib:Peng10}
{Peng}, C.~Y., {Ho}, L.~C., {Impey}, C.~D., \& {Rix}, H.-W. 2010, \aj, 139,
  2097, \adsurl{http://adsabs.harvard.edu/abs/2010AJ....139.2097P},
  \eprint{0912.0731}

\bibitem[{{Perez} {et~al.}(2011){Perez}, {Michel-Dansac}, \&
  {Tissera}}]{Perez11}	\linkup{autobib:Perez11}
{Perez}, J., {Michel-Dansac}, L., \& {Tissera}, P. 2011, ArXiv e-prints,
  \adsurl{http://adsabs.harvard.edu/abs/2011arXiv1106.4556P},
  \eprint{1106.4556}

\bibitem[{{P{\'e}rez-Gonz{\'a}lez} {et~al.}(2005){P{\'e}rez-Gonz{\'a}lez},
  {Rieke}, {Egami}, {Alonso-Herrero}, {Dole}, {Papovich}, {Blaylock}, {Jones},
  {Rieke}, {Rigby}, {Barmby}, {Fazio}, {Huang}, \& {Martin}}]{Perez-Gonzalez05}	\linkup{autobib:Perez-Gonzalez05}
{P{\'e}rez-Gonz{\'a}lez}, P.~G. {et~al.} 2005, \apj, 630, 82,
  \adsurl{http://adsabs.harvard.edu/abs/2005ApJ...630...82P},
  \eprint{arXiv:astro-ph/0505101}

\bibitem[{{Peterson} {et~al.}(2009){Peterson}, {Struck}, {Smith}, \&
  {Hancock}}]{Peterson09}	\linkup{autobib:Peterson09}
{Peterson}, B.~W., {Struck}, C., {Smith}, B.~J., \& {Hancock}, M. 2009, \mnras,
  400, 1208, \adsurl{http://adsabs.harvard.edu/abs/2009MNRAS.400.1208P},
  \eprint{0908.2619}

\bibitem[{{Pilyugin} {et~al.}(2004){Pilyugin}, {V{\'{\i}}lchez}, \&
  {Contini}}]{Pilyugin04}	\linkup{autobib:Pilyugin04}
{Pilyugin}, L.~S., {V{\'{\i}}lchez}, J.~M., \& {Contini}, T. 2004, \aap, 425,
  849, \adsurl{http://adsabs.harvard.edu/abs/2004A\%26A...425..849P},
  \eprint{arXiv:astro-ph/0407014}

\bibitem[{{Pirogov}(2009)}]{Pirogov09}	\linkup{autobib:Pirogov09}
{Pirogov}, L.~E. 2009, Astronomy Reports, 53, 1127,
  \adsurl{http://adsabs.harvard.edu/abs/2009ARep...53.1127P},
  \eprint{0911.4421}

\bibitem[{{Piskunov} {et~al.}(2008){Piskunov}, {Kharchenko}, {Schilbach},
  {R{\"o}ser}, {Scholz}, \& {Zinnecker}}]{Piskunov08}	\linkup{autobib:Piskunov08}
{Piskunov}, A.~E., {Kharchenko}, N.~V., {Schilbach}, E., {R{\"o}ser}, S.,
  {Scholz}, R., \& {Zinnecker}, H. 2008, \aap, 487, 557,
  \adsurl{http://adsabs.harvard.edu/abs/2008A\%26A...487..557P},
  \eprint{0806.2217}

\bibitem[{{Porras} {et~al.}(2003){Porras}, {Christopher}, {Allen}, {Di
  Francesco}, {Megeath}, \& {Myers}}]{Porras03}	\linkup{autobib:Porras03}
{Porras}, A., {Christopher}, M., {Allen}, L., {Di Francesco}, J., {Megeath},
  S.~T., \& {Myers}, P.~C. 2003, \aj, 126, 1916,
  \adsurl{http://adsabs.harvard.edu/abs/2003AJ....126.1916P},
  \eprint{arXiv:astro-ph/0307510}

\bibitem[{{Press} {et~al.}(1992){Press}, {Teukolsky}, {Vetterling}, \&
  {Flannery}}]{Press92}	\linkup{autobib:Press92}
{Press}, W.~H., {Teukolsky}, S.~A., {Vetterling}, W.~T., \& {Flannery}, B.~P.
  1992, {Numerical recipes in FORTRAN. The art of scientific computing}, ed.
  {Press, W.~H., Teukolsky, S.~A., Vetterling, W.~T., \& Flannery, B.~P. },
  \adsurl{http://adsabs.harvard.edu/abs/1992nrfa.book.....P}

\bibitem[{{Rela{\~n}o} \& {Beckman}(2005)}]{Relanyo05b}	\linkup{autobib:Relanyo05b}
{Rela{\~n}o}, M., \& {Beckman}, J.~E. 2005, \aap, 430, 911,
  \adsurl{http://adsabs.harvard.edu/abs/2005A\%26A...430..911R},
  \eprint{arXiv:astro-ph/0410415}

\bibitem[{{Rela{\~n}o} {et~al.}(2005){Rela{\~n}o}, {Beckman}, {Zurita},
  {Rozas}, \& {Giammanco}}]{Relanyo05}	\linkup{autobib:Relanyo05}
{Rela{\~n}o}, M., {Beckman}, J.~E., {Zurita}, A., {Rozas}, M., \& {Giammanco},
  C. 2005, \aap, 431, 235,
  \adsurl{http://adsabs.harvard.edu/abs/2005A\%26A...431..235R},
  \eprint{arXiv:astro-ph/0410484}

\bibitem[{{Renaud} {et~al.}(2008){Renaud}, {Boily}, {Fleck}, {Naab}, \&
  {Theis}}]{Renaud08a}	\linkup{autobib:Renaud08a}
{Renaud}, F., {Boily}, C.~M., {Fleck}, J., {Naab}, T., \& {Theis}, C. 2008,
  \mnras, 391, L98, \adsurl{http://adsabs.harvard.edu/abs/2008MNRAS.391L..98R},
  \eprint{0809.2927}

\bibitem[{{Renaud} {et~al.}(2009){Renaud}, {Boily}, {Naab}, \&
  {Theis}}]{Renaud09}	\linkup{autobib:Renaud09}
{Renaud}, F., {Boily}, C.~M., {Naab}, T., \& {Theis}, C. 2009, \apj, 706, 67,
  \adsurl{http://adsabs.harvard.edu/abs/2009ApJ...706...67R},
  \eprint{0910.0196}

\bibitem[{{Risaliti} {et~al.}(2006){Risaliti}, {Maiolino}, {Marconi}, {Sani},
  {Berta}, {Braito}, {Ceca}, {Franceschini}, \& {Salvati}}]{Risaliti06}	\linkup{autobib:Risaliti06}
{Risaliti}, G. {et~al.} 2006, \mnras, 365, 303,
  \adsurl{http://adsabs.harvard.edu/abs/2006MNRAS.365..303R},
  \eprint{arXiv:astro-ph/0510282}

\bibitem[{{Rodighiero} {et~al.}(2011){Rodighiero}, {Daddi}, {Baronchelli},
  {Cimatti}, {Renzini}, {Aussel}, {Popesso}, {Lutz}, {Andreani}, {Berta}, \&
  {Cava}}]{Rodighiero11}	\linkup{autobib:Rodighiero11}
{Rodighiero}, G. {et~al.} 2011, ArXiv e-prints,
  \adsurl{http://adsabs.harvard.edu/abs/2011arXiv1108.0933R},
  \eprint{1108.0933}

\bibitem[{{Rodr{\'{\i}}guez-Zaur{\'{\i}}n}
  {et~al.}(2011){Rodr{\'{\i}}guez-Zaur{\'{\i}}n}, {Arribas}, {Monreal-Ibero},
  {Colina}, {Alonso-Herrero}, \& {Alfonso-Garz{\'o}n}}]{Rodriguez-Zaurin10}	\linkup{autobib:Rodriguez-Zaurin10}
{Rodr{\'{\i}}guez-Zaur{\'{\i}}n}, J., {Arribas}, S., {Monreal-Ibero}, A.,
  {Colina}, L., {Alonso-Herrero}, A., \& {Alfonso-Garz{\'o}n}, J. 2011, \aap,
  527, A60+, \adsurl{http://adsabs.harvard.edu/abs/2011A\%26A...527A..60R},
  \eprint{1009.0112}

\bibitem[{{Roth} {et~al.}(2005){Roth}, {Kelz}, {Fechner}, {Hahn}, {Bauer},
  {Becker}, {B{\"o}hm}, {Christensen}, {Dionies}, {Paschke}, {Popow}, {Wolter},
  {Schmoll}, {Laux}, \& {Altmann}}]{Roth05}	\linkup{autobib:Roth05}
{Roth}, M.~M. {et~al.} 2005, \pasp, 117, 620,
  \adsurl{http://adsabs.harvard.edu/abs/2005PASP..117..620R},
  \eprint{arXiv:astro-ph/0502581}

\bibitem[{{Rozas} {et~al.}(2006{\natexlab{a}}){Rozas}, {Richer}, {L{\'o}pez},
  {Rela{\~n}o}, \& {Beckman}}]{Rozas06}	\linkup{autobib:Rozas06}
{Rozas}, M., {Richer}, M.~G., {L{\'o}pez}, J.~A., {Rela{\~n}o}, M., \&
  {Beckman}, J.~E. 2006{\natexlab{a}}, \aap, 455, 539,
  \adsurl{http://adsabs.harvard.edu/abs/2006A\%26A...455..539R}

\bibitem[{{Rozas} {et~al.}(2006{\natexlab{b}}){Rozas}, {Richer}, {L{\'o}pez},
  {Rela{\~n}o}, \& {Beckman}}]{Rozas06b}	\linkup{autobib:Rozas06b}
{Rozas}, M., {Richer}, M.~G., {L{\'o}pez}, J.~A., {Rela{\~n}o}, M., \&
  {Beckman}, J.~E. 2006{\natexlab{b}}, \aap, 455, 549,
  \adsurl{http://adsabs.harvard.edu/abs/2006A\%26A...455..549R}

\bibitem[{{Rupke} {et~al.}(2010){Rupke}, {Kewley}, \& {Barnes}}]{Rupke10}	\linkup{autobib:Rupke10}
{Rupke}, D.~S.~N., {Kewley}, L.~J., \& {Barnes}, J.~E. 2010, \apjl, 710, L156,
  \adsurl{http://adsabs.harvard.edu/abs/2010ApJ...710L.156R},
  \eprint{1001.1728}

\bibitem[{{Rupke} {et~al.}(2008){Rupke}, {Veilleux}, \& {Baker}}]{Rupke08}	\linkup{autobib:Rupke08}
{Rupke}, D.~S.~N., {Veilleux}, S., \& {Baker}, A.~J. 2008, \apj, 674, 172,
  \adsurl{http://adsabs.harvard.edu/abs/2008ApJ...674..172R},
  \eprint{0708.1766}

\bibitem[{{Sanders} {et~al.}(2003){Sanders}, {Mazzarella}, {Kim}, {Surace}, \&
  {Soifer}}]{Sanders03}	\linkup{autobib:Sanders03}
{Sanders}, D.~B., {Mazzarella}, J.~M., {Kim}, D., {Surace}, J.~A., \& {Soifer},
  B.~T. 2003, \aj, 126, 1607,
  \adsurl{http://adsabs.harvard.edu/abs/2003AJ....126.1607S},
  \eprint{arXiv:astro-ph/0306263}

\bibitem[{{Sanders} \& {Mirabel}(1996)}]{Sanders96}	\linkup{autobib:Sanders96}
{Sanders}, D.~B., \& {Mirabel}, I.~F. 1996, \araa, 34, 749,
  \adsurl{http://adsabs.harvard.edu/abs/1996ARA\%26A..34..749S}

\bibitem[{{Sanders} {et~al.}(1988){Sanders}, {Soifer}, {Elias}, {Madore},
  {Matthews}, {Neugebauer}, \& {Scoville}}]{Sanders88b}	\linkup{autobib:Sanders88b}
{Sanders}, D.~B., {Soifer}, B.~T., {Elias}, J.~H., {Madore}, B.~F., {Matthews},
  K., {Neugebauer}, G., \& {Scoville}, N.~Z. 1988, \apj, 325, 74,
  \adsurl{http://adsabs.harvard.edu/abs/1988ApJ...325...74S}

\bibitem[{{Santiago-Cort{\'e}s} {et~al.}(2010){Santiago-Cort{\'e}s}, {Mayya},
  \& {Rosa-Gonz{\'a}lez}}]{Santiago-Cortes10}	\linkup{autobib:Santiago-Cortes10}
{Santiago-Cort{\'e}s}, M., {Mayya}, Y.~D., \& {Rosa-Gonz{\'a}lez}, D. 2010,
  \mnras, 496, \adsurl{http://adsabs.harvard.edu/abs/2010MNRAS.tmp..496S},
  \eprint{1002.3370}

\bibitem[{{Schechter}(1976)}]{Schechter76}	\linkup{autobib:Schechter76}
{Schechter}, P. 1976, \apj, 203, 297,
  \adsurl{http://adsabs.harvard.edu/abs/1976ApJ...203..297S}

\bibitem[{{Schlegel} {et~al.}(1998){Schlegel}, {Finkbeiner}, \&
  {Davis}}]{Schlegel98}	\linkup{autobib:Schlegel98}
{Schlegel}, D.~J., {Finkbeiner}, D.~P., \& {Davis}, M. 1998, \apj, 500, 525,
  \adsurl{http://adsabs.harvard.edu/abs/1998ApJ...500..525S},
  \eprint{arXiv:astro-ph/9710327}

\bibitem[{{Schweizer}(1978)}]{Schweizer78}	\linkup{autobib:Schweizer78}
{Schweizer}, F. 1978, in IAU Symposium, Vol.~77, Structure and Properties of
  Nearby Galaxies, ed. {E.~M.~Berkhuijsen \& R.~Wielebinski}, 279--284,
  \adsurl{http://adsabs.harvard.edu/abs/1978IAUS...77..279S}

\bibitem[{{Schweizer} {et~al.}(1996){Schweizer}, {Miller}, {Whitmore}, \&
  {Fall}}]{Schweizer96}	\linkup{autobib:Schweizer96}
{Schweizer}, F., {Miller}, B.~W., {Whitmore}, B.~C., \& {Fall}, S.~M. 1996,
  \aj, 112, 1839, \adsurl{http://adsabs.harvard.edu/abs/1996AJ....112.1839S}

\bibitem[{{Schweizer} \& {Seitzer}(1998)}]{Schweizer98}	\linkup{autobib:Schweizer98}
{Schweizer}, F., \& {Seitzer}, P. 1998, \aj, 116, 2206,
  \adsurl{http://adsabs.harvard.edu/abs/1998AJ....116.2206S},
  \eprint{arXiv:astro-ph/9809026}

\bibitem[{{Scoville} {et~al.}(2000){Scoville}, {Evans}, {Thompson}, {Rieke},
  {Hines}, {Low}, {Dinshaw}, {Surace}, \& {Armus}}]{Scoville00}	\linkup{autobib:Scoville00}
{Scoville}, N.~Z. {et~al.} 2000, \aj, 119, 991,
  \adsurl{http://adsabs.harvard.edu/abs/2000AJ....119..991S},
  \eprint{arXiv:astro-ph/9912246}

\bibitem[{{Sheen} {et~al.}(2009){Sheen}, {Jeong}, {Yi}, {Ferreras}, {Lotz},
  {Olsen}, {Dickinson}, {Barnes}, {Park}, {Ree}, {Madore}, {Barlow}, \&
  {Conrow}}]{Sheen09}	\linkup{autobib:Sheen09}
{Sheen}, Y. {et~al.} 2009, \aj, 138, 1911,
  \adsurl{http://adsabs.harvard.edu/abs/2009AJ....138.1911S},
  \eprint{0910.2351}

\bibitem[{{Simon} \& {Geha}(2007)}]{Simon07}	\linkup{autobib:Simon07}
{Simon}, J.~D., \& {Geha}, M. 2007, \apj, 670, 313,
  \adsurl{http://adsabs.harvard.edu/abs/2007ApJ...670..313S},
  \eprint{0706.0516}

\bibitem[{{Sirianni} {et~al.}(2005){Sirianni}, {Jee}, {Ben{\'{\i}}tez},
  {Blakeslee}, {Martel}, {Meurer}, {Clampin}, {De Marchi}, {Ford}, {Gilliland},
  {Hartig}, {Illingworth}, {Mack}, \& {McCann}}]{Sirianni05}	\linkup{autobib:Sirianni05}
{Sirianni}, M. {et~al.} 2005, \pasp, 117, 1049,
  \adsurl{http://adsabs.harvard.edu/abs/2005PASP..117.1049S},
  \eprint{arXiv:astro-ph/0507614}

\bibitem[{{Smail} {et~al.}(1997){Smail}, {Ivison}, \& {Blain}}]{Smail97}	\linkup{autobib:Smail97}
{Smail}, I., {Ivison}, R.~J., \& {Blain}, A.~W. 1997, \apjl, 490, L5+,
  \adsurl{http://adsabs.harvard.edu/abs/1997ApJ...490L...5S},
  \eprint{arXiv:astro-ph/9708135}

\bibitem[{{Smail} {et~al.}(2002){Smail}, {Ivison}, {Blain}, \&
  {Kneib}}]{Smail02}	\linkup{autobib:Smail02}
{Smail}, I., {Ivison}, R.~J., {Blain}, A.~W., \& {Kneib}, J. 2002, \mnras, 331,
  495, \adsurl{http://adsabs.harvard.edu/abs/2002MNRAS.331..495S},
  \eprint{arXiv:astro-ph/0112100}

\bibitem[{{Soifer} {et~al.}(1989){Soifer}, {Boehmer}, {Neugebauer}, \&
  {Sanders}}]{Soifer89}	\linkup{autobib:Soifer89}
{Soifer}, B.~T., {Boehmer}, L., {Neugebauer}, G., \& {Sanders}, D.~B. 1989,
  \aj, 98, 766, \adsurl{http://adsabs.harvard.edu/abs/1989AJ.....98..766S}

\bibitem[{{Soifer} {et~al.}(1987){Soifer}, {Sanders}, {Madore}, {Neugebauer},
  {Danielson}, {Elias}, {Lonsdale}, \& {Rice}}]{Soifer87}	\linkup{autobib:Soifer87}
{Soifer}, B.~T., {Sanders}, D.~B., {Madore}, B.~F., {Neugebauer}, G.,
  {Danielson}, G.~E., {Elias}, J.~H., {Lonsdale}, C.~J., \& {Rice}, W.~L. 1987,
  \apj, 320, 238, \adsurl{http://adsabs.harvard.edu/abs/1987ApJ...320..238S}

\bibitem[{{Solomon} {et~al.}(1987){Solomon}, {Rivolo}, {Barrett}, \&
  {Yahil}}]{Solomon87}	\linkup{autobib:Solomon87}
{Solomon}, P.~M., {Rivolo}, A.~R., {Barrett}, J., \& {Yahil}, A. 1987, \apj,
  319, 730, \adsurl{http://adsabs.harvard.edu/abs/1987ApJ...319..730S}

\bibitem[{{Spergel} {et~al.}(2007){Spergel}, {Bean}, {Dor{\'e}}, {Nolta},
  {Bennett}, {Dunkley}, {Hinshaw}, {Jarosik}, {Komatsu}, {Page}, {Peiris},
  {Verde}, {Halpern}, {Hill}, \& {Kogut}}]{Spergel07}	\linkup{autobib:Spergel07}
{Spergel}, D.~N. {et~al.} 2007, \apjs, 170, 377,
  \adsurl{http://adsabs.harvard.edu/abs/2007ApJS..170..377S},
  \eprint{arXiv:astro-ph/0603449}

\bibitem[{{Spitzer}(1987)}]{Spitzer87}	\linkup{autobib:Spitzer87}
{Spitzer}, L. 1987, {Dynamical evolution of globular clusters}, ed. {Spitzer,
  L.}, \adsurl{http://adsabs.harvard.edu/abs/1987degc.book.....S}

\bibitem[{{Spoon} {et~al.}(2007){Spoon}, {Marshall}, {Houck}, {Elitzur}, {Hao},
  {Armus}, {Brandl}, \& {Charmandaris}}]{Spoon07}	\linkup{autobib:Spoon07}
{Spoon}, H.~W.~W., {Marshall}, J.~A., {Houck}, J.~R., {Elitzur}, M., {Hao}, L.,
  {Armus}, L., {Brandl}, B.~R., \& {Charmandaris}, V. 2007, \apjl, 654, L49,
  \adsurl{http://adsabs.harvard.edu/abs/2007ApJ...654L..49S},
  \eprint{arXiv:astro-ph/0611918}

\bibitem[{{Springel}(2000)}]{Springel00}	\linkup{autobib:Springel00}
{Springel}, V. 2000, \mnras, 312, 859,
  \adsurl{http://adsabs.harvard.edu/abs/2000MNRAS.312..859S}

\bibitem[{{Springel} {et~al.}(2005){Springel}, {Di Matteo}, \&
  {Hernquist}}]{Springel05a}	\linkup{autobib:Springel05a}
{Springel}, V., {Di Matteo}, T., \& {Hernquist}, L. 2005, \mnras, 361, 776,
  \adsurl{http://adsabs.harvard.edu/abs/2005MNRAS.361..776S},
  \eprint{arXiv:astro-ph/0411108}

\bibitem[{{Springel} \& {Hernquist}(2005)}]{Springel05b}	\linkup{autobib:Springel05b}
{Springel}, V., \& {Hernquist}, L. 2005, \apjl, 622, L9,
  \adsurl{http://adsabs.harvard.edu/abs/2005ApJ...622L...9S},
  \eprint{arXiv:astro-ph/0411379}

\bibitem[{{Stanford} \& {Bushouse}(1991)}]{Stanford91}	\linkup{autobib:Stanford91}
{Stanford}, S.~A., \& {Bushouse}, H.~A. 1991, \apj, 371, 92,
  \adsurl{http://adsabs.harvard.edu/abs/1991ApJ...371...92S}

\bibitem[{{Surace} {et~al.}(2000){Surace}, {Sanders}, \& {Evans}}]{Surace00}	\linkup{autobib:Surace00}
{Surace}, J.~A., {Sanders}, D.~B., \& {Evans}, A.~S. 2000, \apj, 529, 170,
  \adsurl{http://adsabs.harvard.edu/abs/2000ApJ...529..170S},
  \eprint{arXiv:astro-ph/9909085}

\bibitem[{{Surace} {et~al.}(1998){Surace}, {Sanders}, {Vacca}, {Veilleux}, \&
  {Mazzarella}}]{Surace98}	\linkup{autobib:Surace98}
{Surace}, J.~A., {Sanders}, D.~B., {Vacca}, W.~D., {Veilleux}, S., \&
  {Mazzarella}, J.~M. 1998, \apj, 492, 116,
  \adsurl{http://adsabs.harvard.edu/abs/1998ApJ...492..116S}

\bibitem[{{Tacconi} {et~al.}(2002){Tacconi}, {Genzel}, {Lutz}, {Rigopoulou},
  {Baker}, {Iserlohe}, \& {Tecza}}]{Tacconi02}	\linkup{autobib:Tacconi02}
{Tacconi}, L.~J., {Genzel}, R., {Lutz}, D., {Rigopoulou}, D., {Baker}, A.~J.,
  {Iserlohe}, C., \& {Tecza}, M. 2002, \apj, 580, 73,
  \adsurl{http://adsabs.harvard.edu/abs/2002ApJ...580...73T},
  \eprint{arXiv:astro-ph/0207405}

\bibitem[{{Tacconi} {et~al.}(2006){Tacconi}, {Neri}, {Chapman}, {Genzel},
  {Smail}, {Ivison}, {Bertoldi}, {Blain}, {Cox}, {Greve}, \&
  {Omont}}]{Tacconi06}	\linkup{autobib:Tacconi06}
{Tacconi}, L.~J. {et~al.} 2006, \apj, 640, 228,
  \adsurl{http://adsabs.harvard.edu/abs/2006ApJ...640..228T},
  \eprint{arXiv:astro-ph/0511319}

\bibitem[{{Temporin} {et~al.}(2003){Temporin}, {Weinberger}, {Galaz}, \&
  {Kerber}}]{Temporin03}	\linkup{autobib:Temporin03}
{Temporin}, S., {Weinberger}, R., {Galaz}, G., \& {Kerber}, F. 2003, \apj, 587,
  660, \adsurl{http://adsabs.harvard.edu/abs/2003ApJ...587..660T},
  \eprint{arXiv:astro-ph/0301105}

\bibitem[{{Terlevich} \& {Melnick}(1981)}]{Terlevich81}	\linkup{autobib:Terlevich81}
{Terlevich}, R., \& {Melnick}, J. 1981, \mnras, 195, 839,
  \adsurl{http://adsabs.harvard.edu/abs/1981MNRAS.195..839T}

\bibitem[{{Teyssier} {et~al.}(2010){Teyssier}, {Chapon}, \&
  {Bournaud}}]{Teyssier10}	\linkup{autobib:Teyssier10}
{Teyssier}, R., {Chapon}, D., \& {Bournaud}, F. 2010, \apjl, 720, L149,
  \adsurl{http://adsabs.harvard.edu/abs/2010ApJ...720L.149T},
  \eprint{1006.4757}

\bibitem[{{Tissera} {et~al.}(2002){Tissera}, {Dom{\'{\i}}nguez-Tenreiro},
  {Scannapieco}, \& {S{\'a}iz}}]{Tissera02}	\linkup{autobib:Tissera02}
{Tissera}, P.~B., {Dom{\'{\i}}nguez-Tenreiro}, R., {Scannapieco}, C., \&
  {S{\'a}iz}, A. 2002, \mnras, 333, 327,
  \adsurl{http://adsabs.harvard.edu/abs/2002MNRAS.333..327T},
  \eprint{arXiv:astro-ph/0202160}

\bibitem[{{Tody}(1993)}]{Tody93}	\linkup{autobib:Tody93}
{Tody}, D. 1993, in Astronomical Society of the Pacific Conference Series,
  Vol.~52, Astronomical Data Analysis Software and Systems II, ed. R.~J.
  {Hanisch}, R.~J.~V. {Brissenden}, \& J.~{Barnes}, 173--+,
  \adsurl{http://adsabs.harvard.edu/abs/1993ASPC...52..173T}

\bibitem[{{Tolstoy} {et~al.}(2009){Tolstoy}, {Hill}, \& {Tosi}}]{Tolstoy09}	\linkup{autobib:Tolstoy09}
{Tolstoy}, E., {Hill}, V., \& {Tosi}, M. 2009, \araa, 47, 371,
  \adsurl{http://adsabs.harvard.edu/abs/2009ARA\%26A..47..371T},
  \eprint{0904.4505}

\bibitem[{{Toomre}(1977)}]{Toomre77}	\linkup{autobib:Toomre77}
{Toomre}, A. 1977, in Evolution of Galaxies and Stellar Populations, ed.
  {B.~M.~Tinsley \& R.~B.~Larson}, 401--+,
  \adsurl{http://adsabs.harvard.edu/abs/1977egsp.conf..401T}

\bibitem[{{Toomre} \& {Toomre}(1972)}]{Toomre72}	\linkup{autobib:Toomre72}
{Toomre}, A., \& {Toomre}, J. 1972, \apj, 178, 623,
  \adsurl{http://adsabs.harvard.edu/abs/1972ApJ...178..623T}

\bibitem[{{Torres-Peimbert} {et~al.}(1989){Torres-Peimbert}, {Peimbert}, \&
  {Fierro}}]{Torres-Peimbert89}	\linkup{autobib:Torres-Peimbert89}
{Torres-Peimbert}, S., {Peimbert}, M., \& {Fierro}, J. 1989, \apj, 345, 186,
  \adsurl{http://adsabs.harvard.edu/abs/1989ApJ...345..186T}

\bibitem[{{Trager} {et~al.}(1993){Trager}, {Djorgovski}, \& {King}}]{Trager93}	\linkup{autobib:Trager93}
{Trager}, S.~C., {Djorgovski}, S., \& {King}, I.~R. 1993, in Astronomical
  Society of the Pacific Conference Series, Vol.~50, Structure and Dynamics of
  Globular Clusters, ed. {S.~G.~Djorgovski \& G.~Meylan}, 347--+,
  \adsurl{http://adsabs.harvard.edu/abs/1993ASPC...50..347T}

\bibitem[{{Tremonti} {et~al.}(2004){Tremonti}, {Heckman}, {Kauffmann},
  {Brinchmann}, {Charlot}, {White}, {Seibert}, {Peng}, {Schlegel}, {Uomoto},
  {Fukugita}, \& {Brinkmann}}]{Tremonti04}	\linkup{autobib:Tremonti04}
{Tremonti}, C.~A. {et~al.} 2004, \apj, 613, 898,
  \adsurl{http://adsabs.harvard.edu/abs/2004ApJ...613..898T},
  \eprint{arXiv:astro-ph/0405537}

\bibitem[{{Tuffs} \& {Popescu}(2005)}]{Tuffs05}	\linkup{autobib:Tuffs05}
{Tuffs}, R.~J., \& {Popescu}, C.~C. 2005, in American Institute of Physics
  Conference Series, Vol. 761, The Spectral Energy Distributions of Gas-Rich
  Galaxies: Confronting Models with Data, ed. {C.~C.~Popescu \& R.~J.~Tuffs},
  344--363, \eprint{arXiv:astro-ph/0502179},
  \adsurl{http://adsabs.harvard.edu/abs/2005AIPC..761..344T}

\bibitem[{{Vacca} {et~al.}(1996){Vacca}, {Garmany}, \& {Shull}}]{Vacca96}	\linkup{autobib:Vacca96}
{Vacca}, W.~D., {Garmany}, C.~D., \& {Shull}, J.~M. 1996, \apj, 460, 914,
  \adsurl{http://adsabs.harvard.edu/abs/1996ApJ...460..914V}

\bibitem[{{van den Bergh} {et~al.}(1996){van den Bergh}, {Abraham}, {Ellis},
  {Tanvir}, {Santiago}, \& {Glazebrook}}]{Vandenbergh96}	\linkup{autobib:Vandenbergh96}
{van den Bergh}, S., {Abraham}, R.~G., {Ellis}, R.~S., {Tanvir}, N.~R.,
  {Santiago}, B.~X., \& {Glazebrook}, K.~G. 1996, \aj, 112, 359,
  \adsurl{http://adsabs.harvard.edu/abs/1996AJ....112..359V},
  \eprint{arXiv:astro-ph/9604161}

\bibitem[{{van Zee} \& {Haynes}(2006)}]{VanZee06}	\linkup{autobib:VanZee06}
{van Zee}, L., \& {Haynes}, M.~P. 2006, \apj, 636, 214,
  \adsurl{http://adsabs.harvard.edu/abs/2006ApJ...636..214V},
  \eprint{arXiv:astro-ph/0509677}

\bibitem[{{V{\'a}zquez} \& {Leitherer}(2005)}]{Vazquez05}	\linkup{autobib:Vazquez05}
{V{\'a}zquez}, G.~A., \& {Leitherer}, C. 2005, \apj, 621, 695,
  \adsurl{http://adsabs.harvard.edu/abs/2005ApJ...621..695V},
  \eprint{arXiv:astro-ph/0412491}

\bibitem[{{Veilleux} {et~al.}(2006){Veilleux}, {Kim}, {Peng}, {Ho}, {Tacconi},
  {Dasyra}, {Genzel}, {Lutz}, \& {Sanders}}]{Veilleux06}	\linkup{autobib:Veilleux06}
{Veilleux}, S. {et~al.} 2006, \apj, 643, 707,
  \adsurl{http://adsabs.harvard.edu/abs/2006ApJ...643..707V},
  \eprint{arXiv:astro-ph/0601565}

\bibitem[{{Veilleux} {et~al.}(1999){Veilleux}, {Kim}, \&
  {Sanders}}]{Veilleux99}	\linkup{autobib:Veilleux99}
{Veilleux}, S., {Kim}, D., \& {Sanders}, D.~B. 1999, \apj, 522, 113,
  \adsurl{http://adsabs.harvard.edu/abs/1999ApJ...522..113V},
  \eprint{arXiv:astro-ph/9904149}

\bibitem[{{Veilleux} {et~al.}(1995){Veilleux}, {Kim}, {Sanders}, {Mazzarella},
  \& {Soifer}}]{Veilleux95}	\linkup{autobib:Veilleux95}
{Veilleux}, S., {Kim}, D., {Sanders}, D.~B., {Mazzarella}, J.~M., \& {Soifer},
  B.~T. 1995, \apjs, 98, 171,
  \adsurl{http://adsabs.harvard.edu/abs/1995ApJS...98..171V}

\bibitem[{{Veilleux} {et~al.}(2002){Veilleux}, {Kim}, \&
  {Sanders}}]{Veilleux02}	\linkup{autobib:Veilleux02}
{Veilleux}, S., {Kim}, D.-C., \& {Sanders}, D.~B. 2002, \apjs, 143, 315,
  \adsurl{http://adsabs.harvard.edu/abs/2002ApJS..143..315V},
  \eprint{arXiv:astro-ph/0207401}

\bibitem[{{Vesperini}(2010)}]{Vesperini10}	\linkup{autobib:Vesperini10}
{Vesperini}, E. 2010, Royal Society of London Philosophical Transactions Series
  A, 368, 829, \adsurl{http://adsabs.harvard.edu/abs/2010RSPTA.368..829V},
  \eprint{0911.0793}

\bibitem[{{V\'ilchez} \& {Esteban}(1996)}]{Vilchez96}	\linkup{autobib:Vilchez96}
{V\'ilchez}, J.~M., \& {Esteban}, C. 1996, \mnras, 280, 720,
  \adsurl{http://adsabs.harvard.edu/abs/1996MNRAS.280..720V}

\bibitem[{{Wada} {et~al.}(2000){Wada}, {Spaans}, \& {Kim}}]{Wada00}	\linkup{autobib:Wada00}
{Wada}, K., {Spaans}, M., \& {Kim}, S. 2000, \apj, 540, 797,
  \adsurl{http://adsabs.harvard.edu/abs/2000ApJ...540..797W},
  \eprint{arXiv:astro-ph/0005330}

\bibitem[{{Weilbacher} {et~al.}(2003){Weilbacher}, {Duc}, \&
  {Fritze-v.~Alvensleben}}]{Weilbacher03}	\linkup{autobib:Weilbacher03}
{Weilbacher}, P.~M., {Duc}, P.-A., \& {Fritze-v.~Alvensleben}, U. 2003, \aap,
  397, 545, \adsurl{http://adsabs.harvard.edu/abs/2003A\%26A...397..545W},
  \eprint{arXiv:astro-ph/0210393}

\bibitem[{{Weilbacher} {et~al.}(2000){Weilbacher}, {Duc}, {Fritze
  v.~Alvensleben}, {Martin}, \& {Fricke}}]{Weilbacher00}	\linkup{autobib:Weilbacher00}
{Weilbacher}, P.~M., {Duc}, P.-A., {Fritze v.~Alvensleben}, U., {Martin}, P.,
  \& {Fricke}, K.~J. 2000, \aap, 358, 819,
  \adsurl{http://adsabs.harvard.edu/abs/2000A\%26A...358..819W},
  \eprint{arXiv:astro-ph/0004405}

\bibitem[{{Weilbacher} \& {Fritze-v.~Alvensleben}(2001)}]{Weilbacher01}	\linkup{autobib:Weilbacher01}
{Weilbacher}, P.~M., \& {Fritze-v.~Alvensleben}, U. 2001, \aap, 373, L9,
  \adsurl{http://adsabs.harvard.edu/abs/2001A\%26A...373L...9W},
  \eprint{arXiv:astro-ph/0105282}

\bibitem[{{Weilbacher} {et~al.}(2002){Weilbacher}, {Fritze-v.~Alvensleben},
  {Duc}, \& {Fricke}}]{Weilbacher02}	\linkup{autobib:Weilbacher02}
{Weilbacher}, P.~M., {Fritze-v.~Alvensleben}, U., {Duc}, P., \& {Fricke}, K.~J.
  2002, \apjl, 579, L79,
  \adsurl{http://adsabs.harvard.edu/abs/2002ApJ...579L..79W},
  \eprint{arXiv:astro-ph/0210173}

\bibitem[{{Weisz} {et~al.}(2008){Weisz}, {Skillman}, {Cannon}, {Dolphin},
  {Kennicutt}, {Lee}, \& {Walter}}]{Weisz08}	\linkup{autobib:Weisz08}
{Weisz}, D.~R., {Skillman}, E.~D., {Cannon}, J.~M., {Dolphin}, A.~E.,
  {Kennicutt}, Jr., R.~C., {Lee}, J., \& {Walter}, F. 2008, \apj, 689, 160,
  \adsurl{http://adsabs.harvard.edu/abs/2008ApJ...689..160W},
  \eprint{0809.5059}

\bibitem[{{Wen} {et~al.}(2011){Wen}, {Zheng}, {Zhao}, \& {Gao}}]{Wen11}	\linkup{autobib:Wen11}
{Wen}, Z.-Z., {Zheng}, X.-Z., {Zhao}, Y.-H., \& {Gao}, Y. 2011, ArXiv e-prints,
  \adsurl{http://adsabs.harvard.edu/abs/2011arXiv1103.2546W},
  \eprint{1103.2546}

\bibitem[{{Wetzstein} {et~al.}(2007){Wetzstein}, {Naab}, \&
  {Burkert}}]{Wetzstein07}	\linkup{autobib:Wetzstein07}
{Wetzstein}, M., {Naab}, T., \& {Burkert}, A. 2007, \mnras, 375, 805,
  \adsurl{http://adsabs.harvard.edu/abs/2007MNRAS.375..805W},
  \eprint{arXiv:astro-ph/0510821}

\bibitem[{{Whitmore} {et~al.}(2007){Whitmore}, {Chandar}, \&
  {Fall}}]{Whitmore07}	\linkup{autobib:Whitmore07}
{Whitmore}, B.~C., {Chandar}, R., \& {Fall}, S.~M. 2007, \aj, 133, 1067,
  \adsurl{http://adsabs.harvard.edu/abs/2007AJ....133.1067W},
  \eprint{arXiv:astro-ph/0611055}

\bibitem[{{Whitmore} {et~al.}(2010){Whitmore}, {Chandar}, {Schweizer},
  {Rothberg}, {Leitherer}, {Rieke}, {Rieke}, {Blair}, {Mengel}, \&
  {Alonso-Herrero}}]{Whitmore10}	\linkup{autobib:Whitmore10}
{Whitmore}, B.~C. {et~al.} 2010, \aj, 140, 75,
  \adsurl{http://adsabs.harvard.edu/abs/2010AJ....140...75W},
  \eprint{1005.0629}

\bibitem[{{Whitmore} {et~al.}(2005){Whitmore}, {Gilmore}, {Leitherer}, {Fall},
  {Chandar}, {Blair}, {Schweizer}, {Zhang}, \& {Miller}}]{Whitmore05}	\linkup{autobib:Whitmore05}
{Whitmore}, B.~C. {et~al.} 2005, \aj, 130, 2104,
  \adsurl{http://adsabs.harvard.edu/abs/2005AJ....130.2104W},
  \eprint{arXiv:astro-ph/0507706}

\bibitem[{{Whitmore} \& {Schweizer}(1995)}]{Whitmore95}	\linkup{autobib:Whitmore95}
{Whitmore}, B.~C., \& {Schweizer}, F. 1995, \aj, 109, 960,
  \adsurl{http://adsabs.harvard.edu/abs/1995AJ....109..960W}

\bibitem[{{Whitmore} {et~al.}(1993){Whitmore}, {Schweizer}, {Leitherer},
  {Borne}, \& {Robert}}]{Whitmore93}	\linkup{autobib:Whitmore93}
{Whitmore}, B.~C., {Schweizer}, F., {Leitherer}, C., {Borne}, K., \& {Robert},
  C. 1993, \aj, 106, 1354,
  \adsurl{http://adsabs.harvard.edu/abs/1993AJ....106.1354W}

\bibitem[{{Whitmore} {et~al.}(1999){Whitmore}, {Zhang}, {Leitherer}, {Fall},
  {Schweizer}, \& {Miller}}]{Whitmore99}	\linkup{autobib:Whitmore99}
{Whitmore}, B.~C., {Zhang}, Q., {Leitherer}, C., {Fall}, S.~M., {Schweizer},
  F., \& {Miller}, B.~W. 1999, \aj, 118, 1551,
  \adsurl{http://adsabs.harvard.edu/abs/1999AJ....118.1551W},
  \eprint{arXiv:astro-ph/9907430}

\bibitem[{{Williams} {et~al.}(1995){Williams}, {Blitz}, \&
  {Stark}}]{Williams95}	\linkup{autobib:Williams95}
{Williams}, J.~P., {Blitz}, L., \& {Stark}, A.~A. 1995, \apj, 451, 252,
  \adsurl{http://adsabs.harvard.edu/abs/1995ApJ...451..252W}

\bibitem[{{Williams} \& {McKee}(1997)}]{Williams97}	\linkup{autobib:Williams97}
{Williams}, J.~P., \& {McKee}, C.~F. 1997, \apj, 476, 166,
  \adsurl{http://adsabs.harvard.edu/abs/1997ApJ...476..166W}

\bibitem[{{Wilson} {et~al.}(2003){Wilson}, {Scoville}, {Madden}, \&
  {Charmandaris}}]{Wilson03}	\linkup{autobib:Wilson03}
{Wilson}, C.~D., {Scoville}, N., {Madden}, S.~C., \& {Charmandaris}, V. 2003,
  \apj, 599, 1049, \adsurl{http://adsabs.harvard.edu/abs/2003ApJ...599.1049W},
  \eprint{arXiv:astro-ph/0308545}

\bibitem[{{Wright}(2006)}]{Wright06}	\linkup{autobib:Wright06}
{Wright}, E.~L. 2006, \pasp, 118, 1711,
  \adsurl{http://adsabs.harvard.edu/abs/2006PASP..118.1711W},
  \eprint{arXiv:astro-ph/0609593}

\bibitem[{{Wu} {et~al.}(2005){Wu}, {Cao}, {Hao}, {Liu}, {Wang}, {Xia}, {Deng},
  \& {Young}}]{Wu05}	\linkup{autobib:Wu05}
{Wu}, H., {Cao}, C., {Hao}, C.-N., {Liu}, F.-S., {Wang}, J.-L., {Xia}, X.-Y.,
  {Deng}, Z.-G., \& {Young}, C.~K.-S. 2005, \apjl, 632, L79,
  \adsurl{http://adsabs.harvard.edu/abs/2005ApJ...632L..79W},
  \eprint{arXiv:astro-ph/0509281}

\bibitem[{{Wu} {et~al.}(1998){Wu}, {Zou}, {Xia}, \& {Deng}}]{Wu98}	\linkup{autobib:Wu98}
{Wu}, H., {Zou}, Z.~L., {Xia}, X.~Y., \& {Deng}, Z.~G. 1998, \aaps, 132, 181,
  \adsurl{http://adsabs.harvard.edu/abs/1998A\%26AS..132..181W}

\bibitem[{{Yan} {et~al.}(2005){Yan}, {Chary}, {Armus}, {Teplitz}, {Helou},
  {Frayer}, {Fadda}, {Surace}, \& {Choi}}]{Yan05}	\linkup{autobib:Yan05}
{Yan}, L. {et~al.} 2005, \apj, 628, 604,
  \adsurl{http://adsabs.harvard.edu/abs/2005ApJ...628..604Y},
  \eprint{arXiv:astro-ph/0504336}

\bibitem[{{Yoshida} {et~al.}(1994){Yoshida}, {Taniguchi}, \&
  {Murayama}}]{Yoshida94}	\linkup{autobib:Yoshida94}
{Yoshida}, M., {Taniguchi}, Y., \& {Murayama}, T. 1994, \pasj, 46, L195,
  \adsurl{http://adsabs.harvard.edu/abs/1994PASJ...46L.195Y}

\bibitem[{{Yuan} {et~al.}(2010){Yuan}, {Kewley}, \& {Sanders}}]{Yuan10}	\linkup{autobib:Yuan10}
{Yuan}, T., {Kewley}, L.~J., \& {Sanders}, D.~B. 2010, \apj, 709, 884,
  \adsurl{http://adsabs.harvard.edu/abs/2010ApJ...709..884Y},
  \eprint{0911.3728}

\bibitem[{{Zenner} \& {Lenzen}(1993)}]{Zenner93}	\linkup{autobib:Zenner93}
{Zenner}, S., \& {Lenzen}, R. 1993, \aaps, 101, 363,
  \adsurl{http://adsabs.harvard.edu/abs/1993A\%26AS..101..363Z}

\bibitem[{{Zepf} {et~al.}(1999){Zepf}, {Ashman}, {English}, {Freeman}, \&
  {Sharples}}]{Zepf99}	\linkup{autobib:Zepf99}
{Zepf}, S.~E., {Ashman}, K.~M., {English}, J., {Freeman}, K.~C., \& {Sharples},
  R.~M. 1999, \aj, 118, 752,
  \adsurl{http://adsabs.harvard.edu/abs/1999AJ....118..752Z},
  \eprint{arXiv:astro-ph/9904247}

\bibitem[{{Zhang} \& {Fall}(1999)}]{Zhang99}	\linkup{autobib:Zhang99}
{Zhang}, Q., \& {Fall}, S.~M. 1999, \apjl, 527, L81,
  \adsurl{http://adsabs.harvard.edu/abs/1999ApJ...527L..81Z},
  \eprint{arXiv:astro-ph/9911229}

\bibitem[{{Zhang} {et~al.}(2001){Zhang}, {Fall}, \& {Whitmore}}]{Zhang01}	\linkup{autobib:Zhang01}
{Zhang}, Q., {Fall}, S.~M., \& {Whitmore}, B.~C. 2001, \apj, 561, 727,
  \adsurl{http://adsabs.harvard.edu/abs/2001ApJ...561..727Z},
  \eprint{arXiv:astro-ph/0105174}

\bibitem[{{Zwicky}(1956)}]{Zwicky56}	\linkup{autobib:Zwicky56}
{Zwicky}, F. 1956, Ergebnisse der exakten Naturwissenschaften, 29, 344,
  \adsurl{http://adsabs.harvard.edu/abs/1956ErNW...29..344Z}

\end{thebibliography}

\thispagestyle{frontmatter}

\end{document}